# Atlas of Nuclear Isomers - Second Edition

Swati Garg[a], Bhoomika Maheshwari[b,c,d], Balraj Singh[e,*], Yang Sun[a], Alpana Goel[d], Ashok Kumar Jain[d]

[a] *School of Physics and Astronomy, Shanghai Jiao Tong University Shanghai-200240, China*
[b] *Department of Physics, Faculty of Science, University of Zagreb, HR-10000 Zagreb, Croatia*
[c] *Department of Physics, Indian Institute of Technology Ropar, Rupnagar-140001, India*
[d] *Amity Institute of Nuclear Science and Technology, Amity University UP, Noida-201313, India*
[e] *Department of Physics and Astronomy, McMaster University, Hamilton, Ontario-L8S 4M1, Canada*

---

**Abstract**

We present an updated version of the 2015-Atlas of Nuclear Isomers [1], compiling and evaluating experimental data for the isomers with half-life $\geq 10$ $ns$, together with their spectroscopic properties such as excitation-energies, half-lives, decay modes, spins and parities, energies and multipolarities of isomeric transitions, along with the relevant original references in literature. The current version of Atlas presents many re-evaluated half-lives as compared to the 2015 edition, where values were referred to Nuclear Data Sheets publications, when no new data existed. The ENSDF database [2], together with the XUNDL [3] and the NUBASE2020 [4] databases have been consulted for completeness, yet, data from original papers from journals were considered in the present evaluation, and the NSR bibliographic database [5] has been searched to ensure that this work is as complete and current as possible. Several useful systematic features of nuclear isomers covered in this Atlas have been discussed. Literature cutoff date for the extraction of data is October 31, 2022.

---

*corresponding author: balraj@mcmaster.ca

*Email addresses:* swatgarg@sjtu.edu.cn (Swati Garg), bhoomika.physics@gmail.com (Bhoomika Maheshwari), sunyang@sjtu.edu.cn (Yang Sun), agoel1@amity.edu (Alpana Goel), ashkumarjain@yahoo.com (Ashok Kumar Jain)

## Contents

## List of Figures

## 1. INTRODUCTION

The nuclear isomers are excited states of nuclei, whose decay is hindered due to various physical reasons [6]. There has been a rapid growth in the experimental studies of isomers over the past two decades, as these quantum states offer a unique opportunity to investigate nuclear structure in different scenarios. New isomers in different regions of nuclear landscape, excitation energies and spins continue to be discovered regularly with the availability of new state-of-the-art experimental techniques at various laboratories around the world. Besides their role in fundamental research in nuclear physics and astrophysics, such nuclear objects have been gaining prominence due to the exciting possibilities and applications in energy storage devices, possible nuclear clocks, applications in medical diagnostics and therapeutics, gamma ray lasers, etc. We start with a brief glimpse of their history.

Soddy in 1917 [7] remarked, "We can have isotopes with the identity of atomic weight, as well as of chemical character, which are different in their stability and mode of breaking up." This statement is often cited as the first hint of the existence of isomers. It is also generally claimed that Otto Hahn made the first experimental observation of an isomer in $^{234}$Pa while working on uranium salts [8]. It may be noted that Fajans and Gohring [9] had already observed the 1.1 $min$ activity in $UX_2$ isotope in a new chemical element brevium, as they named it, which was to be identified later as an isomeric state of $^{234}$Pa [10]. Hahn in his 1921 paper [8] observed the ground state activity of 6.7 $h$ in $^{234}$Pa and also the short-lived activity populated in the beta decay of $^{234}$Th, whose half life was not measured [11]. This other activity was the 1.1 $min$ isomer, whose link with the ground state remained unobserved. The levels above the 1.1 $min$ isomer, therefore, did not have known energy values. Only recently, Korsakov *et al.* [12] confirmed the 1.1 $min$ isomeric activity with a tentative spin of $(0^-)$ that lies only 2.6 keV above the 73.9 keV, $(3^+)$ state causing its decay to be almost impossible. As pointed out by Walker and Carroll [13], the word "isomer" itself was used, in the nuclear context, for the first time by George Gamow in 1934 after the discovery of neutron by Chadwick in 1932. More confirmed observations of isomers were made by Kurtchatov *et al.* [14] and, Szilard and Chalmers [15] in the Br and In isotopes in the year 1935.

Present day experimental techniques can measure a broad range of nuclear lifetimes, ranging from pico-seconds to years. As a result, the limit of the half-life used to define isomers has been significantly lowered over the years. While a precise definition of lifetime of an isomer is lacking, generally a nuclear excited state of half-life of 1 $ns$ or greater is considered by several researchers as an isomer. In the current evaluation, we continue to maintain the limit as 10 $ns$ or greater, as used in the earlier Atlas of 2015. Work on isomers of 1 $ns$ to 10 $ns$ is progressing and will be presented separately. ENSDF [2], XUNDL [3], NSR [5], and NUBASE2020 databases [4] were extensively consulted for a complete coverage and information about isomers.

This "Atlas of Nuclear Isomers" presents data for about 2623 isomers, all having a half-life of 10 $ns$ or longer. The isomers in odd-odd nuclei are the most abundant (866) as compared to those in even-even (445); even-$N$, odd-$Z$ (679); and odd-$N$, even-$Z$ (633) nuclei. There are about 1312 isomers in the odd-A nuclei compared to about 1311 isomers in the even-A nuclei. These meta-stable nuclear states (isomers) mostly decay to one or more of the lower energy states via electromagnetic transition, which includes gamma-decay and internal conversion; both are together known as isomeric transitions (IT). The isomers away from the line of stability may also decay via beta-decay ($\beta$), beta-delayed neutron decay ($\beta^- n$), beta- and electron capture delayed proton decay ($\beta^+ p$ and/or $\epsilon p$), and proton-decay ($p$). For isomers in heavier nuclei, alpha-decay ($\alpha$), and spontaneous fission (SF) are more likely decay modes.

We discuss important features of the various types of isomers in section II. The policies and explanation of the Table are presented after the bibliography for the introduction. All the compiled and evaluated data are presented in Table I, which include key properties of the isomers, such as their excitation energies, spin-parities, half-lives, emitted gamma-ray energies, multipolarities, and the decay modes. The key-numbers of the references used to extract the data are given in the last column in the style of NSR database [5]. A collection of tentative isomers has also been presented in Table II, which have been proposed in secondary references such as theses or conference reports, but have not yet been published or confirmed in peer reviewed journals. It is useful to identify and classify various regions of the nuclear chart where isomers exist by considering their properties and the reasons for hindrance. We, therefore, present a few systematics of the isomeric properties in section III. These graphs and tables reveal systematic and novel features across the whole nuclear landscape. Section IV summarizes the present work.

## 2. TYPES OF ISOMERS

First physics interpretation of isomers in terms of hindered gamma transition was proposed by Weizsacker in 1936 [16]. The year 1949 seems to be very important in the growth of nuclear isomer research. Occurrence of isomers due to large angular momentum change during their decay was observed to be correlated with the shell structure by Feenberg [17], Feenberg and Hammack [18] and Nordheim [19]. This was also the year when the first review article on nuclear isomers was published by Segre and Helmholtz [20]. Also, the first classification of isomers was attempted by Axel and Dancoff in 1949 [21]. Goldhaber and Hill [22] again reviewed the progress in isomers, and presented the first compilation of isomers and their interpretation in terms of the shell structure. Later on, Feenberg [23] presented a detailed discussion of isomers in his book on the basis of the data of nearly 100 isomers known at that time and provided an interpretation for the majority of them in terms of the shell structure. The gamma-decay probability of an isomeric transition may be represented by the general formula given by,

$$\Lambda \propto |<f|T_\lambda|i>|^2 (E_\gamma)^{2\lambda+1} \tag{1}$$

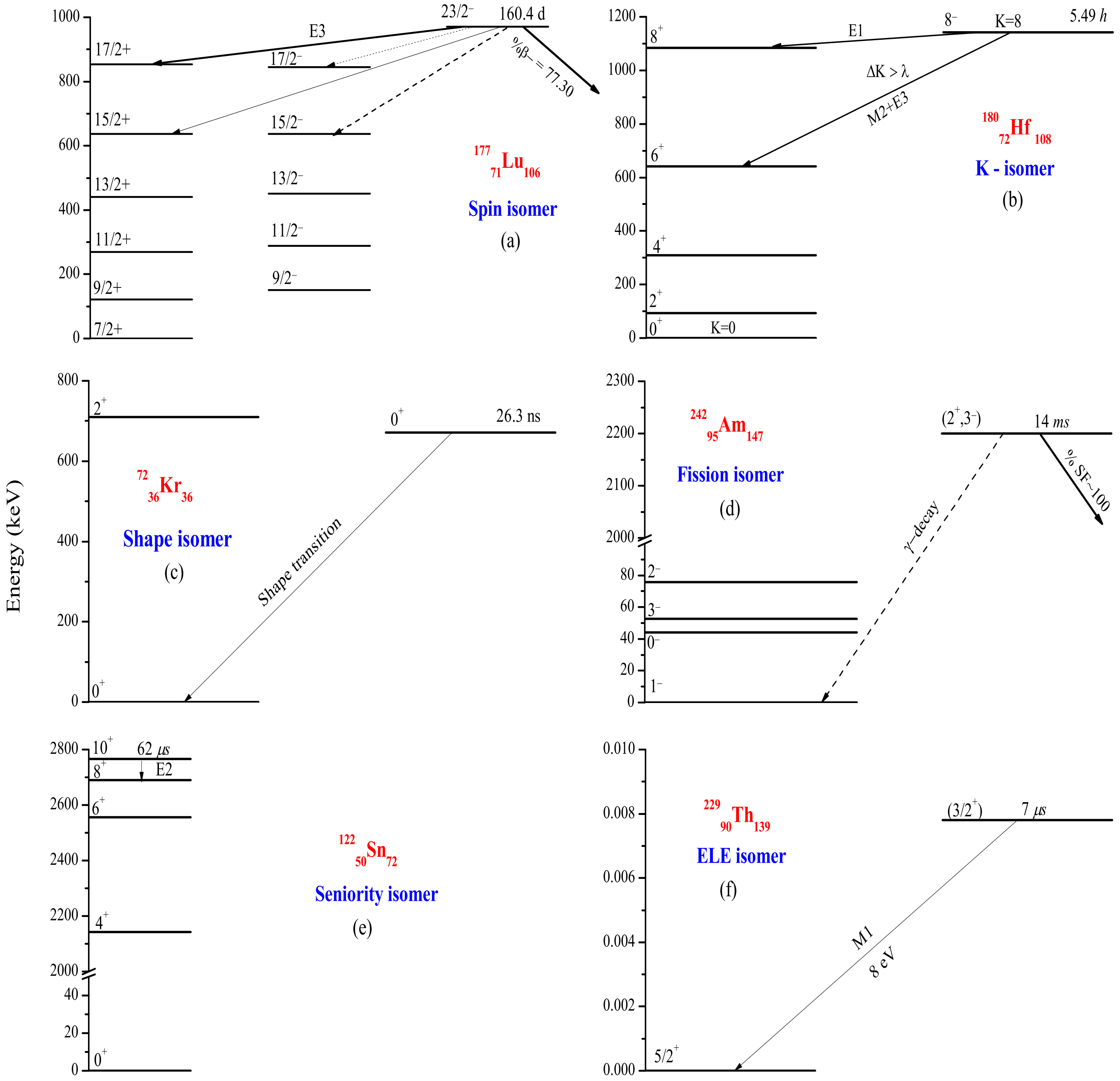

**Fig. 1**: Examples of different types of isomers.

It implies that the decay probability is directly proportional to square of the transition matrix element of the operators causing the transition. This is where the nuclear structure information enters into the decay probability, which requires a knowledge of the nuclear wave-functions in the initial and the final states, which may be quite limited at times. Larger the matrix element, larger the decay probability and smaller the lifetime. An isomer will occur if the matrix element is small and/or, the decay energy is small. The physical reasons for the matrix element to be small can be varied and this is what leads to the broad classification of isomers into spin isomers, seniority isomers, $K$-isomers, shape isomers, fission isomers, etc. Further, the decay probability is also proportional to the transition energy (the energy difference between the initial and final states) raised to the power of $2\lambda + 1$. Therefore, larger the energy of decay, higher will be the decay probability. Only one of the two factors may also control the decay probability provided the other factor is inconsequential. For example, when the decay energy is very small and the matrix element does not hinder the transition, an isomer may still arise. It is also possible to have more than one physical reason behind their long half-lives and the isomer may have a mixed origin. We will now discuss these physics aspects very briefly.

### *2.1. Spin isomers*

Spin isomers are most common type of isomers and the first one to be identified. The matrix elements of electromagnetic transitions lead to the well known spin-parity selection rules. As a result, higher the multi-polarity $\lambda$ of the gamma-transition, smaller is the matrix element causing a hindrance to decay. This leads to the occurrence of spin isomers. This is very well reflected in the well-known case of $^{180}$Ta [24–35]. Here, the multi-polarity of the allowed transitions is so large ($\lambda = 7, 8$) that the decay is nearly forbidden giving rise to the longest-lived isomer with a half-life $T_{1/2} > 4.5 \times 10^{16} years$, the only isomer found in nature. However, in Figure 1, we show the example of $^{177}$Lu, which has a mixed character of spin and $K$-isomer. Here the decay of $23/2^-$ state is hindered due to large multipolarity ($\lambda = 3$) transitions like $E3$ and $M3$. At the same time, the decay is also $K$-hindered due to a transition from $K = 23/2$ to $K = 9/2$ or, $7/2$, as discussed in the following.

### *2.2. K-isomers*

The $K$-isomers represent another class of isomers arising from a sudden change in the $K$-quantum number during a decay [36]. The quantity $K$ is a good quantum number in axially symmetric deformed nuclei and represents the projection of the total angular momentum of the nucleus on its symmetry axis. A change in $K$ implies a change in the inclination of the angular momentum with the symmetry axis. When the change in the $K$-quantum number, $\Delta K$, exceeds the multipolarity $\lambda$ of the decaying transition, it becomes hindered. As a consequence, a $K$-isomer is formed, whose degree of forbiddenness is given by $\nu = \Delta K - \lambda$. As an example, $^{180}$Hf has a $K = 8$, $8^-$ isomer [37–42] with a half-life of 5.5 $h$ as shown in Fig. 1. The nearest level available for decay is the $8^+$ level of the $K = 0$ band. The change in $K$, i.e. $\Delta K = 8$ is very large as compared to the multipolarity of the decaying transition, which is $\lambda = 1$. The high degree of forbiddenness, $\nu = 7$, leads to a very long-lived isomer.

### *2.3. Shape isomers*

A given nucleus may often have several minima in its potential energy surface at deformations different from the ground state. The ground state generally corresponds to the lowest and the deepest minimum. Most of the other minima are usually quite shallow. If one of these minima is deep enough to trap the nucleus, it may lead to a shape isomer. This is the situation in many $0^+$ isomers. For example, $^{72}$Kr, a self conjugate nucleus, has a well known $0^+$ shape isomer with a half-life of 26 $ns$ [43], which exists due to the $0^+ \to 0^+$ shape hindered transition, as shown in Figure 1.

### 2.4. Fission isomers

Fission isomers are also shape isomers but with a difference [44]. These are observed in heavy nuclei where spontaneous fission becomes possible. A double minimum and a double hump barrier in the potential energy surface is a common feature in the trans-actinides [45]. The nucleus trapped in the second minimum has a chance to undergo spontaneous fission besides the possibility of decaying to the ground state in the first minimum. Its half-life, therefore, depends on several factors. The barrier penetration through the outer barrier and the decay to the states in the first minimum both combine together to decide the half-life of the isomer. As an example, the longest-lived fission isomer is observed in $^{242}$Am, which has a half-life of 14 *ms* (see Figure 1) [46, 47]. A table of fission isomers, due to Singh *et al.* [48], had listed 46 such isomers with no restriction on the half-life. The isomer book by Jain *et al.* [6], lists 48 cases of possible fission isomers from the actinide region.

### 2.5. Seniority isomers

Seniority is a quantum number related to the pairing interaction [49]. The seniority selection rules of transitions between good seniority states demand that the decay probability vanishes or becomes very small in seniority preserving transitions near the middle of a shell. As a result, isomers may arise near the middle of the shell filling. These isomers are generally observed in the semi-magic nuclei, where either proton or neutron shell is closed and the other kind of particles are filling their shell. When the filling reaches the middle of the shell, seniority isomers may arise. The seniority remains to be an exactly conserved quantum number up to $j \leq \frac{7}{2}$. Though, for higher-j values, the seniority may still be a nearly good quantum number for some isomers and other excited states [50–59]. More specifically, the high-j intruder orbits near the magic numbers play an important role in the structure of seniority isomers.

The generalization of seniority to multi-j orbits, known as the generalized seniority, is more suitable for the real life situation of nucleons distributed among nearby multi-j orbits in a given shell. For example, the $10^+$ and the $27/2^-$ isomers in the $Z = 50$ isotopic and the $N = 82$ isotonic chains were identified as seniority isomers near the mid-shell, having seniority $v = 2$ and 3 respectively [60, 61]. These are recently understood as generalized seniority isomers [50] which supports the configuration mixing from neighboring orbits. An example of the $10^+$ seniority isomer in $^{122}$Sn has been shown in Figure 1. More recently, $6^+$ isomers have been identified in the extremely neutron-rich Sn-isotopes lying beyond $N = 82$ magic number [62]. These isomers have also been interpreted as generalized seniority isomers, and constitute an important set of data which sheds light on effective interactions in neutron-rich systems [53, 63]. Generalized seniority has also led to the emergence of a new kind of seniority isomer decaying by odd-tensor transitions ($E1, E3$, etc.), which were discovered recently [50]. For example, the $13^-$, $v = 4$ isomers in Sn isotopes decay by seniority preserving $E1$ transitions to the lower-lying $12^+$, $v = 4$ states, and the corresponding $B(E1)$ rates follow a parabolic trend with a minimum at the middle of the active valence space consisting of $h_{11/2} \otimes d_{3/2} \otimes s_{1/2}$ [63].

### 2.6. Extremely Low Energy (ELE) isomers

The ELE isomer is an exciting possibility brought into focus after the world-wide attention on the 8 eV isomer in $^{229}$Th [64–74], see Figure 1. As already discussed, the decay probability depends on two factors: the matrix element and the decay energy dependence. All the previously discussed categories of isomers arise because of the hindrance created by the matrix element becoming small due to various physics reasons. However, if the decay energy becomes very small and the matrix element is not hindering the transition, the decay probability may still become small and an isomer may be created. Such extremely low energy isomers will have their own characteristic features and will eventually be observed in more numbers in the coming years. Nuclei in the heavy mass region are more likely to have ELE isomers. These isomers may be of particular interest due to their potential applications in atomic-nuclear interface phenomena, and may be also in nuclear astrophysics.

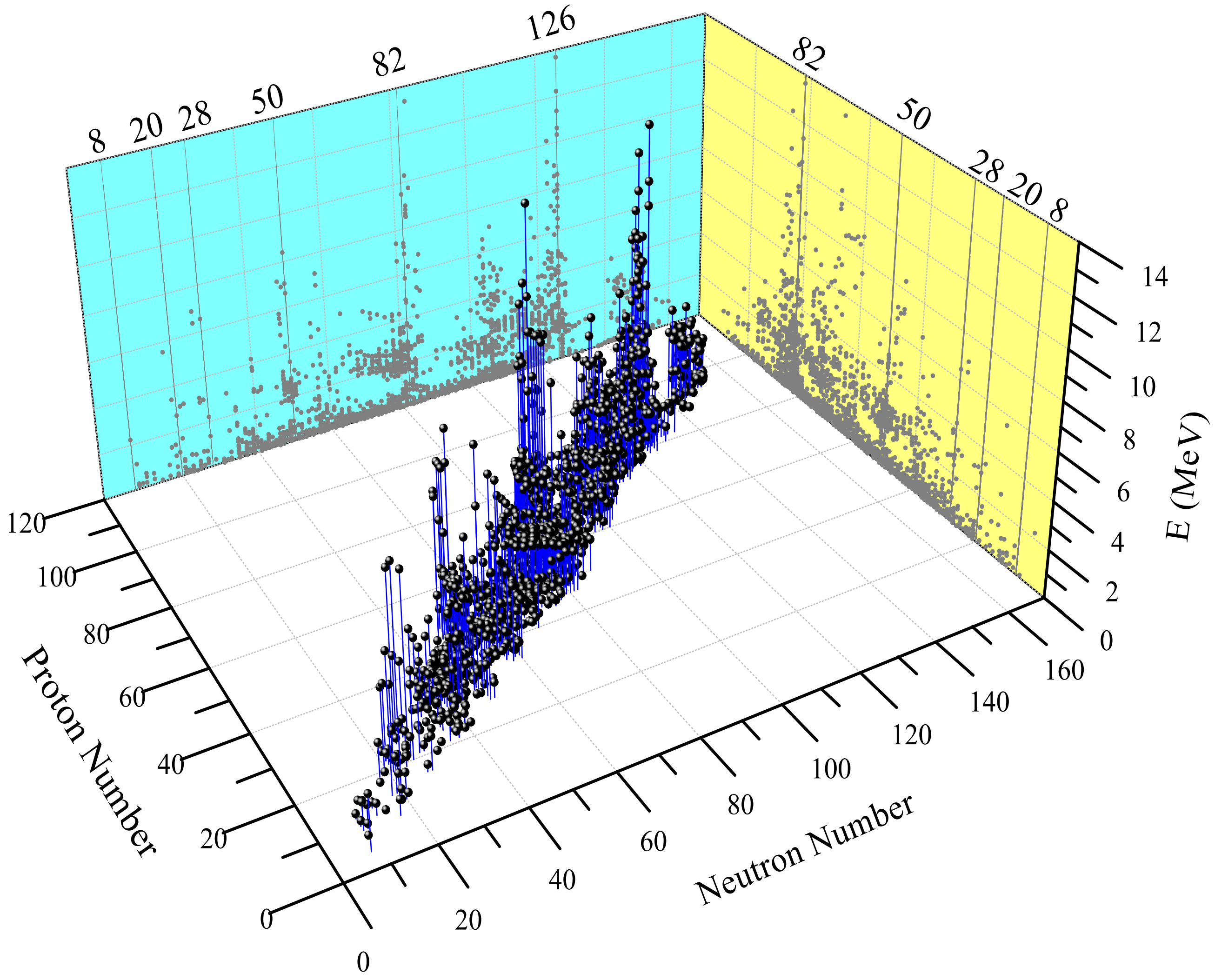

**Fig. 2**: A 3D-nuclear landscape for the variation of isomeric excitation energies with neutron number ($N$) and proton number ($Z$). The two projected panels show the variation of isomeric energies with neutron and proton numbers, respectively, along with the solid lines corresponding to their magic configurations.

### 3. SYSTEMATICS OF ISOMERS

We have plotted various important properties of isomers and presented their global systematics in this section. We also point out the emerging features and the physics behind them very briefly. Many of these systematic features also appeared in the 2015 version of the Atlas, and all of them have been updated with the new data and are plotted again. A fresh look at these systematic features is presented below.

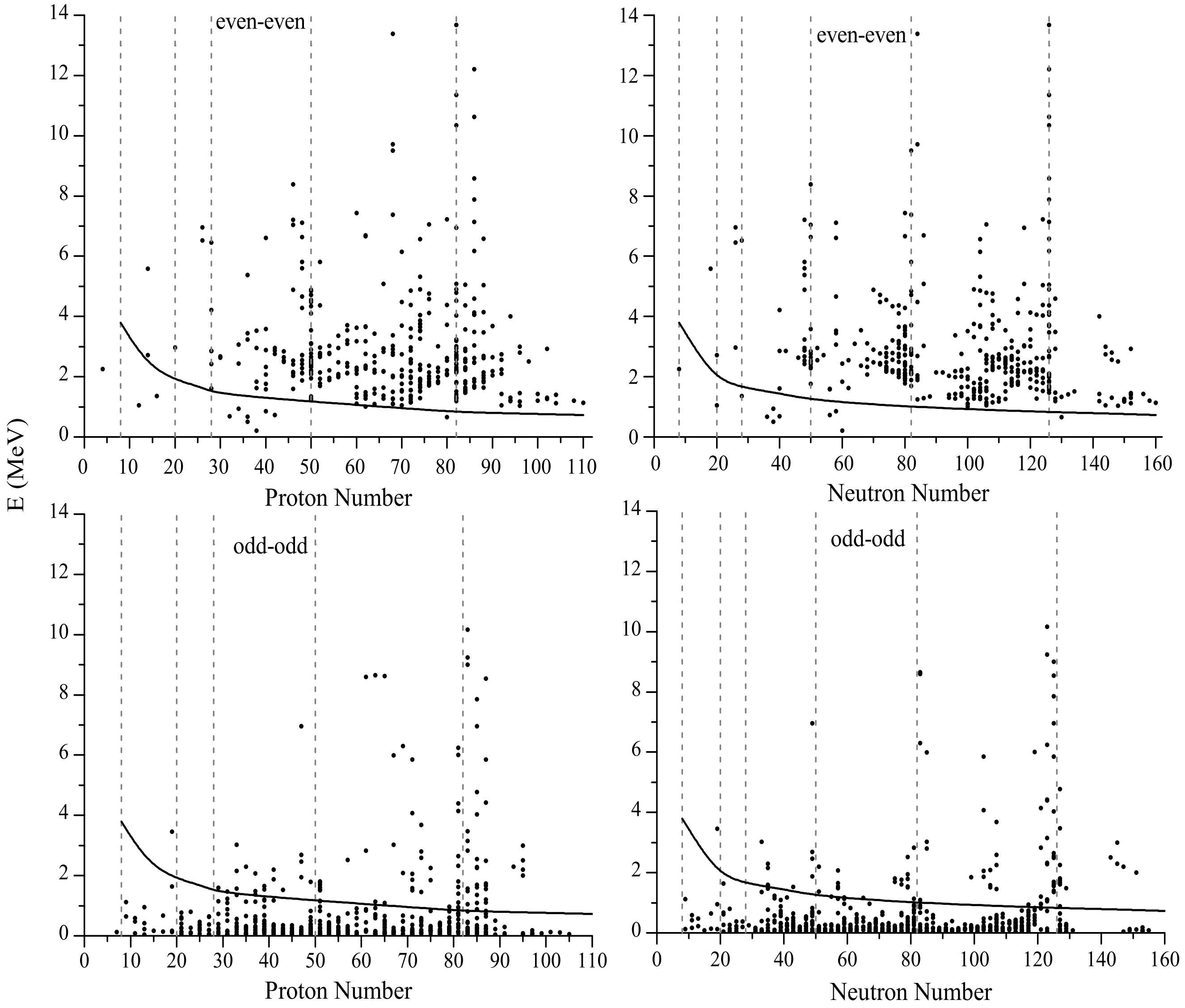

**Fig. 3**: Variation of isomeric energies for even-even (upper panels) and odd-odd (lower panels) nuclei with proton/ neutron number. The dashed lines belong to the respective magic numbers $(2, 8, 20, 50, 82, 126)$ for protons and neutrons. The smooth curve corresponds to an average pairing energy $\Delta = 12/\sqrt{A}$ in MeV.

### *3.1. Excitation energy*

We have plotted in Figure 2, the excitation energy $E$(MeV) of the isomers plotted vertically and $Z$ and $N$ in the horizontal plane. The projections of the 3D-plot on the front and side vertical planes bring out the excitation energy patterns for protons (right side plane) and neutrons (front plane), respectively. The vertical lines in the projected shadows have been drawn at the magic numbers. It is obvious that the excitation energy of isomers rise sharply near the magic numbers, as is expected because of the large shell gaps encountered at the magic numbers. Number of isomers is also seen to cluster in islands near the magic numbers because of the occurrence of high-j intruder orbits near the shell closures, which leads to low-lying large angular momentum excitation. It may also be seen that several isomers with large excitation occur near $Z \approx 70$ and $N \approx 100 - 110$, a region of deformation, where high-$K$ orbits are observed. This could be the island of $K$-isomers.

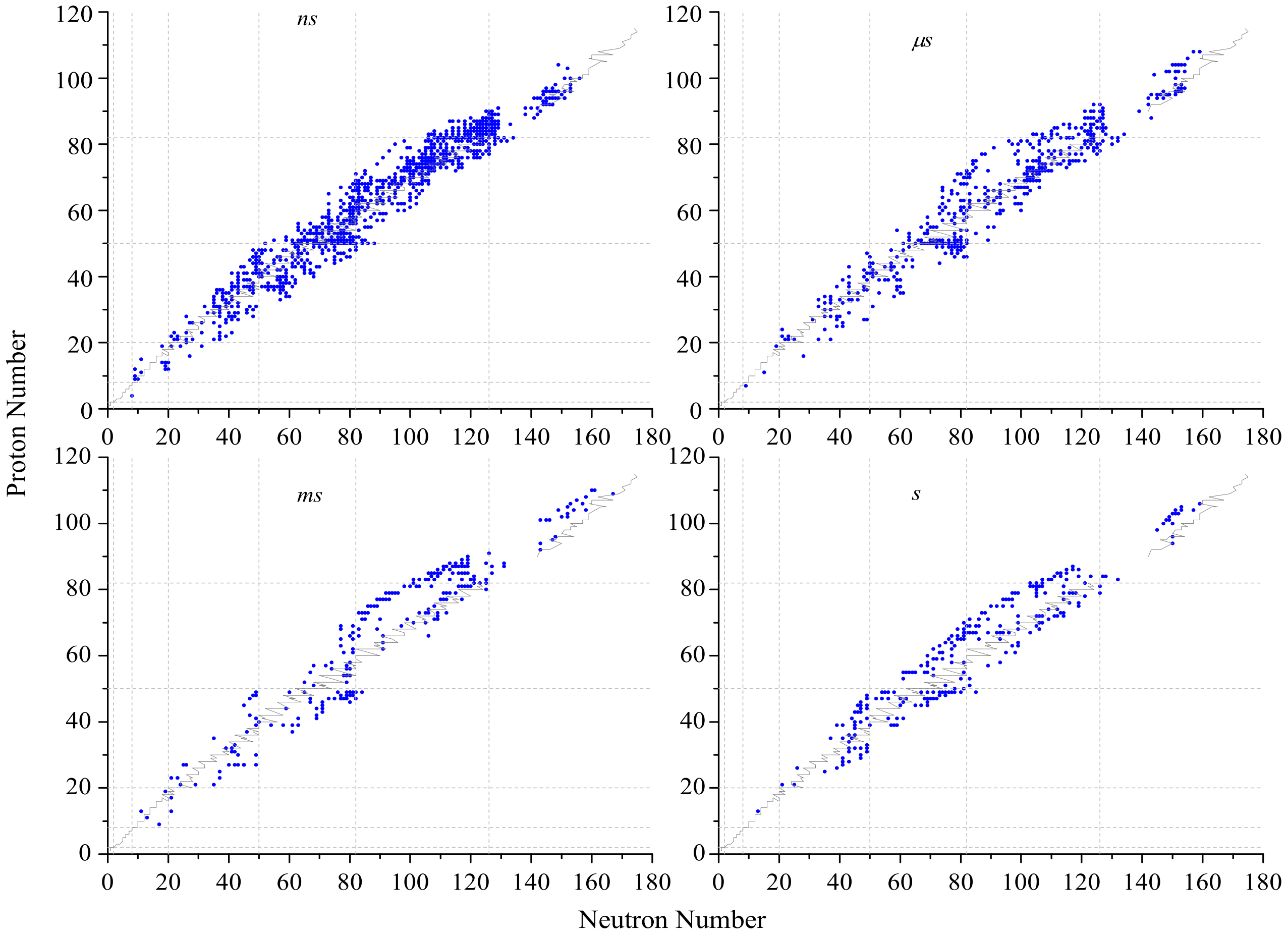

**Fig. 4**: The four-panel nuclear chart representation for the $ns$, $\mu s$, $ms$ and $s$ isomers which include the respective half-life range of $1 - 999$ $ns$, $1 - 999$ $\mu s$, $1 - 999$ $ms$ and $1 - 59$ $s$. The zig-zag black line exhibits the $\beta$-stability line for normal nuclei. The dashed lines belong to the respective magic numbers $(2, 8, 20, 50, 82, 126)$ for protons and neutrons.

The isomers may also be classified into three categories according to the proton and neutron numbers: $Z$-even $N$-even, $Z$-odd $N$-odd, and odd-$A$ (comprising $Z$-even $N$-odd, and $Z$-odd $N$-even). This is quite helpful in understanding their structure and also the occurrence. We have pointed out that the number of odd-odd isomers is the largest, almost double of the even-even isomers. The behavior of isomers in different categories (odd-odd, even-even, even-odd and odd-even) is also quite distinct from each other.

We have plotted the isomeric excitation energies of the odd-odd and even-even isomers as a function of $Z$ and $N$ in Figure 3, to exhibit the role of pairing in isomeric energies. The solid line represents the variation of the average pairing energy $\Delta = 12/\sqrt{A}$ in MeV [75]. The vertical dashed lines again denote the spherical magic numbers. It is rather revealing that majority of the odd-odd isomers lie below the average pairing energy line; these are obviously the two-quasiparticle (2-qp) excitations due to the coupling of odd-proton and odd-neutron without involving any pair break up. The Gallagher-Moszkowski splitting of the 2-qp multiplet due to the n-p interaction may often lead to an easy formation of isomers. This is evident from the large number of odd-odd isomers observed for all $N$ and $Z$ values. The higher lying isomers are obviously 4-qp or higher multi-quasiparticle isomers. The even-even isomers display the opposite feature, where most of the isomers are observed above the pairing energy line; these are again mostly 2-qp excitations, and some higher-qp excitations formed after the breakup of one or more pairs of nucleons. Breaking of each pair requires energy and hence the distinction from the odd-odd isomers arises. The isomers, which lie below the pairing energy gap, are mostly the $0^+$ isomers, and are quite interesting. These are briefly discussed below.

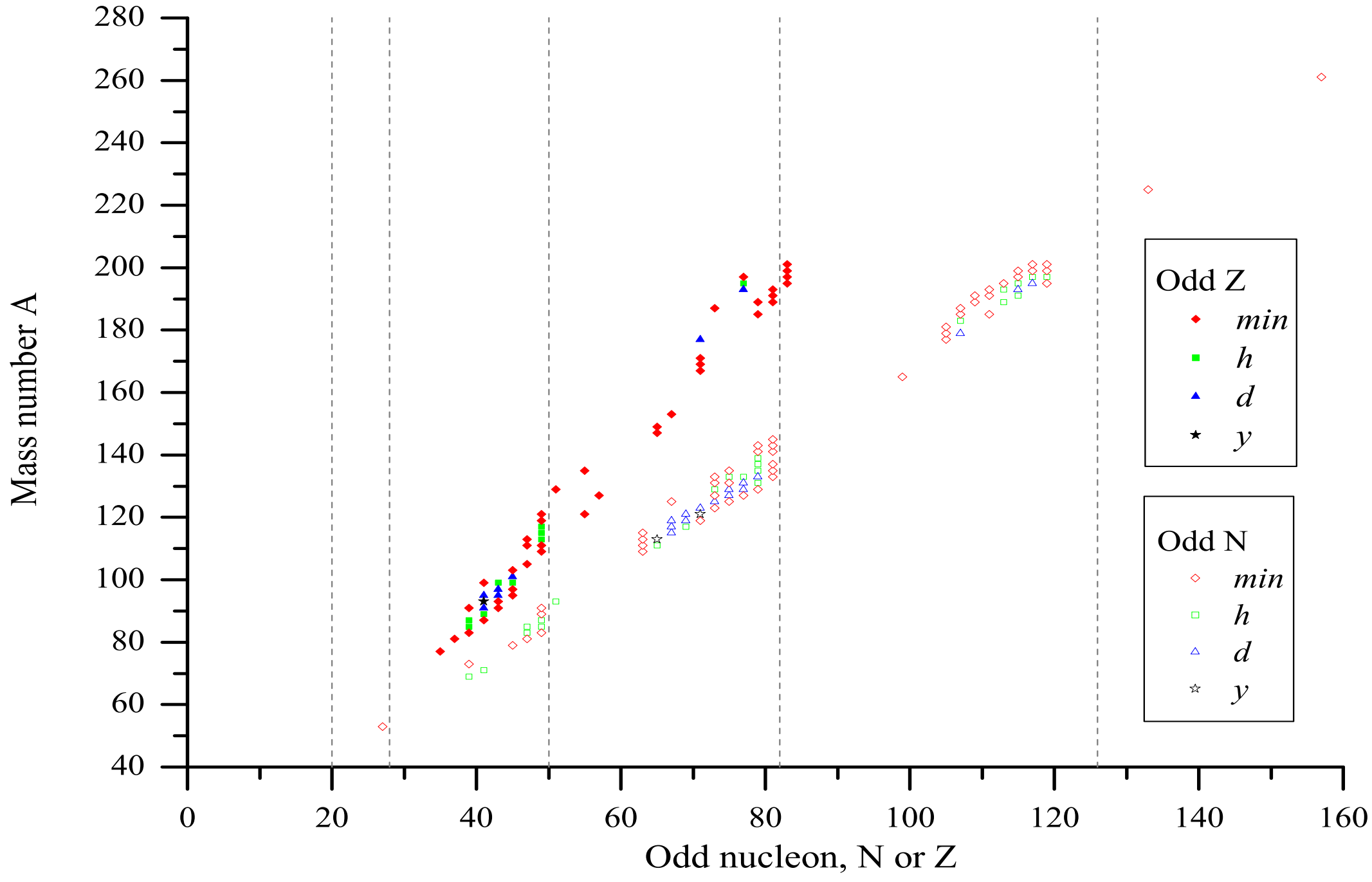

**Fig. 5**: Mass Number $A$ versus odd nucleon number ($N$ or $Z$) for longer lived odd-$A$ isomers. Odd-$Z$ isomers with square symbols and odd-$N$ isomers with triangle symbols are shown for the half-lives in the range of $(1-59)$ $min$, $(1-23)$ $h$, $(1-364)$ $d$ and $(\geq 1)$ $y$.

### *3.2. E0 isomers in even-even nuclei*

The electric monopole ($E0$) transitions connect quantum states of same spin-parity in atomic nuclei and are associated with several shape related phenomena such as shape coexistence and shape isomer. The first suggestion of the importance of $E0$ transition was made by Ellis and Aston in 1930 [76], through the 1.426 MeV transition in Radium C ($^{214}$Po) isotope. Characterization of $E0$ transitions remains a challenge for both experimental nuclear spectroscopy and nuclear theory [77–80].

Transitions between two $0^+$ states can occur only by conversion electrons, electron-positron pair formation, or two-photon emission which is generally negligible. While the active region of multipole order transitions ($E1, E2\ldots, M1, M2..$) is dominantly outside the nucleus, the active region of $E0$ transitions is inside the nuclear volume. Therefore, the monopole $E0$ strength carries vital information about the nuclear structure due to its direct link with the mean-squared charge radius and quadrupole deformation. Pure $E0$ transitions between $0^+$ states, including branching ratios and absolute transition rates, $B(E0)$, are identified in many even-even nuclei throughout the nuclear chart. For $J \rightarrow J$ transitions, where $J > 0$, $E0$ transitions mix with $E2$ and $M1$ transitions. A review of the $E0$ transitions in nuclei has recently been published by Kibedi *et al.* [81], where recent data on 187 pure $E0$ and 95 mixed $(E0 + E2 + M1)$ transitions, with $E0/E2$ and $E2/M1$ mixing ratios, in even-even nuclei have been evaluated. $E0$ isomers in even-even nuclei are of special interest and reveal the shape-coexistence in those nuclei. In the entire nuclear chart, there are 20 confirmed $0^+$ isomers with half-lives $> 10$ $ns$, 15 of them arising from the $0^+ \rightarrow 0^+$ isomeric transitions in even-even nuclei, which can proceed only by conversion electrons, and by pair formation (when transition energy $> 1.02$ MeV), not by gamma emission. Seven $E0$ isomers are known in the even-even nuclei between $Z = 32 - 42$, which lie below the pairing gap line, as shown in Fig. 3. All the seven cases correspond to $0^+ \rightarrow 0^+$ monopole transitions except for the cases of $^{74}$Kr and $^{98}$Sr [1], where $E2$ transitions have also been detected. In fact, most of the low-lying $0^+$ isomers are strong candidates of shape coexistence. A well known example is the $0^+$ shape isomer in $^{72}$Kr [1].

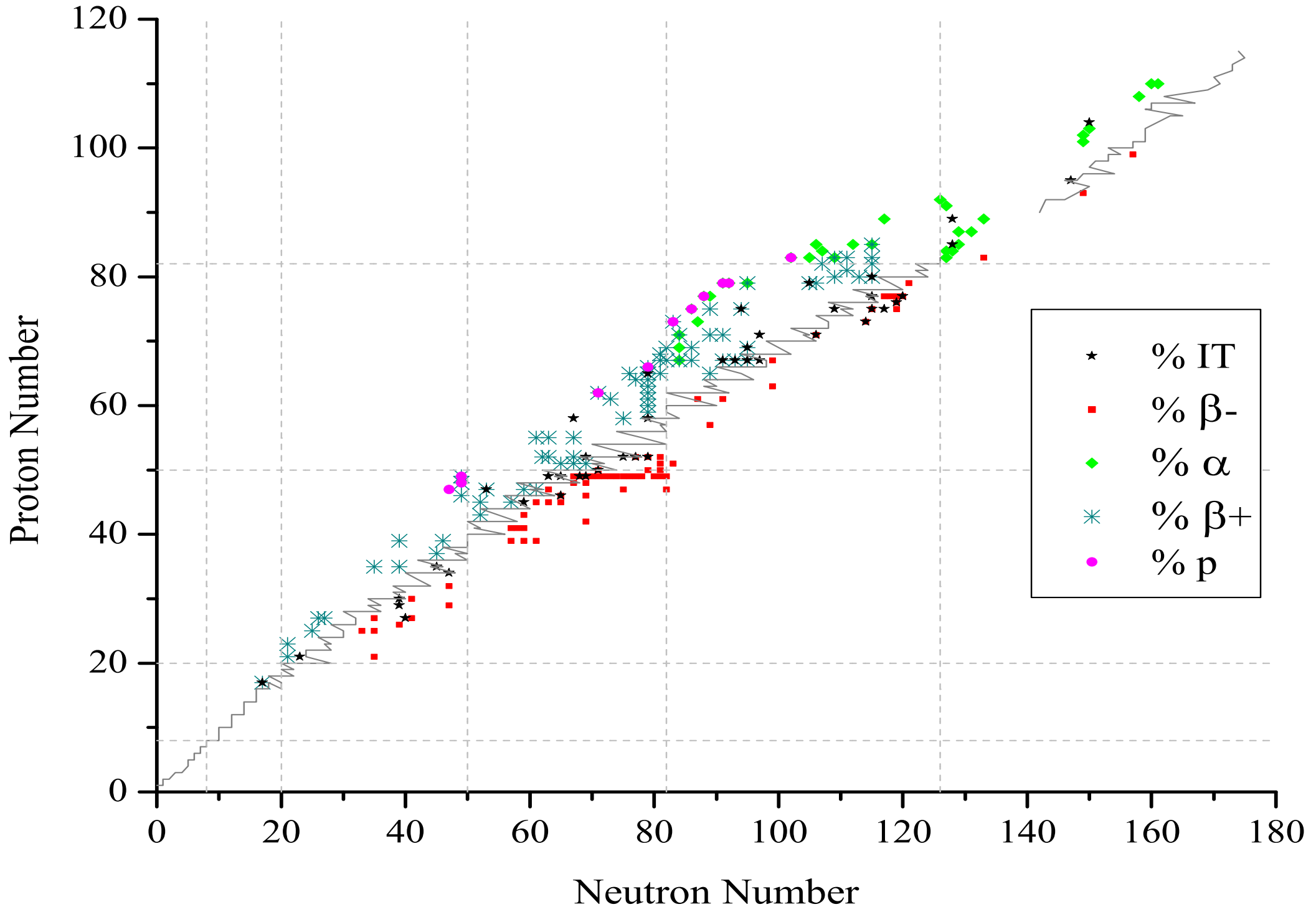

**Fig. 6**: Nuclear isomers having half-lives longer than their respective ground states. The color coding represents their different decay modes such as isomeric transitions (IT), beta($\beta$)-decay, alpha($\alpha$)-decay and proton($p$)-decay. No neutron decaying isomer is known till date.

*3.3. Half-life*

We now discuss briefly some systematic features emerging from the half-life data on isomers. Figure 4 plots the isomers on an $N - Z$ plot, having half-lives in $ns$ (nano-seconds), $\mu s$ (micro-seconds), $ms$ (milli-seconds) and $s$ (seconds) range. The beta stability region having the stable or the longest-lived nuclides with a half-life of $> 10^{10}$ $s$ are plotted in the background. A discontinuity is noticed between $Z \approx 84 - 94$, and $N \approx 127 - 137$, which is the region of $\alpha$-decaying radioactive nuclei. Occurrence of isomers also exhibits a gap in the same range. Only recently, few new isomers have been observed in this region. The latest example is the observation of two new isomers in $^{216}$Fr [82]. We note from Figure 4 that maximum number of isomers are found in the *ns* range. We also notice a distinct shift towards the neutron-deficient side in the occurrence of isomers in the $\mu s$ to *s* range.

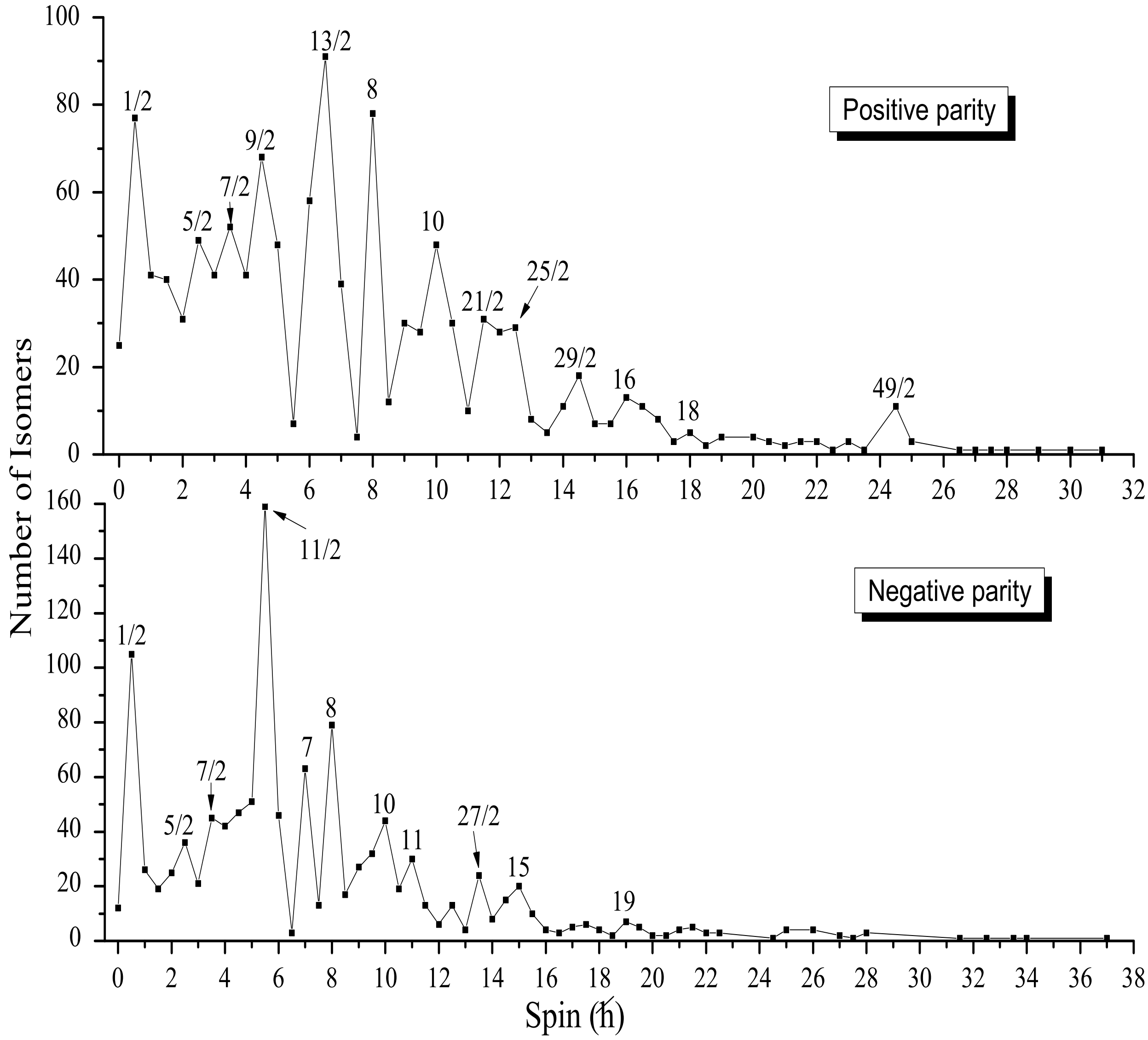

**Fig. 7**: Number of isomers corresponding to each spin. The upper and lower panels belong to the positive and negative parities, respectively. The highest peaks corresponding to the $11/2^-$ and $13/2^+$ spins can be understood in terms of the high-j intruder $h_{11/2}$ and $i_{13/2}$ orbits, respectively.

The longer-lived isomers in the *min* to *y* range mostly lie close to the line of stability. We present another graph for these longer-lived isomers having half-lives in the range of *min* to *y*. Taking a cue from the 1952 review of Goldhaber and Hill [22], we have plotted in Figure 5 the mass number $A$ of these longer-lived isomers versus odd-nucleon number($Z$ or $N$). The vertical dashed lines again depict the magic number configurations for $Z$ and $N$. Such plot naturally separates the odd-$Z$ and odd-$N$ isomers, and also brings out the islands of isomers just before each of the magic numbers. Quite interesting are the isomers, which fall just outside the territory of magic numbers and do not fit into the normal explanation of spin isomers. As an example, $^{93}$Mo ($N = 51$) has an isomer having high-spin ($21/2^+$) and 6.85 $h$ half-life [1]. It was shown that this isomer has pure single-particle character consisting of three particles, $\pi g_{9/2}{}^2 \otimes \nu d_{5/2}$, where the two protons couple to $J = 8$ and combine with neutron $J = 5/2$ to give a total $J = 21/2$ in a fully aligned configuration [83].

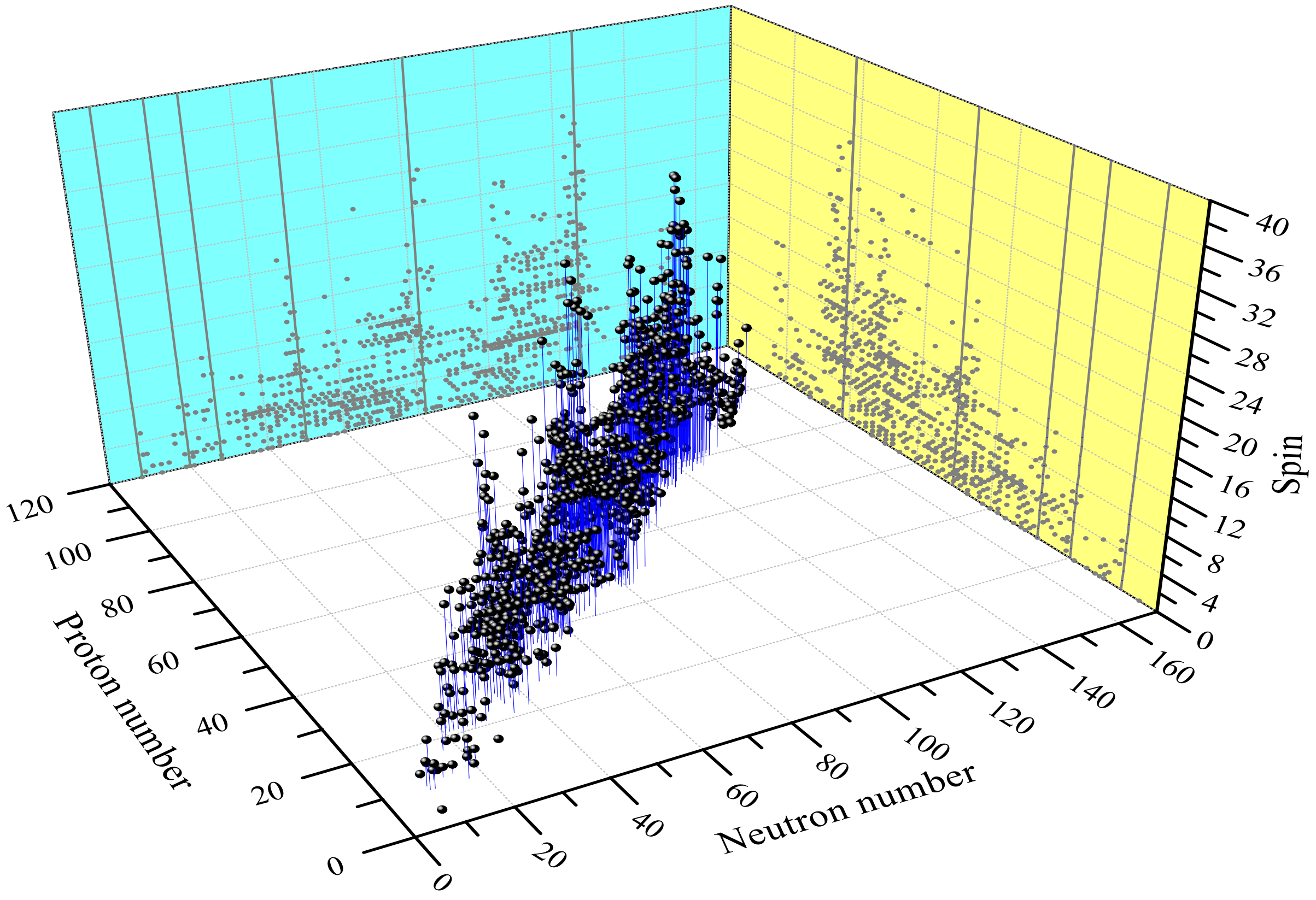

**Fig. 8**: A 3D-plot of isomeric spins for positive parity states vs. $N$ and $Z$. The panels at the back and right side are the projections of the 3D plot.

We have also plotted in Figure 6, isomers having half-lives longer than their respective ground state. The color coding in the figure denotes their different decay modes. There are more than 200 such isomers, of which only few decay via 100% IT mode with high multipolarity transition(s), while others mostly decay via beta-decay. Alpha-decay is also observed for the isomers in heavier nuclei. Such isomeric states offer avenues to study nuclear structure in regions not normally accessible and also in producing heavy/super-heavy elements. We have already pointed out that the number of isomers in odd-odd nuclei is almost double of that in even-even nuclei. This may be correlated with the absence, or unusually large hindrance of isomeric transitions in most odd-odd nuclei as pointed out by Jain *et al.* [84]. This may also be responsible for the observation that many odd-odd isomers have a half-life longer than their respective ground state. In particular, the absence of $\Delta I = 3$ transitions in a number of isomeric state to ground state pairs in many nuclei in $A = 150 - 190$ region still remains a puzzle [84].

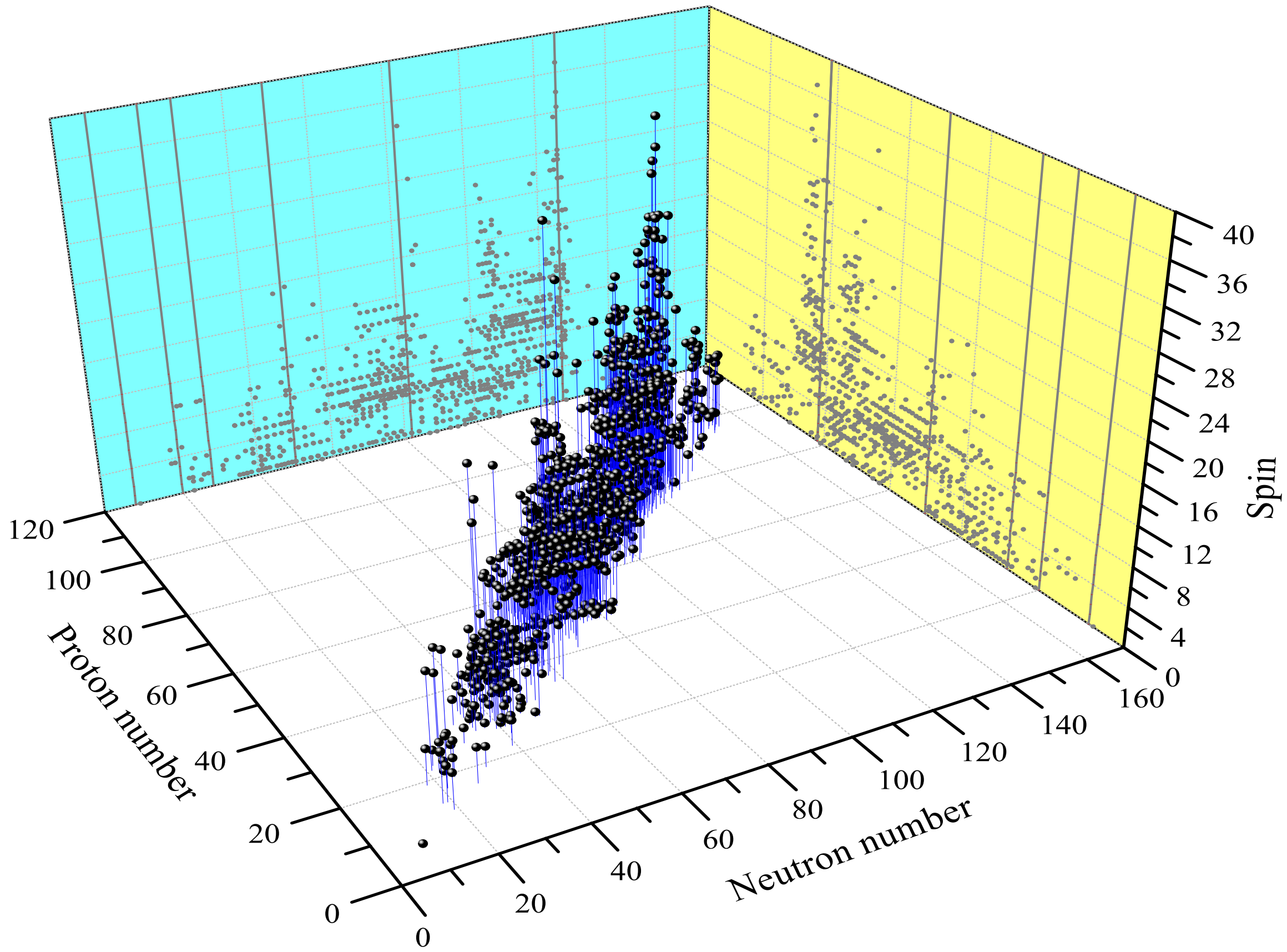

**Fig. 9**: Similar to Fig. 8 but for isomeric spins of negative parity.

### *3.4. Spin*

We show the occurrence of isomers for the negative and the positive parity in the two panels of Fig. 7. This graph is quite revealing as it exhibits distinct peaks at certain spins, which can be correlated with the shell model single particle structure. Interestingly a very high peak is observed at spin 1/2, the lowest spin possible in odd-A nuclei. This peak is present in both the positive and the negative-parity isomers. A simple explanation could be based on the observation that the spin 1/2 is the most common spin in the spectrum of odd-A nuclei. Hence, spin 1/2 is more likely to give rise to an isomer provided a higher spin state exists close to it. This may explain the peak at spin 1/2. Further, the spins $2, 3$, and 4 mostly occur in even-even nuclei where the number of isomers is the lowest. This results in the smaller peaks at the intermediate spins of $5/2, 7/2$, and $9/2$.

Similarly, we notice peaks in the number of isomers at integer spins $7, 8, 10$, and 11 in both the positive as well as negative-parity states, with occasional missing peaks in either of the two parities. The integer high spin isomers mostly arise in odd-odd and even-even nuclei and generally have two quasi-particle configurations.

The peaks at $9/2^+$ and $13/2^+$ in the upper panel are due to the intruder and unique-parity $g_{9/2}$ and $i_{13/2}$ orbits in the respective valence spaces of odd-A, odd-N isomers. Similarly, the highest peak, with more than 160 isomers, is observed at the spin $11/2^-$, as shown in the lower panel. This peak is due to the $h_{11/2}$ orbit in the odd-$A$, odd-$Z$ isomers. We also notice small peaks at 27/2 and 29/2, which are due to three quasi-particle isomers.

We have further plotted the spin values of isomers vs. $N$ and $Z$ in 3D plots, in Fig. 8 and 9. Positive parity isomers are plotted in Figure 8 and the negative parity isomers are plotted in Figure 9. There is no specific reason to separate the positive and negative parity isomeric states except for reducing the crowding in the plots. Both the plots again correlate very well with the shell structure and exhibit peaks near the magic numbers where the high-j orbits are located.

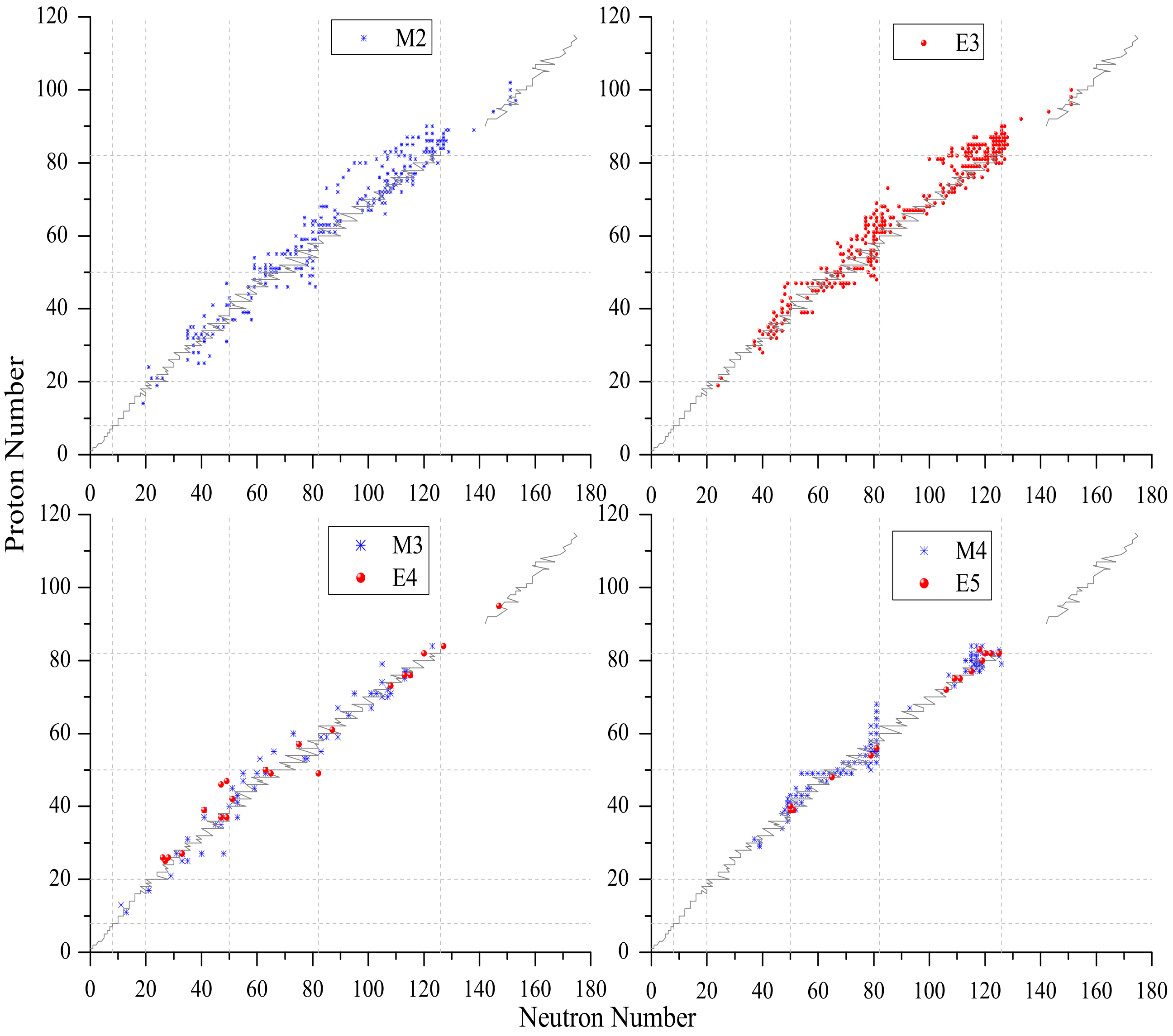

**Fig. 10**: Nuclear isomers for the higher multipolarity transitions across the nuclear landscape. The dashed lines exhibit the respective magic configurations for protons and neutrons. The highest multipolarity isomers are mostly situated along the $\beta$-stability line for normal nuclei.

### *3.5. Multipolarity*

The isomeric transitions are observed to have multipolarities from $E1$, $M1$ to $M4$ and $E5$. Decay by lower multipolarity transitions is quite common and equally distributed across the nuclear landscape. We have plotted in Fig. 10, the occurrence of only those isomers, which decay by higher multipolarity transitions like $M2$, $E3$, $M3$, $E4$, $M4$, and $E5$. We can make several observations: one, the $M2$, $E3$ and $M4$ decaying isomers lie in clusters, mostly around the magic numbers; two, the number of $M3$ and $E4$ decaying isomers seems to be particularly small and these are uniformly spread out. Number of E5 decays is really small. It would be interesting to further analyze the decay patterns and look for the physical reasons behind it.

### *3.6. Mirror isomers and Self-conjugate isomers*

A pair of isomers, for which the proton/neutron number of one member is the same as the neutron/proton number of the other member, are known as mirror isomers. We have listed seven such pairs. It is interesting to note that the pairs have the same spin-parity and similar excitation energies. It is obvious that most such cases will be found in the light mass region. For example, $^{19}$F and $^{19}$Ne both have a $5/2^+$ isomer with an excitation energy of about 200 keV [1]. This may be related to the isospin independence of nuclear force. The self-conjugate isomers are also mirror isomers but have the additional constraint of $N = Z$. We note that of the 23 self-conjugate isomers, 20 are odd-odd and 3 are even-even in nature. For example, the $5^+$ isomer in $^{18}$F is a self-conjugate isomer [1]. Such isomers may be useful to explore the isospin dependence of nuclear force.

### *3.7. List of excluded isomers*

We have excluded a number of isomers from the new edition of the atlas, which were present in the 2015 version of the atlas. The list includes the following cases with the references given as NSR key-numbers and cited in the main reference list at the end.

- $^{73}$Zn: The isomer with $T_{1/2}$ =5.8 $s$ is removed based on 2017VE05.
- $^{105}$Ru: The 159.51 keV isomer with $T_{1/2}$ =0.055(7) $\mu s$ is removed being identical to the 55(7) $ns$ isomer.
- $^{124}$I: The 286.8 keV isomer with $T_{1/2}$ =10(2) $ns$ from 2012MOZZ is removed based on 2021MO09.
- $^{125}$Sb: The 1971.25+X keV isomer with $T_{1/2}$ =25(4) $\mu s$ is removed being identical to the 2112.1 keV isomer.
- $^{130}$La: 110.4 keV isomer with $T_{1/2}$=17(5) $ns$ from 1996XU04 is removed based on 2014IO01, who did not observe this isomer and found only a prompt transition with an upper limit of half-life set at $< 10$ $ns$.
- $^{180}$Os: The 1862.50 keV isomer with $T_{1/2}$ =17(3) $ns$ is removed as 2005MO33 measured its mean life $\tau < 0.3$ $ns$ which is beyond the scope of our atlas.
- $^{180}$Hf: Due to the low energy of the decaying transition (52 keV), and the presence of the 0.94 $\mu s$ isomer at 2485 keV, the half-life for 2537.4 keV level could not be precisely determined. A half-life of $T_{1/2} << 1$ $\mu s$ is assigned to this level. It is also possible that this level may not be isomeric, so we have excluded it.
- $^{190}$Tl: The 161.9+Y keV isomer with $T_{1/2}$ =0.75 $ms$ is removed as it is not confirmed by 1991VA04 and 2005XI06.
- $^{195}$Au: The 2460.84 keV isomer with $T_{1/2}$ =16 $\mu s$ seems to be misquoted. The half-life looks similar to the 31/2- isomer with no information in ENSDF or XUNDL, hence removed.
- $^{201}$Pt: The 1544.4+X keV isomer with $T_{1/2}$ =18.4 $ns$ is removed as it seems to be the duplicate of 21 $ns$ isomer.
- $^{214}$Fr: The 3566+X keV isomer with $\sim 10$ $ns$ half life is removed as there is no information in 1994BY01, ENSDF, or XUNDL.

- $^{245}$Md: The isomer at $\sim$ 300 keV with 0.35 *s* half life is removed since 2020KH08 suggests the reverse ordering of ground state and isomeric state. The previous ground state is now included as an isomeric state.

- $^{254}$No: The isomer at 0+X keV with 0.28 *s* half life is removed as it is the same as the 263 *ms* isomer.

Following is a list of those isomers, which were not present in the Atlas-2015 but discussed in the literature, and not included in the current Atlas due to certain reasons:

- $^{67}$As: The 12 *ns* isomer from 2001JE10 has not been confirmed by 2009OR02, and is not included.

- $^{68}$Ni: 2012DI03 proposed a second excited $0^+$ state at 2202(1) keV decaying through $E_\gamma$ = 168(1) keV, with a half-life of 216 (+66-50) *ns*. In later works from Argonne in 2012CH39 and also 2012BR15, no evidence was found for the isomer at 2202 keV. However, they could not identify the origin of the 168-keV $\gamma$-ray reported by 2012DI03.

- $^{93}$Rb: The 57 $\mu s$ isomer at 253.39 keV from 1970GR38 has not been confirmed by 2014MI12, and is not included.

- $^{117}$La: The 10 *ms* isomer from 2001SO02 has not been observed by 2001MA69 and 2011LI28. This isomer, therefore, is not included.

- $^{123}$Pd,$^{125}$Pd: 2019CH24 suggest $11/2^-$ isomers in $^{123}$Pd and $^{125}$Pd but no data are available, hence not included.

- $^{145}$Cs,$^{147}$Cs: The isomers listed in 2015YAZW and NUBASE2020 were not confirmed in author's later analysis (priv. comm. of Sept 7, 2019).

- Single-level nuclei: Many nuclei in which only single level information is known, but not confirmed as either the ground state or an isomer, are not included. These nuclei are: $^{120}$La, $^{144}$I, $^{148}$Tm, $^{155}$Ta, $^{177}$Tm, $^{181}$Pb, $^{235}$Am, $^{246}$Bk and $^{246}$Es.

- $^{161}$Tm: 7.51 keV state with $1/2^+$ spin-parity and 5 *min* half-life from 1968GRZX, is expected to be an isomer, if the decay is only by an isomeric transition. No confirmed observation till date.

- $^{174}$W: Isomers at 1672.0(5) keV and 1919.7(5) keV with respective half-lives of $\geq$ 187 *ns* and 187(25) *ns* from 1978DR04 are removed as they seem to correspond to the 158(3) *ns* isomer from 2006TA13.

- $^{187}$Po: The 0.5 *ms* isomer with $(13/2^+)$ spin-parity proposed in 2006AN11 is highly tentative, hence not included in our table.

- $^{194}$Os: A high-spin isomer at high excitation is implied from out-of-beam gamma-ray data in $^{186}$W, $^{187}$Re, $^{192}$Os($^{136}$Xe,X$\gamma$), $E$=6 MeV/nucleon reaction studied by 2014DRZZ. The details of this work are still not available, hence not included.

- $^{195}$Au: Isomer at 2418.2(5) keV with half-life $\sim$ 69 *ns* from 2013DR01 is removed as per our policy since authors quote the half-life with a upper limit $<$ (or $\sim$).

- $^{216}$At: There are two isomers, one at 57 keV with spin-parity $(4^-)$ and the other at 413 keV with spin-parity $(9^-)$ and half-life 0.1 *ms* from the systematics in 1994LI10 and 1971BR13, respectively. Both are very tentative according to 1994LI10, hence not included.

- $^{262}$Rf: 0+X keV isomer and with half-life 47 $ms$ from 2017HE08 and also reported in NUBASE2020, is not adopted as there is as yet no evidence for an isomer in $^{262}$Rf.

- $^{263}$Hs,$^{273}$Ds: The isomers as listed in 2021KO07 (NUBASE2020) seem erroneous as no source reference could be found.

## 4. APPLICATIONS OF NUCLEAR ISOMERS

### *4.1. Nuclear Astrophysics*

Although the existence of isomeric states in nuclei has been known for a century (2021 marks the year to celebrate the nuclear isomers centenary), there remains much to learn about their influence on the creation of elements in astrophysical nucleosynthesis; they are expected to have a significant impact due to their unique lifetimes [85]. The usual description of nuclear isomers is based on their transition energies and decay properties, which determine their half-lives. In a stellar environment with high temperature, however, the meaning of isomer may be vastly different from its terrestrial definition. For example, in explosive astrophysical environments such as novae, supernovae, and neutron star mergers, nuclei are expected to exist in excited states including isomers. Inhibited communication between the ground state and isomer can hinder thermal equilibration, as well as trap the nucleus in an excited state that might behave radically differently from the ground state [86]. Further, one finds that although many nuclei possess isomeric states, not all isomers are of astrophysical interest. This unique aspect, recently, led to a new nomenclature of astrophysical isomers, coined as "astromers" [87], which are nuclear isomers that can have influence, different from their ground states, in an astrophysical environment of interest, and according to [88], "which retain their metastable characteristics in a hot environment". For an isomer to be an astromer, it must behave in a significantly different way as compared to the ground state, either survive long enough to undergo a reaction, or undergo $\beta-$decay at a rate different from that of the ground state [88–90].

This has opened a new direction for the research on nuclear isomers in the realm of nuclear astrophysics. Interestingly, theorists have started considering astromers in the r-process nucleosynthesis by connecting them to a light curve of kilonova from the ejecta of neutron star mergers [91]. Let us take a well-known example [86]. $^{26}$Al has a $T_{1/2} = 0.717$ Myr, $5^+$ ground-state, and a $0^+$ isomer at 228 keV, with half-life of 6.35 s, both decaying 100% through $EC$ and $\beta^+$ modes. At high temperatures, the effective lifetime of $^{26}$Al in the stellar plasma can be significantly shorter due to thermally mediated transitions between the two states. At low temperatures, without reaching thermal equilibrium, the two states operate totally differently. The ground-state decay of $^{26}$Al feeds dominantly (97.2%) the first excited $2^+$ state of $^{26}$Mg, and the de-excitation produces 1.8 MeV $\gamma$-rays. The 228 keV isomer, $^{26m}$Al, feeds 100% the ground state of $^{26}$Mg through super-allowed Fermi $\beta$-decay, thus does not contribute to the $\gamma$-ray flux. Therefore, the effective $\beta$-decay rates and the resultant $\gamma$-ray flux change with temperature, causing observational effects. While the astrophysical importance of $^{26m}$Al and $^{180m}$Ta has been well known, G.W. Misch *et al.* [88–90] have analyzed 178 neutron-rich isomers of terrestrial half-life $> 100$ $\mu$s, from $^{69}$Zn to $^{212}$Bi centered around three main abundance peaks at $A \sim 80$, $A \sim 130$ and $A \sim 195$, to compute effective thermal transition rates, thermalization temperatures, and unknown transition rates, with pertinent data taken primarily from the ENSDF database, and deficiencies in available nuclear data pointed out. They classified the isomers, relevant to r, s, and rp-processes, in three categories: astromers which accelerate the decay process, those which slow down the decay and act as storage batteries, and the ones with no significant effect. Based on a detailed analysis of thermal equilibria, Misch *et al.* [88, 89] narrowed the list down to 24 isomers from $^{69}$Zn to $^{195}$Pt in their Table I, which are expected to be important as 'astromers'. Relevant 'Astromers' for which required nuclear data are lacking, will hopefully get studied at facilities such as ATLAS, ANL; ISAC, TRIUMF; ISOLDE, CERN; and upcoming FRIB, MSU, for example, see presentation by K. Kolos *et al.* [92]. Rapid progress is also being made on the use of isomeric ion beams to study important resonance strengths and astrophysical reaction rates, for example first use [93] of 6.35 $s$ isomeric $^{26m}$Al beam has been made to study $^{26m}$Al$(p,\gamma)^{27}$Si reaction, and use [94] of 1.7 $min$ $^{130m}$Sn ion-beam in $^{9}$Be($^{130m}$Sn,$^{8}$Be)$^{131}$Sn and $^{13}$C($^{130m}$Sn,$^{12}$C)$^{131}$Sn reactions of relevance in r-process nucleosynthesis.

*4.2. Diagnostic and therapeutic uses of Nuclear Isomers*

Many radioisotopes are in active use worldwide in medical diagnostic and therapeutic procedures, using SPECT (single-photon emission computerized tomography) and PET (positron emission tomography) techniques, the latter often combined with CT scans (X-ray computerized tomography) and MRI (magnetic resonance imaging). Details of uses and production of these isotopes are provided in recent publications of the IAEA-Coordinated Research project [95–97], in a 2022 report of the Department of Energy of the United States [98], and in a 2014 report of NuPECC of European Science Foundation [99]. Another technique based on Auger electrons, still in an early experimental stage, is promising for the medical therapeutic use in localized cancers [100].

While the number of nuclear isomers as compared to the number of isotopes used in medicine is limited, yet, the application of $^{99m}$Tc isomer by itself dwarfs the usage of all the other isotopes, in the sense, that out of 40 million nuclear medicine procedures performed each year worldwide [101], about 80% (i.e. 32 million) involve $^{99m}$Tc isomer. Due to the center stage taken by this isomer in medical physics, governmental agencies in various countries (for example see refs. [98, 99]) have established dedicated programs to secure availability of this isomeric activity through the production of its parent generator 66 $h$ $^{99}$Mo, which dominantly $\beta$-decays to the 6 $h$ activity of $^{99m}$Tc. First application of this isomer as a medical tracer was reported [102] in the USA, which rapidly expanded worldwide during the 1960s, as gamma cameras. For dose calculations, it is relevant to use most precise and accurate half-life of $^{99m}$Tc, but it is known that its half-life depends on the electronic environment of the nucleus due to a dominant low-energy (2.17 keV) isomeric transition which is heavily internally converted and with an overlap of nuclear and atomic wavefunctions. In Table 1, we have adopted, essentially, the most precisely measured half-life of 6.00660(18) $h$ in ref. [103], where the authors used physiological saline solution sodium pertechnetate ($TcO_4Na$), commonly used in hospitals for nuclear imaging.

Other nuclear isomers [95–97] used are: $^{81m}$Kr (13.10 $s$) and potential diagnostic candidate $^{178m}$Ta (9.31 $min$) as gamma emitters; $^{52m}$Mn (21.1 $min$), $^{82m}$Rb (6.472 $h$), $^{94m}$Tc (52.0 $min$), $^{110m}$In (69.1 $min$), and possible $^{34m}$Cl (31.99 $min$) as positron emitters; and $^{177m}$Lu (160.4 $d$) as parent generator for $^{177}$Lu (6.644 $d$), and $^{178m}$Ta (9.31 $min$) for therapeutic purpose. Soon after the discovery of Auger effect in 1922-23, low-energy electrons emitted in this effect were proposed as a possible therapy for cancer. For targeted AUGER therapy, the following isomers have been considered as potential candidates [100, 104]: $^{58m}$Co (8.853 $h$), $^{60m}$Co (10.467 $min$), $^{99m}$Tc (6.0066 $h$), $^{103m}$Rh (56.114 $min$), $^{117m}$Sn (116.2 $min$), $^{125m}$Te (57.40 $d$), $^{189m}$Os (5.81 $h$), $^{193m}$Ir (10.53 $d$), $^{193m}$Pt (4.33 $d$), $^{195m}$Pt (4.010 $d$) and $^{197m}$Hg (23.82 $h$).

### 4.3. Other applications of Nuclear Isomers

M. Pospelov *et al.* [105] and D. Buker [106] discuss the use of easily available nuclear isomer, in combination with current dark matter detector technology, to search for a class of dark matter in collisional deexcitation of nuclear isomers such as $^{180m}$Ta ($> 4.5 \times 10^{16}$ $y$), $^{137m}$Ba (2.5223 $min$), $^{177m}$Lu (160.4 $d$), and $^{178m}$Hf (31 $y$). For application of isomers in the development of nuclear gamma-ray lasers by pumping methods, 29 outstanding cases have been identified [106, 107] which can store exceptionally high density of energy, out of which most promising are: $^{60m}$Co (10.467 $min$), $^{169m}$Yb (46 $s$), $^{177m}$Hf (51.4 $min$), $^{178m}$Hf (31 $y$), $^{179m}$Hf (25.00 $d$), $^{180m}$Ta ($> 4.5 \times 10^{16}$ $y$), and $^{192m}$Ir (1.45 $min$). J. Feng *et al.* [108] have presented a first report of experimental femtosecond pumping of nuclear isomeric state (in 1.83 $h$ $^{83m}$Kr) by the Coulomb excitation of ions with the quivering electrons induced by laser fields. The $^{83m}$Kr isomeric activity is used for calibration of liquid Xe and Ar detectors such as PANDAX-4T at China Jinping Underground Laboratory (CJPL) for dark matter searches [109]. A well-known promising example [69, 110, 111] of a most precise nuclear optical clock involves the lowest energy (8.12(11) eV) nuclear isomer in $^{229m}$Th (7(1) $\mu s$ for neutral atom). A recent review article by H. J. Pant [112] describes in detail extensive use of radioisotopes in industrial applications in troubleshooting and measuring hydrodynamic parameters, analyzing general aspects of fluid dynamics such as leaks, mixing time, flow rate, residence time distribution, flow pattern, local velocity, and turbulence-related parameters, and in downhole investigations in oil fields for erosion of drill bits and assessment of other conditions of oil wells. In such investigations, $^{99m}$Tc is again among the commonly used 12 or so radiotracers listed in Table 1 of ref. [112], others being $^{137m}$Ba and $^{113m}$In (99.47 $min$).

## 5. Addendum

$^{44}$V, $^{52}$Co: level energies of the $(6)^+$ isomer in $^{44}$V, and the $2^+$ isomer in $^{52}$Co have been measured using mass spectrometry by M. Wang et al., Phys, Rev. C 106, L051301 (2022) as 266(10) keV and 380(12) keV, respectively, compared to 270(10) keV and 381(13) keV in Table 1. $^{73m}$Br: a new isomer at 997.6 keV with spin-parity of $(9/2^+)$ and half-life of 52(2) ns, decaying by 524.7-keV, (M1+E2) and 711.7-keV transitions is proposed by S. Bhattacharya et al., Phys. Rev. C 106, 044312 (2022). $^{101}$In: level energy of the $(1/2^-)$ isomer has been measured, more precisely, using mass spectrometry by M. Mougeot et al., Nature Phys. 17, 1099 (2021) as 668(11) keV, compared to 637(50) keV in Table 1. $^{212m}$Po: two new isomers have been proposed by L. Zago et al., Phys. Lett. B 834, 137457 (1022): 3455 keV with spin-parity of $(21^-)$ and half- life of 1.9(3) $\mu$s, decaying by a 533-keV, E3 transition, and at either 3319 keV, if in parallel or at 3852 keV, if in a cascade, with spin-parity of $(23^+)$ and half-life of 0.20(8) $\mu$s, decaying by a 397- keV, M2 transition. $^{248}$Cf: two new isomers have been identified by R. Orlandi et al., Phys. Rev. C (accepted October 24, 2022): one at 0.95(30) MeV with half-life of 11.64(35) ns, decaying by a 48-keV, E1 transition; and another at 0.90(30) MeV with half-life of $\approx$139 ns isomer, with K$\geq$ 5. $^{253m}$Lr: half-life of an isomer has been measured by T. Huang et al., Phys. Rev. C (accepted October 11, 2022) as 2.46(32) s from decay curve of an $\alpha$-line at 8660(20) keV.

## 6. SUMMARY

We have presented an updated Atlas of Nuclear Isomers comprising the experimental data for the isomers with a half-life $\geq$ 10 ns along with their various spectroscopic properties such as excitation-energy, half-life, decay-mode(s), spin-parity, gamma energy and multipolarity of the isomeric transitions. The most relevant original references are also listed in the Table and were consulted along with ENSDF, XUNDL and NUBASE2020 databases. The present version of Atlas contains a much larger set of evaluated data as compared to the 2015 edition. Many systematic isomeric properties are also presented and discussed. Current and potential applications of nuclear isomers with selected examples are pointed out. The literature cutoff date of the data is October 31, 2022.

**Acknowledgments**

SG and YS acknowledge support from the National Natural Science Foundation of China (Grant Nos. 12235003 and U1932206) and by the National Key Program for S & T Research and Development (Grant No. 2016YFA0400501). BM acknowledges the support received from the Amity University UP and IIT Ropar (India) during her stays in the crucial phases of this work, as well as the financial support from the Croatian Science Foundation and the École Polytechnique Fédérale de Lausanne, under the project TTP-2018-07-3554 "Exotic Nuclear Structure and Dynamics", with funds of the Croatian-Swiss Research Programme. AKJ and AG acknowledge the financial support from the Science and Engineering Research Board (Govt. of India) for the grant no. CRG/2020/000770, and also the support received from the Amity University to carry out this work. AKJ acknowledges the IAEA-NDS for award of Special Service Agreement: TAL-NAPC20190423-001 for this project.

We are grateful to a large number of researchers worldwide with whom we have corresponded over years, who have kindly and promptly provided constructive comments, advice, details of data, and in many cases new data for isomers, although, it is difficult to list all the corresponding researchers. We are indebted to Jun Chen (Michigan State Univ., USA) for help in the conversion of file formats of Tables 1 and 2. We thank Alan Nichols (Surrey, UK); Phil Walker (Surrey, UK); Wendell Misch and Matthew Mumpower (Los Alamos) for consultations, and Steffen Turkat (Dresden, Germany) for supplying and translating 1913 and 1914 articles and a thesis from Karlsruhe work. We also express thanks to Boris Pritychenko and (Ms.) JoAnn Totans at NNDC (BNL) for promptly helping us with entries of new articles, as well as with general bibliographic help in the NSR database.

## 7. POLICIES AND EXPLANATION OF THE TABLE

This Atlas is an update of our previous edition of Atlas of Nuclear Isomers in 2015 [1]. For the earlier work, we have used the experimental data from the ENSDF database [2] as the main source. Other databases such as XUNDL (Experimental Unevaluated Nuclear Data List) [3], NSR (Nuclear Science References) [5] and bibliographic database [5] have been used for completeness. In the update, we have tried to include all the new information which has appeared in the past seven years with a literature cutoff date of 21st July 2022. The NUBASE2020 [4] has also been consulted for completeness. A significant number of half-lives have been re-evaluated, based on original and new references. Finally, the isomers listed in this Atlas are based on the experimental data only, not including certain isomers expected from nuclear models.

### Explanation of Tables

| | |
|---|---|
| $Z$ | Atomic number |
| $N$ | Neutron number |
| $X^A$ | Nuclei with mass number A |
| $E$(level) | Excitation energy of isomer in keV |
| $J^\pi$ | Spin parity of the isomer |
| $T_{1/2}$ | Half-life of the isomeric state |
| $E_\gamma$ | Gamma-ray energy in keV |
| $\lambda$ | Multipolarity corresponding to $E_\gamma$ |
| Decay mode | Decay mode for the isomeric state |
| Key No. | NSR keynumbers corresponding to references for each isomer [5] |

The flagged footnotes are given at the bottom of each page. We have adopted some general policies in this work as listed below:

1. **Energy -** The energies are given in keV with uncertainty in brackets. The unknown level energies listed as 0+X, 0+Y are expected to lie near the ground state. The isomeric energy levels which are expected to be at much higher energy above the ground state have been quoted as X, Y etc. The cases where the ordering of two lowest but unknown energy levels is involved, have been flagged as "†", with a note stating that "the ordering of the ground state and the isomeric state is unknown". We give only the experimental data, not the systematic values from any database.
2. **Spin and parity -** The spin and parities without parentheses are firm assignments based on the strong arguments. The spin and parity within the parentheses are tentative assignments based either on experiments with uncertain results, systematic or, theoretical predictions.
3. **Half-life -** The lower limit on the half-life has been kept at 10 $ns$. The limit of 10 $ns$ represents the central value, not considering the upper or lower bounds of the uncertainties.
   - The isomers reported in literature with upper limit on half-lives have not been included. For example, the 4250 keV level with $< 210$ $ns$ half-life in $^{34}$Si has not been included from 1989BA50.

- The half-lives listed in the table are the weighted average of the most precise values from references in the ENSDF database and from the latest measurements, when available. However, in many cases, previous references are listed for the sake of completeness.
- In general, the uncertainty in the evaluated half-life is not dropped below the minimum uncertainty in a dataset of independent measurements that is being averaged.
- The half-lives measured for bare or, ionized atoms have been identified and flagged as @.

4. **Gamma energy** - The values given without "parentheses or ?" are confirmed values of gamma from the corresponding isomeric state. The values placed in "parentheses" point to gamma rays expected but not observed, generally when these are low energies and highly converted. The values with "?" refers to the $E_\gamma$ whose placement in the level scheme is uncertain.
5. **Multipolarity** - The multipolarities are taken mostly from the Adopted datasets in the ENSDF database or the latest measurements if any. The values within the parentheses show the tentative assignments.
6. **Decay mode** - Nuclear isomers generally decay to a lower energy state via an electromagnetic transition, which is also known as the isomeric transition (IT). The isomers away from the line of stability often decay via beta decay ($\beta$). They may also decay via the beta delayed neutron decay ($\beta^- n$), beta delayed proton decay ($\beta^+ p$), alpha decay ($\alpha$), and spontaneous fission (SF), proton decay ($p$), double proton decay ($2p$), etc. We have included this information for various decay branches from the ENSDF database [2] or XUNDL [3] unless a later reference, not updated in the ENSDF or XUNDL, has become available.
7. **Reference** - The references in the last column are given in the same manner as NSR database [5] key numbers. We have tried to cover almost all the references important for the various quantities mentioned in the table. The references mainly refer to the measured half-life data and new measurements of other quantities mentioned in the table especially where ENSDF has not yet been updated to include that information.
8. We have not listed or discussed any isomers that were proposed in earlier literature but rejected in later studies.
9. **Tentative Isomers** - We have further prepared a list of tentative isomers in Table II which have been proposed in secondary references such as theses or conference reports, but have not yet been published or confirmed. General policies for Table II are the same as those for Table I.
10. Table I lists isomers which have been reported from experimental evidence, and generally in peer-reviewed journals. In this respect, our approach is closer to that of the policies of the ENSDF database. Although, for consistency we have made detailed checks with isomers in NUBASE2020 [4], but have not included several isomers which are proposed either from systematics or from model considerations with no confirmed experimental evidence. In a few other cases, we seem to disagree with the assignments in NUBASE2020.

**Table 1**

Nuclear Isomers with half-life of ≥10 *ns*. @: Half-life for bare or highly-ionized nucleus. †: Energy ordering of the ground state and the isomeric state is unknown.

| Z | N | $^{A}$X | E(keV) | J$^{\pi}$ | $T_{1/2}$ | E$\gamma$(keV) | $\lambda$ | Decay mode | Reference |
|---|---|---|---|---|---|---|---|---|---|
| 4 | 8 | $^{12}$Be | 2251 (1) | 0+ | 233(8) *ns* | 142(2)<br>2251(1) | E2<br>E0 | %IT = 100 | 2018CH38<br>2013JO06<br>2007SH34 |
| 7 | 9 | $^{16}$N | 120.42 (12) | 0− | 5.306(28) $\mu s$ | 120.42(12) | | %IT = 99.99960(4)<br>%$\beta$− = 0.00040(4) | 2013SO20<br>1983MI20<br>1983GA18<br>1975PA01<br>1967BE14 |
| 9 | 9 | $^{18}$F | 1121.36 (15) | 5+ | 160(4) *ns* | 184 | E2 | %IT = 100 | 1979HA60<br>1972AD01<br>1972BE37<br>1967BE14<br>1967PO09<br>1959AL99 |
| 9 | 10 | $^{19}$F | 197.143 (4) | 5/2+ | 88.7(14) *ns* | 87.3<br>197.1 | E2 | %IT = 100 | 1968KL05<br>1968ST25<br>1967BE14<br>1964SU01<br>1956JO35<br>1955FI38 |
| 10 | 9 | $^{19}$Ne | 238.27 (11) | 5/2+ | 18.0(7) *ns* | 238.3 | E2 | %IT = 100 | 1969BL02<br>1967BE14<br>1957BA09 |
| 12 | 9 | $^{21}$Mg | 205.6 (1) | 1/2+ | 11.7(5) *ns* | 205.6(1) | E2 | | 2019RU02 |
| 11 | 11 | $^{22}$Na | 583.05 (10) | 1+ | 243(2) *ns* | 583.04(10) | E2 | %IT = 100 | 1966SU07<br>1958TE15 |
| 11 | 13 | $^{24}$Na | 472.2074 (8) | 1+ | 20.18(10) *ms* | 472.2024(8) | M3 | %IT = 99.95<br>%$\beta$−≈ 0.05 | 1980JO11<br>1972BR53<br>1970CH37<br>1961SC09 |
| 13 | 11 | $^{24}$Al | 425.8 (1) | 1+ | 130.7(13) *ms* | 425.8(1) | M3 | %IT = 69.6(7)<br>%$\epsilon$+%$\beta$+ = 30.4(7)<br>%$\epsilon$ $\alpha$ = 0.028(6) | 2011NI18<br>1988BU12<br>1979SH11<br>1979HO08<br>1971TO12<br>1968AR03 |
| 9 | 17 | $^{26}$F | 643.4 (1) | (4+) | 2.2(1) *ms* | 643.4(1) | | %IT = 82(11)<br>%$\beta$−= 18(11)<br>%$\beta$–n = 12(8) | 2013LE03 |
| 11 | 15 | $^{26}$Na | 82.40 (4) | 1+ | 4.35(16) $\mu s$ | 82.40(4) | (E2) | %IT = 100 | 2014NIZZ<br>1990EN08<br>1987DUZU |
| 13 | 13 | $^{26}$Al | 228.305 (13) | 0+ | 6346.02(54) *ms* | | | %$\epsilon$+%$\beta$+ = 100 | 2020HA30<br>2013CH51<br>2011SC22<br>2011FI01<br>1983KO22<br>1977AL11<br>1975AZ01<br>1969FR08 |

*Continued...*

Table 1 contd...

| Z | N | $^{A}$X | E(keV) | J$^{\pi}$ | $T_{1/2}$ | E$\gamma$(keV) | $\lambda$ | Decay mode | Reference |
|---|---|---|---|---|---|---|---|---|---|
| 15 | 11 | $^{26}$P | 164.4 (1) | | 115(9) *ns* | 164.4(1) | | %IT = 100 | 2017PE09<br>2014NIZZ |
| 12 | 19 | $^{31}$Mg | 49.93 (6) | 3/2+ | 12.0(3) *ns* | 49.9(1) | (M1) | %IT = 100 | 2019NI04<br>2017NI02<br>2005MA96<br>1993KL02 |
| 12 | 19 | $^{31}$Mg | 461.0 (6) | (7/2−) | 10.5(8) *ns* | 239.9(5) | | %IT = 100 | 2007KI08<br>2005MA96 |
| 12 | 20 | $^{32}$Mg | 1057 | 0+ | 7-26 *ns* | 172 | | | 2019EL09<br>2010WI11 |
| 13 | 19 | $^{32}$Al | 956.6 (5) | (4+) | 200(20) *ns* | 221.9(3) | | %IT = 100 | 1996RO02 |
| 14 | 18 | $^{32}$Si | 5581 (4) | (5−) | 27(2) *ns* | 79(1) | | %IT = 100 | 1998FO07<br>1997FO01 |
| 14 | 19 | $^{33}$Si | 1434.9 (5) | 7/2− | 10.2(3) *ns* | 1434.9(5) | M2 | %IT = 100 | 2002ASZY |
| 13 | 21 | $^{34}$Al | 46.47 (17) | 1+ | 22.1(2) *ms* | | | %$\beta-\approx$ 100<br>%$\beta-$n = 11(4) | 2019LI41<br>2017LI03<br>2017HA23<br>2012RO25 |
| 14 | 20 | $^{34}$Si | 2718.52(10) | 0+ | 19.4(5) *ns* | 2718.4(1) | E0 | %IT = 100 | 2019LI41<br>2012RO25 |
| 17 | 17 | $^{34}$Cl | 146.36 (3) | 3+ | 31.99(3) *min* | 146.36(3) | | %IT = 44.6(6)<br>%$\epsilon$+%$\beta$+ = 55.4(6) | 1982GR07<br>1980WI13<br>1975VA02<br>1965EB01 |
| 19 | 18 | $^{37}$K | 1380.25 (3) | 7/2− | 10.4(5) *ns* | 1380.2 | | %IT = 100 | 1971RA22<br>1967GO18 |
| 17 | 21 | $^{38}$Cl | 671.365 (8) | 5− | 715(3) *ms* | 671.360(8) | M3 | %IT = 100 | 1972BR53 |
| 19 | 19 | $^{38}$K | 130.22 (16) | 0+ | 924.33(27) *ms** | 130.1(2) | | %$\epsilon$+%$\beta$+ = 99.9670(43)<br>%IT = 0.0330(43) | 2020HA30<br>2010BA43<br>2000BB01<br>1983KO22<br>1978WI04<br>1978TH02<br>1976WI08<br>1975SQ01<br>1972HA82 |
| 19 | 19 | $^{38}$K | 3458.10 (17) | (7)+ | 21.95(11) $\mu s$ | 38.03(3)<br>811.9(2)<br>3457.7(4) | (E1) | %IT = 100 | 1980JO11 |
| 19 | 21 | $^{40}$K | 1643.638 (11) | 0+ | 0.336(13) $\mu s$ | 843.478(16)<br>1613.84(4) | | %IT = 100 | 1972AD01<br>1968MA09 |
| 21 | 21 | $^{42}$Sc | 616.782 (46) | 7+ | 61.7(4) *s* | | | %$\epsilon$+%$\beta$+ = 100 | 2017ER01<br>1978BE61<br>1974WI14<br>1965NE02<br>1963RO10 |

*Continued...*

*: Seven other values from 1954-1978, mostly with large uncertainties or discrepant.

Table 1 contd...

| Z | N | $^{A}$X | E(keV) | J$^{\pi}$ | $T_{1/2}$ | E$\gamma$(keV) | $\lambda$ | Decay mode | Reference |
|---|---|---|---|---|---|---|---|---|---|
| 16 | 27 | $^{43}$S | 320.7 (5) | (7/2−) | 415(3) $ns$ | 320.7(5) | | %IT = 100 | 2020MO32<br>2012CH16<br>2012KA36<br>2009GA05 |
| 19 | 24 | $^{43}$K | 738.30 (6) | 7/2− | 200(5) $ns$ | 738.23(6) | M2+E3 | %IT = 100 | 1984RA23<br>1977PO07<br>1976DE41<br>1975BO30 |
| 21 | 22 | $^{43}$Sc | 151.79 (8) | 3/2+ | 438(7) $\mu s$ | 151.65(17) | M2 | %IT = 100 | 1965DE15<br>1964HO14 |
| 21 | 22 | $^{43}$Sc | 3123.73 (15) | 19/2− | 472(4) $ns$ | 135.6(1) | E2 | %IT = 100 | 2008FE02<br>2007CH40<br>1981DA06<br>1978HA07 |
| 22 | 21 | $^{43}$Ti | 313.0 (10) | (3/2+) | 11.9(3) $\mu s$ | 312.7(2) | | %IT = 100 | 2011HO02<br>1978ME15 |
| 22 | 21 | $^{43}$Ti | 3066.4 (10) | (19/2−) | 556(6) $ns$ | 114.7 | | %IT = 100 | 2011HO02<br>1981DA06<br>1978HA07<br>1978ME09 |
| 16 | 28 | $^{44}$S | 1365.0 (8) | 0+ | 2.619(26) $\mu s$ | (36(1))<br>1365(1) | E0 | %IT = 100 | 2010FO04 |
| 21 | 23 | $^{44}$Sc | 67.8679 (14) | 1− | 154.8(8) $ns$ | 67.8679(14) | E1 | %IT = 100 | 1988AL27<br>1975GU24<br>1967RI06<br>1963KL06<br>1962TH12<br>1959CY90 |
| 21 | 23 | $^{44}$Sc | 146.1914 (20) | 0− | 51.0(3) $\mu s$ | 78.3234(14)<br>146.22 | M1 | %IT = 100 | 1988AL27<br>1964BR27<br>1963KL06 |
| 21 | 23 | $^{44}$Sc | 271.240 (10) | 6+ | 58.61(10) $h$* | 271.24(1) | | %IT = 98.80(7)<br>%$\epsilon$+%$\beta$+ = 1.20(7) | 1969RA16 |
| 23 | 21 | $^{44}$V | 270 (10) | (6)+ | 150(3) $ms$ | | | %$\epsilon$+%$\beta$+ $\approx$ 100 | 2020PU02<br>2018ZH29<br>1997HA04 |
| 21 | 24 | $^{45}$Sc | 12.40 (5) | 3/2+ | 0.326(4) $s$ | 12.4 | (M2) | %IT = 100 | 1967BL17 |
| 22 | 23 | $^{45}$Ti | 36.53 (15) | 3/2− | 3.0(2) $\mu s$ | 36.69(21) | (E2) | %IT = 100 | 1971BL14<br>1970LY02 |
| 22 | 23 | $^{45}$Ti | 39.39 (23) | 5/2− | 11.29(9) $ns$ | 40.15(30) | (M1(+E2)) | %IT = 100 | 1977BR15<br>1977ST12<br>1970LY02 |
| 23 | 22 | $^{45}$V | 56.8 (6) | (3/2−) | 465(22) $ns$ | (0.8)<br>57.1(8) | <br>(E2) | %IT = 100 | 2011HO02<br>1980GR04 |
| 24 | 21 | $^{45}$Cr | 107 (1) | (3/2+) | >80 $\mu s$ | 107(1) | (M2) | %IT = 100 | 2011HO02 |
| 20 | 26 | $^{46}$Ca | 2973.9 (6) | 6+ | 10.4(5) $ns$ | 399.2(3) | | %IT = 100 | 1975BI01<br>1975KU17 |
| 21 | 25 | $^{46}$Sc | 52.011 (1) | 6+ | 9.4(8) $\mu s$ | 52.011(1) | E2 | %IT = 100 | 1968FO01<br>1966KA20 |

*Continued...*

*: Five values from 1945-1956, with large uncertainties or discrepant. $T_{1/2}$=58.7(3) $h$; M. Teresa Duran et al., App. Rad. Isot. 190, 110507 (2022).

Table 1 contd...

| Z | N | $^{A}$X | E(keV) | $J^{\pi}$ | $T_{1/2}$ | Eγ(keV) | λ | Decay mode | Reference |
|---|---|---|---|---|---|---|---|---|---|
| 21 | 25 | $^{46}$Sc | 142.528 (7) | 1− | 18.75(4) *s* | 142.528(8) | E3 | %IT = 100 | 1972BEWN<br>1967YU01 |
| 23 | 23 | $^{46}$V | 801.46 (10) | 3+ | 1.02(7) *ms* | 801.5(1) | | %IT = 100 | 1967CO12<br>1962MO19 |
| 21 | 26 | $^{47}$Sc | 766.83 (9) | (3/2)+ | 272(8) *ns* | 767.1(3) | (M2) | %IT = 100 | 1968FO02<br>1966BA40 |
| 21 | 28 | $^{49}$Sc | 2228.63 (22) | 1/2+ | 29.9(11) *ns* | 2228.75(35) | | %IT = 100 | 1968CH13 |
| 23 | 26 | $^{49}$V | 152.9282 (17) | 3/2− | 19.90(24) *ns* | 62.289(2)<br>152.928(2) | (M1)<br>E2 | %IT = 100 | 1976WH01 |
| 19 | 31 | $^{50}$K | 172.0 (4) | (2−) | 131(40) *ns* | 101(1)<br>128.1(5)<br>172.0(5) | <br>(E2)<br>(E2) | %IT = 100 | 2012KA36<br>2010DA06<br>2009CR03<br>1999DAZQ |
| 21 | 29 | $^{50}$Sc | 256.895 (10) | 2+ | 0.35(4) *s* | 256.894(10) | (M3) | %IT = 99.5(5)<br>%β−= 0.5(5) | 2017GA25<br>1984AL18<br>1964SH14 |
| 25 | 25 | $^{50}$Mn | 225.28 (9) | 5+ | 1.75(3) *min* | | | %ϵ+%β+ = 100 | 2013SU07<br>1972RA14<br>1962SU10 |
| 25 | 27 | $^{52}$Mn | 377.749 (5) | 2+ | 21.1(2) *min*<br>22.7(30) *min* @ | 377.748(5) | E4 | %IT = 1.78(5)<br>%ϵ+%β+ = 98.22(5) | 1959JU40<br>1995IR01@ |
| 26 | 26 | $^{52}$Fe | 6958.0 (4) | 12+ | 45.9(6) *s* | 465.0(3)<br>597.1(3) | E4<br>E4 | %ϵ+%β+ = 100<br>%IT = 0.021(5) | 2005GA20<br>1979GE02 |
| 27 | 25 | $^{52}$Co | 381 (13) | 2+ | 102(6) *ms* | | | %ϵ+%β+ ≈ 100<br>%IT = ? | 2018ZH29<br>2017NE05<br>2016OR08<br>1997HA04 |
| 26 | 27 | $^{53}$Fe | 741.11 (10) | 3/2− | 63.5(14) *ns* | 741.1(1) | E2 | %IT = 100 | 1973SA10<br>1971CO29 |
| 26 | 27 | $^{53}$Fe | 3040.4 (3) | 19/2− | 2.54(2) *min*<br>2.48(5) *min* @ | 701.1(1)<br>1712.6(3)<br>3040.6(5) | | %IT = 100 | 1974FIZI<br>1968DE27<br>1967ES06<br>1995IR01@ |
| 27 | 26 | $^{53}$Co | 3174.3 (10) | (19/2−) | 247(12) *ms* | | | %ϵ+%β+ ≈ 98.5<br>%p ≈ 1.5 | 2015SH16<br>2010KA26<br>1976VI02<br>1972CE01<br>1970CE04 |
| 21 | 33 | $^{54}$Sc | 110.0 (10) | (4, 5) + | 2.77(2) *μs* | 110 | E2 | %IT = 100 | 2010CR02 |
| 23 | 31 | $^{54}$V | 108.0 (10) | (5)+ | 0.9(5) *μs* | 108 | E2 | %IT = 100 | 1998GR14 |
| 26 | 28 | $^{54}$Fe | 6527.1 (11) | 10+ | 364(7) *ns* | 146.2 (2)<br>3577.6 | E2<br>E4 | %IT = 100 | 1978DA09 |
| 27 | 27 | $^{54}$Co | 197.1 (4) | 7+ | 1.48(2) *min* | | | %ϵ+%β+ = 100 | 1967WE01<br>1962SU10 |
| 28 | 26 | $^{54}$Ni | 6457.4 (9) | 10+ | 155.6(29) *ns* | 146.1(2)<br>3386.2(9) | | %IT = 50.5(23)<br>%p = 49.5(23) | 2021GI18<br>2020ST02<br>2008RU09 |
| 21 | 35 | $^{56}$Sc | 0+X | (5+, 6+) | 75(6) *ms* | | | %β−= 100<br>%β−n > 12 | 2015BL05<br>2010CR02 |

*Continued...*

Table 1 contd…

| Z | N | $^{A}X$ | E(keV) | J$^{\pi}$ | $T_{1/2}$ | Eγ(keV) | $\lambda$ | Decay mode | Reference |
|---|---|---|---|---|---|---|---|---|---|
| 21 | 35 | $^{56}$Sc | 775.0 (1) | (4+) | 0.29(2) $\mu s$ | 47.7(3) | | %IT = 100 | 2020MI13 |
| | | | | | | 187.8(3) | (E2) | | 2012KA36 |
| | | | | | | | | | 2010CR02 |
| 26 | 31 | $^{57}$Fe | 14.4129 (6) | 3/2− | 98.04(20) $ns$* | 14.4129(6) | M1+E2 | %IT = 100 | 2021DU16 |
| | | | | | 100(5) $ns$*[@] | | | | 2006MO26 |
| | | | | | | | | | 1995AH04 |
| | | | | | | | | | 1978ALZX |
| | | | | | | | | | 1969HO28 |
| | | | | | | | | | 1966EC05 |
| | | | | | | | | | 1965KI03 |
| | | | | | | | | | 1961CL11 |
| | | | | | | | | | 1955MI87 |
| | | | | | | | | | 1955LE30 |
| | | | | | | | | | 1950DE06 |
| | | | | | | | | | 1989PH01* |
| | Scenario i[#] | | | | | | | | |
| 21 | 37 | $^{58}$Sc | 580.0 (5) | | 1123(187) $ns$ | 580.0(5) | | | 2021WI05 |
| | Scenario ii[#] | | | | | | | | |
| 21 | 37 | $^{58}$Sc | 412.5 (3) | | 1185(243) $ns$ | 412.5(3) | | %IT = 100 | 2021WI05 |
| 21 | 37 | $^{58}$Sc | 580.0 (5) | | 776(208) $ns$ | 580.0(5) | | %IT = 100 | 2021WI05 |
| 21 | 37 | $^{58}$Sc | X[$] | | 0.60(13) $\mu s$ | | | %IT = 100 | 2020MI13 |
| 25 | 33 | $^{58}$Mn | 71.77 (5) | 4+ | 65.4(5) $s$ | 71.78(5) | M3 | %$\beta-\approx$ 90 | 2015HE28 |
| | | | | | | | | %IT $\approx$ 10 | 1993SCZS |
| | | | | | | | | | 1978WY02 |
| | | | | | | | | | 1971DY01 |
| | | | | | | | | | 1969WA10 |
| | | | | | | | | | 1961CH04 |
| 27 | 31 | $^{58}$Co | 24.95 (6) | 5+ | 8.853(23) $h$ ‡ | 24.889(21) | M3 | %$\epsilon$+%$\beta$+ = 0.00120(5) | 2019MO11 |
| | | | | | | | | %IT = 99.99880(5) | 1970CA19 |
| | | | | | | | | | 1968WI10 |
| | | | | | | | | | 1968RH01 |
| | | | | | | | | | 1967ST23 |
| | | | | | | | | | 1960PR05 |
| | | | | | | | | | 1952AV17 |
| | | | | | | | | | 1952HO58 |
| | | | | | | | | | 1950CH62 |
| 27 | 31 | $^{58}$Co | 53.15 (7) | 4+ | 10.5(3) $\mu s$ | 28.30(15) | E2+M1 | %IT = 100 | 1972HA61 |
| | | | | | | 52.96(13) | | | 1971RO08 |
| | | | | | | | | | 1964BR27 |
| 22 | 37 | $^{59}$Ti | 108.7(4) | (1/2−, 5/2−) | 616(13) $ns$ | 108.5(5) | | %IT = 100 | 2020MI13 |
| | | | | | | | | | 2019WI04 |
| | | | | | | | | | 2012KA36 |
| | | | | | | | | | 2005GA01 |

*Continued…*

*: 1989PH01 measured the internal conversion decay of this level in different ionic charge states; The $T_{1/2}$ for two-electron ion is 100(5) $ns$ and for the total spin f=1 state of the one-electron ion is 79(6) $ns$.

[#]: Two different level schemes are proposed in 2021WI05. In 2020MI13, the transitions at 413 and 580 keV have been observed, albeit with a larger relative intensity for the 580-keV $\gamma$ ray. This would favor scenario (ii) with two different isomers. Two other transitions at 181 and 247 keV claimed in 2020MI13, are not observed in 2021WI05.

[$]: If 180.5(0.6), 247(2), 412.3(0.6), 580.9(0.4) gamma rays form a cascade, and there is no additional gamma ray, then the energy could be 1420.7 keV.

‡: Seven other values from 1950-1968, mostly with less precise values.

Table 1 contd...

| Z | N | $^{A}$X | E(keV) | J$^{\pi}$ | $T_{1/2}$ | E$\gamma$(keV) | $\lambda$ | Decay mode | Reference |
|---|---|---|---|---|---|---|---|---|---|
| 24 | 35 | $^{59}$Cr | 502.7 (11) | (9/2+) | 96(20) $\mu s$ | 193(1) | | %IT = 100 | 1998GR14 |
| 23 | 37 | $^{60}$V$^{\dagger}$ | 0+X<br>0+Y* | | 122(18) $ms$<br>40(15) $ms$ | | | %$\beta$−= 100<br>%$\beta$−= 100<br>%IT = ? | 1999SO20<br>2003SO02 |
| 23 | 37 | $^{60}$V | 103.2 (1) | $ | 204(4) $ns$ | 99.1(1)<br>103.2(1) | | %IT = 100 | 2021WI05#<br>2012KA36$<br>2010DA06$ |
| 25 | 35 | $^{60}$Mn | 271.80 (10) | 4+ | 1.77(2) $s$ | 271.9(1) | M3 | %$\beta$−= 88.5(8)<br>%IT = 11.5(8) | 2015HE28<br>2006LI15<br>1988BO06<br>1985RU05 |
| 27 | 33 | $^{60}$Co | 58.59 (1) | 2+ | 10.467(6) $min$ | 58.603(7) | M3+(E4) | %IT = 99.75(3)<br>%$\beta$−= 0.25(3) | 1990AB02<br>1963SC14 |
| 22 | 39 | $^{61}$Ti | 125.1 (5) | (5/2−) | 200(28) $ns$ | 125.0(5) | | %IT = 100 | 2020MI13<br>2019WI04 |
| 22 | 39 | $^{61}$Ti | 700.7 (7) | (9/2+) | 354(69) $ns$ | 575.1(5) | | %IT = 100 | 2020MI13<br>2019WI04 |
| 26 | 35 | $^{61}$Fe | 861.67 (11) | (9/2+) | 238(5) $ns$ | 654.75(10) | (M2) | %IT = 100 | 2015WI02<br>2004MA80<br>1998GR14 |
| 25 | 37 | $^{62}$Mn | 346 (+3−8) | 4(+) | 671(5) $ms$ | | | %$\beta$−= 100 | 2015HE28<br>2015GA38‡<br>2010CH51<br>1999HA05 |
| 25 | 37 | $^{62}$Mn | 113.8+Y ? | (4) | 95(2) $ns$ | 113.8(3) | (M1+E2) | %IT = 100 | 2010DA06<br>1999DAZQ |
| 27 | 35 | $^{62}$Co | 22 (5) | (5)+ | 13.86(9) $min$ | | | %$\beta$−= 99.5(5)<br>%IT = 0.5(5) | 1970JO12<br>1969MO04<br>1969WA16<br>1962VA23<br>1960PR05<br>1957GA15<br>1949PA01 |
| 29 | 33 | $^{62}$Cu | 390.12 (6) | 4+ | 11(1) $ns$ | 146.81(10)<br>349.25(10) | E2<br>E2 | %IT = 100 | 1977CH31<br>1976CH17 |
| 28 | 35 | $^{63}$Ni | 87.15 (11) | 5/2− | 1.67(3) $\mu s$ | 87.13(11) | | %IT = 100 | 2013AL19<br>1978HO06<br>1975RO05<br>1970BL06 |
| 31 | 32 | $^{63}$Ga | 75.32 (18) | 5/2− | ~ 25 $ns$ | 75.2(2) | M1+E2 | %IT = 100 | 1991BA20 |
| 23 | 41 | $^{64}$V | 82.0 (3) | | 571(58) $ns$ | 82.0 (3) | E2 | %IT ≈ 100 | 2021WI05<br>2014SU11♦ |

*Continued...*

*: Isomer in $^{60}$V is proposed at 4 keV in 2021WI05.

#: We have omitted a 13-$ns$ isomer proposed by 2010DA06, as the 103.2-keV $\gamma$ is reassigned in 2021WI05 to the 204-$ns$ isomer.

$: J$^{\pi}$ = (4+) in 2010DA06 is no longer valid according to level scheme in 2021WI05, where the 99- and 133-keV $\gamma$ rays are in parallel rather than in a cascade as given by 2012KA36.

‡: Energy is estimated by 2015GA38 based on several assumptions.

♦: 2014SU11 reports the half-life as < 1 $\mu s$.

Table 1 contd...

| Z | N | $^{A}$X | E(keV) | J$^{\pi}$ | $T_{1/2}$ | E$\gamma$(keV) | $\lambda$ | Decay mode | Reference |
|---|---|---|---|---|---|---|---|---|---|
| 25 | 39 | $^{64}$Mn | 175.5 (4) | (4+) | 400(49) $\mu s$ | 135.3(5) | (M2) | %IT = 100 | 2011LI50<br>2005GAZR<br>1998GR14 |
| 29 | 35 | $^{64}$Cu | 1594.23 (3) | 6− | 20.4(6) $ns$ | 1019.59(3) | | %IT = 100 | 1997IS13<br>1976CH36<br>1972BL16<br>1971SUZR |
| 31 | 33 | $^{64}$Ga | 42.85 (8) | (2+) | 21.9(7) $\mu s$ | 42.89(10) | (E2) | %IT = 100 | 1999TA29 |
| 26 | 39 | $^{65}$Fe | 393.7 (2) | (9/2+) | 1.12(15) $s$ | 393.7? | | %$\beta-\approx$ 100 | 2013OL06<br>2009PA16<br>2008BL05 |
| 26 | 39 | $^{65}$Fe | 397.56 (7) | (5/2+) | 419(13) $ns$ | 33.9(2) | (E1) | %IT = 100 | 2018ST18<br>2013OL06<br>2010DA06<br>2009PA16<br>1998GR14 |
| 28 | 37 | $^{65}$Ni | 63.37 (5) | 1/2− | 69(3) $\mu s$ | 63.36(5) | (E2) | %IT = 100 | 1978HO06<br>1978MU05 |
| 28 | 37 | $^{65}$Ni | 1017.01 (10) | 9/2+ | 25.6(11) $ns$ | 1017.0(1) | (M2) | %IT = 100 | 2005GE09<br>1997IS13<br>1995BL01<br>1994PA20 |
| 30 | 35 | $^{65}$Zn | 53.928 (10) | (1/2)− | 1.59(5) $\mu s$ | 53.93(1) | E2 | %IT = 100 | 2013RU10<br>1975RO25<br>1960AU03 |
| 25 | 41 | $^{66}$Mn | 464.5 (4) | (5−) | 780(40) $\mu s$ | 169.8(3) | (M2) | %IT = 100 | 2011LI50 |
| 27 | 39 | $^{66}$Co | 175.5 (1) | (3+) | 823(+22−21) $ns$ | 175.5 | E2 | %IT = 100 | 2018ST18<br>2012LI02<br>1998GR14 |
| 27 | 39 | $^{66}$Co | 642 (5) | (8−) | > 100 $\mu s$ | 252 | | %IT = 100 | 1998GR14 |
| 29 | 37 | $^{66}$Cu | 1154.2 (14) | (6)− | 600(20) $ns$ | 563.34 | | %IT = 100 | 2011LO01<br>1972BL16 |
| 31 | 35 | $^{66}$Ga | 43.812 (16) | 1+ | 18.0(9) $ns$ | 43.81(3) | M1 | %IT = 100 | 1975RO25 |
| 31 | 35 | $^{66}$Ga | 66.139 (19) | (2)+ | 23.0(14) $ns$ | 22.33(5) | M1+E2 | %IT = 100 | 1976LE03 |
| 31 | 35 | $^{66}$Ga | 162.472 (20) | (3)+ | 13(5) $ns$ | 53.39(20)<br>96.34(2)<br>118.80(20) | <br>(M1+E2)<br>E2(+M3) | %IT = 100 | 1978MO21 |
| 31 | 35 | $^{66}$Ga | 1464.33 (15) | 7+ | 57.3(14) $ns$ | 113.7(2)<br>600.9(2) | E2 | %IT = 100 | 2017BH05<br>1978FI03 |
| 33 | 33 | $^{66}$As | 1356.63 (17) | 5+ | 1.15(4) $\mu s$ | 124.4(2) | E2 | %IT = 100 | 2013RU10<br>2001GR07 |
| 33 | 33 | $^{66}$As | 3023.8 (3) | 9+ | 7.9(3) $\mu s$ | 114.4(2) | E2 | %IT = 100 | 2013RU10<br>2001GR07 |
| 26 | 41 | $^{67}$Fe | 388+X | | 64(14) $\mu s$ | 388(1) | (E2) | %IT = 100 | 2012KA36<br>2011DA08<br>2003SA02<br>1998GR14 |
| 27 | 40 | $^{67}$Co | 491.54 (10) | (1/2−) | 496(33) $ms$ | 491.55(11) | (M3) | %IT > 80<br>%$\beta-<$ 20 | 2008PA33 |

*Continued...*

Table 1 contd...

| Z | N | $^{A}X$ | E(keV) | J$^{\pi}$ | $T_{1/2}$ | E$\gamma$(keV) | $\lambda$ | Decay mode | Reference |
|---|---|---|---|---|---|---|---|---|---|
| 28 | 39 | $^{67}$Ni | 1006.6 (2) | 9/2+ | 13.3(2) $\mu s$ | 313.8(2) | (M2) | %IT = 100 | 2014DI08<br>2012ZH08<br>2002GE16<br>1998GR14 |
| 30 | 37 | $^{67}$Zn | 93.312 (5) | 1/2− | 9.142(24) $\mu s$* | 93.311(5) | E2 | %IT = 100 | 2015CH57<br>2014DI03<br>1998AT04<br>1996HW03<br>1975RO25<br>1973LE18 |
| 30 | 37 | $^{67}$Zn | 604.48 (5) | 9/2+ | 333(14) $ns$ | 604.48(5) | M2+E3 | %IT = 100 | 1973BE56 |
| 31 | 36 | $^{67}$Ga | 166.98 (3) | 1/2− | 42(21) $ns$ | 167.01(4) | | %IT = 100 | 1978AL32 |
| 32 | 35 | $^{67}$Ge | 18.20 (5) | 5/2− | 13.7(9) $\mu s$ | 18.20(5) | E2 | %IT = 100 | 1983MU15<br>1979AL04<br>1978MU05 |
| 32 | 35 | $^{67}$Ge | 751.70 (6) | 9/2+ | 105(3) $ns$ | 733.50(3) | M2 | %IT = 100 | 2000CH07<br>1979AL04<br>1978NA10 |
| 27 | 41 | $^{68}$Co | 0+X | (2−) | 1.6(3) $s$ | | | %$\beta$−= 100<br>%$\beta$− n ≥ 2.9(4) | 2015FL01<br>2000MU10 |
| 27 | 41 | $^{68}$Co | 48.4+Y | | 101(10) $ns$ | 48.4 | | %IT = 100 | 2010DA06 |
| 28 | 40 | $^{68}$Ni | 1603.5 (3) | 0+ | 270(6) $ns$ | 1604 | E0 | %IT = 100 | 2015FL01<br>2014SU05<br>2013RE18<br>2002SO03<br>1998GR14 |
| 28 | 40 | $^{68}$Ni | 2849.1 (3) | 5− | 0.85(4) $ms$ | 815.0(2) | E3 | %IT = 100 | 2015WI02<br>1998GR14<br>1995BR10 |
| 28 | 40 | $^{68}$Ni | 4210.0 (7) | 8+ | 23(1) $ns$ | 209.3<br>274.7<br>652.0 | E2 | %IT = 100 | 2012BR15<br>2000IS01 |
| 29 | 39 | $^{68}$Cu | 721.26 (8) | 6− | 3.75(5) $min$ | 110.74(6)<br>637.14(6) | E3<br>M4 | %IT = 84(1)<br>%$\beta$−= 16(1) | 2011RA42<br>2010VI07<br>1971SI19 |
| 31 | 37 | $^{68}$Ga | 1229.87 (4) | 7− | 62.0(14) $ns$ | 126.35(3)<br>733.76(34) | E2<br>(E3+M4) | %IT = 100 | 1997IS13<br>1978FI03<br>1973BAYF |
| 33 | 35 | $^{68}$As | 425.1 (2) | 1+ | 107(+23−16) $ns$ | 72.6(5)<br>111.4(2)<br>265.0(3)<br>426.1(4) | <br>E2<br><br>E2 | %IT = 100 | 1994BA50 |
| 33 | 35 | $^{68}$As | 1571.2 (2) | (6−) | 19(3) $ns$ | 248.4(2)<br>266.6(5) | | %IT = 100 | 1998SO23 |
| 33 | 35 | $^{68}$As | 2157.8 (2) | 9(+) | 36(2) $ns$ | 63.9(1)<br>854.2(3)<br>943.6(5)? | <br>M2 | %IT = 100 | 1998SO23 |
| 26 | 43 | $^{69}$Fe | 222 (14) | (9/2+) | 100-200 $ms$ | | | | 2022PO02 |

*Continued...*

*: Seven older references for half-lives from 1953-1965.

Table 1 contd…

| Z | N | $^{A}$X | E(keV) | J$^{\pi}$ | $T_{1/2}$ | E$\gamma$(keV) | $\lambda$ | Decay mode | Reference |
|---|---|---|---|---|---|---|---|---|---|
| 27 | 42 | $^{69}$Co | 182 (100) | (1/2−) | 750(250) *ms* | | | %$\beta$−= 100 | 2020CA08<br>2015LI33 |
| 28 | 41 | $^{69}$Ni | 321 (2) | (1/2−) | 3.5(4) *s* | | | %$\beta$−≈ 100<br>%IT < 0.01 | 1999MU17<br>1999PR10 |
| 28 | 41 | $^{69}$Ni | 2700.0 (10) | (17/2−) | 0.439(3) $\mu s$ | 148 | | %IT = 100 | 1998GR14 |
| 29 | 40 | $^{69}$Cu | 2742.0 (7) | (13/2+) | 357(2) *ns* | 74<br>189.9 | | %IT = 100 | 2016KU11<br>2012DI03<br>2002GE16<br>1998GR14<br>1997IS13 |
| 29 | 40 | $^{69}$Cu | 3692.0 (13) | (19/2−) | 22(1) *ns* | 208.8 | | %IT = 100 | 2000IS01 |
| 29 | 40 | $^{69}$Cu | 3828.0 (10) | (17/2+) | 39(6) *ns* | 613.6<br>1085.8 | | %IT = 100 | 2000IS01 |
| 30 | 39 | $^{69}$Zn | 438.636 (18) | 9/2+ | 13.747(14) *h* | 438.634(18) | M4 | %IT = 99.967(3)<br>%$\beta$−= 0.033(3) | 2017KR01<br>2017WR01<br>2006AB30<br>1977HE20<br>1974RO18<br>1970RA08 |
| 32 | 37 | $^{69}$Ge | 86.76 (2) | 1/2− | 5.1(2) $\mu s$ | 86.78(2) | E2 | %IT = 100 | 1970MU03<br>1964BR27 |
| 32 | 37 | $^{69}$Ge | 397.94 (2) | 9/2+ | 2.81(5) $\mu s$ | 397.95(2) | M2 | %IT = 100 | 1983FU21<br>1983SH47<br>1980KIZT<br>1970CH05 |
| 34 | 35 | $^{69}$Se | 38.85 (22) | 5/2− | 2.0(2) $\mu s$ | 39.4 | E2 | %IT = 100 | 1995PO01 |
| 34 | 35 | $^{69}$Se | 574.0 (4) | 9/2+ | 955(21) *ns* | 534.8(8) | M2 | %IT = 100 | 2000CH07<br>1995PO01 |
| 27 | 43 | $^{70}$Co$^{\dagger}$ | 0+X | | 112(7) *ms* | | | %$\beta$−= 100<br>%$\beta$–n = ? | 2015PR10<br>2011DA08<br>2003SO21<br>2003SA40<br>2000MU10<br>1998AM04 |
| | | | 0+Y | | 508(7) *ms* | | | %$\beta$−= 100 | 2017MO02<br>2015PR10<br>2003SA40<br>2000MU10 |
| 27 | 43 | $^{70}$Co | 437.3+X | (4−) | 54(10) *ns* | 155.8<br>164.1<br>273.2 | E1+M2<br><br>E2 | %IT ≈ 100 | 2010DA06<br>1999DAZQ |
| 28 | 42 | $^{70}$Ni | 2860.93 (7) | 8+ | 0.232(1) $\mu s$ | 183.11(2) | E2 | %IT = 100 | 1999LE68 |
| 29 | 41 | $^{70}$Cu | 101.1 (3) | 3− | 33(2) *s* | 101.1(3) | | %$\beta$−= 52(9)<br>%IT = 48(9) | 2020DE21<br>2010VI07<br>2004VA08<br>2004VA07<br>2002WE03 |
| 29 | 41 | $^{70}$Cu | 242.6 (5) | 1+ | 6.6(2) *s* | 141.3 | | %$\beta$−= 93.2(9)<br>%IT = 6.8(9) | 2020DE21<br>2010VI07<br>2004VA08<br>2004VA07 |

*Continued…*

Table 1 contd...

| Z | N | $^{A}X$ | E(keV) | $J^{\pi}$ | $T_{1/2}$ | Eγ(keV) | λ | Decay mode | Reference |
|---|---|---|---|---|---|---|---|---|---|
| 31 | 39 | $^{70}$Ga | 879.01 (13) | 4− | 22.7(5) *ns* | 188.1(1) | E2 | %IT = 100 | 1975HU06 |
| 33 | 37 | $^{70}$As | 32.046 (23) | 2+ | 96(3) *μs* | 32.05(3) | E2 | %IT = 100 | 1995PO03<br>1979TE06 |
| 35 | 35 | $^{70}$Br | 2292.3 (8) | 9+ | 2.157(51) *s* | | | %ϵ+%β+ = 100 | 2017MO18<br>2002RO05<br>1981VO04 |
| 28 | 43 | $^{71}$Ni | 498.5 (7) | (1/2−) | 2.3(3) *s* | | | %β−= 100 | 2012RA10<br>2009ST07 |
| 29 | 42 | $^{71}$Cu | 2755.7 (6) | (19/2−) | 0.271(14) *μs* | 133.0(3) | | %IT = 100 | 1998GR14<br>1998IS11 |
| 30 | 41 | $^{71}$Zn | 157.7 (13) | 9/2+ | 4.140(15) *h** | 386<br>487<br>620 | | %β−= 100<br>%IT ≤ 0.05 | 2017KR01<br>2012RE05 |
| 32 | 39 | $^{71}$Ge | 174.943 (9) | 5/2− | 79(2) *ns* | 174.956(9) | E2 | %IT = 100 | 1968MO12 |
| 32 | 39 | $^{71}$Ge | 198.354 (14) | 9/2+ | 20.22(12) *ms* | 23.438(15) | M2 | %IT = 100 | 2014DE19<br>1980JO11<br>1976GA33<br>1974BU14<br>1971GO21<br>1971MU14<br>1970RU08<br>1966ME02<br>1963AL32<br>1962RE09<br>1962MO19<br>1961MO06<br>1961SC11 |
| 33 | 38 | $^{71}$As | 143.50 (7) | (1/2)− | 59(10) *ns* | 143.2(3) | E2 | %IT = 100 | 1980TE01 |
| 33 | 38 | $^{71}$As | 1000.31 (13) | 9/2+ | 19.8(3) *ns* | 1000.26(14) | (M2) | %IT = 100 | 1971BEWR |
| 34 | 37 | $^{71}$Se | 48.79 (5) | 1/2− | 5.6(7) *μs* | 48.78(5) | E2 | %IT = 100 | 2012HO16<br>1982HA32 |
| 34 | 37 | $^{71}$Se | 260.48 (10) | 9/2+ | 19.0(5) *μs* | 260.5(1) | | %IT = 100 | 2012HO16<br>2000CH07 |
| 35 | 36 | $^{71}$Br | 759.06 (19) | (9/2+) | 32.5(25) *ns* | 89.3(4)<br>759.1(3) | (E2)<br>(M2) | %IT = 100 | 1990AR23 |
| 27 | 45 | $^{72}$Co† | 0+X | (6−, 7−) | 51.5(3) *ms* | | | %β−= 100 | 2016MO07<br>2014XU07<br>2014RA20<br>2005MA95 |
| | | | 0+Y | (0+, 1+) | 47.8(5) *ms* | | | %β−= 100 | 2016MO07<br>2005MA95 |
| 27 | 45 | $^{72}$Co# | X | (3+) | 0.18(10) *s* | | | %β−= ? | 2014RA20 |
| 29 | 43 | $^{72}$Cu | 137.32 (8) | (3−) | 17.6(7) *ns* | 137.4(1) | E1 | %IT = 100 | 2006TH12<br>2003STZX |
| 29 | 43 | $^{72}$Cu | 270.3 (10) | (6−) | 1.76(3) *μs* | 51 | E2 | %IT = 100 | 2006TH12<br>2003STZX |
| 31 | 41 | $^{72}$Ga | 16.44 (4 ) | 2− | 39.2(7) *ns* | 16.4(3) | (M1) | %IT = 100 | 1969KU08 |
| 31 | 41 | $^{72}$Ga | 119.66 (5) | (0+) | 39.68(13) *ms* | 103.14(17) | (M2) | %IT = 100 | 1972BR53 |

*Continued...*

*: Four earlier values from 1958-1964, mostly with large uncertainties or discrepant.

#: Questionable isomer.

Table 1 contd…

| Z | N | $^{A}$X | E(keV) | J$^{\pi}$ | $T_{1/2}$ | Eγ(keV) | λ | Decay mode | Reference |
|---|---|---|---|---|---|---|---|---|---|
| 32 | 40 | $^{72}$Ge | 691.43 (4) | 0+ | 444.2(8) *ns* | 689.6(5) | E0 | %IT = 100 | 1984BR24 |
| 33 | 39 | $^{72}$As | 46.025 (18) | 1+ | 10.7(3) *ns* | 45.89(4) | E1 | %IT = 100 | 1965HU02 |
| 33 | 39 | $^{72}$As | 213.563 (22) | 3+ | 85(5) *ns* | 166.77(5)<br>213.68(4) | E2<br>E1 | %IT = 100 | 1979TE06<br>1975BE32 |
| 33 | 39 | $^{72}$As | 309.741 (23) | 4− | 27(5) *ns* | 96.09(5)<br>309.77(3) | E1<br>E2 | %IT = 100 | 1996PA17<br>1979TE06<br>1976MA20 |
| 33 | 39 | $^{72}$As | 318.19 (4) | 4+ | 27(1) *ns* | 104.58(11)<br>318.39(25) | (M1) | %IT = 100 | 1979TE06 |
| 33 | 39 | $^{72}$As | 562.85 (8) | 7− | 87.9(17) *ns* | 200.02(9) | E2 | %IT = 100 | 1996PA17<br>1977RA03 |
| 34 | 38 | $^{72}$Se | 937.22 (15) | 0+ | 19.22(21) *ns* | 75.06(20)<br>937 | <br>E0 | %IT = 100 | 2022SM02<br>2011MC01<br>1975RA40<br>1974DR02<br>1974HA04 |
| 35 | 37 | $^{72}$Br | 101.2 (2) | 3− | 10.6(3) *s* | 101.3(3) | M2 | %ϵ+%β+ = ?<br>%IT ≈ 100 | 2022BR01<br>2015VA05<br>1982GA06<br>1980DAZO |
| 36 | 36 | $^{72}$Kr | 671.0 (10) | 0+ | 26.3(21) *ns* | 671 | E0 | %IT = 100 | 2005CLZZ |
| 30 | 43 | $^{73}$Zn | 195.5 (2) | 5/2+ | 13.0(2) *ms* | 195.5(2) | | %IT = 100 | 2017WR01<br>2017VE05<br>1998HU20<br>1985RU05 |
| 32 | 41 | $^{73}$Ge | 13.2845 (15) | 5/2+ | 2.91(3) *µs* | 13.2845(15) | E2 | %IT = 100 | 2002MO46<br>1993CO17<br>1978TA08<br>1971RA10<br>1971VE04<br>1970DO01<br>1970KY01 |
| 32 | 41 | $^{73}$Ge | 66.726 (9) | 1/2− | 0.499(11) *s* | 53.440(9) | M2 | %IT = 100 | 2008LI25<br>1974BU14<br>1970KY01 |
| 33 | 40 | $^{73}$As | 427.906 (21) | 9/2+ | 5.7(2) *µs* | 360.86(2)<br>428.3(3) | M2+E3 | %IT = 100 | 1977KEZY<br>1972REZN<br>1969IV02<br>1969MA21<br>1963BO16<br>1956HA10 |
| 34 | 39 | $^{73}$Se | 25.71 (4) | 3/2− | 39.8(13) *min* | 25.71(4) | E3 | %IT = 72.6(3)<br>%ϵ+%β+ = 27.4(3) | 1976BO19<br>1969MA21 |
| 35 | 38 | $^{73}$Br | 240.48 (14) | (3/2,<br>(5/2)− | 35.0(14) *ns* | 62.4(1)<br>213.5(2) | M1<br>E2 | %IT = 100 | 1987HE27 |
| 36 | 37 | $^{73}$Kr | 433.66 (12) | (9/2+) | 107(10) *ns* | 40.8<br>65.8 | | %IT = 100 | 2000CH07 |
| 31 | 43 | $^{74}$Ga | 56.550 (9) | (2−) | 31(5) *ns* | 56.559(10) | (M1(+E2)) | %IT ≈ 100 | 1977VA01 |
| 31 | 43 | $^{74}$Ga | 59.571 (14) | (0+) | 9.5(10) *s* | 3.2(2)<br>(59.7) | | %IT = 75(25)<br>%β−= 25(25) | 1974VA08 |

*Continued…*

Table 1 contd…

| Z | N | $^{A}$X | E(keV) | J$^{\pi}$ | $T_{1/2}$ | E$\gamma$(keV) | $\lambda$ | Decay mode | Reference |
|---|---|---|---|---|---|---|---|---|---|
| 33 | 41 | $^{74}$As | 259.187 (13) | 4+ | 26.8(5) *ns* | 76.13(1)<br>259.2(3)? | | %IT = 100 | 2014HU09<br>1971CH10 |
| 35 | 39 | $^{74}$Br | 13.58 (21) | 4(+) | 46(2) *min* | | | %$\epsilon$+%$\beta$+ = 100 | 1984MA35<br>1981GA11 |
| 35 | 39 | $^{74}$Br | 85.71 (21) | (3−) | 13.3(4) *ns* | 72.1(1) | | %IT = 100 | 1981WI05 |
| 36 | 38 | $^{74}$Kr | 509.25 (7) | 0+ | 13.4(7) *ns* | 53.23(14)<br>509.40(17) | E2<br>E0 | %IT = 100 | 2013DU14<br>2003BO05 |
| 27 | 48 | $^{75}$Co | 1914 (2) | (1/2−) | 13(6) $\mu s$ | 1914 | (M3) | | 2021ES05 |
| 29 | 46 | $^{75}$Cu | 61.7 (4) | 1/2− | 310(8) *ns* | 61.7(4) | | %IT = 100 | 2019IC02<br>2016PE14 |
| 29 | 46 | $^{75}$Cu | 66.2 (4) | 3/2− | 149(5) *ns* | 66.2(4) | | %IT = 100 | 2019IC02<br>2016PE14 |
| 32 | 43 | $^{75}$Ge | 139.69 (3) | 7/2+ | 47.7(5) *s* | 77.86(15)?<br>139.68(3) | <br>E3 | %IT = 99.970(6)<br>%$\beta$−= 0.030(6) | 1976BH04<br>1974CH22 |
| 32 | 43 | $^{75}$Ge | 192.19 (6) | 5/2+ | 216(5) *ns* | 52.5(1) | M1+E2 | %IT = 100 | 1982IS07 |
| 33 | 42 | $^{75}$As | 303.9243 (8) | 9/2+ | 17.62(23) *ms* | 24.38<br>303.9236(10) | M2(+E3)<br>E3 | %IT = 100 | 2014DE19<br>1998HW05<br>1994SM09<br>1984BR30<br>1980JO11<br>1969KU08<br>1961SC09<br>1957SC11 |
| 34 | 41 | $^{75}$Se | 293.106 (3) | 1/2− | 30.0(4) *ns* | 6.5(1) | | %IT = 100 | 1968RI14 |
| 35 | 40 | $^{75}$Br | 220.80 (7) | (9/2)+ | 31.7(3) *ns* | 88.29(6) | E2 | %IT = 100 | 1995MA97 |
| 27 | 49 | $^{76}$Co | 0+X | (8−) | 21.7(+69−49) *ms* | | | %$\beta$−= 100 | 2015SO23<br>2014XU07 |
| 27 | 49 | $^{76}$Co | 638.4 (4) | (3+) | 2.96(+29−25) $\mu s$ | 192.02(30) | | %IT ≈ 100 | 2015SO23 |
| 28 | 48 | $^{76}$Ni | 2418.00 (50) | (8+) | 547.8(33) *ns* | 142.56(25) | | %IT = 100 | 2015SO23<br>2014RA20<br>2012KA36<br>2005MA59 |
| 29 | 47 | $^{76}$Cu | 64.1 (16)* | (1,2) | 1.27(30) *s* | | | %$\beta$−= 100 | 2022GI08<br>2021GIAA<br>1990WI12 |
| 30 | 46 | $^{76}$Zn | 2634 | | 25.4(4) *ns* | 1337 | | %IT = 100 | 2021CH56 |
| 31 | 45 | $^{76}$Ga | 199 | 1+ | 34(8) *ns* | 199.2(5) | | | 2022CH09 |
| 33 | 43 | $^{76}$As | 44.425 (1) | (1)+# | 1.84(6) $\mu s$ | 44.425(1) | (E1) | %IT = 100 | 2015COZV#<br>1975RE06<br>1968HE15 |
| 35 | 41 | $^{76}$Br | 102.2 (6) | 4+ | 1.31(2) *s* | 57.11(2)<br>102.6? | M2 | %IT = 99.7(3)<br>%$\epsilon$+%$\beta$+ = 0.3(3) | 1997PA25<br>1980HA23 |
| 37 | 39 | $^{76}$Rb | 316.93 (8) | 4(+) | 3.050(7) $\mu s$ | 70.55(5) | | %IT = 100 | 2011WA12<br>2000CH07<br>1986HO22 |

*Continued…*

*: Energy from mass measurements reported by 2021GIAA (Ph.D. thesis). Private communication of July 19, 2022 with the first author of 2022GI08 suggested that the energy of this isomer is subject to some revision in their forthcoming article.

#: 2015COZV suggest 2+ from $\gamma$−$\gamma$-angular correlation data in (p, n$\gamma$), but details of this study are not yet available.

Table 1 contd…

| Z | N | $^{A}$X | E(keV) | J$^{\pi}$ | $T_{1/2}$ | E$\gamma$(keV) | $\lambda$ | Decay mode | Reference |
|---|---|---|---|---|---|---|---|---|---|
| 30 | 47 | $^{77}$Zn | 772.440 (15) | 1/2− | 1.05(10) *s* | 772.43(2) | | %$\beta$−= 66(7)<br>%IT = 34(7) | 2017WR01<br>2009PA35<br>2009IL01<br>1986EK01 |
| 32 | 45 | $^{77}$Ge | 159.71 (6) | 1/2− | 53.7(6) *s* | 159.66(10) | (E3) | %$\beta$−= 81(2)<br>%IT = 19(2) | 1974GR29<br>1970ME20<br>1970OSZZ<br>1969IM02<br>1968MA12<br>1965VA12<br>1957LY49<br>1954BU94<br>1947AR01 |
| 33 | 44 | $^{77}$As | 475.48 (4) | 9/2+ | 114.0(25) $\mu s$ | 211.03(4)<br>475.46(10) | (M2+E3) | %IT = 100 | 1980JO11 |
| 34 | 43 | $^{77}$Se | 161.9223 (10) | 7/2+ | 17.36(5) *s* | 161.9224(11) | E3 | %IT = 100 | 1986NE05<br>1980JO11<br>1967YU01 |
| 35 | 42 | $^{77}$Br | 105.86 (8) | 9/2+ | 4.28(10) *min* | 105.87(10) | E3 | %IT = 100 | 1980EK02<br>1961GO39 |
| 36 | 41 | $^{77}$Kr | 66.50 (5) | 3/2− | 118(12) *ns* | 66.52(5) | (E1) | %IT = 100 | 1975NO11 |
| 30 | 48 | $^{78}$Zn | 2675.3 (10) | (8+) | 0.320(+9−8) $\mu s$ | 144.7(5) | (E2) | %IT = 100 | 2012KA36<br>2000DA07 |
| 31 | 47 | $^{78}$Ga | 498.9 (5) | | 110(3) *ns* | 498.9 | | %IT ≈ 100 | 2010DA06 |
| 35 | 43 | $^{78}$Br | 32.32 (8) | (2−) | 11.3(30) *ns* | 32.3(1) | | %IT = 100 | 1979KL05<br>1972CH34 |
| 35 | 43 | $^{78}$Br | 180.89 (13) | (4+) | 119.4(10) $\mu s$ | 148.55(10) | | %IT = 100 | 1982BE03<br>1970DE46<br>1969RU10<br>1968IO01<br>1967IV03<br>1961SC11<br>1958DU80 |
| 35 | 43 | $^{78}$Br | 227.67 (18) | (5+) | 84(8) *ns* | 46.8(2) | (M1+E2) | %IT = 100 | 1982BE03 |
| 35 | 43 | $^{78}$Br | 242.82 (17) | (2, 3)− | 17(2) *ns* | 46.3(5)<br>242.85(2) | | %IT = 100 | 1982BE03 |
| 37 | 41 | $^{78}$Rb | 46.84 (14) | (1−) | 0.91(4) $\mu s$ | 46.8(2) | (E1) | %IT = 100 | 1997MU02<br>1996KA24 |
| 37 | 41 | $^{78}$Rb | 111.19 (22) | 4(−) | 5.74(3) *min* | 8.6<br>64.4 | <br>(M3) | %IT = 9(2)<br>%$\epsilon$+%$\beta$+ = 91(2) | 1981BA40<br>1979HE18<br>1974SA32<br>1973BA03<br>1972NO14<br>1972DE54<br>1969CH18<br>1968TO05 |
| 39 | 39 | $^{78}$Y | 0+X | (5+) | 5.8(6) *s* | | | %$\epsilon$+%$\beta$+ = 100<br>%$\epsilon$p = ? | 2007WEZX<br>2002FA13<br>2001KI13<br>1998UU01 |

*Continued…*

Table 1 contd...

| Z | N | $^{A}X$ | E(keV) | J$^{\pi}$ | $T_{1/2}$ | E$\gamma$(keV) | $\lambda$ | Decay mode | Reference |
|---|---|---|---|---|---|---|---|---|---|
| 30 | 49 | $^{79}$Zn | 942 (11)* | 1/2+ | ≥ 200 *ms* | | | %β−= ?<br>%IT = ? | 2022GI08<br>2021GIAA<br>2017WR01<br>2016YA02<br>2015OR01 |
| 32 | 47 | $^{79}$Ge | 185.95 (4) | (7/2+) | 39.0(10) *s* | 186.02(7) | | %IT = 4(1)<br>%β−= 96(1) | 1982FOZZ<br>1981HO24<br>1974KRZG<br>1972DE43<br>1970VA31<br>1970KA04 |
| 33 | 46 | $^{79}$As | 772.81 (6) | (9/2)+ | 1.21(1) *μs* | 542.27(7) | | %IT = 100 | 2013RUZX<br>1998GR14<br>1998HO15 |
| 34 | 45 | $^{79}$Se | 95.77 (3) | 1/2− | 3.885(9) *min* | 95.73(3) | E3 | %IT = 99.944(11)<br>%β−= 0.056(11) | 2019DE24<br>1990AB02<br>1988KL03<br>1969MU03<br>1954YT03<br>1953CU33<br>1952RU10<br>1950FL62 |
| 35 | 44 | $^{79}$Br | 207.61 (9) | 9/2+ | 4.85(4) *s* | 207.5(1) | E3 | %IT = 100 | 2009MU15<br>1986AL11<br>1972JO05<br>1968BO52<br>1967YU01<br>1967SC14<br>1963KA34<br>1962AN13<br>1961GO39<br>1954SC37 |
| 36 | 43 | $^{79}$Kr | 129.77 (5) | 7/2+ | 50(3) *s* | 129.76(10) | E3 | %IT = 100 | 1969HA03<br>1940CR06 |
| 36 | 43 | $^{79}$Kr | 147.06 (6) | 5/2− | 78.7(10) *ns* | 17.3(1)<br>147.08(10) | <br>E2 | %IT = 100 | 1978LI28<br>1975BU10<br>1972BR31<br>1968BL04 |
| 37 | 42 | $^{79}$Rb | 39.37 (5) | (3/2−) | 20.5(25) *ns* | 39.41(7) | (E1) | %IT = 100 | 1982DE36<br>1981LI12 |
| 37 | 42 | $^{79}$Rb | 96.76 (7) | 9/2+ | 18.6(5) *ns* | 96.7(1) | | %IT = 100 | 1994IO02<br>1982PA20 |
| 38 | 41 | $^{79}$Sr | 177.29 (6) | (5/2+) | 21(1) *ns* | 177.29(6) | (E1(+M2)) | %IT = 100 | 1992MU12<br>1990CH07<br>1973BOXS |
| 31 | 49 | $^{80}$Ga | 22.45 (10) | 3(−) | 1.3(2) *s* | | | %β−= 100<br>%IT = ? | 2014LI32<br>2013VE03 |

*Continued...*

*: Energy from mass measurements reported by 2021GIAA (Ph.D. thesis). Private communication of July 19, 2022 with the first author of 2022GI08 suggested that the energy of this isomer is subject to some revision in their forthcoming article.

Table 1 contd...

| Z | N | $^{A}X$ | E(keV) | J$^{\pi}$ | $T_{1/2}$ | E$\gamma$(keV) | $\lambda$ | Decay mode | Reference |
|---|---|---|---|---|---|---|---|---|---|
| 31 | 49 | $^{80}$Ga | 707.90 (11) | (1+) | 18.3(5) *ns* | 685.4(1) | (M2) | %IT = 100 | 2014LI32 |
| 35 | 45 | $^{80}$Br | 85.843 (4) | 5− | 4.4205(8) *h* | 48.786(5) | M3 | %IT = 100 | 2011WA21<br>1990AB06 |
| 37 | 43 | $^{80}$Rb | 493.9 (5) | 6+ | 1.63(4) $\mu s$ | (8.0)<br>(21.4) | | | 2013HE08<br>1996IO01 |
| 37 | 43 | $^{80}$Rb | 650.7 (6) | (8+) | 13.9(14) *ns* | 156.8 | | %IT = 100 | 1992DO10 |
| 39 | 41 | $^{80}$Y | 228.5 (1) | (1−) | 4.8(3) *s* | 228.5(1) | M3(+E4) | %$\epsilon$+%$\beta$+ = 19(2)<br>%IT = 81(2) | 2001NO07<br>2000DO10<br>1998DO04 |
| 39 | 41 | $^{80}$Y | 312.6 (9) | (2+) | 4.7(3) $\mu s$ | 84 | | %IT = 100 | 2000CH07 |
| 32 | 49 | $^{81}$Ge | 679.14 (4) | (1/2+) | 7.6(6) *s** | | | %$\beta$−= 99.5(5)<br>%IT = 0.5(5) | 1981HO24 |
| 34 | 47 | $^{81}$Se | 102.966 (10) | 7/2+ | 57.28(2) *min* | 102.966(10) | E3(+M4) | %IT = 99.949(14)<br>%$\beta$−= 0.051(14) | 2015KR02<br>2009PO04<br>1989AB18 |
| 35 | 46 | $^{81}$Br | 536.20 (9) | 9/2+ | 34.6(28) $\mu s$ | 260.21(9) | M2 | %IT = 100 | 1996JA09<br>1971CH28<br>1968IV02<br>1967IV03 |
| 36 | 45 | $^{81}$Kr | 190.64 (4) | 1/2− | 13.10(3) *s* | 190.46(16) | E3 | %IT = 99.9975(4)<br>%$\epsilon$ = 2.5E−3(4) | 1987LO06<br>1987DA06 |
| 37 | 44 | $^{81}$Rb | 86.31 (7) | 9/2+ | 30.5(3) *min* | 86.26(19) | E3 | %IT = 97.6(6)<br>%$\epsilon$+%$\beta$+ = 2.4(6) | 1981FRZY<br>1980HO28<br>1977LI14<br>1956DO52 |
| 38 | 43 | $^{81}$Sr | 79.23 (4) | (5/2)− | 0.39(5) $\mu s$ | 79.23(4) | (E2) | %IT = 100 | 1989WU01<br>1985LI12<br>1983AR16<br>1982DE36 |
| 38 | 43 | $^{81}$Sr | 89.05 (7) | (7/2+) | 6.4(5) $\mu s$ | | (E1) | %IT = 100 | 1989WU01 |
| 38 | 43 | $^{81}$Sr | 119.76 (4) | (1/2+) | 24(4) *ns* | 119.76(4) | (E1) | %IT = 100 | 1983AR16 |
| 31 | 51 | $^{82}$Ga | 140.7 (3) | (4−) | 94(9) $\mu s$ | 140.7(3) | | %IT = 100 | 2016AL10<br>2012KA36 |
| 33 | 49 | $^{82}$As | 131.6 (5) | (5−) | 13.6(5) *s* | | | %$\beta$−= 100 | 2014MI16<br>2008HA23<br>1975KR08<br>1970KA04<br>1970VA31 |
| 35 | 47 | $^{82}$Br | 45.9492 (10) | 2− | 6.13(5) *min* | 45.949(1) | M3 | %IT = 97.6(3)<br>%$\beta$−= 2.4(3) | 1965AN01<br>1965EM02 |
| 37 | 45 | $^{82}$Rb | 69.0 (15) | 5− | 6.472(6) *h* | | | %$\epsilon$+%$\beta$+ = 100<br>%IT < 0.33 | 1982GR07<br>1981TH04 |
| 37 | 45 | $^{82}$Rb | 192.2 (16) | 6+ | 12.3(6) *ns* | 123.2(3) | E1 | %IT = 100 | 1999IO01<br>1996IO01<br>1991DO05 |
| 39 | 43 | $^{82}$Y | 401.10 (10) | 4+ | 11(3) *ns* | 64.3(3)<br>87.5(1)<br>258.7(1) | | %IT = 100 | 1993WO04 |

*Continued...*

*: 1981HO24 assigned the same half-life of 7.6(6) *s* for the g.s. and the isomer, as the two activities could not be separated in time in their experiment. 2022DE07 have measured the half-life of 6.4(2) *s* for only the ground state.

Table 1 contd...

| Z | N | $^{A}X$ | E(keV) | J$^{\pi}$ | $T_{1/2}$ | E$\gamma$(keV) | $\lambda$ | Decay mode | Reference |
|---|---|---|---|---|---|---|---|---|---|
| 39 | 43 | $^{82}$Y | 402.63 (14) | 4− | 268(25) *ns* | 65.7(2)<br>88.9(2) | | %IT = 100 | 1993WO04 |
| 39 | 43 | $^{82}$Y | 405.76 (24) | 4− | 35.3(21) *ns* | 68.2(3) | | %IT = 100 | 1993WO04 |
| 39 | 43 | $^{82}$Y | 507.50 (13) | 6+ | 147(7) *ns* | 106.4(1) | E2 | %IT = 100 | 1993WO04 |
| 41 | 41 | $^{82}$Nb | 1180 | (5+) | 92(17) *ns* | 124 | (M1+E2) | %IT = 100 | 2009GA40<br>2008GA04 |
| 34 | 49 | $^{83}$Se | 228.92 (7) | 1/2− | 70.1(4) *s* | | | %$\beta$−= 100 | 1974KR27<br>1974KRZG<br>1969PH03 |
| 35 | 48 | $^{83}$Br | 3069.2 (4) | (19/2−) | 0.7(1) $\mu s$ | 303.4(4) | E1(+M2) | %IT = 100 | 2013RUZX<br>1997IS13<br>1989WI01 |
| 36 | 47 | $^{83}$Kr | 9.4057 (6) | 7/2+ | 156.8(5) *ns* | 9.4057(6) | M1+E2 | %IT = 100 | 2009KA30<br>1995AH04<br>1963RU03 |
| 36 | 47 | $^{83}$Kr | 41.5575 (7) | 1/2− | 1.83(2) *h* | 32.1516(5) | E3 | %IT = 100 | 2022FE02<br>2010LI13<br>2009KA30<br>1971RU17 |
| 37 | 46 | $^{83}$Rb | 5.2357 (8) | 3/2− | 71.5(8) *ns* | 5.2357(8) | M1+E2 | %IT = 100 | 1972MO16 |
| 37 | 46 | $^{83}$Rb | 42.0780 (20) | 9/2+ | >7.8 *ms** | 42.078(2) | M2 | %IT = 100 | 1967MOZZ |
| 38 | 45 | $^{83}$Sr | 259.15 (9) | 1/2− | 4.95(12) *s* | 259.1(1) | E3 | %IT = 100 | 1972TU07 |
| 39 | 44 | $^{83}$Y | 62.04 (10) | 3/2− | 2.85(2) *min* | (62.1(3)) | E3 | %$\epsilon$+%$\beta$+ = 60(5)<br>%IT = 40(5) | 1987RA06<br>1972TU07 |
| 40 | 43 | $^{83}$Zr | 52.72 (5) | (5/2−) | 0.53(12) $\mu s$ | 52.70(5) | E2 | %IT = 100 | 1988HU01 |
| 40 | 43 | $^{83}$Zr | 77.04 (7) | (7/2+) | 1.8(1) $\mu s$ | 24.30(5) | E1 | %IT = 100 | 1988SU15<br>1988HU01<br>1988KU14 |
| 35 | 49 | $^{84}$Br | 3.2E+2 (10) | (6)− | 6.0(2) *min* | | | %$\beta$−= 100 | 1970HA21<br>1960SA05 |
| 36 | 48 | $^{84}$Kr | 3236.07 (18) | 8+ | 1.83(4) $\mu s$ | 63.5(1) | E2 | %IT = 100 | 2006SC22<br>1982ZA04 |
| 36 | 48 | $^{84}$Kr | 5373.4 (4) | 12+ | 43.7(21) *ns* | 169.3 | E2 | %IT = 100 | 2006SC22<br>1990RO10 |
| 37 | 47 | $^{84}$Rb | 463.59 (8) | 6− | 20.26(4) *min* | 215.61(10)<br>463.62(10) | M3+E4<br>E4 | %IT = 100 | 2010SH12<br>1982GR07 |
| 37 | 47 | $^{84}$Rb | 543.28 (12) | (5+) | 11(1) *ns* | 70.7(2)<br>76.4(2)<br>79.8(1) | | %IT = 100 | 1991DO04 |
| 39 | 45 | $^{84}$Y | 67.0 (2) | 1+ | 4.6(2) *s* | | | %$\epsilon$+%$\beta$+ = 100 | 1976IA01 |
| 39 | 45 | $^{84}$Y | 112.40 (15) | (4+) | 79(2) *ns* | 112.4(2) | E2 | %IT = 100 | 2005IO02 |
| 39 | 45 | $^{84}$Y | 156.70 (9) | (8+) | 14.6(7) *ns* | 156.7(1) | E2 | %IT = 100 | 1994CH01 |
| 39 | 45 | $^{84}$Y | 162.88 (10) | (5−) | 32(4) *ns* | 162.9(1) | (E1) | %IT = 100 | 1994CH01 |
| 39 | 45 | $^{84}$Y | 210.42 (16) | (4−) | 292(10) *ns* | 61.7(2)<br>80.0(2)<br>98.1(2) | (E1)<br>E2<br>(E1) | %IT = 100 | 2005IO02 |
| 39 | 45 | $^{84}$Y | 216.11 (9) | (5−) | 18.7(21) *ns* | 216.1(1) | (E1) | %IT = 100 | 1994CH01 |
| 41 | 43 | $^{84}$Nb | 48.0 (3) | (3+) | 176(46) *ns* | 47.9 | E2 | %IT = 100 | 2009GA40 |

*Continued...*

*: Uncertainty of 0.7 *ms* seems estimated by 1968ET01 from measurement by 1967MOZZ.

Table 1 contd…

| Z | N | $^{A}X$ | E(keV) | J$^{\pi}$ | $T_{1/2}$ | E$\gamma$(keV) | $\lambda$ | Decay mode | Reference |
|---|---|---|---|---|---|---|---|---|---|
| 41 | 43 | $^{84}$Nb | 305.2 (4) | (5+) | 50(8) *ns* | 143.2<br>257.2 | M1<br>E2 | %IT = 100 | 2009GA40 |
| 41 | 43 | $^{84}$Nb | 337.0 (3) | (5−) | 92(5) *ns* | 132.4<br>174.8 | E2<br>E1 | %IT = 100 | 2009GA40<br>2000CH07 |
| 36 | 49 | $^{85}$Kr | 304.871 (20) | 1/2− | 4.480(8) *h* | 304.87(2) | M4 | %IT = 21.2(5)<br>%$\beta$−= 78.8(5) | 1970WO08<br>1949KO13<br>1948WO07<br>1937SN02 |
| 36 | 49 | $^{85}$Kr | 1991.8 (2) | (17/2+) | 1.82(6) $\mu s$ | 60.2(2) | E2 | %IT = 100 | 2013RUZX<br>1989WI01 |
| 37 | 48 | $^{85}$Rb | 514.0065 (22) | 9/2+ | 1.015(1) $\mu s$ | 233?<br>362.81(4)<br>514.0048(22) | <br>(E3)<br>M2 | %IT = 100 | 2019TA19<br>1972MI23<br>1971DEZB<br>1964LO02 |
| 37 | 48 | $^{85}$Rb | 2826.55 (14) | (19/2)− | 12.5(6) *ns* | 349.8(1) | E1 | %IT = 100 | 1989WI01 |
| 38 | 47 | $^{85}$Sr | 238.79 (5) | 1/2− | 67.63(4) *min* | 238.78(5) | M4 | %IT = 86.6(4)<br>%$\epsilon$ = 13.4(4) | 2021HA10<br>1982GR07<br>1972EM01<br>1971BU08<br>1970LYZZ<br>1966KA24<br>1964GU08<br>1940DU05 |
| 39 | 46 | $^{85}$Y | 19.68 (17) | (9/2)+ | 4.86(20) *h* | | | %$\epsilon$+%$\beta$+ = 100<br>%IT < 0.002 | 1966HO04<br>1965NI02<br>1963DO07 |
| 39 | 46 | $^{85}$Y | 266.18 (10) | (5/2)− | 178(7) *ns* | 266.2(1) | E2 | %IT = 100 | 1982RAZY<br>1977IA01 |
| 40 | 45 | $^{85}$Zr | 292.2 (3) | (1/2−) | 10.9(3) *s* | 292.2(3) | | %$\epsilon$+%$\beta$+ > 0.0<br>%IT < 100 | 2005KA39<br>1976IA01 |
| 41 | 44 | $^{85}$Nb | 69+Y | (1/2−,<br>3/2−) | 3.3(9) *s* | 69 | (E2, M2) | %$\epsilon$+%$\beta$+ = ?<br>%IT = ? | 2005KA39 |
| 37 | 49 | $^{86}$Rb | 556.05 (18) | 6− | 1.017(3) *min* | 556.07(18) | E4 | %IT = 100<br>%$\beta$−< 0.3 | 1981TH04<br>1967YU01 |
| 38 | 48 | $^{86}$Sr | 2956.09 (12) | 8+ | 0.455(7) $\mu s$ | 98.68(3) | E2 | %IT = 100 | 1997IS13<br>1978HA52<br>1975MA02<br>1971IS04 |
| 39 | 47 | $^{86}$Y | 208.04 (7) | (5)− | 70(7) *ns* | 208.06(7) | E2(+M1) | %IT=100 | 2010RU07 |
| 39 | 47 | $^{86}$Y | 218.21 (9) | (8+) | 47.4(4) *min* | 10.22 (8) | (E3) | %IT = 99.31(4)<br>%$\epsilon$+%$\beta$+ = 0.69(4) | 2010RU07<br>1972SI11<br>1962KI04<br>1961HA17 |
| 39 | 47 | $^{86}$Y | 242.80 (10) | 2− | 28.6(21) *ns* | 242.80(10) | E2 | %IT = 100 | 2010RU07<br>1968TR11 |
| 39 | 47 | $^{86}$Y | 302.18 (9) | (6+) | 127(14) *ns* | 84.0(1)<br>94.11(7) | <br>(E1) | %IT = 100 | 2010RU07 |
| 41 | 45 | $^{86}$Nb | 49.96+X | (1+,<br>2+) | 68(2) *ns* | 49.95(15) | (E1) | %IT = 100 | 1994SH07 |

*Continued…*

Table 1 contd…

| Z | N | $^{A}$X | E(keV) | J$^{\pi}$ | $T_{1/2}$ | E$\gamma$(keV) | $\lambda$ | Decay mode | Reference |
|---|---|---|---|---|---|---|---|---|---|
| 43 | 43 | $^{86}$Tc | 1524 (10) | (6+) | 1.10(14) $\mu s$ | 81 | (E2) | %IT = 100 | 2009GA40<br>2008GA04<br>2000CH07 |
| 38 | 49 | $^{87}$Sr | 388.5287 (23) | 1/2− | 2.805(1) $h$* | 388.5276(23) | M4 | %IT = 99.70(8)<br>%$\epsilon$ = 0.30(8) | 2021KR05<br>1992AN19<br>1970LE07<br>1968GO30 |
| 39 | 48 | $^{87}$Y | 380.82 (7) | 9/2+ | 13.37(3) $h$ | 380.79(7) | M4 | %IT = 98.43(11)<br>%$\epsilon$+%$\beta$+ = 1.57(11) | 1984PR01<br>1975RU06 |
| 40 | 47 | $^{87}$Zr | 335.84 (19) | 1/2− | 14.0(2) $s$ | 134.93(15) | E3 | %IT = 100 | 2002FO12<br>1972TU03 |
| 41 | 46 | $^{87}$Nb | 3.9 (1) | (9/2)+ | 2.6(1) $min$ | | | %$\epsilon$+%$\beta$+ = 100 | 1974VO03<br>1972TU03 |
| 41 | 46 | $^{87}$Nb | 334.0 (1) | (5/2−) | 28(2) $ns$ | 67.0(3)<br>334.0(1) | <br>(E2) | %IT = 100 | 1995KA06 |
| 43 | 44 | $^{87}$Tc | 71+X | (7/2+) | 647(24) $ns$ | 64?<br>71? | | %IT = 100 | 2009GA40 |
| 35 | 53 | $^{88}$Br | 270.1 (5) | (4−) | 5.50(4) $\mu s$ | 110.9(5) | E2 | %IT = 100 | 2018RZ01<br>2017CZ01<br>2015CZ01<br>2013RUZX<br>2009FO05<br>1999GE01<br>1976SEZN<br>1970GR38 |
| 37 | 51 | $^{88}$Rb | 1373.8 (3) | (7+) | 123(13) $ns$ | 647.2(3)<br>1105.9(7) | (M2) | %IT = 100 | 2009PO10 |
| 39 | 49 | $^{88}$Y | 392.86 (9) | 1+ | 0.304(3) $ms$ | 392.87(9) | E3 | %IT = 100 | 2014DE19<br>1974BA88<br>1967IV04<br>1959LI44<br>1955HY29 |
| 39 | 49 | $^{88}$Y | 674.55 (4) | 8+ | 14.0(4) $ms$ | 442.62(3) | E3 | %IT = 100 | 2014DE19<br>1976GA33<br>1975VA16<br>1974BA06<br>1974BA88<br>1974FI18<br>1967IV04<br>1966ME02<br>1962MO19 |
| 40 | 48 | $^{88}$Zr | 2887.79 (6) | 8+ | 1.337(25) $\mu s$ | 76.99(1) | E2 | %IT = 100 | 2019HA26<br>2017PA35<br>2004CH35<br>1978HA52<br>1978KI06 |
| 41 | 47 | $^{88}$Nb | 0+X | (4−) | 7.7(1) $min$ | | | %$\epsilon$+%$\beta$+ = 100 | 1984OX01<br>1972IA01 |

*Continued…*

*: Precise but discrepant half-life of 2.827(1) $h$ in 1992AN19 is not used in averaging.

Table 1 contd...

| Z | N | $^{A}$X | E(keV) | J$^{\pi}$ | $T_{1/2}$ | E$\gamma$(keV) | $\lambda$ | Decay mode | Reference |
|---|---|---|---|---|---|---|---|---|---|
| 43 | 45 | $^{88}$Tc | 70.4 (31)* | (6+) | 6.4(8) *s** | | | %$\epsilon$+%$\beta$+ = 100<br>%$\epsilon$p = ? | 2019VI05*<br>1996OD01 |
| 43 | 45 | $^{88}$Tc | 95* | (4+) | 146(12) *ns** | 95 | | %IT = 100 | 2019VI05*<br>2009GA40 |
| 36 | 53 | $^{89}$Kr | 28.59 (3) | (5/2+) | 21.7(13) *ns* | 28.51(10) | | %IT = 100 | 1987KA02<br>1974CLZX |
| 37 | 52 | $^{89}$Rb | 1195.48 (21) | 9/2(+) | 10(3) *ns* | 198.10(20)<br>264.10(20)<br>974.6(3) | (E1)<br><br>(M2) | %IT = 100 | 2019TO10<br>2009PA20 |
| 39 | 50 | $^{89}$Y | 908.97 (3) | 9/2+ | 15.663(5) *s* | 908.960(25) | M4+E5 | %IT = 100 | 1995ITZY<br>1968BO52<br>1967YU01<br>1966DU07<br>1962BR42<br>1955SW92<br>1951GO42 |
| 40 | 49 | $^{89}$Zr | 587.82 (10) | 1/2− | 4.161(10) *min* | 587.8(1) | (M4) | %IT = 93.77(12)<br>%$\epsilon$+%$\beta$+ = 6.23(12) | 1992KAZM<br>1969RO02<br>1964VA03<br>1953SH48<br>1953KA11<br>1951SH89<br>1940DU05 |
| 41 | 48 | $^{89}$Nb | <35 | (1/2)− | 66(2) *min* | | | %$\epsilon$+%$\beta$+ = 100 | 1975HAYQ<br>1971AR16<br>1969HAZP<br>1966HA45<br>1964BU11<br>1954DI16 |
| 41 | 48 | $^{89}$Nb | 2192.92 (16) | (21/2+) | 13.8(4) *ns* | 257.7(1) | E2 | %IT = 100 | 1995KA06<br>1994KR01<br>1982DI09<br>1977SP03 |
| 42 | 47 | $^{89}$Mo | 387.5 (2) | (1/2−) | 190(15) *ms* | 268.6(2) | (E3) | %IT = 100 | 1993WE04<br>1980GA16 |
| 43 | 46 | $^{89}$Tc | 62.6 (5) | (1/2−) | 12.9(8) *s* | | | %$\epsilon$+%$\beta$+ = 100<br>%IT < 0.01 | 1995RU03<br>1991HE04<br>1983OXZZ |
| 33 | 57 | $^{90}$As | 124.5+X | | 0.20(+12−9) $\mu s$ | 124.5(5) | | %IT = 100 | 2012KA36 |

*Continued...*

*: 2019VI05 extracted precise energy difference between the two lowest states in $^{88}$Tc from their mass measurements, and considered four different scenarios for the ordering of the (2+), (6+) and a low-lying (4+), together with shell-model calculations using five different interactions, and concluded that (2+) was most likely the g.s., while the (6+) was the long-lived isomeric state. Then based on the Coulomb excitation experiment of 2009GA40, 95-keV gamma can proceed to the g.s., defining the isomer with (4+), 146 *ns* half-life and energy of 95 keV. However, 2019VI05 also noted that this ordering was contrary to the expected production yields of the two activities in their experiment, where high-spin activity should be populated more strongly than the low-spin. In shell-model calculations by 1997HE24, 2+ was predicted as the ground state, with predicted half-life of 10.9 *s*. Considering the available experimental and theoretical results, it is our opinion, that the ordering of the g.s. and the isomer in terms of their spins remains unsettled.

Table 1 contd…

| Z | N | $^{A}X$ | E(keV) | J$^{\pi}$ | $T_{1/2}$ | E$\gamma$(keV) | $\lambda$ | Decay mode | Reference |
|---|---|---|---|---|---|---|---|---|---|
| 37 | 53 | $^{90}$Rb | 106.90 (3) | 3− | 258(4) $ns$ | 106.92(15) | (M3) | %$\beta$−= 97.4(4)<br>%IT = 2.6(4) | 1981TA05<br>1979EK02<br>1977HU03<br>1967AM01 |
| 39 | 51 | $^{90}$Y | 681.67 (10) | 7+ | 3.196(12) $h$* | 479.532(10)<br>682.03(4) | M4(+E5)<br>E5 | %IT = 99.9982(2)<br>%$\beta$ = 0.0018(2) | 2020KR06<br>2014LI50<br>1992AN19<br>1967GR02<br>1962AB03<br>1961CA12<br>1961HE09 |
| 40 | 50 | $^{90}$Zr | 1760.74 (14) | 0+ | 61.3(25) $ns$ | 1760.70(20) | E0 | %IT = 100 | 1972BU18 |
| 40 | 50 | $^{90}$Zr | 2319.000 (9) | 5− | 809.2(20) $ms$ | 132.716(18)<br>2318.959(25) | E3(+M4)<br>E5 | %IT = 100 | 1972BR53 |
| 40 | 50 | $^{90}$Zr | 3589.418 (15) | 8+ | 131(4) $ns$ | 141.178(15)<br>1270.396(18) | E2(+M3)<br>E3 | %IT = 100 | 1977HA49<br>1964LO02 |
| 41 | 49 | $^{90}$Nb | 122.370 (22) | 6+ | 63(2) $\mu s$ | 122.370(22) | E2 | %IT = 100 | 1963GH01<br>1967IV04<br>1971HO27<br>1978BA18 |
| 41 | 49 | $^{90}$Nb | 124.67 (25) | 4− | 18.88(3) $s$ | | | %IT = 100 | 2011KI45<br>1978ME03<br>1974CO33<br>1971SM07<br>1969GE03<br>1968WE03 |
| 41 | 49 | $^{90}$Nb | 382.01 (25) | 1+ | 6.22(8) $ms$ | 257.34(4) | E3(+M4) | %IT = 100 | 1977DEXV<br>1974HA50<br>1974CO33<br>1967IV04 |
| 41 | 49 | $^{90}$Nb | 1880.21 (20) | 11− | 0.458(12) $\mu s$ | 71.1(2)<br>1067.3(2) | E2<br>M2 | %IT = 100 | 2019HA26<br>2017PA35<br>2005CH65<br>1981FI02 |
| 42 | 48 | $^{90}$Mo | 2874.73 (15) | 8+ | 1.14(5) $\mu s$ | 63.15(9) | | %IT = 100 | 2019HA26<br>2017PA35<br>1978HA52<br>1971IS04 |
| 43 | 47 | $^{90}$Tc | 144.1 (17) | (1+) | 8.7(2) $s$ | | | %$\epsilon$+%$\beta$+ = 100 | 2012KA11<br>1981OX01 |
| 45 | 45 | $^{90}$Rh | 0+X | (7+) | 0.56(2) $s$ | | | %$\epsilon$+%$\beta$+ ≈ 100<br>%$\beta$+p = 9.6(10) | 2019PA16<br>2002FA13<br>2001KI13 |
| 36 | 55 | $^{91}$Kr | 144.3 (6) | (3/2+) | 56 $ns$ | 144.1 | E2 | %IT = 100 | 1995KE04<br>1990WOZZ |

*Continued…*

*: Precise but discrepant half-life of 3.244(5) $h$ in 1992AN19 is not used in averaging.

Table 1 contd...

| Z | N | $^{A}$X | E(keV) | J$^{\pi}$ | $T_{1/2}$ | E$\gamma$(keV) | $\lambda$ | Decay mode | Reference |
|---|---|---|---|---|---|---|---|---|---|
| 37 | 54 | $^{91}$Rb | 1133.79 (6) | (9/2+) | 16.6(6) *ns* | 141.5(10)?<br>412.04(8)<br>1024.91(15)<br>1134.1 | | %IT = 100 | 2019TO10<br>2010SI17<br>2009HW03<br>2009PA20<br>1986SI20 |
| 38 | 53 | $^{91}$Sr | 93.628 (4) | (3/2)+ | 89.4(16) *ns* | 93.628(4) | E2(+M1) | %IT = 100 | 1993WO07<br>1976GL02<br>1970MA53<br>1970MA07 |
| 39 | 52 | $^{91}$Y | 555.58 (5) | 9/2+ | 49.71(4) *min* | 555.57(5) | M4 | %IT = 99.2(8)<br>%$\beta$−= 0.8(8) | 1969KN01 |
| 40 | 51 | $^{91}$Zr | 2287.4 (4) | 15/2− | 29.0(8) *ns* | (28.0(2))<br>117.4(4) | (M1) | %IT = 100 | 2014WA12<br>1990AN34<br>1976BA02 |
| 40 | 51 | $^{91}$Zr | 3166.6 (5) | 21/2+ | 4.35(14) $\mu s$ | 879.2(3) | | %IT = 100 | 2014WA12<br>1976BR14 |
| 41 | 50 | $^{91}$Nb | 104.62 (5) | 1/2− | 60.86(22) *d* | 104.62(5) | M4 | %IT = 96.6(5)<br>%$\epsilon$+%$\beta$+ = 3.4(5) | 2014LU05<br>1986WA34 |
| 41 | 50 | $^{91}$Nb | 1984.26 (11) | 13/2− | 10.0(4) *ns* | 193.63(13)<br>1984.15(15) | E2<br>(M2+E3) | %IT = 100 | 2014LU05<br>1976BA02 |
| 41 | 50 | $^{91}$Nb | 2034.42 (20) | 17/2− | 3.76(12) $\mu s$ | 50.1(2) | (E2) | %IT = 100 | 2014LU05<br>1976BR14<br>1975BR01 |
| 42 | 49 | $^{91}$Mo | 653.01 (9) | 1/2− | 64.6(10) *s* | 652.9(1) | M4 | %IT = 50.0(16)<br>%$\epsilon$+%$\beta$+ = 50.0(16) | 1990KAZW<br>1976DE37<br>1965CR10<br>1957PR44<br>1955AX02<br>1953KA11 |
| 42 | 49 | $^{91}$Mo | 2267.4 (4) | 21/2(+) | 47(1) *ns* | 199.5(2) | E2 | %IT = 100 | 1978HA52 |
| 42 | 49 | $^{91}$Mo | 2279.6 (4) | (17/2−) | 38(4) *ns* | 211.7(2) | | %IT = 100 | 1973NI04 |
| 43 | 48 | $^{91}$Tc | 139.3 (3) | (1/2)− | 3.3(1) *min* | | | %$\epsilon$+%$\beta$+ = 99.5(5)<br>%IT = 0.5(5) | 1994RU01<br>1976DE37 |
| 44 | 47 | $^{91}$Ru | 0+X | (1/2−) | 7.6(8) *s* | | | %IT = ?<br>%$\beta$+p > 0<br>%$\epsilon$+%$\beta$+ > 0 | 1983HA06 |
| 45 | 46 | $^{91}$Rh | 172.9 (4) | (1/2−) | 1.60(3) *s** | | | %IT = ?<br>%$\epsilon$+%$\beta$+ = ? | 2019PA16<br>2004DE40 |
| 34 | 58 | $^{92}$Se | 3072 (2) | (9−) | 15.7(7) $\mu s$ | 67 | E2 | %IT = 100 | 2020LI15<br>2012KA36 |
| 35 | 57 | $^{92}$Br | 662.4 (8) | | 0.089(+7−8) $\mu s$ | 106.3(5) | E2 | %IT = 100 | 2012KA36 |
| 35 | 57 | $^{92}$Br | 1137.9 (5) | | 0.084(+10−9) $\mu s$ | 778.8(5)<br>898.3(5)<br>1039.8(5) | | %IT = 100 | 2012KA36 |
| 37 | 55 | $^{92}$Rb | 284.31 (20) | 3− | 54(3) *ns* | 142.0(2) | E2 | %IT = 100 | 2012UR01 |
| 39 | 53 | $^{92}$Y | 0+X | | 3.8(6) $\mu s$ | | | | 2013RUZX<br>2009FO05 |

*Continued...*

*: This half-life is composite of the values for the isomer and the g.s.

Table 1 contd...

| Z | N | $^{A}$X | E(keV) | J$^{\pi}$ | $T_{1/2}$ | Eγ(keV) | λ | Decay mode | Reference |
|---|---|---|---|---|---|---|---|---|---|
| 41 | 51 | $^{92}$Nb | 135.5 (4) | (2)+ | 10.13(2) *d* | | | %ϵ+%β+ = 100 | 2019KR13<br>1985HE18<br>1968RE04<br>1962BU16<br>1959WE30 |
| 41 | 51 | $^{92}$Nb | 225.8 (4) | (2)− | 5.9(2) *μs* | 90.37(9) | E1 | %IT = 100 | 1978BA18 |
| 41 | 51 | $^{92}$Nb | 2203.3 (4) | 11− | 167(4) *ns* | 115.8(2) | E2 | %IT = 100 | 2014WU02<br>1977BR12 |
| 42 | 50 | $^{92}$Mo | 2760.52 (14) | 8+ | 192(5) *ns* | 148.14(13) | E2 | %IT = 100 | 2019HA26<br>2017PA35<br>1979HA60<br>1977HA49<br>1972AD01<br>1971LE19<br>1971CO08<br>1964LO02 |
| 43 | 49 | $^{92}$Tc | 270.09 (8) | (4+) | 0.99(7) *μs* | 56.34(2) | E2 | %IT = 100 | 2019HA26<br>2017PA35<br>1971HO27* |
| 44 | 48 | $^{92}$Ru | 2535.2 (13) | (5−) | 16(2) *ns* | 680.1(10)<br>1670.7(10) | <br>(E3) | %IT = 100 | 1996GO15 |
| 44 | 50 | $^{92}$Ru | 2837.7 (4) | (8+) | 99(10) *ns* | 163.1(2) | E2 | %IT = 100 | 2019HA26<br>2017PA35<br>1980NO06 |
| 45 | 47 | $^{92}$Rh | 0+X | (2+) | 3.18(22) *s* | | | %ϵ = 100<br>%ϵp = 1.7(3) | 2019PA16<br>2004DE40 |
| 45 | 47 | $^{92}$Rh | 55.3+X (3) | (4+) | 0.232(15) *μs* | 55.3(3) | | %IT = 100 | 2019HA26<br>2017PA35 |
| 34 | 59 | $^{93}$Se | 0+X | | 0.39(+12−8) *μs* | 208.3(5)<br>469.9(5) | | %IT = 100 | 2012KA36 |
| 36 | 57 | $^{93}$Kr | 354.85 (25) | (7/2+) | 10(2) *ns* | 237.4(2) | (E2) | %IT = 100 | 2010HW03 |
| 37 | 56 | $^{93}$Rb | 4423.1 (15) | (27/2−) | 111(11) *ns* | 100.9 | | %IT = 100 | 2010SI17 |
| 39 | 54 | $^{93}$Y | 758.719 (21) | 9/2+ | 0.82(4) *s* | 168.499(4) | E3 | %IT = 100 | 2012FO25<br>2007BU35<br>1975CA01 |
| 41 | 52 | $^{93}$Nb | 30.760 (5) | 1/2− | 16.12(12) *y* | 30.760(5) | M4 | %IT = 100 | 2020HO10<br>1983VA25<br>1981LL01 |
| 41 | 52 | $^{93}$Nb | 7435.3+X | | 1.5(5) *μs* | | | | 2007WA45 |
| 42 | 51 | $^{93}$Mo | 2424.95 (4) | 21/2+ | 6.85(7) *h* | 263.049(13) | E4 | %IT = 99.88(1)<br>%ϵ+%β+ = 0.12(1) | 1965GR29<br>1952BO62<br>1950KU15 |
| 42 | 51 | $^{93}$Mo | 9670.0+X | (39/2−) | 1.1(+15−4) *μs* | | | %IT = 100 | 2005FU01 |
| 43 | 50 | $^{93}$Tc | 391.84 (8) | 1/2− | 43.5(10) *min* | 391.83(8) | M4 | %IT = 77.4(6)<br>%ϵ+%β+ = 22.6(6) | 1950ME21 |
| 43 | 50 | $^{93}$Tc | 2185.16 (15) | (17/2−) | 10.2(3) *μs* | 0.31(2)<br>39.75(10)<br>750.7(2) | <br><br>M2+E3 | %IT = 100 | 2019HA26<br>1977HA49<br>1976BR08 |

*Continued...*

*: It appears that 1971HO27, mistakenly, assigned this isomer to a 214-keV level.

Table 1 contd...

| Z | N | $^{A}$X | E(keV) | J$^{\pi}$ | $T_{1/2}$ | Eγ(keV) | λ | Decay mode | Reference |
|---|---|---|---|---|---|---|---|---|---|
| 44 | 49 | $^{93}$Ru | 734.40 (10) | (1/2)− | 10.8(3) *s* | 734.4(1) | | %IT = 22.0(23)<br>%εp = 0.027(5)<br>%ε+%β+ = 78.0(23) | 1976DE37 |
| 44 | 51 | $^{93}$Ru | 2082.5 (9) | (21/2)+ | 2.50(5) *μs* | 146.3(10) | E2 | %IT = 100 | 2019HA26<br>2017PA35<br>2009GA40<br>1983GR33<br>1983KO07<br>1978BR25 |
| 44 | 49 | $^{93}$Ru | 2280.5 (7) | (17/2−) | 35(4) *ns* | 151.3(10)<br>167.8(10)<br>343.8(10) | <br>(E1) | %IT = 100 | 1983GR33 |
| 34 | 60 | $^{94}$Se | 2430 (6) | (7−) | 0.68(5) *μs* | 495.4(3) | | %IT = 100 | 2020LI15 |
| 35 | 59 | $^{94}$Br | 294.6 (5) | | 0.530(15) *μs* | 92.1(5)<br>179.0(5) | E2 | %IT = 100 | 2012KA36 |
| 36 | 58 | $^{94}$Kr | 3445.0 (18) | (9−) | 32(3) *ns* | (88)<br>186.8 | | | 2020GE11 |
| 37 | 57 | $^{94}$Rb | 104.01 (20) | (0−) | 130(15) *ns* | 99.2(2) | | %IT = 100 | 2016MI18 |
| 37 | 57 | $^{94}$Rb | 1485.2 (10) | (8+) | 18(1) *ns* | 168.7<br>169.8<br>817.7 | | %IT = 100 | 2008TS03 |
| 37 | 57 | $^{94}$Rb | 2074.9 (14) | (10−) | 107(16) *ns* | 589.7 | | %IT = 100 | 2008TS03 |
| 39 | 55 | $^{94}$Y | 1202.3 (10) | (5+) | 1.30(1) *μs* | 769.9<br>1202.4 | (M2)<br>(E3) | %IT = 100 | 2017KI09<br>2013RUZX<br>1999GE01 |
| 41 | 53 | $^{94}$Nb | 40.892 (12) | 3+ | 6.263(4) *min* | 40.90(5) | M3 | %IT = 99.50(6)<br>%β−= 0.50(6) | 1990AB06 |
| 41 | 53 | $^{94}$Nb | 140.298 (12) | (2)− | 30(5) *ns* | 99.4074(9) | E1 | %IT = 100 | 1977FE14<br>1971GU05 |
| 42 | 52 | $^{94}$Mo | 2955.55 (13) | 8+ | 98(2) *ns* | 83.6(10)<br>532.1(1) | E2<br>E2(+M3) | %IT = 100 | 2009ZH11<br>1975FA04 |
| 43 | 51 | $^{94}$Tc | 76 (3) | (2)+ | 52.0(10) *min* | 76(3) | | %ε+%β+ = 100<br>%IT < 0.1 | 1962MO06<br>1950ME21<br>1948MO19 |
| 44 | 50 | $^{94}$Ru | 2498.0 (3) | 6+ | 64.5(20) *ns* | 311.4(2) | E2 | %IT = 100 | 2022DA06<br>1977HA49<br>1971LE19 |
| 44 | 50 | $^{94}$Ru | 2644.1 (4) | 8+ | 68(4) *μs*<br>102(17) *μs* @ | 146.1(2) | E2 | %IT = 100 | 2019HA26<br>1977HA49<br>1971LE19<br>2017ZE02@ |
| 45 | 49 | $^{94}$Rh | 54.60 (20) | (2+) | 0.48(3) *μs* | 54.6(2) | E2 | %IT = 100 | 2006BA55 |
| 45 | 49 | $^{94}$Rh | X | (8+) | 25.8(2) *s* | | | %ε+%β+ = 100 | 1980OX01 |
| 46 | 48 | $^{94}$Pd | 4883.1 (4) | 14+ | 0.509(7) *μs** | 95.6(2) | E2 | %IT = 100 | 2019HA26<br>2017PA35<br>2011BR01<br>2009GA40<br>1998GRZS |

*Continued...*

*: Weighted average with 50% weight for $T_{1/2}$ = 0.515(1) $\mu s$ in 2019HA26.

Table 1 contd...

| Z | N | $^AX$ | E(keV) | $J^\pi$ | $T_{1/2}$ | Eγ(keV) | λ | Decay mode | Reference |
|---|---|---|---|---|---|---|---|---|---|
| 46 | 48 | $^{94}$Pd | 7209.8 (9) | (19−) | 217(20) *ns* | 106(1)<br>1651(1) | E1<br>E3 | %IT = 100 | 2019HA26<br>2017PA35<br>2011BR01 |
| 47 | 47 | $^{94}$Ag | 0+X | (7+) | 0.49(5) *s* | | | %ϵ = 100<br>%ϵp = 17.0(6) | 2019PA16<br>2004PL01<br>2004MU30<br>2002LA18<br>1994SC35 |
| 47 | 47 | $^{94}$Ag | X | (21+) | 0.40(4) *s* | | | %ϵ = 95.4(7)<br>%ϵp = 27<br>%p = 4.1(6)<br>%2p = 0.5(3) ? | 2012NA23<br>2006MU03<br>2005MU15<br>2004PL01<br>2004MU30<br>2002LA18 |
| 35 | 60 | $^{95}$Br | 537.9 (5) | (3/2−,<br>1/2) | 6.7(+11−9) *μs* | 537.9 (5) | | %IT = 100 | 2012KA36 |
| 36 | 59 | $^{95}$Kr | 195.5 (3) | (7/2+) | 1.579(22) *μs* | 81.7(2) | (E2) | %IT = 100 | 2022GE01<br>2013RUZX<br>2012KA36<br>2006GE05 |
| 37 | 58 | $^{95}$Rb | 810.57 (25) | (9/2+) | 94(7) *ns* | 191.4<br>810.6 | M2 | %IT = 100 | 2010SI17 |
| 38 | 57 | $^{95}$Sr | 556.08 (8) | (7/2)+ | 21.6(3) *ns* | 204.01(9) | E2 | %IT = 100 | 2015CZ01<br>1986KA20<br>1983KR11<br>1981SEZW<br>1974SU04<br>1974CLZX |
| 39 | 56 | $^{95}$Y | 1087.6 (6) | 9/2+ | 48.6(5) *μs* | 260.7<br>401.8 | (M2)<br>(E3) | %IT = 100 | 2013RUZX<br>2009UR02 |
| 39 | 56 | $^{95}$Y | 3142.5 (14) | (17/2−) | 16(3) *ns* | 969.1 | (M2) | %IT = 100 | 2009UR02 |
| 39 | 56 | $^{95}$Y | 5022.2 (21) | (27/2−) | 65(4) *ns* | 101.2<br>120.6<br>682.3<br>1246.3 | (E2)<br>(E1)<br>(M2)<br>(E3) | %IT = 100 | 2009UR02 |
| 41 | 54 | $^{95}$Nb | 235.69 (2) | 1/2− | 3.61(3) *d* | 235.69(2) | M4 | %IT = 94.4(6)<br>%β−= 5.6(6) | 1974AN22<br>1969FO01 |
| 43 | 52 | $^{95}$Tc | 38.91 (4) | 1/2− | 61.96(24) *d* | 38.9(1) | M4 | %IT = 3.88(32)<br>%ϵ+%β+ = 96.12(32) | 2020SZ02<br>1959UN01 |
| 44 | 51 | $^{95}$Ru | 2538.0 (4) | 21/2+ | 10.05(14) *ns* | 255(1) | E2 | %IT = 100 | 1985CH28 |
| 45 | 50 | $^{95}$Rh | 543.3 (3) | (1/2)− | 1.96(4) *min* | 543.3(3) | | %IT = 88(5)<br>%ϵ+%β+ = 12(5) | 1998JU05 |
| 45 | 50 | $^{95}$Rh | 2236.37 (22) | (17/2−) | 18.8(10) *ns* | 168.84(9) | (E1) | %IT = 100 | 1983GR33<br>1980NO06 |
| 46 | 47 | $^{95}$Pd | 1875.13 (14) | (21/2+) | 13.3(3) *s* | 524.0(1) | (E4) | %IT = 11(3)<br>%β+p = 0.71(7)<br>%ϵ+%β+ = 89(3) | 2019PA16<br>2012LO08<br>2012MA40<br>1982KU15 |
| 46 | 49 | $^{95}$Pd | 4330.9 (4) | (31/2−) | 11.2 (7) *ns* | 260.4(2)<br>1635.4 | | %IT = 100 | 2012MA40<br>1996GO15 |
| 35 | 61 | $^{96}$Br | 311.5+X | | 2.7(+11−7) *μs* | 311.5(5) | | %IT = 100 | 2012KA36 |

*Continued...*

Table 1 contd...

| Z | N | $^{A}X$ | E(keV) | J$^{\pi}$ | $T_{1/2}$ | E$\gamma$(keV) | $\lambda$ | Decay mode | Reference |
|---|---|---|---|---|---|---|---|---|---|
| 37 | 59 | $^{96}$Rb | 1134.6 (11) | (10−) | 1.77(6) $\mu s$ | 40(1) | E2 | %IT = 100 | 2013RUZX<br>2012KA36<br>2005PI13<br>1999GE01 |
| 38 | 58 | $^{96}$Sr | 3523.7 (5) | 9− | 10(5) $ns$ | 195.7<br>398.4 | (E2)<br>E1 | %IT = 100 | 2021UR03<br>2009RZ01<br>2001UR01 |
| 39 | 57 | $^{96}$Y | 1540.5 (4) | 8+ | 9.6(3) $s$ | | | %$\beta$−= 100 | 2020IS08<br>2007HA32<br>1975SA15<br>1975KL11<br>1975BA36<br>1974GRZN |
| 39 | 57 | $^{96}$Y | 1655.1 (2) | (6+) | 181(9) $ns$ | 11.2<br>114.6(3)<br>472.2(3) | <br>(E2)<br>(M2) | %IT = 100 | 2020IS08<br>2017IS03 |
| 40 | 56 | $^{96}$Zr | 1581.64 (6) | 0+ | 38.0(7) $ns$ | 1581.6(4) | E0 | %IT = 100 | 1972BU18<br>1972ANZZ<br>1971ANZF |
| 43 | 53 | $^{96}$Tc | 34.23 (4) | 4+ | 51.5(10) $min$ | 34.20(5) | M3 | %IT = 95.9(+39–34)<br>%$\epsilon$+%$\beta$+ = 4.10(+39 −34) | 2021HE16<br>1967CE01<br>1950ME21 |
| 43 | 53 | $^{96}$Tc | 121.24 (7) | (2)− | 25.6(+4−2) $ns$ | 85.1(1)<br>85.9(1) | E1<br>E1 | %IT = 100 | 1976BI13<br>1974MC14 |
| 45 | 51 | $^{96}$Rh | 51.98 (9) | 3+ | 1.51(2) $min$ | 51.98(9) | M3 | %IT = 60(5)<br>%$\epsilon$+%$\beta$+ = 40(5) | 1975GU01 |
| 46 | 50 | $^{96}$Pd | 2530.57 (23) | 8+ | 1.811(14) $\mu s$ | 106.37(17) | E2 | %IT = 100 | 2019HA26<br>2017PA35<br>2009GA40<br>2007MY02<br>1998GRZS<br>1983GR01 |
| 46 | 50 | $^{96}$Pd | 7040.0 (12) | (15+) | 35(4) $ns$ | 311.3(4) | | %IT = 100 | 1989AL05 |
| 46 | 50 | $^{96}$Pd | 8385.1 (8) | (15,16, 17)− | 37.7(11) $ns$ | 685.2(3) | | %IT = 100 | 2012PA19 |
| 47 | 49 | $^{96}$Ag | 0+X | (2+) | 6.9(6) $s$ | | | %$\epsilon$+%$\beta$+ = 100<br>%$\epsilon$p = 14.9(24) | 2019PA16<br>2012LO08<br>2003BA39<br>1997SC30<br>1982KU15 |
| 47 | 49 | $^{96}$Ag | 2461.4 (3) | (13−) | 103(5) $\mu s$ | 485.7(3)<br>742.7(2) | (M2,E3)<br>(E3) | %IT = 100 | 2019HA26<br>2011BO23 |
| 47 | 49 | $^{96}$Ag | 2686.7 (4) | (15+) | 1.56(2) $\mu s$ | 43.7(2) | (E2) | %IT = 100 | 2019HA26<br>2017PA35<br>2011BO23<br>2011BE34 |
| 47 | 49 | $^{96}$Ag | 6951.8 (18) | (19+) | 0.136(20) $\mu s$ | 98(3)<br>4265(2) | (E2)<br>(E4) | %IT = 100 | 2019HA26<br>2017PA35<br>2011BO23 |
| 48 | 48 | $^{96}$Cd | 5605 | (12−, 13−) | 197(+19−17) $ns$ | 307 | | %IT = 100 | 2019DA02 |

*Continued...*

Table 1 contd...

| Z | N | $^{A}$X | E(keV) | J$^{\pi}$ | $T_{1/2}$ | Eγ(keV) | λ | Decay mode | Reference |
|---|---|---|---|---|---|---|---|---|---|
| 48 | 48 | $^{96}$Cd | 5810 (+1560 −1220) | 16+ | 0.50(4) *s* | | | %ϵ+%β+ ≈ 100<br>%β+p = 19.5(29) | 2019PA16<br>2017DA07<br>2011NA34 |
| 37 | 60 | $^{97}$Rb | 76.6 (2) | (1/2, 3/2)− | 5.7(6) *μs* | 76.6(2) | E1 | %IT = 100 | 2013RU07<br>2012KA36 |
| 38 | 59 | $^{97}$Sr | 308.13 (11) | 7/2+ | 173(4) *ns* | 141.3(5) | E2 | %IT = 100 | 2019ES04<br>2015CZ01<br>2013RUZX<br>2006HW01<br>1983KR11 |
| 38 | 59 | $^{97}$Sr | 830.80 (22) | (9/2+) | 513(8) *ns** | 308.3(2)<br>522.7(2) | (E2) | %IT = 100 | 2019ES04<br>2018RZ01<br>2013RU07<br>2005ZL01 |
| 39 | 58 | $^{97}$Y | 667.52 (23) | (9/2)+ | 1.17(3) *s* | 667.5(5) | | %β−= 99.65(35)<br>%IT = 0.35(35)<br>%β−n < 0.08 | 1986WA17<br>1976KAYO<br>1976MOZC<br>1970EI02 |
| 39 | 58 | $^{97}$Y | 3522.6 (4) | (27/2−) | 142(8) *ms* | 162.3(2) | E3 | %IT = 94.8(9)<br>%β−= 5.2(9) | 1996LH03<br>1987BO19<br>1986LH01 |
| 40 | 57 | $^{97}$Zr | 1264.35 (16) | 7/2+ | 105(2) *ns* | 161.2(2)<br>1264.2(5) | | %IT = 100 | 2013RUZX<br>2006HW01<br>1996LH03<br>1985BE20 |
| 41 | 56 | $^{97}$Nb | 743.35 (3) | 1/2− | 58.7(18) *s* | 743.36(3) | | %IT = 100 | 1992KAZM |
| 43 | 54 | $^{97}$Tc | 96.57 (6) | 1/2− | 91.0(6) *d* | 96.5(1) | M4 | %IT = 96.06(18)<br>%ϵ = 3.94(18) | 2010KR05<br>1998KO27<br>1954BO24 |
| 45 | 52 | $^{97}$Rh | 258.76 (18) | 1/2− | 46.2(16) *min* | 258.76(18) | M4 | %IT = 5.6(6)<br>%ϵ+%β+ = 94.4(6) | 1975PL05<br>1974OH07<br>1971LO12 |
| 47 | 50 | $^{97}$Ag# | 618 (38) | (1/2−) | # | | | | 2020HO03 |
| 48 | 49 | $^{97}$Cd | 0+X | (1/2−) | 0.73(7) *s* | | | %β+ = ? | 2019PA16 |
| 48 | 49 | $^{97}$Cd | 2620 (580) | (25/2+) | 3.86(6) *s* | | | %ϵ+%β+ ≈ 100<br>%β+p = 25.1(5) | 2019PA16<br>2012LO08<br>2011LO09 |
| 49 | 48 | $^{97}$In | 0+X | (1/2−) | 1.3-230 *μs* | | | %p = ? | 2018PA20 |
| 37 | 61 | $^{98}$Rb | 178.3 (5) | | 0.358(7) *μs* | (54.5)<br>178.3(5) | E2<br>E2 | %IT = 100 | 2012KA36<br>2009FO05 |
| 37 | 61 | $^{98}$Rb | ~ 270 | (3+,4+) | 96(3) *ms* | | | %β−= 100 | 2015PR03<br>2002LH01<br>1980SC13 |
| 38 | 60 | $^{98}$Sr | 215.35 (7) | 0+ | 22.8(19) *ns* | 71.0(1)<br>215.5(1) | E2<br>E0 | %IT = 100 | 2002LH01<br>1980SC13 |
| 38 | 60 | $^{98}$Sr | 1838.2 (7) | 3+ | 13(3) *ns* | 1403.8<br>1693.3 | | | 2004LI66 |

*Continued...*

*: $T_{1/2}$= 265(27) *ns* in 2003HW03: J. K. Hwang et al., Phys. Rev. C 67, 054304 (2003); half-life in this work, shorter by a factor of ~2, seems discrepant
#: Half-life of this isomer is not measured, but expected to be > few hundred *ns* from time-of-flight of the ions through the beam transport system.

Table 1 contd…

| Z | N | $^{A}$X | E(keV) | J$^{\pi}$ | $T_{1/2}$ | E$\gamma$(keV) | $\lambda$ | Decay mode | Reference |
|---|---|---|---|---|---|---|---|---|---|
| 39 | 59 | $^{98}$Y | 170.78 (5) | 2− | 0.62(1) $\mu s$ | 51.5(1)<br>170.8(1) | M1+E2<br>E2 | %IT = 100 | 2017UR03<br>2013RUZX<br>2004BR14<br>1972GRYM<br>1970GR38<br>1970JO20 |
| 39 | 59 | $^{98}$Y | 374.97 (9) | 4− | 35.2(5) $ns$ | 204.3(1) | E2 | %IT = 100 | 2017UR03<br>2004BR14<br>2002PFZZ<br>1986LHZW |
| 39 | 59 | $^{98}$Y | 465.7 (7) | (6,7)+ | 2.32(8) $s$ | | | %$\beta$− = 90(10)<br>%IT = 10(10)<br>%$\beta$–n = 3.44(95) | 2017UR03<br>1981EN05<br>1977SI05 |
| 39 | 59 | $^{98}$Y | 496.10 (11) | 4− | 6.90(5) $\mu s$ | 49.9(2)<br>121.1(1)<br>325.2(2) | E1<br>M1+E2 | %IT = 100 | 2017UR03<br>2013RUZX<br>2004BR14<br>1999GE01<br>1972GRYM<br>1970GR38 |
| 39 | 59 | $^{98}$Y | 564.0+X | (3−, 4−) | 180(7) $ns$ | | | %IT = 100 | 2017UR03 |
| 39 | 59 | $^{98}$Y | 972.17 (20) | (8+) | 0.45(15) $\mu s$ | 313.9(1) | | %IT = 100 | 2017UR03 |
| 39 | 59 | $^{98}$Y | 1181.50 (18) | 10− | 0.78(3) $\mu s$ | 110.8(1) | E2 | %IT = 100 | 2017UR03<br>2013RUZX<br>2005PI13<br>2004BR14<br>1972GRYM<br>1970GR38 |
| 40 | 58 | $^{98}$Zr | 854.06 (6) | 0+ | 64(7) $ns$ | 854.06(6) | E0 | %IT = 100 | 1977SI05<br>1976PO11 |
| 40 | 58 | $^{98}$Zr | 6601.9 (11) | (17−) | 1.9(2) $\mu s$ | 63.0(1) | (E2) | %IT = 100 | 2013RUZX<br>2006SI36 |
| 41 | 57 | $^{98}$Nb | 84 (4) | (5+) | 51.3(4) $min$ | | | %$\beta$−= 99.9(1)<br>%IT = 0.1(1) | 1976HE10 |
| 42 | 56 | $^{98}$Mo | 734.75 (4) | 0+ | 21.8(9) $ns$ | 734.75(4) | E0 | %IT = 100 | 1972BU18 |
| 43 | 55 | $^{98}$Tc | 90.76 (16)* | (2)− | 14.7(3) $\mu s$* | | | | 1978BA18<br>1976WE06<br>1961SC11 |
| 45 | 53 | $^{98}$Rh | 56.3 (10) | (5+) | 3.6(2) $min$ | | | %IT = 89(5)<br>%$\epsilon$+%$\beta$+ = 11(5) | 1978KI17<br>1974SI18<br>1966AT02 |
| 47 | 51 | $^{98}$Ag | 107.28 (10) | (4+) | 0.161(7) $\mu s$ # | 107.28(10) | E2 | %IT = 100 | 2019HA26<br>2017PA35#<br>1998GRZS |
| 47 | 51 | $^{98}$Ag | 167.83 (15) | (3+) | 220(20) $ns$ | 107.28(10) | E2 | %IT = 100 | 1998GRZS |
| 48 | 50 | $^{98}$Cd | 2281.1 (3) | (6+) | 13(2) $ns$ | 198.3(1) | | %IT = 100 | 2017PA35 |

*Continued…*

*: This half-life is assumed for the 90.76 level, however, experimental data do not exclude its assignment to the 73.3 level in $^{98}$Tc.

#: 2017PA35 quotes 160(10) $ns$ from M. Vencelj et al., Balk. Phys. Lett. Special issue, p298 (2000); 142(13) $ns$ and 158(26) $ns$ from N. Braun, Ph.D. thesis, Cologne (2012). Our adopted $T_{1/2}$ from 2017PA35 is the same as weighted average of the six available values.

Table 1 contd...

| Z | N | $^{A}$X | E(keV) | J$^{\pi}$ | $T_{1/2}$ | E$\gamma$(keV) | $\lambda$ | Decay mode | Reference |
|---|---|---|---|---|---|---|---|---|---|
| 48 | 50 | $^{98}$Cd | 2428.3 (4) | (8+) | 0.154(16) $\mu$s | 147.2(1) | (E2) | %IT = 100 | 2017PA35<br>2004BL10<br>1997GO18 |
| 48 | 50 | $^{98}$Cd | 6635.1 (21) | (12+) | 0.223(5) $\mu s$ | 49.2(2)<br>4207(2) | | %IT = 100 | 2019HA26<br>2017PA35<br>2010BL13<br>2006VE09<br>2004BL10 |
| 49 | 49 | $^{98}$In | 820 (730) | (9+) | 0.89(2) $s$ | | | %$\epsilon$+%$\beta$+ ≈ 100<br>%$\beta$+p = 44(2) | 2019PA16<br>2012LO08<br>2008BA53<br>2001KI13 |
| 39 | 60 | $^{99}$Y | 2141.65 (19) | (17/2+) | 8.0(5) $\mu s$ | 27.7(4)<br>272.90(10)<br>546.2(3)<br>882.5(3)<br>1166.0(4)<br>1435.5(4) | (M1) | %IT = 100 | 2013RUZX<br>1985ME09 |
| 40 | 59 | $^{99}$Zr | 251.96 (9) | 7/2+ | 335(8) $ns$ | 130.2(1) | E2 | %IT = 100 | 2020BO04<br>2004HW02<br>1999GE01<br>1979SE01<br>1970GR38 |
| 40 | 59 | $^{99}$Zr | 1038.5 (5) | (9/2+) | 54(10) $ns$ | 188.5<br>360.1<br>786.8 | | %IT = 100 | 2003UR01 |
| 41 | 58 | $^{99}$Nb | 365.27 (8) | 1/2− | 2.5(2) $min$ | 365.1 | | %$\beta$−= 98.1(19)<br>%IT = 1.9(19) | 1971HA07<br>1971CA18<br>1963TR01<br>1960OR02 |
| 42 | 57 | $^{99}$Mo | 97.785 (3) | 5/2+ | 15.5(2) $\mu s$ | 97.785(3) | E2 | %IT = 100 | 2013RUZX<br>1978BA18 |
| 42 | 57 | $^{99}$Mo | 684.10 (19) | 11/2− | 742(13) $ns$ | 448.6(2) | | %IT = 100 | 2021DA15<br>1978BA18 |
| 43 | 56 | $^{99}$Tc | 142.6836 (11) | 1/2− | 6.00660(18) $h$* | 2.1726(4)<br>142.63(3) | E3<br>M4 | %IT = 99.9963(6)<br>%$\beta$−= 0.0037(6) | 2018PO10<br>2014UN01<br>2004SC04<br>2021DU15<br>2011KI45<br>2004DA05<br>1980HO17<br>1972EM01<br>1970LE07<br>1969VU03<br>1966GO22<br>1958BE92 |

*Continued...*

*: Weighted average of 2018PO10, 2014UN01 (Erratum) and 2004SC04. 2018PO10 used physiological saline solution sodium pertechnetate ($TcO_4Na$), used in nuclear imaging and other applications.

Table 1 contd...

| Z | N | $^{A}X$ | E(keV) | $J^{\pi}$ | $T_{1/2}$ | Eγ(keV) | λ | Decay mode | Reference |
|---|---|---|---|---|---|---|---|---|---|
| 44 | 55 | $^{99}$Ru | 89.57 (6) | 3/2+ | 20.5(1) *ns* | 89.50(20) | E2+M1 | %IT = 100 | 2015KI14<br>1972GU01<br>1966KI02 |
| 45 | 54 | $^{99}$Rh | 64.4 (5) | 9/2+ | 4.7(1) *h* | | | %ϵ+%β+ = 99.92(8)<br>%IT = 0.08(8) | 2014KU20<br>1956KA25 |
| 47 | 52 | $^{99}$Ag | 506.2 (4) | (1/2−) | 10.5(5) *s* | 163.6(3) | E3 | %IT = 100 | 1982KU15 |
| 48 | 51 | $^{99}$Cd | 2057.5 (4) | (17/2+) | 13.1(5) *ns* | 226.3(2) | | %IT = 100 | 1996LI06 |
| 38 | 62 | $^{100}$Sr | 1618.72 (20) | (4−) | 0.122(9) *μs* | 58.3(2)<br>118.0(2)<br>204.4(3)<br>1201.7(2) | | %IT = 100 | 2012KA36<br>1995PF04 |
| 39 | 61 | $^{100}$Y | 145 (15) | 4+ | 0.94(3) *s* | | | %β−= 100 | 2010BA31<br>2007HA32<br>1977KH03 |
| 41 | 59 | $^{100}$Nb | 314 (23) | (5+) | 2.99(11) *s* | | | %β−= 100 | 1987ME06 |
| 41 | 59 | $^{100}$Nb | 34.3+X | (4−,5−,<br>6−) | 0.46(6) *μs* | 34.3 | (E1) | %IT = 100 | 1986LHZX |
| 41 | 59 | $^{100}$Nb | 420.7+X | (8−) | 12.4(3) *μs* | 28 | (E2) | %IT = 100 | 2013RUZX<br>1999GE01 |
| 43 | 57 | $^{100}$Tc | 200.67 (4) | (4)+ | 8.32(14) *μs* | 28.520(2) | E2 | %IT = 100 | 1980BI01<br>1978MA36 |
| 43 | 57 | $^{100}$Tc | 243.95 (4) | (6)+ | 3.2(2) *μs* | 43.2862(10) | E2 | %IT = 100 | 1980BI01 |
| 45 | 55 | $^{100}$Rh | 32.686 (13) | (2)− | 27.6(6) *ns* | 32.68(2) | M1+E2 | %IT = 100 | 1979EN03 |
| 45 | 55 | $^{100}$Rh | 74.782 (14) | (2)+ | 214.3(20) *ns* | 42.09(2)<br>74.78(2) | <br>E1 | %IT = 100 | 1983BI04<br>1979EN03<br>1971RE06 |
| 45 | 55 | $^{100}$Rh | 107.59 (20) | (5+) | 4.6(2) *min* | 32.7(2)?<br>74.9(2) | | %IT ≈ 98.3<br>%ϵ+%β+ ≈ 1.7 | 1982MAZP<br>1978KI07<br>1974SI18 |
| 45 | 55 | $^{100}$Rh | 219.58 (22) | (7+) | 132(6) *ns* | 112.0(1) | E2 | %IT = 100 | 1986DU04<br>1984MA30 |
| 47 | 53 | $^{100}$Ag | 15.52 (16) | (2)+ | 2.24(13) *min* | | | %IT = ?<br>%ϵ+%β+ = ? | 1983RA10<br>1980HA20 |
| 48 | 52 | $^{100}$Cd | 2548.19 (18) | (8+) | 62(6) *ns* | 90.7(1)<br>452.6(1) | | %IT = 100 | 1994GO38<br>1992AL17<br>1988PI03 |
| 39 | 62 | $^{101}$Y* | 1207.0 | | 0.187(+49−38) *μs**<br>0.860(+90−80) *μs** | 480.0 (16) | | %IT = 100 | 2012KA36<br>2009FO05 |
| 40 | 61 | $^{101}$Zr | 941.81 (23) | (9/2+) | 16(2) *ns* | 322.2(5)<br>331.4(5)<br>474.2(4)<br>533.7(4)<br>620.9(4)<br>710.2(4)<br>843.8(4) | E1<br>(M1+E2)<br>E1<br><br>E1<br><br>E2 | %IT = 100 | 2004UR06 |
| 42 | 59 | $^{101}$Mo | 13.497 (9) | 3/2+ | 226(7) *ns* | 13.49(10) | M1(+E2) | %IT = 100 | 1991SE08 |
| 42 | 59 | $^{101}$Mo | 57.015 (11) | 5/2+ | 133(70) *ns* | 43.515(5)<br>56.892(20) | M1+E2<br>E2 | %IT = 100 | 1991SE08 |
| 42 | 59 | $^{101}$Mo | 273 (2) | 11/2− | 95(15) *ns* | | | | 2005RE11 |

*Continued...*

*: Discrepancy in the two measured half-lives of this isomer remains unresolved. It is possible there are two isomers.

Table 1 contd...

| Z | N | $^{A}$X | E(keV) | J$^{\pi}$ | $T_{1/2}$ | E$\gamma$(keV) | $\lambda$ | Decay mode | Reference |
|---|---|---|---|---|---|---|---|---|---|
| 43 | 58 | $^{101}$Tc | 9.320 (9) | 7/2+ | 14.3(3) *ns* | 9.317(10) | M1+E2 | %IT = 100 | 1976SVZY |
| | | | | | | | | | 1974SVZZ |
| 43 | 58 | $^{101}$Tc | 15.602 (10) | 5/2+ | 26.8(6) *ns* | 6.281(7) | M1+E2 | %IT = 100 | 1976SVZY |
| | | | | | | 15.606(15) | E2 | | 1974SVZZ |
| 43 | 58 | $^{101}$Tc | 207.526 (20) | 1/2− | 636(8) $\mu s$ | 191.92(2) | M2 | %IT = 100 | 1978BA18 |
| 44 | 57 | $^{101}$Ru | 527.56 (10) | 11/2− | 17.5(4) $\mu s$ | 220.7(1) | M2 | %IT = 100 | 1978BA18 |
| 45 | 56 | $^{101}$Rh | 157.32 (3) | 9/2+ | 4.34(1) *d* | 157.41(4) | M4 | %IT = 7.20(25) | 1975RU06 |
| | | | | | | | | %$\epsilon$ = 92.80(25) | 1966AR05 |
| 47 | 54 | $^{101}$Ag | 274.1 (3) | (1/2)− | 3.10(10) *s* | 176.2(5) | E3 | %IT = 100 | 1978HA11 |
| | | | | | | | | | 1975CA01 |
| 49 | 52 | $^{101}$In* | 637 (50) | (1/2−) | | | | | 2020HO03 |
| | | | | | | | | | 2019XU13 |
| 39 | 63 | $^{102}$Y$^{\dagger}$ | 0+X | (>5) | 0.36 (4) *s* | | | %$\beta$−= 100 | 2011HA48 |
| | | | | | | | | %$\beta$−n = 4.9(12)$^{\#}$ | 1983SH13 |
| | | | 0+Y | (2−) | 0.298(9) *s* | | | %$\beta$−= 100 | 2011HA48 |
| | | | | | | | | %$\beta$−n = 4.9(12)$^{\#}$ | 1996ME09 |
| | | | | | | | | | 1991HI02 |
| 41 | 61 | $^{102}$Nb | 93 (23) | 1+ | 1.31(20) *s* | | | %$\beta$−= 100 | 2019DO02 |
| | | | | | | | | | 2007RI01 |
| | | | | | | | | | 1976AH06 |
| 43 | 59 | $^{102}$Tc | 0+X | (4,5) | 4.35(7) *min* | | | %IT = 2(2) | 1970HU02 |
| | | | | | | | | %$\beta$−= 98(2) | |
| 45 | 57 | $^{102}$Rh | 41.940 (15) | 2(−) | 18.9(4) *ns* | 41.938(20) | M1 | %IT = 100 | 2014TO01 |
| | | | | | | | | | 1982BI12 |
| 45 | 57 | $^{102}$Rh | 140.73 (9) | 6(+) | 3.742(10) *y* | 98.8(1) | M4 | %IT = 0.233(24) | 2014TO01 |
| | | | | | | | | %$\epsilon$+%$\beta$+ =99.767(24) | 1998SH21 |
| | | | | | | | | | 1998HI12 |
| 46 | 56 | $^{102}$Pd | 1593.16 (22) | 0+ | 14.5(4) *ns* | 58.3 | | %IT = 100 | 1980FAZX |
| | | | | | | 1035.6 | | | 1979CO17 |
| | | | | | | 1592.6(5) | E0 | | 1979BAZW |
| 47 | 55 | $^{102}$Ag | 9.40 (7) | 2+ | 7.7(5) *min* | (9.40(8)) | (M3) | %IT = 49(5) | 1968GR01 |
| | | | | | | | | %$\epsilon$+%$\beta$+ = 51(5) | 1967CH05 |
| 48 | 54 | $^{102}$Cd | 2718.6 (3) | (8+) | 39(3) *ns* | 157.22(5) | | %IT = 100 | 2001LI24 |
| | | | | | | 487.43(4) | (E2) | | 1992AL17 |
| | | | | | | 684.1(3) | | | |
| | | | | | | 1942.2(10) | | | |
| 50 | 52 | $^{102}$Sn | 1969+Y | (6+) | 367(11) *ns* | | | | 2021GR06 |
| | | | | | | | | | 1998LI50 |
| | | | | | | | | | 1996LI50 |
| 44 | 59 | $^{103}$Ru | 238.2 (7) | 11/2− | 1.69(7) *ms* | (24.6(7)) | | %IT = 100 | 1975KL04 |
| | | | | | | | | | 1975BA60 |
| | | | | | | | | | 1970UY01 |
| 45 | 58 | $^{103}$Rh | 39.753 (6) | 7/2+ | 56.114(9) *min* | 39.755(6) | E3 | %IT = 100 | 2010KR05 |
| | | | | | | | | | 1981VA11 |
| | | | | | | | | | 1973GU06 |
| 46 | 57 | $^{103}$Pd | 784.79 (10) | 11/2− | 25(2) *ns* | 66.95(15) | (E1) | %IT = 100 | 1975DI09 |
| | | | | | | 541.0(1) | M2 | | |

*Continued...*

*: Half-life of this isomer, established in mass measurements, is not measured, but expected to be > few hundred *ns* from time-of-flight of the ions through the beam transport system.

#: The decay mode is combined for the two activities.

Table 1 contd...

| Z | N | $^{A}$X | E(keV) | J$^{\pi}$ | $T_{1/2}$ | E$\gamma$(keV) | $\lambda$ | Decay mode | Reference |
|---|---|---|---|---|---|---|---|---|---|
| 47 | 56 | $^{103}$Ag | 134.45 (4) | 1/2− | 5.7(3) *s* | 134.44(4) | E3 | %IT = 100 | 1962WH02 |
| 49 | 54 | $^{103}$In | 631.7 (1) | (1/2−) | 34(2) *s* | 631.7(1) | (M4) | %$\epsilon$+%$\beta$+ = 67<br>%IT = 33 | 2020HO03<br>1997SZ04 |
| 41 | 63 | $^{104}$Nb | 215 (120)* | | 0.98(7) *s** | | | %$\beta$−= 100<br>%$\beta$−n = 0.05(3) | 2019DO02<br>1996ME09<br>1982KE05<br>1976AH06 |
| 43 | 61 | $^{104}$Tc | 69.7 (2) | (+) | 3.5(3) $\mu s$ | 69.7(2) | E2 | %IT = 100 | 1981TIZZ |
| 43 | 61 | $^{104}$Tc | 106.1 (3) | (+) | 0.40(2) $\mu s$ | 36.3(2) | | %IT = 100 | 1999GE01 |
| 45 | 59 | $^{104}$Rh | 128.9679 (5) | 5+ | 4.34(3) *min* | 31.866(2)<br>77.5447(4) | M3<br>E3 | %IT = 99.87(1)<br>%$\beta$−= 0.13(1) | 1966WA11<br>1963CS01<br>1959EL41<br>1939CR03 |
| 45 | 59 | $^{104}$Rh | 344.594 (3) | 6− | 47(3) *ns* | 169.356(3)<br>215.626(10)<br>293.09(7) | E1<br>E1 | %IT = 100 | 1990BI03 |
| 47 | 57 | $^{104}$Ag | 6.90 (22) | 2+ | 33.5(20) *min* | 6.9(4) | | %$\epsilon$+%$\beta$+ = 99.965(35)<br>%IT = 0.035(35) | 1971MU22 |
| 49 | 55 | $^{104}$In | 93.48 (10) | (3+) | 15.7(5) *s* | 93.5(1) | M3 | %$\epsilon$+%$\beta$+ = 20<br>%IT = 80 | 1988BA10 |
| 43 | 62 | $^{105}$Tc | 85.42 (7) | (5/2+) | 20.8(6) *ns* | 85.4(1) | E1 | %IT = 100 | 1989RUZU |
| 44 | 61 | $^{105}$Ru | 20.606 (14) | 5/2+ | 340(15) *ns* | 20.56(2) | M1+E2 | %IT = 100 | 2014LA15<br>1975SU02 |
| 44 | 61 | $^{105}$Ru | 163.821 (16) | (5/2+) | 55(7) *ns* | 55.74(5)<br>143.26(7) | <br>M1+E2 | %IT = 100 | 1978HO06 |
| 45 | 60 | $^{105}$Rh | 129.782 (4) | 1/2− | 42.9(3) *s* | 129.782(4) | E3 | %IT = 100 | 1998KR08<br>1992KAZM |
| 46 | 59 | $^{105}$Pd | 489.14 (4) | 11/2− | 36.1(4) $\mu s$ | 182.89(4) | M2 | %IT = 100 | 1970BLZT |
| 47 | 58 | $^{105}$Ag | 25.479 (16) | 7/2+ | 7.23(16) *min* | 25.48(2) | E3 | %IT = 99.66(7)<br>%$\epsilon$+%$\beta$+ = 0.34(7) | 2007TI07<br>2006DE15<br>1969HO36 |
| 48 | 57 | $^{105}$Cd | 2517.5 (3) | (21/2+) | 4.5(5) $\mu s$ | 126.46(19) | (E2) | %IT = 100 | 2015RA02<br>1976HAXS<br>1976SPZQ |
| 49 | 56 | $^{105}$In | 674.08 (25) | (1/2−) | 48(6) *s* | 674.1(3) | M4 | %IT = 100 | 2020HO03<br>2006KA44<br>1980WI20<br>1975RI06 |
| 41 | 65 | $^{106}$Nb | 204.8 (1) | (3+) | 0.82(6) $\mu s$ | 204.8(1) | | %IT = 100 | 2014LU07<br>2012KA36<br>1999GE01 |
| 45 | 61 | $^{106}$Rh | 137 (13) | (6)+ | 131(2) *min* | | | %$\beta$−= 100 | 1960SE07<br>1958MA39 |
| 47 | 59 | $^{106}$Ag | 89.66 (7) | 6+ | 8.28(2) *d* | | | %$\epsilon$+%$\beta$+ = 100 | 1988RY01 |
| 48 | 58 | $^{106}$Cd | 4659.71 (9) | 12(+) | 62(6) *ns* | 223.61(5)<br>335.36(5)<br>476.8(3) | E2 | %IT = 100 | 1977DA08 |
| 48 | 58 | $^{106}$Cd | 7118.9 (7) | 16+ | 11(+6−3) *ns* | 602.8(2)<br>892.3(2) | E2 | %IT = 100<br>not found in 1995RE07 | 1994JE05 |

*Continued...*

*: Level energy of 9.8(23) keV measured in R. Orford, Ph.D. thesis (2018), McGill. NUBASE2020 suggests this level as (5-) g.s. and 4.9(3) *s* as (0-,1-) isomer.

Table 1 contd...

| Z | N | $^{A}$X | E(keV) | J$^{\pi}$ | $T_{1/2}$ | E$\gamma$(keV) | $\lambda$ | Decay mode | Reference |
|---|---|---|---|---|---|---|---|---|---|
| 49 | 57 | $^{106}$In | 28.6 (3) | (2)+ | 5.2(1) *min* | | | %$\epsilon$+%$\beta$+ = 100 | 2010EK01 |
| | | | | | | | | | 1978HU06 |
| 51 | 55 | $^{106}$Sb | 103.5 (3) | (4+) | 232(21) *ns* | 103.4(3) | E2 | %IT = 100 | 1999SO08 |
| 42 | 65 | $^{107}$Mo | 65.4 (2) | 5/2+ | 427(35) *ns* | 65.4(1) | E2 | %IT = 100 | 2020UR02 |
| | | | | | | | | | 2006PI14 |
| 43 | 64 | $^{107}$Tc | 30.1 (1) | (1/2+) | 3.85(5) $\mu s$ | 30.1(1) | E1 | %IT = 100 | 2007SI06 |
| 43 | 64 | $^{107}$Tc | 65.72 (14) | (5/2+) | 184(3) *ns* | 20.0 | | %IT = 100 | 1986OHZZ |
| | | | | | | 65.77(14) | E1 | | |
| 44 | 63 | $^{107}$Ru | 102.681 (22) | (+) | 10.8(5) *ns* | 102.70(3) | M1+E2 | %IT = 100 | 1995SC24 |
| 44 | 63 | $^{107}$Ru | 106.307 (22) | | 33.4(10) *ns* | 106.31(3) | M1+E2 | %IT = 100 | 1995SC24 |
| 45 | 62 | $^{107}$Rh | 268.36 (4) | 1/2− | > 10 $\mu s$ | 268.36(5) | (E3) | %IT = 100 | 1986KA43 |
| 45 | 62 | $^{107}$Rh | 374.278 (25) | (3/2+) | 15(2) *ns* | 105.92(5) | (E1) | %IT = 100 | 1986KA43 |
| | | | | | | 374.28(5) | (E2) | | |
| 46 | 61 | $^{107}$Pd | 115.74 (12) | 1/2+ | 0.85(10) $\mu s$ | 115.65(20) | E2 | %IT = 100 | 1969GR18 |
| 46 | 61 | $^{107}$Pd | 214.6 (3) | 11/2− | 21.3(5) *s* | 214.9(5) | E3 | %IT = 100 | 1957ST87 |
| 47 | 60 | $^{107}$Ag | 93.125 (19) | 7/2+ | 44.3(2) *s* | 93.124(20) | E3 | %IT = 100 | 1947BR05 |
| 48 | 59 | $^{107}$Cd | 845.54 (6) | 11/2− | 70.1(14) *ns* | 36.5(1) | E1 | %IT = 100 | 2017GR19 |
| | | | | | | 640.58(10) | M2 | | 1974BE17 |
| | | | | | | 845.5(4) | | | 1974HA41 |
| 48 | 59 | $^{107}$Cd | 2678.88 (15) | 21/2+ | 55(4) *ns* | 520.4(1) | E1 | %IT = 100 | 1974HA41 |
| | | | | | | | | | 1974HA48 |
| 49 | 58 | $^{107}$In | 678.5 (3) | 1/2− | 50.4(6) *s* | 678.5(3) | M4 | %IT = 100 | 2020HO03 |
| | | | | | | | | | 1976HS01 |
| 40 | 68 | $^{108}$Zr | 2074.5 (8) | (6+) | 0.536(+26−25) $\mu s$ | 277.9(5) | | %IT = 100 | 2012KA36 |
| | | | | | | 432.3(5) | | | 2011SU11 |
| 41 | 67 | $^{108}$Nb | 166.6 (5) | | 0.109(2) $\mu s$ | 89.0(5) | (E2) | %IT = 100 | 2012KA36 |
| | | | | | | 102.2(5) | | | |
| 43 | 65 | $^{108}$Tc | 176.6+X | | 28(4) *ns* | 70.4 | | %IT = 100 | 1998HW04 |
| | | | | | | 90.4 | | | |
| | | | | | | 176.6 | | | |
| 45 | 63 | $^{108}$Rh | 115 (18) | (5+) | 6.0(3) *min* | | | %IT = ? | 2007HA20 |
| | | | | | | | | %$\beta$−≈ 100 | 1969PI08 |
| 47 | 61 | $^{108}$Ag | 109.466 (7) | 6+ | 438(9) *y* | 30.332(8) | M4 | %IT = 7.58(37) | 2018SH09 |
| | | | | | | | | %$\epsilon$+%$\beta$+ = 92.42(37) | 2014FE03 |
| | | | | | | | | | 2004SC04 |
| 47 | 61 | $^{108}$Ag | 215.382 (4) | 3+ | 45.8(7) *ns* | 136.241(6) | E1 | %IT = 100 | 1976HA57 |
| | | | | | | 215.381(7) | E2 | | 1974BE47 |
| 49 | 59 | $^{108}$In | 29.75 (5) | 2+ | 39.6(7) *min* | | | %$\epsilon$+%$\beta$+ = 100 | 1975FL01 |
| 41 | 68 | $^{109}$Nb | 312.5 (4) | | 132(18) *ns* | 196.2(5) | | %IT = 100 | 2012KA36 |
| | | | | | | 312.2(5) | | | 2011WA03 |
| 42 | 67 | $^{109}$Mo | 69.7 (5) | 5/2+ | 0.194(+76−49) $\mu s$ | 69.7(5) | | %IT = 100 | 2020UR02 |
| | | | | | | | | | 2012KA36 |
| 44 | 65 | $^{109}$Ru | 96.14 (15) | (5/2−) | 0.68(3) $\mu s$ | 96.2(2) | E1 | %IT = 100 | 1999GE01 |
| | | | | | | | | | 1992PEZX |
| | | | | | | | | | 1976CHZD |
| | | | | | | | | | 1974CLZX |
| 45 | 64 | $^{109}$Rh | 225.873 (19) | 3/2+ | 1.66(4) $\mu s$ | 225.98(3) | E2 | %IT = 100 | 1987KA29 |
| 45 | 64 | $^{109}$Rh | 257.66 (3) | (3/2)+ | 28.7(15) *ns* | 31.80(3) | M1+E2 | %IT = 100 | 1987KA29 |
| 45 | 64 | $^{109}$Rh | 373.99 (3) | 1/2− | 34(2) *ns* | 116.32(3) | | %IT = 100 | 1998LH02 |
| | | | | | | 148.12(3) | | | 1987KA29 |

*Continued...*

Table 1 contd...

| Z | N | $^AX$ | E(keV) | J$^{\pi}$ | $T_{1/2}$ | E$\gamma$(keV) | $\lambda$ | Decay mode | Reference |
|---|---|---|---|---|---|---|---|---|---|
| 46 | 63 | $^{109}$Pd | 113.4000 (14) | 1/2+ | 380(50) *ns* | 113.401(2) | E2 | %IT = 100 | 1978KA10 |
| 46 | 63 | $^{109}$Pd | 188.9903 (10) | 11/2− | 4.694(2) *min** | 188.990(1) | E3 | %IT = 100 | 2019KR05<br>1992AN19<br>1992KAZM<br>1990AB06<br>1970BO22<br>1967NA18<br>1959ST28<br>1957ST87 |
| 47 | 62 | $^{109}$Ag | 88.0337 (10) | 7/2+ | 39.7(2) *s* | 88.0336(10) | E3 | %IT = 100 | 2000YO07<br>1973CO10<br>1967AB07<br>1967MI11<br>1951WO15<br>1947BR05<br>1945WI11<br>1940AL01 |
| 48 | 61 | $^{109}$Cd | 59.60 (7) | 1/2+ | 11.8(18) $\mu s$ | 59.6(2) | E2 | %IT = 100 | 1968IV02<br>1956PE56 |
| 48 | 61 | $^{109}$Cd | 463.10 (11) | 11/2− | 10.6(6) $\mu s$ | 259.7(1) | M2 | %IT = 100 | 1975ME22<br>1968IV02<br>1966MC06<br>1964BR27 |
| 49 | 60 | $^{109}$In | 649.79 (10) | 1/2− | 1.34(7) *min* | 649.90(15) | M4 | %IT = 100 | 2020HO03<br>1968SM08<br>1966MA39<br>1956KU51 |
| 49 | 60 | $^{109}$In | 2101.86 (11) | 19/2+ | 210(1) *ms* | 673.5(1) | M3 | %IT = 100 | 2020HO03<br>1994BYZZ<br>1979VA13<br>1966WE01<br>1965AL15<br>1963PO10 |
| 41 | 69 | $^{110}$Nb$^{\dagger}$ | 0+X | (6−) | 75(1) *ms* | | | %$\beta$−= 100<br>%$\beta$−n = 31(15) | 2020HA14 |
| | | | 0+Y | (2−) | 94(9) *ms* | | | %$\beta$− = 100<br>%$\beta$−n < 36 | 2020HA14 |
| 45 | 65 | $^{110}$Rh | 0+Y | (6+) | 28.0(13) *s* | | | %$\beta$−= 100 | 1970PI01 |
| 47 | 63 | $^{110}$Ag | 1.112 (16) | 2− | 660(40) *ns* | 1.112(16) | E1 | %IT = 100 | 1975CL03 |
| 47 | 63 | $^{110}$Ag | 117.59 (5) | 6+ | 249.794(24) *d* | 116.48(5) | M4 | %IT = 1.33(8)<br>%$\beta$−= 98.67(8) | 2014UN01#<br>1983WA26<br>1980HO17<br>1962NI01<br>1957GE07<br>1950GU54<br>1938LI07 |

*Continued...*

*: The half-life is adopted from 2019KR05.

#: In weighted averaging procedure, value is taken from 2014UN01-Erratum: Appl. Radiat. Isot. 159, 108976 (2020).

Table 1 contd...

| Z | N | $^{A}$X | E(keV) | J$^{\pi}$ | $T_{1/2}$ | E$\gamma$(keV) | $\lambda$ | Decay mode | Reference |
|---|---|---|---|---|---|---|---|---|---|
| 47 | 63 | $^{110}$Ag | 118.719 (10) | 3+ | 36.6(7) *ns* | 117.607(17)<br>118.716(17) | E1(+M2) | %IT = 100 | 1976HA57<br>1974BE47<br>1971GU05<br>1967ES03<br>1967WIZZ<br>1963BE51 |
| 49 | 61 | $^{110}$In | 62.08 (4) | 2+ | 69.1(5) *min* | | | %$\epsilon$+%$\beta$+ = 100 | 1969SA20 |
| 51 | 59 | $^{110}$Sb | 1152.9 (10) | (8−) | 24(1) *ns* | 812 | (M2) | %IT = 100 | 1997LA13 |
| 42 | 69 | $^{111}$Mo | 0+X | (7/2−,<br>9/2−) | ≈ 200 *ms** | | | %$\beta$−= 100<br>%$\beta$−n = ? | 2011KU16* |
| 44 | 67 | $^{111}$Ru | 253.97 (14) | 7/2− | 14 *ns* | 103.8(2)<br>254.0(2) | E1 | %IT = 100 | 1970JO20 |
| 45 | 66 | $^{111}$Rh | 394.95 (17) | (3/2+) | 87(8) *ns* | 91.3(3)<br>395.0(3) | M1, E2 | %IT = 100 | 1990RO13 |
| 46 | 65 | $^{111}$Pd | 172.170 (10) | 11/2− | 5.565(13) *h* | 172.170(10) | E3 | %$\beta$−= 23.2(10)<br>%IT = 76.8(10) | 2019KR05<br>2015KR07<br>2014LI17<br>1957DZ14<br>1952MC34 |
| 47 | 64 | $^{111}$Ag | 59.82 (4) | 7/2+ | 64.8(8) *s* | 59.78(4) | E3 | %$\beta$−= 0.7(2)<br>%IT = 99.3(2) | 1974GR29 |
| 47 | 64 | $^{111}$Ag | 376.71 (5) | 3/2+ | 16(1) *ns* | 87.0<br>316.90(9)<br>376.71(6) | (E1)<br>(E2)<br>(E1) | %IT = 100 | 1977GL06 |
| 48 | 63 | $^{111}$Cd | 245.390 (16) | 5/2+ | 84.5(4) *ns* | 245.395(20) | E2 | %IT = 100 | 1968MC04<br>1957MA26<br>1957SI63 |
| 48 | 63 | $^{111}$Cd | 396.214 (21) | 11/2− | 48.53(5) *min* | 150.824(13) | E3 | %IT = 100 | 2013YO02<br>1997WE13<br>1987NE01<br>1968BO28<br>1949HE06<br>1948HO37<br>1945WI11 |
| 49 | 62 | $^{111}$In | 536.99 (7) | 1/2− | 7.7(2) *min* | 537.22(9) | M4 | %IT = 100 | 1969SH11<br>1968SM08<br>1966MA39 |
| 49 | 62 | $^{111}$In | 2716.79 (14) | 21/2+ | 13.7(4) *ns* | 255.3(2) | E2 | %IT = 100 | 1981VA15<br>1980LE05<br>1978HE10 |
| 50 | 61 | $^{111}$Sn | 254.71 (4) | 1/2+ | 12.5(10) *µs* | 100.24(3) | E2 | %IT = 100 | 1978HO06 |
| 50 | 61 | $^{111}$Sn | 978.6 (3) | 11/2− | 10.0(5) *ns* | 978.6(3) | M2 | %IT = 100 | 1984PR06 |
| 52 | 59 | $^{111}$Te | 839.4 (6) | 11/2− | 32.2(14) *ns* | 723 | M2 | %IT = 100 | 2000ST03 |
| 43 | 69 | $^{112}$Tc | 350.0 (15) | (5+) | 150(17) *ns* | 92(1) | | %IT = 100 | 2010BR15 |
| 45 | 67 | $^{112}$Rh | 0+Y | (6+) | 6.76(12) *s* | | | %$\beta$−= 100 | 1999LH01<br>1988AY02 |

*Continued...*

*: The half-life is only an estimate by 2011KU16 since it could not be distinguished from the half-life of the g.s.

Table 1 contd...

| Z | N | $^{A}$X | E(keV) | J$^{\pi}$ | $T_{1/2}$ | Eγ(keV) | λ | Decay mode | Reference |
|---|---|---|---|---|---|---|---|---|---|
| 49 | 63 | $^{112}$In | 156.592 (25) | 4+ | 20.76(8) *min* | 156.61(3) | M3 | %IT = 100 | 1983RY06<br>1980AD04<br>1968RO03<br>1968KO25<br>1962RU05<br>1953BL44 |
| 49 | 63 | $^{112}$In | 350.80 (5) | (7)+ | 0.69(5) *μs* | 187.93(3) | E2 | %IT = 100 | 1976IO04 |
| 49 | 63 | $^{112}$In | 613.82 (6) | (8)− | 2.81(3) *μs* | 263.01(3) | E1+M2 | %IT = 100 | 1976IO04<br>1976IO05 |
| 50 | 62 | $^{112}$Sn | 2549.22 (14) | 6+ | 13.72(10) *ns* | 301.84(13) | E2 | %IT = 100 | 1989AN14<br>1981VA15<br>1981GO17<br>1980VA13<br>1969YA05 |
| 51 | 61 | $^{112}$Sb | 825.9 (4) | (8−) | 536(22) *ns* | 456.4(3) | M2(+E3) | %IT = 100 | 1982MA29 |
| 43 | 70 | $^{113}$Tc | 114.4 (5) | (5/2−) | 0.526(+16−15) *μs* | 114.4(5) | | %IT = 100 | 2012KA36<br>2010BR15 |
| 44 | 69 | $^{113}$Ru | ~120 | (7/2−) | 510(30) *ms* | | | %β−≈ 100 | 2007KU23<br>1998KU17 |
| 46 | 67 | $^{113}$Pd | 81.10 (10) | (9/2−) | 0.3(1) *s* | 81.1(1) | M2 | %IT = 100 | 2014KU28<br>1993PE11 |
| 47 | 66 | $^{113}$Ag | 43.5 (1) | 7/2+ | 68.7(16) *s* | 43.6(2) | E3 | %IT = 64(7)<br>%β−= 36(7) | 1975BRYM<br>1974GR29 |
| 47 | 66 | $^{113}$Ag | 222.08 (13) | 3/2+ | 23(2) *ns* | 222.06(20) | E1 | %IT = 100 | 1988FOZY |
| 47 | 66 | $^{113}$Ag | 273.59 (16) | (1/2) | ~ 30 *ns* | 51.5(2)<br>273.6(2) | | %IT = 100 | 1988FOZY |
| 48 | 65 | $^{113}$Cd | 263.54 (3) | 11/2− | 13.89(16) *y* | 263.7(3) | E5 | %IT = 0.0964(19)<br>%β−= 99.9036(19) | 2011KO01<br>1972WA11<br>1965FL02 |
| 48 | 65 | $^{113}$Cd | 316.206 (15) | 5/2+ | 10.8(3) *ns* | 17.78(9)<br>316.21(2) | M1<br>E2 | %IT = 100 | 1980OH01<br>1972RAZM |
| 49 | 64 | $^{113}$In | 391.699 (3) | 1/2− | 99.47(6) *min* | 391.698(3) | M4 | %IT = 100 | 1997WE13<br>1987NE01<br>1984IW06<br>1982RUZV<br>1982HOZJ<br>1975BU24<br>1972EM01<br>1971OO01<br>1971HA18<br>1970RO29<br>1970LE07<br>1970GO48<br>1969VA04<br>1967OK02<br>1958GI06<br>1940LA07<br>1939BA03 |
| 50 | 63 | $^{113}$Sn | 77.389 (19) | 7/2+ | 21.4(4) *min* | 77.38(2) | M3+E4 | %IT = 91.1(23)<br>%ϵ+%β+ = 8.9(23) | 1974HO17<br>1961SC12 |

*Continued...*

Table 1 contd...

| Z | N | $^{A}$X | E(keV) | J$^{\pi}$ | $T_{1/2}$ | E$\gamma$(keV) | $\lambda$ | Decay mode | Reference |
|---|---|---|---|---|---|---|---|---|---|
| 50 | 63 | $^{113}$Sn | 738.4 (3) | 11/2− | 86(2) *ns* | 661.0(3) | M2(+E3) | %IT = 100 | 1974BR29<br>1974DI18<br>1973ISZQ |
| 54 | 59 | $^{113}$Xe | 404.8 (4) | 11/2− | 6.9(3) $\mu s$ | 277.6(11) | M2 | %IT = 100 | 2013PR01 |
| 43 | 71 | $^{114}$Tc$^{\dagger}$ | 0+X<br>0+Y | (> 4)<br>(1+) | 100(20) *ms*<br>90(20) *ms* | | | %$\beta$−= 100<br>%$\beta$−= 100 | 2011RI01<br>2011RI01 |
| 45 | 69 | $^{114}$Rh | 0+X | (7−) | 1.85(5) *s* | | | %$\beta$−≈ 100 | 2003LH01<br>1988AY02 |
| 47 | 67 | $^{114}$Ag | 199 | (≤ 6+) | 1.50(5) *ms* | | | %IT = 100 | 1990PE10 |
| 49 | 65 | $^{114}$In | 190.2682 (8) | 5+ | 49.51(1) *d* | 190.2684(8) | E4 | %IT = 96.75(24)<br>%$\epsilon$+%$\beta$+ = 3.25(24) | 1972ME01 |
| 49 | 65 | $^{114}$In | 501.948 (3) | 8− | 42.6(9) *ms* | 311.665(6) | E3 | %IT = 100 | 2014DE19<br>1968KO25<br>1968AL08<br>1966ME02<br>1966MOZZ<br>1960MO19<br>1959GL56<br>1958DU80 |
| 50 | 64 | $^{114}$Sn | 3087.37 (7) | 7− | 733(14) *ns* | 272.3(1) | E2 | %IT = 100 | 1980VA13 |
| 51 | 63 | $^{114}$Sb | 45.86 (2) | (2+),<br>4+ | 26(3) *ns* | 45.87(3) | M1,(E2) | %IT = 100 | 1996ZI01 |
| 51 | 63 | $^{114}$Sb | 54.64 (3) | 3+ | 20.4(9) *ns* | 54.65(3) | M1,(E2) | %IT = 100 | 1996ZI01 |
| 51 | 63 | $^{114}$Sb | 495.5 (7) | (8−) | 219(12) $\mu s$ | 321.6 | M2+E3 | %IT = 100 | 1976KA19 |
| 53 | 61 | $^{114}$I | 265.9 | (7) | 6.2(5) *s* | 134.47(16) | M3 | %IT = 9(2)<br>%$\epsilon$+%$\beta$+ = 91(2) | 1995ZIZZ |
| 44 | 71 | $^{115}$Ru | 61.7+X | (9/2−) | 76(6) *ms* | ~20? | | %IT = 100 | 2011RI07<br>2010KU25 |
| 46 | 69 | $^{115}$Pd | 89.21 (16) | (7/2−) | 50(3) *s* | 89.3(2) | E3 | %IT = 8.0(20)<br>%$\beta$−= 92.0(20) | 1990FO07<br>1987FOZY |
| 47 | 68 | $^{115}$Ag | 41.16 (10) | 7/2+ | 18.0(7) *s* | 41.1(2) | E3 | %$\beta$−= 79.0(3)<br>%IT = 21.0(3) | 1974GR29 |
| 48 | 67 | $^{115}$Cd | 181.0 (5) | 11/2− | 44.56(24) *d* | | | %$\beta$−= 100 | 2013YO02<br>1971BA28<br>1969MOZS<br>1959WA13 |
| 49 | 66 | $^{115}$In | 336.244 (17) | 1/2− | 4.486(4) *h* | 336.241(25) | M4 | %IT = 95.0(7)<br>%$\beta$−= 5.0(7) | 1987NE01<br>1974HA39 |
| 50 | 65 | $^{115}$Sn | 612.81 (4) | 7/2+ | 3.26(8) $\mu s$ | 115.46(4) | E2 | %IT = 100 | 1967IV05 |
| 50 | 65 | $^{115}$Sn | 713.64 (12) | 11/2− | 159(1) $\mu s$ | 100.70(21) | M2 | %IT = 100 | 1964IV01 |
| 51 | 64 | $^{115}$Sb | 2796.26 (9) | (19/2)− | 159(3) *ns* | 157.84(6)<br>279.36(9) | E2<br>E2 | %IT = 100 | 1979SH03 |
| 52 | 63 | $^{115}$Te | < 20 | (1/2)+ | 6.7(4) *min* | | | %$\epsilon$+%$\beta$+ ≤ 100<br>%IT = ? | 1976WI11 |
| 52 | 63 | $^{115}$Te | 280.05 (20) | 11/2− | 7.5(2) $\mu s$ | 280.1(2) | M2 | %IT = 100 | 1972VA38 |
| 45 | 71 | $^{116}$Rh | X | (6−) | 0.57(5) *s* | | | %$\beta$−= 100 | 2001WA04 |
| 47 | 69 | $^{116}$Ag | 47.90 (10) | (3+) | 20(1) *s* | 47.9(1) | E3 | %IT = 7(4)<br>%$\beta$−= 93(4) | 2009BA52<br>2005BA94 |

*Continued...*

Table 1 contd. . .

| Z | N | $^{A}$X | E(keV) | J$^{\pi}$ | $T_{1/2}$ | E$\gamma$(keV) | $\lambda$ | Decay mode | Reference |
|---|---|---|---|---|---|---|---|---|---|
| 47 | 69 | $^{116}$Ag | 129.80 (22) | (6−) | 9.4(4) *s* | 81.9(2) | E3 | %IT = 8(4)<br>%$\beta$−= 92(4) | 2005BA94<br>1974GR29<br>1974BJ01<br>1971FO22<br>1970OSZZ |
| 49 | 67 | $^{116}$In | 127.267 (6) | 5+ | 54.66(25) *min* | | | %$\beta$−= 100 | 2006VO12<br>1986NE01<br>1972EM01<br>1972PA13<br>1965BR34<br>1963BE23<br>1957CA71<br>1953LO09<br>1953DO09<br>1949SI02<br>1947GR16<br>1945RU02 |
| 49 | 67 | $^{116}$In | 289.660 (6) | 8− | 2.18(4) *s* | 162.393(7) | E3 | %IT = 100 | 1963AL32<br>1962WH02<br>1961HE08<br>1960AL27 |
| 50 | 66 | $^{116}$Sn | 2365.975 (21) | 5− | 348(19) *ns* | 99.802(11)<br>1072.37(3) | E2<br>E3 | %IT = 100 | 1988VA13<br>1966RG02 |
| 50 | 66 | $^{116}$Sn | 3547.16 (17) | 10+ | 833(30) *ns* | 54.0(5)<br>319.1(1) | E2<br>M2 | %IT = 100 | 1978VAZK |
| 51 | 65 | $^{116}$Sb | 93.99 (5) | 1+ | 194(4) *ns* | 93.88(3) | E2 | %IT = 100 | 1993DI06 |
| 51 | 65 | $^{116}$Sb | 383 (40) | 8− | 60.3(6) *min* | | | %$\epsilon$+%$\beta$+ = 100 | 1972PA13<br>1967HA27<br>1961FI05 |
| 51 | 65 | $^{116}$Sb | 1158.79 (15) | 7+ | 10.6(7) *ns* | 349.7(3)<br>775.8(2) | E1+M2<br>E1 | %IT = 100 | 1992IO01<br>1982DU11 |
| 53 | 63 | $^{116}$I | 430.4 (5) | (7−) | 3.27 $\mu s$ | 44.3(3) | | %IT = 100 | 1990WU01 |
| 55 | 61 | $^{116}$Cs | 0+X | 4+,5,6 | 3.9(2) *s* | | | %$\epsilon$+%$\beta$+ = 100<br>%$\epsilon$p = 0.51(15)<br>%$\epsilon$ $\alpha$ = 0.008(2) | 1985TI02<br>1980MA16<br>1978DA07<br>1977BO28 |
| 44 | 73 | $^{117}$Ru | 185.0 (4) | | 2.487(+58−55) $\mu s$ | 82.5(5)?<br>127.4(5)<br>184.6(5) | (E1)<br><br>(D, E2) | %IT = 100 | 2012KA36 |
| 45 | 72 | $^{117}$Rh | 321.2 (10) | (3/2+) | 138(17) *ns* | 321.2(10) | | %IT = 100 | 2013LA25 |
| 46 | 71 | $^{117}$Pd | 203.3 (3) | (7/2−) | 19.1(7) *ms* | 71.5(3)<br>168.6(3) | M2<br>M2 | %IT = 100 | 2018KU13<br>1991PE10 |
| 47 | 70 | $^{117}$Ag | 28.6 (2) | (7/2+) | 5.34(5) *s* | 28.6(2) | E3 | %IT = 6.0(15)<br>%$\beta$−= 94.0(15) | 1990FO07<br>1974GR29 |

*Continued. . .*

Table 1 contd...

| Z | N | $^{A}$X | E(keV) | J$^{\pi}$ | $T_{1/2}$ | E$\gamma$(keV) | $\lambda$ | Decay mode | Reference |
|---|---|---|---|---|---|---|---|---|---|
| 48 | 69 | $^{117}$Cd | 136.4 (2) | 11/2− | 3.441(9) *h* | | | %$\beta$− = 100 | 2019GI09<br>2013YO02<br>1975Ta06<br>1972PA13<br>1972GR24<br>1969MO21<br>1967SC37<br>1966DE16<br>1963TA16<br>1940LA07 |
| 49 | 68 | $^{117}$In | 315.303 (11) | 1/2− | 116.2(3) *min* | 315.302(13) | M4 | %IT = 47.1(15)<br>%$\beta$−= 52.9(15) | 1982FU07<br>1955MC17<br>1950ME06<br>1940LA07 |
| 49 | 68 | $^{117}$In | 659.765 (13) | 3/2+ | 53.6(7) *ns* | 71.12(2)<br>344.459(10) | E1<br>E1 | %IT = 100 | 1976PI18<br>1968BE58<br>1968CH08<br>1967BA18 |
| 50 | 67 | $^{117}$Sn | 314.58 (4) | 11/2− | 13.94(3) *d* | 156.02(3)<br>314.3(3) | <br>M4 | %IT = 100 | 2020YO07<br>2016DO10<br>2014UN01<br>2003PO21<br>2002UN02<br>1977KA16 |
| 50 | 67 | $^{117}$Sn | 2406.4 (4) | (19/2+) | 1.75(7) $\mu s$ | 781.0(3)<br>813.4(3) | (E3)<br>(M2) | %IT = 100 | 1979HA12 |
| 51 | 66 | $^{117}$Sb | 3130.76 (19) | (25/2)+ | 355(17) $\mu s$ | 58.1(1) | M2 | %IT = 100 | 1972ME15<br>1970HE13 |
| 51 | 66 | $^{117}$Sb | 3230.7 (2) | (23/2−) | 290(5) *ns* | 16.4(2)<br>99.6(2) | (E2) | %IT = 100 | 1987IO01 |
| 52 | 65 | $^{117}$Te | 274.4 (1) | 5/2+ | 19.9(4) *ns* | 274.4(1) | E2 | %IT = 100 | 1981KA10<br>1979HA18 |
| 52 | 65 | $^{117}$Te | 296.1 (5) | (11/2−) | 103(3) *ms* | | (M2) | %IT = 100 | 1969BR02<br>1963DE37 |
| 53 | 64 | $^{117}$I | 353.59 (11) | (9/2)+ | 12.1(7) *ns* | 294.7(1)<br>353.6(2) | M1+(E2) | %IT = 100 | 1982GA21 |
| 53 | 64 | $^{117}$I | 5063.8 (7) | (35/2+) | 50 *ns* | 380.5(6)<br>777.2(6) | M1, E2<br>E2 | %IT = 100 | 1993WA21 |
| 54 | 63 | $^{117}$Xe | 205.52 (14) | (7/2−) | 17(8) *ns* | 205.6(2) | E1 | %IT = 100 | 1998LI06 |
| 54 | 63 | $^{117}$Xe | 229.78 (18) | (11/2−) | 59(20) *ns* | 24 | | %IT = 100 | 1998LI06 |
| 55 | 62 | $^{117}$Cs$^{\dagger}$ | 0+X<br>0+Y | (9/2+)<br>(3/2+) | 8.4(6) *s*<br>6.5(4) *s* | | | %$\epsilon$+%$\beta$+ = 100<br>%$\epsilon$+%$\beta$+ = ? | 1986MA41<br>1986MA41<br>1978DA07 |
| 47 | 71 | $^{118}$Ag | 45.79 (9) | 0(−), 1(−)<br>2(−) | ~ 0.1 $\mu s$ | 45.8(1) | M1 | %IT = 100 | 1989KO22 |
| 47 | 71 | $^{118}$Ag | 127.63 (10) | (4,5)+ | 2.0(2) *s* | 127.6(1) | E3 | %IT = 41<br>%$\beta$−= 59 | 2017KI08<br>1979HIZR |
| 47 | 71 | $^{118}$Ag | 279.37 (20) | (2+,<br>3+) | ~ 0.1 $\mu s$ | 151.6(2) | | %IT = 100 | 1989KO22 |
| 49 | 69 | $^{118}$In | ~ 60 | 5+ | 4.366(12) *min* | | | %$\beta$−= 100 | 1995ITZY<br>1969DO11 |

*Continued...*

Table 1 contd…

| Z | N | $^{A}X$ | E(keV) | $J^{\pi}$ | $T_{1/2}$ | Eγ(keV) | λ | Decay mode | Reference |
|---|---|---|---|---|---|---|---|---|---|
| 49 | 69 | $^{118}$In | ~ 200 | 8− | 8.5(3) *s* | 138.2(5) | (E3) | %IT = 98.6(3)<br>%β−= 1.4(3) | 1969HA08 |
| 50 | 68 | $^{118}$Sn | 2321.23 (4) | 5− | 20.8(6) *ns* | 40.8(1)<br>1091.51(8) | E1<br>E3 | %IT = 100 | 2021DA02<br>1969HA08<br>1962BO16<br>1961BO13 |
| 50 | 68 | $^{118}$Sn | 2574.91 (4) | 7− | 230(10) *ns* | 253.678(10) | E2 | %IT = 100 | 2021DA02<br>1980VA13<br>1961BO13 |
| 50 | 68 | $^{118}$Sn | 3108.06 (22) | 10+ | 2.52(6) *μs* | 55.9(2) | E2 | %IT = 100 | 2014IS01<br>1987LU06 |
| 51 | 67 | $^{118}$Sb | 50.814 (21) | (3)+ | 20.6(6) *μs* | 50.82(5) | E2 | %IT = 100 | 1975PL04 |
| 51 | 67 | $^{118}$Sb | 250 (6) | 8− | 5.08(5) *h* | | | %ϵ+%β+ = 100 | 2021DA02<br>1974CA06<br>1972PA13<br>1968KI06<br>1967HA27 |
| 51 | 67 | $^{118}$Sb | 269.82 (3) | (3)− | 13.3(2) *ns* | 103.65(2)<br>187.85(7)<br>238.56(2) | E1<br>E1 | %IT = 100 | 1985DI07<br>1974CH44 |
| 51 | 67 | $^{118}$Sb | 964.88 (4) | 7+ | 22.6(3) *ns* | 396.65(4)<br>714.9(1) | E1+M2 | %IT = 100 | 1997FA18<br>1985DI07<br>1983VA14 |
| 53 | 65 | $^{118}$I | 188.8 (7) | (7−) | 8.5(5) *min* | 104(2) | | %ϵ+%β+ < 100<br>%IT > 0 | 2003MO36<br>1968LA18 |
| 55 | 63 | $^{118}$Cs | 0+X | (7−) | 17(3) *s* | | M1 | %ϵ+%β+ = 100<br>%ϵp < 4.2E−2(6)<br>%ϵ α < 2.4E−3(4) | 2021ZH57<br>1977GE03 |
| 55 | 63 | $^{118}$Cs | 125.7+X | (7+) | 0.55(6) *μs* | 46.3<br>64.7<br>126 | <br><br>(E1) | | 2021ZH57 |
| 44 | 75 | $^{119}$Ru | 227.1 (7) | | 0.383(+22−21) *μs* | 90.8(5)? | (D, E2) | %IT = 100 | 2012KA36 |
| 47 | 72 | $^{119}$Ag | 33.4 (1) | (7/2+) | 2.1(1) *s* | | | %β−∼ 100 | 2022KU09<br>1975KA09 |
| 48 | 71 | $^{119}$Cd | 146.54 (11) | 11/2(−) | 2.20(2) *min* | | | %β−= 100 | 2013YO02<br>1976SC30 |
| 48 | 71 | $^{119}$Cd | 228.27 (9) | (7/2−,<br>9/2−) | 43(3) *ns* | 14.3<br>81.7(1) | <br>E2 | %IT = 100 | 1975KA09 |
| 49 | 70 | $^{119}$In | 311.37 (3) | 1/2− | 18.0(3) *min* | 311.39(3) | M4 | %IT = 4.4<br>%β−= 95.6 | 1976SC30 |
| 49 | 70 | $^{119}$In | 654.27 (7) | (3/2)+ | 130(15) *ns* | 50.2(1)<br>343.0(1) | <br>E1 | %IT = 100 | 1976SC30<br>1974MC09 |
| 49 | 70 | $^{119}$In | 2656.9 (18) | (25/2+) | 265(11) *ns* | 152 | (E2) | %IT = 100 | 2020BI06<br>2002LU15 |
| 50 | 69 | $^{119}$Sn | 23.871 (8) | 3/2+ | 18.03(7) *ns* | 23.870(8) | M1+E2 | %IT = 100 | 1980LA03 |
| 50 | 69 | $^{119}$Sn | 89.531 (13) | 11/2− | 293.1(7) *d* | 65.66(1) | M4 | %IT = 100 | 2020YO07<br>2003PO21<br>1988GEZS<br>1976MA63 |
| 50 | 69 | $^{119}$Sn | 2127.0 (10) | (19/2+) | 9.6(12) *μs* | 748(1)<br>818(1) | | %IT = 100 | 1992MA27 |

*Continued…*

Table 1 contd. . .

| Z | N | $^{A}X$ | E(keV) | $J^{\pi}$ | $T_{1/2}$ | Eγ(keV) | λ | Decay mode | Reference |
|---|---|---|---|---|---|---|---|---|---|
| 50 | 69 | $^{119}$Sn | 2368.89 (20) | 23/2+ | 0.096(9) $\mu s$ | 241.7(2) | | %IT = 100 | 2016IS03 |
| 50 | 69 | $^{119}$Sn | 3101.75 (19) | 27/2− | 0.039(3) $\mu s$ | 174.7(1) | | %IT = 100 | 2016IS03 |
| | | | | | | | | | 2012AS05 |
| | | | | | | | | | 1994MA48 |
| 51 | 68 | $^{119}$Sb | 2553.6 (3) | (19/2−) | 130(3) $ns$ | 134 | (E1+M2) | %IT = 100 | 1991IO02 |
| | | | | | | 240 | (E2) | | 1987LU06 |
| 51 | 68 | $^{119}$Sb | 2799 (30) | (25/2+) | 835(90) $ms$ | 288.2 | (E3) | %IT = 100 | 2019MI18 |
| | | | | | | | | | 2005PO03 |
| | | | | | | | | | 1987LU06 |
| | | | | | | | | | 1979SH03 |
| 52 | 67 | $^{119}$Te | 260.96 (5) | 11/2− | 4.70(5) $d$ | | | $\%\epsilon+\%\beta+ = 100$ | 1975DU04 |
| | | | | | | | | $\%\beta+ = 0.41(4)$ | 1973KA45 |
| | | | | | | | | %IT < 0.008 | 1963KA04 |
| | | | | | | | | | 1961FI05 |
| | | | | | | | | | 1960KO12 |
| | | | | | | | | | 1960SO02 |
| 53 | 66 | $^{119}$I | 306.66 (14) | 9/2+ | 34.6(5) $ns$ | 207.8(2) | M1+E2 | %IT = 100 | 1982DA17 |
| | | | | | | 306.9(2) | E2 | | |
| 54 | 65 | $^{119}$Xe | 176.08 (5) | (7/2−) | 20(4) $ns$ | 176.05(5) | E1 | %IT = 100 | 1982BA31 |
| 54 | 65 | $^{119}$Xe | 243.39 (11) | (11/2−) | 27(5) $ns$ | 67.3(1) | E2 | %IT = 100 | 1982BA31 |
| 54 | 65 | $^{119}$Xe | 245.99 (11) | (1/2+) | 70 $ns$ | 76.4(2) | E2 | %IT = 100 | 2001GE01 |
| | | | | | | 246.2(2) | E2 | | |
| 55 | 64 | $^{119}$Cs | 0+X | $3/2^{(+)}$ | 30.4(1) $s$ | | | $\%\epsilon+\%\beta+ = 100$ | 1984ICZY |
| 55 | 64 | $^{119}$Cs | 110.0 (7) | 11/2− | 55(5) $\mu$s | (23) | (M2) | | 2021ZH54 |
| | | | | | | (25) | (E1) | | 2021ZH56 |
| 56 | 63 | $^{119}$Ba | 66.0 | (5/2−) | 0.36(2) $\mu$s | 66.0 | | | 2021ZH38 |
| 45 | 75 | $^{120}$Rh | 157.2 (7) | | 0.294(+16−15) $\mu s$ | 59.1(5)? | (D, E2) | %IT = 100 | 2012KA36 |
| 47 | 73 | $^{120}$Ag$^{\dagger}$ | 0+X | (0−, 1−) | 0.94(10) $s$ | | | $\%\beta-\approx 100$ | 2012BA58 |
| | | | 0+Y | (3+, 4+) | 1.52(7) $s$ | | | $\%\beta-\approx 100$ | 2012BA58 |
| 47 | 73 | $^{120}$Ag | 203.2+Y | (6−, 7−) | 0.41(3) $s$ | 203.1(3) | E3 | $\%\beta-\approx 32(5)$ | 2012BA58 |
| | | | | | | | | $\%IT \approx 68(10)$ | 2007RI17 |
| | | | | | | | | | 2003WA13 |
| | | | | | | | | | 1971FO22 |
| 49 | 71 | $^{120}$In | 0+X | (8−) | 47.3(5) $s$ | | | $\%\beta-= 100$ | 1960PO06 |
| 49 | 71 | $^{120}$In | 0+Y | (5)+ | 46.2(8) $s$ | | | $\%\beta-= 100$ | 1978CH25 |
| 50 | 70 | $^{120}$Sn | 2481.63 (6) | 7− | 11.8(5) $\mu s$ | 197.37(2) | E2 | %IT = 100 | 1960IK01 |
| 50 | 70 | $^{120}$Sn | 2902.22 (22) | 10+ | 6.26(11) $\mu s$ | 65.7(2) | (E2) | %IT = 100 | 2012AS05 |
| | | | | | | | | | 1987LU06 |
| 50 | 70 | $^{120}$Sn | 4891.1 (3) | 15− | 30(3) $ns$ | 241.7(1) | | %IT = 100 | 2014IS04 |
| | | | | | | | | | 2012AS05 |
| 51 | 69 | $^{120}$Sb | 0+X | 8− | 5.76(2) $d$ | | | $\%\epsilon+\%\beta+ = 100$ | 1967HA27 |
| 51 | 69 | $^{120}$Sb | 78.16 (5) | 3+ | 246(2) $ns$ | 69.78(4) | E1 | %IT = 100 | 1979AD02 |
| | | | | | | | | | 1976IO03 |
| 51 | 69 | $^{120}$Sb | 2328.3+X | 13+ | 400(8) $ns$ | 204.4(3) | | %IT = 100 | 2014LI31 |
| | | | | | | | | | 1987LU06 |
| | | | | | | | | | 1984QUZX |
| 51 | 69 | $^{120}$Sb | 2884.7+X | 16− | 14(3) $ns$ | 148.3(3) | | %IT = 100 | 2014LI31 |
| | | | | | | | | | 1984QUZX |
| 53 | 67 | $^{120}$I | 25.07 (8) | 1+ | 13.6(7) $ns$ | 25.1(2) | E1 | %IT = 100 | 1974MU10 |
| 53 | 67 | $^{120}$I | 72.3 (3) | 3+ | 244(3) $ns$ | 13.5(3) | | %IT = 100 | 2018MO16 |
| | | | | | | 72.3(3) | E1 | | 1974MU10 |

*Continued. . .*

Table 1 contd...

| Z | N | $^{A}$X | E(keV) | $J^{\pi}$ | $T_{1/2}$ | E$\gamma$(keV) | $\lambda$ | Decay mode | Reference |
|---|---|---|---|---|---|---|---|---|---|
| 53 | 67 | $^{120}$I | 132 (30) | (7−) | 53(4) *min* | | | %$\epsilon$+%$\beta$+ = 100 | 2018MO16<br>1968LA18 |
| 53 | 67 | $^{120}$I | 6418.3+X | (25+) | 49(2) *ns* | | | %IT = 100 | 2018MO16 |
| 55 | 65 | $^{120}$Cs | 0+X | (7−) | 57(6) *s* | | | %$\epsilon$+%$\beta$+ = 100 | 1977GE03 |
| 46 | 75 | $^{121}$Pd | 135.5 (5) | | 0.460(+85−92) $\mu s$ | 135.5(5) | E2 | %IT = 100 | 2012KA36 |
| 46 | 75 | $^{121}$Pd | 135.5+X | | 0.463(+83−94) $\mu s$ | | | | 2012KA36 |
| 48 | 73 | $^{121}$Cd | 214.86 (15) | 11/2(−) | 8.3(8) *s* | | | %$\beta$−= 100 | 2013YO02<br>1982FO04 |
| 48 | 73 | $^{121}$Cd | 314.50 (6) | (5/2+,<br>7/2+) | 19(2) *ns* | 314.55(10) | (E2) | %IT = 100 | 1982FO10<br>1982AL29 |
| 49 | 72 | $^{121}$In | 313.68 (7) | 1/2− | 3.88(10) *min* | 313.60(9) | M4 | %IT = 1.2(2)<br>%$\beta$−= 98.8(2) | 1976FO02<br>1974GR29 |
| 49 | 72 | $^{121}$In* | 2447.0+Y | (25/2+) | 350(50) *ns** | | | %IT = 100 | 2002LU15 |
| 49 | 72 | $^{121}$In* | 2448.0 (10) | (21/2−) | 17(2) $\mu s$* | 214? | | | 2010RE01 |
| 50 | 71 | $^{121}$Sn | 6.31 (6) | 11/2− | 43.9(5) *y* | 6.29(8) | | %IT = 77.6(20)<br>%$\beta$−= 22.4(20) | 2020YO07<br>2002RE18 |
| 50 | 71 | $^{121}$Sn | 1998.3 (11) | 19/2+ | 5.3(5) $\mu s$ | 752<br>841 | | %IT = 100 | 2012AS05<br>1995DA26 |
| 50 | 71 | $^{121}$Sn | 2221.98 (19) | 23/2+ | 0.52(5) $\mu s$ | 223.4(2) | E2 | %IT = 100 | 2016IS03<br>2012AS05 |
| 50 | 71 | $^{121}$Sn | 2833.9 (2) | 27/2− | 0.167(25) $\mu s$ | 175.4(2) | | %IT = 100 | 2016IS03<br>1995DA26 |
| 51 | 70 | $^{121}$Sb | 2721.5+X | (25/2) | 179(6) $\mu s$ | | | %IT = 100 | 2009WA02<br>2008KO03<br>2008JO03 |
| 52 | 69 | $^{121}$Te | 293.974 (22) | 11/2− | 164.7(7) *d* | 81.788(18) | M4 | %IT = 85.9(28)<br>%$\epsilon$+%$\beta$+ = 14.1(28) | 2019JO03<br>2008EA01 |
| 52 | 69 | $^{121}$Te | 443.11 (3) | 7/2+ | 85.3(5) *ns* | 230.95(5) | E2 | %IT = 100 | 1980IO01 |
| 53 | 68 | $^{121}$I | 2353.0 (3) | (21/2) | 80(12) *ns* | 134.8(1) | | %IT = 100 | 1982HA46 |
| 53 | 68 | $^{121}$I | 2376.9 (4) | | 9.0(14) $\mu s$ | 158.7(3) | | %IT = 100 | 1982GA21 |
| 54 | 67 | $^{121}$Xe | 153.99 (11) | (1/2+) | 80(15) *ns* | 153.9(2) | E2 | %IT = 100 | 1991GE02 |
| 55 | 66 | $^{121}$Cs | 68.5 (3) | 9/2(+) | 122(3) *s* | 68.5(3) | M3 | %$\epsilon$+%$\beta$+ = 83<br>%IT = 17 | 1991GE02 |
| 45 | 77 | $^{122}$Rh | 271.0 (7) | | 0.82(+13−11) $\mu s$ | 63.9(5)? | (E2) | %IT = 100 | 2012KA36 |
| 47 | 75 | $^{122}$Ag | 0+X | (1−) | 0.55(5) *s* | | | %$\beta$−= ?<br>%IT = ? | 2000KR18 |
| 47 | 75 | $^{122}$Ag | 0+Y | (9−) | 0.20(5) *s* | | | %$\beta$−= ? | 2000KR18 |
| 47 | 75 | $^{122}$Ag | 91+X | (1+) | 6.3(10) $\mu s$ | 91? | | %IT = 100 | 2013LA11 |
| 49 | 73 | $^{122}$In | 0+X | 5+ | 10.3(6) *s* | | | %$\beta$−= 100 | 1987EB02<br>1979CH10<br>1971TA07 |
| 49 | 73 | $^{122}$In | 0+Y$^{\#}$ | (8−) | 10.8(4) *s* | | | %$\beta$−= 100 | 1979CH10<br>1978AL18 |
| 50 | 72 | $^{122}$Sn | 2409.03 (4) | 7− | 7.5(9) $\mu s$ | 163.22(3) | E2 | %IT = 100 | 2012AS05<br>1979CH10<br>1979FO10 |

*Continued...*

*: Further investigations are required to clarify the situation about the the (25/2+) and (21/2−) isomers.

#: From $\beta$-$\gamma$ coincidence data in 1978AL18 and 1971TA07 for the isomers, and and Q($\beta$) for $^{122}$In g.s. in AME2020, we obtain energies of 172(205) keV and 220(190) keV for the 5+ and (8−) isomers, respectively. Energy of 290(140) keV for the (8−) isomer in NUBASE2020 and in ENSDF database seems erroneous.

Table 1 contd...

| Z | N | $^{A}$X | E(keV) | J$^{\pi}$ | $T_{1/2}$ | E$\gamma$(keV) | $\lambda$ | Decay mode | Reference |
|---|---|---|---|---|---|---|---|---|---|
| 50 | 72 | $^{122}$Sn | 2765.3 (4) | 10+ | 62(3) $\mu s$ | 75.2(5) | E2 | %IT = 100 | 2017KI09<br>2014IS04<br>2012AS05<br>1992BR06 |
| 50 | 72 | $^{122}$Sn | 4478.7 (3) | 13− | 40(3) $ns$ | 70.2(5)<br>264.5(2)<br>609.6(1)<br>1712.8(5) | <br>(E2)<br>(E1)<br> | %IT = 100 | 2014IS04 |
| 50 | 72 | $^{122}$Sn | 4721.2 (3) | 15− | 139(12) $ns$ | 242.5(1) | (E2) | %IT = 100 | 2014IS04<br>2012AS05 |
| 51 | 71 | $^{122}$Sb | 61.4131 (5) | 3+ | 1.86(8) $\mu s$ | 61.4127(5) | E1 | %IT = 100 | 1973HE10 |
| 51 | 71 | $^{122}$Sb | 137.4726 (8) | (5)+ | 0.53(3) $ms$ | 76.0595(7) | E2 | %IT = 100 | 1963DE05 |
| 51 | 71 | $^{122}$Sb | 163.5591 (17) | (8)− | 4.191(3) $min$ | 26.0867(24) | (E3) | %IT = 100 | 1990AB06 |
| 53 | 69 | $^{122}$I | 155.7 (3) | (4+) | 16.6(2) $ns$ | 46.0(3)<br>94.4(3) | <br>E2 | %IT = 100 | 2019MO28<br>2012MOZZ |
| 53 | 69 | $^{122}$I | 315.0 (4) | (7−) | 193.3(9) $ns$ | 159.3(3) | E3 | %IT = 100 | 2019MO28<br>2012MOZZ |
| 53 | 69 | $^{122}$I | 379.5 (5) | (7−) | 79.1(12) $\mu s$ | 64.5(3) | M1+E2 | %IT = 100 | 2019MO28<br>2012MOZZ |
| 53 | 69 | $^{122}$I | 393.9 (5) | 8+ | 78.2(4) $\mu s$ | 51.0(3) | E2 | %IT = 100 | 2019MO28<br>2012MOZZ |
| 53 | 69 | $^{122}$I | 444.2 (5) | 8− | 146.5(12) $ns$ | 64.7(3) | M1+E2 | %IT = 100 | 2019MO28<br>2012MOZZ |
| 55 | 67 | $^{122}$Cs | 45.87 (12) | (3)+ | > 1 $\mu s$ | 45.85(15) | E2 | %IT = 100 | 1987WEZW |
| 55 | 67 | $^{122}$Cs | 127.07 (16) | (5)− | 0.36(2) $s$ | 81.20(10) | M2 | %IT = 100 | 1983WE07 |
| 55 | 67 | $^{122}$Cs | 135 (15) | 8(−) | 3.70(11) $min$ | | | %$\epsilon$+%$\beta$+ = 100 | 1999AM05<br>1993AL03 |
| 47 | 76 | $^{123}$Ag | 1365.2+X | | 202(20) $ns$ | 389.5(5) | | %IT = 100 | 2013LA11 |
| 47 | 76 | $^{123}$Ag | 1472.8 (8) | (17/2−) | 393(16) $ns$ | (48(1))<br>732.1(5) | | %IT = 100 | 2013LA11<br>2009ST28 |
| 48 | 75 | $^{123}$Cd | 144 (4) | 11/2(−) | 1.80(3) $s$ | | | %$\beta$−≤ 100<br>%IT = ? | 2014TEZY<br>2013YO02<br>2013KA08<br>1989HU03<br>1986MA42 |
| 48 | 75 | $^{123}$Cd | 263.867 (20) | (7/2+) | 80(15) $ns$ | 263.87(2) | | %IT = 100 | 1989HU10 |
| 49 | 74 | $^{123}$In | 327.21 (4) | (1/2)− | 47.4(4) $s$ | | | %$\beta$−= 100 | 1986GO10<br>1974GR29<br>1960YU01 |
| 49 | 74 | $^{123}$In | 2078.1 (6) | (17/2−) | 1.4(2) $\mu s$ | 31.5(5) | (E2) | %IT = 100 | 2004SC42 |
| 49 | 74 | $^{123}$In | 2079+X | (21/2−) | ≥ 100 $\mu s$ | | | %IT = 100 | 2010RE01 |
| 50 | 73 | $^{123}$Sn | 24.6 (4) | 3/2+ | 40.06(1) $min$ | | | %$\beta$−= 100 | 2020YO07<br>1990AB06 |
| 50 | 73 | $^{123}$Sn | 1945.0 (10) | 19/2+ | 7.4(26) $\mu s$ | 19<br>728<br>838 | | %IT = 100 | 2012AS05<br>1992MA27 |
| 50 | 73 | $^{123}$Sn | 2153.0 (12) | 23/2+ | 6 $\mu s$ | 208 | | %IT = 100 | 2012AS05<br>1995DA26<br>1994MA48 |

*Continued...*

Table 1 contd...

| Z | N | $^{A}$X | E(keV) | J$^{\pi}$ | $T_{1/2}$ | E$\gamma$(keV) | $\lambda$ | Decay mode | Reference |
|---|---|---|---|---|---|---|---|---|---|
| 50 | 73 | $^{123}$Sn | 2713.0 (14) | 27/2− | 34 $\mu s$ | 170<br>560 | | %IT = 100 | 2012AS05<br>1995DA26<br>1994MA48 |
| 51 | 72 | $^{123}$Sb | 2037.4 (3) | 15/2− | 37.3(8) $ns$ | 381.2(3)<br>949.0<br>1007.3 | E2 | %IT = 100 | 2009WA02 |
| 51 | 72 | $^{123}$Sb | 2237.9 (4) | (19/2−) | 214(3) $ns$ | 200.5(3) | | %IT = 100 | 2019BI04<br>2009WA02 |
| 51 | 72 | $^{123}$Sb | 2613.6 (5) | 23/2+ | 65(1) $\mu$s | 127.6(3)<br>375.7 | (E2) | %IT = 100 | 2019BI04<br>2009WA02 |
| 52 | 71 | $^{123}$Te | 247.45 (4) | 11/2− | 119.3(1) $d$ | 88.46(3)<br>247.5(2) | M4 | %IT = 100 | 1992CO11<br>1987JA13<br>1970EMZY |
| 52 | 71 | $^{123}$Te | 489.78 (5) | 7/2+ | 30.7(4) $ns$ | 330.75(6) | | %IT = 100 | 1981IO05 |
| 53 | 70 | $^{123}$I | 2659.9 (3) | (21/2+) | 26.6(20) $ns$ | 298.07(20) | (M1) | %IT = 100 | 2011MO28<br>1982SH20<br>1977HA36 |
| 54 | 69 | $^{123}$Xe | 185.30 (16) | 7/2(−) | 5.7(3) $\mu s$ | 4.6? | | %IT = 100 | 1982ZE05<br>1981LU01<br>1970KE01 |
| 54 | 69 | $^{123}$Xe | 206.38 (19) | (9/2−) | 11.8(14) $ns$ | 21.1(2) | | %IT = 100 | 1982ZE05 |
| 55 | 68 | $^{123}$Cs | 156.27 (5) | 11/2− | 1.64(12) $s$ | 61.70(5) | E3 | %IT = 100 | 2004SI27<br>1981MA01<br>1972DR06 |
| 55 | 68 | $^{123}$Cs | 231.63+X | (9/2+) | 114(5) $ns$ | | | %IT = 100 | 2000GI12 |
| 56 | 67 | $^{123}$Ba | 92.68 (10) | (7/2−) | ~ 40 $ns$ | 92.7(1) | (E1) | %IT = 100 | 1979YO06 |
| 56 | 67 | $^{123}$Ba | 120.98 (23) | (1/2+) | 0.83(6) $\mu s$ | 120.9(3) | (E2) | %IT = 100 | 1991II02 |
| 46 | 78 | $^{124}$Pd | 62.2+X (17) | | > 20 $\mu s$ | 62.2(17) | | %IT = 100 | 2012KA36 |
| 47 | 77 | $^{124}$Ag* | 0+X | (2−, 3+) | 191(28) $ms$ | | | %$\beta$−= 100<br>%$\beta$−n = 1.3 (9) | 2014BA18<br>2013LA11 |
| | | | 0+Y | (8−) | 144(20) $ms$ | | | %$\beta$−= 100 | 2014BA18 |
| 47 | 77 | $^{124}$Ag | 155.6+X (5)? | (1+) | 0.14(5) $\mu s$ | 155.6(5)? | | %IT = 100 | 2013LA11<br>2012KA36 |
| 47 | 77 | $^{124}$Ag | 231.1+X (7) | (1−) | 1.47(20) $\mu s$ | 75.5(5)? | (E1) | %IT = 100 | 2013LA11<br>2012KA36<br>2005WAZY |
| 49 | 75 | $^{124}$In | ≈ 50 | (8−) | 3.67(3) $s$ | | | %$\beta$−= 100 | 2014LE20<br>1986GO10 |
| 49 | 75 | $^{124}$In | 242.68 (1) | (1)+ | 50(6) $ns$ | 62.80(10) | E1 | %IT = 100 | 1974FO23 |
| 50 | 74 | $^{124}$Sn | 2204.620 (23) | 5− | 0.27(6) $\mu s$ | 102.91(2)<br>1072.88(2) | E1 | %IT = 100 | 2012AS05<br>1979FO10 |
| 50 | 74 | $^{124}$Sn | 2325.01 (4) | 7− | 2.84(12) $\mu s$ | 120.38(3) | E2 | %IT = 100 | 2017KI09<br>2012AS05<br>1979FO10 |
| 50 | 74 | $^{124}$Sn | 2656.6 (5) | 10+ | 51(5) $\mu s$ | 78.2(5) | E2 | %IT = 100 | 2017KI09<br>2012AS05<br>1992BR06 |
| 50 | 74 | $^{124}$Sn | 4551.4 (7) | 15− | 260(25) $ns$ | 228.5(4) | (E2) | %IT = 100 | 2014IS04<br>2012AS05 |

*Continued...*

*: Higher spin level may be the isomer.

Table 1 contd...

| Z | N | $^{A}X$ | E(keV) | J$^{\pi}$ | $T_{1/2}$ | E$\gamma$(keV) | $\lambda$ | Decay mode | Reference |
|---|---|---|---|---|---|---|---|---|---|
| 51 | 73 | $^{124}$Sb | 10.8627 (8) | 5+ | 93(5) *s* | 10.8630(11) | | %IT = 75(5)<br>%$\beta$−= 25(5) | 1962VA18 |
| 51 | 73 | $^{124}$Sb | 36.8440 (14) | (8)− | 20.2(2) *min* | 25.981(3) | E3 | %IT = 100 | 1967HA27 |
| 51 | 73 | $^{124}$Sb | 40.8038 (7) | 3+,4+ | 3.2(3) $\mu s$ | 29.940(1)<br>40.8041(10) | E2<br>E1 | %IT = 100 | 1980AL22<br>1973SHZZ |
| 51 | 73 | $^{124}$Sb | 125.2291 (13) | (6)− | 86(2) *ns* | 37.629(3)<br>88.3852(6)<br>114.3679(13) | E2<br>E2 | %IT = 100 | 1980AL22<br>1973SHZZ |
| 53 | 71 | $^{124}$I | 55.2 (3) | 3+ | 63.5(7) *ns* | 55.2(3) | E1 | %IT = 100 | 2021MO09<br>1982BU12 |
| 53 | 71 | $^{124}$I | 122.8 (3) | 4− | 10.6(3) *ns* | 122.8(3) | E2 | %IT = 100 | 2021MO09<br>1982BU12 |
| 53 | 71 | $^{124}$I | 249.8 (4) | (4+) | 13.4(8) *ns* | 87.0(3)<br>194.6(3) | | %IT = 100 | 2021MO09 |
| 53 | 71 | $^{124}$I | 689.1 (5) | 8− | 13.7(4) *ns* | 130.4(3)<br>132.6(3)<br>235.7(3)<br>378.3(3) | M1+E2<br>M1+E2<br>M1+E2<br>M1+E2 | %IT = 100 | 2021MO09 |
| 55 | 69 | $^{124}$Cs | 301.14 (7) | (4)− | 69(3) *ns* | (30.8)<br>58.20(8)<br>89.50(8) | <br>E1<br>E1 | %IT = 100 | 1983WE07 |
| 55 | 69 | $^{124}$Cs | 462.54 (9) | (7)+ | 6.41(7) *s* | 64.90(5)<br>161.0 | M2<br>(E3) | %$\epsilon$+%$\beta$+ =0.11(2)<br>%IT = 99.89(2) | 2017RA20<br>2014LE20<br>1983WE07 |
| 57 | 67 | $^{124}$La$^{\dagger}$ | 0+X | | 21(4) *s* | | | %$\epsilon$+%$\beta$+ = 100 | 1997AS05 |
| | | | 0+Y | (8−) | 29.21(17) *s* | | | %$\epsilon$+%$\beta$+ = 100 | 1997AS05 |
| 46 | 79 | $^{125}$Pd | 1805.23 (18) | (23/2+) | 144(4) *ns* | 108.4(1) | E2 | %IT = 100 | 2019WA14 |
| 47 | 78 | $^{125}$Ag | 97.1 | (1/2−) | 159(21) *ms* | | | %$\beta$− ≈ 100<br>%$\beta$–n = 4.6(10)<br>%IT = ? | 2021BA34<br>2019CH24 |
| 47 | 78 | $^{125}$Ag | 859.9+X (14) | | 80(17) *ns* | 147.0(10) | | %IT = 100 | 2013LA11 |
| 47 | 78 | $^{125}$Ag | 1501.2 (6) | (17/2−) | 0.491(20) $\mu s$ | 102.9(5)<br>786.9(5) | (E2) | %IT = 100 | 2013LA11<br>2012KA36<br>2009ST28 |
| 48 | 77 | $^{125}$Cd | 188.5 (2) | 11/2(−) | 0.48(3) *s* | | | %$\beta$−= 100 | 2021BA34<br>2017LA16<br>2013YO02<br>2013KA08<br>1989HU03 |
| 48 | 77 | $^{125}$Cd | 1461.6+X (7) | (19/2+) | 19(3) $\mu s$ | 742.2(5) | | %IT = 100 | 2012KA36<br>2011SI32 |
| 49 | 76 | $^{125}$In | 360.12 (9) | 1/2(−) | 12.2(2) *s* | | | %$\beta$−= 100 | 1986GO10<br>1983SH07<br>1974GR29 |
| 49 | 76 | $^{125}$In | 2009.4 (7) | (19/2+) | 9.4(6) $\mu s$ | 56.3 | E2 | %IT = 100 | 1998FOZY |
| 49 | 76 | $^{125}$In | 2161.2 (9) | (23/2−) | 5.0(15) *ms* | 151.8 | M2 | %IT = 100 | 1998FOZY |

*Continued...*

Table 1 contd...

| Z | N | $^{A}$X | E(keV) | J$^{\pi}$ | $T_{1/2}$ | E$\gamma$(keV) | $\lambda$ | Decay mode | Reference |
|---|---|---|---|---|---|---|---|---|---|
| 50 | 75 | $^{125}$Sn | 27.50 (14) | 3/2+ | 9.70(10) *min* | | | %$\beta$−= 100 | 2020YO07<br>2020GU04<br>1968ER03<br>1950NE52<br>1950NOAA<br>1949LE05 |
| 50 | 75 | $^{125}$Sn | 1892.8 (3) | (19/2+) | 6.2(2) $\mu s$ | ~ 10<br>805.5(2) | | %IT = 100 | 2008LO07 |
| 50 | 75 | $^{125}$Sn | 2059.69 (17) | 23/2+ | 0.65(6) $\mu s$ | 166.8(2) | | %IT=100 | 2016IS03<br>2012AS05<br>2008LO07 |
| 50 | 75 | $^{125}$Sn | 2623.5 (5) | 27/2− | 0.23(2) $\mu s$ | 161.3(3) | | %IT = 100 | 2012AS05<br>2008LO07 |
| 51 | 74 | $^{125}$Sb | 1971.25 (20) | 15/2− | 4.1(2) $\mu s$ | 881.8(3)<br>904.0(3) | E3(+M2)<br>E3 | %IT = 100 | 2007JU06 |
| 51 | 74 | $^{125}$Sb | 2112.1 (3) | 19/2− | 28.4(6) $\mu s$ | 140.9(3) | E2 | %IT=100 | 2019BI04<br>2010RE01<br>2007JU06 |
| 51 | 74 | $^{125}$Sb | 2325.0 (3) | (19/2)+ | 31(2) *ns* | 107.9(3)<br>131.59(17)<br>331.36(24) | <br>E2<br>(E2) | %IT = 100 | 2007JU06 |
| 51 | 74 | $^{125}$Sb | 2471.0 (4) | 23/2+ | 277(7) *ns* | 146.02(19) | E2 | %IT=100 | 2019BI04<br>2007JU06<br>2005PO03 |
| 52 | 73 | $^{125}$Te | 144.775 (8) | 11/2− | 57.40(15) *d* | 109.276(15)<br>144.780(25)? | M4 | %IT = 100 | 1978LA21 |
| 54 | 71 | $^{125}$Xe | 252.61 (14) | 9/2− | 57(1) s | 140.5(2) | E3 | %IT = 100 | 2011AL02<br>1968WIZY |
| 54 | 71 | $^{125}$Xe | 295.89 (15) | 7/2(+) | 0.14(3) $\mu s$ | 43.3(3)<br>183.8(2) | (E1)<br>(E2) | %IT = 100 | 1979HE15 |
| 55 | 70 | $^{125}$Cs | 84.82 (24) | (5/2+) | 14.5(15) *ns* | 84.6(3) | E2 | %IT = 100 | 1976BE11 |
| 55 | 70 | $^{125}$Cs | 266.1 (11) | (11/2−) | 0.90(3) *ms* | 13? | (M2) | %IT = 100 | 1998SU16 |
| 56 | 69 | $^{125}$Ba | 67.7+X | (7/2−) | 2.76(14) $\mu s$ | 67.6(5) | E1 | %IT = 100 | 2002SH01 |
| 57 | 68 | $^{125}$La | 107.00 (10) | | 0.39(4) *s* | 107.0(1) | E3 | %IT = 100 | 1999CA21 |
| 58 | 67 | $^{125}$Ce | 92.1 (6) | (1/2+) | 131 (+640−60) *s* @<br>4.4 *s* # | (92.1)* | (E3)* | %IT = 100 | 2007SU07@<br>2002PE15 |
| 46 | 80 | $^{126}$Pd | 2023.5 (7) | (5−) | 0.33(4) $\mu s$ | 542.4<br>1330.2 | | %IT = 100 | 2013WA24 |
| 46 | 80 | $^{126}$Pd | 2109.7 (9) | (7−) | 0.44(3) $\mu s$ | 86.2 | | %IT = 100 | 2013WA24 |
| 46 | 80 | $^{126}$Pd | 2406.4 (10) | (10+) | 23.0(9) *ms* | 296.7 | | %IT = 28(8)<br>%$\beta$−= 72(8) | 2014WA26 |
| 47 | 79 | $^{126}$Ag$ | 0+X<br>0+Y | (3+)<br>(≥ 8) | 52(10) *ms*<br>92(9) *ms* | | | %$\beta$−= 100<br>%$\beta$−= 100 | 2014BA18<br>2014BA18 |
| 47 | 79 | $^{126}$Ag | 254.8+X (5) | (1−) | 27(6) $\mu s$ | 254.8(5) | (M2) | %IT = 100 | 2013LA11<br>2012KA36 |

*Continued...*

*: $\gamma$ not seen, its energy and multipolarity assumed.

#: Estimated half-life for neutral atom.

$: Higher spin level may be the isomer.

Table 1 contd...

| Z | N | $^{A}X$ | E(keV) | $J^{\pi}$ | $T_{1/2}$ | Eγ(keV) | λ | Decay mode | Reference |
|---|---|---|---|---|---|---|---|---|---|
| 49 | 77 | $^{126}$In | 102 (64) | (8−) | 1.64(5) *s* | | | %β−= 100 | 1987SP09<br>1986GO10 |
| 49 | 77 | $^{126}$In | 243.3 (2) | 1− | 22(2) *μs* | 243.3(2) | | %IT = 100 | 2004SC42 |
| 50 | 76 | $^{126}$Sn | 2161.54 (7) | 5− | 10.8(7) *ns* | 111.79(5)<br>1020.41(10) | E1 | %IT = 100 | 1979FO10 |
| 50 | 76 | $^{126}$Sn | 2218.99 (8) | 7− | 6.1(8) *μs* | 57.47(5) | E2 | %IT = 100 | 2010IL01<br>1979FO10 |
| 50 | 76 | $^{126}$Sn | 2564.5 (5) | 10+ | 7.6(3) *μs* | 76.3(5) | | %IT = 100 | 2012AS05<br>2010IL01<br>2000ZH47 |
| 50 | 76 | $^{126}$Sn | 4345.7 (8) | 15− | 126(20) *ns* | 180.5(3) | E2 | %IT = 100 | 2014IS04<br>2012AS05 |
| 51 | 75 | $^{126}$Sb | 17.7 (3) | (5+) | 19.15(8) *min* | 17.7(3) | (E3) | %β−= 86(4)<br>%IT = 14(4) | 1972PA13<br>1971OR04<br>1970MU16<br>1967HA27<br>1962DR01 |
| 51 | 75 | $^{126}$Sb | 40.4 (3) | (3−) | ~ 11 *s* | 22.70(7) | (M2) | %IT = 100 | 1976SM01 |
| 51 | 75 | $^{126}$Sb | 104.6 (3) | (3+) | 553(5) *ns* | 21.646(10)<br>64.281(10)<br>86.938(10) | (E1)<br>(E1)<br>(E2) | %IT = 100 | 1976SM01 |
| 51 | 75 | $^{126}$Sb | 127.9 (3) | (2+) | 78.0(5) *ns* | 23.280(10)<br>87.567(10) | (M1)<br>(E1) | %IT = 100 | 1976SM01 |
| 51 | 75 | $^{126}$Sb | 1689.4 (13) | (11+) | 60(20) *ns* | 644.6(10) | | %IT = 100 | 2019BI04 |
| 51 | 75 | $^{126}$Sb | 1810.7 (17) | (13+) | 90(16) *ns* | 121.3(10) | | %IT = 100 | 2019BI04 |
| 52 | 74 | $^{126}$Te | 2974.4 (4) | 10+ | 10.7(9) *ns* | 208.6(3) | E2 | %IT = 100 | 1998ZH09<br>1983GO02<br>1971KE20 |
| 53 | 73 | $^{126}$I | 56.40 (24) | 1+ | 20 *ns* | 56.4(3) | E1 | %IT = 100 | 2012MOZZ |
| 53 | 73 | $^{126}$I | 111.00 (23) | 3+ | 128 *ns* | 54.6(3)<br>111.0(3) | E2<br>E1 | %IT = 100 | 2012MOZZ |
| 53 | 73 | $^{126}$I | 122.2 (3) | 4− | 11 *ns* | 122.2(3) | E2 | %IT = 100 | 2012MOZZ |
| 53 | 73 | $^{126}$I | 237.4 (4) | 6− | 16 *ns* | 115.2(3) | E2 | %IT = 100 | 2012MOZZ |
| 55 | 71 | $^{126}$Cs | 272.5 (10) | (4−) | ≥ 1 *μs* | 31<br>55 | | %IT = 100 | 2003LI30<br>1991TAZX |
| 55 | 71 | $^{126}$Cs | 596.1 (11) | | 171(14) *μs* | 112<br>223 | | %IT = 100 | 1991TAZX |
| 57 | 69 | $^{126}$La$^{\dagger}$ | 0+X<br>0+Y | (4,5)<br>(0−, 1, 2−) | 54(2) *s*<br>< 50 *s* | | | %ϵ+%β+ > 0.0<br>%ϵ+%β+ = ?<br>%IT = ? | 2002KO02<br>2002KO02 |
| 46 | 81 | $^{127}$Pd | 1717.91 (23) | (19/2+) | 39(6) *μs* | 422.4(1) | M2 | %IT = 100 | 2019WA14 |
| 47 | 79 | $^{127}$Ag | 1942 (+14–20) | (27/2+) | 67.5(9) *ms* | 150(+14–20) | | %β−= 91.2(8)<br>%IT = 8.8(8) | 2021WA49 |
| 48 | 79 | $^{127}$Cd | 283.3 (56) | 11/2(−) | 0.36(4) *s* | | | %β−= 100 | 2020MA09<br>2019LO04<br>2017LA16<br>2016YO01<br>2013YO02 |
| 48 | 79 | $^{127}$Cd | 1559.7 | (19/2+) | 17.5(3) *μs* | 738.8(2)<br>848.9(2) | | %IT = 100 | 2021WA49<br>2020MA09<br>2010NA17 |

*Continued...*

Table 1 contd...

| Z | N | $^{A}$X | E(keV) | J$^{\pi}$ | $T_{1/2}$ | Eγ(keV) | λ | Decay mode | Reference |
|---|---|---|---|---|---|---|---|---|---|
| 49 | 78 | $^{127}$In | 401 (12) | (1/2−) | 3.618(32) *s* | | | %β−= 100<br>%β−n = 0.70(6) | 2021IZ01<br>2018BA08<br>1993RU01<br>1986REZU<br>1983SH07<br>1980DE35 |
| 49 | 78 | $^{127}$In | 1743 (9) | (21/2−) | 1.04(10) *s* | | | %β−= 100 | 2021IZ01<br>2018BA08<br>2004GA24 |
| 49 | 78 | $^{127}$In | 2364 (60)? | (29/2+) | 9(2) *μs* | 47.0(5) | E2 | %IT = 100 | 2004SC42 |
| 50 | 77 | $^{127}$Sn | 5.07 (6) | 3/2+ | 4.13(3) *min* | | | %β−= 100 | 2020YO07<br>2005LE34<br>1974GR29 |
| 50 | 77 | $^{127}$Sn | 1826.67 (16) | (19/2+) | 4.52(15) *μs* | 16.52(11)<br>732.04(11) | E2 | %IT = 100 | 2010AT03<br>2008LO07<br>2004GA24<br>2000PI03 |
| 50 | 77 | $^{127}$Sn | 1930.97 (17) | (23/2+) | 1.19(13) *μs* | 104.30(6) | (E2) | %IT = 100 | 2008LO07<br>2004GA24 |
| 50 | 77 | $^{127}$Sn | 2552.4 (10) | (27/2−) | 0.25(3) *μs* | 142.0(3) | (E2) | %IT = 100 | 2008LO07 |
| 51 | 76 | $^{127}$Sb | 1920.19 (21) | (15/2−) | 11.7(1) *μs* | 805.9(4)<br>824.7(4) | (E3)<br>(M2) | %IT = 100 | 2019BI04<br>1974AP01 |
| 51 | 76 | $^{127}$Sb | 2194.4 (6) | 19/2+ | 14(1) *ns* | 143.3<br>246.9 | E2<br>E2 | %IT = 100 | 2009WA24 |
| 51 | 76 | $^{127}$Sb | 2325.2 (7) | (23/2+) | 269(5) *ns* | 130.8 | E2 | %IT = 100 | 2019BI04<br>2009WA24 |
| 52 | 75 | $^{127}$Te | 88.23 (7) | 11/2− | 106.1(7) *d* | 88.3(1) | M4 | %IT = 97.86(3)<br>%β−= 2.14(3) | 2017NI03<br>2008EA01 |
| 54 | 73 | $^{127}$Xe | 297.10 (8) | 9/2− | 69.2(9) *s* | 172.4(1) | E3 | %IT = 100 | 1967GE15 |
| 54 | 73 | $^{127}$Xe | 342.23 (4) | 7/2+ | 36.7(9) *ns* | 45.1(2)<br>217.48(5) | (E1)<br>E2 | %IT = 100 | 2018CH24<br>1985UR01<br>1984LO07<br>1981HE04 |
| 54 | 73 | $^{127}$Xe | 2730.3 (10) | 23/2+ | 28(1) *ns* | 335.9<br>486.6 | E2<br>E1 | %IT = 100 | 2018CH24<br>1985UR01 |
| 55 | 72 | $^{127}$Cs | 66.09 (11) | (5/2)+ | 24.88(30) *ns* | 66.06(14) | E2 | %IT = 100 | 1999CO22 |
| 55 | 72 | $^{127}$Cs | 452.23 (21) | (11/2)− | 55(3) *μs* | 179.30(20)<br>386.3(3) | M2<br>(E3) | %IT = 100 | 1980DR07 |
| 56 | 71 | $^{127}$Ba | 80.32 (11) | 7/2− | 1.93(7) *s* | 24.2(1)<br>80.2(2) | M2<br>E3 | %IT = 100 | 2002SH01 |
| 56 | 71 | $^{127}$Ba | 81.29 (13) | (5/2)+ | 75(4) *ns* | 25.1(2)<br>81.3(2) | M1 | %IT = 100 | 2002SH01 |
| 57 | 70 | $^{127}$La | 14.2 (4) | (3/2+) | 3.7(4) *min* | | | %ϵ+%β+ = 100 | 1963PR02<br>1963YA05 |
| 58 | 69 | $^{127}$Ce | 7.3 (11) | (5/2+) | 28.6(7) *s* | | | %ϵ+%β+ = 100 | 2005II01 |
| 58 | 69 | $^{127}$Ce | 36.8 (12) | (7/2−) | > 10 *μs* | 29.56(5) | | %IT = 100 | 1995OS03 |
| 46 | 82 | $^{128}$Pd | 2151.0 (10) | (8+) | 5.8(8) *μs* | 75.1 | | %IT = 100 | 2013WA24 |
| 48 | 80 | $^{128}$Cd | 1870.5 (3) | (5−) | 270(7) *ns* | 440.3(3)<br>1224.0(6) | | %IT = 100 | 2009CA02 |
| 48 | 80 | $^{128}$Cd | 2108.3 (4) | (7−) | 12(2) *ns* | 237.9(5) | | %IT = 100 | 2009CA02 |

*Continued...*

Table 1 contd...

| Z | N | $^{A}X$ | E(keV) | J$^{\pi}$ | $T_{1/2}$ | E$\gamma$(keV) | $\lambda$ | Decay mode | Reference |
|---|---|---|---|---|---|---|---|---|---|
| 48 | 80 | $^{128}$Cd | 2714.6 (4) | (10+) | 3.56(6) $\mu s$ | 68.7(1) | E2 | %IT = 100 | 2012KA36<br>2009CA02 |
| 48 | 80 | $^{128}$Cd | 4288 (3) | (15−) | 6.3(8) $ms$ | 309 | | %IT = 100 | 2017JU02 |
| 49 | 79 | $^{128}$In | 247.87 (10) | (1−) | 23(2) $\mu s$ | 247.92(10) | M2, E3 | %IT = 100 | 2004SC42 |
| 49 | 79 | $^{128}$In | 285.1 (25) | (8−) | 0.72(10) $s$ | | | %$\beta$−= 100<br>%$\beta$−n < 0.046 | 2021IZ01<br>2020NE06<br>2018BA08<br>1990ST13<br>1986GO10 |
| 49 | 79 | $^{128}$In | 1797.6 (20) | (16+) | ≥ 0.3 $s$ | | | %$\beta$−≈ 100 | 2021IZ01<br>2020NE06 |
| 50 | 78 | $^{128}$Sn | 2091.50 (11) | (7−) | 6.5(5) $s$ | 91.15(10) | E3 | %IT = 100 | 1979FO10 |
| 50 | 78 | $^{128}$Sn | 2491.89 (17) | (10+) | 2.91(15) $\mu s$ | 79.28(15) | E2 | %IT = 100 | 2011PI05<br>1981FO02 |
| 50 | 78 | $^{128}$Sn | 4099.5 (4) | (15−) | 220(30) $ns$ | 119.7(2) | | %IT = 100 | 2014IS04<br>2011PI05 |
| 51 | 77 | $^{128}$Sb | 0+X | 5+ | 10.4(2) $min$ | | | %IT = 3.6(10)<br>%$\beta$−= 96.4(10) | 1966TO02<br>1962UH01<br>1962HA16<br>1962DR01<br>1962DE11<br>1955FR11 |
| 51 | 77 | $^{128}$Sb | 1617 | (11+) | 217(7) $ns$ | 384.8<br>556.2<br>843.8<br>1617.2 | | | 2019BI04 |
| 51 | 77 | $^{128}$Sb | 1769 | (13+) | 500(20) $ns$ | 152.6 | E2 | %IT = 100 | 2019BI04 |
| 52 | 76 | $^{128}$Te | 2790.8 (3) | 10+ | 236(20) $ns$* | 101.9(3) | E2 | %IT = 100 | 2014AS01<br>2004VA03<br>1998ZH09 |
| 53 | 75 | $^{128}$I | 133.6107 (10) | 2− | 12.3(5) $ns$ | 48.148(7)<br>106.2486(7)<br>133.6106(11) | E1<br>E1<br>E1 | %IT = 100 | 1978BU15 |
| 53 | 75 | $^{128}$I | 137.851 (3) | 4− | 0.845(20) $\mu s$ | 4.240(5)<br>52.378(5) | <br>E1 | %IT = 100 | 1991SA07 |
| 53 | 75 | $^{128}$I | 167.368 (4) | (6)− | 175(15) $ns$ | 29.512(5) | (E2) | %IT = 100 | 1991SA07 |
| 54 | 74 | $^{128}$Xe | 2787.2 (3) | 8− | 78(5) $ns$ | 204.2(5)<br>274.4(4)<br>286.3(5) | (M1)<br><br>E2(+M1) | %IT = 100 | 2006OR10<br>1984LO07 |
| 55 | 73 | $^{128}$Cs | 750+X | (9+) | 50(8) $ns$ | 158.7(3)<br>301.8(3) | (E1)<br>(E1) | %IT = 100 | 2018GR01<br>1989PA09 |
| 47 | 82 | $^{129}$Ag | X | (1/2−) | ∼ 160 $ms$ | | | %$\beta$−= ? | 2000KR18 |

*Continued...*

*: We adopt half-life from 2014AS01 with better statistics than those in 2004VA03 ($T_{1/2}$ = 337(59) $ns$) and 1998ZH09 ($T_{1/2}$ = 370(30) $ns$).

Table 1 contd…

| Z | N | $^{A}$X | E(keV) | J$^{\pi}$ | $T_{1/2}$ | E$\gamma$(keV) | $\lambda$ | Decay mode | Reference |
|---|---|---|---|---|---|---|---|---|---|
| 48 | 81 | $^{129}$Cd | 343 (8) | 3/2+ | 157(8) *ms* | | | %$\beta$−= 100 | 2020MA09<br>2016YO01<br>2016DU13<br>2015TA13<br>2014TA29<br>2013YO02<br>2005KR20<br>2003ARZX |
| 48 | 81 | $^{129}$Cd | 1940 | (21/2+) | 3.6(2) *ms* | 353 | (E3) | %IT = 100 | 2014TA29 |
| 49 | 80 | $^{129}$In | 450.72 (16) | (1/2−) | 1.16(1) *s* | | | %$\beta$−= 99.85(15)<br>%$\beta$−n = 3.6(4)<br>%IT = 0.15(15) | 2021GA10<br>2018BA08<br>2015TA13<br>2013KA08<br>1986GO10<br>1986REZU<br>1980DE35 |
| 49 | 80 | $^{129}$In | 1630 (56) | (23/2−) | 0.65(2) *s* | | | %$\beta$−≈ 100<br>%IT = ? | 2021GA10<br>2004GA24<br>1998FOZY |
| 49 | 80 | $^{129}$In | 1687.97 (25) | (17/2−) | 10.9(1) $\mu s$ | 333.8(2) | (M2) | %IT = 100 | 2020SA31<br>2014TAZV<br>2012KA36<br>2003HEZT<br>2003GE04 |
| 49 | 80 | $^{129}$In | 1911 (56) | (29/2+) | 98(15) *ms* | 281.0(2) | (E3) | %IT = 98.0(5)<br>%$\beta$− = 2.0(5) | 2021GA10<br>2004GA24<br>1998FOZY |
| 50 | 79 | $^{129}$Sn | 35.15 (5) | 11/2− | 6.9(1) *min* | | | %$\beta$−= 100<br>%IT < 2E−3 | 2020YO07<br>1982HU09 |
| 50 | 79 | $^{129}$Sn | 1761.6 (10) | (19/2+) | 3.40(13) $\mu s$ | 19.7(10) | (E2) | %IT = 100 | 2008LO07<br>2004GA24<br>2002GE07 |
| 50 | 79 | $^{129}$Sn | 1802.6 (10) | (23/2+) | 2.22(14) $\mu s$ | 41.0(2) | (E2) | %IT = 100 | 2008LO07<br>2004GA24<br>2002GE07 |
| 50 | 79 | $^{129}$Sn | 2552.9 (11) | (27/2−) | 217(19) *ns* | 145.3(3) | | %IT = 100 | 2011PI05<br>2008LO07 |
| 51 | 78 | $^{129}$Sb | 1851.31 (6) | (19/2−) | 17.7(1) *min* | 722.69(5) | (M4) | %$\beta$−= 85<br>%IT = 15 | 1982HU09 |
| 51 | 78 | $^{129}$Sb | 1861.06 (5) | (15/2−) | 2.2(2) $\mu s$ | (9.76(8))<br>699.64(6)<br>732.48(5) | | %IT = 100 | 2019BI04<br>2003GE04<br>1998GEZX |
| 51 | 78 | $^{129}$Sb | 2139.4 (3) | (23/2+) | 0.91(6) $\mu s$ | 98.6(2) | E2 | %IT = 100 | 2019BI04<br>2003GE04 |
| 52 | 77 | $^{129}$Te | 105.51 (3) | 11/2− | 33.67(16) *d* | 105.50(5) | M4 | %IT = 64(7)<br>%$\beta$−= 36(7) | 1972EM01<br>1971BA28<br>1970BO22<br>1965AN05<br>1963HA23<br>1953PA25<br>1940SE01 |

*Continued…*

Table 1 contd...

| Z | N | $^{A}X$ | E(keV) | J$^{\pi}$ | $T_{1/2}$ | E$\gamma$(keV) | $\lambda$ | Decay mode | Reference |
|---|---|---|---|---|---|---|---|---|---|
| 52 | 77 | $^{129}$Te | 2137.83 (17) | (23/2+) | 33(3) *ns* | 180.4(2)<br>251.1(2) | (E1)<br>(E1) | %IT = 100 | 1998ZH09 |
| 53 | 76 | $^{129}$I | 27.793 (20) | 5/2+ | 16.8(2) *ns* | 27.81(5) | M1+E2 | %IT = 100 | 1966SA06 |
| 54 | 75 | $^{129}$Xe | 236.14 (3) | 11/2− | 8.88(2) *d* | 196.56(3) | M4 | %IT = 100 | 1990TA18<br>1975HO18<br>1973MI08 |
| 55 | 74 | $^{129}$Cs | 6.5450 (10) | 5/2+ | 72(6) *ns* | 6.545(1) | E2 | %IT = 100 | 1976BE11 |
| 55 | 74 | $^{129}$Cs | 575.40 (14) | (11/2−) | 0.718(21) $\mu s$ | 149.1(3)<br>354.8?<br>366.1(2)?<br>386.6(3)<br>569.2(3) | (E1)<br><br><br>(M2)<br> | %IT = 100 | 1978DE29<br>1977CH23 |
| 56 | 73 | $^{129}$Ba | 8.42 (6) | 7/2+ | 2.135(10) *h* | (8.4(2)) | | %IT = ?<br>%$\epsilon$+%$\beta$+ ≈ 100 | 1973IS04<br>1972TA02<br>1966LI05<br>1961AR05 |
| 56 | 73 | $^{129}$Ba | 182.04 (11) | 9/2− | 15.2(10) *ns* | 173.6(1) | E1 | %IT = 100 | 2013KA27<br>1992BY03 |
| 56 | 73 | $^{129}$Ba | 2462.6 (2) | (23/2+) | 47(1) *ns* | 126.0(1)<br>316.3(1)<br>472.8(1) | (E1)<br><br>(E2) | %IT = 100 | 2013KA27<br>1992BY03 |
| 57 | 72 | $^{129}$La | 172.33 (20) | (11/2−) | 0.56(5) *s* | 104.0(3) | E3 | %IT = 100 | 1969AL05 |
| 58 | 71 | $^{129}$Ce | 107.60 (16) | (7/2−) | 60(2) *ns* | 107.7(2) | (E1) | %IT = 100 | 1998IO01 |
| 60 | 69 | $^{129}$Nd | 0+X | (7/2−) | ~ 7 *s** | | | %$\epsilon$+%$\beta$+ = ? | 2010XU12 |
| 60 | 69 | $^{129}$Nd | 0+Z | (1/2+) | 2.6(4) *s* | | | %$\epsilon$+%$\beta$+ = ?<br>%$\epsilon$p = ? | 2010XU12 |
| 48 | 82 | $^{130}$Cd | 2129.6 (10) | (8+) | 0.248(+21−19) $\mu s$ | 138.0(5) | E2 | %IT = 100 | 2012KA36<br>2007JU05 |
| 49 | 81 | $^{130}$In | 58.6 (82) | (10−) | 0.534(6) *s* # | | | %$\beta$−= 100<br>%$\beta$−n = 1.65(15)# | 2020NE06<br>1993RU01<br>1986REZU<br>1983SH07<br>1981FO02<br>1973KE12 |
| 49 | 81 | $^{130}$In | 385.5 (50) | (5+) | 0.534(6) *s* # | | | %$\beta$−= 100<br>%$\beta$−n = 1.65(15)# | 2021IZ01<br>2020NE06<br>2018BA08<br>1993RU01<br>1986REZU<br>1983SH07<br>1981FO02<br>1973KE12 |
| 49 | 81 | $^{130}$In | 388.47 (14) | (3+) | 4.3(6) $\mu s$ | 388.4(2) | | %IT = 100 | 2016JU03<br>2012KA36<br>2004SC42 |
| 50 | 80 | $^{130}$Sn | 1946.88 (10) | (7−) | 1.7(1) *min* | | | %$\beta$−= 100 | 1974KE08 |
| 50 | 80 | $^{130}$Sn | 2084.84 (8) | (5−) | 52(3) *ns* | 89.23(3)<br>137.96(5) | E1<br>E2 | %IT = 100 | 1981FO02 |

*Continued...*

*: In ENSDF database, half-life assumed as comparable to that for the ground state.
#: Combined half-life and decay modes for the (10−) and (5+) isomers.

Table 1 contd...

| Z | N | $^{A}X$ | E(keV) | $J^{\pi}$ | $T_{1/2}$ | Eγ(keV) | λ | Decay mode | Reference |
|---|---|---|---|---|---|---|---|---|---|
| 50 | 80 | $^{130}Sn$ | 2434.79 (12) | (10+) | 1.502(17) $\mu s$ | 96.54(5) | E2 | %IT = 100 | 2019DU12 |
| | | | | | | | | | 2012KA36 |
| | | | | | | | | | 2011PI05 |
| | | | | | | | | | 1981FO02 |
| 51 | 79 | $^{130}Sb$ | 4.8 (2) | (4, 5)+ | 6.3(2) *min* | | | %β−= 100 | 1974ER07 |
| | | | | | | | | | 1974FO06 |
| | | | | | | | | | 1974GR29 |
| | | | | | | | | | 1974KE08 |
| 51 | 79 | $^{130}Sb$ | 84.7 (3) | 6− | 0.8(1) $\mu s$ | 84.7 | E2 | %IT = 100 | 2002GE07 |
| 51 | 79 | $^{130}Sb$ | 1508 | (11+) | 0.600(15) $\mu s$ | 365.2 | | %IT = 100 | 2019BI04 |
| 51 | 79 | $^{130}Sb$ | 1544.7 (5) | (13+) | 1.25(1) $\mu s$ | 36.5 | E2 | %IT = 100 | 2019BI04 |
| | | | | | | | | | 2002GE07 |
| 52 | 78 | $^{130}Te$ | 2146.41 (4) | 7− | 111(5) *ns* | (46) | | %IT = 100 | 2022KUAA |
| | | | | | | 330.7(3) | E1+M2 | | 2014AS01 |
| | | | | | | 330.7(3) | E1+M2 | | 2004VA03* |
| | | | | | | | | | 1972KE28 |
| 52 | 78 | $^{130}Te$ | 2664.7 (9) | 10+ | 1.90(8) $\mu s$ | (17.9) | | %IT = 100 | 2022KUAA |
| | | | | | | | | | 2014AS01 |
| | | | | | | | | | 2004BR19 |
| | | | | | | | | | 2001GE07 |
| 52 | 78 | $^{130}Te$ | 4373.9 (9) | (15−) | 53(8) *ns* | 126.2(3) | E2 | %IT = 100 | 2014AS01 |
| | | | | | | | | | 1998HOZP# |
| 53 | 77 | $^{130}I$ | 39.9525 (13) | 2+ | 8.91(12) *min* | 39.9542(21) | M3 | %IT = 84(2) | 1996NA23 |
| | | | | | | | | %β−= 16(2) | 1989SA11 |
| | | | | | | | | | 1974DI03 |
| | | | | | | | | | 1972BA51 |
| | | | | | | | | | 1970QA03 |
| | | | | | | | | | 1968RE04 |
| | | | | | | | | | 1966WI15 |
| 53 | 77 | $^{130}I$ | 69.5865 (7) | (6)− | 133(7) *ns* | 69.5862(7) | E1 | %IT = 100 | 1989SA11 |
| 53 | 77 | $^{130}I$ | 82.3960 (19) | (8)− | 315(15) *ns* | (12.81) | E2 | %IT = 100 | 1989SA11 |
| 53 | 77 | $^{130}I$ | 82.4+X$ | | 66(8) *ns* | | | | 1989SA11 |
| 53 | 77 | $^{130}I$ | 85.1099 (10) | (6)− | 254(4) *ns* | 85.1104(10) | E1 | %IT = 100 | 1989SA11 |
| | | | | | | | | | 1975BLZY |
| 55 | 75 | $^{130}Cs$ | 163.25 (11) | 5− | 3.46(6) *min* | 14.9(3) | | %IT = 99.84(2) | 1983WE07 |
| | | | | | | 31.5(3) | | %ϵ+%β+ = 0.16(2) | |
| | | | | | | 82.9(1) | E3 | | |
| 56 | 74 | $^{130}Ba$ | 2475.12 (18) | 8− | 9.4(4) *ms* | 80.3(2) | E1 | %IT = 100 | 1999DEZZ |
| | | | | | | 462.3(2) | E3 | | 1969WAZX |
| | | | | | | 882.3(2) | M2+E3 | | 1966BR14 |
| 57 | 73 | $^{130}La$ | 131.01 (8) | 1+ | 77(10) *ns* | 131.1(2) | | %IT = 100 | 1996XU04 |
| 57 | 73 | $^{130}La$ | 214.0 (3) | 5+ | 0.76(9) $\mu s$ | 103.6(2) | (E2) | %IT = 100 | 2014IO01 |
| | | | | | | | | | 2012TA18 |
| 57 | 73 | $^{130}La$ | 319.1 (3) | 6+ | 33(1) *ns* | 82.5(2) | (E1) | %IT = 100 | 2014IO01 |
| | | | | | | 105.1(2) | (M1) | | |

*Continued...*

*: Half-life of 186(11) ns in 2004VA03 seems too high as compared to that in 2014AS01 and 1972KE28.
#: Half-life of 261(33) *ns* in 1998HOZP seems too high as compared to that in 2014AS01.
$: X < 25 keV from 1989SA11.

Table 1 contd. . .

| Z | N | $^{A}$X | E(keV) | J$^{\pi}$ | $T_{1/2}$ | E$\gamma$(keV) | $\lambda$ | Decay mode | Reference |
|---|---|---|---|---|---|---|---|---|---|
| 58 | 72 | $^{130}$Ce | 2453.6 (3) | (7−) | 100(8) *ns* | 72.8<br>120.5<br>401.0(5)<br>499.0(4)<br>556.1(3)<br>1129.7(3) | | %IT = 100 | 1999IO02<br>1984TO10 |
| 59 | 71 | $^{130}$Pr | 0+X | (5+) | 40.0(4) s | | | %$\epsilon$+%$\beta$+ = 100 | 1988BA42 |
| 49 | 82 | $^{131}$In | 375 (18) | (1/2−) | 328(15) *ms* | | | %$\beta$− = 99.991(9)<br>%IT = 0.009(9) | 2021IZ01<br>2019DU12<br>2016TA22<br>2013KA08<br>1984FO19 |
| 49 | 82 | $^{131}$In | 3771 (15) | (21/2+) | 322(55) *ms* | | | %$\beta$− = 99.5(5)<br>%$\beta$− n ≈ 0.03<br>%IT = 0.5(5) | 2021IZ01<br>2019DU12<br>1984FO19 |
| 49 | 82 | $^{131}$In | 3783.6 (5) | (17/2+) | 0.685(+42−39) $\mu s$ | 3783.6(5) | E4 | %IT = 100 | 2012KA36<br>2009GO40 |
| 50 | 81 | $^{131}$Sn | 65.1 (3) | (11/2−) | 58.4(5) s | | | %IT = ?<br>%$\beta$−≈ 100 | 2020YO07<br>2004FO06<br>1986STZZ |
| 50 | 81 | $^{131}$Sn | 4605.7+X | (23/2−) | 316(5) *ns* | 158.8(5) | E2 | %IT = 100 | 2019DU12<br>2012KA36<br>1984FO19 |
| 51 | 80 | $^{131}$Sb | 1676.06 (6) | (15/2−) | 64(3) $\mu s$ | 447.4<br>450.03(5)<br>1676.0? | E3<br>M2 | %IT = 100 | 2019BI04<br>2000GE18<br>1987BO19<br>1982PR01<br>1977SC14<br>1969WA29 |
| 51 | 80 | $^{131}$Sb | 1687.2 (9) | (19/2−) | 4.3(8) $\mu s$ | 11.2 | | %IT = 100 | 2000GE18 |
| 51 | 80 | $^{131}$Sb | 2165.6 (15) | (23/2+) | 0.97(3) $\mu s$ | 96.4 | E2 | %IT = 100 | 2019BI04<br>2000GE18 |
| 52 | 79 | $^{131}$Te | 182.258 (18) | 11/2− | 32.48(11) *h* | 182.25(2) | (M4) | %$\beta$−= 74.1(5)<br>%IT = 25.9(5) | 2008EA01<br>2002RE30 |
| 52 | 79 | $^{131}$Te | 1940.0 (4) | (23/2+) | 93(12) *ms* | 360.3(2) | E3 | %IT = 100 | 1998FOZY<br>1988ZH09 |
| 53 | 78 | $^{131}$I | 1918.4 (4) | 19/2− | 24(1) $\mu s$ | (33)<br>122.0 | <br>E2(+M3) | %IT = 100 | 2009WA11<br>2020BA52 |
| 53 | 78 | $^{131}$I | 2350.3 (5) | 23/2+ | 42(2) *ns* | 116 | E2(+M3) | %IT = 100 | 2009WA11<br>2004VA03 |
| 53 | 78 | $^{131}$I | 4307.4 (6) | (31/2−,<br>33/2−) | 25(3) *ns* | 702.2<br>1265.6 | M2<br>(E3) | %IT = 100 | 2009WA11 |
| 54 | 77 | $^{131}$Xe | 163.930 (8) | 11/2− | 11.939(20) *d* | 163.930(8) | M4 | %IT = 100 | 2014UN01<br>1990TA18<br>1975HO18<br>1974ME21<br>1972EM01<br>1966KN03<br>1965AN05 |
| 54 | 77 | $^{131}$Xe | 1805.3 (3) | 19/2+ | 14(3) *ns* | 189.1(2)<br>204.5(3) | E1<br>E1 | %IT = 100 | 1971KE13 |

*Continued. . .*

Table 1 contd...

| Z | N | $^{A}X$ | E(keV) | J$^{\pi}$ | $T_{1/2}$ | E$\gamma$(keV) | $\lambda$ | Decay mode | Reference |
|---|---|---|---|---|---|---|---|---|---|
| 55 | 76 | $^{131}$Cs | 775.28 (6) | 11/2− | 10.46(14) *ns* | 117.69(13)?<br>159.0(5)<br>279.17(2) | <br><br>E1 | %IT = 100 | 2000FUZM |
| 56 | 75 | $^{131}$Ba | 187.995 (9) | 9/2− | 14.26(9) *min* | 79.918(7) | E3 | %IT = 100 | 2012DA04<br>1963HO05 |
| 57 | 74 | $^{131}$La | 304.60 (24) | 11/2− | 170(7) $\mu s$ | 108.9(3) | M2 | %IT = 100 | 1981AN17<br>1973CO32<br>1973LE09<br>1970CO05 |
| 57 | 74 | $^{131}$La | 2121.8 (3) | 21/2− | 37.2(2) *ns** | 711.3(8)<br>947.6(1) | <br>M1, E2 | %IT = 100 | 2017KA03<br>1989HI02 |
| 58 | 73 | $^{131}$Ce | 63.09 (9) | (1/2+) | 5.4(4) *min* | | | %$\epsilon$+%$\beta$+ = 100<br>%IT = ? | 1983AKZZ<br>1966NO05 |
| 58 | 73 | $^{131}$Ce | 161.98 (5) | 9/2− | 88(2) *ns* | 161.98(5) | E1 | %IT = 100 | 1998IO01 |
| 59 | 72 | $^{131}$Pr | 152.4 (3) | (11/2−) | 5.73(20) *s* | 64.8(3) | E3 | %IT = 96.4(12)<br>%$\epsilon$+%$\beta$+ = 3.6(12) | 1996GE12<br>1983VIZU |
| 50 | 82 | $^{132}$Sn | 4715.91 (17) | (6+) | 20.1(5) *ns* | 299.6 | (E2) | %IT = 100 | 1982KA25 |
| 50 | 82 | $^{132}$Sn | 4848.52 (20) | (8+) | 2.080(17) $\mu s$ | 132.5 | (E2) | %IT = 100 | 2017CH51<br>2012KA36<br>1994FO14 |
| 51 | 81 | $^{132}$Sb | 0+X | (8−) | 4.10(5) *min* | | | %$\beta$−= 100 | 1975BA36 |
| 51 | 81 | $^{132}$Sb | 85.55 (6) | (3)+ | 15.62(13) *ns* | 85.58(8) | M1+E2 | %IT = 100 | 1995MA02 |
| 51 | 81 | $^{132}$Sb | 254.5 (3) | (6−) | 102(4) *ns* | 91.7(2) | (E1) | %IT = 100 | 1989ST06<br>1974CLZX |
| 52 | 80 | $^{132}$Te | 1774.80 (9) | 6+ | 135.0(44) *ns* | 103.519(4) | E2 | %IT = 100 | 2022KUAA<br>1973MC09 |
| 52 | 80 | $^{132}$Te | 1925.47 (9) | (7)− | 28.5(9) $\mu s$ | 150.672(10) | E1 | %IT = 100 | 2022KUAA<br>2017KI09<br>1979SI18 |
| 52 | 80 | $^{132}$Te | 2723.3 (8) | (10+) | 3.62(9) $\mu s$ | 22(1)<br>798.0 | | %IT = 100 | 2022KUAA<br>2017KI09<br>2001GE07<br>1979SI18 |
| 53 | 79 | $^{132}$I | 120 (20) | (8−) | 1.387(15) *h* | 98.0(10) | E3 | %IT = 86(2)<br>%$\beta$−= 14(2) | 1976LA14<br>1974DI03<br>1973DI14 |
| 53 | 79 | $^{132}$I | X | (11+) | 342(12) *ns* | | | | 2020BA52 |
| 54 | 78 | $^{132}$Xe | 2214.01 (14) | (7−) | 87(3) *ns* | 173.7(1) | E2 | %IT = 100 | 2017VO06<br>2004VA03<br>1986VO14<br>1971KE13 |
| 54 | 78 | $^{132}$Xe | 2752.21 (17) | 10+ | 8.37(8) *ms* | 538.2(1) | E3 | %IT = 100 | 2018KA47<br>1976HA50 |
| 57 | 75 | $^{132}$La | 188.20 (11) | 6− | 24.3(5) *min* | 52.8(1)<br>188.5(3) | M3<br>E4 | %IT = 76<br>%$\epsilon$+%$\beta$+ = 24 | 1969GE11 |
| 58 | 74 | $^{132}$Ce | 2341.15 (21) | (8−) | 9.4(3) *ms* | 11.6<br>526.8(3)<br>798.5(2) | | %IT = 100 | 2001MO05 |
| 59 | 73 | $^{132}$Pr | 219.9+X | (8+) | 2.46(4) $\mu s$ | 101.9(1) | | %IT = 100 | 2012TA18 |

*Continued...*

*: Statistical uncertainty of 0.1 *ns* in 2017KA03 doubled to include possible systematic uncertainty.

Table 1 contd...

| Z | N | $^{A}$X | E(keV) | J$^{\pi}$ | $T_{1/2}$ | E$\gamma$(keV) | $\lambda$ | Decay mode | Reference |
|---|---|---|---|---|---|---|---|---|---|
| 59 | 73 | $^{132}$Pr | 273.0+X | (8−) | 486(70) *ns* | 53.1(5) | (E1) | %IT = 100 | 2012TA18 |
| 49 | 84 | $^{133}$In | 642 (60) | (1/2−) | 167(11) *ms** | | | %$\beta$−~ 100<br>%$\beta$−n = 93(3) | 2021IZ01<br>2020BE16<br>2019PI04<br>2000HO32<br>1996HO16 |
| 51 | 82 | $^{133}$Sb | 4526+X# | (21/2+) | 16.5(3) $\mu s$ # | | | | 2016BO19<br>2010SU11<br>2010SU18<br>2009UR01<br>2000GE08 |
| 52 | 81 | $^{133}$Te | 334.26 (4) | (11/2−) | 55.4(4) *min* | 334.27(4) | M4 | %IT = 16.5(20)<br>%$\beta$−= 83.5(20) | 1968BE64 |
| 52 | 81 | $^{133}$Te | 1610.4 (5) | (19/2−) | 100(5) *ns* | 125.5(3) | | %IT = 100 | 2005HW06<br>2001BH06 |
| 53 | 80 | $^{133}$I | 1634.148 (10) | (19/2−) | 9(2) *s* | 74.05(1) | (M2) | %IT = 100 | 1970OSZZ |
| 53 | 80 | $^{133}$I | 1729.137 (10) | (15/2−) | ~ 170 *ns* | 94.989(2)<br>169.025(5) | E2<br>(E1) | %IT = 100 | 1984WA04 |
| 53 | 80 | $^{133}$I$ | 2493.7 (4) | (23/2+) | 474(16) *ns* | 58.7(3) | E2 | %IT = 100 | 2020BA52<br>2009WA11<br>2004VA03 |
| 54 | 79 | $^{133}$Xe | 233.221 (15) | 11/2− | 2.212(17) *d* | 233.221(15) | M4+E5 | %IT = 100 | 1975HO18<br>1974FOZY<br>1968AL16<br>1961ER04<br>1951BE11 |
| 54 | 79 | $^{133}$Xe | 2107 | 23/2+ | 8.64(13) *ms* | 231 | (M2+E3) | %IT = 100 | 2018KA47<br>2017VO06 |
| 56 | 77 | $^{133}$Ba | 288.252 (9) | 11/2− | 38.91(8) *h* | 275.925(7)<br>288(1) | M4 | %IT = 99.9896(5)<br>%$\epsilon$ = 0.0104(5) | 2012DA04<br>2011GR01<br>1960WI10<br>1941CO03 |
| 56 | 77 | $^{133}$Ba | 1942.1 | 19/2+ | 66.6(20) *ns* | 83.1 | | | 2019KA36<br>1975GI11 |
| 57 | 76 | $^{133}$La | 535.595 (21) | 11/2− | 62(4) *ns* | 58.39(3)<br>404.78(4)<br>535 | E1<br>M2 | %IT = 100 | 2020LA06<br>1975BU10<br>1973LE09 |
| 58 | 75 | $^{133}$Ce | 37.2 (7) | 9/2− | 5.326(11) *h* | | | %$\epsilon$+%$\beta$+ < 100<br>%IT = ? | 2011FA01<br>1976GE10<br>1967GE08 |
| 59 | 74 | $^{133}$Pr | 192.12 (14) | (11/2−) | 1.1(2) *s* | 130.4(2) | E3 | %IT = 100 | 2001XU04 |
| 60 | 73 | $^{133}$Nd | 127.97 (12) | (1/2+) | ~ 70 *s* | 127.9(6) | M3 | %IT = ?<br>%$\epsilon$+%$\beta$+ = ? | 1995BR21 |
| 60 | 73 | $^{133}$Nd | 176.10 (10) | (9/2−) | 291(4) *ns* | 176.1(1) | E1 | %IT = 100 | 2012TA18<br>1998BA81 |
| 60 | 73 | $^{133}$Nd | 353.62 (12) | (3/2−) | 46(9) *ns* | 62.3(3)<br>180.6(1) | <br>E1 | %IT = 100 | 1995BR21 |

*Continued...*

*: This half-life assigned to the g.s. in 1996HO16.

#: X < 30 keV from 2016BO19. 2010SU11 gives half-life for bare ion as 17 *ms*.

$: We omit a 780-*ns* isomer at 2435 keV proposed by 2004VA03 as according to 2009WA11, it gives unrealistic B(E2) values.

Table 1 contd...

| Z | N | $^{A}$X | E(keV) | J$^{\pi}$ | $T_{1/2}$ | Eγ(keV) | λ | Decay mode | Reference |
|---|---|---|---|---|---|---|---|---|---|
| 62 | 71 | $^{133}$Sm | 0+Y | (1/2−) | 3.5(4) *s* | | | %ϵ+%β+ = ?<br>%IT = ?<br>%ϵp = ? | 2001XU04<br>1993BRZU |
| 49 | 85 | $^{134}$In | 56.7 (1) | (5−) | 3.5(4) μs | | | %IT = 100 | 2019PH02 |
| 50 | 84 | $^{134}$Sn | 1246.3 (9) | 6+ | 81.7(12) *ns* | 174.1(5) | (E2) | %IT = 100 | 2021PI11<br>2012KA36<br>2000KO15 |
| 51 | 83 | $^{134}$Sb | 279 (1) | (7−) | 9.97(10) *s* | | | %β−= 100<br>%β−n = 0.088(4) | 2018SI28<br>1993RU01 |
| 52 | 82 | $^{134}$Te | 1691.34 (16) | 6+ | 164.8(10) *ns* | 115.2(1) | E2 | %IT = 100 | 2017UR03<br>2017KI09<br>2004HW02<br>2001MI22<br>1995OM01<br>1976CHZD<br>1974SU04<br>1974CLZX<br>1974BL03<br>1970JO20 |
| 52 | 82 | $^{134}$Te | 5804.0 (3) | (12+) | 17.1(17) *ns* | 182.6(2) | | %IT = 100 | 2021HA13<br>2002SA02 |
| 53 | 81 | $^{134}$I | 316.49 (22) | (8)− | 3.52(4) *min* | 272.1(1)<br>316.3(10)? | E3 | %IT = 97.7(10)<br>%β−= 2.3(10) | 1976LA14<br>1974DI03<br>1972CO04<br>1971AC01<br>1970CA16 |
| 54 | 80 | $^{134}$Xe | 1965.5 (5) | 7− | 290(17) *ms* | 234.3(5) | E3 | %IT = 100 | 1968WIZY |
| 54 | 80 | $^{134}$Xe | 3025.2 (15) | (10+) | 5(1) *μs* | 28(1) | E2 | %IT = 100 | 2001GE07 |
| 55 | 79 | $^{134}$Cs | 11.2442 (18) | 5+ | 46.6(6) *ns* | 11.242(7) | M1+E2 | %IT = 100 | 1972TUZV<br>1971DRZX<br>1970BLZT<br>1969LY08 |
| 55 | 79 | $^{134}$Cs | 138.7441 (26) | 8− | 2.912(2) *h* | 127.5021(28)<br>138.733(11) | E3<br>M4 | %IT = 100 | 1999NA39<br>1973MA68<br>1970QA02<br>1968RE04<br>1964WA10<br>1964FR02<br>1961KE03<br>1960BA49 |
| 55 | 79 | $^{134}$Cs | 176.4044 (16) | 3−,4− | 49.7(8) *ns* | 116.3755(26)<br>176.4047(24) | E1<br>E1 | %IT = 100 | 1975AL21 |
| 55 | 79 | $^{134}$Cs | 257.1075 (26) | 6− | 12.3(11) *ns* | 63.490(4)<br>118.367(4)<br>245.861(4) | E2<br>E2<br>E1 | %IT = 100 | 1975AL21 |
| 56 | 78 | $^{134}$Ba | 1986.35 (21) | 5− | 50(5) *ns* | 16<br>585.5(3)<br>1382.0(3) | <br>E1 | %IT = 100 | 2019KA36<br>1980MO27 |
| 56 | 78 | $^{134}$Ba | 2957.2 (5 ) | 10+ | 2.61(14) *μs* | 121.3(3) | E2 | %IT = 100 | 2019KA36<br>1982BEZY |
| 57 | 77 | $^{134}$La | 336.44+X | | 29(4) *μs* | | | %IT = 100 | 1985MO01 |

*Continued...*

Table 1 contd...

| Z | N | $^{A}X$ | E(keV) | $J^{\pi}$ | $T_{1/2}$ | Eγ(keV) | λ | Decay mode | Reference |
|---|---|---|---|---|---|---|---|---|---|
| 58 | 76 | $^{134}$Ce | 3208.6 (4) | 10+ | 308(5) *ns* | 190.8(3)<br>397.4(3) | E2<br>E2 | %IT = 100 | 1983DA29 |
| 59 | 75 | $^{134}$Pr | 67.7 (4) | (6−) | ~ 11 *min* | | | %ϵ+%β+ = 100 | 2011TI10<br>1973AR13 |
| 60 | 74 | $^{134}$Nd | 2293.3 (5) | 8− | 0.39(2) *ms* | 166.5(3)<br>595.5(6)<br>873.1(6) | E1<br><br>(M2+E3) | %IT = 100 | 2017PE03<br>1972PA26 |
| 61 | 73 | $^{134}$Pm | 0+X | (5+) | 22(1) *s* | | | %ϵ+%β+ = 100 | 1990KO25<br>1988BEYG<br>1977BO02 |
| 61 | 73 | $^{134}$Pm | 70.7+X | (7−) | 20(1) *µs* | 70.7 (2) | E1 | %IT = 100 | 2009CU02 |
| 51 | 84 | $^{135}$Sb | 1343.2 (2) | (19/2+) | ~ 20 *ns* | 225.1(1) | (E2) | %IT = 100 | 1998BH09 |
| 52 | 83 | $^{135}$Te | 1554.89 (16) | (19/2−) | 0.511(20) *µs* | 50.0(1) | E2 | %IT = 100 | 2001MI22<br>1980KA30 |
| 54 | 81 | $^{135}$Xe | 526.551 (13) | 11/2− | 15.29(5) *min* | 526.561(17) | M4 | %IT = 99.7(3)<br>%β−= 0.3(3) | 1975FU12 |
| 55 | 80 | $^{135}$Cs | 1632.9 (15) | 19/2− | 53(2) *min* | 846.1 | M4 | %IT = 100 | 1964HA18<br>1962WA22 |
| 56 | 79 | $^{135}$Ba | 268.218 (20) | 11/2− | 28.7(2) *h* | 268.218(20) | M4 | %IT = 100 | 2012DA04<br>1960WI10 |
| 56 | 79 | $^{135}$Ba | 2388 | 23/2+ | 1.06(4) *ms* | 254 | (M2+E3) | %IT = 100 | 2018KA47 |
| 57 | 78 | $^{135}$La | 2738.43 (24) | (23/2+) | 25.3(3) *ns* | 470.6(1) | | %IT = 100 | 2019LA01<br>2013LE27 |
| 58 | 77 | $^{135}$Ce | 445.81 (21) | (11/2−) | 20(1) *s* | 149.7(2) | E3 | %IT = 100 | 1971VA22<br>1970DR04<br>1963BR30 |
| 59 | 76 | $^{135}$Pr | 358.06 (6) | (11/2−) | 105(10) *µs* | 112.60(5)<br>316.6(1) | M2<br>E3 | %IT = 100 | 1973CO32 |
| 60 | 75 | $^{135}$Nd | 64.95 (24) | (1/2+) | 5.5(5) *min** | | | %ϵ+%β+ = 99.985(15)<br>%IT = 0.015(15) | 1989KO07<br>1989VI04<br>1970AB07 |
| 61 | 74 | $^{135}$Pm# | 0+X | (3/2+,<br>5/2+) | 49(3) *s* | | | %ϵ+%β+ = 100 | 1989KO07<br>1989VI04 |
| 61 | 74 | $^{135}$Pm | 68.7+Y | (11/2−) | 45(4) *s* | | | %ϵ+%β+ = 100 | 1989KO07<br>1989VI04 |
| 62 | 73 | $^{135}$Sm | ~ 100+Y | (9/2−) | 21(14) *ns* | ~ 100 | | %IT = 100 | 1996MUZZ |
| 50 | 86 | $^{136}$Sn | 1295 (2) | (6+) | 46(7) *ns* | 216(1) | | %IT = 100 | 2014SI18 |
| 51 | 85 | $^{136}$Sb | 269.3 (5) | (6−) | 536(40) *ns* | 53.4(3) | (E2) | %IT = 100 | 2015LO08<br>2012KA36<br>2007SI27<br>2001MI22 |
| 53 | 83 | $^{136}$I | 201 (26) | (6−) | 46.6(11) *s* | | | %β−= 100 | 2007FO02<br>1977WE04<br>1971LU02<br>1970CA25 |

*Continued...*

*: Evidence for this isomer is still lacking since no decay data are available.
#: ENSDF evaluators suggest that this level may be the ground state.

Table 1 contd…

| Z | N | $^{A}$X | E(keV) | $J^{\pi}$ | $T_{1/2}$ | Eγ(keV) | λ | Decay mode | Reference |
|---|---|---|---|---|---|---|---|---|---|
| 54 | 82 | $^{136}$Xe | 1891.703 (14) | 6+ | 2.92(3) $\mu s$ | 197.316(7) | E2 | %IT = 100 | 2017KI09<br>1974CLZX<br>1970JO20<br>1970GR38<br>1970CA25<br>1969WA29 |
| 55 | 81 | $^{136}$Cs | 517.9 (1) | 8− | 17.5(2) $s$ | 413.1(3)<br>517.9 | M4<br>E3 | %IT > 0<br>%β−= ? | 2011WI09 |
| 56 | 80 | $^{136}$Ba | 2030.535 (18) | 7− | 0.3075(31) $s$ | 163.920(2) | E3 | %IT = 100 | 2018KA47<br>1972BR53 |
| 56 | 80 | $^{136}$Ba | 3357.19 (25) | 10(+) | 91(2) $ns$ | 363.0(2) | E2 | %IT = 100 | 2004VA03<br>2004SH15 |
| 57 | 79 | $^{136}$La | 259.5 (3)? | (7−) | 114(5) $ms$ | 87.5(2)? | (M4) | %IT = 100 | 2005BH06<br>1985MO01<br>1980SUZY<br>1966GR19 |
| 57 | 79 | $^{136}$La | 270.13 (25) | (3) | 17(4) $ns$ | 98.0(2)<br>130.2(2)<br>248.4(2) | | %IT = 100 | 2005BH06<br>1985MO01 |
| 57 | 79 | $^{136}$La | 2520.6 (4) | (14+) | 187(27) $ns$ | 149.0(2)<br>408.1(2) | | %IT = 100 | 2015NI05 |
| 58 | 78 | $^{136}$Ce | 3095.0 (6) | 10+ | 2.0(1) $\mu s$ | 105.7(5) | E2 | %IT = 100 | 2013VA10<br>1975YO01 |
| 59 | 77 | $^{136}$Pr | 594.60 (20) | (6)+ | 91(1) $ns$ | 46.8(2)<br>420.1(2) | E2<br>M2 | %IT = 100 | 1996PE12<br>1993BA42<br>1992OL03<br>1975BR16 |
| 61 | 75 | $^{136}$Pm$^{\dagger}$ | 0+X | (2) | 300(50) $s$ | | | %ϵ+%β+ = 100 | 1989VI04<br>1988KE04 |
| | | | 0+Y | (5−) | 107(6) $s$ | | | %ϵ+%β+ = 100 | 1973PAZV |
| 61 | 75 | $^{136}$Pm | 70.0+X (6) | (8+) | 1.5(1) $\mu s$ | 42.7(2) | (E1) | %IT = 100 | 2008RI05 |
| 62 | 74 | $^{136}$Sm | 2264.7 (11) | (8−) | 15(1) $\mu s$ | 465.9 | | %IT = 100 | 1994BR15 |
| 63 | 73 | $^{136}$Eu$^{\dagger}$ | 0+X | (7+) | 3.3(3) $s$ | | | %ϵ+%β+ = 100<br>%β+p = 0.09 | 1989VI04 |
| | | | 0+Y | (3+) | 3.8(3) $s$ | | | %ϵ+%β+ = 100<br>%β+p = 0.09 | 1989VI04<br>1987KE05 |
| 54 | 83 | $^{137}$Xe | 1935.2 (3) | 19/2− | 10.1(9) $ns$ | 314.10(10) | (E2) | %IT = 100 | 2005HW06<br>2004HW02 |
| 56 | 81 | $^{137}$Ba | 661.659 (3) | 11/2− | 2.5223(10) $min$ | 378.0(4)<br>661.657(3) | E5<br>M4 | %IT = 100<br>%2γ = 0.000205(37) | 2015WA29<br>2014MO32<br>1973LEZJ<br>1967MI11<br>1966MA28<br>1965ME03 |
| 56 | 81 | $^{137}$Ba | 2349.1 (5) | (19/2−) | 589(20) $ns$ | 120.2(3) | E2 | %IT = 100 | 2017VO01<br>1973KE07<br>1973KE22 |
| 57 | 80 | $^{137}$La | 10.59 (4) | 5/2+ | 89(4) $ns$ | 10.61(5) | M1 | %IT = 100 | 1963RU03 |
| 57 | 80 | $^{137}$La | 1869.50 (21) | 19/2− | 342(25) $ns$ | 83.2 | E2 | %IT = 100 | 2004VA03 |
| 58 | 79 | $^{137}$Ce | 254.29 (5) | 11/2− | 34.80(3) $h$ | 254.29(5) | M4 | %IT = 99.21(4)<br>%ϵ+%β+ = 0.79(4) | 2012TO09<br>1958DA13 |

*Continued…*

Table 1 contd…

| Z | N | $^{A}X$ | E(keV) | $J^{\pi}$ | $T_{1/2}$ | Eγ(keV) | λ | Decay mode | Reference |
|---|---|---|---|---|---|---|---|---|---|
| 59 | 78 | $^{137}$Pr | 561.22 (23) | 11/2− | 2.66(7) $\mu s$ | 331.3(2) | M2 | %IT = 100 | 1992DR04 |
| | | | | | | 563.4(2) | | | 1987DR12 |
| 60 | 77 | $^{137}$Nd | 519.43 (20) | 11/2− | 1.60(15) $s$ | 233.5(2) | E3 | %IT = 100 | 1971VA22 |
| | | | | | | | | | 1970DR04 |
| 50 | 88 | $^{138}$Sn | 1344 (2) | (6+) | 210(45) $ns$ | 168(1) | | | 2014SI18 |
| 53 | 85 | $^{138}$I | 67.9 (3)? | (3−) | 1.07(12) $\mu s$ | 67.9? | E2 | %IT = 100 | 2022MO11 |
| | | | | | | | | | 2007RZ01 |
| 55 | 83 | $^{138}$Cs | 79.9 (3) | 6− | 2.91(10) $min$ | 79.9(3) | M3 | %β−= 19(3) | 1978AU08 |
| | | | | | | | | %IT = 81(3) | 1971CA21 |
| 56 | 82 | $^{138}$Ba | 2090.536 (21) | 6+ | 0.85(10) $\mu s$ | 191.95(2) | | %IT = 100 | 2004VA03 |
| | | | | | | | | | 1978AU08 |
| | | | | | | | | | 1973KE07 |
| | | | | | | | | | 1971CA21 |
| 57 | 81 | $^{138}$La | 72.57 (3) | (3)+ | 116(5) $ns$ | 72.57(3) | (E2) | %IT = 100 | 1975ISZY |
| 57 | 81 | $^{138}$La | 738.80 (20) | (7)− | 2.0(3) $\mu s$ | 738.8(2) | (M2) | %IT = 100 | 2014AS02 |
| 58 | 80 | $^{138}$Ce | 2129.28 (12) | 7− | 8.73(20) $ms$ | 302.8(1) | E3 | %IT = 100 | 1977GO15 |
| | | | | | | | | | 1960MO19 |
| 58 | 80 | $^{138}$Ce | 3539.21 (15) | 10+ | 81(2) $ns$ | 31.9(2)? | (E1) | %IT = 100 | 1984LO14 |
| | | | | | | 109.0(2)? | | | 1983DA29 |
| | | | | | | 430.2(1) | E2 | | 1980ME11 |
| | | | | | | 1409.9(2) | (E3) | | 1978MU09 |
| | | | | | | | | | 1976LU07 |
| 59 | 79 | $^{138}$Pr | 364 (23) | 7− | 2.07(4) $h$ | | | %ϵ+%β+ = 100 | 1972EK04 |
| | | | | | | | | | 1971JU01 |
| | | | | | | | | | 1970HO28 |
| | | | | | | | | | 1964FU08 |
| | | | | | | | | | 1958DA13 |
| | | | | | | | | | 1951ST03 |
| 60 | 78 | $^{138}$Nd | 3174.5 (4) | 10+ | 370(5) $ns$ | 66.6(3) | E2 | %IT = 100 | 2013VA10 |
| | | | | | | | | | 1975YO01 |
| 61 | 77 | $^{138}$Pm | 0+X | (5−) | 3.24(5) $min$ | | | %ϵ+%β+ = 100 | 1981DE38 |
| 61 | 77 | $^{138}$Pm | 584.26+X (15) | (8+) | 21(5) $ns$ | 173.6(1) | | %IT = 100 | 1990BE28 |
| 64 | 74 | $^{138}$Gd | 2232.6 (11) | (8−) | 6.2(2) $\mu s$ | 583.2 | | %IT = 100 | 2011PR02 |
| | | | | | | | | | 1997BR02 |
| 57 | 82 | $^{139}$La | 1800.4 (4) | (17/2+) | 315(35) $ns$ | 88.7(5) | | %IT = 100 | 2012AS06 |
| 58 | 81 | $^{139}$Ce | 754.24 (8) | 11/2− | 57.58(32) $s$ | 754.24(8) | M4 | %IT = 100 | 2012TO09 |
| | | | | | | | | | 1967GE09 |
| 58 | 81 | $^{139}$Ce | 2631.9 (3) | (19/2−) | 70(5) $ns$ | 270.7(2) | E2 | %IT = 100 | 1977LU04 |
| | | | | | | | | | 1976LU07 |
| 59 | 80 | $^{139}$Pr | 821.98 (7) | 11/2− | 43.4(16) $ns$ | 708.06(6) | M2+E3 | %IT = 100 | 1980PI03 |
| | | | | | | 821.9(2) | E3 | | 1979KE07 |
| | | | | | | | | | 1972KR24 |
| 60 | 79 | $^{139}$Nd | 231.16 (5) | 11/2− | 5.50(20) $h$ | 231.15(5) | M4 | %ϵ+%β+ = 87.0(10) | 1951ST03 |
| | | | | | | | | %IT = 13.0(10) | |
| 60 | 79 | $^{139}$Nd | 2616.9 (6) | (23/2+) | 277(2) $ns$ | (45) | | | 2013VA10 |
| | | | | | | | | | 2011BH07 |
| | | | | | | | | | 2008FE02 |
| 61 | 78 | $^{139}$Pm | 188.7 (3) | (11/2)− | 180(20) $ms$ | 188.7(3) | E3 | %IT = 100 | 2011ZH47 |
| | | | | | | | | %ϵ+%β+ < 0.05 | 2010ZH12 |
| | | | | | | | | | 2009DH01 |
| | | | | | | | | | 1975VA14 |

*Continued…*

Table 1 contd…

| Z | N | $^{A}$X | E(keV) | J$^{\pi}$ | $T_{1/2}$ | E$\gamma$(keV) | $\lambda$ | Decay mode | Reference |
|---|---|---|---|---|---|---|---|---|---|
| 62 | 77 | $^{139}$Sm | 457.38 (23) | 11/2− | 10.7(6) *s* | 190.1(2) | E3 | %$\epsilon$+%$\beta$+ = 6.3(5)<br>%IT = 93.7(5) | 1986DE35<br>1975VA14 |
| 63 | 76 | $^{139}$Eu | 148.3 (3) | (7/2+) | 10(2) $\mu s$ | 26.7(2) | (E1) | %IT = 100 | 2011CU01 |
| 64 | 75 | $^{139}$Gd | 0+X ? | | 4.8(9) *s* | | | %$\epsilon$+%$\beta$+ = 100<br>%$\epsilon$p > 0 | 1999XI04 |
| 51 | 89 | $^{140}$Sb | 298.2+X | (6−, 7−) | 41(8) $\mu s$ | | | %IT = 100 | 2016LO01 |
| 53 | 87 | $^{140}$I | 0+X | (0−, 1) | 0.91(5) *s* | | | %$\beta$−≈ 100<br>%IT = ? | 2022YA13 |
| 53 | 87 | $^{140}$I | 0+Y | (4−, 5) | 0.47(4) *s* | | | %$\beta$−≈ 100<br>%IT = ? | 2022YA13 |
| 55 | 85 | $^{140}$Cs | 13.931 (21) | (2)− | 471(51) *ns* | 13.93(5) | M1 | %IT = 100 | 1975MO03 |
| 58 | 82 | $^{140}$Ce | 2107.854 (24) | 6+ | 7.3(15) $\mu s$ | 24.595(4) | E2 | %IT = 100 | 1969IV02 |
| 58 | 82 | $^{140}$Ce | 3714.3 (3) | 10+ | 23.1(4) *ns* | 202.0(3)<br>222.0(3) | E2<br>E1 | %IT = 100 | 1984EN01 |
| 59 | 81 | $^{140}$Pr | 127.8 (3) | 5+ | 0.35(2) $\mu s$ | 98.5(3) | E2 | %IT = 100 | 1975SC17 |
| 59 | 81 | $^{140}$Pr | 763.7 (5) | (7)− | 3.05(20) $\mu s$ | 635.9(3) | (M2, E3) | %IT = 100 | 1964KR02 |
| 60 | 80 | $^{140}$Nd | 2221.65 (9) | 7− | 0.60(5) *ms* | 419.81(1) | E3 | %IT = 100 | 1962RE04 |
| 60 | 80 | $^{140}$Nd | 3621.52 (13) | 10+ | 22.4(7) *ns* | 166.57(4) | E1 | %IT = 100 | 2006PE25<br>1987GU22<br>1981ME09<br>1980ME11 |
| 60 | 80 | $^{140}$Nd | 7435.1 (4) | (20+) | 1.22(7) $\mu s$ | 36.8<br>227.5(2)<br>1028.0(5) | | %IT = 100 | 2013VA10<br>2008FE02 |
| 61 | 79 | $^{140}$Pm | 0+X | 8− | 5.95(5) *min* | | | %$\epsilon$+%$\beta$+ = 100 | 1975KE09 |
| 62 | 78 | $^{140}$Sm | 3172.1 (4) | 10+ | 19.4(7) *ns* | (44)<br>202.6(2) | <br>E2 | %IT = 100 | 1988BA22<br>1988ST02 |
| 62 | 78 | $^{140}$Sm | 3652.8 (3) | 12+ | 15.2(21) *ns* | 441.9(1) | E2 | %IT = 100 | 1991CA17 |
| 63 | 77 | $^{140}$Eu | 0+X | (5−) | 125(2) *ms* | <48?<br><59? | (M2)<br>(E3) | %IT = 100<br>%$\epsilon$+%$\beta$+ < 1 | 2006TA08<br>1991FI03 |
| 63 | 77 | $^{140}$Eu | 459.5+X (3) | (8+) | 299.8(21) *ns* | 37.1(3)<br>98.1(3) | E1<br>E1 | %IT = 100 | 2006TA08<br>2002CU05 |
| 66 | 74 | $^{140}$Dy | 2166.1 (5) | (8−) | 7.0(5) $\mu s$ | 573.8(2)? | (E1) | %IT = 100 | 2002CU01<br>2002KR04 |
| 55 | 86 | $^{141}$Cs | 69.05 (3) | (3/2)+ | 23.3(7) *ns* | 69.05(3) | (E2) | %IT = 100 | 1975MO03 |
| 60 | 81 | $^{141}$Nd | 756.51 (5) | 11/2− | 62.0(10) *s* | 756.51(5) | M4 | %IT = 99.975(25)<br>%$\epsilon$+%$\beta$+ = 0.025(25) | 1988CH39<br>1969JA02<br>1967GE09<br>1966GR05<br>1960KO02<br>1960JA06 |
| 60 | 81 | $^{141}$Nd | 2886+X? | | 26(5) *ns* | | | %IT = 100 | 1976LU07 |
| 61 | 80 | $^{141}$Pm | 628.62 (7) | 11/2− | 0.63(2) $\mu s$ | 431.6(1)<br>628.6(1) | M2<br>E3 | %IT = 100 | 1985AR19 |
| 61 | 80 | $^{141}$Pm | 2349.52 (16) | 19/2(−) | 54(5) *ns* | 110.5(1) | | %IT = 100 | 2004BH01 |
| 61 | 80 | $^{141}$Pm | 2530.75 (17) | | > 2 $\mu s$ | 639.0(1)<br>1020.3(3) | | %IT = 100 | 1985AR19 |
| 61 | 80 | $^{141}$Pm | 2574.4 (4)? | | ≥ 2 $\mu s$ | 44.0(2)? | | %IT = 100 | 2004BH01<br>2000BH08 |

*Continued…*

Table 1 contd…

| Z | N | $^{A}$X | E(keV) | J$^{\pi}$ | $T_{1/2}$ | E$\gamma$(keV) | $\lambda$ | Decay mode | Reference |
|---|---|---|---|---|---|---|---|---|---|
| 62 | 79 | $^{141}$Sm | 175.9 (3) | 11/2− | 22.6(2) *min* | 174.2(3) | M4 | %IT = 0.31(3)<br>%$\epsilon$+%$\beta$+ = 99.69(3) | 1993AL03<br>1972EP01<br>1972DE23<br>1970AR17<br>1968BL13<br>1967HE23 |
| 63 | 78 | $^{141}$Eu | 96.45 (7) | 11/2− | 2.7(3) *s* | 96.4(1) | E3 | %$\epsilon$+%$\beta$+ = 13 (+4−2)<br>%IT = 87(+2−4) | 1989GI06 |
| 64 | 77 | $^{141}$Gd | 377.76 (9) | 11/2− | 24.5(5) *s* | 119.6(1) | E3 | %IT = 11(2)<br>%$\epsilon$+%$\beta$+ = 89(2) | 1989GI06 |
| 67 | 74 | $^{141}$Ho | 66 (12) | 1/2+ | 7.3(3) $\mu s$ | | | %p = 100 | 2008KA16<br>2001SE03<br>1999RY04 |
| 55 | 87 | $^{142}$Cs | 122.9+X | | 11(3) *ns* | 26.4<br>97.3 | <br>(E2) | %IT = 100 | 2009RZ02 |
| 57 | 85 | $^{142}$La | 145.82 (8) | (4)− | 0.87(17) $\mu s$ | 68.3(1) | E2 | %IT = 100 | 1983CH39 |
| 59 | 83 | $^{142}$Pr | 3.694 (3) | 5− | 14.6(5) *min* | (3.683(4)) | M3 | %IT = 100 | 1968KE08<br>1967KE05 |
| 59 | 83 | $^{142}$Pr | 911.4 (13) | (9+) | 61(6) *ns* | 553 | | %IT = 100 | 1975SC17 |
| 60 | 82 | $^{142}$Nd | 2100.787 (13) | 4+ | 28(2) *ns* | 16.9<br>525.009(10) | | %IT = 100 | 1975KL01 |
| 60 | 82 | $^{142}$Nd | 2209.303 (21) | 6+ | 18.6(37) $\mu s$ | 108.52(2) | | %IT = 100 | 1969IV02<br>1964KR02 |
| 61 | 81 | $^{142}$Pm | 449.47 (13) | (5)+ | 16.5(15) *ns* | 37.5<br>208.5(1) | | %IT = 100 | 1976FU07 |
| 61 | 81 | $^{142}$Pm | 883.17 (16) | (8)− | 2.0(2) *ms* | 433.7(1) | E3 | %IT = 100 | 1976FU07 |
| 61 | 81 | $^{142}$Pm | 2828.7 (6) | (13−) | 67(5) $\mu s$ | 638.6(3)<br>1019.4 | | %IT = 100 | 2004LI49 |
| 62 | 80 | $^{142}$Sm | 2372.1 (4) | 7− | 142.1(28) *ns* | 24.1(3)<br>580.7(4)? | E2 | %IT = 100 | 2014RA03<br>1975KE08 |
| 62 | 80 | $^{142}$Sm | 3662.2 (7) | 10+ | 480(60) *ns* | 275.1<br>336.0<br>1290.3 | | %IT = 100 | 2014RA03<br>1984LAZU |
| 62 | 80 | $^{142}$Sm | 6657.9 (5) | 17+ | 41.6(14) *ns* | 351.8(1) | M1 | | 2014RA03 |
| 63 | 79 | $^{142}$Eu | 0+X | 8− | 1.223(8) *min* | | | %$\epsilon$+%$\beta$+ = 100 | 1993AL03 |
| 65 | 77 | $^{142}$Tb | 279.7 (4) | 5− | 303(17) *ms* | 68.2(3)<br>98.3(9) | M2<br>E3 | %IT = 100 | 2006TA08<br>1991FI03 |
| 65 | 77 | $^{142}$Tb | 652.1 (6) | 8+ | 26(1) $\mu s$ | 37.2(3)<br>340.1(6) | E1<br>M2 | %IT = 100 | 2009MA06 |
| 60 | 83 | $^{143}$Nd | 8987.7 (9) | 49/2+ | 35(8) *ns* | 300.6(5)<br>338.6(5) | | %IT = 100 | 2000ZH03 |
| 61 | 82 | $^{143}$Pm | 959.73 (13) | 11/2− | 24.0(7) *ns* | 687.7(4)<br>959.8(2) | M2<br>E3 | %IT = 100 | 1981KO16<br>1980PR02 |
| 61 | 82 | $^{143}$Pm | 1898.36 (14) | 15/2+ | 10.5(4) *ns* | 234.9(1) | E2 | %IT = 100 | 1981KO16<br>1980PR02 |
| 62 | 81 | $^{143}$Sm | 753.99 (16) | 11/2− | 66(2) *s* | 754.0(2) | (M4) | %IT = 99.76(5)<br>%$\epsilon$+%$\beta$+ = 0.24(5) | 1969JA02<br>1967GO06<br>1963AL05<br>1960KO02 |
| 62 | 81 | $^{143}$Sm | 2793.8 (13) | 23/2(−) | 30(3) *ms* | 208 | E3 | %IT = 100 | 1969NE04 |

*Continued…*

Table 1 contd...

| Z | N | $^{A}X$ | E(keV) | $J^{\pi}$ | $T_{1/2}$ | Eγ(keV) | λ | Decay mode | Reference |
|---|---|---|---|---|---|---|---|---|---|
| 63 | 80 | $^{143}$Eu | 389.51 (4) | 11/2− | 50.0(5) *μs* | 117.57(5)<br>389.47(5) | M2<br>E3 | %IT = 100 | 1978FI02 |
| 64 | 79 | $^{143}$Gd | 152.6 (5) | (11/2−) | 110.0(14) *s* | | | %ε+%β+ = 100 | 1978FI02<br>1976WI09<br>1973VAYZ |
| 66 | 77 | $^{143}$Dy | 310.7 (6) | (11/2−) | 3.0(3) *s* | | | %ε+%β+ = 100<br>%εp = ? | 2003XU04 |
| 66 | 77 | $^{143}$Dy | 405.7 | | 1.2(3) *μs* | 86<br>95 | (E2)<br>(E2) | %IT = 100 | 2005RI17 |
| 55 | 89 | $^{144}$Cs | 92.2 (5) | (4−) | 1.1(1) *μs* | 49.4 | (E2) | %IT = 100 | 2009RZ02 |
| 59 | 85 | $^{144}$Pr | 59.03 (3) | 3− | 7.2(3) *min* | 59.03(3) | M3 | %IT = 99.95(3)<br>%β−= 0.05(3) | 1985DA14<br>1976CH33<br>1970FA03 |
| 61 | 83 | $^{144}$Pm | 840.90 (5) | (9)+ | 0.78(20) *μs* | 326.55(5)<br>608.7<br>669.15(8) | M2<br>E3<br>E3 | %IT = 100 | 1993MU03 |
| 61 | 83 | $^{144}$Pm | 8595.8 (22) | (25 to 35) | ~ 2.7 *μs* | 85.5 | | %IT = 100 | 1994ZHZW |
| 62 | 82 | $^{144}$Sm | 2323.60 (8) | 6+ | 880(25) *ns* | 132.6(1) | E2 | %IT = 100 | 1972KO42 |
| 63 | 81 | $^{144}$Eu | 926.0 (4) | (6−) | 27.8(16) *ns* | 38.7(3)<br>163.1(5) | (M1+E2) | %IT = 100 | 1981HA25<br>1976FU07 |
| 63 | 81 | $^{144}$Eu | 1127.6 (6) | (8−) | 1.0(1) *μs* | 7.5<br>201.6(5) | <br>E2 | %IT = 100 | 1976FU07 |
| 64 | 80 | $^{144}$Gd | 2471.6 (4) | (7−) | 13(2) *ns* | 169.0(2) | E2 | %IT = 100 | 1978MA43 |
| 64 | 80 | $^{144}$Gd | 3433.1 (5) | (10+) | 145(30) *ns* | 87.4(3)<br>415.2(3) | E1<br>M2 | %IT = 100 | 1978MA43 |
| 65 | 79 | $^{144}$Tb | 396.9 (5) | 6− | 4.25(15) *s*<br>12(2) *s* @ | 113.0(3) | E3 | %IT = 66<br>%ε+%β+ = 34 | 2014CH22<br>1986RE11<br>2003LI42@ |
| 65 | 79 | $^{144}$Tb | 476.2 (5) | 8− | 2.8(3) *μs* | 79.3(2) | E2 | %IT = 100 | 2014CH22<br>1996SF01 |
| 65 | 79 | $^{144}$Tb | 517.1 (5) | 9+ | 0.67(6) *μs* | 40.8(2) | E1 | %IT = 100 | 2014CH22<br>1996SF01 |
| 67 | 77 | $^{144}$Ho | 265.3(3) | (8+) | 519(5) *ns* | 56.4(2) | E1 | %IT = 100 | 2010MA08<br>2006TA08<br>2001SC09 |
| 58 | 87 | $^{145}$Ce | 64.3 (2) | (−) | 13(3) *ns* | 64.3(2) | E2 | %IT = 100 | 1978PF02 |
| 60 | 85 | $^{145}$Nd | 67.167 (7) | 3/2− | 29.4(10) *ns* | 67.10(1) | E2 | %IT = 100 | 1970KA36<br>1967MY01 |
| 61 | 84 | $^{145}$Pm | 794.6 (4) | 11/2− | 17.1(10) *ns* | (44.2)<br>80.7(3)<br>733.4(3)<br>794.6(5) | <br>E1<br>M2<br>E3 | %IT = 100 | 1996UR03<br>1980KO16<br>1976SH05 |
| 62 | 83 | $^{145}$Sm | 1105.03 (16) | 13/2+ | 13.1(18) *ns* | 1105.0(2) | E3 | %IT = 100 | 2020GE08<br>1977HA04<br>1975CL01 |
| 62 | 83 | $^{145}$Sm | 8815 | 49/2+ | 3.52(16) *μs* | 30.2(8)<br>(236)<br>(743) | E1<br>(E2)<br>(E2) | %IT = 100 | 2020GE08<br>1993FE14 |
| 63 | 82 | $^{145}$Eu | 716.0 (3) | 11/2− | 490(30) *ns* | 386.6(3)<br>716.0(3) | M2<br>E3 | %IT = 100 | 1975FR18 |

*Continued...*

Table 1 contd...

| Z | N | $^{A}$X | E(keV) | J$^{\pi}$ | $T_{1/2}$ | E$\gamma$(keV) | $\lambda$ | Decay mode | Reference |
|---|---|---|---|---|---|---|---|---|---|
| 64 | 81 | $^{145}$Gd | 27.3 (11) | 3/2+ | 11.5(3) *ns* | 27.3(1) | M1+E2 | %IT = 100 | 1975FI02 |
| 64 | 81 | $^{145}$Gd | 749.1 (2) | 11/2− | 85(3) *s* | 721.8(1) | M4 | %IT = 94.3(5) | 1974KO29 |
| | | | | | | | | %$\epsilon$+%$\beta$+ = 5.7(5) | 1970EP02 |
| | | | | | | | | | 1970SEZP |
| | | | | | | | | | 1969JA02 |
| 64 | 81 | $^{145}$Gd | 2200.1 (2) | 13/2+ | 20.4(16) *ns* | 927.2 | E3 | %IT = 100 | 1982PA04 |
| | | | | | | 1451.3 | E1+M2+E3 | | |
| 65 | 80 | $^{145}$Tb* | 0+Y* | (11/2−) | 30.9(6) *s* | | | %$\epsilon$+%$\beta$+ = 100 | 1993AL03 |
| | | | | | | | | | 1982NO08 |
| | | | | | | | | | 1982SO02 |
| | | | | | | | | | 1982AL07 |
| | | | | | | | | | 1982STZU |
| 66 | 79 | $^{145}$Dy | 118.2 (2) | (11/2−) | 14.1(7) *s* | | | %$\epsilon$+%$\beta$+ = 100 | 1993TO04 |
| | | | | | | | | %$\epsilon$p ≈ 50 | 1982NO08 |
| 68 | 77 | $^{145}$Er | 253 | (11/2−) | 0.9(3) *s* | | | %$\epsilon$+%$\beta$+ = 100 | 2010MA20 |
| | | | | | | | | %$\epsilon$p = ? | 2006TA08 |
| | | | | | | | | | 1989VI02 |
| 55 | 91 | $^{146}$Cs$^{\#}$ | | | 1.7(1) $\mu s$ | | | | 2015YAZW |
| 57 | 89 | $^{146}$La | 141.4 (26) | (6−) | 9.8(4) *s* | | | %$\beta$−= 100 | 2020OR02 |
| | | | | | | | | %IT = ? | 1979EN02 |
| | | | | | | | | | 1978MOYW |
| | | | | | | | | | 1974AR17 |
| 57 | 89 | $^{146}$La | 619.9+X (2) | (3+) | 14(1) *ns* $^{\$}$ | 43.0? | | %IT = 100 | 2017WA50 |
| | | | | | | 148.3(1) | | | 1998HW08 |
| | | | | | | | | | 1981SEZW |
| 63 | 83 | $^{146}$Eu | 666.33 (11) | 9+ | 235(3) $\mu s$ | 18.8 | (M2) | %IT = 100 | 1980LEZN |
| | | | | | | 293.72(9) | E3 | | 1980ER04 |
| | | | | | | 377.00(9) | E3 | | 1971HAXM |
| | | | | | | | | | 1962RE04 |
| 63 | 83 | $^{146}$Eu | 8649.9 (7) | (27) | 10.0(6) *ns* | 204 | D, E2 | %IT = 100 | 1999ID01 |
| 65 | 81 | $^{146}$Tb | 0+X | 5− | 24.0(5) *s* | | | %$\epsilon$+%$\beta$+ = 100 | 1993AL03 |
| | | | | | | | | | 1974NE01 |
| 65 | 81 | $^{146}$Tb | 779.57+X (16) | 10+ | 1.20(3) *ms* | 417.7(1) | E3 | %IT = 100 | 2011KO08 |
| | | | | | | | | | 1989BR22 |
| 66 | 80 | $^{146}$Dy | 2934.5 (4) | 10+ | 150(20) *ms* | 127.0(3) | E3 | %IT = 100 | 1982GU07 |
| | | | | | | 416.5(3) | E3 | | |
| 69 | 77 | $^{146}$Tm | 182 (4) | (10+) | 198(3) *ms* | | | %p = ? | 2006TA08 |
| | | | | | | | | %$\epsilon$+%$\beta$+ = ? | 2005RO40 |
| 59 | 88 | $^{147}$Pr | 93.29 (9) | (7/2+) | 12 *ns* | 65.21 | | %IT = 100 | 1981SCZM |
| | | | | | | 90.44 | | | |
| | | | | | | 93.17 | M1 | | |
| 61 | 86 | $^{147}$Pm | 649.14 (6) | 11/2− | 27(3) *ns* | 117.98(5)? | E3 | %IT = 100 | 1995UR01 |
| | | | | | | 159.7(2)? | M2 | | |
| | | | | | | 241.2(2) | E1 | | |
| | | | | | | 649.04(8) | M2 | | |
| 63 | 84 | $^{147}$Eu | 625.27 (5) | 11/2− | 0.765(15) $\mu s$ | 396.00(10) | M2 | %IT = 100 | 1970KL07 |
| | | | | | | 625.18(10) | E3 | | |

*Continued...*

*: This isomer could correspond to the g.s. as no other long-lived activity has been reported.

#: Based on private communication to B. Singh from the first author A. Yagi of 2015YAZW, Sept 7, 2019.

$: The uncertainty could also be 6 *ns* as stated by 2017WA50.

Table 1 contd...

| Z | N | $^{A}$X | E(keV) | J$^{\pi}$ | $T_{1/2}$ | E$\gamma$(keV) | $\lambda$ | Decay mode | Reference |
|---|---|---|---|---|---|---|---|---|---|
| 64 | 83 | $^{147}$Gd | 997.10 (10) | 13/2+ | 21.4(11) *ns* | 997.1(1) | E3 | %IT = 100 | 1983KO42<br>1979HA15 |
| 64 | 83 | $^{147}$Gd | 3581.9 (4) | 27/2− | 26.8(7) *ns* | 182.9(1)<br>543.7(1)<br>821.3(2) | E1 | %IT = 100 | 1979HA15 |
| 64 | 83 | $^{147}$Gd | 8587.8 (5) | 49/2+ | 510(20) *ns* | 254.4(1)<br>434.5(1)<br>593.7(1)<br>623.8(3) | E2<br><br>(E3) | %IT = 100 | 2020BR06<br>1982HA22<br>1982BA46 |
| 65 | 82 | $^{147}$Tb | 50.6 (9) | (11/2−) | 1.87(6) *min* | | | %$\epsilon$+%$\beta$+ = 100 | 1993AL03<br>1973BO13 |
| 66 | 81 | $^{147}$Dy | 750.5 (4) | (11/2−) | 55.2(5) *s* | 678.4(3) | (M4) | %IT = 31.1(23)<br>%$\epsilon$+%$\beta$+ = 68.9(23) | 1993AL03<br>1985AL08<br>1984SCZT<br>1983BYZZ |
| 66 | 81 | $^{147}$Dy | 3407.2 (8) | (27/2−) | 0.40(1) $\mu s$ | 81.9(1) | E2 | %IT = 100 | 1985BR07 |
| 66 | 81 | $^{147}$Dy | 3650.6 (9) | (27/2−) | 16(3) *ns* | 92.3(3)<br>243.4(3) | (M1) | %IT = 100 | 1985BR07 |
| 67 | 80 | $^{147}$Ho | 2687.1 (4) | (27/2−) | 315(30) *ns* | 32.3 | | %IT = 100 | 2001RO15<br>1982NO07 |
| 68 | 79 | $^{147}$Er | 0+X | (11/2−) | 1.6(2) *s* | | | %$\epsilon$+%$\beta$+ = 100<br>%$\beta$+p > 0 | 2010MA27<br>1993TO02<br>1984SCZT |
| 69 | 78 | $^{147}$Tm | 68 (6) | 3/2+ | 0.36(4) *ms* | | | %p = 100 | 1993SE04<br>1993TO02 |
| 55 | 91 | $^{148}$Cs* | 0+X | | 4.8(2) $\mu s$ | | | | 2015YAZW<br>2015ODZZ |
| 57 | 91 | $^{148}$La | 56.034 (25) | (1+) | 67(4) *ns* | 56.08(4) | E1 | %IT = 100 | 1984CH02 |
| 59 | 89 | $^{148}$Pr | 76.80 (20) | 4− | 2.01(7) *min* | 76.8(2) | M3 | %IT = 36(10)<br>%$\beta$−= 64(10) | 1986WA06<br>1979IK06 |
| 61 | 87 | $^{148}$Pm | 137.9 (3) | 5−, 6− | 41.29(11) *d* | 62.2(5) | E4 | %IT = 4.2(6)<br>%$\beta$−= 95.8(6) | 1971WA05 |
| 62 | 86 | $^{148}$Sm | 6694.32 (21) | 21(−) | 32(3) *ns* | 101.5(1)<br>217.3(1)<br>302.0(2) | | %IT = 100 | 1998URZZ |
| 63 | 85 | $^{148}$Eu | 720.4 (3) | 9+ | 162(8) *ns* | (12.0(4))<br>201.9(3) | <br>M2 | %IT = 100 | 1995JO04<br>1981PI10<br>1980BA67 |
| 64 | 84 | $^{148}$Gd | 2694.67 (13) | 9− | 16.6(3) *ns* | 130.8(3)<br>883.6(3) | E2<br>E3 | %IT = 100 | 1990PI17<br>1979HA15<br>1973KR10<br>1972HAXQ |
| 65 | 83 | $^{148}$Tb | 90.1 (3) | (9)+ | 2.20(5) *min* | | | %$\epsilon$+%$\beta$+ = 100 | 1973BO13 |
| 65 | 83 | $^{148}$Tb | 109.6 (2) | 4− | 80(4) *ns* | 109.6(2) | E2 | %IT = 100 | 1987STZV |
| 65 | 83 | $^{148}$Tb | 1095.8 | (11)− | 22(1) *ns* | 690.2<br>1006.2 | <br>E3+M2 | %IT = 100 | 1980BO07<br>1980JA16<br>1979BR28<br>1979SI08 |

*Continued...*

*: Isomer confirmed but half-life is subject to revision based on private communication of Sept 7, 2019 from A. Yagi to B. Singh.

Table 1 contd...

| Z | N | $^{A}$X | E(keV) | J$^{\pi}$ | $T_{1/2}$ | Eγ(keV) | λ | Decay mode | Reference |
|---|---|---|---|---|---|---|---|---|---|
| 65 | 83 | $^{148}$Tb | 8618.6 (10) | (27+) | 1.310(7) $\mu s$ | 280.6<br>784.8<br>857.7 | | %IT = 100 | 1995ID01 |
| 66 | 82 | $^{148}$Dy | 2832.2 (10) | 8+ | 65(10) $ns$ | 94.1(2)<br>101.2(6) | E1<br>E2 | %IT = 100 | 1980KL09 |
| 66 | 82 | $^{148}$Dy | 2919.1 | 10+ | 471(20) $ns$ | 86.9 | E2 | %IT = 100 | 1981HA17 |
| 67 | 81 | $^{148}$Ho | 0+X | (5)− | 9.59(15) $s$ | | | %$\epsilon$+%$\beta^{+}$ = 100<br>%$\epsilon$p = 0.08(1) | 2010KO12<br>1989TA11<br>1988TO03 |
| 67 | 81 | $^{148}$Ho | 694.4+X | (10)+ | 2.36(6) $ms$ | 373.1(1) | E3 | %IT = 100 | 2010KO12<br>1989BR22<br>1984BR07 |
| 68 | 80 | $^{148}$Er | 2913.2 (4) | (10+) | 13(3) $\mu s$ | 131.2(2) | E2 | %IT = 100 | 1982NO07 |
| 59 | 90 | $^{149}$Pr | 58.5 (3) | (7/2−) | 23.4(18) $ns$ | 58.5 | E1 | %IT = 100 | 2010RZ02<br>1996YAZV*<br>1974CLZX |
| 61 | 88 | $^{149}$Pm | 240.214 (7) | 11/2− | 35(3) $\mu s$ | 240.220(7) | M2 | %IT = 100 | 1967BA27<br>1966HE04 |
| 63 | 86 | $^{149}$Eu | 496.386 (2) | 11/2− | 2.45(5) $\mu s$ | 346.651(3)<br>496.383(2) | M2+E3<br>E3 | %IT = 100 | 1970KL07<br>1962PR06<br>1961BE08<br>1969IV02<br>1961SO04 |
| 65 | 84 | $^{149}$Tb | 35.78 (13) | 11/2− | 4.17(5) $min$ | | | %$\epsilon$+%$\beta$+ = 99.978(4)<br>%$\alpha$ = 0.022(4) | 1973BI06<br>1973BO13<br>1971AR31<br>1969CH32<br>1964MA19<br>1962MA14 |
| 66 | 83 | $^{149}$Dy | 1073.24 (9) | (13/2)+ | 12.5(15 ) $ns$ | 1073.2(1) | E3 | %IT = 100 | 1980DA18<br>1976ST01 |
| 66 | 83 | $^{149}$Dy | 2661.1 (4) | (27/2−) | 0.490(15) $s$<br>11(1) $s$ @ | 110.7(3) | E3 | %IT = 99.3(3)<br>%$\epsilon$+%$\beta$+ = 0.7(3) | 1987BAZV<br>1980JA16<br>1976ST08<br>2003LI42@ |
| 66 | 83 | $^{149}$Dy | 8520.5 (12) | (49/2+) | 28(2) $ns$ | 249?<br>475? | | %IT = 100 | 1981HA17<br>1980DA18<br>1980JA16<br>1979HA29 |
| 67 | 82 | $^{149}$Ho | 48.8 (2) | (1/2+) | 56(3) s | | | %$\epsilon$+%$\beta$+ = 100 | 1994ME13 |
| 67 | 82 | $^{149}$Ho | 2735.9 (5) | (27/2−) | 59(4) $ns$ | 144 | (E2) | %IT = 100 | 1980WI11 |
| 67 | 82 | $^{149}$Ho | 7.20E+3? (35) | | ≥ 100 $ns$ | | | | 1980BO07 |
| 68 | 81 | $^{149}$Er | 741.69 (23) | (11/2−) | 9.6(6) $s$ | 630.5(2) | M4 | %IT = 3.5(7)<br>%$\epsilon$+%$\beta$+ = 96.5(7)<br>%$\epsilon$p = 0.18(7) | 1989FI01<br>1984TO07<br>1984SCZT |
| 68 | 81 | $^{149}$Er | 2611.1 (3) | (19/2+) | 0.61(8) $\mu s$ | 68.9(1)<br>132.3(1) | E2<br>E2 | %IT = 100 | 1987BR14 |
| 68 | 81 | $^{149}$Er | 3242.7+X | (27/2−) | 4.8(1) $\mu s$ | 55.1(3) | E2 | %IT = 100 | 1987BR14 |

*Continued...*

*: 1996YAZV reports $T_{1/2}$=10(2) $ns$ from beta-gamma-coin timing in the decay of $^{149}$Ce.

Table 1 contd…

| Z | N | $^{A}X$ | E(keV) | J$^{\pi}$ | $T_{1/2}$ | E$\gamma$(keV) | $\lambda$ | Decay mode | Reference |
|---|---|---|---|---|---|---|---|---|---|
| 59 | 91 | $^{150}$Pr | 109.84 (5) | 1+ | 46.4(21) *ns* | 109.84 (5) | E1 | %$\beta$−= ?<br>%IT = ? | 2015KO23<br>1996YAZV |
| 63 | 87 | $^{150}$Eu | 41.7 (10) | 0− | 12.8(1) *h* | (~1) | | %$\beta$−= 89(2)<br>%$\epsilon$+%$\beta$+ = 11(2)<br>%IT ≤ 5E+(−8) | 1972EK05<br>1963YO07 |
| 63 | 87 | $^{150}$Eu | 588.81 (7) | 8+ | 45(3) *ns* | 26.8(1)?<br>171.50(9) | (E2) | %IT = 100 | 1983SO13 |
| 65 | 85 | $^{150}$Tb | 461 (27) | 9+ | 5.8(2) *min* | | | %$\epsilon$+%$\beta$+ ≈ 100 | 1972HA18 |
| 67 | 83 | $^{150}$Ho | X | (9)+ | 23.5(3) *s* | | | %$\epsilon$+%$\beta$+ = 100 | 1993AL03<br>1990SA32<br>1982MO19<br>1980LI18 |
| 67 | 83 | $^{150}$Ho | 1096.03+X | (11−) | 18(2) *ns* | 879.4(2)<br>1096.0(1) | | %IT = 100 | 1981WI08 |
| 67 | 83 | $^{150}$Ho | 2625.44+X | (17+) | 84(8) *ns* | 98.2(1) | (E2) | %IT = 100 | 1986MC14 |
| 67 | 83 | $^{150}$Ho | 7912.1+X | (28) | 787(36) *ns* | 776(1) | | %IT = 100 | 2006FU06 |
| 68 | 82 | $^{150}$Er | 2733.3 (4) | 8+ | ~ 20 *ns* | 100.52(9)<br>112.6(3) | E1 | %IT = 100 | 1984CH11 |
| 68 | 82 | $^{150}$Er | 2796.5 (5) | 10+ | 2.55(10) $\mu s$ | 63.2(3) | E2 | %IT = 100 | 1984CH11<br>1981LA26 |
| 68 | 82 | $^{150}$Er | 7371.9 (6) | | 15(4) *ns* | 39.4(4)<br>219.0(2)<br>1013.1(3) | E1 | %IT = 100 | 1984CH11 |
| 68 | 82 | $^{150}$Er | 9508.6 (7) | | 43(3) *ns* | 360.1(2) | | %IT = 100 | 1984CH11 |
| 69 | 81 | $^{150}$Tm | 671.3 (10) | (10+) | 5.2(3) *ms* | 331.2 | (E3) | %IT = 100 | 1989BR22 |
| 71 | 79 | $^{150}$Lu | 22 (6) | (1−,<br>2−) | 39(+8−6) $\mu s$ | | | %p = 100 | 2003GI10 |
| 58 | 93 | $^{151}$Ce | 0+X | | 1.02(6) *s* | | | %$\beta$−= ? | 1970WIZN<br>1969WIZX |
| 59 | 92 | $^{151}$Pr | 35.10 (10) | (1/2+,<br>3/2+,<br>7/2+) | 50(8) $\mu s$ | 35.1(1) | E1 | %IT = 100 | 2012MA03<br>2006KO25 |
| 59 | 92 | $^{151}$Pr | 96.0+X | | 20(8) *ns* | 96 | | %IT = 100 | 1974CLZW |
| 62 | 89 | $^{151}$Sm | 4.821 (3) | 3/2− | 35(1) *ns* | 4.821(3) | M1+E2 | %IT = 100 | 1971DR05 |
| 62 | 89 | $^{151}$Sm | 91.532 (9) | (9/2)+ | 78(1) *ns* | 25.71(1) | E1 | %IT = 100 | 1971HO09 |
| 62 | 89 | $^{151}$Sm | 261.13 (4) | (11/2)− | 1.4(1) $\mu s$ | 85.7(1)<br>113.21(5)<br>169.57(7)<br>195.26(5) | | %IT = 100 | 1973CO34 |
| 63 | 88 | $^{151}$Eu | 196.245 (10) | 11/2− | 58.9(5) $\mu s$ | 174.70(1)<br>196 | M2 | %IT = 100 | 1994SI11<br>1969FAZY<br>1968IO01<br>1967CO20<br>1960BE27<br>1958SH61 |
| 65 | 86 | $^{151}$Tb | 99.53 (5) | (11/2−) | 25(3) *s* | 27.1(1) | E3 | %IT = 93.4(20)<br>%$\epsilon$+%$\beta$+ = 6.6(20) | 1978KE12 |
| 66 | 85 | $^{151}$Dy | 6032.2 (8) | (49/2+) | 11.9(8) *ns* | 25 | E2 | %IT = 100 | 1981HA17<br>1979LI14<br>1979PI07<br>1979HA29 |

*Continued…*

Table 1 contd…

| Z | N | $^{A}X$ | E(keV) | $J^{\pi}$ | $T_{1/2}$ | Eγ(keV) | λ | Decay mode | Reference |
|---|---|---|---|---|---|---|---|---|---|
| 67 | 84 | $^{151}$Ho | 41.0 (2) | (1/2+) | 47.2(13) *s* | | | %ϵ+%β+ = 20(+20−15)<br>%α = 80(+15−20) | 1982BO04<br>1970TO16<br>1963MA17 |
| 68 | 83 | $^{151}$Er | 1140.35 (14) | (13/2+) | 10(3) *ns* | 338.5(2)<br>1140.2(2) | (M2)<br>E3 | %IT = 100 | 1980JA16 |
| 68 | 83 | $^{151}$Er | 2586.0 (5) | (27/2−) | 0.58(2) *s*<br>19(3) *s* @ | 57.7 | E3 | %IT = 95.3(4)<br>%ϵ+%β+ = 4.7(4) | 1988BA02<br>1980JA16<br>2003LI42@ |
| 68 | 83 | $^{151}$Er | 10286.6 (10) | (67/2−) | 0.42(5) *µs* | 894.6 | (E3) | %IT = 100 | 1990AN25 |
| 69 | 82 | $^{151}$Tm | 0+X | (1/2+) | 6.6(20) *s* | | | %ϵ+%β+ = 100 | 1990AK01<br>1988BA02 |
| 69 | 82 | $^{151}$Tm | 2655.67 (22) | (27/2−) | 0.451(34) *µs* | 140.4(1) | E2 | %IT = 100 | 1987MCZZ<br>1982HE08<br>1982NO13 |
| 69 | 82 | $^{151}$Tm | 6908.5 (4) | | 24(4) *ns* | 1050.2(2) | | %IT = 100 | 1987MCZZ |
| 70 | 81 | $^{151}$Yb | 679 (105) | | 1.6(1) *s* | | | %ϵp = ?<br>%IT ≈ 0.4<br>%ϵ+%β+ ≈ 100 | 2021BE23<br>1990AK01<br>1985KL10 |
| 70 | 81 | $^{151}$Yb | 1791.2+Y | | 2.6(7) *µs* | 57.1(2)<br>259.4(3) | | %IT ≈ 100 | 1993NI05 |
| 70 | 81 | $^{151}$Yb | 2448+Z | (3/2+) | 20(1) *µs* | 57(2) | | %IT = 100 | 1993NI05<br>1987BR14 |
| 71 | 80 | $^{151}$Lu | 61 (5) | 3/2+ | 16.0(8) *µs* | | | %p = 100 | 2017WA18<br>2015TA12<br>1999BI14 |
| 59 | 93 | $^{152}$Pr | 115.1 (3) | (1+) | 4.2(2) *µs* | 17.0(4)<br>115.1 | (E1)<br>E2 | %IT = 100 | 2018AL14<br>1995YA21 |
| 61 | 91 | $^{152}$Pm | 1.5E+2 (9) | 4− | 7.52(8) *min* | | | %β−= 100 | 1977YA07 |
| 61 | 91 | $^{152}$Pm | 150+X | (8) | 13.8(2) *min* | | | %IT ≥ 0<br>%β−≤ 100 | 1985YAZV |
| 63 | 89 | $^{152}$Eu | 45.5998 (4) | 0− | 9.3116(13) *h* | | | %ϵ+%β+ = 27(3)<br>%β−= 73(3) | 1990AB06 |
| 63 | 89 | $^{152}$Eu | 65.2969 (4) | 1− | 0.94(8) *µs* | 19.695(7)<br>65.2965(5) | M1<br>E2 | %IT = 100 | 1978VO05 |
| 63 | 89 | $^{152}$Eu | 77.2593 (4) | 3− | 38(4) *ns* | 77.2583(6) | M1+E2 | %IT = 100 | 1978VO05 |
| 63 | 89 | $^{152}$Eu | 78.2331 (4) | 1+ | 165(10) *ns* | 12.965(15)<br>32.6341(3) | <br>E1 | %IT = 100 | 1978VO05 |
| 63 | 89 | $^{152}$Eu | 89.8496 (4) | 4+ | 384(10) *ns* | 12.598(15)<br>89.8492(7) | <br>E1 | %IT = 100 | 1965TA03 |
| 63 | 89 | $^{152}$Eu | 147.86 (10) | 8− | 95.8(4) *min* | 39.75(10) | E3 | %IT = 100 | 2015HU02<br>1975PR05 |
| 65 | 87 | $^{152}$Tb | 342.15 (16) | 5− | 0.96 *µs* | 58.9(2)<br>65.0(3)<br>106.6(3) | M1 | %IT = 100 | 1980ZO02<br>1972BOYE |
| 65 | 87 | $^{152}$Tb | 501.74 (19) | 8+ | 4.2(1) *min* | 159.59(10) | E3 | %IT = 78.9(6)<br>%ϵ+%β+ = 21.1(6) | 1971BO12 |
| 66 | 86 | $^{152}$Dy | 5088.1 (3) | 17+ | 60(4) *ns* | 53.3(2) | E2 | %IT = 100 | 1981HA17 |

*Continued…*

Table 1 contd…

| Z | N | $^{A}X$ | E(keV) | J$^{\pi}$ | $T_{1/2}$ | E$\gamma$(keV) | $\lambda$ | Decay mode | Reference |
|---|---|---|---|---|---|---|---|---|---|
| 67 | 85 | $^{152}$Ho | 160 (1) | 9+ | 49.8(3) *s* | | | $\%\alpha$ = 10.8(17)<br>$\%\epsilon$+$\%\beta$+ = 89.2(17) | 1993AL03<br>1990SA32<br>1989GA11<br>1982BA75<br>1982BO04<br>1963MA17 |
| 67 | 85 | $^{152}$Ho | 3019.59 (19) | 19− | 8.4(3) $\mu s$ | 734.1(1) | E3 | %IT = 100 | 1994AN13 |
| 67 | 85 | $^{152}$Ho | 5997.9 (3) | 28− | 47(7) *ns* | 105.5(1) | E2 | %IT = 100 | 1980JA16 |
| 68 | 84 | $^{152}$Er | 9711.8 (5) | 28+ | 35(4) *ns* | 848.3(3)<br>1021.0(7)<br>1051.9(6)<br>1184.1(6) | <br><br>M2+E3<br> | %IT = 100 | 1992KU13 |
| 68 | 84 | $^{152}$Er | 13386.9 (9) | | 11(1) *ns* | 357.9<br>1265.5(6) | | %IT = 100 | 1992KU13 |
| 69 | 83 | $^{152}$Tm | 0+X | (9)+ | 5.2(6) *s* | | | $\%\epsilon$+$\%\beta$+ = 100 | 1980LI18 |
| 69 | 83 | $^{152}$Tm | 2554.98+X<br>(20) | 17+ | 301(8) *ns* | 103.5(1) | E2 | %IT = 100 | 2018NA20<br>1986MC14 |
| 69 | 83 | $^{152}$Tm | ~ 6300 | | 42(5) *ns* | | | %IT ≤ 100 | 1986MC14 |
| 70 | 82 | $^{152}$Yb | 2744.5 (10) | (10+) | 30(1) $\mu s$ | 54.6(10) | (E2) | %IT = 100 | 1995NI10 |
| 60 | 93 | $^{153}$Nd | 191.71 (16) | (5/2+) | 1.10(5) $\mu s$ | 71.5(2)<br>141.8(2)<br>191.7(2) | (E1)<br>E1<br>E1 | %IT = 100 | 2010SI03<br>1996YA12 |
| 62 | 91 | $^{153}$Sm | 98.39 (10) | 11/2− | 10.6(3) *ms* | 32.9(1) | | %IT = 100 | 1971KIZC |
| 62 | 91 | $^{153}$Sm | 182.902 (4) | 5/2− | 17(7) *ns* | 55.61(1)<br>92.03(1)<br>129.36(1)<br>147.06(1)<br>175.370(11)<br>182.900(8) | <br><br>E1<br>M1,E2<br>E1<br>E1 | %IT = 100 | 1968NA21 |
| 63 | 90 | $^{153}$Eu | 1771.0 (4) | 19/2− | 475(10) *ns* | 477.0(4)<br>709.4(4)<br>919.4(4) | <br><br>E1 | %IT = 100 | 2000SM09 |
| 64 | 89 | $^{153}$Gd | 95.1737 (8) | 9/2+ | 3.5(4) $\mu s$ | (1.8307(14))<br>53.60(2)? | <br>M2 | %IT = 100 | 1979KA16 |
| 64 | 89 | $^{153}$Gd | 171.188 (4) | (11/2−) | 76.0(14) $\mu s$ | (2.8)<br>76.015(4)<br>77.9 | | %IT = 100 | 1979KA16<br>1970BO02<br>1967CO20 |
| 65 | 88 | $^{153}$Tb | 163.175 (5) | 11/2− | 186(4) $\mu s$ | 82.464(4) | M2 | %IT = 100 | 1977KOZH<br>1968IO01<br>1967CO20<br>1965GR04 |
| 67 | 86 | $^{153}$Ho | 68.7 (3) | 1/2+ | 9.3(5) *min* | | | $\%\epsilon$+$\%\beta$+ = 99.82(8)<br>$\%\alpha$ = 0.18(8) | 1967HA34 |
| 67 | 86 | $^{153}$Ho | 2771.9 (7) | 31/2+ | 229(2) *ns* | 36<br>475 | | %IT = 100 | 2016PR06<br>1983RA19 |
| 68 | 85 | $^{153}$Er | 2798.2 (10) | (27/2−) | 373(9) *ns* | (46.4) | | | 1982CA09 |
| 68 | 85 | $^{153}$Er | 2992.7 | (29/2+) | ~ 10 *ns* | 83.9<br>194.5 | <br>E1 | %IT = 100 | 1982FO06 |
| 68 | 85 | $^{153}$Er | 3939.5 | (33/2+) | ~ 10 *ns* | 287.8<br>946.7 | | %IT = 100 | 1982FO06 |

*Continued…*

Table 1 contd...

| Z | N | $^{A}X$ | E(keV) | J$^{\pi}$ | $T_{1/2}$ | E$\gamma$(keV) | $\lambda$ | Decay mode | Reference |
|---|---|---|---|---|---|---|---|---|---|
| 68 | 85 | $^{153}$Er | 5248.1 (10) | (41/2−) | 248(32) *ns* | 356.3<br>403.7<br>428.7 | M2<br>M2 | %IT = 100 | 1981HO05<br>1980JA16 |
| 69 | 85 | $^{153}$Tm | 43.2 (2) | (1/2+) | 2.5(2) *s* | | | $\%\epsilon+\%\beta+$ = 8(3)<br>$\%\alpha$ = 92(3) | 1989KO02<br>1988SCZV |
| 70 | 83 | $^{153}$Yb | 2578.2+X | 27/2− | 15(1) $\mu s$ | | | %IT = 100 | 1993MC03<br>1989MC01 |
| 71 | 82 | $^{153}$Lu | 2502.5 (4) | 23/2− | > 0.1 $\mu s$ | (21)<br>291.0(3)<br>355.4(3) | | %IT = 100 | 1993MC03 |
| 71 | 82 | $^{153}$Lu | 2632.9 (5) | 27/2− | 15(3) $\mu s$ | 130.4(2) | E2 | %IT = 100 | 1993MC03 |
| 60 | 94 | $^{154}$Nd | 233.2+X | | 1.3(5) $\mu s$ | | | | 1974CLZX |
| 60 | 94 | $^{154}$Nd | 1297.9 (4) | (4−) | 3.2(3) $\mu s$ | 134.8(3)<br>169.8(3)<br>270.3(3)<br>295.3(3) | | %IT = 100 | 2009SI21 |
| 61 | 93 | $^{154}$Pm | X | (0−, 1−) | 1.73(10) *min* | | | $\%\beta-$= 100 | 1974YA07<br>1973PR05<br>1972TA13<br>1971DA28 |
| 63 | 91 | $^{154}$Eu | 68.1702 (4) | 2+ | 2.2(1) $\mu s$ | 68.1711(5) | E1 | %IT = 100 | 1988KA01<br>1977ST14 |
| 63 | 91 | $^{154}$Eu | 82.8200 (6) | 1− | 20(5) *ns* | 10.905(14)<br>14.634(17) | (E1)<br>(E1) | %IT = 100 | 1977ST14 |
| 63 | 91 | $^{154}$Eu | 100.8612 (4) | 4+ | 54(3) *ns* | (0.91)<br>32.700(4)<br>100.8592(10) | <br>E2<br>E1 | %IT = 100 | 1984RO06<br>1977ST14<br>1975CA22 |
| 63 | 91 | $^{154}$Eu | 145.3 (3) | 8− | 46.3(4) *min* | (8.6) | | %IT = 100 | 1988KA01<br>1976CH08<br>1976ZO01<br>1975CA22 |
| 64 | 90 | $^{154}$Gd | 2137.48 (4) | 7− | 68 *ns* | 225.94(3)<br>992.92(12)<br>1419.81(8) | E1<br>E1(+M2)<br>E1 | %IT = 100 | 1978WE08 |
| 65 | 89 | $^{154}$Tb | 0+X | 3− | 9.994(39) *h* | | | %IT = 21.8(7)<br>$\%\epsilon+\%\beta+$ = 78.2(7)<br>$\%\beta-$< 0.1 | 2009GY01<br>1983BE03<br>1976NEZT<br>1975SO03<br>1973LA20<br>1972VY04<br>1970AD09 |
| 65 | 89 | $^{154}$Tb | 0+Y | 7− | 22.7(5) *h* | | | %IT = 1.8(6)<br>$\%\epsilon+\%\beta+$ = 98.2(6) | 1975SO03<br>1973LA20<br>1972VY04 |
| 65 | 89 | $^{154}$Tb | 0+Z | | 513(42) *ns* | | | | 1982BE46 |
| 67 | 87 | $^{154}$Ho | 0+X | 8+ | 3.10(14) *min* | | | $\%\epsilon+\%\beta+$ = 100<br>%IT ≈ 0<br>$\%\alpha$ < 0.001 | 1993AL03<br>1971TO01<br>1968WA12 |
| 68 | 86 | $^{154}$Er | 3025 | 11− | 39(4) *ns* | (9)<br>(11) | | %IT ≈ 100<br>$\%\alpha$ ≈ 0 | 1980BO07<br>1979BA03<br>1978AG01 |

*Continued...*

Table 1 contd…

| Z | N | $^{A}$X | E(keV) | J$^{\pi}$ | $T_{1/2}$ | Eγ(keV) | λ | Decay mode | Reference |
|---|---|---|---|---|---|---|---|---|---|
| 69 | 85 | $^{154}$Tm | 0+X | 9+ | 3.30 (7) *s* | | | %IT = ?<br>%ϵ+%β+ = 42(5)<br>%α = 58(5) | 1982BO04<br>1978AFZZ<br>1964MA45 |
| 70 | 84 | $^{154}$Yb | 2046.2 (4) | (8+) | 28(2) *ns* | 96.6(2) | E2 | %IT = 100 | 1993ZH10 |
| 70 | 84 | $^{154}$Yb | 4479.1 | (16+) | 18.6(15) *ns* | 160.3 | E2 | %IT = 100 | 1993ZH10 |
| 71 | 83 | $^{154}$Lu | 59 (9) | (9+) | 1.12(8) *s* | | | %ϵ+%β+ ≈ 98.6<br>%α = 1.4(2) | 2014CA03<br>1997DA07<br>1988VI02<br>1981HO10 |
| 71 | 83 | $^{154}$Lu | 130.4+Y | (17+) | 35(3) *μs* | 130.4(3) | E2 | %IT = 100 | 1993MC03<br>1990MC02 |
| 72 | 82 | $^{154}$Hf | 2671+X | (10+) | 9(4) *μs* | | | %IT = 100 | 1993MC03<br>1989MC07 |
| 62 | 93 | $^{155}$Sm | 16.547 (2) | 5/2+ | 2.8(5) *μs* | 16.547(10) | E1 | %IT = 100 | 2010SI03 |
| 62 | 93 | $^{155}$Sm | 538.4 (11) | 11/2− | 1.00(8) *μs* | 179.2(2)<br>287.5(2)<br>385.6(2) | E1<br>E1<br>E1 | %IT = 100 | 2010SI03 |
| 64 | 91 | $^{155}$Gd | 121.05 (19) | 11/2− | 31.97(27) *ms* | 13.47(19) | E1 | %IT = 100 | 1972BR53 |
| 66 | 89 | $^{155}$Dy | 132.195 (22) | 9/2+ | 51(3) *ns* | 45.38(5) | E1 | %IT = 100 | 1982KA36 |
| 66 | 89 | $^{155}$Dy | 234.33 (3) | 11/2− | 6(1) *μs* | 9.1(1)<br>79.72(5)<br>102.16(3)<br>147.63(6) | M1+E2<br>E1+M2<br>E1+M2 | %IT = 100 | 1970BO02 |
| 67 | 88 | $^{155}$Ho | 141.97 (11) | 11/2− | 0.88(8) *ms* | 31.7(1) | M2 | %IT = 100 | 1979FO11 |
| 68 | 87 | $^{155}$Er | 563.3 (3) | 13/2+ | 34.8(6) *ns* | 31.5(1) | E1 | %IT = 100 | 1987RA20 |
| 69 | 86 | $^{155}$Tm | 41 (6) | 1/2+ | 45(3) *s* | | | %ϵ+%β+ = 99(1)<br>%α = 1(1) | 1991TO08<br>1990PO13 |
| 71 | 84 | $^{155}$Lu | 20 (6) | 1/2+ | 138(8) *ms* | | | %ϵ+%β+ = 24(16)<br>%α = 76(16) | 1997DA07<br>1996PA01<br>1991TO09 |
| 71 | 84 | $^{155}$Lu | 1781 (2) | (25/2−) | 2.69(3) *ms* | | | %α ≈ 100 | 1996PA01<br>1989HO12 |
| 60 | 96 | $^{156}$Nd | 1431.3 (4) | (5−) | 0.36(15) *μs* | 970.6(2)<br>1209.0(2) | E1<br>E1 | %IT = 100 | 2009SI21 |
| 61 | 95 | $^{156}$Pm | 150.3 (1) | 1(+) | 2.3(20) *s** | 150.3(1) | | %IT ≈ 98<br>%β−≈ 2 | 2007SH05 |
| 62 | 94 | $^{156}$Sm | 1397.55 (9) | 5− | 185(7) *ns* | 376.75(10)<br>880.39(10)<br>1147.84(10) | | %IT = 100 | 1990HE11 |
| 63 | 93 | $^{156}$Eu | 87.4897 (3) | 1− | 12.0(3) *ns* | 39.7805(4)<br>64.9725(4)<br>87.4897(3) | <br>E1<br>E1 | %IT = 100 | 1968AN09 |
| 64 | 92 | $^{156}$Gd | 2137.60 (5) | 7− | 1.3(1) *μs* | 228.35(4)<br>383.93(6) | E1 | %IT = 100 | 1981KO03 |
| 65 | 91 | $^{156}$Tb | 49.630 (10) | 4+ | 49(7) *ns* | 49.630(10) | E1 | %IT = 100 | 1982BE46 |
| 65 | 91 | $^{156}$Tb | 49.630+X | (7−) | 24.4(10) *h* | | | %IT = 100 | 1970TO11 |
| 65 | 91 | $^{156}$Tb | 88.4 (2) | (0+) | 5.3(2) *h* | 88.4 | E3 | %ϵ+%β+ > 0<br>%IT < 100 | 1970TO11<br>1955HA52<br>1950WI13 |

*Continued…*

*: 2007SH05 also suggest < 5 *s*.

Table 1 contd…

| Z | N | $^{A}X$ | E(keV) | $J^{\pi}$ | $T_{1/2}$ | Eγ(keV) | λ | Decay mode | Reference |
|---|---|---|---|---|---|---|---|---|---|
| 67 | 89 | $^{156}$Ho | 52.37 (30) | 1− | 9.5(15) *s* | 52.37 | M3 | %IT = 100 | 1995KAZS |
| 67 | 89 | $^{156}$Ho | 52.37+X | 9+ | 7.6(3) *min* | | | %IT = 25<br>%ϵ+%β+ = 75 | 2003KAZP<br>2002KAZO<br>1999KAZV |
| 67 | 89 | $^{156}$Ho | 117.58 | 1+ | 58(3) *ns* | 26.55(10)<br>35.37<br>65.16? | M1+E2<br>E1<br>E1 | %IT = 100 | 2005KAZY |
| 69 | 87 | $^{156}$Tm | 203.6+X | (11−) | ~ 400 *ns* | 203.6 | | %IT = 100 | 1985KO30 |
| 71 | 85 | $^{156}$Lu | 0+X | 10+ | 198(2) *ms* | | | %α = 100 | 2018LE10<br>1996PA01 |
| 71 | 85 | $^{156}$Lu | 2601.0+X (14) | 19− | 179(4) *ns* | 307.1(10)<br>580.5(10)<br>765.0(10)<br>923.9(10) | | %IT = 100 | 2018LE10 |
| 72 | 84 | $^{156}$Hf | 1959 (6) | 8+ | 0.52(1) *ms* | | | %α = 100 | 1996PA01 |
| 73 | 83 | $^{156}$Ta | 102 (7) | 9+ | 0.36(4) *s* | | | %ϵ+%β+ = 95.8(9)<br>%p = 4.2(9) | 1996PA01<br>1993LI34 |
| 64 | 93 | $^{157}$Gd | 63.916 (5) | 5/2+ | 0.46(4) *μs* | 9.365(12)<br>63.929(8) | E1<br>E1 | %IT = 100 | 1964KA04 |
| 64 | 93 | $^{157}$Gd | 426.539 (23) | 11/2− | 18.5(23) *μs* | 199<br>246.31(2) | | %IT = 100 | 1967BO05<br>1961KR01 |
| 66 | 91 | $^{157}$Dy | 161.99 (3) | 9/2+ | 1.3(2) *μs* | 14.23(5) | E1 | %IT = 100 | 1974AN11 |
| 66 | 91 | $^{157}$Dy | 199.38 (7) | 11/2− | 21.6(16) *ms* | 37.36(8)<br>51.7(1) | (E1)<br>(E2) | %IT = 100 | 1973KL03<br>1971KIZQ<br>1970BO02 |
| 67 | 90 | $^{157}$Ho | 53.048 (20) | 5/2+ | 20(1) *ns* | 53.05(2) | (E1) | %IT = 100 | 1979AL33 |
| 68 | 89 | $^{157}$Er | 155.4 (3)? | (9/2+) | 76(6) *ms* | 155.4(3) | E3 | %IT = 100 | 1971LEYU |
| 70 | 87 | $^{157}$Yb | 528.8 (3) | 13/2+ | 45 *ns* | (34.4)<br>323.3(3) | | %IT = 100 | 1984RA11 |
| 71 | 86 | $^{157}$Lu | 20.9 (20) | (11/2−) | 4.62(17) *s* | | | %ϵ+%β+ = 94(2)<br>%α = 6(2) | 1992HA10<br>1991LE15<br>1991TO09<br>1983TO01<br>1979BE52<br>1972GAZR |
| 72 | 85 | $^{157}$Hf | 2875.4 | (29/2+) | 52(12) *ns* | 70.3 | E1 | %IT = 100 | 1995SA31 |
| 73 | 84 | $^{157}$Ta | 22 (5) | 11/2− | 4.3(1) *ms* | | | %α = 100 | 1996PA01 |
| 73 | 84 | $^{157}$Ta | 1589 (10)? | (25/2−) | 1.7(1) *ms* | | | %α = 100 | 1996PA01 |
| 60 | 98 | $^{158}$Nd | 1648.1 (14) | (6−) | 339(20) *ns* | 1197.1(5) | | %IT = 100 | 2016ID02<br>2015TAZX |
| 62 | 96 | $^{158}$Sm | 1279.70 (17) | (5−) | 74(6) *ns* | 781.3(2)<br>1039.4(1) | | %IT = 100 | 2014WA53<br>2009SI21<br>1998GA12 |
| 65 | 93 | $^{158}$Tb | 110.3 (12) | 0− | 10.70(17) *s* | 110.3(12) | M3 | %IT = 100<br>%β−< 0.6<br>%ϵ+%β+ < 0.01 | 1965SC11<br>1958GO78<br>1957HA12 |
| 65 | 93 | $^{158}$Tb | 388.39 (11) | 7− | 0.40(4) *ms* | 65.76(10)<br>171.07(10) | E1 | %IT = 100 | 1961KR01 |
| 67 | 91 | $^{158}$Ho | 67.20 (1) | 2− | 28(2) *min* | 67.200(10) | E3 | %IT = 90(10)<br>%ϵ+%β+ = 10(10) | 1962SC10<br>1960DN01 |

*Continued…*

Table 1 contd...

| Z | N | $^{A}$X | E(keV) | J$^{\pi}$ | $T_{1/2}$ | E$\gamma$(keV) | $\lambda$ | Decay mode | Reference |
|---|---|---|---|---|---|---|---|---|---|
| 67 | 91 | $^{158}$Ho | 156.9 (1) | (5−) | 29(3) *ns* | 156.9(1) | | %IT = 100 | 1986SO02<br>1976RI09 |
| 67 | 91 | $^{158}$Ho | X* | (9+) | 21.3(23) *min* | | | %$\epsilon$+%$\beta$+ = 96.5(35)<br>%IT = 3.5(35) | 1975AL13<br>1974AL30 |
| 69 | 89 | $^{158}$Tm | 0+Y | (9−) | 16(4) *ns* | | | | 1981DR07 |
| 73 | 85 | $^{158}$Ta | 141 (9) | (9+) | 36.7(15) *ms* | | | %$\epsilon$+%$\beta$+ = 5(5)<br>%$\alpha$ = 95(5) | 1997DA07<br>1996PA01<br>1979HO10 |
| 73 | 85 | $^{158}$Ta | 2805.5 (4) | (19−) | 6.1(1) $\mu s$ | 418.5(7)<br>708.1(9)<br>1001.6(11) | (M2)<br>(E3)<br>(E3) | %IT = 98.6(2)<br>%$\alpha$ = 1.4(2) | 2014CA03 |
| 74 | 84 | $^{158}$W | 1888 (8) | (8+) | 0.143(19) *ms* | | | %$\alpha$ ≈ 100<br>%2p ≤ 0.17 | 2017JO09<br>2000MA95<br>1996PA01 |
| 61 | 98 | $^{159}$Pm | 1495 | (17/2+) | 4.97(12) $\mu s$ | 330.2<br>644.4<br>841.0<br>999.6 | | %IT = 100 | 2021YO08<br>2015YOZX |
| 62 | 97 | $^{159}$Sm | 1275.9 (14) | (15/2+) | 83(33) *ns* | 869.6(8) | | %IT = 100 | 2017PA25<br>2009UR04# |
| 64 | 95 | $^{159}$Gd | 67.829 (24) | 5/2+ | 26.2(8) *ns* | 17.1<br>67.8(1) | <br>E1 | %IT = 100 | 1968BO10 |
| 66 | 93 | $^{159}$Dy | 352.77 (14) | 11/2− | 122(3) $\mu s$ | 113.3(2)<br>116.9(2)<br>218(1) | E1<br>M1<br>E2 | %IT = 100 | 1967CO26 |
| 67 | 92 | $^{159}$Ho | 205.91 (5) | 1/2+ | 8.30(8) *s* | 40.0(1)<br>205.92(5) | <br>E3 | %IT = 100 | 1971GE01 |
| 68 | 91 | $^{159}$Er | 182.602 (24) | 9/2+ | 0.337(14) $\mu s$ | 38.32(3) | E1 | %IT = 100 | 1975BU10 |
| 68 | 91 | $^{159}$Er | 429.05 (3) | 11/2− | 0.59(6) $\mu s$ | 170.75(9)<br>247<br>284.84(3) | M1<br><br>(E2) | %IT = 100 | 1975ST07<br>1971LEYU |
| 69 | 90 | $^{159}$Tm | 166.17 (5) | (7/2−) | 37.5(13) *ns* | 113.18(6)<br>166.16(5) | E1 | %IT = 100 | 1992TL01<br>1985AN09 |
| 73 | 86 | $^{159}$Ta | 64 (5) | 11/2− | 0.56(6) *s* | | | %$\epsilon$+%$\beta$+ = 45(1)<br>%$\alpha$ = 55(1) | 2002RO17<br>1997DA07<br>1996PA01<br>1979HO10 |
| 75 | 84 | $^{159}$Re | 0+X | 11/2− | 20(4) $\mu s$ | | | %$\alpha$ = 7.5(35)<br>%p = 92.5(35) | 2007PA27<br>2006JO10 |
| 60 | 100 | $^{160}$Nd | 1107.9 (9) | (4−) | 1.63(21) $\mu s$ | 892.8(5) | | %IT = 100 | 2016ID02 |
| 61 | 99 | $^{160}$Pm$ | 191 (11) | (1−) | | | | | 2020OR03 |
| 62 | 98 | $^{160}$Sm | 1361.3 (4) | (5−) | 120(46) *ns* | 878.0(2)<br>1128.2(2) | | %IT = 100 | 2009SI21 |
| 62 | 98 | $^{160}$Sm | 2757.6 (12) | (11+) | 1.8(4) $\mu s$ | 432.1(4)<br>641.1(3) | (E1)<br>(E1) | %IT = 100 | 2016PA01 |

*Continued...*

*: The theoretical calculated energy is 180 keV in 1986SO02.

#: 2009UR04 measured the half-life of the isomer as 116(8) *ns* with spin parity (11/2–). 2017PA25 mentioned this discrepancy and indicated the need for additional experimental data.

$: Half-lives are not measured in 2020OR03 but all have to be >100 *ms*, from time-of-flight through the beam lines.

Table 1 contd...

| Z | N | $^{A}$X | E(keV) | J$^{\pi}$ | $T_{1/2}$ | E$\gamma$(keV) | $\lambda$ | Decay mode | Reference |
|---|---|---|---|---|---|---|---|---|---|
| 63 | 97 | $^{160}$Eu | 93.0 (12) | (0−) | 30.8(5) *s* | | | %$\beta$−≈ 100 | 2020HA13<br>2018HA19 |
| 65 | 95 | $^{160}$Tb | 63.6856 (20) | 1− | 60(5) *ns* | 63.6859(20) | E2 | %IT = 100 | 1978SC10 |
| 67 | 93 | $^{160}$Ho | 59.98 (3) | 2− | 5.02(5) *h* | 59.98(3) | E3(+M4) | %IT = 73(3)<br>%$\epsilon$+%$\beta$+ = 27(3) | 1965ST08 |
| 67 | 93 | $^{160}$Ho | 107.27 (2) | 6+ | 48(10) *ns* | 107.28(2) | M1+E2 | %IT = 100 | 1990SA19 |
| 67 | 93 | $^{160}$Ho | 118.441 (18) | 6− | 56(8) *ns* | (11.13(7))<br>118.44(2) | <br>E1 | %IT = 100 | 1990SA19 |
| 67 | 93 | $^{160}$Ho | 169.56+X* | (9+) | 3.2(2) *s* | | | %IT = 100 | 2005KAZX<br>1990SA19<br>1988BH05 |
| 69 | 91 | $^{160}$Tm | 70.9 (5) | 5 | 74.5(15) *s* | 28.85 | | %IT = 85(5)<br>%$\epsilon$+%$\beta$+ = 15(5) | 2008SU08<br>1983SI20 |
| 69 | 91 | $^{160}$Tm | 98.2+X | (8) | ~ 200 *ns* | 98.2 | | %IT = 100 | 1986DR06 |
| 69 | 91 | $^{160}$Tm | 174.38 (6) | 1+ | 17(1) *ns* | 34.18(10)<br>132.23(5)<br>174.40(10) | M1<br>E1<br>E1 | %IT = 100 | 1978AD03 |
| 71 | 89 | $^{160}$Lu | 0+X | | 40(1) *s* | | | %$\alpha$ = ?<br>%$\epsilon$+%$\beta$+ ≤ 100 | 1984AU13 |
| 73 | 87 | $^{160}$Ta | X | | 1.7(2) *s* | | | %$\alpha$ = 100 | 1996PA01 |
| 75 | 85 | $^{160}$Re | 184 (4) | (9+) | 2.8(1) $\mu s$ | 96(1) | (E2) | %IT = 100 | 2014DR02<br>2011DA01 |
| 61 | 100 | $^{161}$Pm | 966 | (13/2+) | 0.79(4) $\mu s$ | 609.2<br>727.5 | | %IT = 100 | 2021YO08<br>2015YOZX |
| 62 | 99 | $^{161}$Sm | 1337.8 (6) | (17/2−) | 2.6(4) $\mu s$ | 882.8(4) | | %IT = 100 | 2017PA25 |
| 66 | 95 | $^{161}$Dy | 25.65136 (3) | 5/2− | 29.10(11) *ns* | 25.65135(3) | E1 | %IT = 100 | 2013LO07<br>1978ALZC<br>1977PE20<br>1969VE05<br>1969BE54<br>1965ME08<br>1959FA06<br>1958HA13<br>1957VE17 |
| 66 | 95 | $^{161}$Dy | 485.7 (2) | 11/2− | 0.76(17) $\mu s$ | 165.0(1)<br>217.8(1)<br>284.3(1)<br>301.4(1)<br>382.6(1)<br>385.3(1) | | %IT = 100 | 2012SW01 |
| 67 | 94 | $^{161}$Ho | 211.15 (3) | 1/2+ | 6.76(7) *s* | 211.15(3) | E3 | %IT = 100 | 1971GE01<br>1965ST08 |
| 68 | 93 | $^{161}$Er | 189.42 (3) | 9/2+ | 84(10) *ns* | 45.54(3) | E1 | %IT = 100 | 1979ALZU<br>1975BU10 |
| 68 | 93 | $^{161}$Er | 396.44 (4) | 11/2− | 7.5(7) $\mu s$ | 99.76(4)<br>128.90(7)<br>146.65(8)<br>207.12(6)<br>252.50(10) | E1<br>E1<br>M1+E2<br>E1<br>E2 | %IT = 100 | 1970BO02 |

*Continued...*

*: X < 55 keV from 1988BH05.

Table 1 contd...

| Z | N | $^{A}X$ | E(keV) | J$^{\pi}$ | $T_{1/2}$ | Eγ(keV) | λ | Decay mode | Reference |
|---|---|---|---|---|---|---|---|---|---|
| 69 | 92 | $^{161}$Tm | 78.20 (3) | 7/2− | 110(3) *ns* | 78.20(3) | E1 | %IT = 100 | 1984FO04<br>1981AD02 |
| 71 | 90 | $^{161}$Lu | 166.5+X | (9/2−) | 7.3(4) *ms* | (30.9) | | %IT ≈ 100 | 1973AN10 |
| 72 | 89 | $^{161}$Hf | 329.0 (5) | (13/2+) | 4.8(2) *μs* | 202.2(4) | M2 | %IT = 100 | 2014MA91 |
| 73 | 88 | $^{161}$Ta | X | (11/2−) | 3.09(11) *s* | | | %ϵ+%β+ = 93(7)<br>%α = 7(3) | 2012TH13<br>2005SC22<br>1992HA10<br>1986RU05 |
| 75 | 86 | $^{161}$Re | 123.8 (13) | 11/2− | 14.7(3) *ms* | | | %p = 7.0(3)<br>%α = 93.0(3) | 2006LA16 |
| 62 | 100 | $^{162}$Sm | 1010.7 (6) | (4−) | 1.78(7) *μs* | 774.8(3) | E1 | %IT = 100 | 2017YO01<br>2017PA25 |
| 63 | 99 | $^{162}$Eu | 158.4 (24) | (6+) | 15.0(5) *s* | | | %β−≈ 100 | 2020VI04<br>2018HA19 |
| 64 | 98 | $^{162}$Gd | 1448.6 | (6−) | 99(3) *μs* | 205.2<br>329.9 | | | 2021WA04 |
| 65 | 97 | $^{162}$Tb* | 285.5 (32) | (4−) | | | | | 2020OR03 |
| 66 | 96 | $^{162}$Dy | 2187.9 (10) | 8+ | 8.3(3) *μs* | 146.4(1)<br>228.6(1)<br>248.0(1)<br>300.3(1)<br>341.8(1)<br>380.2(1)<br>435.4(1)<br>504.3(1)<br>550.3(1)<br>1266.5(2)<br>1639.2(2) | M1<br>E1<br>E1<br>M1<br>E1<br>E1<br>E2<br>E1<br>E1<br>M1<br>E2 | %IT = 100 | 2011SW02 |
| 67 | 95 | $^{162}$Ho | 105.87 (6) | 6− | 67.1(10) *min* | 9.80(5) | E3 | %IT = 62<br>%ϵ+%β+ = 38 | 1973BA21<br>1971WO09<br>1969HO17<br>1969AK01<br>1965GRZZ<br>1964MA10 |
| 68 | 94 | $^{162}$Er | 2026.7 (6) | 7(−) | 77(4) *ns* | 930.1(4)<br>1359.6(2) | | %IT = 100 | 2020KN03<br>2012SW01 |
| 69 | 93 | $^{162}$Tm | 0+X | 5+ | 24.3(17) *s* | (< 125) | | %IT = 81(4)<br>%ϵ+%β+ = 19(4) | 1974DE47 |
| 71 | 91 | $^{162}$Lu | 0+X | (4−) | 1.5 *min* | | | %ϵ+%β+ ≤ 100 | 1980BEYG |
| 71 | 91 | $^{162}$Lu | 0+Y | | 1.9 *min* | | | %ϵ+%β+ ≤ 100 | 1980BEYG |
| 75 | 87 | $^{162}$Re | 173 (13) | (9+) | 77(9) *ms* | | | %ϵ+%β+ = 9(5)<br>%α = 91(5) | 1997DA07<br>1996PA01 |
| 63 | 100 | $^{163}$Eu | 964.5 (5) | (13/2−) | 930(61) *ns* | 256.1#<br>674.9 | E1<br>E1 | %IT = 100 | 2017YO01<br>2017PA25 |
| 64 | 99 | $^{163}$Gd | 138.2 (2) | (1/2−) | 23.5(10) *s* | 138.2 | | %IT = ?<br>%β−= ? | 2020VI04<br>2020ZA04<br>2020OR02<br>2014HA38 |

*Continued...*

*: Isomer established in mass measurement, half-life of >100 *ms* only an estimate from time-of-flight through the beam transport system.

#: Isomer may decay by 537 and 675 keV gammas due to reversing reported by 2017PA25.

Table 1 contd…

| Z | N | $^{A}X$ | E(keV) | J$^{\pi}$ | $T_{1/2}$ | E$\gamma$(keV) | $\lambda$ | Decay mode | Reference |
|---|---|---|---|---|---|---|---|---|---|
| 67 | 96 | $^{163}$Ho | 297.88 (7) | 1/2+ | 1.09(3) *s* | 297.88(10) | E3 | %IT = 100 | 1967GE09 |
| 67 | 96 | $^{163}$Ho | 1506.4 (4) | (17/2+) | 41(11) *ns* | 581.1(1)<br>973.8(1) | | %IT = 100 | 2012SW01<br>2004HO19 |
| 67 | 96 | $^{163}$Ho | 2109.4 (4) | (23/2+) | 0.80(15) $\mu s$ | 183.3(1)<br>340.2(1) | | %IT = 100 | 2012SW01 |
| 68 | 95 | $^{163}$Er | 445.5 (6) | (11/2−) | 0.58(10) $\mu s$ | 125.8<br>198.5?<br>246.2<br>255.8<br>325.1 | <br><br><br>(M1) | %IT = 100 | 1974AN04 |
| 69 | 94 | $^{163}$Tm | 86.92 (5) | (7/2)− | 0.38(3) $\mu s$ | 63.62(3) | E1 | %IT = 100 | 1975AD09 |
| 69 | 94 | $^{163}$Tm | 164.69 (11) | (9/2+) | ~ 43 *ns* | 141.4(1) | (M1, E2) | %IT = 100 | 1977ROYU |
| 69 | 94 | $^{163}$Tm | 290.30 (13) | (11/2−) | ~ 43 *ns* | 115.7(2)<br>203.4(2) | | %IT = 100 | 1977ROYU |
| 70 | 93 | $^{163}$Yb | 58.1 | (3/2−,<br>5/2,<br>7/2−) | > 10 *ns* | 4.2<br>58 | | %IT = 100 | 1994MOZY |
| 70 | 93 | $^{163}$Yb | 99.2 (2) | (5/2+) | ~ 10 *ns* | 41<br>45.3 | | %IT = 100 | 1983KO05 |
| 70 | 93 | $^{163}$Yb | 124.0 (3) | (9/2+) | ~ 10 *ns* | (24.7) | | %IT = 100 | 1983KO05 |
| 74 | 89 | $^{163}$W | 480.3 (7) | 13/2+ | 154(3) *ns* | 38.4(7)<br>378.3(6) | E1<br>M2 | %IT = 100 | 2010SC02 |
| 75 | 88 | $^{163}$Re | 115 (4) | 11/2− | 214(5) *ms* | | | %$\alpha$ = 66(4)<br>%$\epsilon$+%$\beta$+ = 34(4) | 1997DA07 |
| 62 | 102 | $^{164}$Sm | 1485.5 (12) | (6−) | 0.60(14) $\mu s$ | 349.4(2) | (E1) | %IT = 100 | 2014PA55 |
| 64 | 100 | $^{164}$Gd | 1095.8 (4) | (4−) | 0.580(23) $\mu s$ | 60.2<br>854.7(5) | | %IT = 100 | 2021WA04<br>2018GA18<br>2017YO01<br>2017PA25 |
| 65 | 99 | $^{164}$Tb* | 145 (12) | (2+) | | | | | 2020OR03 |
| 67 | 97 | $^{164}$Ho | 139.78 (7) | 6− | 36.6(3) *min* | 45.79(6) | E3 | %IT = 100 | 2008HA21<br>1972DR04<br>1972KA19 |
| 68 | 96 | $^{164}$Er | 1985.06 (6) | 7− | 23.0(12) *ns* | 139.44(8)<br>240.49(3)<br>626.4(8)<br>960.5(2)<br>1370.73(10) | E2+M1<br>M1 | %IT = 100 | 1977DR03<br>1973CH28<br>1971DE22 |
| 68 | 96 | $^{164}$Er | 3377.57 (11) | (12+) | 68(2) *ns* | 156.4(1)<br>427.3(1)<br>555.0(1)<br>1294.8(3)<br>1859.5(6)? | (E1) | %IT = 100 | 2012SW02<br>1997BA63 |
| 69 | 95 | $^{164}$Tm | 0+X | 6− | 5.1(1) *min* | | | %IT ≈ 80<br>%$\epsilon$+%$\beta$+ ≈ 20 | 1971DE22<br>1971EK01 |
| 69 | 95 | $^{164}$Tm | 124.04+X (3) | (6−) | 36(5) *ns* | 124.04(3) | | %IT = 100 | 1987DR07 |
| 75 | 89 | $^{164}$Re | 0+Y | (9+) | 0.86(+15−11) *s* | | | %$\epsilon$+%$\beta$+ = 97(1)<br>%$\alpha$ = 3(1) | 2009HA42 |

*Continued…*

*: Isomer established in mass measurement, half-life of >100 *ms* only an estimate from time-of-flight through the beam transport system.

Table 1 contd...

| Z | N | $^{A}$X | E(keV) | J$^{\pi}$ | $T_{1/2}$ | E$\gamma$(keV) | $\lambda$ | Decay mode | Reference |
|---|---|---|---|---|---|---|---|---|---|
| 77 | 87 | $^{164}$Ir* | 0+X | (9+) | 70(10) $\mu s$ | | | %p >0 | 2014DR02 |
| | | | | | | | | $\%\alpha$ = 4(2) | 2002MA61 |
| | | | | | | | | $\%\epsilon+\%\beta+$ = ? | 2001KE05 |
| 65 | 100 | $^{165}$Tb | 207 | (7/2−) | 0.81(8) $\mu s$ | 76(2) | (E1) | %IT = 100 | 2017GU08 |
| | | | | | | 152(2) | (E1) | | |
| 66 | 99 | $^{165}$Dy | 108.1552 (13) | 1/2− | 1.257(6) *min* | 108.160(3) | E3 | $\%\beta-$= 2.24(11) | 1972MA06 |
| | | | | | | | | %IT = 97.76(11) | 1964HA19 |
| | | | | | | | | | 1960HO16 |
| | | | | | | | | | 1960WI10 |
| | | | | | | | | | 1959CR80 |
| 67 | 98 | $^{165}$Ho | 361.675 (11) | 3/2+ | 1.512(4) $\mu s$ | 266.80(15) | | %IT = 100 | 1959CR80 |
| | | | | | | 361.68(2) | M2+E3 | | 1958KA10 |
| 68 | 97 | $^{165}$Er | 551.3 (6) | 11/2− | 0.25(3) $\mu s$ | 314.8 | | %IT = 100 | 1974AN04 |
| | | | | | | 375 | | | |
| | | | | | | 383.7 | | | |
| | | | | | | 473.7 | | | |
| 68 | 97 | $^{165}$Er | 1823.0 (4) | (19/2) | 0.37(4) $\mu s$ | 317.0(1) | | %IT = 100 | 2012SW01 |
| | | | | | | 1050.6(1) | | | |
| | | | | | | 1144.6(1) | | | |
| 69 | 96 | $^{165}$Tm | 80.37 (6) | 7/2+ | 80(3) $\mu s$ | 68.86(5) | E2 | %IT = 100 | 1967CO26 |
| 69 | 96 | $^{165}$Tm | 160.47 (6) | 7/2− | 9.0(5) $\mu s$ | 30.80(10) | E1 | %IT = 100 | 1978AD06 |
| | | | | | | 80.11(2) | E1+M2 | | 1968TA05 |
| 70 | 95 | $^{165}$Yb | 126.80 (9) | 9/2+ | 300(30) *ns* | 39.23(8) | E1 | %IT = 100 | 1980ADZO |
| 75 | 90 | $^{165}$Re | 48 (26) | (11/2−) | 1.74(6) *s* | | | $\%\epsilon+\%\beta+$ = 87(1) | 2012TH13 |
| | | | | | | | | $\%\alpha$ = 13(1) | 2005SC22 |
| | | | | | | | | | 1999PO09 |
| | | | | | | | | | 1996PA01 |
| | | | | | | | | | 1984SC06 |
| | | | | | | | | | 1981HO10 |
| | | | | | | | | | 1978SC26 |
| 77 | 88 | $^{165}$Ir | X | (11/2−) | 0.34(4) *ms* | | | %p = 88(2) | 2014DR02 |
| | | | | | | | | $\%\alpha$ = 12(2) | 1997DA07 |
| 64 | 102 | $^{166}$Gd | 1601.5 (11) | (6−) | 950(60) *ns* | 146.3(2) | (E1) | %IT = 100 | 2014PA55 |
| | | | | | | 183.1(2) | (E1) | | |
| 65 | 101 | $^{166}$Tb | 159.0 (15) | | 3.5(4) $\mu s$ | 119(1) | (E1) | %IT = 100 | 2017GUZW |
| 67 | 99 | $^{166}$Ho | 5.969 (12) | 7− | 1132.6(39) *y* | | | $\%\beta-$= 100 | 2012NE05 |
| | | | | | | | | | 1965FA01 |
| 67 | 99 | $^{166}$Ho | 190.9021 (20) | 3+ | 185(15) $\mu s$ | 10.43(2) | | %IT = 100 | 1965BJ03 |
| | | | | | | 19.840(6) | E1 | | |
| | | | | | | 136.662(2) | E1 | | |
| 69 | 97 | $^{166}$Tm | 109.338+X | (6−) | 340(25) *ms* | 34.418(1) | E1 | %IT = 100 | 2002CA46 |
| | | | | | | | | | 1996DR07 |
| 69 | 97 | $^{166}$Tm | 231.053+X | (6−) | 36(2) *ns* # | 59.488(2) | E1 | %IT = 100 | 2002CA46 |
| | | | | | | 121.710(5) | M1 | | |
| 71 | 95 | $^{166}$Lu | 34.37 (22) | 3(−) | 1.41(10) *min* | 34.37(22) | (M3) | %IT = 42(5) | 1974DE09 |
| | | | | | | | | $\%\epsilon$ = 58(5) | |
| 71 | 95 | $^{166}$Lu | 43.0 (4) | 0− | 2.12(10) *min* | | | $\%\epsilon+\%\beta+$ = 90(10) | 1974DE09 |
| | | | | | | | | %IT = 10(10) | |

*Continued...*

*: Expected low-spin ground state of $^{164}$Ir has not yet been identified.
#: Half-life of ~ 2(1) $\mu s$ in 1996DR07 seems too long as compared to 36 *ns* in 2002CA46.

Table 1 contd...

| Z | N | $^{A}$X | E(keV) | J$^{\pi}$ | $T_{1/2}$ | E$\gamma$(keV) | $\lambda$ | Decay mode | Reference |
|---|---|---|---|---|---|---|---|---|---|
| 71 | 95 | $^{166}$Lu | 83.50 (10) | (5, 6, 7)+ | 92(7) *ns* | 83.5(1) | E1 | %IT = 100 | 1992HO02 |
| 77 | 89 | $^{166}$Ir | 172 (6) | (9+) | 15.1(9) *ms* | | | %$\alpha$ = 98.2(6)<br>%p = 1.8(6) | 1997DA07 |
| 65 | 102 | $^{167}$Tb | 200 | (7/2−) | 2.1(1) $\mu s$* | 73(4)<br>147(4) | (E1)<br>(E1) | %IT = 100 | 2017GU08 |
| 67 | 100 | $^{167}$Ho | 259.34 (11) | 3/2+ | 6.0(10) $\mu s$ | 259.33(13) | M2 | %IT = 100 | 1977TU01 |
| 68 | 99 | $^{167}$Er | 207.801 (5) | 1/2− | 2.269(6) *s* | 207.801(5) | E3 | %IT = 100 | 1986NE05 |
| 69 | 98 | $^{167}$Tm | 179.480 (19) | (7/2)+ | 1.16(6) $\mu s$ | 37.05(2)<br>62.91(1)<br>169.04(3) | M1+E2<br>M1+E2<br>E2 | %IT = 100 | 1964LO04 |
| 69 | 98 | $^{167}$Tm | 292.820 (20) | 7/2− | 0.9(1) $\mu s$ | (6.93(4))<br>105.19(2)<br>113.34(1)<br>150.40(3)<br>176.25(2) | <br>M1<br>E1<br>E1<br>E1 | %IT = 100 | 1965TA01 |
| 70 | 97 | $^{167}$Yb | 188.754 (18) | 1/2− | ~ 23 *ns* | 188.66(4) | E2 | %IT = 100 | 1976GR06 |
| 70 | 97 | $^{167}$Yb | 571.548 (22) | (11/2)− | ~ 180 *ns* | 151.960(17)<br>160.490(17)<br>254.0(2)<br>270.00(9)<br>385.55(11)<br>392.61(9)<br>445.56(12)<br>513.10(10) | M1(+E2)<br><br><br><br>(E1)<br>M1+E2<br>E1(+M2)<br>(E1) | %IT = 100 | 1976GR06 |
| 71 | 96 | $^{167}$Lu | 0+X | 1/2+ | ≥ 1 *min* | | | %$\epsilon$+%$\beta$+ = ?<br>%IT = ? | 1998GE13 |
| 75 | 92 | $^{167}$Re | X | (1/2+) | 5.9(3) *s* | | | | 2013AN10<br>1992ME10<br>1984SC06<br>1978CA11 |
| 76 | 91 | $^{167}$Os | 435.1 (10) | 13/2+ | 672(7) *ns* | 348 | M2 | %IT = 100 | 2010SC02<br>2009OD02 |
| 77 | 90 | $^{167}$Ir | 175.3 (22) | 11/2− | 30.0(6) *ms* | | | %p = 0.4(1)<br>%$\epsilon$+%$\beta$+ = 20(10)<br>%$\alpha$ = 80(10) | 1997DA07<br>1996PA01 |
| 65 | 103 | $^{168}$Tb | 211 (1) | (6+) | 0.69(3) $\mu s$ | 114(1) | (E1) | %IT = 100 | 2017GU24 |
| 66 | 102 | $^{168}$Dy | 1378.2 (6) | (4−) | 0.57(7) $\mu s$ | 151.7(3) | (E1) | %IT = 100 | 2019ZH49 |
| 67 | 101 | $^{168}$Ho | ~ 59 | (6+) | 132(4) *s* | (~ 59) | (M3) | %IT = 99.75(25)<br>%$\beta$−= 0.25(25) | 1990CH37 |
| 67 | 101 | $^{168}$Ho | 143.43 (17) | (1)− | > 4 $\mu s$ | 143.5(2) | M2 | %IT = 100 | 1990CH37 |
| 67 | 101 | $^{168}$Ho | 192.57 (20) | 1+ | 108(11) *ns* | 192.5(2) | E2 | %IT = 100 | 1990CH37 |
| 68 | 100 | $^{168}$Er | 1094.0383 (16) | 4− | 109.0(7) *ns* | 99.289(2)<br>198.241(1)<br>272.876(2)<br>829.958(7)<br>1014.11(4) | E1+M2<br>E1+M2<br>M2<br>E1+M2<br>M2+E3 | %IT = 100 | 2010DR02<br>1981IW04<br>1973KI09<br>1967GU04<br>1966JU02<br>1959KO64<br>1957MI01 |

*Continued...*

*: Half-life of bare or H-like ions.

Table 1 contd…

| Z | N | $^{A}$X | E(keV) | J$^{\pi}$ | $T_{1/2}$ | E$\gamma$(keV) | $\lambda$ | Decay mode | Reference |
|---|---|---|---|---|---|---|---|---|---|
| 70 | 98 | $^{168}$Yb | 1998.74 (6) | (5)− | 82(5) *ns* | 156.6(2)<br>179.6(2)<br>228.6(2)<br>324.7(2)<br>348.3(2)<br>401.1(3)<br>1413.5(3)<br>1712.0(5) | M1<br>(E1)<br>(M1)<br>(E1+M2)<br>E2<br>M1 | %IT = 100 | 1973CH28 |
| 70 | 98 | $^{168}$Yb | 2222.37 (20) | (−) | 62(8) *ns* | 111.4<br>223.59(19) | | %IT = 100 | 1973CH28 |
| 71 | 97 | $^{168}$Lu | 160 (40) | 3+ | 6.5(3) *min* | | | %IT = 0.4(4)<br>%$\epsilon$+%$\beta$+ = 99.6(4) | 2019HU15<br>1999BA65<br>1981BY04<br>1974EK03<br>1972CH44<br>1970CHYY<br>1960WI09 |
| 77 | 91 | $^{168}$Ir | 0+Y | | 159(+16–13) *ms* | | | %p = ?<br>%$\alpha$ = 77(9)<br>%$\epsilon$+%$\beta$+ ≤ 23(9) | 2009HA42<br>1996PA01 |
| 66 | 103 | $^{169}$Dy | 166 | (1/2−) | 1.26(17) $\mu s$ | 166.1 | (E2) | %IT = 100 | 2019ZH49 |
| 67 | 102 | $^{169}$Ho | 1386.2 (4) | (19/2+) | 118(6) $\mu s$ | 483.6<br>685.4<br>867.3 | | %IT = 100 | 2010DR05 |
| 68 | 101 | $^{169}$Er | 92.05 (10) | (5/2)− | 285(20) *ns* | (17.46(12))<br>27.6(2) | | %IT = 100 | 1969BOZP |
| 68 | 101 | $^{169}$Er | 243.69 (17) | 7/2+ | 200(10) *ns* | 67.3(3)<br>151.5(2) | E1<br>E1 | %IT = 100 | 1969BOZP |
| 69 | 100 | $^{169}$Tm | 316.14633 (11) | 7/2+ | 659.9(23) *ns* | 177.21307(6)<br>197.95675(7)<br>307.73586<br>(10) | M1+E2<br>M1+E2<br>E2 | %IT = 100 | 1994DE01<br>1974EN09<br>1950FU63 |
| 69 | 100 | $^{169}$Tm | 379.26678 (12) | 7/2− | 52.2(8) *ns* | 63.12044(4)<br>240.331(3)<br>261.07712(9)<br>370.854(8)<br>379.284(18) | E1<br>E1(+M2)<br>E1+M2 | %IT = 100 | 1974BO30<br>1974EN09<br>1974VI05 |
| 70 | 99 | $^{169}$Yb | 24.1999 (16) | 1/2− | 46(2) *s* | 24.20(2) | E3 | %IT = 100 | 1960HO10 |
| 71 | 98 | $^{169}$Lu | 29.0 (5) | 1/2− | 160(10) *s* | 29.0(5) | E3 | %IT = 100 | 1965BJ01 |
| 72 | 97 | $^{169}$Hf | 28.80 (4) | (7/2)+ | 82(+40−15) *ns* | 28.80(4) | E1 | %IT = 100 | 1975RE05 |
| 73 | 96 | $^{169}$Ta | 180.1 (3) | (5/2+) | 17(4) *ns* | 152.6(2)<br>168.7(2) | | %IT = 100 | 1998ZH03 |
| 73 | 96 | $^{169}$Ta | 219.72 (19) | (9/2−) | 28(5) *ns* | 123.0(2) | E1 | %IT = 100 | 1998ZH03 |
| 75 | 94 | $^{169}$Re | 187(17) | (1/2+,<br>3/2+) | 15.1(15) *s* | | | %IT = ?<br>%$\alpha$ ≈ 0.2<br>%$\epsilon$+%$\beta$+ = ? | 2021HA32<br>1992ME10<br>1984SC06 |
| 77 | 92 | $^{169}$Ir | 153 (24) | (11/2−) | 0.281(4) *s* | | | %p = ?<br>%$\alpha$ = 72(7)<br>%$\epsilon$+%$\beta$+ = ? | 2005SC22<br>1996PA01 |

*Continued…*

Table 1 contd…

| Z | N | $^{A}$X | E(keV) | J$^{\pi}$ | $T_{1/2}$ | Eγ(keV) | λ | Decay mode | Reference |
|---|---|---|---|---|---|---|---|---|---|
| 66 | 104 | $^{170}$Dy | 1643.8 (3) | (6+) | 990(40) *ns* | 255 (1) | | %IT = 100 | 2016SO13 |
| | | | | | | 386.33(15) | | | |
| | | | | | | 496.64(14) | | | |
| | | | | | | 527.28(22) | | | |
| | | | | | | 1148.9(7) | | | |
| | | | | | | 1406(1) | | | |
| 67 | 103 | $^{170}$Ho | 120 (70) | (1+) | 43(2) *s* | | | %β−= 100 | 1978TU04 |
| | | | | | | | | | 1974KA21 |
| 68 | 102 | $^{170}$Er | 1268.68 (3) | (4−) | 42.8(17) *ns* | 51.30(10) | E1 | %IT = 100 | 2003WU07 |
| | | | | | | 141.50(9) | | | 2000WUZY |
| | | | | | | 165.33(4) | (E1) | | |
| | | | | | | 258.136(20) | | | |
| | | | | | | 1008.3(3) | | | |
| 69 | 101 | $^{170}$Tm | 183.197 (4) | (3)+ | 4.12(13) *μs* | 68.6491(4) | E1 | %IT = 100 | 1967AN04 |
| | | | | | | 144.4797(5) | E1 | | |
| 70 | 100 | $^{170}$Yb | 1258.46 (14) | 4− | 370(15) *ns* | 981.1(2) | E1 | %IT = 100 | 1981WA14 |
| 71 | 99 | $^{170}$Lu | 92.91 (9) | (4)− | 0.67(10) *s* | 48.42(10) | M2 | %IT = 100 | 1988SO04 |
| | | | | | | | | | 1969TR02 |
| | | | | | | | | | 1965BJ01 |
| 72 | 98 | $^{170}$Hf | 2182.70 (16) | (8−) | 23(2) *ns* | 216.6(1) | (E1) | %IT = 100 | 1979DR08 |
| | | | | | | 1141(1) | | | |
| 77 | 93 | $^{170}$Ir | 0+X | (8+) | 811(18) *ms* | | | %α = 38(5) | 2007HA45 |
| | | | | | | | | %IT ≤ 62(5) | |
| | | | | | | | | %ϵ+%β+ ≤ 62(5) | |
| 79 | 91 | $^{170}$Au | 275 (14) | (9+) | 0.617(+50−40) *ms* | | | %α = 42(5) | 2004KE06 |
| | | | | | | | | %p = 58(5) | 2002KEZZ |
| 68 | 103 | $^{171}$Er | 198.6 (1) | 1/2− | 210(10) *ns* | 198.6(1) | E2 | %IT = 100 | 1969BOZL |
| 69 | 102 | $^{171}$Tm | 424.9560 (15) | 7/2− | 2.60(2) *μs* | 295.9010(12) | E1 | %IT = 100 | 1972GR09 |
| | | | | | | 308.2996(16) | E1 | | |
| | | | | | | 419.9(3) | M2 | | |
| | | | | | | 424.9(5) | E3 | | |
| 69 | 102 | $^{171}$Tm | 1674.5 (3) | 19/2+ | 1.7(2) *μs* | 219.2(1) | | %IT = 100 | 2009WA06 |
| | | | | | | 428.0(1) | E2 | | |
| | | | | | | 558.1(3) | | | |
| 70 | 101 | $^{171}$Yb | 95.282 (2) | 7/2+ | 5.25(24) *ms* | 19.394(2) | E1 | %IT = 100 | 1974HA50 |
| 70 | 101 | $^{171}$Yb | 122.416 (2) | 5/2− | 265(20) *ns* | 27.133(1) | E1 | %IT = 100 | 1968LO10 |
| | | | | | | 46.543(5) | M1+E2 | | |
| | | | | | | 55.689(2) | M1+E2 | | |
| | | | | | | 122.37(5) | | | |
| 71 | 100 | $^{171}$Lu | 71.13 (8) | 1/2− | 79(2) *s* | 71.10(9) | E3 | %IT = 100 | 1967GI10 |
| 71 | 100 | $^{171}$Lu | 208.15 (9) | (1/2)+ | 29.7(11) *ns* | 137.01(3) | E1 | %IT = 100 | 1972LO22 |
| 72 | 99 | $^{171}$Hf | 21.93 (9) | 1/2− | 29.5(9) *s* | (21.93(9)) | | %IT ≤ 100 | 2012ZH22 |
| | | | | | | | | %ϵ+%β+ = ? | 1997CA39 |
| 72 | 99 | $^{171}$Hf | 49.60 (8) | 5/2− | 64(4) *ns* | 49.6(1) | (E1) | %IT = 100 | 2012ZH22 |
| | | | | | | | | | 1979DR08 |
| 72 | 99 | $^{171}$Hf | 1984.4 (3) | 23/2− | 18(2) *ns* | 190.3(2) | E1 | %IT = 100 | 2012ZH22 |
| | | | | | | | | | 1979DR08 |
| 73 | 98 | $^{171}$Ta | 235.90 (20) | (9/2−) | 46(3) *ns* | 53.1(3) | E1 | %IT = 100 | 1985BA48 |
| | | | | | | 184.2(2) | E1 | | |
| | | | | | | 236.3(3) | | | |

*Continued…*

Table 1 contd...

| Z | N | $^{A}$X | E(keV) | J$^{\pi}$ | $T_{1/2}$ | Eγ(keV) | λ | Decay mode | Reference |
|---|---|---|---|---|---|---|---|---|---|
| 77 | 94 | $^{171}$Ir | 0+X | (11/2−) | 1.2(1) *s* | | | %α = 54(5) | 2014PE02 |
| | | | | | | | | %ϵ+%β+ ≤ 46(5) | 2010AN01 |
| | | | | | | | | %p ≤ 46(5) | 2002RO17 |
| | | | | | | | | | 1996PA01 |
| | | | | | | | | | 1992SC16 |
| | | | | | | | | | 1978CA11 |
| | | | | | | | | | 1967SI02 |
| 78 | 93 | $^{171}$Pt | 412.6 (10) | (13/2+) | 901(9) *ns* | 323.1(6) | M2 | %IT = 100 | 2010SC02 |
| 79 | 92 | $^{171}$Au | 250 (16) | (11/2−) | 1.014(19) *ms* | | | %α = 54(4) | 2004KE06 |
| | | | | | | | | %p = 46(4) | 2003BA20 |
| | | | | | | | | | 1999PO09 |
| | | | | | | | | | 1997DA07 |
| 66 | 106 | $^{172}$Dy | 1277.9 (5) | (8−) | 0.71(5) *s* | (45.0(7)) | E1 | %IT = 81(3) | 2017WU04 |
| | | | | | | 400.4(2) | E1 | %β−= 19(3) | 2016WA19 |
| | | | | | | 758(1) | M2 | | |
| 68 | 104 | $^{172}$Er | 1263.3 (3) | (4−) | 39.5(21) *ns* | 137.8(6) | | %IT = 100 | 2010DR02 |
| | | | | | | 229.4(6) | | | |
| | | | | | | 1008.1(5) | | | |
| | | | | | | 1186.1(6) | | | |
| 68 | 104 | $^{172}$Er | 1500.9 (3) | (6+) | 0.57(6) *µs* | 133.6(2) | E1 | %IT = 100 | 2010DR02 |
| | | | | | | 249.6(2) | | | |
| | | | | | | 369.7(2) | | | |
| | | | | | | 970.5(2) | | | |
| 69 | 103 | $^{172}$Tm | 476.2 (2) | 6+ | 132(7) *µs* | 34.6(3) | E1 | %IT = 100 | 2008HU05 |
| | | | | | | 101.8(1) | E1 | | |
| | | | | | | 226.3(2) | E1 | | |
| 70 | 102 | $^{172}$Yb | 1550.43 (6) | 6− | 3.6(1) *µs* | 174.7(10) | (E1) | %IT = 100 | 1969NO05 |
| | | | | | | 197.6(3) | | | |
| | | | | | | 1010.45(6) | E1+M2 | | |
| 71 | 101 | $^{172}$Lu | 41.86 (4) | 1− | 3.7(5) *min* | 41.86(4) | M3 | %IT = 100 | 1962VA07 |
| | | | | | | | | %ϵ+%β+ < 0.18 | |
| 71 | 101 | $^{172}$Lu | 65.79 (4) | (1)+ | 0.332(20) *µs* | 23.9331(2) | E1 | %IT = 100 | 1965BR26 |
| 71 | 101 | $^{172}$Lu | 109.41 (10) | (1)+ | 440(12) *µs* | 67.35(10) | E1 | %IT = 100 | 1967CO26 |
| | | | | | | | | | 1966GR22 |
| | | | | | | | | | 1965BJ01 |
| 71 | 101 | $^{172}$Lu | 213.57 (17) | (6−) | 150 *ns* | 102.5(2) | | %IT = 100 | 1974BEXY |
| | | | | | | 213.5(2) | | | |
| 72 | 100 | $^{172}$Hf | 2005.84 (11) | (8−) | 163(3) *ns* | 127.67(10) | (E1) | %IT = 100 | 1980WA23 |
| | | | | | | 149.4 | | | |
| | | | | | | 278.2 | | | |
| | | | | | | 321.0 | | | |
| | | | | | | 408.4 | | | |
| | | | | | | 968.2 | | | |
| 73 | 99 | $^{172}$Ta | 73.0+X (5) | (6+) | 50(2) *ns* | 49.3(3) | (E1) | %IT = 100 | 2000HO16 |
| 73 | 99 | $^{172}$Ta | 253.0+X (5) | (7+) | 37(3) *ns* | 180.2(3) | M1+E2 | %IT = 100 | 2000HO16 |
| 75 | 97 | $^{172}$Re$^{\dagger}$ | 0+X | (5) | 15(3) *s* | | | %ϵ+%β+ = 100 | 1977BE66 |
| | | | 0+Y | (2) | 55(5) *s* | | | %ϵ+%β+ = 100 | 1977BE66 |

*Continued...*

Table 1 contd...

| Z | N | $^{A}$X | E(keV) | J$^{\pi}$ | $T_{1/2}$ | E$\gamma$(keV) | $\lambda$ | Decay mode | Reference |
|---|---|---|---|---|---|---|---|---|---|
| 77 | 95 | $^{172}$Ir$^{\dagger}$ | 0+X | (3−,4−) | 4.4(3) *s* | | | %$\epsilon$+%$\beta$+ ≈ 98 | 2021HA32 |
| | | | | | | | | %$\alpha$ ≈ 2 | 2014AN10 |
| | | | | | | | | %IT = ? | 2004GOZZ |
| | | | | | | | | | 1992SC16 |
| | | | 0+Y | (7+,8+) | 2.19(7) *s* | | | %$\epsilon$+%$\beta$+ = 90.5(11) | 2021HA32 |
| | | | | | | | | %$\alpha$ = 9.5(11) | 2014AN10 |
| | | | | | | | | %IT = ? | 2004GOZZ |
| | | | | | | | | | 1992SC16 |
| | | | | | | | | | 1978SC26 |
| 79 | 93 | $^{172}$Au | 0+X | | 7.7(14) *ms* | | | %$\alpha$ ≈ 100 | 2009HA42 |
| | | | | | | | | %p < 0.02 | 1996PA01 |
| | | | | | | | | %$\epsilon$+%$\beta$+ = ? | |
| 67 | 106 | $^{173}$Ho | 405 | (3/2+) | 3.7(12) $\mu s$ | | | | 2020LI28 |
| 69 | 104 | $^{173}$Tm | 317.73 (20) | 7/2− | 10.4(21) $\mu s$ | 192.8(2) | (E1) | %IT = 100 | 2012HU10 |
| | | | | | | 199.2(2) | (E1) | | 1972PU02 |
| | | | | | | 315.2 | | | |
| 69 | 104 | $^{173}$Tm | 1905.7 (4) | 19/2− | 250(69) *ns* | 43.4(6) | M1 | %IT = 100 | 2012HU10 |
| | | | | | | 124(1) | E2 | | |
| | | | | | | 154.3(2) | E1 | | |
| | | | | | | 214.3(4) | E2 | | |
| 69 | 104 | $^{173}$Tm | 2061.1 (4) | 21/2+ | 15.2(55) *ns* | 155.4(1) | E1 | %IT = 100 | 2012HU10 |
| 69 | 104 | $^{173}$Tm | 4047.9 (5) | 35/2− | 121(28) *ns* | 412(1) | M1 | %IT = 100 | 2012HU10 |
| | | | | | | 655(1) | E1 | | |
| | | | | | | 703.5(1) | E2 | | |
| 70 | 103 | $^{173}$Yb | 398.9 (5) | 1/2− | 2.9(1) $\mu s$ | 398.9(6) | E2 | %IT = 100 | 1969BOZP |
| | | | | | | | | | 1967BLZY |
| | | | | | | | | | 1963OR01 |
| 71 | 102 | $^{173}$Lu | 123.672 (13) | 5/2− | 74.2(10) $\mu s$ | 123.672(13) | E1 | %IT = 100 | 1969FAZY |
| 72 | 101 | $^{173}$Hf | 107.16 (5) | 5/2− | 180(8) *ns* | 25.70(5) | M1 | %IT = 100 | 1973RE03 |
| | | | | | | 37.40(5) | M1+E2 | | 1971BOZG |
| | | | | | | 107.2(2) | (E2) | | |
| 72 | 101 | $^{173}$Hf | 197.47 (10) | 7/2+ | 160(40) *ns* | 90.3(1) | E1 | %IT = 100 | 1973RE03 |
| | | | | | | | | | 1971BOZG |
| 72 | 101 | $^{173}$Hf | 1982.1 (5) | (23/2)− | 19.5(6) *ns* | 164.8(3) | E1 | %IT = 100 | 1980WA23 |
| 73 | 100 | $^{173}$Ta | 173.05 (24) | 9/2− | 205(8) *ns* | 35.4(2) | E1+M2 | %IT = 100 | 2017WO02 |
| | | | | | | 63.0(2) | | | 1995CA27 |
| | | | | | | 166.0(2) | (M2) | | 1991KU12 |
| 73 | 100 | $^{173}$Ta | 198.16 (20) | 3/2(+) | 13.5(58) *ns* | 174.21(10) | | | 1995CA27 |
| | | | | | | 196.61(10) | | | |
| 73 | 100 | $^{173}$Ta | 1717.2 (4) | 21/2− | 137(7) *ns* | 351.3(2) | M1 | %IT = 100 | 2017WO02 |
| | | | | | | 604.1(2) | M1 | | 2006TH07 |
| | | | | | | 835.3(2) | E2 | | 1997AN04 |
| 74 | 99 | $^{173}$W | 85.37 (10) | (7/2)+ | 14(4) *ns* | 85.37(10) | E1 | %IT = 100 | 1978WA16 |
| 74 | 99 | $^{173}$W | 148.0 (5) | 1/2− | 1.31(15) $\mu s$ | 148.0(5) | | | 2012WA31 |
| 76 | 97 | $^{173}$Os | 141.2 (2) | (9/2+) | > 28 *ns* | 49.6(2) | E1 | %IT = 100 | 1992BO21 |
| | | | | | | | | | 1990BA29 |
| 77 | 96 | $^{173}$Ir | 219 (16) | (11/2−) | 2.20(5) *s* | | | %$\epsilon$+%$\beta$+ = 34(10) | 2021HA32 |
| | | | | | | | | %$\alpha$ = 66(10) | 2009AN14 |
| | | | | | | | | | 1992SC16 |

*Continued...*

Table 1 contd…

| Z | N | $^{A}$X | E(keV) | J$^{\pi}$ | $T_{1/2}$ | E$\gamma$(keV) | $\lambda$ | Decay mode | Reference |
|---|---|---|---|---|---|---|---|---|---|
| 79 | 94 | $^{173}$Au | 214 (23) | (11/2−) | 12.2(1) *ms* | | | %$\alpha$ = 92(13) | 2012TH13 |
| | | | | | | | | %$\epsilon$+%$\beta$+ = ? | 2001KO44 |
| | | | | | | | | %p = ? | 1999PO09 |
| | | | | | | | | | 1996PA01 |
| 68 | 106 | $^{174}$Er | 0+X? | | 37.5(375) *ns* | | | | 2005CA02 |
| 68 | 106 | $^{174}$Er | 1111.5 | 8− | 3.90(35) *s* | 162.9 | E1 | %IT = 100 | 2017WU04 |
| | | | | | | 551? | | | 2009DR06 |
| 69 | 105 | $^{174}$Tm | 191.50 (19) | (4+) | 15.2(21) *ns* | 91.2(6) | E1 | %IT = 100 | 2013HU08 |
| | | | | | | 191.5(2) | E1 | | |
| 69 | 105 | $^{174}$Tm | 252.4 (5) | 0+ | 2.29(1) *s* | 152.1 | E3 | | 2006CH10 |
| 69 | 105 | $^{174}$Tm | 496.7 (3) | (7+) | 10.4(14) *ns* | 90.5(5) | M1 | %IT = 100 | 2013HU08 |
| 69 | 105 | $^{174}$Tm | 2091.7 (3) | 14− | 106.1(69) $\mu s$ | 37(1) | E1 | %IT = 100 | 2013HU08 |
| | | | | | | (39(1)) | E2 | | |
| | | | | | | 145(1) | M1 | | |
| | | | | | | 168.1(3) | M2 | | |
| | | | | | | 298.3(2) | M2 | | |
| | | | | | | 397.7(3) | M1 | | |
| | | | | | | 401.3(2) | M1 | | |
| | | | | | | 440.0(5) | E3 | | |
| | | | | | | 637.1(4) | E2 | | |
| 70 | 104 | $^{174}$Yb | 1518.148 (13) | 6+ | 830(40) $\mu s$ | 628.37(4) | | %IT = 100 | 1967BO08 |
| | | | | | | 992.128(13) | (M1+E2) | | 1964KA15 |
| | | | | | | 1265.18(10) | | | |
| 70 | 104 | $^{174}$Yb | 1765.2 (5) | 7− | 256(11) *ns* | 193 | E2 | | 2005DR05 |
| | | | | | | 247.1 | E1 | | |
| | | | | | | 875.5 | E1+M2 | | |
| | | | | | | 1239.2 | E1+M2 | | |
| 70 | 104 | $^{174}$Yb | 3698.8 (6) | 14+ | 55(4) *ns* | 409.6 | M1+E2 | | 2005DR05 |
| | | | | | | 695.6 | M1+E2 | | |
| | | | | | | 785.5 | E1 | | |
| | | | | | | 964.1 | E2 | | |
| 70 | 104 | $^{174}$Yb | 6146.6 (9) | (22,23) | 28.4(14) *ns* | 172.5? | | | 2005DR05 |
| | | | | | | 417.3? | | | |
| 71 | 103 | $^{174}$Lu | 170.83 (5) | 6− | 144(2) *d* | 59.08(2) | M3 | %IT = 99.38(2) | 1998GE13 |
| | | | | | | 126.2 | | %$\epsilon$ = 0.62(2) | 1975KI06 |
| | | | | | | | | | 1973NE03 |
| | | | | | | | | | 1969KA19 |
| | | | | | | | | | 1967GI06 |
| | | | | | | | | | 1965FU01 |
| | | | | | | | | | 1964BA25 |
| | | | | | | | | | 1962BO12 |
| | | | | | | | | | 1960WI10 |
| 71 | 103 | $^{174}$Lu | 240.818 (4) | (3+) | 395(15) *ns* | 129.065(2) | (E1) | %IT = 100 | 1980KE08 |
| | | | | | | 196.112(10) | (E1) | | 1973ANYG |
| 71 | 103 | $^{174}$Lu | 365.183 (6) | (4−) | 145(3) *ns* | 105.654(9) | (E1) | %IT = 100 | 1980KE08 |
| | | | | | | 124.360(8) | (E1) | | 1973ANYG |
| | | | | | | 164.885(10) | (M1,E2) | | |
| | | | | | | 253.435(10) | (M1,E2) | | |

*Continued…*

Table 1 contd. . .

| Z | N | $^{A}$X | E(keV) | J$^{\pi}$ | $T_{1/2}$ | E$\gamma$(keV) | $\lambda$ | Decay mode | Reference |
|---|---|---|---|---|---|---|---|---|---|
| 71 | 103 | $^{174}$Lu | 1855.7 (5) | 13+ | 194(24) $ns$ | 267.4(3) | M1 | %IT = 100 | 2009KO19 |
| | | | | | | 373.3(3) | M1 | | 2006DR07 |
| | | | | | | 427.1(3) | E2 | | |
| | | | | | | 484.8(3) | E1 | | |
| | | | | | | 608.1(3) | M1 | | |
| | | | | | | 821.6(3) | E2 | | |
| 71 | 103 | $^{174}$Lu | 2062.5 (6) | 14− | 38(4) $ns$ | 206.8(3) | E1 | %IT = 100 | 2009KO19 |
| 71 | 103 | $^{174}$Lu | 4068.6 (8) | (21+) | 97(10) $ns$ | 328.1(3) | E2 | %IT = 100 | 2009KO19 |
| 71 | 103 | $^{174}$Lu | 5849.6 (9) | (26−) | 242(19) $ns$ | 299.2(3) | E1 | %IT = 100 | 2009KO19 |
| | | | | | | 432.7(3) | E2 | | |
| 72 | 102 | $^{174}$Hf | 1549.3 (18) | (6+) | 138(4) $ns$ | 100.10(22) | | %IT = 100 | 1980WA23 |
| | | | | | | 154.71(13) | | | |
| | | | | | | 241.97(19) | | | |
| | | | | | | 486.61(25) | | | |
| | | | | | | 539.67(25) | | | |
| | | | | | | 941.02(5) | | | |
| | | | | | | 1251.81(7) | | | |
| 72 | 102 | $^{174}$Hf | 1797.5 (18) | (8−) | 2.39(4) $\mu s$ | 60.18(13) | | %IT = 100 | 1974KHZW |
| | | | | | | 248.3(5) | | | |
| | | | | | | 788.0(12) | | | |
| 72 | 102 | $^{174}$Hf | 3311.7 (18) | (14+) | 3.7(2) $\mu s$ | 10.3(10) | | %IT = 100 | 1974KHZW |
| | | | | | | 11.87(25) | | | |
| | | | | | | 15.7(7) | | | |
| | | | | | | 31.9(5) | | | |
| | | | | | | 42.69(14) | | | |
| | | | | | | 54(5) | | | |
| | | | | | | 82.0(3) | | | |
| | | | | | | 132.4(14) | | | |
| | | | | | | 155.09(16) | | | |
| | | | | | | 194.8(12) | | | |
| | | | | | | 221.97(22) | | | |
| | | | | | | 318.8(3) | | | |
| | | | | | | 328.36(5) | | | |
| | | | | | | 339.7(5) | | | |
| | | | | | | 379.38(12) | | | |
| | | | | | | 457.70(14) | | | |
| | | | | | | 539.3(6) | | | |
| | | | | | | 627.22(14) | | | |
| | | | | | | 714.2(3) | | | |
| | | | | | | 822.70(15) | | | |
| | | | | | | 1291.32(24) | | | |
| 74 | 100 | $^{174}$W | 2267.8 (4) | 8− | 158(3) $ns$ | 247.0(2) | | %IT = 100 | 2006TA13 |
| | | | | | | | | | 1978DR04 |
| 74 | 100 | $^{174}$W | 3515.6 (4) | 12+ | 126(8) $ns$ | 306.8(2) | | %IT = 100 | 2020RO25 |
| | | | | | | 557.1(4) | | | 2006TA13 |
| | | | | | | 685.6(5) | | | |
| | | | | | | 1327.7(4) | | | |
| | | | | | | 1879.3(5) | | | |
| 75 | 99 | $^{174}$Re | 152.9+Y | (8−) | ~ 91 $ns$ | 115.5(1) | (E1) | %IT = 100 | 2012GU14 |
| 75 | 99 | $^{174}$Re | 177.74+Z | | ≥ 1 $\mu s$ | 177.7(1) | | %IT = 100 | 2012GU14 |

*Continued. . .*

Table 1 contd...

| Z | N | $^{A}$X | E(keV) | J$^{\pi}$ | $T_{1/2}$ | E$\gamma$(keV) | $\lambda$ | Decay mode | Reference |
|---|---|---|---|---|---|---|---|---|---|
| 75 | 99 | $^{174}$Re | 343.6+X | (6+) | 23 *ns* | 137.5 | | %IT = 100 | 1989MCZT |
| | | | | | | | | | 1988CHZQ |
| 75 | 99 | $^{174}$Re | 1846.9 (20) | (14−) | 53(5) *ns* | 875.1(3) | M1 | | 2020CA11 |
| | | | | | | 925.8(3) | M1 | | |
| | | | | | | (1127) | E2 | | |
| | | | | | | 1143.6(3) | E2 | | |
| 77 | 97 | $^{174}$Ir | 129 (17) | (6,7,8,9) | 4.9(3) *s* | | | $\%\epsilon$ = 97.5(3) | 2020CU04 |
| | | | | | | | | $\%\alpha$ = 2.5(3) | 1992BO21 |
| | | | | | | | | | 1992SC16 |
| 79 | 95 | $^{174}$Au | 0+X | (9+) | 162(3) *ms* | | | $\%\alpha$ = ? | 2004GOZZ |
| | | | | | | | | $\%\epsilon+\%\beta+$ = ? | 1996PA01 |
| 69 | 106 | $^{175}$Tm | 440.0 (11) | 7/2− | 319(35) *ns* | 87.4(4) | E1 | %IT = 100 | 2012HU10 |
| | | | | | | 311.0(2) | E1 | | |
| | | | | | | 313.0(2) | E1 | | |
| | | | | | | 440(1) | M2 | | |
| 69 | 106 | $^{175}$Tm | 946.7 (11) | 15/2− | 44.4(21) *ns* | (5.2) | | %IT = 100 | 2012HU10 |
| | | | | | | 163.8(1) | M1 | | |
| | | | | | | 300.6(1) | E2 | | |
| 69 | 106 | $^{175}$Tm | 1517.7 (12) | 23/2+ | 21(14) $\mu s$ | 108.3(1) | E1 | %IT = 100 | 2012HU10 |
| | | | | | | 321.7(2) | | | |
| | | | | | | 513(1) | | | |
| 70 | 105 | $^{175}$Yb | 514.866 (4) | 1/2− | 68.2(3) *ms* | 514.862(11) | (M3) | %IT = 100 | 1972BR53 |
| 71 | 104 | $^{175}$Lu | 353.48 (13) | 5/2− | 1.49(7) $\mu s$ | 353.3(2) | E1 | %IT = 100 | 1969JO16 |
| 71 | 104 | $^{175}$Lu | 626.53 (15) | (1/2+) | 10.6(5) *ns* | 111.9(4) | | %IT = 100 | 1974FO01 |
| | | | | | | 255.72(10) | | | 1974WI06 |
| 71 | 104 | $^{175}$Lu | 1392.2 (6) | (19/2+) | 984(30) $\mu s$ | 368 | | %IT = 100 | 2004GA04 |
| | | | | | | 592 | | | 1998WH02 |
| | | | | | | 797 | | | |
| 72 | 103 | $^{175}$Hf | 125.89 (12) | 1/2− | 53.7(15) $\mu s$ | 125.9(2) | | %IT = 100 | 1967CO20 |
| 72 | 103 | $^{175}$Hf | 1433.41 (12) | (19/2)+ | 1.10(8) $\mu s$ | 357.40(20) | | %IT = 100 | 2002PF01 |
| | | | | | | 536.42(19) | | | 1980DR06 |
| | | | | | | 614.8(3) | E1 | | |
| | | | | | | 722.23(3) | M1 | | |
| | | | | | | 867.16(8) | E2 | | |
| 72 | 103 | $^{175}$Hf | 3015.6 (4) | (35/2−) | 1.21(15) $\mu s$ | 74.5(8) | | %IT = 100 | 2002PF01 |
| | | | | | | 82.7(4) | | | 1980DR06 |
| | | | | | | 381.3(4) | | | |
| 72 | 103 | $^{175}$Hf | 4636.2 (12) | (45/2+) | 2.8 $\mu s$ | 131 | | %IT = 100 | 2001DRZZ |
| | | | | | | 291 | | | 1990GJ01 |
| | | | | | | 479 | | | |
| 73 | 102 | $^{175}$Ta | 131.41 (17) | (9/2−) | 222(8) *ns* | 131.6(2) | | %IT = 100 | 1996KO17 |
| 73 | 102 | $^{175}$Ta | 339.2 (13) | (1/2+) | 0.17(2) $\mu s$ | 121.16 | (E1) | %IT = 100 | 1969ADZY |
| | | | | | | 270.25 | E1 | | |
| 73 | 102 | $^{175}$Ta | 1567.6 (3) | (21/2−) | 1950(150) *ns* | 216.8(2) | | %IT = 100 | 1996KO17 |
| | | | | | | 473.8(2) | | | |
| | | | | | | 709.6 | (M1) | | |
| 74 | 101 | $^{175}$W | 104.03 (17) | (5/2−) | 45(12) *ns* | | | | 1978WA16 |
| 74 | 101 | $^{175}$W | 234.96 (15) | (7/2+) | 216(6) *ns* | 38.69(10) | E1 | %IT=100 | 1978WA16 |
| | | | | | | 130.92(10) | | | |
| | | | | | | 145.74(12) | | | |

*Continued...*

Table 1 contd…

| Z | N | $^{A}$X | E(keV) | J$^{\pi}$ | $T_{1/2}$ | E$\gamma$(keV) | $\lambda$ | Decay mode | Reference |
|---|---|---|---|---|---|---|---|---|---|
| 75 | 100 | $^{175}$Re | 1793.8 (11) | (19/2) | ~ 27.7 *ns* | 179.5 | | %IT = 100 | 1992KI06 |
| | | | | | | 428.7 | | | |
| | | | | | | 684.4 | | | |
| | | | | | | 693.7 | | | |
| 76 | 99 | $^{175}$Os | 105.7 (2) | (7/2+) | 10(2) *ns* | 105.7(2) | (E1) | %IT = 100 | 1990FA02 |
| 77 | 98 | $^{175}$Ir | 71.3 | (5/2−) | 6.58(15) $\mu s$ | 45.2 | E1 | %IT = 100 | 2019GI11 |
| 79 | 96 | $^{175}$Au | 0+X | (11/2−) | 137(1) *ms* | | | $\%\alpha$ = 90(3) | 2017BA46 |
| | | | | | | | | $\%\epsilon+\%\beta+$ = 10(3) | 2011WA37 |
| | | | | | | | | | 2010AN01 |
| | | | | | | | | | 2001KO44 |
| | | | | | | | | | 1983SC24 |
| 80 | 95 | $^{175}$Hg | 494 (2) | (13/2+) | 0.34(3) $\mu s$ | 414 | M2 | %IT=100 | 2009OD01 |
| 70 | 106 | $^{176}$Yb | 1049.8 (6) | 8− | 11.4(3) *s* | 95.92(9) | E1 | %IT=100 | 1967BO08 |
| | | | | | | | | | 1965VE01 |
| | | | | | | | | | 1962KA24 |
| 71 | 105 | $^{176}$Lu | 122.845 (4) | 1− | 3.68(4) *h* | | | $\%\beta-$= 99.905(16) | 1990AB02 |
| | | | | | | | | $\%\epsilon$ = 0.095(16) | 1981LO12 |
| | | | | | | | | | 1978HE06 |
| | | | | | | | | | 1965WH03 |
| | | | | | | | | | 1965AV01 |
| | | | | | | | | | 1963SC22 |
| | | | | | | | | | 1960WI10 |
| | | | | | | | | | 1958BE41 |
| | | | | | | | | | 1945AT02 |
| | | | | | | | | | 1935MC06 |
| | | | | | | | | | 1935MA03 |
| 71 | 105 | $^{176}$Lu | 194.358 (4) | 1+ | 35.0(10) *ns* | 71.516(1) | E1+M2 | %IT = 100 | 1974AN12 |
| 71 | 105 | $^{176}$Lu | 1514.5 (5) | 12+ | 312(69) *ns* | 162.4 | (E2) | %IT = 100 | 2000MC03 |
| | | | | | | 200.3 | M1 | | |
| | | | | | | 355.0 | | | |
| | | | | | | 382.3 | | | |
| | | | | | | 396.0 | | | |
| | | | | | | 454.2 | | | |
| | | | | | | 617.0 | (E2) | | |
| | | | | | | 658.0 | | | |
| | | | | | | 687.1 | | | |
| | | | | | | 1126.5 | | | |
| 71 | 105 | $^{176}$Lu | 1587.5 (11) | (14+) | 40(3) $\mu s$ | 73.0 | (E2) | %IT = 100 | 2010DR01 |
| | | | | | | 190.0 | | | 2000MC03 |
| 72 | 104 | $^{176}$Hf | 1333.07 (7) | 6+ | 9.6(3) $\mu s$ | 736.20(7) | E2 | %IT = 100 | 1973KH02 |
| | | | | | | 1043.0(1) | E2 | | 1964BR27 |
| 72 | 104 | $^{176}$Hf | 1559.31 (9) | 8− | 9.9(2) $\mu s$ | 53.49(7) | (E1) | %IT = 100 | 1973KH02 |
| | | | | | | 226.25(6) | M2 | | 1967BO08 |
| 72 | 104 | $^{176}$Hf | 2865.8 (7) | 14− | 401(6) $\mu s$ | 38.7 | (M1) | %IT = 100 | 1975KH04 |
| | | | | | | 227.9 | (E2) | | |
| | | | | | | 302.2 | (E2) | | |
| 72 | 104 | $^{176}$Hf | 4376.6 (16) | 19+ | 34 *ns* | 529.1 | (M1) | %IT = 100 | 2010MU13 |
| | | | | | | 836.5 | | | 1976KH03 |
| 72 | 104 | $^{176}$Hf | 4863.5 (16) | 22− | 45(13) $\mu s$ | 37 | E2 | %IT = 100 | 2010MU13 |
| | | | | | | 97.1 | | | 2001CH89 |

*Continued…*

Table 1 contd…

| Z | N | $^{A}X$ | E(keV) | $J^{\pi}$ | $T_{1/2}$ | Eγ(keV) | λ | Decay mode | Reference |
|---|---|---|---|---|---|---|---|---|---|
| 73 | 103 | $^{176}$Ta | 90.4+X | (4,5)+ | 27(8) *ns* | 90.4(8) | E1 | %IT = 100 | 1998KO09<br>1994DA11 |
| 73 | 103 | $^{176}$Ta | 99.9+X | (3+) | 38(6) *ns* | 99.9(8) | E1 | %IT = 100 | 1998KO09<br>1994DA11 |
| 73 | 103 | $^{176}$Ta | 100.2 (10) | (0+) | 30.5(10) *ns* | 100.2 | E1 | %IT = 100 | 1978DU06 |
| 73 | 103 | $^{176}$Ta | 103.0 (18) | (+) | 1.1(1) *ms* | 33.5 | E1 | %IT = 100 | 1971GO21 |
| 73 | 103 | $^{176}$Ta | 193.8? | (+) | 13.3(10) *ns* | 90.8? | (M1, E2) | %IT = 100 | 1978DU06 |
| 73 | 103 | $^{176}$Ta | 1371+W | 14− | 3.8(4) *μs* | 393.6(6)<br>618.4(6) | M1+E2<br>E2 | %IT = 100 | 1978BU16 |
| 73 | 103 | $^{176}$Ta | 1432+W | (13+) | 25(8) *ns* | 61.4(8) | E1 | %IT = 100 | 1998KO09<br>1994DA11 |
| 73 | 103 | $^{176}$Ta | 2771+W | 20− | 0.97(7) *ms* | 6.9(6)<br>11.2(6)<br>345.6(6) | <br><br>E2 | %IT = 100 | 1994DA11 |
| 74 | 102 | $^{176}$W | 3747.0 (8) | 14+ | 41(1) *ns* | 445.0(1)<br>490.9(2)<br>518.6(3)<br>714.1(1)<br>884.5(2)<br>916.8(3)<br>945.0(2) | | %IT = 100 | 2000IO03 |
| 74 | 102 | $^{176}$W | 4894.7 (8) | | ~ 10 *ns* | 315.8(1) | | %IT = 100 | 1996CR02 |
| 75 | 101 | $^{176}$Re | 114.8+X | (8−) | 30(3) *ns* | 99.5 | E1 | %IT = 100 | 1999CA08 |
| 79 | 97 | $^{176}$Au | 0+X | (2−, 3−) | 1.05(1) *s* | | | %α = 58(5)<br>%ϵ+%β+ = 42(5)<br>%IT = ? | 2021HA32<br>2014AN10<br>2013KOZR<br>2004GOZZ |
| 79 | 97 | $^{176}$Au | 0+Y | (7+,8+,<br>9+) | 1.36(2) *s* | | | %α = 29(5)<br>%ϵ+%β+ = 71(5)<br>%IT = ? | 2021HA32<br>2014AN10<br>2004GOZZ |
| 70 | 107 | $^{177}$Yb | 331.5 (3) | 1/2− | 6.41(2) *s* | 227.0(2) | M3 | %IT = 100 | 2012FL05<br>1965BUZZ<br>1962FE02 |
| 71 | 106 | $^{177}$Lu | 150.3984 (10) | 9/2− | 133.1(44) *ns* | 150.399(1) | E1 | %IT = 100 | 2002DRZZ<br>2002MCZY<br>1955DE18<br>1949MC41 |
| 71 | 106 | $^{177}$Lu | 569.6721 (15) | 1/2+ | 155(7) *μs* | 111.715(1)<br>569.680(9) | E2 | %IT = 100 | 1970FL09<br>1965HE06 |
| 71 | 106 | $^{177}$Lu | 761.7063 (14) | 5/2− | 33(2) *ns* | 52.085(2)<br>188.086(1)<br>209.610(1)<br>303.751(4)<br>761.708(5) | <br>E1<br><br><br>E1 | %IT = 100 | 2016DE30<br>2002DRZZ<br>2002MCZY<br>1972MA54 |
| 71 | 106 | $^{177}$Lu | 970.1757 (24) | 23/2− | 160.4(3) *d* | 115.8682(23)<br>125.3(2)?<br>333.1(2)<br>334? | E3 | %β−= 77.30(8)<br>%IT = 22.70(8) | 2008CA13<br>1975WA19<br>1973CH18<br>1967NE05 |

*Continued…*

Table 1 contd...

| Z | N | $^{A}$X | E(keV) | J$^{\pi}$ | $T_{1/2}$ | Eγ(keV) | λ | Decay mode | Reference |
|---|---|---|---|---|---|---|---|---|---|
| 71 | 106 | $^{177}$Lu | 1356.860 (7) | 15/2+ | 10.8(5) *ns* | 502.54(6) | | %IT=100 | 2016DE30 |
| | | | | | | 720.721(18) | M1 | | 2002DRZZ |
| | | | | | | 916.25(16) | M1 | | 2002MCZY |
| | | | | | | 1067.0(5) | | | 1996PE05 |
| | | | | | | 1088.129(10) | (E2) | | |
| | | | | | | 1206.0(5)? | | | |
| 71 | 106 | $^{177}$Lu | 2771.7 (5) | 33/2+ | 625(62) *ns* | 542.6(5) | | %IT=100 | 2004DR06 |
| | | | | | | 586.5(5) | | | |
| | | | | | | 864.4(5) | | | |
| 71 | 106 | $^{177}$Lu* | 3530.4 (6) | 39/2− | 6(2) *μs** | 226.7(5) | M1 | %IT=100 | 2015KO14 |
| | | | | | | 402.2(5) | | | 2004DR06 |
| | | | | | | 618.7(5) | | | 2004AL04 |
| | | | | | | 758.8(5) | | | |
| 72 | 105 | $^{177}$Hf | 1315.4502 (8) | 23/2+ | 1.09(5) *s* | (14.050(10)) | | %IT=100 | 2013LA08 |
| | | | | | | 55.15(2) | E1+M2 | | 1971GL09 |
| | | | | | | 228.4838(6) | E2 | | 1966BO01 |
| 72 | 105 | $^{177}$Hf | 1342.4 (10) | (19/2−) | 55.9(12) *μs* | 548 | M1+E2 | %IT=100 | 1976REZH |
| 72 | 105 | $^{177}$Hf | 2740.02 (15) | 37/2− | 51.4(5) *min* # | 214 | E3 | %IT=100 | 1972CH38 |
| | | | | | | | | | 1971WA16 |
| 73 | 104 | $^{177}$Ta | 70.48 (5) | 5/2+ | 70.2(19) *ns* | 70.45(5) | M1+E2 | %IT=100 | 2000DA09 |
| | | | | | | | | | 1976AO02 |
| | | | | | | | | | 1973SC20 |
| 73 | 104 | $^{177}$Ta | 73.16 (7) | 9/2− | 410(7) *ns* | 73.15(9) | E1 | %IT=100 | 2000DA09 |
| 73 | 104 | $^{177}$Ta | 186.16 (6) | 5/2− | 3.62(10) *μs* | (13.90(9)) | | %IT=100 | 2000DA09 |
| | | | | | | (113.0(1)) | | | 1972AD12 |
| | | | | | | 115.65(5) | E1 | | 1972RO05 |
| | | | | | | 186.2(2) | E1 | | 1971HU14 |
| 73 | 104 | $^{177}$Ta | 487.63 (6) | 1/2+ | 26(3) *ns* | 115.05(5) | E1 | %IT=100 | 1972AD12 |
| | | | | | | 271.02(9) | E1 | | |
| | | | | | | 417.16(5) | E2 | | |
| 73 | 104 | $^{177}$Ta | 1354.8 (3) | 21/2− | 5.30(20) *μs* | 53(1) | M1 | %IT=100 | 2002PF01 |
| | | | | | | 311.3(3) | M1+E2 | | 2000DA09 |
| | | | | | | 549.9(3) | E2 | | 1982AO04 |
| | | | | | | | | | 1972RO05 |
| | | | | | | | | | 1971HU14 |
| 73 | 104 | $^{177}$Ta | 2826.3 (4) | 31/2+ | 23(4) *ns* | 155.2(3) | M1 | %IT=100 | 2000DA09 |
| | | | | | | 244(1)? | | | |
| | | | | | | 257(1) | | | |
| | | | | | | 296.5(3) | (M1) | | |
| | | | | | | 502(1) | | | |
| | | | | | | 552(1) | | | |
| | | | | | | 555.3(3) | M1 | | |
| | | | | | | 789.3(3) | E2 | | |
| 73 | 104 | $^{177}$Ta | 2852.5 (5) | 33/2− | 46(4) *ns* | 26.2(3) | E1 | %IT=100 | 2000DA09 |
| 73 | 104 | $^{177}$Ta | 4656.3 (8) | 49/2− | 133(4) *μs* | 86.1(3) | E2 | %IT=100 | 2000DA09 |

*Continued...*

*: The isomer half-life and decay mode are from a priv. comm. as cited in 2015KO14 and the beta-decaying, 39/2−isomer with $T_{1/2}$ = 7(2) *min* proposed by 2004AL04 was not confirmed by 2004DR06.

#: 2004AL04 gives half-life as 76(+16−9) *min*.

Table 1 contd...

| Z | N | $^{A}$X | E(keV) | J$^{\pi}$ | $T_{1/2}$ | E$\gamma$(keV) | $\lambda$ | Decay mode | Reference |
|---|---|---|---|---|---|---|---|---|---|
| 74 | 103 | $^{177}$W | 101.23 (8) | 5/2− | 38(8) *ns* | (6.25(11)) | | %IT=100 | 1997SH36 |
| | | | | | | (21.74(11)) | | | |
| | | | | | | 101.2(1) | | | |
| 74 | 103 | $^{177}$W | 185.31 (17) | 7/2+ | 13(3) *ns* | 49.8(2) | | %IT=100 | 1997SH36 |
| | | | | | | 84.3(2) | E1 | | |
| 75 | 102 | $^{177}$Re | 84.71 (10) | 5/2+ | 50(10) $\mu s$ | 84.7(1) | | %IT=100 | 1972LE04 |
| 76 | 101 | $^{177}$Os | 152.30 (24) | 5/2− | 40(3) *ns* | 61.7(3)? | | %IT=100 | 1983DR05 |
| | | | | | | 152.3(3)? | | | |
| 76 | 101 | $^{177}$Os | 300.6 (4) | 7/2+ | 46.3(3) *ns* | 60.2(3) | | %IT=100 | 1983DR05 |
| | | | | | | 148.3(3) | (E1) | | |
| 78 | 99 | $^{177}$Pt | 147.4 (4) | 1/2− | 2.35(4) $\mu s$ | 147.4(4) | | %IT=100 | 2019GI11 |
| | | | | | | | | | 1979HA10 |
| 79 | 98 | $^{177}$Au | 182.7 (5) | (11/2−) | 1.193(13) *s* | | | $\%\alpha$ = 60(8) | 2021HA32 |
| | | | | | | | | $\%\epsilon+\%\beta+$ = 40(8) | 2020VE03 |
| | | | | | | | | | 2014ALZX |
| | | | | | | | | | 2009AN14 |
| | | | | | | | | | 2001KO44 |
| | | | | | | | | | 1996PA01 |
| | | | | | | | | | 1991SE01 |
| | | | | | | | | | 1975CA06 |
| | | | | | | | | | 1968SI01 |
| 80 | 97 | $^{177}$Hg | 323.0 (10) | (13/2+) | 1.50(15) $\mu s$ | 246 | M2 | %IT = 100 | 2003ME20 |
| 81 | 96 | $^{177}$Tl | 807 (18) | (11/2−) | 230(40) $\mu s$ | | | $\%\alpha$ = 49(8) | 1999PO09 |
| | | | | | | | | %p = 51(8) | 1998AK04 |
| 71 | 107 | $^{178}$Lu | 123.8 (26) | (9−) | 23.1(3) *min* | | | $\%\beta-$= 100 | 1975KA15 |
| | | | | | | | | | 1975WA24 |
| | | | | | | | | | 1973OR03 |
| | | | | | | | | | 1967TA09 |
| 72 | 106 | $^{178}$Hf | 1147.416 (6) | 8− | 4.0(2) *s* | 88.8667(10) | E1 | %IT = 100 | 1965BUZZ |
| | | | | | | | | | 1962AL08 |
| 72 | 106 | $^{178}$Hf | 1553.997 (4) | 6+ | 77.5(7) *ns* | 40(1) | | %IT = 100 | 1980WA23 |
| | | | | | | 169.537(2) | E2 | | 1977KH01 |
| | | | | | | 406.579(6) | (M2) | | |
| | | | | | | 921.827(13) | E2 | | |
| | | | | | | 1247.391(18) | (E2) | | |
| 72 | 106 | $^{178}$Hf | 2446.09 (8) | 16+ | 31(1) *y* | 12.7(2) | | %IT = 100 | 1973HE19 |
| | | | | | | 309.50(15) | M4(+E5) | | 1968HE10 |
| | | | | | | 587.0(1) | E5 | | |
| 72 | 106 | $^{178}$Hf | 2572.4 (3) | 14− | 68(2) $\mu s$ | 126.1(3) | | %IT = 100 | 1977KH01 |
| | | | | | | 140.3 | M1 | | |
| | | | | | | 437.0 | | | |
| 73 | 105 | $^{178}$Ta$^{\dagger *}$ | 0+X | 7− | 2.36(8) *h* | | | $\%\epsilon+\%\beta+$ = 100 | 1975WA24 |
| | | | | | | | | | 1963RA14 |
| | | | | | | | | | 1958CA10 |
| | | | | | | | | | 1950WI67 |
| | | | 0+Y | (1+) | 9.31(3) *min* | | | $\%\epsilon+\%\beta+$ = 100 | 1967NI02 |
| | | | | | | | | | 1958CA10 |
| | | | | | | | | | 1950WI67 |
| 73 | 105 | $^{178}$Ta | 1467.82+X | 15− | 58(4) *ms* | 227.3(1) | E2 | %IT = 100 | 1998KO09 |

*Continued...*

*:1979DU02 and 1998KO09 assumed 7− as the g.s., however, from analysis of particle-transfer data, 2006BU19 concluded that the energy ordering is unsettled.

Table 1 contd…

| Z | N | $^{A}$X | E(keV) | J$^{\pi}$ | $T_{1/2}$ | E$\gamma$(keV) | $\lambda$ | Decay mode | Reference |
|---|---|---|---|---|---|---|---|---|---|
| 73 | 105 | $^{178}$Ta | 1551.92+X | 14+ | 43(8) *ns* | 84.1(1) | E1 | %IT = 100 | 1998KO09 |
| 73 | 105 | $^{178}$Ta | 2902.2+X | (21−) | 290(12) *ms* | 431.1(8) | E3 | %IT = 100 | 1998KO09 |
| 73 | 105 | $^{178}$Ta | 447.7+Y | (4+) | 60(5) *ns* | 118.4(3)<br>329.2(1)<br>401.7(8) | | %IT = 100 | 1998KO09 |
| 74 | 104 | $^{178}$W | 3654.93 (19) | 15+ | 30(1) *ns* | 61.4(1) | (E1) | %IT = 100 | 1999CU02<br>1998PU01 |
| 74 | 104 | $^{178}$W | 5313.7 (3) | 21− | 64(2) *ns* | 43.8(1)<br>251.0(5) | (M1) | %IT = 100 | 1999CU02<br>1998PU01 |
| 74 | 104 | $^{178}$W | 6572.7 (3) | 25+ | 220(10) *ns* | 182.9(1)<br>946(1) | (E1)<br>(E3) | %IT = 100 | 1999CU02<br>1998PU01 |
| 79 | 99 | $^{178}$Au | 173.6 (3) | | 294(7) *ns* | 63.4(2)<br>113.4(4) | M1+E2<br>E2 | | 2021GI08 |
| 79 | 99 | $^{178}$Au | 189 (14) | (7+,8−) | 2.7(5) *s* | | | %$\alpha$ = 18(1)<br>%$\epsilon$+%$\beta$+ = 82 (1) | 2020CU04 |
| 79 | 99 | $^{178}$Au | 246 (14) | | 373(9) *ns* | 56.7(2) | (E1 or M1) | | 2021GI08 |
| 71 | 108 | $^{179}$Lu | 592.4 (4) | 1/2+ | 3.1(9) *ms* | 592.1(4) | M3 | %IT = 100 | 1993BO14 |
| 72 | 107 | $^{179}$Hf | 375.0352 (25) | 1/2− | 18.67(4) *s* | 160.696(2)<br>~ 375 | M3 | %IT = 100 | 1972BEWN<br>1972BF03<br>1972JO05<br>1967YU01 |
| 72 | 107 | $^{179}$Hf | 1105.74 (16) | 25/2− | 25.00(27) *d* | 21.01(12)<br>257.37(15) | M2<br>E3 | %IT = 100 | 2019KR06<br>1973CH18<br>1970KAZV |
| 72 | 107 | $^{179}$Hf | 1404.5+X | (21/2+) | 14(2) *ns* | | | %IT = 100 | 2000MU06 |
| 72 | 107 | $^{179}$Hf | 2549.6 (13) | (33/2−) | 30(10) *ns* | 847.1 | | %IT = 100 | 2000MU06 |
| 72 | 107 | $^{179}$Hf | 3439.2 (18) | (39/2−) | 12(6) *ns* | 171 | (M1, E2) | %IT = 100 | 2000MU06 |
| 72 | 107 | $^{179}$Hf | 3775.2 (21) | (43/2+) | 15(5) $\mu s$ | 336 | (M2) | %IT = 100 | 2000MU06 |
| 73 | 106 | $^{179}$Ta | 30.7 (1) | 9/2− | 1.42(8) $\mu s$ | 30.7(1) | E1 | %IT = 100 | 1964LO04 |
| 73 | 106 | $^{179}$Ta | 238.56 (9) | 5/2+ | 65(10) *ns* | 238.56(9) | M1+E2 | %IT = 100 | 1997KO13 |
| 73 | 106 | $^{179}$Ta | 520.23 (18) | (1/2)+ | 0.28(8) $\mu s$ | 281.69(16) | (E2) | %IT = 100 | 1997KO13<br>1974MA26 |
| 73 | 106 | $^{179}$Ta | 627.98 (15) | 5/2− | 80(7) *ns* | 100.50(9)<br>283.99(21)<br>389.32(21) | E1<br>(E1) | %IT = 100 | 1997KO13<br>1974MA26 |
| 73 | 106 | $^{179}$Ta | 1252.60 (23) | (21/2−) | 322(16) *ns* | 232.4(2)<br>474.92(18) | M1(+E2) | %IT = 100 | 1997KO13<br>1982BA21 |
| 73 | 106 | $^{179}$Ta | 1317.2 (4) | (25/2+) | 9.0(2) *ms* | 64.7(3) | M2 | %IT = 100 | 1982BA21 |
| 73 | 106 | $^{179}$Ta | 1328.0 (4) | (23/2−) | 1.6(4) $\mu s$ | 75.3(3) | M1(+E2) | %IT = 100 | 1982BA21 |
| 73 | 106 | $^{179}$Ta | 2639.5 (5) | (37/2+) | 54.1(17) *ms* | 108.8(3) | E2 | %IT = 100 | 1997KO13<br>1982BA21 |
| 73 | 106 | $^{179}$Ta | 2792.8 (4) | (33/2−) | 17(3) *ns* | 232.0(8)<br>262.2(3)<br>573.45(23)<br>595? | M1 | %IT = 100 | 1997KO13<br>1982BA21 |
| 73 | 106 | $^{179}$Ta | 5391.8 (8) | (49/2+) | 53(+3−7) *ns* | 500.0(5)<br>988.8(5) | | %IT = 100 | 2004KO58 |
| 74 | 105 | $^{179}$W | 221.91 (3) | 1/2− | 6.40(7) *min* | 101.6(5)<br>221.93(5) | <br>M3 | %IT = 99.71(4)<br>%$\epsilon$+%$\beta$+ = 0.29(4) | 1969BI10 |

*Continued…*

Table 1 contd…

| Z | N | $^{A}$X | E(keV) | J$^{\pi}$ | $T_{1/2}$ | Eγ(keV) | λ | Decay mode | Reference |
|---|---|---|---|---|---|---|---|---|---|
| 74 | 105 | $^{179}$W | 1631.90 (8) | (21/2+) | 390(30) $ns$ | 99.5(2) | | %IT = 100 | 1994WA05 |
| | | | | | | 114.8(2) | | | 1978BE15 |
| | | | | | | 206.7(2) | | | |
| | | | | | | 415.6(2) | | | |
| | | | | | | 508.6(2) | | | |
| | | | | | | 567.3(2) | | | |
| | | | | | | 671.1(2) | | | |
| | | | | | | 883.8(2) | | | |
| 74 | 105 | $^{179}$W | 3348.41 (14) | (35/2−) | 750(80) $ns$ | 266.3(2) | | %IT = 100 | 1994WA05 |
| | | | | | | 316.4(2) | | | |
| | | | | | | 575.7(2) | | | |
| | | | | | | 609.8(2) | E2 | | |
| | | | | | | 625.0(2) | | | |
| 75 | 104 | $^{179}$Re | 65.35 (9) | (5/2−) | 95(25) $\mu s$ | 65.35(9) | (E1) | %IT = 100 | 1972LE04 |
| 75 | 104 | $^{179}$Re | 1202.2 (6) | (21/2) | ~ 21 $ns$ | 295.9(5) | | %IT = 100 | 1976ROYE |
| 75 | 104 | $^{179}$Re | 1772.20+X | (23/2+) | 0.408(12) $\mu s$ | | | %IT = 100 | 2002TH12 |
| 75 | 104 | $^{179}$Re | 5408.0 (5) | (47/2, 49/2+) | 0.466(15) $ms$ | 244.9(2) | | %IT = 100 | 2002TH12 |
| 76 | 103 | $^{179}$Os | 145.41 (12) | (7/2)− | 0.50 $\mu s$ | 45.2(1) | M1 | %IT = 100 | 1983DR05 |
| 76 | 103 | $^{179}$Os | 243.0 (8) | (9/2)+ | 0.783(14) $\mu s$ | 97.5(1) | E1 | %IT = 100 | 1983DR05 |
| 79 | 100 | $^{179}$Au | 89.5 (5) | (3/2−) | 328(2) $ns$ | 62.4(2) | E1 | %IT = 100 | 2019MOAA |
| | | | | | | 89.5(3) | E1 | | 2011VE01 |
| | | | | | | | | | 2004RA28 |
| 80 | 99 | $^{179}$Hg | 171.4 (4) | (13/2+) | 6.4(9) $\mu s$ | 110.8(3) | M2 | %IT = 100 | 2002JE09 |
| 81 | 98 | $^{179}$Tl | 0+X | (11/2−) | 1.43(3) $ms$ | | | %α ≈ 100 | 2017BA46 |
| | | | | | | | | %IT = ? | 2010AN01 |
| | | | | | | | | %ϵ+%β+ = ? | |
| | | | | | | | | %p = ? | |
| 81 | 98 | $^{179}$Tl | 904.5 (9) | (9/2−) | 114(+18–10) $ns$ | 94.0(5) | E1 | %IT = 100 | 2017BA46 |
| 71 | 109 | $^{180}$Lu | 13.9 (3) | 3− | ~ 1 $s$ | 14? | | %IT =? | 1995ME03 |
| | | | | | | | | %β−= ? | |
| 71 | 109 | $^{180}$Lu | 624.0 (5) | (9−) | ≥ 1 $ms$ | 128 | E1 | %IT = 100 | 2001WH02 |
| 72 | 108 | $^{180}$Hf | 1141.552 (15) | 8− | 5.49(3) $h$ | 57.538(17) | E1 | %IT = 99.69(8) | 2006VO12 |
| | | | | | | 500.697(15) | M2+E3 | %β = 0.31(8) | 2003AL36 |
| | | | | | | | | | 2001AL23 |
| | | | | | | | | | 1980HO17 |
| | | | | | | | | | 1963RA14 |
| | | | | | | | | | 1951BU50 |
| 72 | 108 | $^{180}$Hf | 1374.36 (4) | (4−) | 0.57(2) $\mu s$ | 113.66(2) | | %IT = 100 | 2016TA23 |
| | | | | | | 1065.77(5) | E1(+M2) | | 1999DA09 |
| | | | | | | 1281.7(2) | | | 1990GRZS |
| 72 | 108 | $^{180}$Hf | 2485.5 (5) | 12+ | 0.94(11) $\mu s$ | 223.3(2) | | %IT = 100 | 2016TA23 |
| | | | | | | 539.5(3) | | | 2001CH10 |
| | | | | | | 832.4(15) | | | 1999DA09 |
| | | | | | | 1101.1(7) | | | |
| 72 | 108 | $^{180}$Hf | 3597.5 (10) | (18−) | 90(10) $\mu s$ | 150.6(5) | (M2) | | 2016TA23 |
| | | | | | | | | | 2001CH10 |
| | | | | | | | | | 1999DA09 |

*Continued…*

Table 1 contd...

| Z | N | $^{A}$X | E(keV) | J$^{\pi}$ | $T_{1/2}$ | Eγ(keV) | λ | Decay mode | Reference |
|---|---|---|---|---|---|---|---|---|---|
| 73 | 107 | $^{180}$Ta | 76.79 (55) | 9− | > 4.5E+16 *y* | | | %β−= ?<br>%ϵ = ? | 2022NE10<br>2017LE01<br>2009HU21<br>2006HU15<br>2006BI14<br>1985CU03<br>1981NO09<br>1977AR11<br>1967SA05<br>1958MI90<br>1958BA51<br>1958EB09<br>1955EB14 |
| 73 | 107 | $^{180}$Ta | 107.78 (4) | 0− | 19.2(7) *ns* | 107.69(5) | E1 | %IT = 100 | 1999SA59<br>1998DR07 |
| 73 | 107 | $^{180}$Ta | 177.87 (11) | 8+ | 70.0(14) *ns* | 100.71(5) | E1 | %IT = 100 | 1998DR07 |
| 73 | 107 | $^{180}$Ta | 357.01 (12) | 7+ | 42(3) *ns* | 179.08(9) | M1 | %IT = 100 | 1999SA59<br>1998DR07 |
| 73 | 107 | $^{180}$Ta | 463.62 (13) | 7− | 31.2(19) *ns* | 285.54(12) | E1 | %IT = 100 | 1999SA59<br>1998DR07 |
| 73 | 107 | $^{180}$Ta | 520.04 (9) | 4+ | 37.4(20) *ns* | (71.8)<br>209.2(1)<br>335.2(1)<br>409.07(14) | <br>M1+E2<br><br>M1 | %IT = 100 | 1999SA59<br>1998DR07 |
| 73 | 107 | $^{180}$Ta | 594.43 (16) | (5) | 16.1(19) *ns* | 72.2 | | %IT = 100 | 1999SA59<br>1998DR07 |
| 73 | 107 | $^{180}$Ta | 1452.39 (22) | 15− | 31.2(1) *µs* | 142.7(3)<br>431.6(2) | M1 | %IT = 100 | 1996DR02 |
| 73 | 107 | $^{180}$Ta | 2588.3 (3) | (18+) | 22(2) *ns* | 431.07(9)<br>1136.8 | (E1) | %IT = 100 | 1999SA59<br>1996DR02 |
| 73 | 107 | $^{180}$Ta | 3678.9 (10) | (22−) | 2.0(5) *µs* | 369.8 | (E2) | %IT = 100 | 2000WH04 |
| 73 | 107 | $^{180}$Ta | 4169.9+X (18) | (23, 24, 25) | 17(5) *µs* | | | | 2000WH04 |
| 74 | 106 | $^{180}$W | 1529.05 (4) | 8− | 5.47(9) *ms* | 67<br>390.581(15) | <br>E1 | %IT = 100 | 1981AV04<br>1967CO20<br>1967CO26<br>1966BU08 |
| 74 | 106 | $^{180}$W | 1639.80 (3) | 5 | 19.2(3) *ns* | 179.1<br>279.31(4)<br>332.24(3)<br>454.88(3)<br>951.25(12) | | %IT = 100 | 1979MA08 |
| 74 | 106 | $^{180}$W | 3264.9 (3) | 14− | 2.3(2) *µs* | 222.3(3)<br>298.4<br>528.0(3)<br>813.4(3) | (M1) | %IT = 100 | 2002PF01<br>1979FAZR |
| 75 | 105 | $^{180}$Re | 20.10 (10) | 1+ | 88(3) *ns* | 20.1(1) | E1 | %IT = 100 | 1968KO10 |
| 75 | 105 | $^{180}$Re | 284.2+X (6) | (9−) | 75.1(14) *ns* | 78.6(3)<br>121.3(3) | E1<br>E1 | %IT = 100 | 2005EL10<br>1990VE07<br>1987KR20 |

*Continued...*

Table 1 contd...

| Z | N | $^{A}$X | E(keV) | J$^{\pi}$ | $T_{1/2}$ | E$\gamma$(keV) | $\lambda$ | Decay mode | Reference |
|---|---|---|---|---|---|---|---|---|---|
| 75 | 105 | $^{180}$Re | 1566.8+X (6) | (13+) | 73.4(16) $ns$ | 141.3(2) | M1 | %IT = 100 | 2005EL10 |
| | | | | | | 163.9(1) | E1 | | 1990VE07 |
| | | | | | | 262.4(1) | E1 | | 1987KR20 |
| | | | | | | 416.8(1) | M1 | | |
| | | | | | | 524.0(2) | E1 | | |
| | | | | | | 678.5(1) | E2 | | |
| | | | | | | 761.7(2) | E1 | | |
| 75 | 105 | $^{180}$Re | 3471.8+X (6) | 21− | 9.0(7) $\mu s$ | 62.9? | | %IT = 100 | 2005EL10 |
| | | | | | | 102.2(1) | E2 | | |
| | | | | | | 456.8(6) | E1 | | |
| 76 | 104 | $^{180}$Os | 1928.76 (20) | 7− | 15.8(14) $ns$ | 51.6(2) | E1 | %IT = 100 | 2005MO33 |
| | | | | | | 301.6(5) | | | 1993VE01* |
| | | | | | | 324.0(7) | (E2) | | 1988LI02 |
| | | | | | | 550.0(3) | | | 1982DR03 |
| | | | | | | 670.9(4) | (E1) | | |
| | | | | | | 1133.8(4) | E1+(M2) | | |
| 72 | 109 | $^{181}$Hf | 595.27 (4) | 9/2+ | 80(5) $\mu s$ | 129.36(4) | | %IT = 100 | 2001SH36 |
| | | | | | | 154.56(3) | | | 1999DA09 |
| | | | | | | 291.39(1) | | | |
| | | | | | | 391.19(12) | | | |
| 72 | 109 | $^{181}$Hf | 1043.5 (8) | (17/2+) | ~ 100 $\mu s$ | 115.3 | M1 | %IT = 100 | 2001SH36 |
| | | | | | | 284.9 | E2 | | |
| 72 | 109 | $^{181}$Hf | 1741.9 (13) | (25/2−) | 1.5(5) $ms$ | 357 | E3 | %IT = 100 | 2001SH36 |
| | | | | | | 499.2 | E3 | | |
| 73 | 108 | $^{181}$Ta | 6.237 (20) | 9/2− | 6.05(12) $\mu s$ | 6.240(20) | E1 | %IT = 100 | 1981MO15 |
| 73 | 108 | $^{181}$Ta | 482.168 (23) | 5/2+ | 10.8(1) $ns$ | 345.97(4) | E2 | %IT = 100 | 1971BO13 |
| | | | | | | 475.99(9) | M2+E3 | | 1969HR02 |
| | | | | | | 482.17(3) | M1+E2 | | 1966HO13 |
| | | | | | | | | | 1965SO02 |
| | | | | | | | | | 1965ME08 |
| | | | | | | | | | 1963MA10 |
| | | | | | | | | | 1961NA06 |
| | | | | | | | | | 1961MA47 |
| 73 | 108 | $^{181}$Ta | 615.19 (3) | 1/2+ | 18.5(4) $\mu s$ | 133.027(18) | E2 | %IT = 100 | 1972RI14 |
| | | | | | | 615.17(11) | M3(+E4) | | 1970GL02 |
| | | | | | | | | | 1969FAZY |
| | | | | | | | | | 1967CO20 |
| | | | | | | | | | 1960BE19 |
| | | | | | | | | | 1959LI44 |
| | | | | | | | | | 1957DR13 |
| | | | | | | | | | 1956VE03 |
| | | | | | | | | | 1955GO30 |
| | | | | | | | | | 1953BR30 |
| | | | | | | | | | 1953MU60 |
| | | | | | | | | | 1953BA60 |
| | | | | | | | | | 1948BU14 |
| 73 | 108 | $^{181}$Ta | 1403.2+X | (19/2+) | 140(36) $ns$ | | | | 1998SA60 |

*Continued...*

*: In 1993VE01, it is not clear whether the listed value is half-life or mean lifetime, thus not considered in averaging.

Table 1 contd…

| Z | N | $^{A}$X | E(keV) | J$^{\pi}$ | $T_{1/2}$ | E$\gamma$(keV) | $\lambda$ | Decay mode | Reference |
|---|---|---|---|---|---|---|---|---|---|
| 73 | 108 | $^{181}$Ta | 1483.43 (21) | 21/2− | 25(2) $\mu s$ | 177<br>455.3<br>710.6 | | %IT = 100 | 1998WH02 |
| 73 | 108 | $^{181}$Ta | 2227.9 (9) | | 210(20) $\mu s$ | 130<br>295 | | %IT = 100 | 2001OL03<br>1998WH02 |
| 74 | 107 | $^{181}$W | 365.55 (13) | 5/2− | 14.6(2) $\mu s$ | 252.2(3)<br>365.5(3) | E3<br>M2 | %IT = 100 | 2002PF01<br>1994SI11<br>1968IV02<br>1967CO20<br>1958DU80<br>1957BU39<br>1956VE03 |
| 74 | 107 | $^{181}$W | 1653.1 (6) | 21/2+ | 140(20) $ns$ | 838.9<br>1053.7 | | %IT = 100 | 2002PF01 |
| 74 | 107 | $^{181}$W | 1744.9 (8) | (23/2−) | ~ 50 $ns$ | 91.8 | (E1) | %IT = 100 | 1973LI17 |
| 74 | 107 | $^{181}$W | 3943.9 (16) | 37/2 | ~ 20 $ns$ | 351<br>712 | | %IT = 100 | 1992YEZW |
| 75 | 106 | $^{181}$Re | 262.91 (11) | 9/2− | 156.7(19) $ns$ | 144.76(11) | E1 | %IT = 100 | 2000PE18<br>1974SI14<br>1967GO25 |
| 75 | 106 | $^{181}$Re | 356.72 (7) | 5/2− | 87.6(12) $ns$ | 238.75(7)<br>356.7(2) | E1<br>(E1) | %IT = 100 | 2000PE18 |
| 75 | 106 | $^{181}$Re | 1656.37 (14) | 21/2− | 250(10) $ns$ | 328.84(8)<br>584.23(9) | <br>(E2) | %IT = 100 | 2000PE18 |
| 75 | 106 | $^{181}$Re | 1880.57 (16) | 25/2+ | 11.5(9) $\mu s$ | 224.25(12) | M2 | %IT = 100 | 2000PE18<br>1969CO13 |
| 75 | 106 | $^{181}$Re | 3869.40 (18) | (35/2−) | 1.2(2) $\mu s$ | 356.6(1) | | %IT = 100 | 2000PE18 |
| 75 | 106 | $^{181}$Re | 3990.02 (17) | (37/2−) | 22.2(5) $ns$ | 22.6(1)<br>120.6(1)<br>278.6(1) | <br>M1 | %IT = 100 | 2000PE18 |
| 76 | 105 | $^{181}$Os | 49.20 (14) | 7/2− | 2.7(1) $min$ | (49.2) | | %$\epsilon$+%$\beta$+ = 98.5(15)<br>%IT = 1.5(15) | 1967GO25 |
| 76 | 105 | $^{181}$Os | 156.91 (15) | 9/2+ | 262(6) $ns$ | 107.73(7) | E1 | %IT = 100 | 2003CU03 |
| 76 | 105 | $^{181}$Os | 3738.67 (18) | 35/2− | 24(4) $ns$ | 159.4(1) | M1 | %IT = 100 | 2003CU03 |
| 77 | 104 | $^{181}$Ir | 289.33 (13) | 5/2+ | 298 $ns$ | 289.34(13) | E1 | %IT = 100 | 1993DR02 |
| 77 | 104 | $^{181}$Ir | 366.30 (22) | 9/2− | 126(6) $ns$ | 341.5<br>366.5 | M1 | %IT = 100 | 1992KA01 |
| 77 | 104 | $^{181}$Ir | 2034.3 (4) | 23/2+ | 22(4) $ns$ | 54.9<br>151.8<br>547.6 | | %IT = 100 | 1992KA01 |
| 78 | 103 | $^{181}$Pt | 116.65 (8) | (7/2)− | > 300 $ns$ | 22.8(1) | (M1) | %IT = 100 | 1992SA03 |
| 78 | 103 | $^{181}$Pt | 275.99 (10) | (9/2)+ | 99(9) $ns$ | 19.7(3)<br>40.5(1)<br>159.42(9) | E1<br>E1<br>(E1) | %IT = 100 | 1990DE03 |
| 81 | 100 | $^{181}$Hg | 161.7+Y | (13/2+) | 0.48(2) $ms$ | | | %IT = ? | 2009AN17 |
| 81 | 100 | $^{181}$Tl | 835.9 (5) | (9/2−) | 1.40(3) $ms$ | 577.9(3) | (E3) | %$\alpha$ = 0.40(6)<br>%IT = 99.60(4) | 2009AN14 |
| 72 | 110 | $^{182}$Hf | 1172.87 (18) | (8−) | 61.5(15) $min$ | 50.80(8)<br>506.60(8) | (E1) | %IT = 46(2)<br>%$\beta$−= 54(2) | 1974WA14 |
| 72 | 110 | $^{182}$Hf | 2571.3 (12) | (13+) | 40(10) $\mu s$ | 264 | | %IT ≈ 100 | 1999DA09 |
| 73 | 109 | $^{182}$Ta | 16.273 (4) | 5+ | 283(3) $ms$ | (16.273(4)) | (M2) | %IT = 100 | 1968CL06 |

*Continued…*

Table 1 contd…

| Z | N | $^{A}$X | E(keV) | J$^{\pi}$ | $T_{1/2}$ | E$\gamma$(keV) | $\lambda$ | Decay mode | Reference |
|---|---|---|---|---|---|---|---|---|---|
| 73 | 109 | $^{182}$Ta | 519.577 (16) | 10− | 15.84(10) *min* | 184.951(15)<br>356.47(10) | E3<br>M4 | %IT = 100 | 1966BI10 |
| 74 | 108 | $^{182}$W | 2230.65 (14) | (10+) | 1.3(1) $\mu s$ | 518.5(1)<br>1086.5(1) | (M1) | %IT = 100 | 2002PF01<br>1969NO05 |
| 74 | 108 | $^{182}$W | 3754.89 (21) | (15+) | 37(2) $ns$ | 19?<br>356.5(1)<br>676.8(2) | <br>(M1+E2)<br>(E2) | %IT = 100 | 1995SH27 |
| 74 | 108 | $^{182}$W | 4040.6 (3) | (17−) | 20(1) $ns$ | 146.9(1) | (E1) | %IT = 100 | 1995SH27 |
| 75 | 107 | $^{182}$Re | 0+X | 2+ | 14.14(45) $h$ | | | %$\epsilon$+%$\beta$+ = 100 | 2014MA43<br>2011BO01<br>1963BA37<br>1950WI14 |
| 75 | 107 | $^{182}$Re | 235.732+X | (2)− | 585(30) $ns$ | 180.20(3)<br>235.75(6) | E1<br>(E1+M2) | %IT = 100 | 1973BU08<br>1969AN13 |
| 75 | 107 | $^{182}$Re | 461.3+X | (4−) | 0.78(9) $\mu s$ | 461.3(1) | | %IT = 100 | 1984SL01 |
| 75 | 107 | $^{182}$Re | 2256.48 (19) | (16−) | 82(1) $ns$ | 344.54(8)<br>647.36(8) | (M1)<br>(E2) | %IT = 100 | 1988JA02 |
| 76 | 106 | $^{182}$Os | 1831.4 (3) | (8)− | 0.78(7) $ms$ | 553.5(2) | E1 | %IT = 100 | 1966BU08 |
| 76 | 106 | $^{182}$Os | 7049.5 (4) | (25+) | 150(10) $ns$ | 148.9(5)<br>187.5(2)<br>505.6(2)<br>566.0(2)<br>726.9(2)<br>1061.6(5) | <br>(M1+E2)<br><br><br><br>(M1+E2) | %IT = 100 | 1989AL19 |
| 77 | 105 | $^{182}$Ir | 71.02 (17) | (5)+ | 170(40) $ns$ | 45.3(3)<br>71.1(3) | M1+E2 | %IT = 100 | 1990KR06 |
| 77 | 105 | $^{182}$Ir | 176.4 (3) | (6−) | 130(50) $ns$ | 105.4(3) | (E1) | %IT = 100 | 1990KR06 |
| 81 | 101 | $^{182}$Tl$^{\dagger}$ | 0+X | (4−) | 1.9(1) $s$ | | | %$\epsilon$+%$\beta$+ = 99.75(25)<br>%$\alpha$ = 0.25(25) | 2017BA04<br>2016VA01 |
| | | | 0+Y | (7+) | 3.1(10) $s$ | | | %$\epsilon$+%$\beta$+ = 97.5(25)<br>%$\alpha$ = 2.5(25) | 2017BA04<br>1997BA21<br>1991BO22 |
| 72 | 111 | $^{183}$Hf | 1464 (64) | (27/2−) | 10(+48−5) $s$ $^{@}$ | | | %IT > 0.0<br>%$\beta$−= ? | 2012RE19$^{@}$<br>2010RE07$^{@}$ |
| 73 | 110 | $^{183}$Ta | 73.164 (14) | (9/2)− | 109(3) $ns$ | 73.174(12) | E1 | %IT = 100 | 2016PAAA<br>2009SH17<br>1967MO13 |
| 73 | 110 | $^{183}$Ta | 1332 | 19/2+ | 832(97) $ns$ | 22.7(2)?<br>465.4(3) | E1<br>E1 | %IT = 100 | 2016PAAA<br>2009SH17 |
| 74 | 109 | $^{183}$W | 309.492 (4) | 11/2+ | 5.32(10) $s$ | 102.481(3) | M2 | %IT = 100 | 1961GA06<br>1963KA34<br>1965BUZZ<br>1972JO05 |
| 74 | 109 | $^{183}$W | 453.0695 (17) | 7/2− | 18.9(4) $ns$ | 40.976(1)<br>144.1217(24)<br>161.3439(27)<br>244.263(4)<br>246.059(4)<br>353.989(6)<br>406.599(13) | M1<br>M1+E2<br>M1+E2<br>E2<br>M1+E2<br>M1+E2<br>(E2) | %IT = 100 | 1966HO13<br>1967ME01<br>1967MA28<br>1971HO14<br>1999SA60 |

*Continued…*

Table 1 contd...

| Z | N | $^{A}$X | E(keV) | J$^{\pi}$ | $T_{1/2}$ | E$\gamma$(keV) | $\lambda$ | Decay mode | Reference |
|---|---|---|---|---|---|---|---|---|---|
| 74 | 109 | $^{183}$W | 1746.11 (7) | 19/2− | 12.7(20) *ns* | 306.5(1)<br>307.0(1)<br>555.6(1)<br>577.1(1)<br>830.9(1) | | %IT = 100 | 1999SA60 |
| 75 | 108 | $^{183}$Re | 1907.21 (15) | (25/2)+ | 1.04(4) *ms* | 193.84(8) | E2 | %IT = 100 | 1981AV04<br>1966EM02 |
| 75 | 108 | $^{183}$Re | 1936.66+X | (21/2) | 10(4) *ns* | | | %IT = 100 | 2000PU01 |
| 76 | 107 | $^{183}$Os | 170.73 (7) | 1/2− | 9.9(3) *h* | 170.7(1) | M4 | %$\epsilon$+%$\beta$+ = 85(2)<br>%IT = 15(2) | 1960NE03 |
| 76 | 107 | $^{183}$Os | 4181.78+X | | ~ 30 *ns* | | | | 1985PE07 |
| 76 | 107 | $^{183}$Os | 5067.68 (15) | (43/2−) | 27(3) *ns* | 133.0(1)<br>351.3(1) | (E1) | %IT = 100 | 2001SH41 |
| 76 | 107 | $^{183}$Os | 5167.61 (15) | (43/2+) | 24(2) *ns* | 451.3(1)<br>1079.0(1) | | %IT = 100 | 2001SH41 |
| 78 | 105 | $^{183}$Pt | 34.50 (8) | (7/2)− | 43(5) *s* | | | %$\epsilon$+%$\beta$+ ≈ 100<br>%IT = ?<br>%$\alpha$ < 4E−4 | 1979VI02 |
| 78 | 105 | $^{183}$Pt | 195.68 (11) | (9/2)+ | > 150 *ns* | 46.03(9)<br>161.18(7) | <br>E1 | %IT = 100 | 1990NY02 |
| 79 | 104 | $^{183}$Au | 73.10 (1) | (1/2)+ | > 1 $\mu s$ | 60.5(3) | E1 | %IT = 100 | 2017VE04<br>1984MA41 |
| 80 | 103 | $^{183}$Hg | 204 (14) | (13/2+) | 290(30) $\mu s$ | ≤81 | | | 2022HU09<br>1981MI12 |
| 81 | 102 | $^{183}$Tl | 628.7 (3) | (9/2−) | 53.3(3) *ns* | 356.5(2) | E3 | %IT = 98.55(42)<br>%$\alpha$ = 1.45(42) | 2022VE01<br>2004RA28<br>1980SC09 |
| 81 | 102 | $^{183}$Tl | 976.8 (3) | (13/2+) | 1.48(10) $\mu s$ | 69.3(5)<br>346.8(3) | | %IT = 100 | 2001MU26 |
| 82 | 101 | $^{183}$Pb | 97 (9) | (13/2+) | 415(20) *ms* | | | %$\alpha$ ≈ 100 | 2002JE09 |
| 72 | 112 | $^{184}$Hf | 1272.2 (4) | (8−) | 48(10) *s*<br>113(+60−47) *s* @ | 72.7(2)<br>555.0(2) | | %IT ≈ 100<br>%$\beta$−= ? | 2012RE19@<br>2010RE07@<br>1995KR04 |
| 72 | 112 | $^{184}$Hf | 2477 (10) | (15+) | 12(+8−6) *min* @ | | | %$\beta$−= ?<br>%IT = ? | 2012RE19@<br>2010RE07@ |
| 74 | 110 | $^{184}$W | 1284.997 (8) | 5− | 8.33(18) $\mu s$ | 63.6890(14)<br>151.134(20)<br>381.82(14)<br>536.674(15)<br>920.933(21)<br>1173.77(3) | E2<br><br><br>E1+M2+E3<br>E1+M2+E3<br>(E3) | %IT = 100 | 1969GL04 |
| 74 | 110 | $^{184}$W | 3863.2 (25) | (14−,15,<br>17−)* | 188(38) *ns** | 148 | (M1) | %IT = 100 | 2004WH02 |
| 75 | 109 | $^{184}$Re | 188.0463 (17) | 8(+) | 177.25(7) *d* # | 83.3067(8)<br>188.0462(17) | M4<br>(E5) | %$\epsilon$ = 25.5(8)<br>%IT = 74.5(8) | 1963JO03 |
| 76 | 108 | $^{184}$Os | 2366.81 (19) | 10+ | 23.6(14) *ns* | 145.0(2)<br>495.8(2)<br>1092.1(2) | E1<br>M1(+E2)<br>E2 | %IT = 100 | 2002WH01 |

*Continued...*

*: J$^{\pi}$ = (14+), $T_{1/2}$ = 156(3) *ns*, and E$\gamma$ = 273.2 keV reported by J. Werner in B.Sc. Hons. thesis (2005, ANU, Canberra).
#: $T_{1/2}$ from 2022JA13; L. Janiak et al., Phys. Rev. C 106, 044303 (2022). $T_{1/2}$ = 169(8) *d* in 1963JO03.

Table 1 contd...

| Z | N | $^{A}X$ | E(keV) | J$^{\pi}$ | $T_{1/2}$ | E$\gamma$(keV) | $\lambda$ | Decay mode | Reference |
|---|---|---|---|---|---|---|---|---|---|
| 76 | 108 | $^{184}$Os | 4756.71 (24) | (18−) | 48(5) *ns* | 280.9(2) | | %IT = 100 | 2002SH21 |
| | | | | | | 289.1(2) | | | |
| | | | | | | 554.0(2) | | | |
| | | | | | | 634.4(2) | | | |
| 77 | 107 | $^{184}$Ir | 225.65 (11) | 3+ | 470(30) $\mu s$ | 154.8(1) | E1+M2 | %IT = 100 | 1988KR17 |
| | | | | | | 225.8(1) | M2 | | |
| 77 | 107 | $^{184}$Ir | 328.40 (24) | (7)+ | 350(90) *ns* | 286.5(3) | (E1(+M2)) | %IT = 100 | 1988KR17 |
| | | | | | | 328.5(3) | M2 | | |
| 77 | 107 | $^{184}$Ir | 432.49 (11) | (2)+ | > 10 *ns* | 89.8(1) | | %IT = 100 | 1988BE16 |
| | | | | | | 139.1(1) | M1(+E2) | | |
| | | | | | | 169.8(1) | | | |
| | | | | | | 206.9(1) | | | |
| 78 | 106 | $^{184}$Pt | 1842.6 (8) | 8− | 0.98(6) *ms* | 49(1) | | %IT = 100 | 2017PE03 |
| | | | | | | 112(1) | E1 | | 1966BU08 |
| | | | | | | 610.1(20) | E1 | | |
| 79 | 105 | $^{184}$Au | 68.46 (4) | 2+ | 46.8(10) *s* | 68.46(4) | M3 | $\%\epsilon+\%\beta+$ = 70(10) | 1997ZA03 |
| | | | | | | | | %IT = 30(10) | 1997LE22 |
| | | | | | | | | $\%\alpha \leq 0.016$ | 1995BI01 |
| | | | | | | | | | 1994ROZY |
| | | | | | | | | | 1992RO21 |
| | | | | | | | | | 1972FI12 |
| | | | | | | | | | 1970HA18 |
| | | | | | | | | | 1969HA03 |
| 79 | 105 | $^{184}$Au | 228.40 (7) | 3− | 69(6) *ns* | 81.9(1) | E1 | %IT = 100 | 1994IB01 |
| | | | | | | 141.8(1) | (E1+M2) | | |
| | | | | | | 156.5(1) | E1 | | |
| | | | | | | 160.0(1) | (E1) | | |
| 81 | 103 | $^{184}$Tl | 0+X | (7+) | 10.1(6) *s* | | | $\%\alpha$ = 0.047(6) | 2017BA04 |
| | | | | | | | | $\%\epsilon+\%\beta+$ = 99.953(6) | 2016VA01 |
| | | | | | | | | | 1993BOZK |
| | | | | | | | | | 1976TO06 |
| | | | | | | | | | 1976CO24 |
| 81 | 103 | $^{184}$Tl | 506.09+X (10) | (10−) | 47.1(7) *ms* | 61.29(14) | (M2) | $\%\alpha$ = 0.089(19) | 2016VA01 |
| | | | | | | 186.2(3) | E3 | %IT = 99.911(19) | 2015VA10 |
| | | | | | | 506.1(1) | E3 | | 2003AN26 |
| 83 | 101 | $^{184}$Bi$^{\dagger}$ | 0+X | | 13(2) *ms* | | | $\%\alpha \approx 100$ | 2003AN27 |
| | | | 0+Y | | 6.6(15) *ms* | | | $\%\alpha \approx 100$ | 2003AN27 |
| 73 | 112 | $^{185}$Ta | 175.5 (2) | 9/2− | 11.8(21) *ns* | (22.6) | E1 | %IT = 100 | 2009LA17 |
| | | | | | | 175.5(2) | E1 | | |
| 73 | 112 | $^{185}$Ta | 406.0 (10) | (3/2+) | 0.9(3) $\mu s$ | 406 | | %IT = 100 | 2007SH42 |
| 73 | 112 | $^{185}$Ta | 1273.4 (4) | 21/2− | 11.8(14) *ms* | (14.5) | M1 | %IT = 100 | 2009LA17 |
| | | | | | | 280.4(2) | E2 | | 1999WH03 |
| 74 | 111 | $^{185}$W | 197.383 (23) | 11/2+ | 1.67(3) *min* | (9.53(6)) | | %IT = 100 | 1970PA32 |
| | | | | | | 23.54(5) | | | 1970GU02 |
| | | | | | | 131.55(2) | E3 | | 1969DA01 |
| 74 | 111 | $^{185}$W | 243.62 (5) | 7/2− | 19.3(5) *ns* | 69.7(3) | M1+E2 | %IT = 100 | 1969KU07 |
| | | | | | | 150.3(2) | | | |
| | | | | | | 177.59(8) | M1(+E2) | | |
| | | | | | | 243.7(3) | | | |
| 75 | 110 | $^{185}$Re | 368.3 (5) | 9/2− | 33(3) *ns* | 242.9 | | %IT = 100 | 1971EV02 |

*Continued...*

Table 1 contd...

| Z | N | $^{A}$X | E(keV) | J$^{\pi}$ | $T_{1/2}$ | E$\gamma$(keV) | $\lambda$ | Decay mode | Reference |
|---|---|---|---|---|---|---|---|---|---|
| 75 | 110 | $^{185}$Re | 2123.8 (11) | (21/2) | 121(13) *ns* | 115 | E1(+M2) | %IT = 100 | 2002PF01<br>1997SH37 |
| 76 | 109 | $^{185}$Os | 102.37 (11) | 7/2− | 3.0(4) $\mu s$ | (4.9)<br>(64.9) | | | 1970FIZZ |
| 76 | 109 | $^{185}$Os | 275.53 (12) | 11/2+ | 0.78(5) $\mu s$ | (15.2) | | %IT = 100 | 1970FIZZ |
| 76 | 109 | $^{185}$Os | 5007.1 (3) | (41/2) | 18(2) *ns* | 426.0(1) | | %IT = 100 | 2004SH08 |
| 77 | 108 | $^{185}$Ir | 646.6 (3) | (11/2−) | 21.5(20) *ns* | 488.0(5)<br>640.8(2) | <br>(M1) | %IT = 100 | 1979AN20 |
| 77 | 108 | $^{185}$Ir | 2157.2+X | | 120(20) *ns* | | | %IT = 100 | 1979AN20 |
| 77 | 108 | $^{185}$Ir | 2614.0+X | | 40(10) *ns* | 99.7(5)<br>317.6(5) | | %IT = 100 | 1979AN20 |
| 78 | 107 | $^{185}$Pt | 103.41 (5) | 1/2− | 33.0(8) *min* | | | $\%\epsilon+\%\beta+$ = 99(1)<br>%IT < 2 | 1970FIZZ |
| 78 | 107 | $^{185}$Pt | 200.89 (4) | 5/2− | 728(20) *ns* | 19.8(91)<br>97.5(1) | M1<br>E2 | %IT = 100 | 1996OM01 |
| 79 | 106 | $^{185}$Au | 0+X | | 6.8(3) *min* | | | $\%\epsilon+\%\beta+$ < 100<br>%IT = ? | 1970FIZZ |
| 79 | 106 | $^{185}$Au | 220.1 (1) | (11/2−) | 26(2) *ns* | 211.2(1) | M1 | %IT = 100 | 1983BE48 |
| 80 | 105 | $^{185}$Hg | 103.7 (4) | 13/2+ | 21.6(15) *s* | 65.3(5) | E3 | $\%\epsilon+\%\beta+$ = 46(10)<br>%IT = 54(10)<br>$\%\alpha \approx 0.03$ | 2013SA43<br>1982BO27<br>1970FIZZ<br>1970HA18 |
| 81 | 104 | $^{185}$Tl | 454.8 (15) | (9/2−) | 1.93(8) *s* | 168.8 | (E3) | %IT = ?<br>$\%\alpha$ = ? | 1993BOZK<br>1980SC09<br>1977SC03<br>1976TO06 |
| 82 | 103 | $^{185}$Pb | 0+X | 13/2+ | 4.3(2) *s* | | | $\%\alpha$ = 50(25)<br>$\%\epsilon+\%\beta+$ = ? | 2002AN15 |
| 83 | 102 | $^{185}$Bi | X | (7/2−,<br>9/2−) | 58(2) $\mu s$ | | | %p = ?<br>$\%\alpha$ = ? | 2021DO08<br>2004AN07<br>2001PO05 |
| 72 | 114 | $^{186}$Hf | 2968 (43) | (17+) | > 20 *s* @ | | | $\%\beta-$= ?<br>%IT = ? | 2012RE19@<br>2010RE07@ |
| 73 | 113 | $^{186}$Ta | 336 (20) | (2, 8−) | 1.54(5) *min*<br>3.0(+15−8) *min* @ | | | %IT > 0.0<br>$\%\beta-$= ? | 2014SO21<br>2004XU08<br>2012RE19@<br>2010RE07@ |
| 73 | 113 | $^{186}$Ta | 347.9 (3) | (9+) | 17(2) *s* | 161.1(2) | E3 | | 2021WA39 |
| 74 | 112 | $^{186}$W | 1517.2 (6) | (7−) | 18(1) $\mu s$ | 119<br>195<br>708 | | %IT = 100 | 1998WH02 |
| 74 | 112 | $^{186}$W* | 3542.8 (21) | (16+) | 2.0(2) *s** | 9.6<br>219<br>300.5<br>399.4 | | $\%\beta-$= ?<br>%IT = ? | 2015KO14<br>2012LAAA<br>2012RE19<br>1998WH02<br>1999WHZZ |

*Continued...*

*: The $T_{1/2}$ for He-like $^{186}$W ion is 7.5(+48–33) *s* from 2012RE19. The half-life for neutral state of isomer is given as ≥ 3 *ms* or ≤ 30 *s* in 1998WH02. All the four gamma rays are from 2012LAAA, lowest energy to (14−), other three to (13−) states. 1998WH02 and 1999WHZZ had proposed two gamma rays (180.0 keV and 399.5 keV), one of which was the same as in 2012LAAA. The uncertainty of 0.2 *s* is assigned by 2015KO14.

Table 1 contd...

| Z | N | $^{A}$X | E(keV) | J$^{\pi}$ | $T_{1/2}$ | Eγ(keV) | λ | Decay mode | Reference |
|---|---|---|---|---|---|---|---|---|---|
| 75 | 111 | $^{186}$Re | 99.361 (3) | (3)− | 25.5(25) *ns* | 40.350(3)<br>99.362(4) | M1+E2<br>E2 | %IT = 100 | 1978SC10 |
| 75 | 111 | $^{186}$Re | 148.2 (5) | (8+) | 2.0E+5 *y* | 50(7) | (E5) | %IT = 100 | 2015MA60<br>1972SE06 |
| 75 | 111 | $^{186}$Re | 314.009 (5) | (3)+ | 24.1(11) *ns* | 103.310(6)<br>167.737(8)<br>214.648(8)<br>254.995(15) | <br>(E1)<br>E1<br>E1 | %IT = 100 | 1978SC10<br>1973GL06 |
| 75 | 111 | $^{186}$Re | 324.4 (9) | 5+ | 17.3(6) *ns* | 144.152(5) | E1 | %IT = 100 | 2017MA39<br>1978SC10<br>1973GI06 |
| 77 | 109 | $^{186}$Ir | 0+X | 2− | 1.90(5) *h* | (≤1.5) | | %ϵ+%β+ ≈ 75<br>%IT ≈ 25 | 1991BE25 |
| 79 | 107 | $^{186}$Au | 227.77 (7) | 2+ | 110(10) *ns* | 191.6(1)<br>227.7(1) | E1<br>E1 | %IT = 100 | 1985AB03 |
| 79 | 107 | $^{186}$Au | 455.1+X | (11−) | 39(4) *ns* | 57? | (M1) | %IT = 100 | 1992JA01 |
| 80 | 106 | $^{186}$Hg | 2217.3 (4) | (8−) | 82(5) *µs* | (31.6)<br>241.5<br>628.1 | | %IT = 100 | 1984JAZS |
| 81 | 105 | $^{186}$Tl | 0+X* | 7(+) | 27.5(10) *s* | | | %ϵ+%β+ = 100<br>%α ≈ 0.006 | 2020ST11<br>1977CO21 |
| 81 | 105 | $^{186}$Tl | 293.7+Y | (3+) | 11(4) *ns* | 78.5<br>293.7(3) | <br>E1 | %IT = 100 | 1991VA04 |
| 81 | 105 | $^{186}$Tl | 374+X* | 10(−) | 3.40(9) *s* | (18.2(4))<br>374.2(1) | E3<br>E3 | %IT < 94.1(3)<br>%β−> 5.9(3) | 2020ST11<br>2014BO26<br>1981KR20 |
| 83 | 103 | $^{186}$Bi | 0+X | (10−) | 9.8(13) *ms* | | | %α ≈ 100 | 1997BA21<br>1973TA30 |
| 72 | 115 | $^{187}$Hf | 0+X | | 270(80) *ns* | | | | 2012AL05<br>2009AL30 |
| 73 | 114 | $^{187}$Ta | 1778.1 (10) | (25/2−) | 7.3(9) *s* | 191.7(5) | (E2) | %IT > 60<br>%β−< 40 | 2020WA29 |
| 73 | 114 | $^{187}$Ta | 1793 (10) | (27/2−) | 22(9) *s* @ | | | %IT > 0.0<br>%β−= ? | 2012RE19@<br>2010RE07@ |
| 73 | 114 | $^{187}$Ta | 2933 (14) | (41/2+) | > 5 *min* @ | | | %β−= ?<br>%IT = ? | 2012RE19@<br>2010RE07@ |
| 74 | 113 | $^{187}$W | 410.06 (4) | 11/2+ | 1.38(7) *µs* | 45.8(3) | (E1) | %IT = 100 | 2008BO26<br>2005SH26 |
| 75 | 112 | $^{187}$Re | 206.2473 (10) | 9/2− | 555.3(17) *ns* | 72.002(4)<br>206.247(1) | E1(+M2)<br>M2+E3 | %IT = 100 | 1972GU03 |
| 75 | 112 | $^{187}$Re | 1681.8 (4) | 21/2+ | 354(62) *ns* | 207.1<br>298.0 | E1 | %IT = 100 | 2016RE02<br>2003SH13 |
| 76 | 111 | $^{187}$Os | 100.45 (4) | 7/2− | 112(6) *ns* | 25.62(5)<br>90.37(10) | M1+E2<br>E2 | %IT = 100 | 1971MA24 |
| 76 | 111 | $^{187}$Os | 257.10 (7) | 11/2+ | 231(2) *µs* | 156.63(7) | M2+E3 | %IT = 100 | 1967CO20 |
| 77 | 110 | $^{187}$Ir | 106.478 (24) | 1/2+ | 11.5(3) *ns* | 106.44(3) | E2 | %IT = 100 | 1973SE13 |
| 77 | 110 | $^{187}$Ir | 186.16 (4) | 9/2− | 30.3(6) *ms* | 76.09(3)<br>186.2(1) | M2+E3<br>E3 | %IT = 100 | 1974RO05 |
| 77 | 110 | $^{187}$Ir | 433.75 (6) | 11/2− | 152(12) *ns* | 247.61(6) | M1 | %IT = 100 | 1973SE13 |

*Continued...*

*: X is about 77(56) keV from 2020ST11.

Table 1 contd...

| Z | N | $^{A}$X | E(keV) | J$^{\pi}$ | $T_{1/2}$ | E$\gamma$(keV) | $\lambda$ | Decay mode | Reference |
|---|---|---|---|---|---|---|---|---|---|
| 77 | 110 | $^{187}$Ir | 2487.7 (4) | 29/2− | 1.8(5) $\mu s$ | 109.8 | | %IT = 100 | 2010MO09 |
| 78 | 109 | $^{187}$Pt | 57.11 (14) | (7/2−) | 18(2) $ns$ | 31.6(4) | M1 | %IT = 100 | 1979BE51 |
| 78 | 109 | $^{187}$Pt | 174.38 (22) | 11/2+ | 311(15) $\mu s$ | 117.2(4) | M2+E3 | %IT = 100 | 1976PI03 |
| 79 | 108 | $^{187}$Au | 120.33 (14) | 9/2− | 2.3(1) $s$ | 101.0(2) | E3 | %IT = 100 | 2020BA29<br>1983BR26 |
| 79 | 108 | $^{187}$Au | 223.90 (14) | (11/2−) | 48(2) $ns$ | 103.4(2) | M1+E2 | %IT = 100 | 1983BE48 |
| 79 | 108 | $^{187}$Au | 2431.3 (14) | (25/2+) | 24(3) $ns$ | 149.6 | (M1+E2) | %IT = 100 | 1997PE26<br>1989JO02 |
| 79 | 108 | $^{187}$Au | 2669.4 (18) | (31/2−) | 100(5) $ns$ | 508.8 | (M1+E2) | %IT = 100 | 1997PE26<br>1989JO02 |
| 80 | 107 | $^{187}$Hg | 54 (7) | 13/2(+) | 2.4(3) $min$ | | | %$\epsilon$+%$\beta$+ = 100<br>%$\alpha$ < 3.7E−4(10) | 2001SC41<br>1970HA18 |
| 80 | 107 | $^{187}$Hg | 161.57 (24) | (9/2+) | 33 $ns$ | 161.5(3) | | %IT = 100 | 2008WOZY<br>1983COZP |
| 81 | 106 | $^{187}$Tl | 334 (4) | (9/2−) | 15.60(12) $s$ | (35(4)) | | %$\alpha$ = 0.15(5)<br>%IT < 99.9<br>%$\epsilon$+%$\beta$+ < 99.9 | 1992SC25<br>1981MI12 |
| 81 | 106 | $^{187}$Tl | 1434.23+X | | 1.11 $\mu s$ | | | | 2000BY02 |
| 81 | 106 | $^{187}$Tl | 2582.5 (3) | (25/2−,<br>27/2,<br>29/2+) | 0.69(4) $\mu s$ | 478.2(2) | (E1, M1, E2) | %IT = 100 | 2000BY02 |
| 82 | 105 | $^{187}$Pb | 33 (13) | (13/2+) | 18.0(2) $s$ | | | %$\alpha$ = 12(2)<br>%$\epsilon$+%$\beta$+ = 88(2) | 2022HU09<br>2005WE11<br>1981MI12<br>1974LE02<br>1972GA27 |
| 82 | 105 | $^{187}$Pb | 307.9 (1) | | 5.15(15) $\mu s$ | 307.9(1) | E2 | %IT = 100 | 2022ZH22 |
| 83 | 104 | $^{187}$Bi | 112 (20) | (1/2+) | 0.370(20) $ms$ | | | %$\alpha$ = 100 | 2006AN11 |
| 83 | 104 | $^{187}$Bi | 252 (18) | (13/2+) | 3.2(+76−20) $\mu s$ | 252 | | %IT = 100 | 1984SC13 |
| 73 | 115 | $^{188}$Ta | 291.9+X ? | | 3.5(4) $\mu s$ | 291.9(5)? | | %IT = 100 | 2011ST21<br>2009AL30<br>2005CA02 |
| 74 | 114 | $^{188}$W | 1929.3 (16) | 8− | 109.5(35) $ns$ | 184<br>198 | M1<br>M1 | %IT = 100 | 2010LA16 |
| 75 | 113 | $^{188}$Re | 172.0848 (24) | 6− | 18.59(4) $min$ | (2.636(3))<br>15.93(10) | (M3)<br>M3 | %IT = 100 | 2010BA48<br>1989AB18 |
| 75 | 113 | $^{188}$Re | 182.7594 (24) | (4−) | 20.3(18) $ns$ | (13.310(3))<br>(26.704(3)) | | | 1978SC10 |
| 75 | 113 | $^{188}$Re | 230.9213 (12) | 3+ | 21.2(15) $ns$ | 61.1(3)<br>74.864(3)<br>167.3258(6) | <br>(E1)<br>E1 | %IT = 100 | 1978SC10 |
| 76 | 112 | $^{188}$Os | 1771.0 (4) | 7− | 14.0(2) $ns$ | 102.4(5)<br>830.7(5) | E2 | %IT = 100 | 2009MO05<br>1984GO06<br>1978SH21 |
| 76 | 112 | $^{188}$Os | 2144.2 (7) | (10−) | 12.3(1) $ns$ | 89.2(5) | (M1) | | 2009MO05 |
| 77 | 111 | $^{188}$Ir | 923.53+X | (11−) | 4.2(2) $ms$ | | | | 2008JU02<br>1984KR18<br>1975AN08<br>1971GO21 |
| 77 | 111 | $^{188}$Ir | 2642.64+X | (18−) | 12.27(14) $ns$ | 88.3(1) | (M1) | %IT = 100 | 2008JU02 |

*Continued...*

Table 1 contd...

| Z | N | $^{A}X$ | E(keV) | $J^{\pi}$ | $T_{1/2}$ | Eγ(keV) | λ | Decay mode | Reference |
|---|---|---|---|---|---|---|---|---|---|
| 80 | 108 | $^{188}$Hg | 2724.3 (4) | (12+) | 134(15) *ns* | 61.6(7)<br>233.3(3) | | %IT = 100 | 1983HA15 |
| 81 | 107 | $^{188}$Tl$^{\dagger}$ | 0+X<br>0+Y | (2−)<br>(7+) | 71(2) *s*<br>71(1) *s* | | | %ϵ+%β+ = 100<br>%ϵ+%β+ = 100 | 1984CO17<br>1984CO17 |
| 81 | 107 | $^{188}$Tl | 184.6+X | (1+,3+) | 34 *ns* | 184.6(3) | E1 | %IT = 100 | 1991VA04 |
| 81 | 107 | $^{188}$Tl | 268.8+Y | (9−) | 41(4) *ms* | 268.8(2) | M2 | %IT ≈ 100<br>%ϵ+%β+ = ? | 1981KR20 |
| 82 | 106 | $^{188}$Pb | 2577.2 (4) | 8− | 820(60) *ns* | 103.0(3)<br>129(1)<br>278.2(3)<br>360.2(3)<br>709.9(3) | M1<br>E2<br>E1<br>E1 | %IT = 100 | 2010IO01<br>2004DR04<br>1999DR10 |
| 82 | 106 | $^{188}$Pb | 2701.6 (5) | 11− | 27(5) *ns* | 335.4(3) | E1 | %IT = 100 | 2010IO01<br>2004DR04<br>1999DR10 |
| 82 | 106 | $^{188}$Pb | 2709.8 (5) | 12+ | 99(10) *ns* | 343.5(3) | E2 | %IT = 100 | 2010IO01<br>2004DR04<br>1999DR10 |
| 82 | 106 | $^{188}$Pb | 4783.4 (7) | (19−) | 0.44(6) *µs* | 217.8(3) | (E2) | %IT = 100 | 2004DR04 |
| 83 | 105 | $^{188}$Bi | 0+X | (10−) | 265(15) *ms* | | | %α ≈ 100<br>%β+F ≈ 0.046(9)<br>%β+ = ? | 2020AN12<br>2006AN04<br>2003AN26 |
| 83 | 105 | $^{188}$Bi | 0+Y* | | 0.25(5) *µs* | | | | 2022ZH46 |
| 83 | 105 | $^{188}$Bi | 65 (25) | | > 5 *µs* | | | %α = 100 | 2006AN04 |
| 73 | 116 | $^{189}$Ta | X | | 0.58(22) *µs* | 153.9(5)?<br>283.7(5)?<br>342.5(5)?<br>388.7(5)?<br>481.6(5)? | | | 2011ST21<br>2009AL30 |
| 75 | 114 | $^{189}$Re | 1692.9 (5) | (25/2−) | 51(17) *ns* | (14.0)<br>102.6 | | %IT = 100 | 2016RE02 |
| 75 | 114 | $^{189}$Re | 1770.9 (6) | (29/2+) | 223(14) *µs* | 78 | (M2) | %IT = 100 | 2016RE02 |
| 76 | 113 | $^{189}$Os | 30.812 (15) | 9/2− | 5.81(6) *h* | 30.810(21) | M3+E4 | %IT = 100 | 2000AH03<br>1963PR12<br>1958SC30 |
| 77 | 112 | $^{189}$Ir | 94.34 (3) | 1/2+ | 11.4(3) *ns* | 94.34(4) | M1+E2 | %IT = 100 | 1972BA21<br>1969HA03 |
| 77 | 112 | $^{189}$Ir | 372.17 (4) | 11/2− | 13.3(3) *ms* | 71.69(4)<br>258.37(6) | M2(+E3)<br>E3 | %IT = 100 | 1973ROYQ<br>1967CO20<br>1963RE13 |
| 77 | 112 | $^{189}$Ir | 2333.8 (3) | (25/2)+ | 3.7(2) *ms* | 84.5(3)<br>224(1)<br>247.6(3) | (E1+M2)<br>(M2)<br>(E1) | %IT = 100 | 1975KE06<br>1975AN08 |
| 78 | 111 | $^{189}$Pt | 172.80 (6) | 9/2− | 464(25) *ns* | 166.40(5) | E2 | %IT = 100 | 1970FI16 |
| 78 | 111 | $^{189}$Pt | 191.5 (7) | (13/2+) | 143(5) *µs* | (18.7(9)) | | %IT = 100 | 1976PI03 |
| 79 | 110 | $^{189}$Au | 9.94 (11) | 3/2+ | 30(4) *ns* | 9.9(2) | | %IT = 100 | 1975BE17 |

*Continued...*

*: Isomer above 0+X, (10−) isomer; 58-, 81-, 143- and 243-keV γ rays associated with its decay.

Table 1 contd...

| Z | N | $^{A}X$ | E(keV) | J$^{\pi}$ | $T_{1/2}$ | E$\gamma$(keV) | $\lambda$ | Decay mode | Reference |
|---|---|---|---|---|---|---|---|---|---|
| 79 | 110 | $^{189}$Au | 247.25 (16) | 11/2− | 4.59(11) *min* | | | %$\epsilon$+%$\beta$+ = 100 | 1970FI16<br>1967HE06<br>1967AL17<br>1966FO13 |
| 79 | 110 | $^{189}$Au | 325.12 (16) | 9/2− | 190(15) *ns* | 77.9(2) | M1+E2 | %IT = 100 | 1975BE17 |
| 79 | 110 | $^{189}$Au | 2554.8 (8) | 31/2+ | 242(10) *ns* | 38.8(3) | (E2) | | 1997PE26 |
| 80 | 109 | $^{189}$Hg | 80 (30) | 13/2+ | 8.6(2) *min* | | | %$\epsilon$+%$\beta$+ = 100<br>%$\alpha$ < 0.00003 | 2001SC41<br>1979DA06<br>1975BE17<br>1970ERZX<br>1967NA02<br>1966FO13 |
| 81 | 108 | $^{189}$Tl | 281 (7) | (9/2−) | 1.4(1) *min* | | | %$\epsilon$+%$\beta$+ = 98(2)<br>%IT = 2(2) | 1976HA25<br>1974HA10 |
| 82 | 107 | $^{189}$Pb | 40 (4) | (13/2+) | 50(3) *s* | | | %$\epsilon$+%$\beta$+ = 99.5(5)<br>%$\alpha$ = 0.5(5) | 2013SA43<br>2009SA09 |
| 82 | 107 | $^{189}$Pb | 2474.5 (40) | (31/2−) | 22.2(+69−14) $\mu s$ | 193.9(3)<br>336.7(1)<br>609.3(3) | M2<br>E3<br>(E3) | %IT = 100 | 2013SA43<br>2005BA51 |
| 83 | 106 | $^{189}$Bi | 184 (8) | (1/2+) | 5.0(1) *ms* | (84) | | %$\alpha$ = 83(5)<br>%IT = 17(5) | 2007DOZW<br>2003KE08<br>2001AN11<br>1999AN52 |
| 83 | 106 | $^{189}$Bi | 357.6 (5) | (13/2+) | 880(5) *ns* | 357.6(5) | M2 | %IT ≈ 100 | 2007DOZW<br>2003KE08<br>2002HU14<br>2001AN11 |
| 73 | 117 | $^{190}$Ta | 175+X | | 42(7) *ns* | 175 | | %IT ≈ 100 | 2009AL30 |
| 74 | 116 | $^{190}$W | 1742.0 (20) | 8+ | 111(17) *ns* | 102<br>694 | M1<br>E2 | %IT = 100 | 2010LA16 |
| 74 | 116 | $^{190}$W | 1839.0 (22) | 10− | 166(6) $\mu s$ | 97 | M2 | %IT = 100 | 2011ST21*<br>2010LA16 |
| 74 | 116 | $^{190}$W | X | | 0.35(4) $\mu s$ | | | | 2011ST21 |
| 74 | 116 | $^{190}$W | 2394.5 (14) | (10−) | 0.06(+150−3) *ms* | 58.5(5) | (E1) | %IT = 100 | 2005CA02 |
| 75 | 115 | $^{190}$Re | 204 (10) | (6−) | 3.2(2) *h* @ | | | %IT = 45.6(20)<br>%$\beta$−= 54.4(20) | 2012RE19@<br>1974YA02<br>1972RU06 |
| 76 | 114 | $^{190}$Os | 1705.4 (2) | (10)− | 9.86(3) *min* | 38.9(1) | M2+E3 | %IT = 100 | 2012KR05<br>1958SC30 |
| 77 | 113 | $^{190}$Ir | 26.1 (1) | (1−) | 1.120(3) *h* | 26.3(1) | M3 | %IT = 100 | 1996GA30<br>1964HA06 |
| 77 | 113 | $^{190}$Ir | 36.154 (25) | (4)+ | > 2 $\mu s$ | 36.155(25) | E1 | %IT = 100 | 1996GA30 |
| 77 | 113 | $^{190}$Ir | 148.696+X | | 90(3) *ns* | 148.696(20) | | %IT = 100 | 2000GA03 |
| 77 | 113 | $^{190}$Ir | 376.4 (1) | (11)− | 3.087(12) *h* | 148.7(1) | M4 | %IT = 8.6(2)<br>%$\epsilon$+%$\beta$+ = 91.4(2) | 1996GA30<br>1970BO22<br>1964HA06<br>1963GR22<br>1950CH11 |

*Continued...*

*: 2011ST21 reports lower half-life of 108(9) $\mu s$.

Table 1 contd...

| Z | N | $^{A}$X | E(keV) | J$^{\pi}$ | $T_{1/2}$ | E$\gamma$(keV) | $\lambda$ | Decay mode | Reference |
|---|---|---|---|---|---|---|---|---|---|
| 78 | 112 | $^{190}$Pt | 2297.45 (17) | (10−) | 48(5) *ns* | 75.0(5)<br>219.14(14) | <br>(E2) | %IT = 100 | 2014LI21<br>1976HJ01<br>1976CU02 |
| 79 | 111 | $^{190}$Au | 0+X | (11−) | 125(20) *ms* | | | %IT ≈ 100 | 1982NE05 |
| 80 | 110 | $^{190}$Hg | 2620.7 (5) | (12+) | 23(1) *ns* | 23.9(5) | | %IT = 100 | 1980HJ01<br>1975LI16<br>1972IN02 |
| 81 | 109 | $^{190}$Tl | 83 (10) | 7+ | 3.6(3) *min* | | | %$\epsilon$+%$\beta$+ = 100 | 2019GH11<br>2014BO26<br>2013ST25<br>1976BI09 |
| 81 | 109 | $^{190}$Tl | 151.3 (3) | 1+, 2+,<br>3+ | > 34 *ns* | 151.3 | (E1) | %IT = 100 | 1991VA04 |
| 81 | 109 | $^{190}$Tl | 325.2 (5) | (9−) | >1 $\mu s$ | | | %IT = 100 | 1991VA04 |
| 82 | 108 | $^{190}$Pb | 2614.8 (8) | (10)+ | 150 *ns* | 338.6<br>362.9 | E2 | %IT = 100 | 1998DR06 |
| 82 | 108 | $^{190}$Pb | 2615+X | (12+) | 25 $\mu s$ | | | | 1998DR06 |
| 82 | 108 | $^{190}$Pb | 2658.2 (8) | (11)− | 7.2(6) $\mu s$ | 43.2<br>382<br>406.5 | <br>(E3)<br>E3 | %IT = 100 | 2001DR05 |
| 82 | 108 | $^{190}$Pb | 4516.8 (11) | (16+) | ~ 14 *ns* | 188.7 | | %IT = 100 | 1998DR06 |
| 83 | 107 | $^{190}$Bi | 121.5 (3) | (5−) | 175(8) *ns* | 76.5(3) | E1 | %IT ≈ 100 | 2009AN11 |
| 83 | 107 | $^{190}$Bi | 191 (65) | (10−) | 6.2(1) *s* | | | %$\alpha$ = 70(9)<br>%$\epsilon$+%$\beta$+ = 30(9)<br>%$\beta$+F=0.000013<br>(+39−11) | 2020ST11<br>2020AN12<br>2001AN11<br>1988HU03 |
| 83 | 107 | $^{190}$Bi | 465.0 (5) | (–) | 1.0(+10–5) $\mu s$ | 274.0(5) | (E2+M1) | %IT ≈ 100 | 2009AN11<br>2001AN11 |
| 74 | 117 | $^{191}$W | X | | 0.34(2) $\mu s$ | 67.5(5)?<br>167.4(5)? | | %IT = 100 | 2011ST21<br>2009AL30 |
| 75 | 116 | $^{191}$Re | 1507.6 (3) | 21/2+ | 70(33) *ns* | 157.9<br>418.9 | (E1)<br>(E1) | %IT = 100 | 2016RE02<br>2011ST21 |
| 75 | 116 | $^{191}$Re | 1601.5 (4) | 25/2− | 50.6(35) $\mu s$ | 93.8<br>115.9 | (M2)<br>(E2) | %IT = 100 | 2016RE02<br>2011ST21 |
| 75 | 116 | $^{191}$Re | 1678.6 (4) | 23/2+ | 33.3(28) *ns* | 171.1<br>192.8 | <br>(E1) | %IT = 100 | 2016RE02 |
| 76 | 115 | $^{191}$Os | 74.382 (3) | 3/2− | 13.10(5) *h* | 74.379(9) | M3+E4 | %IT = 100 | 1975CA03 |
| 76 | 115 | $^{191}$Os | 2640 | (29/2,<br>31/2) | 61(4) *ns* | 453 | | %IT = 100 | 2004VA03 |
| 77 | 114 | $^{191}$Ir | 171.29 (4) | 11/2− | 4.899(23) *s* | 41.89(4) | E3 | %IT = 100 | 1972JO05<br>1970JO16<br>1967AB09<br>1963KA34 |
| 77 | 114 | $^{191}$Ir | 2101.1 (9) | 31/2(+) | 5.8(6) *s* | (54.0(10))<br>308.5(5) | <br>M3 | %IT = 100 | 2012DR02 |
| 78 | 113 | $^{191}$Pt | 100.663 (20) | (9/2)− | > 1 $\mu s$ | 91.11(2) | E2 | %IT = 100 | 1976PI03 |
| 78 | 113 | $^{191}$Pt | 149.035 (22) | (13/2)+ | 95(5) $\mu s$ | 48.37(1) | M2 | %IT = 100 | 1976PI03 |
| 79 | 112 | $^{191}$Au | 11.1 (5) | (1/2+) | 15.5(15) *ns* | 11.2(6) | | %IT = 100 | 1986BE07 |
| 79 | 112 | $^{191}$Au | 266.2 (7) | (11/2−) | 0.92(11) *s* | 13.7(6) | (E3) | %IT = 100 | 1971BE61 |
| 79 | 112 | $^{191}$Au | 540.4 (8) | (9/2−) | 10(2) *ns* | 274.1(3) | M1+E2 | %IT = 100 | 1979GO15 |

*Continued...*

Table 1 contd…

| Z | N | $^{A}X$ | E(keV) | J$^{\pi}$ | $T_{1/2}$ | E$\gamma$(keV) | $\lambda$ | Decay mode | Reference |
|---|---|---|---|---|---|---|---|---|---|
| 79 | 112 | $^{191}$Au | 2489.6 (9) | (31/2+) | 402(20) *ns* | 67.0(3) | (E2) | %IT = 100 | 1997PE26<br>1985KO13 |
| 80 | 111 | $^{191}$Hg | 128 (8) | 13/2(+) | 50.8(15) *min* | | | $\%\epsilon$+$\%\beta$+ = 100 | 2001SC41<br>1971BE61 |
| 81 | 110 | $^{191}$Tl | 0+X | 9/2(−) | 5.22(16) *min* | | | $\%\epsilon$+$\%\beta$+ = 100<br>$\%\epsilon$ = 98.2(+6−17)<br>$\%\beta$+ = 1.8(+17−6) | 1974VA19 |
| 82 | 109 | $^{191}$Pb | 55 (12) | (13/2+) | 2.18(8) *min* | | | $\%\alpha \approx$ 0.02<br>$\%\epsilon$+$\%\beta$+ = 100 | 2017AL34<br>1998FO02<br>1981MI11 |
| 82 | 109 | $^{191}$Pb | 2417.7+X | (29/2−) | 15(4) *ns* | 200.1(6) | E2 | %IT = 100 | 1999LA06 |
| 82 | 109 | $^{191}$Pb | 2440.7+X | (27/2−,<br>29/2−) | 17(4) *ns* | ($\approx$ 23.6)<br>149.19(5) | <br>E1 | %IT = 100 | 1999LA06 |
| 82 | 109 | $^{191}$Pb | 2602.31+X | (33/2+) | 0.15(+10−5) $\mu s$ | 106.85(15) | E2 | %IT = 100 | 1999LA06 |
| 83 | 108 | $^{191}$Bi | 240 (4) | (1/2+) | 125(8) *ms* | (92(4)) | (E3) | $\%\alpha$ = 68(5)<br>$\%\epsilon$+$\%\beta$+ = ?<br>%IT = 32(5) | 2003KE04<br>1999AN36<br>1999TA20<br>1985CO06 |
| 83 | 108 | $^{191}$Bi | 429.7 (5) | (13/2+) | 562(10) *ns* | 429.7(5) | M2 | %IT = 100 | 2004NI06<br>2001NI04 |
| 84 | 107 | $^{191}$Po | 40 (15) | (13/2+) | 93(3) *ms* | | | $\%\alpha$ = 96(4) | 2002AN19 |
| 85 | 106 | $^{191}$At | 55 (30) | (7/2−) | 2.1(+4−3) *ms* | | | $\%\alpha \approx$ 100 | 2003KE08 |
| 75 | 117 | $^{192}$Re | X | | 85(10) $\mu s$ | 159.3(5)? | | %IT = 100 | 2011ST21 |
| 75 | 117 | $^{192}$Re | 267 (10) | | 61(+40−20) *s* @ | 267(10) | | %IT = 100 | 2012RE19@ |
| 76 | 116 | $^{192}$Os | 2015.40 (11) | 10− | 5.9(1) *s*<br>10.5(+10−9) *s* @ | 302.48(6)<br>307.02(9) | (E3)<br>(M2) | %IT = 93.5(65)<br>$\%\beta$−= 6.5(65) | 2013DR05<br>1979KAYT<br>1973PA21<br>2015AK02@ |
| 76 | 116 | $^{192}$Os | 4113.8 (20) | (16+) | 0.19(10) $\mu s$ | 446 | E2 | %IT = 100 | 2004VA03<br>2004RE11 |
| 76 | 116 | $^{192}$Os | 4580.3 (1) | 20+ | 205(7) *ns* | | | | 2013DR05 |
| 77 | 115 | $^{192}$Ir | 56.720 (5) | 1− | 1.45(5) *min* | 56.71(3) | E3 | %IT = 99.9825<br>$\%\beta$−= 0.0175 | 1954WE10<br>1953WE02 |
| 77 | 115 | $^{192}$Ir | 66.830 (20) | (4)− | 15(4) *ns* | 50.780(10)?<br>66.83(2) | E2 | %IT = 100 | 1997BAZV<br>1994GA05<br>1991KE10 |
| 77 | 115 | $^{192}$Ir | 104.776 (5) | (1)− | 17.4(26) *ns* | 48.0568(8) | M1+E2 | %IT = 100 | 1991KE10 |
| 77 | 115 | $^{192}$Ir | 118.7824 (18) | 3− | > 15 *ns* | 34.520(10)<br>118.7817(18) | M1<br>E1 | %IT = 100 | 1991KE10 |
| 77 | 115 | $^{192}$Ir | 168.14 (12) | 11− | 241(9) *y* | 155.16(12) | (E5) | %IT = 100 | 2015BA01<br>1970HA32 |
| 78 | 114 | $^{192}$Pt | 2172.37 (13) | (10)− | 272(23) *ns* | 69.12(10)<br>207.93(15) | M1<br>(E2) | %IT = 100 | 2004VA03<br>2004RE11<br>1976CU02<br>1976HJ01 |
| 79 | 113 | $^{192}$Au | 135.41 (25) | (5)+ | 29 *ms* | 62.8(3)<br>103.8(3) | M2<br>E3 | %IT = 100 | 1980ROZT<br>1976ROZE |
| 79 | 113 | $^{192}$Au | 431.6 (5) | 11− | 160(20) *ms* | 59.8(3) | E3 | %IT = 100 | 2012OK04<br>1982NE05 |
| 80 | 112 | $^{192}$Hg | 2535.5 (4) | (12+) | 11.1(5) *ns* | 28.4 | (E2) | %IT = 100 | 1983GU05 |

*Continued…*

Table 1 contd...

| Z | N | $^{A}$X | E(keV) | J$^{\pi}$ | $T_{1/2}$ | E$\gamma$(keV) | $\lambda$ | Decay mode | Reference |
|---|---|---|---|---|---|---|---|---|---|
| 81 | 111 | $^{192}$Tl | 138 (45) | (7+) | 10.8(2) *min* | | | $\%\epsilon$+$\%\beta$+ = 100 | 1981SO09 |
| | | | | | | | | | 1975VA20 |
| 81 | 111 | $^{192}$Tl | 388 (45) | (8−) | 296(5) *ns* | 250.6(2) | (E1) | %IT = 100 | 1982DA17 |
| 82 | 110 | $^{192}$Pb | 2581.1 (4) | (10)+ | 166(6) *ns* | (19.0) | | %IT = 100 | 2007IO03 |
| | | | | | | 60.8(4) | (E2) | | 2003WIZZ |
| | | | | | | 66.9(4) | E1 | | 2001DR05 |
| | | | | | | 277.3(6) | E2 | | |
| 82 | 110 | $^{192}$Pb | 2624.0 (8) | (12+) | 1.09(4) $\mu s$ | 44.0(10) | (E2) | %IT = 100 | 2010KM01 |
| | | | | | | | | | 1985ST16 |
| | | | | | | | | | 1983ST16 |
| 82 | 110 | $^{192}$Pb | 2743.5 (4) | (11)− | 0.756(14) $\mu s$ | 120.6 | | %IT = 100 | 2007LO03 |
| | | | | | | 162.5(3) | E1 | | 2001DR05 |
| | | | | | | 229 | | | |
| | | | | | | 439.7(3) | (E3) | | |
| 83 | 109 | $^{192}$Bi | 147 (34) | (10−) | 39.6(4) *s* | | | $\%\epsilon$+$\%\beta$+ = 90(3) | 1988HU03 |
| | | | | | | | | $\%\alpha$ = 10(3) | |
| 84 | 108 | $^{192}$Po | 2294.6 | (11−) | 0.58(10) $\mu s$ | 153.9(3) ? | (E1) | %IT = 100 | 2016AN10 |
| | | | | | | 733.1(4)? | (E3) | | 2003VA16 |
| 85 | 107 | $^{192}$At$^{\dagger}$ | 0+X | (9−, | 88(6) *ms* | | | $\%\alpha$ = 99.58(9)* | 2013AN03 |
| | | | | 10−) | | | | $\%\beta$+F = 0.42(9) | 2006AN04 |
| | | | | | | | | | 1973TA30 |
| | | | 0+Y | | 11.5(6) *ms* | | | $\%\alpha$ = 99.58(9)* | 2013AN03 |
| | | | | | | | | $\%\beta$+F = 0.42(9) | 2006AN04 |
| 75 | 118 | $^{193}$Re | 146.1+X (3) | (9/2−) | 69(8) $\mu s$ | 146.1(3) | | %IT = 100 | 2011ST21 |
| | | | | | | | | | 2009AL30 |
| | | | | | | | | | 2009AL16 |
| 76 | 117 | $^{193}$Os | 315.6 (3) | (9/2−) | 121(28) *ns* | 242.7(3) | | %IT = 100 | 2014GA14 |
| | | | | | | | | | 2011ST21 |
| 77 | 116 | $^{193}$Ir | 80.239 (6) | 11/2− | 10.53(4) *d* | 80.234(7) | M4 | %IT = 100 | 1987LI16 |
| 77 | 116 | $^{193}$Ir | 2277.4 (7) | 31/2+ | 124.8(21) $\mu s$ | 226.7(5) | M2 | %IT = 100 | 2012DR02 |
| | | | | | | 334.5(5) | (E3) | | |
| | | | | | | 385.0(5) | | | |
| 78 | 115 | $^{193}$Pt | 149.78 (4) | 13/2+ | 4.33(3) *d* | 135.50(3) | M4 | %IT = 100 | 1949WI08 |
| 79 | 114 | $^{193}$Au | 290.20 (4) | 11/2− | 3.9(3) *s* | 32.21(3) | E3 | %IT = 99.97 | 1955FI30 |
| | | | | | | 289.8 | | $\%\epsilon$+$\%\beta$+ ≈ 0.03 | |
| 79 | 114 | $^{193}$Au | 1947.10 (25) | (21/2)+ | 10.4(8) *ns* | 528.0(3) | E1 | %IT = 100 | 1985KO13 |
| | | | | | | 1249.3(3) | (E3) | | |
| 79 | 114 | $^{193}$Au | 2486.7 (6) | (31/2+) | 150(50) *ns* | 161.8(3) | | %IT = 100 | 1985KO13 |
| 80 | 113 | $^{193}$Hg | 140.76 (5) | 13/2(+) | 11.8(2) *h* | 101.25(4) | M4 | %IT = 7.2(5) | 1974VIZS |
| | | | | | | | | $\%\epsilon$+$\%\beta$+ = 92.8(5) | |
| 81 | 112 | $^{193}$Tl | 373 (9) | (9/2−) | 2.11(15) *min* | | | $\%\epsilon$+$\%\beta$+ ≥ 25 | 2014BO26 |
| | | | | | | | | %IT ≤ 75 | 1963DI10 |
| 82 | 111 | $^{193}$Pb | 95 (28) | (13/2+) | 5.8(2) *min* | | | $\%\epsilon$+$\%\beta$+ = 100 | 2017AL34 |
| | | | | | | | | | 1976HA25 |
| 82 | 111 | $^{193}$Pb | 1585.9+X | (21/2−) | 20.5(4) *ns* | (66.5) | | %IT = 100 | 2004IO01 |
| | | | | | | 184.0(4) | E1+M2 | | |
| 82 | 111 | $^{193}$Pb | 2612.5+X | (33/2+) | 180(15) *ns* | 85.6(4) | (E2) | %IT = 100 | 2004IO01 |

*Continued...*

*: The branching ratios are combined for both the isomers.

Table 1 contd…

| Z | N | $^{A}$X | E(keV) | J$^{\pi}$ | $T_{1/2}$ | Eγ(keV) | λ | Decay mode | Reference |
|---|---|---|---|---|---|---|---|---|---|
| 83 | 110 | $^{193}$Bi | 308 (7) | 1/2+ | 3.20(20) *s* | | | %α = 84(16)<br>%ϵ+%β+ = 16(16) | 2015HE27<br>2005UU02<br>1985CO06<br>1974LE02 |
| 83 | 110 | $^{193}$Bi | 605.53 (18) | 13/2+ | 153(10) *ns* | 604.7(3) | M2 | %IT = 100 | 2015HE27<br>2004NI06 |
| 83 | 110 | $^{193}$Bi | 2349.6 (6) | 29/2+ | 85(3) *µs* | 84.0(6) | E2 | %IT = 100 | 2015HE27<br>2004NI06 |
| 83 | 110 | $^{193}$Bi | 2405.1 (7) | (29/2−) | 3.02(8) *µs* | 48.8(6) | (E2) | %IT = 100 | 2015HE27<br>2004NI06 |
| 84 | 109 | $^{193}$Po | 95 (7) | (13/2+) | 244(10) *ms* | | | %α ≤ 100 | 2013SA43<br>1996EN02<br>1995MO14<br>1993WA04<br>1981LE23<br>1977DE32 |
| 85 | 108 | $^{193}$At | 5 (10) | (7/2−) | 21(5) *ms* | | | %α ≈ 100 | 2003KE08 |
| 85 | 108 | $^{193}$At | 39 (7) | (13/2+) | 27(+4−3) *ms* | | | %α = 24(10)<br>%IT = 76(10) | 2003KE08 |
| 75 | 119 | $^{194}$Re | X | | 45(18) *µs* | | | %IT ≈ 100 | 2011ST21<br>2005CA02 |
| 75 | 119 | $^{194}$Re | 285 (40) | (11−) | 25(8) *s* @ | | | %β−≈ 100 | 2012RE19@<br>2012AL05 |
| 75 | 119 | $^{194}$Re | 833 (33) | | 100(10) *s* @ | | | %β−≈ 100 | 2012RE19@<br>2012AL05 |
| 77 | 117 | $^{194}$Ir | 147.072 (2) | 4+ | 31.85(24) *ms* | 34.829(10)<br>62.793(3) | M2<br>M2 | %IT = 100 | 1972BR53<br>1968LU01<br>1962FE02<br>1961AL21<br>1959CA13 |
| 77 | 117 | $^{194}$Ir | 190+X | (10, 11) | 171(11) *d* | | | %β−= 100 | 1968SU02 |
| 79 | 115 | $^{194}$Au | 107.4 (5) | 5+ | 600(8) *ms* | 26.9(5) | (M2) | %IT = 100 | 2012GA46<br>1975YA14 |
| 79 | 115 | $^{194}$Au | 475.8 (6) | 11− | 420(10) *ms* | 69.0(3) | (E3) | %IT = 100 | 2012GA46<br>1982NE05<br>1975YA14<br>1953HE57 |
| 81 | 113 | $^{194}$Tl | 260 (15) | (7+) | 32.8(2) *min* | | | %ϵ+%β+ = 100 | 2014BO26<br>2013ST25<br>2013MA03<br>1972AM03<br>1960JU01 |
| 82 | 112 | $^{194}$Pb | 2407.46 (20) | (9)− | 17(3) *ns* | 166.2(1) | E2 | %IT = 100 | 2004VY01<br>2001GU31<br>1985ST16<br>1972AL49 |
| 82 | 112 | $^{194}$Pb | 2437.53 (20) | (8)+ | 17(4) *ns* | 196.2(2)<br>302.5(1) | (E1)<br>E2 | %IT = 100 | 1987VA09 |
| 82 | 112 | $^{194}$Pb | 2581.18 (22) | (10)+ | 17.2(5) *ns* | 55?<br>173.7(1) | <br>E1 | %IT = 100 | 1986PA18 |

*Continued…*

Table 1 contd...

| Z | N | $^{A}$X | E(keV) | J$^{\pi}$ | $T_{1/2}$ | E$\gamma$(keV) | $\lambda$ | Decay mode | Reference |
|---|---|---|---|---|---|---|---|---|---|
| 82 | 112 | $^{194}$Pb | 2628.33 (25) | (12+) | 370(13) *ns* | 46.8(4) | | %IT = 100 | 2009KU03<br>2004GL04<br>1985ST16<br>1977RO15<br>1972AL49 |
| 82 | 112 | $^{194}$Pb | 2933.21 (25) | (11)− | 133(7) *ns* | 304.9(1)<br>352.1(3)<br>495.7(10)<br>526 | E1<br>E1 | %IT = 100 | 2009KU03<br>2005DR11<br>2004VY01<br>1986VA03<br>1986PA18 |
| 83 | 111 | $^{194}$Bi | 145 (50) | (6+, 7+) | 125(2) *s* | | | %$\epsilon$+%$\beta$+ = 100 | 2008WE02<br>1987VA09<br>1976CH30 |
| 83 | 111 | $^{194}$Bi | 161 (8) | (10−) | 115(4) *s* | | | %$\epsilon$+%$\beta$+ = 99.80(7)<br>%$\alpha$ = 0.20(7) | 2019GH11<br>2013ST25<br>1991VA04<br>1988HU03<br>1985HUZY |
| 84 | 110 | $^{194}$Po | 2313.6 (3) | (10−) | 12.9(5) $\mu s$ | (33)<br>248.0(1) | <br>(M2) | %IT = 100 | 2016AN10<br>2001JU09<br>1999HE32 |
| 85 | 109 | $^{194}$At$^{\dagger}$ | 0+X | (4−,<br>5−) | 286(7) *ms* | | | %$\alpha$ ≈ 99.2<br>%$\beta$+F ≈ 0.8 | 2013AN03<br>2013NY01<br>2013UU01<br>2013KA16<br>2009NA11<br>1995LE15<br>1984YAZY |
| | | | 0+Y | (9−,<br>10−) | 323(7) *ms* | | | %$\alpha$ ≈ 99.2<br>%$\beta$+F ≈ 0.8 | 2013AN03<br>2013NY01<br>2009NA11<br>1995LE15 |
| 76 | 119 | $^{195}$Os | 454 (10) | (13/2+) | > 9 *min* @<br>47(3) *s* | 148.8 | E3 | %IT > 0.0<br>%$\beta$−= ? | 2012RE19@<br>2020WA12 |
| 76 | 119 | $^{195}$Os | 1464.7 (4)? | (15/2+) | 26(4) *ns* | 533.1(2) | | %IT = 100 | 2005CA02 |
| 76 | 119 | $^{195}$Os | 2178.7+X | (25/2+,<br>29/2−) | 34.0(23) *ns* | 714.0(5) | | %IT = 100 | 2011ST21 |
| 77 | 118 | $^{195}$Ir | 100 (5) | 11/2− | 3.67(8) *h* | (100(5)) | (M4) | %IT = 5(5)<br>%$\beta$−= 95(5) | 1973JA10<br>1968HO01<br>1968JA06 |
| 77 | 118 | $^{195}$Ir | 2354.0 (11) | (27/2+) | 4.4(6) $\mu s$ | 268.4(5) | | %IT = 100 | 2011ST21 |
| 78 | 117 | $^{195}$Pt | 259.077 (23) | 13/2+ | 4.010(5) *d* | (19.8)<br>129.5(2) | <br>M4 | %IT = 100 | 2000MO05 |
| 79 | 116 | $^{195}$Au | 318.58 (4) | 11/2− | 30.5(2) *s* | 56.80(3)<br>318.60(10) | E3<br>M4 | %IT = 100 | 1967FR05<br>1955FI30 |
| 79 | 116 | $^{195}$Au | 2460.84+X | 31/2(−) | 12.89(21) $\mu s$ | | | %IT = 100 | 2013DR01 |
| 80 | 115 | $^{195}$Hg | 176.07 (4) | 13/2+ | 41.6(2) *h* | 122.78(3) | M4 | %IT = 54.2(20)<br>%$\epsilon$+%$\beta$+ = 45.8(20) | 2015DO01<br>1973VI09 |
| 81 | 114 | $^{195}$Tl | 482.63 (17) | 9/2− | 3.6(4) *s* | 98.97(12) | E3 | %IT = 100 | 1963DI10 |
| 82 | 113 | $^{195}$Pb | 202.9 (7) | 13/2+ | 15.0(12) *min* | | | %$\epsilon$+%$\beta$+ = 100 | 1982HI04 |

*Continued...*

Table 1 contd...

| Z | N | $^{A}$X | E(keV) | J$^{\pi}$ | $T_{1/2}$ | E$\gamma$(keV) | $\lambda$ | Decay mode | Reference |
|---|---|---|---|---|---|---|---|---|---|
| 82 | 113 | $^{195}$Pb | 1759.0 (7) | 21/2− | 10.0(7) $\mu s$ | (5.1)<br>586.5(2) | <br>E3(+M2) | %IT = 100 | 1981HE07<br>1977HE06 |
| 82 | 113 | $^{195}$Pb | 2901.7 (8) | 33/2+ | 95(20) $ns$ | 86.8(3) | E2 | %IT = 100 | 1982ALZQ |
| 83 | 112 | $^{195}$Bi | 401 (7) | (1/2+) | 87(1) $s$ | | | %$\epsilon$+%$\beta$+ = 67(17)<br>%$\alpha$ = 33(17) | 1985CO06 |
| 83 | 112 | $^{195}$Bi | 886.7 (1) | 13/2+ | 32(2) $ns$ | 886.7(1) | M2 | %IT = 100 | 1986LO05 |
| 83 | 112 | $^{195}$Bi | 2194.3 (3) | 23/2+ | 80(10) $ns$ | 150.7(2) | E2 | %IT = 100 | 1986LO05 |
| 83 | 112 | $^{195}$Bi | 2380.1 (6) | (29/2−) | 0.614(5) $\mu s$ | 46.0(6) | (E2) | %IT = 100 | 2017HE12<br>2015RO20<br>1986LO05 |
| 83 | 112 | $^{195}$Bi | 2615.9 (5) | 29/2+ | 1.490(10) $\mu s$ | 235.8(4)<br>457.1(4) | (E1)<br>E2 | %IT = 100 | 2017HE12 |
| 83 | 112 | $^{195}$Bi | 3335.0 (7) | (31/2−) | 1.6(1) $\mu s$ | 238.0(4) | (M2) | %IT = 100 | 2015RO20 |
| 84 | 111 | $^{195}$Po | 150 (10) | (13/2+) | 1.92(2) $s$ | | | %$\epsilon$+%$\beta$+ ≈ 10<br>%$\alpha$ ≈ 90<br>%IT < 0.01 | 2017AL34<br>1993WA04 |
| 85 | 110 | $^{195}$At | 33.0 (10) | (7/2−) | 143(3) $ms$ | (33) | | %$\alpha$ = 88(4)<br>%IT = 12(4) | 2013NY01<br>2003KE04 |
| 86 | 109 | $^{195}$Rn | (59) | 13/2+ | 5(+3−2) $ms$ | | | %$\alpha$ ≈ 100 | 2001KE06<br>2001UU01 |
| 75 | 121 | $^{196}$Re | 0+X | | 3.6(6) $\mu s$ | | | %IT = 100 | 2011ST21 |
| 77 | 119 | $^{196}$Ir | 4.1E+2 (11) | (10,<br>11−) | 1.40(2) $h$ | | | %$\beta$−= 99.85(15)<br>%IT = 0.15(15) | 1968JA06 |
| 79 | 117 | $^{196}$Au | 84.656 (20) | 5+ | 8.1(2) $s$ | 84.66(2) | E3 | %IT = 100 | 1972GLZX<br>1971RO16 |
| 79 | 117 | $^{196}$Au | 595.66 (4) | 12− | 9.605(22) $h$ | 174.91(2) | M4 | %IT = 100 | 2020MO24<br>2019JA03<br>1982HA04<br>1963TI02<br>1963KA24<br>1962CH13<br>1960KA21<br>1960BA63 |
| 81 | 115 | $^{196}$Tl | 394.2 (5) | (7+) | 1.41(2) $h$ | 120.1(3) | M4 | %IT = 3.8(4)<br>%$\epsilon$+%$\beta$+ = 96.2(4) | 1960JU01 |
| 81 | 115 | $^{196}$Tl | 738.1 (6) | (8−) | 21.3(5) $ns$ | 343.9(3) | E1+M2 | %IT = 100 | 1978KR12 |
| 82 | 114 | $^{196}$Pb | 1797.51 (14) | 5− | 140(14) $ns$ | 59.23(9)<br>748.4(3) | E1<br>E3 | %IT = 100 | 1985ST16<br>1973PA03<br>1973DJ01 |
| 82 | 114 | $^{196}$Pb | 2307.84 (18) | 9− | 52(5) $ns$ | 138.41(7) | E2 | %IT = 100 | 1973PA03<br>1973DJ01 |
| 82 | 114 | $^{196}$Pb | 2621.9 (9) | (8+) | 50(15) $ns$ | 198.0(5) | | %IT = 100 | 1987PE13 |
| 82 | 114 | $^{196}$Pb | 2694.6 (3) | 12+ | 270(4) $ns$ | 47.7(5) | (E2) | %IT = 100 | 1989SU12<br>1983ST15<br>1977RO15 |
| 82 | 114 | $^{196}$Pb | 3192.67 (25) | 11− | 72(4) $ns$ | 497.8(2)<br>547.8(2)<br>562<br>884 | E1<br>E1<br>E3<br>E2 | %IT = 100 | 1986VA03 |
| 83 | 113 | $^{196}$Bi | 169 (4) | (7+) | 0.6(5) $s$ | 11(4) | | %IT = ?<br>%$\epsilon$+%$\beta$+ = ? | 1987VA09 |

*Continued...*

Table 1 contd…

| Z | N | $^{A}$X | E(keV) | J$^{\pi}$ | $T_{1/2}$ | E$\gamma$(keV) | $\lambda$ | Decay mode | Reference |
|---|---|---|---|---|---|---|---|---|---|
| 83 | 113 | $^{196}$Bi | 271 (5) | (10−) | 240(3) *s* | 102.0(20) | (E3) | %IT = 25.8(25)<br>%$\epsilon$+%$\beta$+ = 74.2(25)<br>%$\alpha$ = 0.00038(10) | 1987VA09 |
| 84 | 112 | $^{196}$Po | 2493.9 (4) | 11− | 856(17) *ns* | 198<br>552 | <br>E3 | %IT = 100 | 1998CIZY |
| 85 | 111 | $^{196}$At | 157.9 (1) | (5+) | 11(2) $\mu s$ | 157.9(1) | E2 | %IT = 100 | 2000SM06 |
| 76 | 121 | $^{197}$Os | 1735.6+X | (25/2−,<br>27/2−) | 78.2(66) *ns* | 204.4(5) | | %IT = 100 | 2011ST21 |
| 77 | 120 | $^{197}$Ir | 0+X | | 30(8) $\mu s$ | 161.0(5)?<br>278.5(2)?<br>378.8(2)? | | | 2005CA02 |
| 77 | 120 | $^{197}$Ir | 0+Y | | 15(9) $\mu s$ | 458.3(5)?<br>495.0(3)?<br>567.1(3)?<br>609.1(5)? | | | 2005CA02 |
| 77 | 120 | $^{197}$Ir | 115 (5) | 11/2− | 8.9(3) *min* | | | %IT = 0.25(10)<br>%$\beta$−= 99.75(10) | 1978PEZJ |
| 78 | 119 | $^{197}$Pt | 53.088 (19) | 5/2− | 16.58(17) *ns* | 53.10(2) | E2 | %IT = 100 | 1982SO05 |
| 78 | 119 | $^{197}$Pt | 399.59 (20) | 13/2+ | 95.41(18) *min* | 346.5(2) | M4 | %IT = 96.7(4)<br>%$\beta$−= 3.3(4) | 1982SO05 |
| 78 | 119 | $^{197}$Pt | 1753.2+X | (25/2−,<br>27/2−) | 9.5(5) *ns* | | | | 2019WA22<br>2011ST21 |
| 79 | 118 | $^{197}$Au | 409.15 (8) | 11/2− | 7.73(6) *s* | 130.2(1)<br>409.1(1) | E3<br>M4 | %IT = 100 | 1986NE05 |
| 79 | 118 | $^{197}$Au | 2532.5 (1) | (25/2,<br>27/2,<br>29/2) | 150(5) *ns* | 435.6 | | %IT = 100 | 2006WH02 |
| 80 | 117 | $^{197}$Hg | 298.93 (8) | 13/2+ | 23.82(4) *h* | 164.97(7) | M4 | %IT = 94.68(9)<br>%$\epsilon$ +%$\beta$+ = 5.32(9) | 2020LE04<br>1978ME11<br>1966EL09 |
| 81 | 116 | $^{197}$Tl | 608.22 (8) | 9/2− | 0.54(1) *s* | 222.45(5) | E3 | %IT = 100 | 2018HI07<br>1978LI10<br>1975TO08<br>1953HE57 |
| 81 | 116 | $^{197}$Tl | 2529.7 (4) | 19/2− | 18(4) *ns* | 263.1(2) | | %IT = 100 | 2013PA42<br>1978LI10 |
| 82 | 115 | $^{197}$Pb | 319.31 (11) | 13/2+ | 42.9(9) *min* | 234.4(1) | M4 | %IT = 19(2)<br>%$\epsilon$+%$\beta$+ = 81(2) | 1980HI04<br>1979RA04<br>1957AN53 |
| 82 | 115 | $^{197}$Pb | 1914.10 (25) | 21/2− | 1.15(20) $\mu s$ | 32.4(1)<br>57.35(10)<br>589.0(2) | (E1)<br>(E1)<br>M2+E3 | %IT = 100 | 1985ST16 |
| 82 | 115 | $^{197}$Pb | 3168.9 (3) | (33/2)+ | 55(5) *ns* | 88.7(1) | E2 | %IT = 100 | 1985PA22<br>1984ALZA<br>1978RI01 |
| 83 | 114 | $^{197}$Bi | 0+X | (1/2+) | 5.04(16) *min* | | | %$\epsilon$+%$\beta$+ = 45(40)<br>%$\alpha$ = 55(40)<br>%IT < 0.3 | 1991VA09<br>1985CO06 |
| 83 | 114 | $^{197}$Bi | 1600.95 (25) | (17/2+) | 15.3(30) *ns* | 404.7(2) | (E2) | %IT = 100 | 1995ZH36<br>1995ZH56<br>1986CH01 |

*Continued…*

Table 1 contd...

| Z | N | $^{A}X$ | E(keV) | $J^{\pi}$ | $T_{1/2}$ | Eγ(keV) | λ | Decay mode | Reference |
|---|---|---|---|---|---|---|---|---|---|
| 83 | 114 | $^{197}$Bi | 1968+X | (25/2+) | 18.0(31) *ns* | | | | 2005MA51<br>1995ZH36<br>1995ZH56<br>1986CH01 |
| 83 | 114 | $^{197}$Bi | 2064.7 (10) | 25/2+ | 36.7(70) *ns* | 96.9(2) | (E2) | %IT = 100 | 2005MA51<br>1995ZH36<br>1995ZH56 |
| 83 | 114 | $^{197}$Bi | 2088.6 (10) | (25/2+) | 19.3(49) *ns* | 121.1(5) | (E2) | %IT = 100 | 1995ZH36<br>1995ZH56 |
| 83 | 114 | $^{197}$Bi | 2129.3+X | | 204(18) *ns* | | | | 1986CH01 |
| 83 | 114 | $^{197}$Bi | 2357.4 (11) | (27/2+) | 53(21) *ns* | 292.7(5) | (E2) | %IT = 100 | 1995ZH36<br>1995ZH56 |
| 83 | 114 | $^{197}$Bi | 2383.1+X | (29/2−) | 253(39) *ns* | 255.2(5) | | %IT = 100 | 1995ZH36<br>1995ZH56<br>1986CH01 |
| 83 | 114 | $^{197}$Bi | 2929.5 (5) | (31/2−) | 209(30) *ns* | 864.0(2) | (E3) | %IT = 100 | 1986CH01 |
| 84 | 113 | $^{197}$Po | 199 (11) | 13/2+ | 30.5(20) *s* | | | %ϵ+%β+ = 16(9)<br>%α = 84(9) | 2017AL34<br>2014SE07<br>1996TA18<br>1993WA04<br>1982BO04<br>1971HO06<br>1967SI09<br>1967LE21 |
| 85 | 112 | $^{197}$At | 52 (10) | (1/2+) | 2.0(2) *s* | | | %ϵ+%β+ = ?<br>%IT ≤ 0.004<br>%α ≈ 100 | 2014KA23<br>1999SM07 |
| 85 | 112 | $^{197}$At | 310.70 (20) | 13/2+ | 1.3(2) *μs* | 310.7(2) | M2 | %IT = 100 | 2008AN11 |
| 86 | 111 | $^{197}$Rn | 194 (12) | (13/2+) | 25(+3−2) *ms* | | | %α ≈ 100 | 2013SA43<br>2008AN05<br>1996EN02 |
| 76 | 122 | $^{198}$Os | 1680.8+X | (7−) | 16.1(8) *ns* | | | %IT = 100 | 2011ST21<br>2009PO02 |
| 76 | 122 | $^{198}$Os | 3198.6+Y | (12+) | 18.0(28) *ns* | | | %IT = 100 | 2011ST21<br>2009PO02 |
| 77 | 121 | $^{198}$Ir | 116.4 (2)? | | 75(9) *ns* | 116.4(2) | | %IT = 100 | 2011ST21<br>2005CA02 |
| 78 | 120 | $^{198}$Pt | 3017.0 (17) | (12−) | 36(2) *ns* | 104?<br>337 | | %IT = 100 | 2004VA03<br>2004RE11 |
| 79 | 119 | $^{198}$Au | 312.2200 (20) | 5+ | 124(4) *ns* | 97.249(2) | E1 | %IT = 100 | 1975MI05 |
| 79 | 119 | $^{198}$Au | 811.7 (15) | (12−) | 2.272(16) *d* | 115.2(15) | (M4) | %IT = 100 | 1992ZHZG<br>1973PA08<br>1972CU06 |
| 81 | 117 | $^{198}$Tl | 543.6 (4) | 7+ | 1.87(3) *h* | 260.9(3) | M4 | %IT = 44.1(23)<br>%ϵ+%β+ = 55.9(23) | 1960JU01 |
| 81 | 117 | $^{198}$Tl | 686.8 (5) | (5+) | 150(40) *ns* | (13) | | %IT = 100 | 1977KR04 |
| 81 | 117 | $^{198}$Tl | 742.4 (4) | (10−) | 32.1(10) *ms* | 198.8(2) | (E3) | %IT = 100 | 1975SE12 |
| 81 | 117 | $^{198}$Tl | 934.5 (4) | (8)− | 12.3(3) *ns* | 391.0(3) | E1+M2 | %IT = 100 | 1977KR04 |
| 82 | 116 | $^{198}$Pb | 1823.5 (4) | (5)− | 50.4(5) *ns* | 197.6(2)<br>760 | E1 | %IT = 100 | 1987CA23 |
| 82 | 116 | $^{198}$Pb | 2141.4 (4) | (7)− | 4.19(10) *μs* | 317.9(2) | E2 | %IT = 100 | 1987CA23 |

*Continued...*

Table 1 contd. . .

| Z | N | $^{A}$X | E(keV) | J$^{\pi}$ | $T_{1/2}$ | E$\gamma$(keV) | $\lambda$ | Decay mode | Reference |
|---|---|---|---|---|---|---|---|---|---|
| 82 | 116 | $^{198}$Pb | 2231.4 (5) | (9)− | 137(10) *ns* | 90.0(2) | E2 | %IT = 100 | 1989SU12 |
| 82 | 116 | $^{198}$Pb | 2821.7 (6) | (12)+ | 212(4) *ns* | 49.2(5) | E2 | %IT = 100 | 1987CA23 |
| 83 | 115 | $^{198}$Bi | 0+X | (7+) | 11.6(3) *min* | | | %$\epsilon$+%$\beta$+ = 100 | 1992HU04 |
| 83 | 115 | $^{198}$Bi | 248.5+X (5) | (10−) | 7.7(5) *s* | 248.5(5) | (E3) | %IT = 100 | 1992HU04<br>1972HA73 |
| 84 | 114 | $^{198}$Po | 1853.63 (18) | 8+ | 29(2) *ns* | 136.1(2) | E2 | %IT = 100 | 1991AL15<br>1990MA14<br>1986MA31 |
| 84 | 114 | $^{198}$Po | 2565.92 (20) | 11− | 200(20) *ns* | 241.2(2)<br>712.3(1) | E2<br>E3 | %IT = 100 | 1991AL15<br>1990MA14<br>1986MA31 |
| 84 | 114 | $^{198}$Po | 2691.86+X | 12+ | 0.75(5) $\mu s$ | | | | 1990MA14 |
| 85 | 113 | $^{198}$At | 266 (5) | (10−) | 1.23(6) *s* | | | %$\alpha$ = 84(16)<br>%$\epsilon$+%$\beta$+ = 16(16) | 2019GH11<br>2014KA23<br>2013ST25<br>2005UU02 |
| 87 | 111 | $^{198}$Fr$^{\dagger}$ | 0+X | | 15(3) *ms* | | | %$\alpha$ ≈100 | 2013KA16<br>2013UU01* |
| | | | 0+Y | | 1.1(7) *ms* | | | %$\alpha$ ≈100 | 2013KA16 |
| 76 | 123 | $^{199}$Os | 0+X | | 25.2(20) *ns* | | | | 2011ST21 |
| 77 | 122 | $^{199}$Ir | 0+X | | 80-390 *ns* | | | | 2005CA02 |
| 78 | 121 | $^{199}$Pt | 424 (2) | (13/2)+ | 13.5(3) *s* | 391.93(14) | E3 | %IT = 100 | 2018MU19<br>2017HI05<br>1973UR01<br>1959WA15 |
| 78 | 121 | $^{199}$Pt | 1760+X | (25/2−,<br>27/2−) | 18.6(34) *ns* | | | | 2011ST21 |
| 79 | 120 | $^{199}$Au | 548.9405 (21) | (11/2)− | 0.44(3) *ms* | 55.150(19) | M2+E3 | %IT = 100 | 1968BO22 |
| 80 | 119 | $^{199}$Hg | 532.48 (10) | 13/2+ | 42.67(9) *min* | (118.6)<br>374.1(1) | M4+E5 | %IT = 100 | 2001LI17<br>1969KL06<br>1968BO28<br>1955BO29<br>1948MO33 |
| 81 | 118 | $^{199}$Tl | 748.87 (6) | 9/2− | 28.4(2) *ms* | 29(2)<br>381.98(5) | <br>E3 | %IT = 100 | 1977KOZH<br>1967CO20<br>1965GR04<br>1963DI10<br>1963DE38 |
| 82 | 117 | $^{199}$Pb | 424.8+X | (13/2+) | 12.2(3) *min* | 424.8(2) | M4 | %IT ≈ 93<br>%$\epsilon$+%$\beta$+ ≈ 7 | 1999PO13<br>1994BA43<br>1978LEZA<br>1962JU05<br>1956ST05<br>1955AN01 |
| 82 | 117 | $^{199}$Pb | 2559.1+X | (29/2−) | 10.1(2) $\mu s$ | 59.1(3) | E2 | %IT = 100 | 1989SU12<br>1988PA12<br>1988RO08<br>1981HE07 |

*Continued. . .*

*: 2013UU01 suggest a 2− and 6+, 7+ isomers with respective half-lives of 15(+12−5) *ms* and 16(+13−5) *ms*, but do not rule out just one activity.

Table 1 contd…

| Z | N | $^{A}X$ | E(keV) | J$^{\pi}$ | $T_{1/2}$ | E$\gamma$(keV) | $\lambda$ | Decay mode | Reference |
|---|---|---|---|---|---|---|---|---|---|
| 82 | 117 | $^{199}$Pb | 3490.1+X | (33/2+) | 63(4) *ns* | 88.7(2) | E2 | %IT = 100 | 1989SU12<br>1988PA12<br>1988RO08<br>1985ST16<br>1981HE07 |
| 82 | 117 | $^{199}$Pb | 4474.7+X | (41/2+) | 40(10) *ns* | 217.2(3) | (E2) | %IT = 100 | 1988PA12 |
| 83 | 116 | $^{199}$Bi | 667 (4) | (1/2+) | 24.70(15) *min* | (667) | | %$\epsilon$+%$\beta$+ = 99(1)<br>%$\alpha$ ≈ 0.01<br>%IT = 1(1) | 1970DAZM<br>1966MA51<br>1964SI11<br>1950NE77 |
| 83 | 116 | $^{199}$Bi | 1647.5 (3) | 17/2+ | 34.1(24) *ns* | 145.70(20)<br>251.13(25) | E1<br>(E2) | %IT = 100 | 1985PI05 |
| 83 | 116 | $^{199}$Bi | 1922.3+X | (25/2+) | 0.10(3) $\mu s$ | | | %IT = 100 | 1985PI05 |
| 83 | 116 | $^{199}$Bi | 2523.2+X | 29/2− | 168(13) *ns* | (80.0)<br>(87.6)<br>285.16(25)<br>343.4(3)<br>493.2(3) | | %IT = 100 | 2003GL05<br>1985PI05 |
| 84 | 115 | $^{199}$Po | 310 (2) | (13/2+) | 4.17(5) *min* | 238 | M4 | %IT = 2.5(10)<br>%$\epsilon$+%$\beta$+ = 73.5(10)<br>%$\alpha$ = 24(1) | 1996TA18<br>1985ST02<br>1976KO13<br>1971HO01<br>1967LE08<br>1967SI09<br>1967TI04<br>1965BR17 |
| 85 | 114 | $^{199}$At | 244.0 (20) | 1/2+ | 273(9) *ms* | 103(2) | (E3) | %$\alpha$ ≈ 1<br>%IT = 100 | 2014AU03<br>2013JA06 |
| 85 | 114 | $^{199}$At | 572.90 (10) | 13/2+ | 70(20) *ns* | 572.9(1) | M2 | %IT = 100 | 2010JA05<br>2000LA36 |
| 85 | 114 | $^{199}$At | 2293.4 (5) | (29/2+) | 0.80(5) $\mu s$ | 162.9(2) | E2 | %IT = 100 | 2010JA05 |
| 86 | 113 | $^{199}$Rn | 222 (13) | (13/2+) | 0.31(2) *s* | | | %$\alpha$ ≈ 97<br>%$\epsilon$+%$\beta$+ = 3 | 2017AL34<br>2014KA23<br>2005UU02<br>1999TA03<br>1984CA32<br>1982HI14<br>1981EN02 |
| 87 | 112 | $^{199}$Fr* | X | | 6.30(+105−71) *ms* | | | %$\alpha$ ≈ 100 | 2013UU01<br>2013KA16 |
| 87 | 112 | $^{199}$Fr* | Y | | 4.59 (+278−87) *ms* | | | %$\alpha$ ≈ 100 | 2013UU01<br>2013KA16 |
| 78 | 122 | $^{200}$Pt | 1566.9+X | (7−) | 14.2(4) *ns* | | | %IT = 100 | 2011ST21<br>2005CA02<br>1988YA03 |
| 78 | 122 | $^{200}$Pt | 3136.2+X | (12+) | 13.4(13) *ns* | | | %IT = 100 | 2011ST21<br>2005CA02 |
| 79 | 121 | $^{200}$Au | 962 (88) | 12− | 18.7(5) *h* | | | %IT = 16(1)<br>%$\beta$−= 84(1) | 1968SA08 |

*Continued…*

*: Only three events for one isomer and four for the other were assigned in 2013UU01.

Table 1 contd...

| Z | N | $^{A}$X | E(keV) | J$^{\pi}$ | $T_{1/2}$ | E$\gamma$(keV) | $\lambda$ | Decay mode | Reference |
|---|---|---|---|---|---|---|---|---|---|
| 81 | 119 | $^{200}$Tl | 753.60 (24) | 7+ | 34.0(9) *ms* | 212.7(2) | E3 | %IT = 100 | 1967CO20<br>1963DE38<br>1963DI10 |
| 81 | 119 | $^{200}$Tl | 762.00 (24) | 5+ | 390(20) *ns* | 221.1(2) | E1 | %IT = 100 | 2019RO12<br>1981KR03<br>1972IS01 |
| 81 | 119 | $^{200}$Tl | 6007 | (26−) | 57(2) *ns* | 99 | E2 | %IT = 100 | 2019RO12 |
| 82 | 118 | $^{200}$Pb | 2153.8 (3) | 7− | 45.2(10) *ns* | 245.15(13) | E2 | %IT = 100 | 1989SU12<br>1987FA15<br>1978MC03<br>1974LU03<br>1973PA04<br>1972KR08<br>1972IS01 |
| 82 | 118 | $^{200}$Pb | 2183.3 (11) | (9−) | 478(12) *ns**  | 29.5(10) | | %IT = 100 | 2018LA03<br>1989SU12<br>1988PA12<br>1978MC03<br>1974LU03<br>1973PA04 |
| 82 | 118 | $^{200}$Pb | 3005.8 (12) | (12+) | 199(3) *ns* | 45.5(5) | | %IT = 100 | 1989SU12<br>1987FA15<br>1979MA37 |
| 82 | 118 | $^{200}$Pb | 5075.9 (16) | (19−) | 72(3) *ns* | 67.7(8) | E2 | %IT = 100 | 1989SU12<br>1988PA12<br>1987FA15 |
| 82 | 118 | $^{200}$Pb | 6948.3 (17) | (25−) | 58(4) *ns* | 148.3(5) | (E2) | %IT = 100 | 1988PA12 |
| 83 | 117 | $^{200}$Bi | 0+X | (2+) | 31(2) *min* | | | $\%\epsilon+\%\beta+ \leq 100$ | 1978LIZM |
| 83 | 117 | $^{200}$Bi | 428.20 (10) | (10−) | 0.40(5) *s* | 428.2(1) | (E3) | %IT = 100 | 1972HA73 |
| 84 | 116 | $^{200}$Po | 1773.6 (3) | 8+ | 61(3) *ns* | (12.3(4)) | | %IT = 100 | 1990MA14 |
| 84 | 116 | $^{200}$Po | 2596.1 (3) | 11− | 100(10) *ns* | 334.8(1)<br>822.5(1) | E2<br>E3 | %IT = 100 | 1990MA14 |
| 84 | 116 | $^{200}$Po | 2804.5+X | 12+ | 268(3) *ns* | | | | 1990MA14<br>1985WE05 |
| 85 | 115 | $^{200}$At | 112 (5) | (7+) | 47(1) *s* | | | $\%\alpha = 43(7)$<br>$\%\epsilon+\%\beta+ = 57(7)$ | 1992HU04 |
| 85 | 115 | $^{200}$At | 344 (5) | (10−) | 7.3(+26−15) *s* | 230.9(2) | E3 | $\%\alpha \approx 10.5(3)$<br>$\%IT+\%\epsilon+\%\beta+ =$<br>89.5(3) | 2005UU02 |
| 86 | 114 | $^{200}$Rn | 2300.5+X | | 25(+11−6) $\mu s$ | | | | 2002DO19 |
| 87 | 113 | $^{200}$Fr | > 101? | | 0.6(+5−2) $\mu s$ | | | $\%IT \approx 100$ | 2014KA23 |
| 77 | 124 | $^{201}$Ir | 0+X | | 10.5(17) *ns* | | | | 2011ST21 |
| 78 | 123 | $^{201}$Pt | 1455.5+X# | (19/2+) | 18.4(13) *ns* | | | %IT = 100 | 2005CA02<br>2011ST21 |
| 79 | 122 | $^{201}$Au | X | | 5.6(24) $\mu s$ | 378.2(5)<br>638.0(5) | | %IT = 100 | 2011ST21 |
| 79 | 122 | $^{201}$Au | 594 (5) | 11/2− | 0.34(+90−29) *ms* | 41(6) | | %IT = 100 | 2011ST21 |

*Continued...*

*: Low value of 424(10) *ns* from 1987FA15 is omitted in the averaging procedure.
#: X < 90 keV in both 2005CA02 and 2011ST21.

Table 1 contd…

| Z | N | $^{A}X$ | E(keV) | $J^{\pi}$ | $T_{1/2}$ | Eγ(keV) | λ | Decay mode | Reference |
|---|---|---|---|---|---|---|---|---|---|
| 80 | 121 | $^{201}$Hg | 1.5648 (10) | 1/2− | 81(5) *ns* | 1.5648(10) | M1+E2 | %IT = 100 | 2007ME12 |
| 80 | 121 | $^{201}$Hg | 766.22 (15) | 13/2+ | 94(2) *μs* | 218.9(1) | M2(+E3) | %IT = 100 | 1976UY01 |
| | | | | | | | | | 1962EU01 |
| | | | | | | | | | 1961KR01 |
| 81 | 120 | $^{201}$Tl | 919.4 (2) | (9/2−) | 2.12(12) *ms* | 588.0(2) | (E3) | %IT = 100 | 1975UY01 |
| | | | | | | | | | 1967CO20 |
| | | | | | | | | | 1965GR04 |
| | | | | | | | | | 1964BR27 |
| | | | | | | | | | 1963DE38 |
| | | | | | | | | | 1962MO19 |
| 81 | 120 | $^{201}$Tl | 2747 | (33/2) | 0.095(+39−21) *μs* | | | | 2013BO18 |
| 82 | 119 | $^{201}$Pb | 629.1 (3) | 13/2+ | 60.8(18) *s* | 629.1(5) | M4 | %IT ≈ 100 | 1956ST05 |
| | | | | | | | | | 1955FI30 |
| 82 | 119 | $^{201}$Pb | 2719.6 (9) | 25/2− | 63(3) *ns* | 222.4(5) | E2 | %IT = 100 | 1988RO08 |
| 82 | 119 | $^{201}$Pb | 2719.6+X | (29/2−) | 508(3) *ns* | | | %IT ≈ 100 | 1989SU12 |
| | | | | | | | | | 1988RO08 |
| 82 | 119 | $^{201}$Pb | 4640.9+X | 41/2+ | 48(3) *ns* | 79.5(5) | E2 | %IT = 100 | 2019RO12 |
| | | | | | | | | | 1989SU12 |
| | | | | | | | | | 1988RO08 |
| 83 | 118 | $^{201}$Bi | 846.35 (18) | 1/2+ | 57.5(21) *min* | 846.3(3) | M4(+E5) | %ϵ = 95.4(43) | 1970DAZM |
| | | | | | | | | %IT = 4.3(43) | 1966MA51 |
| | | | | | | | | %α ≈ 0.3 | 1964SI11 |
| | | | | | | | | | 1950NE77 |
| 83 | 118 | $^{201}$Bi | 1933.3+X | (25/2+) | 118(28) *ns* | (<80) | | | 1985PI05 |
| 83 | 118 | $^{201}$Bi | 1972.3+X | (27/2+) | 105(75) *ns* | (39.0(8)) | | %IT = 100 | 1985PI05 |
| 83 | 118 | $^{201}$Bi | 2741.0+X | (29/2−) | 124(4) *ns* | (71.7(5)) | | %IT = 100 | 1985PI05 |
| | | | | | | 88.88(12) | | | |
| | | | | | | 150.5(6) | (E1) | | |
| | | | | | | 190.49(25) | (E1) | | |
| | | | | | | 440.9(4) | (E1) | | |
| 84 | 117 | $^{201}$Po | 423.4 (3) | 13/2+ | 8.96(20) *min* | 417.8(2) | M4 | %IT = 55 | 1963HO18 |
| | | | | | | | | %ϵ+%β+ = 43 | 1967LE08 |
| | | | | | | | | %α = 2.4(5) | 1967LE21 |
| | | | | | | | | | 1967TI04 |
| | | | | | | | | | 1970DAZM |
| | | | | | | | | | 1970JO26 |
| | | | | | | | | | 1970RA14 |
| | | | | | | | | | 1971JO19 |
| | | | | | | | | | 1976KO13 |
| | | | | | | | | | 1986BR28 |
| | | | | | | | | | 2010DE04 |
| 85 | 116 | $^{201}$At | 459.20 (15) | 1/2+ | 45(3) *ms* | 269.1(1) | E3 | %IT = 100 | 2014AU03 |
| 85 | 116 | $^{201}$At | 749.0 (3) | (13/2+) | 15.9(14) *ns* | 114.1 | | %IT = 100 | 1983DY02 |
| | | | | | | 749.0(3) | (M2) | | |
| 85 | 116 | $^{201}$At | 2319.0 (10) | 29/2+ | 3.39(9) *μs* | 269 | E2 | %IT = 100 | 2015AU01 |
| | | | | | | 339 | E3 | | |
| 86 | 115 | $^{201}$Rn | 248 (12) | (13/2+) | 3.8(1) *s* | | | %ϵ+%β+ = ? | 2017AL34 |
| | | | | | | | | %α = ? | 1996TA18 |
| | | | | | | | | | 1971HO01 |

*Continued…*

Table 1 contd…

| Z | N | $^{A}X$ | E(keV) | J$^{\pi}$ | $T_{1/2}$ | E$\gamma$(keV) | $\lambda$ | Decay mode | Reference |
|---|---|---|---|---|---|---|---|---|---|
| 87 | 114 | $^{201}$Fr | 130 (14) | (1/2+) | 37(+14–8) *ms* | | | %$\alpha$ = 100 | 2020AU01<br>2014KA23<br>2005UU02 |
| 87 | 114 | $^{201}$Fr | 289.5 (4) | (13/2+) | 720(40) *ns* | 289.5(4) | M2 | %IT = 100 | 2020AU01<br>2014KA23 |
| 88 | 113 | $^{201}$Ra | 260 (30) | (13/2+) | 1.6(+77–7) *ms* | | | %$\alpha$ ≈100 | 2014KA23<br>2005UU02 |
| 77 | 125 | $^{202}$Ir | X | | 3.4(6) $\mu s$ | 311.5(5)?<br>655.9(5)?<br>737.2(5)?<br>889.2(5)?<br>967.6(5)? | | | 2011ST21 |
| 78 | 124 | $^{202}$Pt | 1788.5 (4) | (7−) | 141(7) $\mu s$ | 534.9(2) | | %IT ≈ 100 | 2011ST21<br>2005CA02 |
| 80 | 122 | $^{202}$Hg | 2059 | (7−) | 10.4(4) *ns* | (70.6)<br>94.2 | | | 2021SU02 |
| 81 | 121 | $^{202}$Tl | 950.19 (10) | 7+ | 591(3) $\mu s$ | 459.72(7) | E3 | %IT = 100 | 2007FO06 |
| 81 | 121 | $^{202}$Tl | 4148 | (20+) | 215(10) $\mu s$ | 286 | | %IT = 100 | 2020WA24 |
| 82 | 120 | $^{202}$Pb | 2169.85 (8) | 9− | 3.54(2) *h* | 129.1(2)<br>547.4(2)<br>786.99(6) | E4<br>E5<br>E5 | %IT = 90.5(5)<br>%$\epsilon$+%$\beta$+ = 9.5(5) | 1981AN11<br>1957AS65<br>1954MA78 |
| 82 | 120 | $^{202}$Pb | 2208.44 (8) | 7− | 65.4(2) *ns* | 168.11(4)<br>825.4(3) | E2<br>E3 | %IT = 100 | 1986JA13 |
| 82 | 120 | $^{202}$Pb | 3237.7 (5) | 12+ | 24.1(3) *ns* | 46.0(5)<br>179.7(5) | <br>E1 | %IT = 100 | 2019RO12<br>1986JA13 |
| 82 | 120 | $^{202}$Pb | 4091.0+X | 16+ | 100(8) *ns* | | | | 2019RO12<br>1986JA13 |
| 82 | 120 | $^{202}$Pb | 5251.0+X | 19− | 108(3) *ns* | 1160.0(5) | E2 | %IT = 100 | 2019RO12<br>1987FA15 |
| 83 | 119 | $^{202}$Bi | 605+X | (10−) | 3.04(6) $\mu s$ | | | | 1981TH03 |
| 83 | 119 | $^{202}$Bi | 2597.07+X | (17+) | 310(50) *ns* | (50.7(3)) | | %IT = 100 | 1981TH03 |
| 84 | 118 | $^{202}$Po | 1691.5+X | 8+ | 110(15) *ns* | | | | 1976HA56 |
| 84 | 118 | $^{202}$Po | 2604.1+X | 11− | 85(10) *ns* | 386.0(5)<br>912.5(5) | E2<br>E3 | %IT = 100 | 1976HA56 |
| 84 | 118 | $^{202}$Po | 3040.5+X | 12+ | 19(4) *ns* | 143.1(5)<br>436.4(5) | <br>E1 | %IT = 100 | 1990FA03<br>1976BE12 |
| 84 | 118 | $^{202}$Po | 3573.5+X | 15− | 11(3) *ns* | 138.3(5) | E2 | %IT = 100 | 1990FA03<br>1976BE12 |
| 85 | 117 | $^{202}$At | 0+X | 7+ | 182(2) *s* | | | %$\epsilon$+%$\beta$+ = 91.3(15)<br>%$\alpha$ = 8.7(15) | 2018CU02<br>2016LY01<br>1992HU04 |
| 85 | 117 | $^{202}$At | 391.7+X (2) | 10− | 0.46(5) *s* | 391.7(2) | E3 | %IT = 99.90(1)<br>%$\alpha$ = 0.096(11) | 2016LY01<br>1992HU04 |
| 86 | 116 | $^{202}$Rn | 2260.4+X | | 2.22(7) $\mu s$ | | | %IT = 100 | 2002DO19 |
| 87 | 115 | $^{202}$Fr | 258 (11) | (10−) | 0.29(5) *s* | | | %$\alpha$ ≈ 100 | 2019GH11<br>2013ST25<br>2005UU02 |
| 77 | 126 | $^{203}$Ir | 1943.0+X | (23/2+) | 0.80(35) $\mu s$ | 207.0(5)? | | %IT = 100 | 2011ST21 |
| 78 | 125 | $^{203}$Pt | 1104.0+X | (33/2+) | 641(55) *ns* | 1104.0(5) | | %IT = 100 | 2011ST21 |
| 79 | 124 | $^{203}$Au | 641 (3) | 11/2− | 140(44) $\mu s$ | (78(3)) | | %IT = 100 | 2011ST21<br>1994GR07 |

*Continued…*

Table 1 contd…

| Z | N | $^{A}$X | E(keV) | J$^{\pi}$ | $T_{1/2}$ | E$\gamma$(keV) | $\lambda$ | Decay mode | Reference |
|---|---|---|---|---|---|---|---|---|---|
| 80 | 123 | $^{203}$Hg | 932.1 (7) | (13/2+) | 21.9(10) $\mu s$ | 341.0(5) | | %IT = 100 | 2011ST21<br>1972MO12 |
| 80 | 123 | $^{203}$Hg | 8281.0 (5) | (53/2+) | 146(30) $ns$ | 753.9(5)<br>887.4(4)<br>1320.3(2) | | %IT = 100 | 2011SZ01 |
| 81 | 122 | $^{203}$Tl | 3265.6 (4) | (29/2+) | 6.9(5) $\mu s$ | (5.7(5))<br>333.7(3) | | %IT = 100 | 2022BO04<br>1998PF02<br>1977SL01 |
| 82 | 121 | $^{203}$Pb | 126.5 (3) | 1/2− | 75(3) $ns$ | 126.5(10) | E2 | %IT = 100 | 1961BE29 |
| 82 | 121 | $^{203}$Pb | 825.11 (10) | 13/2+ | 6.21(11) $s$ | (4.9(3))<br>825.1(1) | <br>M4 | %IT = 100 | 1977LI04<br>1958FR53<br>1957AS65<br>1956ST05<br>1955FI30 |
| 82 | 121 | $^{203}$Pb | 1921.99 (17) | 21/2+ | 42(3) $ns$ | 258.4(1) | E2 | %IT = 100 | 1977SA18 |
| 82 | 121 | $^{203}$Pb | 2923.3+X | (25/2−) | 122(4) $ns$ | | | | 1988RO08 |
| 82 | 121 | $^{203}$Pb | 2949.12 (24) | 29/2− | 480(7) $ms$ | 153.4(2)<br>1027.0(3) | E3<br>M4 | %IT = 100 | 1977LI04 |
| 83 | 120 | $^{203}$Bi | 1098.21 (9) | 1/2+ | 305(5) $ms$ | 189.5(1)<br>204.7(1) | E3<br>M2+E3 | %IT = 100 | 1984LO16 |
| 83 | 120 | $^{203}$Bi | 1990.6 (3) | 21/2+ | 90(7) $ns$ | 87.1(2) | E2 | %IT = 100 | 1982LO14 |
| 83 | 120 | $^{203}$Bi | 2041.5 (6) | 25/2+ | 194(30) $ns$ | 50.9(5) | E2 | %IT = 100 | 1978HU02 |
| 83 | 120 | $^{203}$Bi | 3032.2 (7) | 29/2− | 22.4(9) $ns$ | 176.7(2)<br>301.3(2) | E1<br>E1 | %IT = 100 | 1978HU02 |
| 84 | 119 | $^{203}$Po | 641.64 (14) | 13/2+ | 45(2) $s$ | (2.3(2))<br>641.5(2) | <br>M4 | %IT = 100<br>%$\epsilon$+%$\beta$+ = ? | 1986FA04 |
| 84 | 119 | $^{203}$Po | 2158.3 (6) | | > 200 $ns$ | 182.5 | | %IT = 100 | 1986FA04 |
| 84 | 119 | $^{203}$Po | 2789.1 (2) | 25/2− | 12(2) $ns$ | 289.3(1)<br>303.5(1)<br>385.1(1) | (E1)<br>E1<br>E1 | %IT = 100 | 2022CHAA*<br>1986FA04 |
| 85 | 118 | $^{203}$At | 683.4 (3) | 1/2+ | 3.5(6) $ms$ | 462.0(2) | E3 | %IT = 100 | 2017AU05 |
| 85 | 118 | $^{203}$At | 1942.03+X | | 16.6(7) $ns$ | ≤100 | | %IT = 100 | 1983DY02 |
| 85 | 118 | $^{203}$At | 2044.2 (3) | 25/2+ | 13.9(14) $ns$ | 63.5(3)<br>80.0(2) | E2<br>(E1) | %IT = 100 | 2018AU01<br>1983DY02 |
| 85 | 118 | $^{203}$At | 2330.0 (3) | 29/2+ | 9.77(21) $\mu s$ | 285.8(2)<br>365.8(2) | E2<br>E3 | %IT = 100 | 2018AU01 |
| 86 | 117 | $^{203}$Rn | 362 (5) | (13/2+) | 26.9(5) $s$ | | | %$\epsilon$+%$\beta$+ = 25(10)<br>%$\alpha$ = 75(10) | 1996TA18<br>1996LE09<br>1971HO01<br>1967VA17 |
| 87 | 116 | $^{203}$Fr | 356.9 (9) | 1/2+ | 43(4) $ms$ | (≈ 20)<br>≈ 195? | (M2)<br>(E3) | %IT = 80(20)<br>%$\alpha$ = 20(4) | 2013JA06 |
| 87 | 116 | $^{203}$Fr | 426 (1) | 13/2+ | 0.37(5) $\mu s$ | 426(1) | M2 | %IT = 100 | 2013JA06 |
| 88 | 115 | $^{203}$Ra | 246 (17) | (13/2+) | 24.2(78) $ms$ | | | %$\alpha$ ≈ 100 | 2017AL34<br>2014KA23<br>2005UU02 |
| 78 | 126 | $^{204}$Pt | 1995.0 (15) | (5−) | 5.5(7) $\mu s$ | 1123(1) | | %IT = 100 | 2011ST21<br>2008ST20 |

*Continued…*

*: 2022CHAA report half-life of 7.1(1) $ns$. Unweighted average of the two values is 9.6(25) $ns$.

Table 1 contd...

| Z | N | $^{A}X$ | E(keV) | $J^{\pi}$ | $T_{1/2}$ | Eγ(keV) | λ | Decay mode | Reference |
|---|---|---|---|---|---|---|---|---|---|
| 78 | 126 | $^{204}$Pt | 1995.0+X | (7−) | 55(3) *μs* | | | %IT = 100 | 2011ST21<br>2008ST20 |
| 78 | 126 | $^{204}$Pt | 3153+X | (10+) | 146(14) *ns* | 97(1)<br>1158(1) | | %IT = 100 | 2011ST21<br>2008ST20 |
| 79 | 125 | $^{204}$Au | 1815.6+X | (16+) | 2.1(3) *μs* | 976.6(5) | | %IT = 100 | 2011ST21 |
| 80 | 124 | $^{204}$Hg | 4388.71 (13) | 14+ | 21.4(20) *ns* | 101.8(1) | E2 | %IT = 100 | 2015WR02<br>2011ST21 |
| 80 | 124 | $^{204}$Hg | 7226.08 (17) | 22+ | >485 *ns* | 921.03(5) | E3 | %IT = 100 | 2015WR02 |
| 81 | 123 | $^{204}$Tl | 145.89 (5) | (0−) | 18.7(14) *ns* | 145.88(10) | | %IT = 100 | 1975RAYX |
| 81 | 123 | $^{204}$Tl | 1103.9 (3) | 7+ | 61.7(12) *μs* | 690.1(1) | E3 | %IT = 100 | 2011BR12<br>2008FO03<br>1975UY01<br>1966MOZZ<br>1958DU80 |
| 81 | 123 | $^{204}$Tl | 2319.07 (20) | 12− | 2.6(2) *μs* | 668.9(1) | E3 | %IT = 100 | 2011BR12<br>1998PF02 |
| 81 | 123 | $^{204}$Tl | 4391.62 (24) | 18+ | 420(30) *ns* | 754.3(1)<br>1036.3(1) | (M2)<br>(E3) | %IT = 100 | 2011BR12<br>1998PF02 |
| 81 | 123 | $^{204}$Tl | 6239.4 (3) | 22− | 90(3) *ns* | 954.1(3)<br>1084.8(2)<br>1542.4(1) | <br><br>(M2) | %IT = 100 | 2011BR12 |
| 82 | 122 | $^{204}$Pb | 1274.13 (5) | 4+ | 265(6) *ns* | 374.76(7)<br>1274 | E2 | %IT = 100 | 1978SO02<br>1967LI12<br>1963SA19 |
| 82 | 122 | $^{204}$Pb | 2185.88 (8) | 9− | 66.93(10) *min* | 622.2(2)<br>911.74(15) | E5<br>E5 | %IT = 100 | 2001LI17<br>1977SMZV<br>1972SI22<br>1958BA04<br>1956HE50 |
| 82 | 122 | $^{204}$Pb | 2264.42 (6) | 7− | 0.45(+10−3) *μs* | (6.26(3))<br>78.54(8)<br>990.4(2) | <br>E2<br>E3 | %IT = 100 | 1978SO02 |
| 82 | 122 | $^{204}$Pb | 8349 | (22+) | 0.22(2) *ms* | 481 | | | 2022WA20 |
| 83 | 121 | $^{204}$Bi | 805.5 (3) | 10− | 13.0(1) *ms* | 752.1(2) | E3 | %IT = 100 | 1974RA25 |
| 83 | 121 | $^{204}$Bi | 2833.4 (11) | 17+ | 1.07(3) *ms* | 181.8(2)<br>918.1(2) | M2<br>E3 | %IT = 100 | 1974RA25 |
| 84 | 120 | $^{204}$Po | 1639.03 (6) | 8+ | 157(3) *ns* | (12.1) | | %IT = 100 | 2010KA29<br>1990FA03<br>1987RA04<br>1983HE08<br>1971HA01<br>1970YA03<br>1970BRZO |
| 84 | 120 | $^{204}$Po | 2227.33 (6) | 9− | 15.6(5) *ns* | 588.30(2) | E1 | %IT = 100 | 1987RA04 |
| 84 | 120 | $^{204}$Po | 3564.1 (5) | 15− | 11.5(9) *ns* | 124.8(3) | (E2) | %IT = 100 | 1990FA03<br>1982HA16<br>1976BE12 |
| 85 | 119 | $^{204}$At | 587.30 (20) | 10− | 108(10) *ms* | 587.3(2) | E3 | %IT = 100 | 1969MOZU |
| 85 | 119 | $^{204}$At | 4016.8 (15) | 16+ | 36.0(35) *ns* | 718.3(10)<br>975.6(10) | M2<br>M2 | %IT = 100 | 2022KA24 |

*Continued...*

Table 1 contd...

| Z | N | $^{A}X$ | E(keV) | J$^{\pi}$ | $T_{1/2}$ | Eγ(keV) | λ | Decay mode | Reference |
|---|---|---|---|---|---|---|---|---|---|
| 86 | 118 | $^{204}$Rn | 2461.9 (4) | 10− | 33.4(24) *ns* | 242.8(4) | M1 | %IT = 100 | 2002DO19<br>1981HO29 |
| 86 | 118 | $^{204}$Rn | 3035.3 (3) | 12+ | 14(4) *ns* | 438.0(4)<br>582.6(2) | E1<br>E2 | %IT = 100 | 2002DO19 |
| 87 | 117 | $^{204}$Fr | 41 (7) | 7(+) | 2.3(3) *s* | | | %α = 90(2)<br>%ϵ+%β+ = 10(2) | 2015VO05<br>2013JA06<br>2005UU02<br>1992HU04<br>1974HO27<br>1967VA20 |
| 87 | 117 | $^{204}$Fr | 316 (7) | 10(−) | 1.65(15) *ms* | 275 | (E3) | %α = 53(10)<br>%IT = 47(10) | 2015VO05<br>2014LY01<br>2013JA06<br>1992HU04 |
| 79 | 126 | $^{205}$Au | 907 (5) | 11/2− | 6(2) *s* | 907(5) | M4 | %β−> 0.0<br>%IT > 0.0 | 2009PO01 |
| 79 | 126 | $^{205}$Au | 2849.7 (6) | 19/2+ | 163(5) *ns* | (34.2)<br>962.5(5) | | %IT = 100 | 2011ST21<br>2009PO01 |
| 80 | 125 | $^{205}$Hg | 1556.40 (17) | 13/2+ | 1.09(4) *ms* | 161.4(2)<br>210.3(2) | M2<br>E3 | %IT = 100 | 1986ZE03<br>1985MA48 |
| 80 | 125 | $^{205}$Hg | 3315.8 (9) | (23/2−) | 5.89(18) *μs* | 722.6(5)<br>950.2(5) | | %IT = 100 | 2011ST21 |
| 81 | 124 | $^{205}$Tl | 3290.60 (17) | 25/2+ | 2.6(2) *μs* | 739.16(10) | E3 | %IT = 100 | 1984BE14<br>1982MA05 |
| 81 | 124 | $^{205}$Tl | 4835.6 (15) | (35/2−) | 235(10) *ns* | 1217 | | %IT = 100 | 2004WR01 |
| 82 | 123 | $^{205}$Pb | 2.329 (7) | 1/2− | 24.2(4) *μs* | 2.328(7) | E2 | %IT = 100 | 1994KR11 |
| 82 | 123 | $^{205}$Pb | 1013.85 (3) | 13/2+ | 5.55(2) *ms* | 26.220(11)<br>310.35(5)<br>1013.8(1) | M2<br>E3<br>M4 | %IT = 100 | 1971MA59 |
| 82 | 123 | $^{205}$Pb | 3195.8 (6) | 25/2− | 217(5) *ns* | 27.8(12)<br>1175.1(5) | <br>E3 | %IT = 100 | 1976LI09 |
| 82 | 123 | $^{205}$Pb | 5161.9 (7) | (33/2+) | 63(3) *ns* | 97.1(5)<br>1251.7(5)<br>1535.6(5) | E2 | %IT = 100 | 1983ST15 |
| 83 | 122 | $^{205}$Bi | 1497.17 (9) | 1/2+ | 7.9(7) *μs* | (10.6)<br>24.9(2)<br>495.9(2)<br>624.8(1) | <br>(E1)<br><br>M2 | %IT = 100 | 1972AL25 |
| 83 | 122 | $^{205}$Bi | 2064.7 (4) | 21/2+ | 100(6) *ns* | 23.8(5)<br>720.5(5) | <br>M2 | %IT = 100 | 1982LO14 |
| 83 | 122 | $^{205}$Bi | 2139.0 (7) | 25/2+ | 220(25) *ns* | 74.3(5) | E2 | %IT = 100 | 1978HU02 |
| 83 | 122 | $^{205}$Bi | 7913+X* | (51/2−) | 8(2) *ms* | <150 | | | 2022WA20 |
| 84 | 121 | $^{205}$Po | 143.166 (17) | 1/2− | 310(60) *ns* | 143.166(17) | E2 | %IT = 100 | 1971JO19 |
| 84 | 121 | $^{205}$Po | 880.31 (7) | 13/2+ | 0.645(20) *ms* | 161.30(17) | M2 | %IT = 100 | 1973FO07 |
| 84 | 121 | $^{205}$Po | 1461.21 (21) | 19/2− | 57.4(9) *ms* | 580.9(2) | E3 | %IT = 100 | 1974RO36<br>1974OH06<br>1973FO07 |

*Continued...*

*: X < 150 keV from 2022WA20.

Table 1 contd. . .

| Z | N | $^{A}$X | E(keV) | J$^{\pi}$ | $T_{1/2}$ | E$\gamma$(keV) | $\lambda$ | Decay mode | Reference |
|---|---|---|---|---|---|---|---|---|---|
| 84 | 121 | $^{205}$Po | 3087.2 (4) | 29/2− | 115(10) *ns* | 106.5(2) | | %IT = 100 | 1985RA18 |
| | | | | | | 260.3(2) | E1 | | |
| | | | | | | 374.8(2) | | | |
| 85 | 120 | $^{205}$At | 2062.61 (23) | 25/2+ | 66.6(19) *ns* | (8.30(13)) | | %IT = 100 | 1984DA19 |
| | | | | | | 126.7(1) | E1 | | 1982SJ01 |
| 85 | 120 | $^{205}$At | 2339.64 (23) | 29/2+ | 7.76(14) $\mu s$ | 277.1(1) | E2 | %IT = 100 | 1984DA19 |
| | | | | | | 403.6(1) | E3 | | |
| 86 | 119 | $^{205}$Rn | 657.1 (5)? | (13/2+) | > 10 *s* | 657.1? | | | 2010DE04 |
| 87 | 118 | $^{205}$Fr | 544.0 (10) | 13/2+ | 80(20) *ns* | 544(1) | (M2) | %IT = 100 | 2012JA01 |
| 87 | 118 | $^{205}$Fr | 609 (6) | 1/2+ | 1.15(4) *ms* | (165(5)) | | %IT = 100 | 2012JA01 |
| 88 | 117 | $^{205}$Ra | 278 (31) | (13/2+) | 170(+60−40) *ms* | | | %$\alpha$ ≈ 100 | 2017AL34 |
| | | | | | | | | | 1996LE06 |
| 80 | 126 | $^{206}$Hg | 2102.4 (3) | 5− | 2.09(2) $\mu s$ | 1034.2(2) | E3 | %IT = 100 | 2018LA03 |
| | | | | | | | | | 2011ST21 |
| | | | | | | | | | 1982BE38 |
| 80 | 126 | $^{206}$Hg | 3723.4 (14) | (10+) | 107(6) *ns* | 100(1) | | %IT = 100 | 2018LA03 |
| | | | | | | 1257(1) | | | 2011ST21 |
| | | | | | | | | | 2009AL29 |
| | | | | | | | | | 2001FO08 |
| 81 | 125 | $^{206}$Tl | 1405.47 (7) | (5)+ | 78(2) *ns* | 453.28(5) | | %IT = 100 | 2011ST21 |
| | | | | | | 604.3(2) | | | 1976HA44 |
| | | | | | | 1139.9(3) | | | |
| 81 | 125 | $^{206}$Tl | 1621.70 (11) | (7)+ | 10.1(6) *ns* | 216.26(8) | | %IT = 100 | 1976HA44 |
| 81 | 125 | $^{206}$Tl | 2643.10 (18) | (12−) | 3.74(3) *min* | 316.8(2) | | %IT = 100 | 1978UR01 |
| | | | | | | 564.2(1) | M4 | | 1976BE44 |
| | | | | | | 1021.5(2) | | | 1976HA44 |
| 82 | 124 | $^{206}$Pb | 2200.16 (4) | 7− | 125(2) $\mu s$ | 202.44(3) | | %IT = 100 | 1995AN36 |
| | | | | | | 516.18(4) | E3 | | 1973SA22 |
| | | | | | | | | | 1973DAZL |
| | | | | | | | | | 1968TA13 |
| | | | | | | | | | 1967CO20 |
| | | | | | | | | | 1966MOZZ |
| | | | | | | | | | 1962TH12 |
| | | | | | | | | | 1960BE36 |
| | | | | | | | | | 1957TO22 |
| | | | | | | | | | 1957AS65 |
| | | | | | | | | | 1953AL47 |
| 82 | 124 | $^{206}$Pb | 4027.3 (3) | 12+ | 202(3) *ns* | 69.7(5) | E2 | %IT = 100 | 1993BL02 |
| | | | | | | 1369.0(3) | E3 | | 1983ST15 |
| | | | | | | | | | 1979MA37 |
| | | | | | | | | | 1971BE37 |
| 83 | 123 | $^{206}$Bi | 59.897 (17) | 4+ | 7.7(2) $\mu s$ | 59.908(18) | E2 | %IT = 100 | 1975KA13 |
| | | | | | | | | | 1957AR61 |
| 83 | 123 | $^{206}$Bi | 1044.8 (7) | 10− | 0.89(1) *ms* | 229.0(5) | M2 | %IT = 100 | 1974RA25 |
| | | | | | | 903.6(5) | E3 | | |
| 83 | 123 | $^{206}$Bi | 3147.1 (12) | 15+ | 15.6(3) *ns* | 543.5(5) | E1 | %IT = 100 | 1978LO09 |
| 83 | 123 | $^{206}$Bi | 9233.3 (8) | (28−) | 155(15) *ns* | 713.2(1) | M2(+E3) | %IT = 100 | 2012CI05 |
| 83 | 123 | $^{206}$Bi | 10170.5 (8) | (31+) | 27(2) $\mu s$ | 937.2(1) | | %IT = 100 | 2022WA20 |
| | | | | | | | | | 2012CI05 |
| 84 | 122 | $^{206}$Po | 1585.90 (11) | 8+ | 232(4) *ns* | (12.50(10)) | | %IT = 100 | 1990BA31 |

*Continued. . .*

Table 1 contd...

| Z | N | $^{A}$X | E(keV) | J$^{\pi}$ | $T_{1/2}$ | E$\gamma$(keV) | $\lambda$ | Decay mode | Reference |
|---|---|---|---|---|---|---|---|---|---|
| 84 | 122 | $^{206}$Po | 2262.09 (12) | 9− | 1.05(6) $\mu s$ | 61.766(19)<br>676.4(2) | E1<br>(E1) | %IT = 100 | 1990BA31 |
| 85 | 121 | $^{206}$At | 806.7+X | (10−) | 813(21) $ns$* | 121(1) | E1 | | 2009DR08 |
| 86 | 120 | $^{206}$Rn | 1924.7 (6) | (8)+ | 19(3) $ns$ # | 161.3 | E2 | %IT = 100 | 1981HO29 |
| 86 | 120 | $^{206}$Rn | 2475.55 (25) | 10− | 65(5) $ns$ | 205.7(2)<br>551.2(2) | M1 | %IT = 100 | 1981HO29 |
| 86 | 120 | $^{206}$Rn | 4129.8 (9) | (15) | 11(2) $ns$ | 242(1)<br>768(1) | E2 | %IT = 100 | 1981HO29 |
| 87 | 119 | $^{206}$Fr | 0+X | 7(+) | ≈ 16.0(1) $s$ $ | | | %$\alpha$ ≈ 84<br>%$\epsilon$ ≈ 16 | 2015VO05<br>1981RI04<br>1974HO27<br>1967VA20<br>1964GR04 |
| 87 | 119 | $^{206}$Fr | 531+X | 10(−) | 0.7(1) $s$ | 531 | | %$\alpha$ = ?<br>%IT = ? | 2016LY01<br>2015VO05<br>1981RI04 |
| 89 | 117 | $^{206}$Ac | 226 (37) | (10−) | 33(+22−9) $ms$ | | | %$\alpha$ ≈ 100 | 2019GH11<br>2013ST25<br>1998ES02 |
| 81 | 126 | $^{207}$Tl | 1348.18 (16) | 11/2− | 1.33(11) $s$<br>1.47(32) $s$ @ | 997.1(3)<br>1348.1(3) | | %IT = 100 | 2000RE12<br>1965EC02<br>2006BO41@ |
| 82 | 125 | $^{207}$Pb | 1633.356 (4) | 13/2+ | 0.805(8) $s$ | 1063.656(3) | M4+E5 | %IT = 100 | 1986AL11<br>1973SA22<br>1971GL09<br>1967YU01<br>1961GL16<br>1956CA50<br>1956VE10<br>1955BE24<br>1954RE33<br>1953FR17<br>1952HO41<br>1951LA18 |
| 83 | 124 | $^{207}$Bi | 2101.61 (16) | 21/2+ | 182(6) $\mu s$ | 456.1(1)<br>743.30(15) | E3<br>E3 | %IT = 100 | 1969BE47 |
| 83 | 124 | $^{207}$Bi | 3887.4 (9) | 29/2− | 12.7(9) $ns$ | 386.8(5) | E1+M2 | %IT = 100 | 1979LO04 |
| 84 | 123 | $^{207}$Po | 68.557 (14) | 1/2− | 205(10) $ns$ | 68.55(2) | E2 | %IT = 100 | 1963AS02 |
| 84 | 123 | $^{207}$Po | 1115.076 (17) | 13/2+ | 49(4) $\mu s$ | 300.648(13) | M2 | %IT = 100 | 1973CO30<br>1962HA26 |
| 84 | 123 | $^{207}$Po | 1383.16 (7) | 19/2− | 2.79(8) $s$ | 109.1<br>268.08(6) | M3<br>E3 | %IT = 100 | 1978SC12 |
| 84 | 123 | $^{207}$Po | 2379.7 (3) | 25/2+ | 43.0(3) $ns$ | 66.2(2) | E2 | %IT = 100 | 1985RO07 |
| 85 | 122 | $^{207}$At | 2117.3 (6) | 25/2+ | 108(2) $ns$ | 219.6(3) | E1 | %IT = 100 | 1981SJ01 |
| 86 | 121 | $^{207}$Rn | 899.1 (10) | 13/2+ | 184.5(9) $\mu s$ | 234(1) | M2 | %IT = 100 | 2006HA17 |

*Continued...*

*: Low values of 410(80) $ns$ in 1999FE10 and 377(44) $ns$ in 2010KA29 are reported.
#: Low values of 6.3(24) $ns$ in 1981RI02 and 13.5(10) $ns$ in 1981MA28 are reported.
$: Composite half-life for g.s.+ 0+X isomer.

Table 1 contd...

| Z | N | $^{A}$X | E(keV) | J$^{\pi}$ | $T_{1/2}$ | E$\gamma$(keV) | $\lambda$ | Decay mode | Reference |
|---|---|---|---|---|---|---|---|---|---|
| 88 | 119 | $^{207}$Ra | 554 (15) | (13/2+) | 59(4) *ms* | | | %IT = 74(20)<br>%$\alpha$ = 26(20) | 2021NI08<br>1996LE09<br>1987HE10 |
| 80 | 128 | $^{208}$Hg | 1296.9+X | (8+) | 99(14) *ns* | | | %IT = 100 | 2009AL29 |
| 81 | 127 | $^{208}$Tl | 1806.7 (3) | (0−) | 1.3(1) $\mu s$ | 220.7(2)<br>1314.0(3) | E2<br>E3 | %IT = 100 | 2020CA25 |
| 82 | 126 | $^{208}$Pb | 4895.23 (5) | 10+ | 535(35) *ns* | (34.4)<br>284.5(1)<br>857.7(1) | E2<br>E2<br>E3 | %IT = 100 | 2017BR08<br>1989RO04 |
| 82 | 126 | $^{208}$Pb | 10341.87 (20) | 20− | 22(3) *ns* | 145.8(2)<br>206.1(2)<br>947.6(1) | E1<br>(E1, E2)<br>E1 | %IT = 100 | 2017BR08 |
| 82 | 126 | $^{208}$Pb | 11360.87 (23) | 23+ | 12.7(15) *ns* | 1019.0(1) | E3 | %IT = 100 | 2017BR08 |
| 82 | 126 | $^{208}$Pb | 13674.7 (3) | 28− | 60(6) *ns* | 138.7(2)<br>725.4(1) | E2<br>E3 | %IT = 100 | 2017BR08 |
| 83 | 125 | $^{208}$Bi | 1571.1 (4) | 10− | 2.60(5) *ms* | 920.5(3) | E3(+M4) | %IT = 100 | 2014DE19<br>1995AN36<br>1976GA33<br>1973SA22<br>1967HI08<br>1966ME02<br>1962MO19 |
| 83 | 125 | $^{208}$Bi | ≈ 9000 | | ≈ 40 *ns* | | | | 2003FO03 |
| 84 | 124 | $^{208}$Po | 1528.22 (4) | 8+ | 373(9) *ns* | 4.02(3) | E2 | %IT = 100 | 2021BR06<br>2020BRZZ<br>1985RA21<br>1976HA56<br>1968TR06 |
| 85 | 123 | $^{208}$At | 1090.5 | 10− | 47.8(10) *ns* | 186.5<br>1018.5 | E1 | %IT = 100 | 1985NO09 |
| 85 | 123 | $^{208}$At | 2276.4 (18) | 16− | 1.5(2) $\mu s$ | 472.2<br>750.9 | E3<br>E3 | %IT = 100 | 1984FA10 |
| 86 | 122 | $^{208}$Rn | 1828.3 (4) | 8+ | 487(12) *ns* | 88.9(1) | E2 | %IT = 100 | 2010KA29<br>1983TR03<br>1981HO29<br>1981MA28 |
| 86 | 122 | $^{208}$Rn | 2618.1 (4) | 10− | 11.8(7) *ns* | 153.0(1)<br>298.7(1)<br>789.7(1) | <br>M1<br>M2 | %IT = 100 | 1983TR03 |
| 86 | 122 | $^{208}$Rn | 4066.4 (5) | 16− | 18.3(4) *ns* | 141.2(2) | M1 | %IT = 100 | 1983TR03 |
| 87 | 121 | $^{208}$Fr | 826.3 (5) | (9, 10)− | 432(11) *ns** | 194.1 | E1 | %IT = 100 | 2009DR08 |
| 87 | 121 | $^{208}$Fr | 1209 (4) | 11− | 33(7) *ns* | 382.9(18) | E2 | %IT = 100 | 2010KA29 |
| 88 | 120 | $^{208}$Ra | 2147.4 (4) | (8+) | 270(21) *ns* | 130.2(2)<br>392.1(2) | | %IT = 100 | 2005RE02<br>1999CO13 |
| 89 | 119 | $^{208}$Ac | 506 (26) | (10−) | 25(+9−5) *ms* | | | %$\alpha$ ≈ 90<br>%$\epsilon$ ≈ 10 | 1994LE05<br>1973TA30 |
| 81 | 128 | $^{209}$Tl | 1180.7 (4) | 13/2+ | 14(5) *ns* | 137.8(2) | E2 | %IT = 100 | 2017AM01 |
| 81 | 128 | $^{209}$Tl | 1228.1 (20) | 17/2+ | 146(10) *ns* | 47.4(20) | E2 | %IT = 100 | 2017AM01<br>2009AL29 |

*Continued...*

*: 2010KA29 quotes the half-life as 233(18) *ns* for the same state.

Table 1 contd…

| Z | N | $^{A}$X | E(keV) | J$^{\pi}$ | $T_{1/2}$ | E$\gamma$(keV) | $\lambda$ | Decay mode | Reference |
|---|---|---|---|---|---|---|---|---|---|
| 83 | 126 | $^{209}$Bi | 2442.92 (6) | 1/2+ | 10.2(11) *ns* | 1546.52(9) | E3 | %IT = 100 | 2016RO02<br>1996DE48<br>1978EL07 |
| 83 | 126 | $^{209}$Bi | 2986.80 (5) | 19/2+ | 17.9(5) *ns* | 245.73(2)<br>385.9? | E2 | %IT = 100 | 1996DE48<br>1978BE17 |
| 84 | 125 | $^{209}$Po | 1417.66 (9) | 13/2− | 24.2(4) *ns* | 90.8(1) | E2 | %IT = 100 | 2009NI05<br>1976HA56<br>1974BE74<br>1971AL31 |
| 84 | 125 | $^{209}$Po | 1472.56 (19) | 17/2− | 89.3(5) *ns* | 54.9(2) | E2 | %IT = 100 | 2009NI05<br>1976HA56<br>1974BE74 |
| 84 | 125 | $^{209}$Po | 4265.4 (3) | 31/2− | 119(4) *ns* | 96.9(1)<br>1289.1(2) | E1<br>E3 | %IT = 100 | 1974BE74<br>2000PO03 |
| 85 | 124 | $^{209}$At | 1427.67 (16) | 21/2− | 25.0(10) *ns* | 106.1(1) | E2 | %IT = 100 | 1985RA21<br>1983MA08<br>1976SJ01<br>1975BE39 |
| 85 | 124 | $^{209}$At | 2429.32 (22) | 29/2+ | 0.893(32) $\mu s$ | 577.60(10) | E3 | %IT = 100 | 1987DR01<br>1985RA21<br>1983MA08<br>1975BE39 |
| 86 | 123 | $^{209}$Rn | 1174.01 (13) | 13/2+ | 13.4(13) $\mu s$ | 376.2(1) | M2 | %IT = 100 | 1985PO08 |
| 86 | 123 | $^{209}$Rn | 3157.51 (21) | 29/2− | 13.9(21) *ns* | 199.9(1) | M1+E2 | %IT = 100 | 1985PO08 |
| 86 | 123 | $^{209}$Rn | 3636.81 (23) | 35/2+ | 3.0(3) $\mu s$ | 479.3(1) | E3 | %IT = 100 | 1985PO08 |
| 86 | 123 | $^{209}$Rn | 4833.7 (3) | 41/2− | 10.0(4) *ns* | 1196.9(2) | E3 | %IT = 100 | 1985PO08 |
| 87 | 122 | $^{209}$Fr | 826.1 (6) | (13/2+) | 446(14) *ns* | 194.1(4) | | %IT = 100 | 2006ME03 |
| 87 | 122 | $^{209}$Fr | 2130.5 (4) | 25/2+ | 33.3(21) *ns* | 201.9(2) | E1 | %IT = 100 | 2009DR04 |
| 87 | 122 | $^{209}$Fr | 3415.9 (6) | 33/2(−) | ~ 62 *ns* | 181.4(3) | E1 | %IT = 100 | 2009DR04 |
| 87 | 122 | $^{209}$Fr | 4659.8 (7) | 45/2− | 420(18) *ns* | 335.5(3)?<br>620.2(3) | <br>E3 | %IT = 100 | 2011KA37<br>2009DR04 |
| 88 | 121 | $^{209}$Ra | 882.4 (7) | 13/2+ | 117(5) $\mu s$ | 238.4(5) | M2 | %IT = 100 | 2008HA12 |
| 90 | 119 | $^{209}$Th* | 0+X | (13/2+) | 2.5(+17–7) *ms* | | | %$\alpha \approx$ 100 | 2010HE25<br>1996IK01 |
| 80 | 130 | $^{210}$Hg | 663.0 (20)? | (3−) | 2.1(7) $\mu s$ | (20)<br>663 | | %IT = 100 | 2013GO10 |
| 80 | 130 | $^{210}$Hg | 1366+X | (8+) | 2(1) $\mu s$ | | | %IT = 100 | 2013GO10 |
| 82 | 128 | $^{210}$Pb | 1194.6 | 6+ | 92(10) *ns* # | 97.6(1) | E2 | %IT = 100 | 2018BR15<br>1983DE34<br>1980SJ01 |
| 82 | 128 | $^{210}$Pb | 1278 (5) | 8+ | 166(15) *ns* | 83(3) | | %IT = 100 | 1983DE34<br>1981BO29<br>1980SJ01 |
| 83 | 127 | $^{210}$Bi | 271.31 (11) | 9− | 3.04E+6(6) *y* | | | %$\alpha$ = 100 | 1976TUZY |
| 83 | 127 | $^{210}$Bi | 433.48 (12) | 7− | 57.5(10) *ns* | 162.19(5) | | %IT = 100 | 1973PR11<br>1972BA65 |
| 83 | 127 | $^{210}$Bi | 439.24 (4) | 5− | 38(1) *ns* | 91.32(8) | | %IT = 100 | 1973PR11 |

*Continued…*

*: The 2.5-*ms* activity is assigned as an isomer, although, the g.s. activity has not yet been identified.

#: Low values of 49(6) *ns* in 1983DE34 and 21(7) *ns* in 1980SJ01 are reported.

Table 1 contd...

| Z | N | $^{A}X$ | E(keV) | J$^{\pi}$ | $T_{1/2}$ | Eγ(keV) | λ | Decay mode | Reference |
|---|---|---|---|---|---|---|---|---|---|
| 83 | 127 | $^{210}$Bi | 3469.2 (13) | (15+) | 11.1(7) *ns* | 175 | (E2) | %IT = 100 | 2014CI03 |
| | | | | | | 744 | (E1) | | |
| 84 | 126 | $^{210}$Po | 1473.357 (21) | 6+ | 42.6(10) *ns* | 46.85(5) | E2 | %IT = 100 | 1976HA56 |
| 84 | 126 | $^{210}$Po | 1556.97 (3) | 8+ | 98.9(25) *ns* | 83.54(8) | E2 | %IT = 100 | 1988MA32 |
| | | | | | | | | | 1976HA56 |
| 84 | 126 | $^{210}$Po | 2849.17 (4) | 11− | 19.6(4) *ns* | 661.17(3) | E3 | %IT = 100 | 1988MA32 |
| | | | | | | 1292.20(1) | E3 | | |
| 84 | 126 | $^{210}$Po | 4371.96 (4) | 13− | 54.4(24) *ns* | 47.8 | | %IT = 100 | 1988MA32 |
| | | | | | | 1522.79(2) | E2 | | 1985BE22 |
| 84 | 126 | $^{210}$Po | 5057.65 (5) | 16+ | 263(5) *ns* | 279.89(10) | M2 | %IT = 100 | 1985BE22 |
| | | | | | | 685.69(2) | E3 | | 1985KA07 |
| 85 | 125 | $^{210}$At | 1363.2 (2) | (11)+ | 23.6(7) *ns* | 111.3 | E2 | %IT = 100 | 2001BA79 |
| | | | | | | | | | 1987DR01 |
| | | | | | | | | | 1983MA08 |
| | | | | | | | | | 1972WI19 |
| 85 | 125 | $^{210}$At | 1688.7 (3) | (10)− | 15.5(10) *ns* | 193.5 | E1 | %IT = 100 | 2001BA79 |
| | | | | | | 325.3 | E1 | | 1982LO18 |
| | | | | | | 436.6 | | | 1980RAZM |
| 85 | 125 | $^{210}$At | 2549.6 (2) | (15)− | 478(6) *ns* | 579.5 | E3 | %IT = 100 | 2001BA79 |
| | | | | | | 644.3 | E3 | | 1987CA23 |
| | | | | | | | | | 1987DR01 |
| | | | | | | | | | 1983MA08 |
| | | | | | | | | | 1978RA03 |
| | | | | | | | | | 1972WI19 |
| 85 | 125 | $^{210}$At | 4027.7 (2) | (19)+ | 5.66(7) *µs* | 372.3 | E3 | %IT = 100 | 2001BA79 |
| | | | | | | 485.2 | M2 | | 1987DR01 |
| | | | | | | 611.8 | | | 1987CA23 |
| | | | | | | 920.4 | | | 1978RA03 |
| 85 | 125 | $^{210}$At | 6959.3 (6) | (26−) | 98(2) *ns* | 435 | | %IT = 100 | 2001BA79 |
| 85 | 125 | $^{210}$At | 7847.3 (6) | (29+) | 36.0(14) *ns* | 888 | | %IT = 100 | 2001BA79 |
| 86 | 124 | $^{210}$Rn | 1664.6+X | (8+) | 644(40) *ns* | | | %IT = 100 | 2005PO10 |
| | | | | | | | | | 1982PO03 |
| | | | | | | | | | 1981MA28 |
| | | | | | | | | | 1980PO07 |
| | | | | | | | | | 1979PO19 |
| 86 | 124 | $^{210}$Rn | 2562.31+X | (11)− | 64(3) *ns* | 185.5(1) | E1 | %IT = 100 | 2005PO10 |
| | | | | | | 897.6(2) | (E3) | | 1982PO03 |
| | | | | | | | | | 1981MA28 |
| | | | | | | | | | 1979PO19 |
| 86 | 124 | $^{210}$Rn | 3248.06+X | (14)+ | 76(7) *ns* | 325.4(1) | E2 | %IT = 100 | 2005PO10 |
| | | | | | | | | | 1982PO03 |
| | | | | | | | | | 1981MA28 |
| | | | | | | | | | 1979PO19 |
| 86 | 124 | $^{210}$Rn | 3812.40+X | (17)− | 1.06(5) *µs* | 564.3(1) | E3 | %IT = 100 | 2005PO10 |
| | | | | | | | | | 1982PO03 |
| | | | | | | | | | 1981MA28 |
| | | | | | | | | | 1979PO19 |
| 86 | 124 | $^{210}$Rn | 4993.43+X | (20)+ | 12.3(9) *ns* | 1181.0(1) | E3 | %IT = 100 | 2005PO10 |
| | | | | | | | | | 1985PO13 |
| | | | | | | | | | 1982PO03 |

*Continued...*

Table 1 contd...

| Z | N | $^{A}$X | E(keV) | J$^{\pi}$ | $T_{1/2}$ | E$\gamma$(keV) | $\lambda$ | Decay mode | Reference |
|---|---|---|---|---|---|---|---|---|---|
| 86 | 124 | $^{210}$Rn | 6469.02+X | 23+ | 1.04(7) $\mu s$ | 433.0(1)<br>602.7(1) | E2<br>E2 | %IT = 100 | 2005PO10<br>1982PO03 |
| 86 | 124 | $^{210}$Rn | 7311.02+X | 26− | 34(2) $ns$ | 415.9(1)<br>842.0(1) | <br>E3 | %IT = 100 | 2005PO10<br>1982PO03 |
| 87 | 123 | $^{210}$Fr | 729.1 (23) | (9−) | 41(2) $ns$ | 203.4(23) | | %IT = 100 | 2011KA37 |
| 87 | 123 | $^{210}$Fr | 1112.6 (5) | 10− | 20.7(5) $ns$ | 271 | E1 | | 2016MA41 |
| 87 | 123 | $^{210}$Fr | 4417.3 (8) | (23)+ | 475(6) $ns$ | 573.5<br>663.3 | (E3)<br>(E3) | | 2016MA41<br>2006ME03 |
| 88 | 122 | $^{210}$Ra | 2050.9 (7) | 8+ | 2.1(1) $\mu s$ | 95.7(5) | E2 | %IT = 100 | 2013BA29<br>2006HA17<br>2004RE04 |
| 81 | 130 | $^{211}$Tl | 144+X | (13/2+) | 0.58(8) $\mu s$ | 144(1) | (E2) | %IT = 100 | 2019GO10 |
| 82 | 129 | $^{211}$Pb | 1193.1 (4) | (21/2+) | 42(7) $ns$ | 137.4(2) | E2 | %IT = 100 | 2005LA01 |
| 82 | 129 | $^{211}$Pb | 1679.1+X | (27/2+) | 159(28) $ns$ | | | %IT = 100 | 2005LA01 |
| 83 | 128 | $^{211}$Bi | 1227.2 (3) | (21/2−) | 70(5) $ns$ | 97.4(3) | (E2) | %IT = 100 | 1989MA12 |
| 83 | 128 | $^{211}$Bi | 1257 (10) | (25/2−) | 1.4(3) $\mu s$ | (30) | | %IT = 100 | 1998PF02 |
| 84 | 127 | $^{211}$Po | 1064.8 (4) | 15/2− | 14.0(2) $ns$ | 377.6(5)<br>1064.9(5) | M2<br>E3 | %IT = 100 | 1998MC03<br>1973FAZD |
| 84 | 127 | $^{211}$Po | 1427.8+X | | 25.0(14) $ns$ | | | %IT = 100 | 1998MC03<br>1981FA01 |
| 84 | 127 | $^{211}$Po | 1462 (5) | (25/2+) | 25.2(6) $s$ | (34(5)) | (E4) | %IT = 0.016(4)<br>%$\alpha$ = 99.984(4) | 1998MC03<br>1989KU08<br>1982BO04<br>1974BA29<br>1962PE15<br>1962KA15<br>1954SP32 |
| 84 | 127 | $^{211}$Po | 2135 (5) | (31/2−) | 243(21) $ns$ | 315.4(5)<br>672.7(5) | (M2)<br>E3 | %IT = 100 | 1998MC03 |
| 84 | 127 | $^{211}$Po | 4872 (6) | (43/2+) | 2.8(7) $\mu s$ | 508.6(5) | | %IT = 100 | 1998MC03<br>1998FO04 |
| 85 | 126 | $^{211}$At | 1270.09 (17) | (15/2)− | 12.1(15) $ns$ | 146.8(2)<br>203.3(2) | E2<br>E2 | %IT = 100 | 1995BA66<br>1975MCZO<br>1972AS04<br>1970BE37 |
| 85 | 126 | $^{211}$At | 1416.27 (25) | (21/2)− | 35.1(7) $ns$ | 95.9(2) | E2 | %IT = 100 | 1995BA66<br>1976HA62<br>1975MCZO<br>1972AS04<br>1970BE37 |
| 85 | 126 | $^{211}$At | 2640.7 (4) | (29/2)+ | 50.8(7) $ns$ | (24.4(5))<br>713.3(2) | <br>E3 | %IT = 100 | 1995BA66<br>1976HA62<br>1971MA36<br>1970BE53 |
| 85 | 126 | $^{211}$At | 4814.5 (5) | (39/2−) | 4.23(7) $\mu s$ | 435.2(2) | E3 | %IT = 100 | 2001BA79<br>1971MA36 |
| 85 | 126 | $^{211}$At | 6567.2 (6) | (49/2+) | 50.6(14) $ns$ | 549.9(2) | E2 | %IT = 100 | 2001BA79 |
| 85 | 126 | $^{211}$At | 7517.4 (6) | (55/2−) | 24.3(14) $ns$ | 950.2(2) | | %IT = 100 | 2001BA79 |
| 86 | 125 | $^{211}$Rn | 1577.8+X | (17/2−) | 596(28) $ns$ | | | %IT = 100 | 1993DA10<br>1981PO08 |

*Continued...*

Table 1 contd…

| Z | N | $^{A}$X | E(keV) | J$^{\pi}$ | $T_{1/2}$ | E$\gamma$(keV) | $\lambda$ | Decay mode | Reference |
|---|---|---|---|---|---|---|---|---|---|
| 86 | 125 | $^{211}$Rn | 3926.0+X | (35/2+) | 40.2(14) *ns* | (52.1)<br>81.7(2) | <br>E2 | %IT = 100 | 1981PO08 |
| 86 | 125 | $^{211}$Rn | 5245.9+Y | (43/2−) | 14(2) *ns* | | | %IT = 100 | 1985PO06 |
| 86 | 125 | $^{211}$Rn | 6100.1+Y | (49/2+) | 28.4(14) *ns* | 366.3(2)<br>854.3(2) | M2<br>E3 | %IT = 100 | 1985PO13 |
| 86 | 125 | $^{211}$Rn | 8854.5+Y | (63/2−) | 201(4) *ns* | 687.0(2) | E3 | %IT = 100 | 1985PO06 |
| 87 | 124 | $^{211}$Fr | 2423.16 (24) | (29/2+) | 146(14) *ns* | 112.9(1)<br>563.3(3) | (E2)<br>(E3) | %IT = 100 | 1986BY01 |
| 87 | 124 | $^{211}$Fr | 4657.3 (4) | (45/2−) | 124.5(12) *ns* | 728.4(2) | E3 | %IT = 100 | 2016MA41<br>1986BY01 |
| 88 | 123 | $^{211}$Ra | 1198.1 (8) | (13/2+) | 9.4(4) $\mu s$ | 396.1(6) | M2 | %IT = 100 | 2013BA29<br>2006HA17 |
| 90 | 121 | $^{211}$Th | 778 (8) | 13/2+ | 83(8) $\mu s$ | 203(5) | M2 | %IT = 96(3)<br>%$\alpha$ = 4(3) | 2021AU03 |
| 82 | 130 | $^{212}$Pb | 1335 (10)? | (8+) | 6.0(8) $\mu s$ | (58(10)) | | | 2012GO19<br>1998PF02 |
| 83 | 129 | $^{212}$Bi | 239 (30) | (8−,<br>9−) | 25.0(2) *min* | | | %$\beta$−= 33(1)<br>%$\beta$−$\alpha$ = 30(1)<br>%$\alpha$ = 67(1) | 2013CH12<br>1984ES01 |
| 83 | 129 | $^{212}$Bi | 1478 (30) | (18−) | 7.0(3) *min** | | | %IT ≈ 75<br>%$\beta$−≈ 25 | 2013CH12<br>1984ES01 |
| 84 | 128 | $^{212}$Po | 1476.39 (17) | 8+ | 14.6(3) *ns* | 120.9(1) | E2 | %IT ≈ 58<br>%$\alpha$ ≈ 42 | 2010AS03<br>1981BO29<br>1978LI14 |
| 84 | 128 | $^{212}$Po | 2930 (10) | (18+) | 45.1(6) *s* | (45(10)) | | %IT = 0.07(2)<br>%$\alpha$ = 99.93(2) | 1962PE15 |
| 85 | 127 | $^{212}$At | 222.9 (4) | (9−) | 0.121(2) *s* | | | %$\alpha$ = 99.5(5)<br>%IT = 0.5(5) | 2007KU30<br>1999HO28<br>1974BA29<br>1970RE02 |
| 85 | 127 | $^{212}$At | 275.2 (10) | (5−) | 32(1) *ns* | 69.9 | (E2) | %IT = 100 | 1999BA30<br>1998BY01<br>1979SJ01 |
| 85 | 127 | $^{212}$At | 885.4 (4) | (11+) | 18.7(7) *ns* | 183.9(2)<br>662.5(2) | (E1)<br>M2 | %IT = 100 | 1999BA30<br>1998BY01<br>1979SJ01 |
| 85 | 127 | $^{212}$At | 1604.3 (9) | (15−) | 35.4(14) *ns* | 63.9 | (E2) | %IT = 100 | 1999BA30<br>1998BY01<br>1979SJ01 |
| 85 | 127 | $^{212}$At | 2250.0 (11) | (18+) | 42(2) *ns* | (37.7)<br>295.4<br>486<br>645.5 | (E2)<br>M2<br><br>E3 | %IT = 100 | 1999BA30<br>1998BY01<br>1979SJ01 |
| 85 | 127 | $^{212}$At | 4771.4 (15) | (25−) | 152(5) $\mu s$ | 224.2<br>1265.4? | E3 | %IT = 100 | 1999BA30<br>1998BY01<br>1979SJ01 |
| 86 | 126 | $^{212}$Rn | 1639.68 (15) | 6+ | 118(14) *ns* | 138.3(1) | E2 | %IT = 100 | 2009DR12<br>1988ST17 |

*Continued…*

*: 2013CH12 gives $T_{1/2} \geq 30$ *min* for H-, He- and Li-like ions of $^{212}$Bi.

Table 1 contd…

| Z | N | $^{A}X$ | E(keV) | J$^{\pi}$ | $T_{1/2}$ | Eγ(keV) | λ | Decay mode | Reference |
|---|---|---|---|---|---|---|---|---|---|
| 86 | 126 | $^{212}$Rn | 1694.1 (3) | 8+ | 0.91(3) $\mu s$ | 54.2(3) | (E2) | %IT = 100 | 2009DR12<br>1988ST17<br>1976HA62<br>1971MAXH |
| 86 | 126 | $^{212}$Rn | 4066.6 (3) | 17− | 28.9(14) $ns$ | 75.7(2)<br>709.1(1) | (E2)<br>E3 | %IT = 100 | 2009DR12<br>1988ST17 |
| 86 | 126 | $^{212}$Rn | 6174.2 (3) | (22+) | 101.8(35) $ns$ | (7.5(6))<br>402.5(3)<br>747.3(1) | E2<br>E3<br>E2 | %IT = 100 | 2009DR12<br>1988ST17<br>1977HO17 |
| 86 | 126 | $^{212}$Rn | 7142.0 (3) | (25−) | 18.0(6) $ns$ | 432.5(2)<br>967.8(1) | (M2)<br>E3 | %IT = 100 | 2009DR12<br>1977HO17 |
| 86 | 126 | $^{212}$Rn | 7878.1 (5) | (27−) | 14(4) $ns$ | (15.5(6))<br>(59.2(6)<br>353.8(3)<br>736.3(1) | M1<br><br><br>E2 | %IT = 100 | 2009DR12<br>1977HO17 |
| 86 | 126 | $^{212}$Rn | 8579.2 (4) | (30+) | 154(14) $ns$ | (21.9(6))<br>(81.9(6))<br>700.9(1) | (E2)<br>E2<br>E3 | %IT = 100 | 2009DR12<br>1977HO17 |
| 86 | 126 | $^{212}$Rn | 10619.5 (4) | (34−) | ~ 20 $ns$ | 923.7(1) | (M1) | %IT = 100 | 2009DR12<br>1990DR07 |
| 86 | 126 | $^{212}$Rn | 12211.3 (4) | (37−) | 17.3(14) $ns$ | 158.4(3)<br>856.7(2)<br>949.5(3) | <br>(E2)<br> | %IT = 100 | 2009DR12<br>2008DR01 |
| 87 | 125 | $^{212}$Fr | 1551.4 (3) | (11+) | 31.9(7) $\mu s$ | 162.0(1) | (E2) | %IT = 100 | 1990BY02 |
| 87 | 125 | $^{212}$Fr | 2492.2 (4) | (15−) | 604(28) $ns$ | 612.4(1) | E3 | %IT = 100 | 1986BY01 |
| 87 | 125 | $^{212}$Fr | 5854.7 (6) | (27−) | 312(21) $ns$ | 588.3(2) | E3 | %IT = 100 | 1986BY01 |
| 87 | 125 | $^{212}$Fr | 8533.4 (11) | | 23.6(21) $\mu s$ | 126<br>278 | <br>(E3) | %IT = 100 | 1990BY02 |
| 88 | 124 | $^{212}$Ra | 1958.4 (20) | (8)+ | 9.3(9) $\mu s$ | 63.3(20) | | %IT = 100 | 2013BA29<br>2006HE17<br>1986KO01 |
| 88 | 124 | $^{212}$Ra | 2613.3 (20) | (11)− | 0.85(13) $\mu s$ | (36.1)<br>504.9(2)<br>654.9(2) | <br>(E3)<br>(E3) | %IT = 100 | 2018PA04<br>2013BA29<br>1986KO01 |
| 88 | 124 | $^{212}$Ra | 5043.5 (20) | (19+) | 21.5(21) $ns$ | 490.9(1) | (E1) | %IT = 100 | 2018PA04 |
| 81 | 132 | $^{213}$Tl | 380+X | | 4.1(5) $\mu s$ | 380(1) | | %IT = 100 | 2019GO10 |
| 81 | 132 | $^{213}$Tl | 698+X | | 0.6(3) $\mu s$ | 698(1) | | %IT = 100 | 2019GO10 |
| 82 | 131 | $^{213}$Pb | 1331 | (21/2+) | 0.26(2) $\mu s$ | 71<br>190 | | %IT = 100 | 2021VA03 |
| 85 | 128 | $^{213}$At | 1318.1+X | | 110(17) $ns$ | | | | 1980SJ01 |
| 85 | 128 | $^{213}$At | 1319+Y | (27/2−) | 85 $ns$ | | | | 2003LAZZ |
| 85 | 128 | $^{213}$At | 1838+Y | (33/2+) | 82 $ns$ | 156<br>520 | | %IT = 100 | 2003LAZZ |
| 85 | 128 | $^{213}$At | 2620+Y | (43/2−) | 34.7 $ns$ | (50) | | | 2003LAZZ |
| 85 | 128 | $^{213}$At | 2926+Y | (49/2+) | 45(4) $\mu s$ | 306 | (E3) | %IT = 100 | 2003LAZZ |
| 86 | 127 | $^{213}$Rn | 896.1 (8) | (15/2−) | 26(1) $ns$ | 191.1<br>896.1 | M2<br>E3 | %IT = 100 | 1988ST10 |
| 86 | 127 | $^{213}$Rn | 1664.0 (10) | (21/2+) | 29(2) ns | 135.0<br>767.9 | E2<br>(E3) | %IT = 100 | 1988ST10 |
| 86 | 127 | $^{213}$Rn | 1664.0+X | (25/2+) | 1.00(21) $\mu s$ | | | | 1988ST10 |

*Continued…*

Table 1 contd...

| Z | N | $^{A}X$ | E(keV) | J$^{\pi}$ | $T_{1/2}$ | E$\gamma$(keV) | $\lambda$ | Decay mode | Reference |
|---|---|---|---|---|---|---|---|---|---|
| 86 | 127 | $^{213}$Rn | 2186.7+X | (31/2−) | 1.36(7) $\mu s$ | (65.1) | | %IT = 100 | 1988ST10 |
| | | | | | | 330.1 | (E3) | | |
| | | | | | | 522.7 | E3 | | |
| 86 | 127 | $^{213}$Rn | 3029.3+X | (37/2+) | 26(1) $ns$ | 45.3 | E2 | %IT = 100 | 1988ST10 |
| | | | | | | 113.5 | E2 | | |
| | | | | | | 842.6? | | | |
| 86 | 127 | $^{213}$Rn | 3495.4+X | (43/2−) | 28(1) $ns$ | 54.3 | E2 | %IT = 100 | 1988ST10 |
| 86 | 127 | $^{213}$Rn | 4505.5+X | (49/2+) | 12(1) $ns$ | 457.6 | | %IT = 100 | 1988ST10 |
| | | | | | | 1010.1 | E3 | | |
| 86 | 127 | $^{213}$Rn | 5929+Y | (55/2+) | 164(11) $ns$ | | | | 1988ST10 |
| 86 | 127 | $^{213}$Rn | 6744.00+Y | | 59 ns | 815.0(2) | | %IT = 100 | 1988ST10 |
| 87 | 126 | $^{213}$Fr | 1411.01 (15) | 17/2− | 18(1) $ns$ | 222.2(1) | E2 | %IT = 100 | 1986BY01 |
| 87 | 126 | $^{213}$Fr | 1590.41 (18) | 21/2− | 505(14) $ns$ | 179.4(1) | E2 | %IT = 100 | 1986BY01 |
| | | | | | | | | | 1976HA37 |
| 87 | 126 | $^{213}$Fr | 2537.62 (23) | 29/2+ | 238(6) $ns$ | 681.3(1) | E3 | %IT = 100 | 1976HA37 |
| 87 | 126 | $^{213}$Fr | 4992.7 (4) | 45/2− | 13(2) $ns$ | 94.4(3) | E2 | %IT = 100 | 1986BY01 |
| | | | | | | 909.8(2) | E3 | | 1979HO06 |
| 87 | 126 | $^{213}$Fr | 8094.8 (7) | (65/2−) | 3.1(2) $\mu s$ | 111.3(2) | | %IT = 100 | 1989BY01 |
| | | | | | | 371.2(2) | (E3) | | |
| 88 | 125 | $^{213}$Ra | 1770 (7) | (17/2−) | 2.20(5) $ms$ | (161(7)) | | %IT ≈ 99 | 2006KU26 |
| | | | | | | | | %$\alpha$ ≈ 1 | 2004HE25 |
| | | | | | | | | | 1976RA37 |
| 88 | 125 | $^{213}$Ra | 2609.7+X (2) | 23/2+ | 18.7(21) $ns$ | 322.4(1) | E1 | %IT = 100 | 2018PA04 |
| 88 | 125 | $^{213}$Ra | 4047.5+X (7) | 33/2+ | 34.7(21) $ns$ | (40.8(9)) | | %IT = 100 | 2018PA04 |
| | | | | | | 169.7(3) | E2 | | |
| | | | | | | 606.3(5) | | | |
| 90 | 123 | $^{213}$Th | 1180.0 (15) | (13/2)+ | 8.30(82) $\mu s$ | 381(1) | M2 | %IT = 100 | 2021ZH24 |
| | | | | | | | | | 2007KH22* |
| 82 | 132 | $^{214}$Pb | 1365+X | (8+) | 6.2(3) $\mu s$ | | | | 2012GO19 |
| 83 | 131 | $^{214}$Bi | 0+X# | (8–) | 9.39(10) min | | | %$\beta$−≈ 100 | 2021AN14 |
| | | | | | | | | %IT = ? | 2008CHZI$ |
| 85 | 129 | $^{214}$At | 59 (9) ? | | 265(30) $ns$ | | | %$\alpha$ < 100 | 1982EW01 |
| 85 | 129 | $^{214}$At | 232 (6) | 9− | 760(15) $ns$ | | | %$\alpha$ ≤ 100 | 1982EW01 |
| 86 | 128 | $^{214}$Rn | 3490 (3) | 18+ | 44(3) $ns$ | 162.2 | E2 | %IT = 100 | 1983LO01 |
| 86 | 128 | $^{214}$Rn | 4595 (4) | (22+) | 245(30) $ns$ | 767.7 | (M2) | %IT = 100 | 1987DR08 |
| | | | | | | | | | 1983LO01 |
| 87 | 127 | $^{214}$Fr | 121 (5) | (8−) | 3.38(5) $ms$ | | | %$\alpha$ = 100 | 2015KH09 |
| | | | | | | | | | 1968TO10 |
| | | | | | | | | | 1968VA18 |
| 87 | 127 | $^{214}$Fr | 638 (5) | (11+) | 103(4) $ns$ | 471.7(1) | M2 | %IT = 100 | 1994BY01 |
| | | | | | | 516.6(2) | | | |
| 87 | 127 | $^{214}$Fr | 1661 (5) | (14−) | 11(2) $ns$ | 63.9(2) | M1 | %IT = 100 | 1994BY01 |
| | | | | | | 580.6(2) | M2 | | |
| | | | | | | 1023.1(2) | E3 | | |
| 87 | 127 | $^{214}$Fr | 1732 (5) | (15−) | 10(2) $ns$ | 71.4(2) | M1 | %IT = 100 | 1994BY01 |
| | | | | | | 135.1(2) | E2 | | |
| 88 | 126 | $^{214}$Ra | 1639.30 (14) | 4+ | 35.1(3) $ns$ | 257.0(1) | E2 | %IT = 100 | 1992ST09 |

*Continued...*

*: 2007KH22 gives a low half-life of 1.4(4) $\mu s$.

#: 2021AN14 gives the energy of ~100 keV from systematics.

$: 2008CHZI gives an isomer at 539(30) keV with half-life of >93 $s$ for a Li-like ion of $^{214}$Bi.

Table 1 contd...

| Z | N | $^{A}X$ | E(keV) | J$^{\pi}$ | $T_{1/2}$ | E$\gamma$(keV) | $\lambda$ | Decay mode | Reference |
|---|---|---|---|---|---|---|---|---|---|
| 88 | 126 | $^{214}$Ra | 1819.70 (17) | 6+ | 118(7) *ns* | 180.4(1) | E2 | %IT = 100 | 1992ST09 |
| 88 | 126 | $^{214}$Ra | 1865.2 (11) | 8+ | 67.8(20) $\mu s$ | 45.5 | E2 | %$\alpha$ = 0.09(7)<br>%IT = 99.91(7) | 2006KU26<br>1992ST09<br>1974YA09<br>1971MAXI |
| 88 | 126 | $^{214}$Ra | 2683.2 (14) | 11− | 295(7) *ns* | 609.3<br>818 | E3<br>E3 | %IT = 100 | 1992ST09 |
| 88 | 126 | $^{214}$Ra | 3478.4 (16) | 14+ | 279(4) *ns* | 148.9<br>222.1 | E2<br>E2 | %IT = 100 | 1992ST09 |
| 88 | 126 | $^{214}$Ra | 4146.8 (17) | 17− | 225(4) *ns* | 156.6<br>668.4 | E2<br>E3 | %IT = 100 | 1992ST09 |
| 88 | 126 | $^{214}$Ra | 6577.0 (24) | (25−) | 128(4) *ns* | 46.8 | E1 | %IT = 100 | 1992ST09 |
| 90 | 124 | $^{214}$Th | 2181.0 (27) | (8+) | 1.21(12) $\mu s$ | 89(2) | (E2) | %IT = 100 | 2021ZH24<br>2007KH22 |
| 83 | 132 | $^{215}$Bi | 1347.50+X | (25/2−,<br>27/2−,<br>29/2−) | 36.9(6) *s* | | | %IT = 76.9(5)<br>%$\beta$−= 23.1(5) | 2003KU26 |
| 86 | 129 | $^{215}$Rn | 1804.8+X | | 57(+21−12) *ns* | | | %IT = 100 | 2013BO18<br>2012BOZU |
| 87 | 128 | $^{215}$Fr | 3068.9 (4) | (39/2)− | 13.0(16) *ns* | 262.01(15)<br>817.53(15) | E2<br>(E3) | %IT = 100 | 2022YA10<br>1984SC25 |
| 87 | 128 | $^{215}$Fr | 3462.3 (6) | (47/2+) | 22.9(21) *ns* | 45.2(3) | (E1) | %IT = 100 | 1984SC25 |
| 88 | 127 | $^{215}$Ra | 773.0 (2) | (15/2−) | 67.2(14) *ns* | 773.0(2) | E3 | %IT = 100 | 1998ST24<br>1989DR02 |
| 88 | 127 | $^{215}$Ra | 1821.2 (3) | (21/2+) | 25.0(14) *ns* | 196.0(2)<br>1048.2(2) | E2<br>E3 | %IT = 100 | 1998ST24 |
| 88 | 127 | $^{215}$Ra | 1877.8 (3) | (25/2+) | 7.29(20) $\mu s$ | 56.5(2) | E2 | %IT = 100 | 2004HE25<br>1998ST24<br>1988FU10 |
| 88 | 127 | $^{215}$Ra | 2246.9 (4) | (29/2−) | 1.39(7) $\mu s$ | (32.5)<br>193.1(2)<br>369.1(2) | <br>M2(+E3)<br>M2+E3 | %IT = 100 | 1998ST24 |
| 88 | 127 | $^{215}$Ra | 3756.6+X | (43/2−) | 555(10) *ns* | (18.0)<br>425.5(2) | <br>E3 | %IT = 100 | 1998ST24 |
| 88 | 127 | $^{215}$Ra | 4567.0+X | (49/2+) | 10.47(14) *ns* | (13.5)<br>200.1(2)<br>810.2(2) | <br>(E2)<br>(E3) | %IT = 100 | 1998ST24 |
| 89 | 126 | $^{215}$Ac | 1621.0 (7) | (17/2−) | 30(10) *ns* | 304.0(5) | (E2) | %IT = 100 | 1983DE08 |
| 89 | 126 | $^{215}$Ac | 1796.0 (9) | (21/2−) | 185(30) *ns* | 175.0(5) | (E2) | %IT = 100 | 1983DE08 |
| 89 | 126 | $^{215}$Ac | 2438+X | (29/2+) | 335(10) *ns* | 642.0(5) | (E3) | %IT = 100 | 1983DE08 |
| 90 | 125 | $^{215}$Th | 1421.3+X ? | | 0.75(6) $\mu s$ | | | %IT ≈ 100 | 2021ZH24<br>2005KU31 |
| 82 | 134 | $^{216}$Pb | 1459+X | (8+) | 0.40(4) $\mu s$ | | | | 2012GO19 |
| 83 | 133 | $^{216}$Bi | X | (3) | 6.6(21) *min* | | | %$\beta$−≤ 100 | 1989BU09 |
| 87 | 129 | $^{216}$Fr | 133.3 (1) | (3−) | 71(5) *ns* | 133.3(1) | E2 | %$\alpha$ > 50 | 1971EPZY |
| 87 | 129 | $^{216}$Fr | 219 (6) | (9−) | 850(30) *ns* | | | %$\alpha$ ≈ 100 | 2007KU30 |
| 87 | 129 | $^{216}$Fr | 2200+X | | 89(9) *ns* | | | | 2022MA16 |
| 89 | 127 | $^{216}$Ac | 48 (7) | (9−) | 441(7) $\mu s$ | | | %$\alpha$ = 100<br>%$\epsilon$+%$\beta$+ ≈ 7E–5 | 2000HE17 |
| 89 | 127 | $^{216}$Ac | 322+X | | ~ 300 *ns* | 322 | | %IT = 100 | 2006POZX |

*Continued...*

Table 1 contd…

| Z | N | $^{A}$X | E(keV) | J$^{\pi}$ | $T_{1/2}$ | Eγ(keV) | λ | Decay mode | Reference |
|---|---|---|---|---|---|---|---|---|---|
| 90 | 126 | $^{216}$Th | 2040 (9) | 8+ | 133(4) $\mu s$ | | | %IT = 97.2(9) | 2019ZH54 |
| | | | | | | | | %α = 2.8(9) | 2005KU31 |
| | | | | | | | | | 2001HA46 |
| | | | | | | | | | 1983HI08 |
| 90 | 126 | $^{216}$Th | 2646.8 (1) | 11− | 0.58(3) $\mu s$ | 516.3(2) | | %IT = 100 | 2005KU31 |
| | | | | | | 606.8(1) | E3 | | 2001HA46 |
| 90 | 126 | $^{216}$Th | 3681.4 (7) | (14+) | 0.74(7) $\mu s$ | 151.2(6) | E2 | %IT = 100 | 2005KU31 |
| 92 | 124 | $^{216}$U | 2240 (42) | (8+) | 0.89(+27−17) $ms$ | | | %α = 100 | 2022ZH45 |
| | | | | | | | | | 2015MA37 |
| 83 | 134 | $^{217}$Bi | 1436+X | (25/2−) | 3.0(2) $\mu s$ | | | %IT = 100 | 2014GO20 |
| 89 | 128 | $^{217}$Ac | 2013 | (29/2)+ | 740(40) $ns$ | 96 | E1+M2 | %IT = 95.7(10) | 1985DE14 |
| | | | | | | 220 | M2 | %α = 4.3(10) | |
| 90 | 127 | $^{217}$Th | 673.8 | (15/2−) | 141(50) $ns$ | 673.8 | (E3) | %IT = 100 | 1989DR02 |
| 90 | 127 | $^{217}$Th | 2251.9+X | (25/2+) | 67(+17−11) $\mu s$ | | | | 2017BRAA |
| | | | | | | | | | 2005KU31 |
| | | | | | | | | | 2003MUZV |
| 91 | 126 | $^{217}$Pa | 1854 (7) | (23/2−) | 1.2(2) $ms$ | | | %IT = 27(4) | 2002HE29 |
| | | | | | | | | %α = 73(4) | 1998IK01 |
| | | | | | | | | | 1995NIZS |
| | | | | | | | | | 1979SC09 |
| 87 | 131 | $^{218}$Fr | 86 (8) | (8−, 9−) | 22.0(5) $ms$ | | | %IT = ? | 2014BU06 |
| | | | | | | | | %α ≤ 100 | 1982EW01 |
| 89 | 129 | $^{218}$Ac | 122.5+Y | (9−) | 32(9) $ns$* | | | %IT = 100 | 1994DE04 |
| 89 | 129 | $^{218}$Ac | 506.99+Y (13) | (11+) | 103(11) $ns$ | 91.0(2) | (E1) | %IT = 100 | 1994DE04 |
| | | | | | | 384.5(2) | M2 | | |
| 91 | 127 | $^{218}$Pa | 83 (6) | (1−) | 135 (+62–32) $\mu s$ | | | %α ≈ 100 | 2020ZH01 |
| 92 | 126 | $^{218}$U | 2105 (19) | (8+) | 0.40(+6−5) $ms$ | | | %α ≈ 100 | 2022ZH45 |
| | | | | | | | | | 2018YA01 |
| | | | | | | | | | 2015MA37 |
| | | | | | | | | | 2007LE14 |
| 86 | 133 | $^{219}$Rn | 4.413 (12) | (9/2)+ | 15.4(13) $ns$ | 4.413(12) | (E2) | %IT = 100 | 1974RI05 |
| 88 | 131 | $^{219}$Ra | 16.6 (2) | (11/2+) | 10(3) $ms$ | | | %α ≈ 100 | 2018SA45 |
| 91 | 129 | $^{220}$Pa | 34 (26) | | 0.31(+25–10) $\mu s$ | | | %α ≈ 100 | 2018HU13 |
| 91 | 129 | $^{220}$Pa | 297 (65) | | 0.07(+33–3) $\mu s$ | | | %α ≈ 100 | 2018HU13 |
| 89 | 133 | $^{222}$Ac | 0+X | | 63(3) $s$ | | | %α = 94(5) | 1982BO04 |
| | | | | | | | | %IT = 5(5) | 1973MO07 |
| | | | | | | | | %ϵ+%β+ = 1.35(65) | 1972ES03 |
| 89 | 138 | $^{227}$Ac | 27.369 (11) | 3/2+ | 38.3(3) $ns$ | 27.36(2) | E1(+M2) | %IT = 100 | 1985IS03 |
| 88 | 141 | $^{229}$Ra | 142.67 (6) | 1/2+ | 17.23(12) $ns$ | 35.6 | (E2) | %IT = 100 | 1999FR33 |
| | | | | | | 142.7(1) | E2 | | 1992BO05 |

*Continued…*

*: No half-life was associated with a level in $^{218}$Ac by 1994DE04. Present assignment is based on systematic occurrence of (1−) ground states and (9−) isomers in $^{216}$Ac, $^{214}$At and possibly in $^{212}$Bi.

Table 1 contd…

| Z | N | $^{A}$X | E(keV) | J$^{\pi}$ | $T_{1/2}$ | E$\gamma$(keV) | $\lambda$ | Decay mode | Reference |
|---|---|---|---|---|---|---|---|---|---|
| 90 | 139 | $^{229}$Th | 0.008338(24)* | (3/2+) | 7.0(10) $\mu s$ # <br> 5.0(11) $\times 10^3$ $s$ | 0.00812(11) | | %IT = 100 | 2020VO15 <br> 2017SE01 <br> 2016WE07 <br> 2007BE16 <br> 2010BEZZ <br> 2021SH28 <br> 2019SH38 <br> 2019YA18 <br> 2019MA63 <br> 2009IN01 <br> 2009KI14 |
| 91 | 138 | $^{229}$Pa | 0+X | | 420(30) ns | | | %IT = 100 | 1982AH08 |
| 88 | 143 | $^{231}$Ra | 66.21 (9) | (1/2+) | ~ 53 $\mu s$ | 66.2(1) | E2 | %IT = 100 | 2001FR05 <br> 2001BO34 |
| 89 | 142 | $^{231}$Ac | 116.00 (7) | (3/2−) | 14.3(11) $ns$ | 41.27(5) <br> 47.45(5) <br> 54.29(5) <br> 77.97(6) | <br> <br> E1 <br> E1 | %IT = 100 | 2008BO29 <br> 2007BO48 |
| 91 | 140 | $^{231}$Pa | 84.2148 (13) | 5/2+ | 45.1(13) $ns$ | 25.64(2) <br> 84.2140(13) | E1 <br> E1 | %IT = 100 | 1999CH12 <br> 1975HO14 |
| 90 | 143 | $^{233}$Th | 1.85E+3 (25) | | 50(+50−49) $ns$ | | | %IT ≈ 100 | 1994OB02 |
| 91 | 142 | $^{233}$Pa | 86.481 (8) | 5/2+ | 36.5(4) $ns$ | 29.373(10) <br> 86.485(10) | E1 <br> E1 | %IT = 100 | 1972MC12 <br> 1972MC29 <br> 1972WI11 <br> 1971GA16 <br> 1968OB02 <br> 1961MA10 <br> 1954EN11 |
| 91 | 143 | $^{234}$Pa$^{\$}$ | 76.5 (5) | (0−) | 1.159(11) $min$ | (2.6(5)) | [E3] | %$\beta$− = 99.84(4) <br> %IT = 0.16(4) | 2016RI06 <br> 1969SAZR <br> 1963BJ02 <br> 1956ON07 <br> 1951BA83 <br> 1921HA01 <br> 1921HA02 <br> 1914GOZZ <br> 1913FA01 |

*Continued…*

*: Level energy from S. Kraemer et al., arXiv: 2209:10276v1(2022); and measured radiative $T_{1/2}$ >670(102) $s$.

#: Half-life of 7.0(10) $\mu s$ is for a neutral atom from electron measurement by 2017SE01 and 2016WE07, who also point out that for doubly-charged ions, half-life is > 60 $s$, as electron conversion process gets shut off. The radiative half-life is estimated to be $5.0(11)\times10^3 s$ by 2021SH28. For the earliest work on this isomer and follow-up papers, see book (A. K. Jain, Bhoomika Maheshwari, Alpana Goel, Nuclear Isomers −A Primer, Springer Nature, Switzerland (2021)).

$: Half-life of short-lived activity of 1.15(1) $min$ was first measured by 1913FA01 and 1914GOZZ, assigned to a new element brevium, later named as protactinium, and the activity now known as its isomer. 1921HA01 and 1921HA02 identified two activities for this element and measured half-life of 6.7(2) $h$, now known as the g.s. of $^{234}$Pa, while the short-lived activity was identified by beta absorption curve, but its half-life was not measured.

Table 1 contd...

| Z | N | $^{A}$X | E(keV) | J$^{\pi}$ | $T_{1/2}$ | E$\gamma$(keV) | $\lambda$ | Decay mode | Reference |
|---|---|---|---|---|---|---|---|---|---|
| 92 | 142 | $^{234}$U | 1421.257 (17) | 6− | 33.5(20) $\mu s$ | 143.78(2) | (M1+E2) | %IT = 100 | 1963HA30 |
| | | | | | | 159.48(2) | | | |
| | | | | | | 226.50(3) | M1+E2 | | |
| | | | | | | 249.22(1) | E1 | | |
| | | | | | | 293.79(5) | M1+E2 | | |
| | | | | | | 295.91(8) | | | |
| | | | | | | 330.40(5) | | | |
| | | | | | | 351.9(1) | E2 | | |
| | | | | | | 397.7(3) | | | |
| | | | | | | 458.68(5) | M1+E2 | | |
| | | | | | | 1125.2(1) | | | |
| | | | | | | 1277.7(2) | | | |
| 92 | 133 | $^{235}$U | 0.076737 (18) | 1/2+ | 26.5(10) $min$* | 0.0765(4) | E3 | %IT = 100 | 2018PO07 |
| | | | | | | | | | 2018SH23 |
| | | | | | | | | | 2015BE11 |
| 92 | 143 | $^{235}$U | 2.5E+3 (3) | | 11(3) $ms$ # | | | %IT = ? | 2021OB01 |
| | | | | | | | | %SF = ? | 2007OB02 |
| 94 | 141 | $^{235}$Pu | 3.0E+3 (2) | | 25(5) $ns$ | | | %SF ≤ 100 | 1972GA42 |
| | | | | | | | | | 1970BU02 |
| | | | | | | | | | 1969ME11 |
| 92 | 144 | $^{236}$U | 1052.5 (6) | (4)− | 100(4) $ns$ | 65(1) | (E2) | %IT = 100 | 1980BU13 |
| | | | | | | 204.6(10) | (E2) | | 1979MCZP |
| | | | | | | 307.9(10) | E2+M1 | | 1973BR05 |
| | | | | | | 903.2(10) | E1 | | |
| 92 | 144 | $^{236}$U | 2750 (3) | (0+) | 121(2) $ns$ $ | 560(10) | | %IT = 87(6) | 2020BA53 |
| | | | | | | 1170(10) | | %SF = 13(6) | 1990MA59 |
| | | | | | | 1783(10) | | %$\alpha$ < 10 | 1989MA57 |
| | | | | | | 2062(10) | | | 1982GO02 |
| | | | | | | ~ 2705 | | | 1978GU02 |
| | | | | | | | | | 1975CH09 |
| | | | | | | | | | 1974HEZE |
| | | | | | | | | | 1972PIZR |
| | | | | | | | | | 1972PE01 |
| | | | | | | | | | 1972HOXQ |
| | | | | | | | | | 1971BO61 |
| | | | | | | | | | 1971BE62 |
| | | | | | | | | | 1971BR38 |
| | | | | | | | | | 1970VI05 |
| | | | | | | | | | 1970RE05 |
| | | | | | | | | | 1970WO06 |
| | | | | | | | | | 1970EL03 |
| | | | | | | | | | 1969LA14 |
| 93 | 143 | $^{236}$Np | 0+X | 1 | 22.5(4) $h$ | | | %$\beta$−= 50(3) | 1969LE05 |
| | | | | | | | | %$\epsilon$ = 50(3) | 1949JA01 |

*Continued...*

*: Due to the low-energy of the electrons emitted by the isomer at 76.7 eV, significant effects on half-life due to the chemical environment are possible as discussed in 2018SH23 and 2015BE11.

#: This half-life for the decay of the shape isomer was determined in 2021OB01, which is about three times longer than reported by the same authors in 2007OB02, difference explained due to an improved analysis of the measured time distribution of the isomeric fission events.

$: The measured half-lives of this fission isomer fall in two groups, one near 120 $ns$, and the other near 70 $ns$. We adopt the value from 1989MA57, as this seems to be best documented. It is possible there are two isomers corresponding to two different half-lives.

Table 1 contd…

| Z | N | $^{A}$X | E(keV) | J$^{\pi}$ | $T_{1/2}$ | Eγ(keV) | λ | Decay mode | Reference |
|---|---|---|---|---|---|---|---|---|---|
| 94 | 142 | $^{236}$Pu | 1185.45 (15) | 5− | 1.2(3) $\mu s$ | 319.50(11)<br>879.7(2)<br>1037.8(2) | M1 | %IT = 100 | 2005AS01 |
| 94 | 142 | $^{236}$Pu | 4.0E+3 (2) | | 34(8) $ns$ | | | %SF ≤ 100 | 1971BR39<br>1969LA14 |
| 95 | 141 | $^{236}$Am | X | (1−) | 2.9(2) $min$ | | | %ϵ = ?<br>%α = ? | 2005AS01 |
| 92 | 145 | $^{237}$U | 274.0 (10) | (7/2)− | 155(6) $ns$ | 114.0(10) | E1 | %IT = 100 | 1968AH01 |
| 93 | 144 | $^{237}$Np | 59.54092 (10) | 5/2− | 67.50(25) $ns$ | 26.3446(2)<br>59.5409(1) | E1<br>E1 | %IT = 100 | 2021DU16<br>1972MI23<br>1972MC12<br>1971GA16<br>1968OB02<br>1952BE47 |
| 93 | 144 | $^{237}$Np | 945.20 (10) | 11/2,<br>13/2 | 0.71(4) $\mu s$ | 640.4(3)<br>675.6(4)<br>719.2(2)<br>753.6(2)<br>786.8(2)<br>815.3(2) | | %IT = 100 | 1990ST29 |
| 93 | 144 | $^{237}$Np | 2.8E+3 (4) | | 45(5) $ns$ | | | %SF ≤ 100 | 1977MI09<br>1973WO03 |
| 94 | 143 | $^{237}$Pu | 145.543 (8) | 1/2+ | 0.18(2) $s$ | 145.542(8) | E3 | %IT = 100 | 1975AH05<br>1957ST67 |
| 94 | 143 | $^{237}$Pu | 26.0E+2 (20) | | 103(7) $ns$ | | | %SF > 0 | 1982RA04<br>1979GU03<br>1978DE07<br>1974BA82<br>1973BR38<br>1972VI10<br>1971RE11<br>1971RU03<br>1971TE07<br>1971BR39<br>1970BU02<br>1969LA14<br>1969VAZX |
| 94 | 143 | $^{237}$Pu | 29.0E+2 (25) | | 1055(80) $ns$ | | | %SF < 100 | 1979GU03<br>1974BA82<br>1973VA16<br>1972VI10<br>1971TE07<br>1971RU03<br>1970PO01 |

*Continued…*

Table 1 contd…

| Z | N | $^{A}$X | E(keV) | J$^{\pi}$ | $T_{1/2}$ | Eγ(keV) | λ | Decay mode | Reference |
|---|---|---|---|---|---|---|---|---|---|
| 92 | 146 | $^{238}$U | 2557.9 (5) | 0+ | 267(13) *ns** | 1879 | | %IT = 97.4(4) | 1992ST05 |
| | | | | | | 2512.7(5) | | %SF = 2.6(4) | 1991KU23 |
| | | | | | | 2558(2) | E0 | | 1985DR01 |
| | | | | | | | | | 1989MA54 |
| | | | | | | | | | 1983DM04 |
| | | | | | | | | | 1983DR14 |
| | | | | | | | | | 1982GO02 |
| | | | | | | | | | 1977ARZZ |
| | | | | | | | | | 1977VOZU |
| | | | | | | | | | 1975RU03 |
| | | | | | | | | | 1974WOZW |
| | | | | | | | | | 1970PO01 |
| | | | | | | | | | 1970WO06 |
| | | | | | | | | | 1970RE05 |
| 93 | 145 | $^{238}$Np | 2300 | | 112(39) *ns* | | | %SF ≤ 100 | 1970VI05 |
| 95 | 143 | $^{238}$Am | ~ 2500 | | 35 *µs* | | | %SF ≤ 100 | 1969JOZU |
| 92 | 147 | $^{239}$U | 133.7991 (10) | 1/2+ | 0.78(4) *µs* | 133.799(1) | E2 | %IT = 100 | 1975YA03 |
| 92 | 147 | $^{239}$U | 0+X | (5/2+) | > 0.25 *µs* | 708.2 | | | 1994OB01 |
| | | | | | | 1600.3 | | | |
| 94 | 145 | $^{239}$Pu | 391.584 (3) | 7/2− | 190.2(2) *ns* | (4.2) | | %IT = 100 | 2022BA06 |
| | | | | | | 61.460(2) | E1 | | 1974PA03 |
| | | | | | | 106.125(2) | E1(+M2) | | 1955EN07 |
| | | | | | | 315.880(3) | E1(+M2) | | |
| | | | | | | 334.310(3) | E1(+M2) | | |
| 94 | 145 | $^{239}$Pu | 31.E+2 (2) | (5/2+) | 7.5(10) *µs* | | | %SF ≤ 100 | 1977GOZH |
| 95 | 144 | $^{239}$Am | 25.E+2 (2) | (7/2+) | 163(12) *ns* | | | %SF ≤ 100 | 1972BR35 |
| | | | | | | | | | 1971BR38 |
| 93 | 147 | $^{240}$Np | 0+X | (1+) | 7.22(2) *min* | | | %IT = 0.12(1) | 1981HS02 |
| | | | | | | | | %β−= 99.88(1) | |
| 94 | 146 | $^{240}$Pu | 1308.74 (5) | (5−) | 165(10) *ns* | 147.20(10) | (M1, E2) | %IT = 100 | 1967WA27 |
| | | | | | | 193.30(10) | (M1, E2) | | |
| | | | | | | 271.30(10) | (M1, E2) | | |
| | | | | | | 306.80(10) | (E2) | | |
| | | | | | | 566.34(6) | (M1, E2) | | |
| | | | | | | 1014.40(10) | | | |
| | | | | | | 1167.10(10) | | | |
| 95 | 145 | $^{240}$Am | 3.0E+3 (2) | | 0.94(4) *ms* | | | %SF ≤ 100 | 1979BE46 |
| 96 | 144 | $^{240}$Cm | ~ 3000 | | 55(12) *ns* | | | %SF ≈ 100 | 1976SL01 |
| 94 | 147 | $^{241}$Pu | 161.6853 (9) | 1/2+ | 0.88(5) *µs* | 161.685(1) | E2 | %IT = 100 | 1975YA03 |
| 94 | 147 | $^{241}$Pu | ~ 2200 | | 21(3) *µs* | | | %SF = 100 | 1981GU04 |
| | | | | | | | | | 1970GA10 |
| | | | | | | | | | 1970PO01 |
| 94 | 147 | $^{241}$Pu | 2200+X | | 32(5) *ns* | | | %SF = 100 | 1981GU04 |
| | | | | | | | | | 1969LA14 |
| 95 | 146 | $^{241}$Am | ~ 2200 | | 1.2(3) *µs* | | | %SF = 100 | 1993KU16 |
| | | | | | | | | | 1969LA14 |

*Continued…*

*: As for $^{236}$U, the measured half-life of this fission isomer fall in two groups, one near 250 *ns*, and the other near 150 *ns*. We adopt the value from 1989MA54, as this seems to be best documented. It is possible there are two isomers corresponding to two different half-lives.

Table 1 contd...

| Z | N | $^{A}X$ | E(keV) | J$^{\pi}$ | $T_{1/2}$ | E$\gamma$(keV) | $\lambda$ | Decay mode | Reference |
|---|---|---|---|---|---|---|---|---|---|
| 96 | 145 | $^{241}$Cm | ~ 2300 | | 15.3(10) *ns* | | | %SF = 100 | 1974SPZS<br>1972GA42<br>1971BR39<br>1971RE11<br>1970PO01<br>1969ME11 |
| 93 | 149 | $^{242}$Np | 0+Y | (6+) | 5.5(1) *min* | | | %$\beta$−= 100 | 1981FR07 |
| 94 | 148 | $^{242}$Pu | ~ 2000+Y | | 28 *ns* | | | %SF ≤ 100 | 1970PO01 |
| 95 | 147 | $^{242}$Am | 48.60 (5) | 5− | 141(2) *y* | 48.63(5) | E4 | %IT = 99.55(2)<br>%$\alpha$ = 0.45(2)<br>%SF < 4.7E−9 | 1979ZE05<br>1959BA22 |
| 95 | 147 | $^{242}$Am | 2200 (80) | (2+,<br>3−) | 14.0(10) *ms* | 2200(80)? | | %SF ≈ 100<br>%IT > 0.0?<br>%$\alpha$ > 0.0? | 1996BA52<br>1992MA34<br>1981VAZQ<br>1975VA21<br>1968ER01<br>1966BR23<br>1965LI05<br>1965FL04<br>1963PE27<br>1963FL08<br>1962PO09 |
| 96 | 146 | $^{242}$Cm | ~ 2800 | | 180(70) *ns* | | | %SF = ?<br>%IT = ? | 1971RE11 |
| 97 | 145 | $^{242}$Bk | 0+Y | | 600(100) *ns* | | | %SF ≤ 100 | 1972WO07 |
| 94 | 149 | $^{243}$Pu | 383.64 (25) | (1/2+) | 0.33(3) $\mu s$ | 96.2(2) | | %IT = 100 | 1975YA03 |
| 94 | 149 | $^{243}$Pu | 1.7E+3 (3) | | 46(13) *ns* | | | %SF = 100 | 1970PO01<br>1970VI05 |
| 95 | 148 | $^{243}$Am | 2.3E+3 (2) | | 5.5(5) $\mu s$ | | | %SF ≤ 100 | 1973NA35<br>1972WO07<br>1970PO01 |
| 96 | 147 | $^{243}$Cm | 87.4 (1) | 1/2+ | 1.08(3) $\mu s$ | 87.4(1) | E2 | %IT = 100 | 1975YA03 |
| 96 | 147 | $^{243}$Cm | 1.9E+3 (3) | | 42(6) *ns* | | | %SF ≤ 100 | 1972WO05 |
| 98 | 145 | $^{243}$Cf | ~315 | | 5.1 *s* | | | | 2021KH07 |
| 94 | 150 | $^{244}$Pu | 1211.2 (8) | 8− | 1.75(12) *s* | (10(1))<br>681.0(1) | | | 2016HO13 |
| 95 | 149 | $^{244}$Am | 89.5 (16)* | 1+ | 26.13(43) min | | | %$\beta$−= 99.9639(13)<br>%$\epsilon$ = 0.0361(13) | 2019TR05<br>1984VO07<br>1954GH24<br>1950ST61 |
| 95 | 149 | $^{244}$Am | 0+X | | 0.90(15) *ms* | | | %SF ≤ 100 | 1972WO07<br>1969BO25<br>1968BJ04<br>1967FL08 |
| 95 | 149 | $^{244}$Am | 0+Y | | ~ 6.5 $\mu s$ | | | %SF ≤ 100 | 1969SLZZ |
| 96 | 148 | $^{244}$Cm | 1040.188 (12) | 6+ | 34(2) *ms* | 538.400(16)<br>743.971(5)<br>897.848(7) | (E2)<br>M1+E2<br>E2 | %IT = 100 | 1963HA29 |
| 96 | 148 | $^{244}$Cm | 0+X | | > 500 *ns* | | | %SF ≤ 100 | 1969MEZX |

*Continued...*

*: Energy of the isomer is from primary E$\gamma$ in 1984VO07, and for S(n) from AME2020.

Table 1 contd...

| Z | N | $^{A}$X | E(keV) | J$^{\pi}$ | $T_{1/2}$ | E$\gamma$(keV) | $\lambda$ | Decay mode | Reference |
|---|---|---|---|---|---|---|---|---|---|
| 97 | 147 | $^{244}$Bk | 0+X | | 820(60) *ns* | | | %SF ≤ 100 | 1972GA42<br>1972WO07 |
| 101 | 143 | $^{244}$Md | 0+X | | 5(+3−2) *ms* | | | %SF = ?<br>%$\alpha$ = ?<br>%$\epsilon$+%$\beta$+ = ? | 1996NI09 |
| 94 | 151 | $^{245}$Pu | ∽ 311 | (1/2+) | 0.33(2) $\mu s$ | (∽ 47) | [E2] | %IT = 100 | 2007MA82 |
| 94 | 151 | $^{245}$Pu | 2.0E+3 (4) | | 90(30) *ns* | | | %SF ≤ 100 | 1980BJ02 |
| 95 | 150 | $^{245}$Am | 2.4E+3 (4) | | 640(60) *ns* | | | %SF ≤ 100 | 1972WO07 |
| 96 | 149 | $^{245}$Cm | 355.92 (10) | 1/2+ | 0.29(2) $\mu s$ | 103.1(1) | E2 | %IT = 100 | 1975YA03 |
| 96 | 149 | $^{245}$Cm | 2.1E+3 (3) | | 13.2(18) *ns* | | | %SF ≤ 100 | 1972WO07<br>1971BR39 |
| 101 | 144 | $^{245}$Md | 0+X | | 0.90(25) *ms* | | | %SF ≈ 100 | 2020KH08<br>1996NI09 |
| 95 | 151 | $^{246}$Am | 0+X | 2(−) | 25.0(2) *min* | | | %$\beta$−= 100<br>%IT < 0.02 | 1955EN16 |
| 95 | 151 | $^{246}$Am | ~ 2000 | | 73(10) $\mu s$ | | | %SF ≤ 100 | 1983PO14 |
| 96 | 150 | $^{246}$Cm | 1179.7 | 8− | 1.12(24) *s* | 128.0 | | | 2019SH34<br>2008RO21 |
| 98 | 148 | $^{246}$Cf | ~ 2500 | | 45(10) *ns* | | | %SF ≤ 100 | 1980GA07<br>1968GA04 |
| 101 | 145 | $^{246}$Md$^{\dagger}$ | 0+X | | 0.9(2) *s* | | | %$\alpha$ = 100<br>%SF = ?<br>%$\epsilon$ = ? | 2010AN08 |
| | | | 0+Y | | 4.4(8) *s* | | | %$\epsilon$ = 88(12)<br>%$\alpha$ = 12(12) | 2010AN08 |
| 96 | 151 | $^{247}$Cm | 227.38 (2) | 5/2+ | 26.3(3) $\mu s$ | 165.70(5)<br>227.38(2) | E3<br>M2+E3 | %IT = 100 | 2003AH07 |
| 96 | 151 | $^{247}$Cm | 404.90 (3) | 1/2+ | 100.6(6) *ns* | 177.52(2) | E2 | %IT = 100 | 2003AH07 |
| 100 | 147 | $^{247}$Fm | 45 (7) | (1/2+) | 5.1(2) *s* | | | %IT = 12(2)<br>%$\alpha$ = 88(2) | 2006HE27 |
| 101 | 146 | $^{247}$Md | 153 | (1/2−) | 0.23(3) *s* | | | %$\alpha$ = 80(2)<br>%SF = 20(2) | 2022HE04<br>2010AN08 |
| 96 | 152 | $^{248}$Cm | 1461 | 8− | 146(18) $\mu s$ | (7)<br>954 | | | 2019SH34 |
| 97 | 151 | $^{248}$Bk | 0+X | 1(−) | 23.7(2) *h* | | | %$\beta$−= 70(5)<br>%$\epsilon$ = 30(5) | 1978GR10 |
| 96 | 153 | $^{249}$Cm | 48.76 (4) | 7/2+ | 23 $\mu s$ | (22.53) | | %IT = ?<br>%$\alpha$ = ? | 1966AS12 |
| 96 | 153 | $^{249}$Cm | 375.23 (13) | (11/2−) | 19(1) *ns* | 192.5(1)<br>265.7(1) | (E1)<br>(E1) | %IT = 100 | 2008IS05 |
| 97 | 152 | $^{249}$Bk | 8.777 (14) | 3/2− | 0.3 *ms* | | | %IT = 100 | 1960AS06 |
| 98 | 151 | $^{249}$Cf | 144.98 (5) | 5/2+ | 45(5) $\mu s$ | 144.99(6) | M2+E3 | %IT = 100 | 1967AH02 |
| 101 | 148 | $^{249}$Md$^{\dagger}$ | 0+X | (7/2−) | 21.7(20) *s* | | | %$\alpha$ = 80(20)<br>%$\epsilon$+%$\beta$+ = 20(20) | 2008GA25<br>2005KUZZ<br>2001HE35<br>1973ES01 |
| | | | | (1/2−) | 1.5(+12−5) *s* | | | %$\alpha$ = 100 | 2001HE35 |
| 101 | 148 | $^{249}$Md | ≥ 910 | | 2.8(5) *ms* | | | | 2021GO26 |
| 97 | 153 | $^{250}$Bk | 35.59 (10) | 4+ | 29(1) $\mu s$ | (1.12)<br>35.59 | <br>(M2) | %IT = 100 | 2008AH02<br>1982KOZZ<br>1966MC02 |

*Continued...*

Table 1 contd...

| Z | N | $^{A}$X | E(keV) | J$^{\pi}$ | $T_{1/2}$ | E$\gamma$(keV) | $\lambda$ | Decay mode | Reference |
|---|---|---|---|---|---|---|---|---|---|
| 97 | 153 | $^{250}$Bk | 83.88 (18) | 7+ | 213(8) $\mu s$ | 6.1? | | %IT = 100 | 2008AH02 |
| | | | | | | | | | 1982KOZZ |
| | | | | | | | | | 1966MC02 |
| 97 | 153 | $^{250}$Bk | 97.48 (10) | 5− | 38(5) $ns$ | 61.89(1) | (E1) | %IT = 100 | 2008AH02 |
| | | | | | | | | | 1982KOZZ |
| | | | | | | | | | 1966MC02 |
| 97 | 153 | $^{250}$Bk | 175.13 | 1+ | 42(2) $ns$ | 50.12 | (E1) | %IT = 100 | 2008AH02 |
| | | | | | | 71.30 | (E1) | | 1973AH04 |
| | | | | | | 175.12 | | | |
| 99 | 151 | $^{250}$Es | 0+X | 1(−) | 2.22(5) $h$ | | | %$\epsilon$+%$\beta$+ ≤ 100 | 1980AH03 |
| 100 | 150 | $^{250}$Fm | 1199.2 (16) | (8−) | 1.92(5) $s$ | 23? | | %IT ≈ 100 | 2008GR17 |
| | | | | | | 682.3(5) | | %$\alpha$ = ? | |
| | | | | | | | | %$\epsilon$ = ? | |
| | | | | | | | | %SF = ? | |
| 101 | 149 | $^{250}$Md | 123 | | 42.4(45) $s$ | | | %$\alpha$ = ? | 2019VO03 |
| 102 | 148 | $^{250}$No | 0+X* | (6+) | 35.5(20) $\mu s$ | | | %IT ≈ 100 | 2022KH08 |
| | | | | | | | | %SF < 3.5 | 2022TE01 |
| | | | | | | | | | 2020KA02 |
| | | | | | | | | | 2017SV02 |
| | | | | | | | | | 2006PE17 |
| | | | | | | | | | 2003BE18 |
| | | | | | | | | | 2001OG08 |
| 102 | 148 | $^{250}$No | 0+Y | | 0.7(+14−3) $\mu s$ | | | %IT ≈ 100 | 2022KH08 |
| 97 | 154 | $^{251}$Bk | 35.5 (13) | (7/2+) | 58(4) $\mu s$ | (35.5) | | %IT = 100 | 1970HOZN |
| 98 | 153 | $^{251}$Cf | 106.309 (18) | 7/2+ | 38(2) $ns$ | 0.57? | | %IT = 100 | 1971AH01 |
| | | | | | | 58.48(2) | M1(+E2) | | |
| | | | | | | 81.48(2) | E2 | | |
| 98 | 153 | $^{251}$Cf | 370.47 (3) | 11/2− | 1.3(1) $\mu s$ | 45.2(1) | | %IT = 100 | 1971AH01 |
| | | | | | | 131.13(5) | E1 | | |
| | | | | | | 204.17(2) | E1 | | |
| | | | | | | 264.15(3) | | | |
| 100 | 151 | $^{251}$Fm | 200.0 (1) | 5/2+ | 23.0(11) $\mu s$ | 200.0(1) | M2+E3 | %IT = 100 | 2018RE07 |
| | | | | | | | | | 2011AS03 |
| 100 | 151 | $^{251}$Fm | 391.8 (2) | 1/2+ | 22(3) $ns$ | 192.0(2) | (E2) | %IT = 100 | 2011AS03 |
| 101 | 150 | $^{251}$Md | ≥ 844 | (23/2+) | 1.4(3) $s$ | | | | 2021GO26 |
| 102 | 149 | $^{251}$No | 106 (6) | (1/2+) | 1.02(3) $s$ | | | %$\alpha$ ≈ 100 | 2006HE27 |
| 102 | 149 | $^{251}$No | 1699.2+X | | ~ 2 $\mu s$ | | | | 2006HE27 |
| 102 | 150 | $^{252}$No | 1253.1 (19) | 8− | 109(3) $ms$ | 710 | | %IT = 100 | 2012SU22 |
| | | | | | | | | | 2011LO06 |
| | | | | | | | | | 2008RO21 |
| | | | | | | | | | 2007SU19 |
| 100 | 153 | $^{253}$Fm | 211+X | 11/2− | 0.56(6) $\mu s$ | 76.8 | (E1) | %IT = 100 | 2011AN13 |
| | | | | | | 150.5(5) | (E1) | | |
| 102 | 151 | $^{253}$No | 167.5 (5) | 5/2+ | 30.3(21) $\mu s$ | 167.5(5) | M2 | %IT = 100 | 2011AN13 |
| | | | | | | | | | 2009HE23 |
| | | | | | | | | | 2007LO11 |
| | | | | | | | | | 1973BE33 |
| 102 | 151 | $^{253}$No | 1440+X | | 627(5) $\mu s$ # | | | | 2011AN13 |

*Continued...*

*: 2006PE17 gives theoretical value of 1050 keV.

#: Another isomer with half-life of 552(15) $\mu s$ may exist as pointed out by 2011AN13 from their (x ray)(ce)-coin data.

Table 1 contd...

| Z | N | $^{A}X$ | E(keV) | $J^{\pi}$ | $T_{1/2}$ | Eγ(keV) | λ | Decay mode | Reference |
|---|---|---|---|---|---|---|---|---|---|
| 102 | 151 | $^{253}$No | X | (19/2+) | 706(24) $\mu s$ | | | | 2011LO06 |
| 103 | 150 | $^{253}$Lr | 0+X | (1/2−) | 1.32(14) $s$ | | | %α = 90(10)<br>%SF = 12(3)<br>%IT = ? | 2009HE20<br>2010HE11<br>2001HE35 |
| 104 | 149 | $^{253}$Rf$^{\dagger}$ | 0+X | (7/2+) | 10.3(12) $ms$ | | | %SF > 0.0<br>%α = ? | 2022LO03<br>2021KH07<br>1997HE29 |
| | | | 0+Y | (1/2+) | 52.0(44) $\mu s$ | | | %SF ≤ 100<br>%α = 17(6) | 2022LO03<br>2021KH07<br>2000HO27<br>1999HE11<br>1997HE29 |
| 104 | 149 | $^{253}$Rf | ≥ 1020+X | | 0.66(+40−18) $ms$ | | | | 2022LO03<br>2021KH07 |
| 99 | 155 | $^{254}$Es | 84.2 (25) | 2+ | 39.3(2) $h$ | | | %β−= 98.5(15)<br>%ϵ = 0.076(7)<br>%α = 0.32(1)<br>%IT = 1.5(15)<br>%SF < 0.045 | 1962UN01 |
| 101 | 153 | $^{254}$Md | 0+X | | 28(8) $min$* | | | %ϵ+%β+ ≈100 | 1970FI12 |
| 102 | 152 | $^{254}$No | 1297.1 (17) | 8− | 263(2) $ms$ | 53(1)<br>778(1) | | %IT = 100 | 2011LO06<br>2010CL01<br>2010HE10<br>2006HE19<br>2006TA19<br>1973GH03 |
| 102 | 152 | $^{254}$No | 2930.2 (21) | 16+ | 184(2) $\mu s$ | 133.4(4)<br>312.4(4) | | %IT = 100 | 2010CL01 |
| 103 | 151 | $^{254}$Lr | 108 (25) | (1−) | 20.3(42) $s$ | | | %α = ?<br>%ϵ+%β+ = ? | 2019VO03<br>2009HE05 |
| 104 | 150 | $^{254}$Rf | X | (8−) | 4.3(10) $\mu s$ | | | %SF < 10<br>%IT ≈ 100 | 2020KH01<br>2015DA12 |
| 104 | 150 | $^{254}$Rf | Y | (16+) | 247(73) $\mu s$ | | | %SF < 40<br>%IT ≈ 100 | 2015DA12 |
| 102 | 153 | $^{255}$No | 240-300 | (11/2−) | 109(9) $\mu s$ | <25 | | %IT = 100 | 2022BR08<br>2010HE10 |
| 102 | 153 | $^{255}$No | 1400-1600 | (19/2,<br>21/2,<br>23/2) | 77(6) $\mu s$ | 657 | | %IT = 100 | 2022BR08 |
| 102 | 153 | $^{255}$No | ≥ 1500 | ≥ (19/2) | 1.2(+6−4) $\mu s$ | 355 | | %IT = 100 | 2022BR08 |
| 103 | 152 | $^{255}$Lr | 38 (10) | [7/2−] | 2.54(5) $s$ | | | %α ≈ 40<br>%IT ≈ 60 | 2008HA31<br>2008AN16<br>2006CH52 |
| 103 | 152 | $^{255}$Lr | 878.8+Y | (19/2−) | ≥ 10 $ns$ | (28) | | %IT = 100 | 2009JE02 |
| 103 | 152 | $^{255}$Lr | 1408.6+Y | (25/2+) | 1.70(3) $ms$ | 243.9(3)<br>300.6(3) | | %IT ≈ 100<br>%α < 0.15 | 2009JE02 |
| 104 | 151 | $^{255}$Rf | (~135) | (5/2+) | 50(17) $\mu s$ # | | | %IT > 0 | 2020KH01<br>2015AN05 |

*Continued...*

*: 2017SO07 suggests from theory that 10 min activity is the K=0, 1− g.s. and 28 *min* activity is the K=3, 3− isomeric state.
#: Half-life in 2015AN05 is stated as tentative. 2020KH01 from the same group gives the half-life as > 30 $\mu s$.

Table 1 contd…

| Z | N | $^{A}X$ | E(keV) | $J^{\pi}$ | $T_{1/2}$ | Eγ(keV) | λ | Decay mode | Reference |
|---|---|---|---|---|---|---|---|---|---|
| 104 | 151 | $^{255}$Rf | 900–1200 | > 17/2 | 15(+6−4) $\mu s$ | | | %IT = ?<br>%SF = ? | 2020MO11 |
| 104 | 151 | $^{255}$Rf | 1150–1450 | > 17/2 | 38(+12−7) $\mu s$ | | | %IT = ?<br>%SF = ? | 2020MO11 |
| 99 | 157 | $^{256}$Es | 0+X | (8+) | 7.6 $h$ | | | %β−= 100<br>%β–F = 0.002 | 1976HOZB |
| 100 | 156 | $^{256}$Fm | 1425.5 | (7−) | 70(5) ns | 96.8<br>211.2?<br>275.3?<br>302.0<br>380.0<br>861.8<br>1092.9 | | %IT = 100 | 1989HA10 |
| 102 | 154 | $^{256}$No | > 1089 | (7−, 5−) | 7.8(+83−26) $\mu s$ | | | %IT = ? | 2021KE10 |
| 104 | 152 | $^{256}$Rf | ~ 1120 | (5−) | 25(2) $\mu s$ | | | %IT = ?<br>%SF = ? | 2021KH03<br>2011RO20<br>2013RI07<br>2009JE01 |
| 104 | 152 | $^{256}$Rf | ~ 1400 | (8−) | 17(2) $\mu s$ | | | %IT = ?<br>%SF = ? | 2021KH03<br>2013RI07<br>2009JE01 |
| 104 | 152 | $^{256}$Rf | > 2200 | | 27(5) $\mu s$ | | | %IT = ?<br>%SF = ? | 2020MO10<br>2009JE01 |
| 104 | 153 | $^{257}$Rf | 73 (11) | (11/2−) | 4.37(5) $s$ | | | %α = 81.0(25)<br>%IT = 14(1)<br>%ϵ = 4.8(6)<br>%SF = 0.4(2) | 2022HA04<br>2013RI07<br>2010ST14<br>2010BE16<br>2009QI04<br>2008DR05<br>1997HE29 |
| 104 | 153 | $^{257}$Rf | 1083.2 (17) | 21/2+ | 106(6) $\mu s$ | 295.6(7)<br>446.8(7)<br>586.1(13) | | %IT = 100 | 2013RI07<br>2010MO10 |
| 105 | 152 | $^{257}$Db | ~ 370 | (1/2−) | 0.67(6) $s$ | | | %α = 93.5(65)<br>%SF = 6.5(65) | 2009HE20 |
| 101 | 157 | $^{258}$Md | 0+X | | 57.0(9) $min$ | | | %ϵ ≥ 70<br>%SF+%β−≤ 13 | 1993MO18 |
| 104 | 154 | $^{258}$Rf | 0+Y | | 2.4(+24−8) $ms$ | | | %SF = ?<br>%α = ?<br>%IT = ? | 2016HE15 |
| 104 | 154 | $^{258}$Rf | 0+Z | | 15(10) $\mu s$ | | | %SF = ?<br>%α = ?<br>%IT = ? | 2016HE15 |
| 105 | 153 | $^{258}$Db | 51 (14) | (5+,<br>6+) | 4.41(21) $s$ | | | %α = 77(8)<br>%ϵ+%β+ = 23(8) | 2019VO03<br>2016HE15<br>2009HE20 |
| 106 | 153 | $^{259}$Sg | ~90 | (1/2+) | 226(27) $ms$* | | | %α ≈ 100 | 2015AN05<br>2013AN08 |

*Continued…*

*: Half-life in 2015AN05 is stated as tentative.

Table 1 contd...

| Z | N | $^{A}$X | E(keV) | J$^{\pi}$ | $T_{1/2}$ | Eγ(keV) | λ | Decay mode | Reference |
|---|---|---|---|---|---|---|---|---|---|
| 104 | 157 | $^{261}$Rf | 0+X | (11/2−) | 74(7) *s* | | | %α ≈ 100 | 2017HE08 |
| | | | | | | | | %SF < 11 | 2013MU08 |
| | | | | | | | | | 2012HA05 |
| | | | | | | | | | 2008GA08 |
| | | | | | | | | | 2008DV02 |
| | | | | | | | | | 2008DU09 |
| | | | | | | | | | 2006DV01 |
| | | | | | | | | | 2002HO11 |
| | | | | | | | | | 2000SY01 |
| | | | | | | | | | 1996KA66 |
| | | | | | | | | | 1994LA22 |
| | | | | | | | | | 1994LO27 |
| | | | | | | | | | 1970GH01 |
| 104 | 157 | $^{261}$Rf | 0+Y | (3/2+) | 2.6(3) *s* | | | %SF = 88(5) | 2017HE08 |
| | | | | | | | | %α = 12(5) | 2013SU04 |
| | | | | | | | | | 2013MU08 |
| | | | | | | | | | 2012HA05 |
| | | | | | | | | | 2011HA13 |
| | | | | | | | | | 2008DV02 |
| | | | | | | | | | 2008DU09 |
| | | | | | | | | | 2007MO09 |
| | | | | | | | | | 2003TU05 |
| | | | | | | | | | 2002HO11 |
| | | | | | | | | | 1996LA11 |
| | | | | | | | | | 1994LA22 |
| 106 | 155 | $^{261}$Sg | ~ 200 | (11/2−) | 9.0(+20−15) *μs* | | | %IT ≈ 100 | 2010BE16 |
| 107 | 155 | $^{262}$Bh | 0+X | | 22(4) *ms* | | | %SF < 24 | 2009HE20 |
| | | | | | | | | | 2008NE08 |
| | | | | | | | | | 2006FO02 |
| | | | | | | | | | 1989MU09 |
| | | | | | | | | | 1984OG03 |
| | | | | | | | | | 1981MU06 |
| 106 | 157 | $^{263}$Sg | 0+Y | | 306(+160−80) *ms* | | | %α = ? | 2004FO08 |
| | | | | | | | | %IT = ? | 2004MO40 |
| | | | | | | | | %SF = ? | 2003GI05 |
| | | | | | | | | | 1998HO13 |
| 106 | 159 | $^{265}$Sg$^{\dagger}$ | 0+X | (11/2−) | 8.8(+26−16) *s* | | | %α ≈ 50 | 2012HA05 |
| | | | | | | | | %SF ≈ 50 | 2008DU09 |
| | | | | | | | | | 2003TU05 |
| | | | | | | | | | 2002HO11 |
| | | | | | | | | | 1998TU01 |
| | | | 0+Y | | 16.0(25) *s* | | | %α ≈ 50 | 2013SU04 |
| | | | | | | | | %SF ≈ 50 | 2012HA05 |
| | | | | | | | | | 2008DU09 |
| | | | | | | | | | 2008DV02 |
| | | | | | | | | | 2006DV01 |
| | | | | | | | | | 2003TU05 |
| | | | | | | | | | 2002HO11 |
| | | | | | | | | | 1998TU01 |

*Continued...*

Table 1 contd...

| Z | N | $^{A}$X | E(keV) | J$^{\pi}$ | $T_{1/2}$ | E$\gamma$(keV) | $\lambda$ | Decay mode | Reference |
|---|---|---|---|---|---|---|---|---|---|
| 108 | 157 | $^{265}$Hs | 0+X | | 0.78(+17−12) *ms* | | | %$\alpha$ = 99.5(5) | 2009HE20 |
| | | | | | | | | %SF = 0.5(5) | 1999HE11 |
| 108 | 158 | $^{266}$Hs | ~ 1200 | (9−) | 74(+354–34) *ms* | | | %$\alpha$ ≈ 100 | 2017AC02 |
| | | | | | | | | | 2012AC04 |
| 108 | 159 | $^{267}$Hs | 0+X | | 940(+120−45) $\mu s$ | | | %$\alpha$ > 0.0 | 2004F008 |
| | | | | | | | | | 2004MO40 |
| 108 | 159 | $^{267}$Hs | 0+Y | | 0.80(+380−27) *s* | | | %$\alpha$ ≈ 100 | 2004MO40 |
| | | | | | | | | %SF = ? | 2015MO25 |
| | | | | | | | | %IT = ? | |
| 110 | 160 | $^{270}$Ds | 1.13+E3 ? | | 6.0(+82−22) *ms* | | | %$\alpha$ = 85(15) | 2017AC02 |
| | | | | | | | | %IT = 15(15) | 2001HO06 |
| 110 | 161 | $^{271}$Ds | 0+X | | 69(+56−21) *ms* | | | %$\alpha$ ≈ 100 | 2015MO25 |
| | | | | | | | | | 2004MO40 |
| | | | | | | | | | 1998HO13 |
| 109 | 167 | $^{276}$Mt† | 0+X | | 0.54(+14−9) *s* | | | | 2013OG01 |
| | | | | | | | | | 2013RU11 |
| | | | | | | | | | 2012OG02 |
| | | | 0+Y | | 6(+8−2) *s* | | | | 2013OG01 |
| | | | | | | | | | 2013RU11 |
| | | | | | | | | | 2012OG02 |

@: Half-life for bare or highly-ionized nucleus.

†: Energy ordering of the ground state and the isomeric state is unknown.

**Table II**

List of tentative isomers proposed either in journal articles or in theses or in conference proceedings, but have not been confirmed as yet.

| Z | N | $^{A}$X | E(keV) | J$^{\pi}$ | $T_{1/2}$ | E$\gamma$(keV) | $\lambda$ | Decay mode | Reference |
|---|---|---|---|---|---|---|---|---|---|
| 27 | 37 | $^{64}$Co | 107 (20) | | > 280 *ms* | | | | 2010FE01 |
| 31 | 39 | $^{70}$Ga* | 1086.31 (16)? | | 24(4) *ns* | 184.9(1)? | | | 1977MO01 |
| 30 | 45 | $^{75}$Zn | 126.94 (9) | 1/2− | 5 *s* SYS | | | %$\beta$−= ?<br>%IT = ? | 2017WR01<br>2011IL01 |
| 35 | 52 | $^{87}$Br$^{\#}$ | 1463.89 | 9/2+ | ~ 20 *ns* | 318.2(2)<br>429.0(2)<br>589.0(2)<br>662.60(8)<br>846.0(2)<br>1457.8(2)<br>1463.8(3) | | | 2019WI11 |
| 44 | 65 | $^{109}$Ru | 68.75 (12) | (1/2+) | 0.50(20) $\mu s$ | 68.8(2) | E2 | %IT = 100 | 1992PEZX |
| 45 | 73 | $^{118}$Rh$^{\$}$ | 0+X ? | | | | | | 2000JO18<br>2006WA10 |
| 46 | 73 | $^{119}$Pd | ~ 340 | | 0.85(1) *s* | 100?$^{‡}$ | | %$\beta$−≈ 100<br>% IT > 0 | 2022KU09 |
| 46 | 77 | $^{123}$Pd | Unknown | (11/2−) | Unknown | | | %$\beta$−= ? | 2019CH24 |
| 47 | 76 | $^{123}$Ag | 59.5 | (1/2−) | Unknown | | | %$\beta$−= ?<br>%IT = ? | 2019CH24 |
| 46 | 79 | $^{125}$Pd | Unknown | (11/2−) | Unknown | | | % $\beta$−= ? | 2019CH24 |
| 59 | 70 | $^{129}$Pr | 382.57 (24) | (11/2−) | 1 *ms* (SYS) | 140.9(4)<br>291.6(4) | | | 1997GI07 |
| 59 | 70 | $^{129}$Pr | 497.3 (4) | (9/2+) | ~ 42 *ns* | 255.5(3) | | | 1990JAZU |
| 60 | 70 | $^{130}$Pr | 139.1 | (7−) | ~ 280 *ns* | 80 | | | 1990JAZU |
| 60 | 71 | $^{131}$Nd | 210.8 (5) | (7/2−) | ~ 50 *ns* | 210.6(5) | (E1) | %IT = 100 | 1990JAZU |
| 59 | 80 | $^{139}$Pr | 3697.90 (15) | (25/2−) | ~ 12 *ns* | 71.1?<br>214.0(1)<br>1330.7(1) | (E1) | | 2012BHZZ |
| 53 | 87 | $^{140}$I | 0.0+X | (2−) | Unknown | | | %$\beta$−= ?<br>%$\beta$−n = ?<br>%IT = ? | 2017MO19 |
| 65 | 76 | $^{141}$Tb$^{♦}$ | 0+X | | 7.9(6) *s* | | | %$\epsilon$+%$\beta$+ = 100 | 1988TUZY |
| 55 | 91 | $^{146}$Cs | 47 | | 1.25(5) $\mu s$ | 47 | | | 2015YAZW |
| 55 | 93 | $^{148}$Cs | | | 4.8(2) $\mu s$ | | | | 2015YAZW |
| 61 | 97 | $^{158}$Pm | 121+X | | > 16 $\mu s$ | 121 | | %IT = 100 | 2015YOZX |
| 69 | 89 | $^{158}$Tm$^{\star}$ | 0+X | (5+) | ~ 20 *s* | | | | 1981DR07 |
| 67 | 91 | $^{158}$Ho | 74.897 (11) | 2+ | 60(10) *ns* | 7.697(4) | E1 | | 2005KAZY |
| 67 | 91 | $^{158}$Ho | 91.595 (12) | 1−,2−,3− | 140(25) *ns* | 24.395(6) | M1+E2 | | 2005KAZY |
| 67 | 93 | $^{160}$Ho | 67.11 (3) | 1+ | 28(2) *ns* | 7.133(10) | E1(+M2) | | 2006KAZX |
| 62 | 101 | $^{163}$Sm | 0+X | | in $\mu s$ range | | | | 2017PA25 |
| 63 | 101 | $^{164}$Eu | 0+X | | in $\mu s$ range | | | | 2017PA25 |

*Continued...*

*: Questionable isomer.

#: possible 20 ns isomer, half-life not directly measured but estimated from transition probabilities.

$: possibly a high-spin (J=4-10) isomer from systematics and from $^{118}$Rh decay to $^{118}$Pd (2006WA10) which populates both low- and high-spin levels.

‡: Authors assign a gamma ray of possible E3 multipolarity with an approximate energy of 100 keV.

♦: Isomer was not confirmed by 1989GI06 and 2001BEZY.

⋆: 1981DR07 quoted ~ 20 *ns* in one place and ~ 20 *s* in another place in their paper, however, the former value is unrealistic as the activity was reported by authors in 'beam-off' mode. Choice of ~ 20 *ns* in the ENSDF database seems erroneous.

Table II contd. . .

| Z | N | $^{A}$X | E(keV) | $J^{\pi}$ | $T_{1/2}$ | E$\gamma$(keV) | $\lambda$ | Decay mode | Reference |
|---|---|---|---|---|---|---|---|---|---|
| 63 | 102 | $^{165}$Eu | 0+X | | in $\mu s$ range | | | | 2017PA25 |
| 75 | 95 | $^{170}$Re | 64 (20) | | | | | | 2020CU04 |
| 74 | 106 | $^{180}$W | 3547.9 (12) | (16+) | 20.3(6) $ns$ | 158.1 | (M1) | | 1979FAZR |
| 76 | 104 | $^{180}$Os | 4540+X | | 41(10) $ns$* | | | | 1993VE01 |
| 76 | 104 | $^{180}$Os | 5848.3 (15) | | 12(4) $ns$* | 286.8 | | %IT = 100 | 1993VE01 |
| 73 | 109 | $^{182}$Ta | 1950 | 14+ | 247(11) $ns$ | 245.3(1)<br>570.2(1) | | | 2016PAAA |
| 73 | 109 | $^{182}$Ta | 2082 | (14−) | < 208 $ns$ | 132.1(1) | | | 2016PAAA |
| 73 | 109 | $^{182}$Ta | 3399 | (18−) | 73.5(25) $ns$ | 58.7(1)?<br>447.6(1)<br>637.5(1) | | | 2016PAAA |
| 73 | 110 | $^{183}$Ta | 2475 | 29/2− | 28(3) $ns$ | 171.1(2) | E1 | | 2016PAAA |
| 73 | 110 | $^{183}$Ta | 3870 | 41/2− | 49(3) $ns$ | 153.0(2) | E1 | | 2016PAAA |
| 73 | 115 | $^{188}$Ta | 99 (33) | | < 10 $s$ | | | | 2012REZZ |
| 83 | 108 | $^{191}$Bi | 1825.1+X | | 400(40) $ns$ | | | %IT ≈ 100 | 2004NI06 |
| 81 | 129 | $^{210}$Tl | 1127 | (11+) | | | | %$\beta$−= 100 | 2018BR15 |
| 83 | 130 | $^{213}$Bi | 1353 (21) | | Unknown | | | | 2012CH19 |
| 83 | 130 | $^{213}$Bi | 1319 (30) | | > 168 $s$ [@] | | | | 2008CHZI[@] |
| 83 | 131 | $^{214}$Bi | 539 (30) | | > 93 $s$ [@] | | | | 2008CHZI[@] |
| 87 | 127 | $^{214}$Fr | 6475+Y | (33+) | 108(7) $ns$ | 114.7(5)<br>296.3(2) | M1+E2<br>M1+E2 | %IT = 100 | 1994BY01 |
| 91 | 129 | $^{220}$Pa | 0+X | (3−) | 233(+108−56) $ns$ | | | %$\alpha$ ≈ 100 | 2021MA66 |
| 87 | 141 | $^{228}$Fr | 1004 (30) | | 94(+170−29) $s$ [@] | | | | 2008CHZI[@] |
| 89 | 139 | $^{228}$Ac | 6.11 (9) | (1−) | 299(11) $ns$ | 6.11(9) | | %IT ≈ 100 | 2021KI08<br>1995SO11 |
| 89 | 139 | $^{228}$Ac | 20.6 (5) | (1−) | 115(25) $ns$ | 13.8(4) | | %IT ≈ 100 | 2021KI08<br>1995SO11 |
| 89 | 139 | $^{228}$Ac | 356 (30) | | 119(+61−30) $s$ [@] | | | | 2008CHZI[@] |
| 89 | 145 | $^{234}$Ac | 140 (30) | | > 93 $s$ [@] | | | | 2008CHZI[@] |
| 89 | 139 | $^{228}$Ac | 620 (30) | | 149(+95−42) $s$ [@] | | | | 2008CHZI[@] |
| 92 | 144 | $^{236}$U | ~ 5500[#] | | 66(3) $ns$ | | | | 2020BA53[#] |
| 101 | 143 | $^{244}$Md | 0+X ? | | 5(+3−2) $ms$ | | | %SF = ?<br>%$\alpha$ = ?<br>%$\epsilon$+%$\beta$+ = ? | 2020KH08<br>2020PO07 |
| 100 | 148 | $^{248}$Fm | > 1074 ? | | 10.1 (6) $ms$ | | | | 2010KEZY |
| 105 | 150 | $^{255}$Db | | | 2.8 $ms$ | | | %SF = 100 | 2017HE08 |
| 105 | 153 | $^{258}$Db | 0+Y | | 20(10) $s$ | | | %$\epsilon$ ≈ 100 | 1985HE22 |
| 108 | 169 | $^{277}$Hs | 0+X | | 34(+164−16) $s$ | | | %SF ≈ 100 | 2012HO12 |
| 110 | 171 | $^{281}$Ds | 0+X | | 0.25(+118−11) $s$ | | | %$\alpha$ ≈ 100 | 2012HO12 |
| 112 | 173 | $^{285}$Cn | 0+X | | 4.0(+191−18) $s$ | | | %$\alpha$ ≈ 100 | 2012HO12 |
| 114 | 175 | $^{289}$Fl | 0+X | | 0.28(+135−13) $s$ | | | %$\alpha$ ≈ 100 | 2012HO12 |

*: The value is either half-life or mean lifetime, confirmation of these isomers is lacking.

#: Communication of June 2, 2021 with the corresponding author, E. Mendoza suggested possibility of two isomers in their work, the well-known fission isomer at 2750 keV, and another isomer at ~5500 keV.

@: Half-life for bare or highly-ionized nucleus.

*Alpha Decay of the Isomers of $^{214}$Fr.*
1968Tr06 W.J.Treytl, E.K.Hyde, T.Yamazaki - Nucl.Phys. A117, 481 (1968).
*Levels of $^{208}$Po from Radioactive Decay and Nuclear Reaction $\gamma$-RaySpectroscopy.*
1968Tr11 W.J.Treytl, E.Matthias - Proc.Intern.Conf.Hyperfine Struct.Nucl.Radiations, Asilomar, Pacific Grove, Calif. (1967), E.Matthias, D.A.Shirley, Eds., North-Holland Publ.Co., Amsterdam, p.145 (1968).
*g-Factor of the 243-keV Level in $^{86}$Y.*
1968Va18 K.Valli, E.K.Hyde - Phys.Rev. 176, 1377 (1968).
*New Isotopes of Thorium Studied with an Improved Helium-Jet Recoil Transport Apparatus.*
1968Wa12 D.Ward, M.Neiman - Nucl.Phys. A115, 529 (1968).
*Energy Levels in $^{154}$Dy.*
1968We03 W.Weirauch, W.-D.Schmidt-Ott, F.Smend, A.Flammersfeld - Z.Physik 209, 289 (1968).
*Suche nach Dem Einfluss der Chemischen Bindung Auf die Halbwertszeit des 90m-Nb ($T_{1/2}$=19,2 sec).*
1968Wi10 B.Wilken - Z.Phys. 213, 56 (1968).
*Konversionskoeffizienten und Fluoreszenzausbeuten beim Zerfall des Ir$^{192m}$ und Co$^{58m}$.*
1968WiZY W.G.Winn - Thesis, Cornell University (1968); NYO-3664-6 (1968).
*Experimental Studies of $^{136m}$Ba, $^{134m}$Xe, $^{127m}$Xe, and $^{125m}$Xe.*
1969AdZY B.Ader - Thesis, Univ.Paris (1969).
*Etude de la Desintegration par Capture Electronique des Isotopes 175 et 177 du Tungstene. Schemas de Niveaux des Isotopes 175 et 177 du Tantale, Interpretes dans le Cadre du Modele de Nilsson.*
1969Ak01 M.R.Akhmed, K.A.Baskova, S.S.Vasilev, L.Y.Shavtvalov - Izv.Akad.Nauk SSSR, Ser.Fiz. 33, 104 (1969); Bull.Acad.Sci.USSR, Phys.Ser. 33, 97 (1970).
*The Radiation from $^{162}$Ho.*
1969Al05 K.F.Alexander, W.Neubert, H.Rotter, S.Chojnacki et al. - Nucl.Phys. A133, 77 (1969).
*Isomerism in $^{129}$La: Evidence for Oblate Nuclear Shape.*
1969An13 W.Andrejtscheff, L.Funke, W.Meiling, F.Stary - Nucl.Phys. A135, 170 (1969).
*±-Forbidden E1 Transitions in $^{182}$Re.*
1969Be47 I.Bergstrom, C.J.Herrlander, P.Thieberger, J.Blomqvist - Phys.Rev. 181, 1642 (1969).
*High-Spin Three-Particle States in $^{207}$Bi.*
1969Be54 R.B.Begzhanov, K.M.Sadykov - Yadern.Fiz. 10, 445 (1969); Soviet J.Nucl.Phys. 10, 254 (1970).
*Some Properties of Excited Levels of Tb$^{159}$, Tb$^{161}$, and Dy$^{161}$.*
1969Bi10 C.Birattari, G.Strini, G.Tagliaferri, G.Bertolini - Nuovo Cimento Lett. 2, 473 (1969).
*Decay of $^{179}$W-g,m.*
1969Bl02 J.Bleck, D.W.Haag, W.Leitz, R.Michaelsen et al. - Nucl.Phys. A123, 65(1969).
*The g-Factor of the First Excited State in $^{19}$Ne.*
1969Bo25 I.Boca, N.Martalogu, M.Sezon, I.Vilcov et al. - Nucl.Phys. A134, 541 (1969).
*Population of the Spontaneously Fissioning Isomer $^{244}$mf-Am Throughthe (n,$\gamma$) Reaction.*
1969BoZL M.Bonitz - ZFK-172 (1969).
*Kernspektroskopie am Deutronenstrahl.*
1969BoZP M.Bonitz - ZFK-172 (1969).
*Kernspektroskopie am Deuteronenstrahl: Habilitationsschrift.*
1969Br02 H.F.Brinckmann, C.Heiser, U.Hagemann - Nucl.Phys. A123, 689 (1969).
*Zur Isomerie in $^{117}$Te.*
1969Ch18 J.Chaumont, E.Roeckl, Y.Nir-El, C.Thibault-Philippe et al. - Phys.Letters 29B, 652 (1969).
*Identification of New Neutron-Deficient Nuclides $^{76}$Rb and $^{118}$Cs. Half-Lives of $^{78}$Rb, $^{119}$-$^{124}$Cs, $^{126}$Cs.*
1969Ch32 Y.Y.Chu, E.M.Franz, G.Friedlander - Phys.Rev. 187, 1529 (1969).
*Isomerism in $^{147}$Tb.*
1969Co13 T.W.Conlon - Nucl.Phys. A136, 70 (1969).
*Three-Quasiparticle States in the Decay of the New Isomer $^{181}$Re.*
1969Da01 P.J.Daly, P.Kleinheinz, R.F.Casten - Nucl.Phys. A123, 186 (1969).
*The 1.7 min Isomer in $^{185}$W.*
1969Do11 L.C.M.do Amaral, C.F.de Barros Leite, J.M.F.Jeronymo, A.G.de Pinho et al. - Z.Naturforsch. 24a, 1196 (1969).
*Decay Scheme of $^{118}$In (4.45 min).*
1969FaZY D.Favret - Thesis, Universite de Lyon (1969); LYCEN-6924 (1969).
*Recherches sur la Mesure Precise de Durees de Vie des Niveaux Isomeres ($10^{-6}$ a $10^{-2}$ s).*
1969Fo01 C.Foin, J.Oms, J.Blachot, J.Crancon - Nucl.Phys. A123, 513 (1969).
*Etude de la Desintegration $^{95}$Zr $\rightarrow$ $^{95}$Nb.*
1969Fr08 J.M.Freeman, J.G.Jenkin, G.Murray, D.C.Robinson, W.E.Burcham - Nucl.Phys. A132, 593 (1969).
*The ft Value of the Superallowed Fermi Beta Decay $^{26m}$Al($\beta^+$)$^{26}$Mg.*
1969Ge03 J.S.Geiger, R.L.Graham, M.W.Johns - Can.J.Phys. 47, 949 (1969).
*The Half-Life of the First Metastable State in $^{90}$Nb.*
1969Ge11 C.Gerschel, N.Perrin - Compt.Rend. 269B, 220 (1969).
*Isomerie dans le Lanthane-132.*
1969Gl04 J.Glatz, K.E.G.Lobner - Z.Physik 227, 83 (1969).

*Activation Cross-Sections for Some 14.8 MeV Neutron-Induced Nuclear Reactions on Caesium: Isomer Ratios of $^{134}$Cs, $^{133}$Xe and $^{130}$I.*
1970Qa03 S.M.Qaim - Nucl.Phys. A154, 145 (1970).
*The Decay of 12.3 h $^{130g}$I, 9.2 min $^{130m}$I and 6.5 d $^{132}$Cs.*
1970Ra08 S.Raman, R.G.Couch - Phys.Rev. C1, 744 (1970).
*Spin and Parity Assignment for the 872-keV Level in $^{69}$Ga.*
1970Ra14 K.Raichev, L.Tron - Acta Phys. 28, 263 (1970).
*Investigation of the Alpha Spectra of Po Isotopes.*
1970Re02 P.L.Reeder - Phys.Rev. C1, 721 (1970).
*Alpha Spectroscopy of $^{212g}$At and $^{212m}$At.*
1970Re05 R.Repnow, V.Metag, J.D.Fox, P.von Brentano - Nucl.Phys. A147, 183 (1970).
*Evidence for a Direct Reaction Mechanism in the Production of FissionIsomers.*
1970Ro29 R.J.Rousselin, C.Gauthier - Int.J.Appl.Radiat.Isotop. 21, 599 (1970).
*Caracteristiques de Deux Generateurs d'Indium 113m.*
1970Ru08 E.Rurarz, Z.Haratym, T.Kozlowski, J.Wojtkowska - Acta Phys.Polon. B1, 97 (1970).
*Excitation of Isomeric Activities in $^{71}$Ge, $^{78}$Br and $^{79}$Br Using 14.8 MeV Neutrons.*
1970SeZP V.Sergejev, Z.Sujkowski, J.Kownacki, H.Ryde - Ann.Rept., Research Inst.Phys., Stockholm, p.33 (1970).
*On the Beta-Decay of an Isomeric State in $^{145}$Gd.*
1970To11 T.Toriyama, M.Fujioka, M.Akiba, K.Hisatake - J.Phys.Soc.Jap. 29, 9 (1970).
*Existence of a New Isomer of $T_{1/2}$ = 24.4 hr in $^{156}$Tb.*
1970To16 K.S.Toth, R.L.Hahn, M.A.Ijaz, W.M.Sample - Phys.Rev. C2, 1480 (1970).
*Production of Rare-Earth α Emitters with Energetic $^{3}$He Particles;New Isotopes: $^{151}$Er, $^{156}$Yb, and $^{157}$Yb.*
1970Uy01 J.Uyttenhove, J.Demuynck, M.Dorikens, L.Dorikens-Vanpraet - Z.Phys. 238, 90 (1970).
*Experimental Study of the Isomeric States $^{101m}$Ru, $^{103m}$Ru and $^{101m}$Tc.*
1970Va31 J.van Klinken, L.M.Taff, H.T.Dijkstra, A.H.de Haan et al. - Nucl.Phys. A157, 385 (1970).
*The Decays of $^{82a+82b}$As and $^{79}$Ge.*
1970Vi05 N.Vilcov - Stud.Cercet.Fiz. 22, 795 (1970).
*Izomeri Spontan Fisionabili Ai Nucleelor Transuraniene.*
1970WiZN J.B.Wilhelmy, S.G.Thompson, J.O.Rasmussen, J.T.Routti, J.E.Phillips - UCRL-19530, p.178 (1970).
*Gamma-Ray Studies on Short-Lived Fission Products.*
1970Wo06 K.L.Wolf, R.Vandenbosch, P.A.Russo, M.K.Mehta, C.R.Rudy - Phys.Rev. C1, 2096 (1970).
*Spontaneous Fission Isomerism in Uranium Isotopes.*
1970Wo08 F.K.Wohn, W.L.Talbert, Jr., J.K.Halbig - Nucl.Phys. A152, 561 (1970).
*The Decay of $^{85m}$Kr.*
1970Ya03 T.Yamazaki - Phys.Rev. C1, 290 (1970).
*Isomeric States of Po Isotopes in the Nanosecond Range Populated by Pb(α,xn) Reactions.*
1971Ac01 E.Achterberg, E.Y.de Aisenberg, F.C.Iglesias, A.E.Jech et al. - Phys.Rev. C4, 188 (1971).
*On-Line Spectroscopy of $^{134m}$,$^{134g}$I.*
1971Ah01 I.Ahmad, F.T.Porter, M.S.Freedman, R.F.Barnes et al. - Phys.Rev. C3, 390 (1971).
*Energy Levels in Cf$^{251}$ via Alpha Decay of Fm$^{255}$.*
1971Al31 M.Alpsten, A.Appelqvist - Phys.Scr. 4, 137 (1971).
*Properties of $^{209}$Po as Observed in the Decay of $^{209}$At.*
1971AnZF H.R.Andrews, J.S.Geiger, R.L.Graham, S.H.Sie, D.Ward - AECL-4068, p.54 (1972).
*Electric Monopole Transition Probabilities from $0^+$ Excited States.*
1971Ar16 R.Arlt, N.G.Zaitseva, B.Kracik, G.Musiol et al. - Izv.Akad.Nauk SSSR, Ser.Fiz. 35, 27 (1971); Bull.Acad.Sci.USSR, Phys.Ser. 35, 25 (1972).
*The Decay of $^{89}$Nb.*
1971Ar31 R.Arlt, H.Bayer, V.V.Kuznetsov, W.Neubert et al. - Izv.Akad.Nauk SSSR, Ser.Fiz. 35, 1612 (1971); Bull.Acad.Sci.USSR, Phys.Ser. 35, 1470 (1972).
*Short-Lived Isomers of $^{148}$Tb, $^{149}$Tb, $^{150}$Tb, and $^{152}$Tb.*
1971Ba28 S.Baba, H.Baba, H.Natsume - J.Inorg.Nucl.Chem. 33, 589 (1971).
*Half-Lives of Some Fission Product Nuclides.*
1971Be37 I.Bergstrom, J.Blomqvist, B.Fant, A.Filevich et al. - Phys.Scr. 3, 11 (1971).
*E2 Core Polarization by Neutrons in the $(i_{13/2})^{-2}$ Configuration in $^{206}$Pb.*
1971Be61 H.Beuscher, P.Jahn, R.M.Lieder, C.Mayer-Boricke - Z.Phys. 247, 383 (1971).
*Electron Capture Decay of 51 min $^{191}$Hg.*
1971Be62 A.G.Belov, Y.P.Gangrskii, B.Dalkhsuren, A.M.Kucher - Yad.Fiz. 14, 685 (1971); Sov.J.Nucl.Phys. 14, 385 (1972).
*Production of the Spontaneously Fissioning U$^{236}$ Isomer in Thermal Neutron Radiative Capture.*
1971BeWR M.Becker, H.Bertschat, U.Leithauser, W.Leitz et al. - HMI-B-112, p.58 (1971).
*Kernphysikalische Experimente zur Festkorperphysik.*
1971Bl14 P.Blasi, M.Morando, P.R.Maurenzig, N.Taccetti - Lett.Nuovo Cim. 2, 63 (1971).
*A K = 3/2$^+$ Rotational Band in $^{45}$Ti.*
1971Bo12 W.W.Bowman, T.T.Sugihara, F.R.Hamiter - Phys.Rev. C3, 1275 (1971).

*Fractional Cumulative Yield of 2.6 min $^{99m}$Nb from Low Energy Neutron Fission of $^{233}$U, $^{235}$U, $^{238}$U and Spontaneous Fission of $^{252}$Cf.*

1971Ha18 H.H.Hansen, E.De Roost, D.Mouchel, R.Vaninbroukx - Int.J.Appl.Radiat.Isotop. 22, 1 (1971).
*Internal Conversion Coefficients of the 392 keV $\gamma$-Ray and Half-Life of the Isomeric State in $^{113}$In.*

1971HaXM U.Hagemann, W.Neubert, L.K.Peker, W.Schulze, F.Stary - JINR-E6-6193 (1971).
*Two Quasi-Particle Excitation in $^{146}$Eu.*

1971Ho01 P.Hornshoj, K.Wilsky, P.G.Hansen, A.Lindahl, O.B.Nielsen - Nucl.Phys. A163, 277 (1971).
*Alpha Decay of Neutron-Deficient Radon and Polonium Isotopes.*

1971Ho06 T.L.Houk - Phys.Rev. C3, 1886 (1971).
*Neutron-Proton Scattering Cross Section at a Few Electron Volts and Charge Independence.*

1971Ho09 S.Hofmann, P.Steiner, G.Weyer - Nucl.Phys. A166, 436 (1971).
*Lifetime of the 91.5 keV State in $^{151}$Sm.*

1971Ho14 A.Hoglund, S.G.Malmskog, A.Marelius, K.G.Valivaara, J.Kozyczkowski - Nucl.Phys. A169, 49 (1971).
*Absolute Transition Rates in $^{182}$W.*

1971Ho27 R.E.Holland, R.D.Lawson, F.J.Lynch - Ann.Phys.(N.Y.) 63, 607 (1971).
*Electric-Quadrupole Isomers in $^{90}$Nb and $^{92}$Tc.*

1971Hu14 H.Hubel, R.A.Naumann, E.H.Spejewski - Phys.Rev. C4, 2272 (1971); Erratum Phys.Rev. C5, 1438 (1972).
*Decay of a Three-Quasiparticle State in $^{177}$Ta.*

1971Is04 M.Ishihara, H.Kawakami, N.Yoshikawa, M.Sakai, K.Ishii - Phys.Lett. 35B, 398 (1971).
*Lifetime Measurements on E2 Isomers in N = 48 Nuclei and E2 EffectiveCharges of $1g_{9/2}$.*

1971Jo19 B.Jonson, M.Alpsten, A.Appelqvist, G.Astner - Nucl.Phys. A174, 225 (1971).
*Properties of Neutron-Deficient Odd-Mass Polonium Isotopes.*

1971Ju01 G.M.Julian, T.E.Fessler - Phys.Rev. C3, 751 (1971).
*Excited States of $^{138}$Ce.*

1971Ke13 A.Kerek, A.Luukko, M.Grecescu, J.Sztarkier - Nucl.Phys. A172, 603 (1971).
*Two- and Tree-Quasiparticle States in $^{132}$Xe and $^{131}$Xe.*

1971Ke20 A.Kerek - Nucl.Phys. A176, 466 (1971).
*Two-Quasiparticle States in $^{122,124,126}$Te. The Level Systematics in Doubly Even Tellurium Nuclei.*

1971KiZC M.B.Kime - Thesis, Cornell Univ. (1971); Diss.Abstr.Int. 31B, 7511 (1971).
*K Forbidden Isomerism of the 11/2$^{-}$[505] Nilsson Orbital in the OddA, N=91 Isotones.*

1971KiZQ M.B.Kime - Thesis, Cornell Univ. (1971); Diss.Abst.Int. 31B, p.7511 (1971).
*K Forbidden Isomerism of the 11/2$^{-}$ [505] Nilsson Orbital in the Odd A,N=91 Isotones.*

1971Le19 C.M.Lederer, J.M.Jaklevic, J.M.Hollander - Nucl.Phys. A169, 449 (1971).
*In-Beam Gamma-Ray Spectroscopy of Even Mo and Ru Isotopes.*

1971LeYU J.R.Leigh, R.M.Diamond, K.H.Maier, R.Nordhagen, F.S.Stephens - UCRL-20426, p.18 (1971); See Keynumber 1971LEZM.
*Levels in $^{157,159}$Er.*

1971Lo12 A.M.Lopez, W.V.Prestwich, B.Arad - Can.J.Phys. 49, 1828 (1971).
*Isomeric State of $^{97}$Rh.*

1971Lu02 A.Lundan - Z.Phys. 242, 107 (1971).
*The Decay of the $^{136}$I Isomers to the Energy Levels of $^{136}$Xe.*

1971Ma24 S.G.Malmskog, V.Berg, B.Fogelberg, A.Backlin - Nucl.Phys. A166, 573 (1971).
*On the Low-Energy Band Structure in $^{187}$Os.*

1971Ma36 K.H.Maier, J.R.Leigh, F.Puhlhofer, R.M.Diamond - Phys.Lett. 35B, 401 (1971).
*A High-Spin Isomer in $^{211}$At.*

1971Ma59 K.H.Maier, J.R.Leigh, R.M.Diamond - Nucl.Phys. A176, 497 (1971).
*Measurement of the Magnetic Moment and Lifetime of the 13/2$^{+}$ Levelin $^{205}$Pb.*

1971MaXH K.H.Maier, J.R.Leigh, F.Puhlhofer, R.M.Diamond - J.Phys.(Paris) 32, Suppl.No.11-12, Colloq.C6-221 (1971).
*Excited Levels in N = 126 Isotones Studied by (HI,xn$\gamma$)-Reactions.*

1971MaXI K.H.Maier, J.R.Leigh, R.M.Diamond, F.S.Stephens - UCRL-20426, p.21 (1971).
*The Ground-State Bands of the N = 126 Nuclei $^{212}$Rn and $^{214}$Ra Studied by (HI,xn) Reactions.*

1971Mu14 G.Murray, N.E.Sanderson, J.C.Willmott - Nucl.Phys. A171, 435 (1971).
*Low-Lying Positive-Parity States in $^{71}$Ge.*

1971Mu22 F.Munnich, A.Kjelberg, D.J.Hnatowich - Nucl.Phys. A178, 97 (1971).
*The Decay of 33 min $^{104m}$Ag and 69 min $^{104g}$Ag.*

1971Oo01 J.M.Oottukulam, M.K.Ramaswamy - Amer.J.Phys. 39, 221 (1971); Addendum Amer.J.Phys. 39, 1282 (1971).
*Radioactive Half-Life Determination with an Isotope Generator.*

1971Or04 C.J.Orth, B.J.Dropesky, N.J.Freeman - Phys.Rev. C3, 2402 (1971).
*Decay of $^{126}$Sn and of the 19-min and 12.4-day Isomers of $^{126}$Sb.*

1971Ra10 R.S.Raghavan - Z.Phys. 243, 441 (1971).
*Decay of the First Excited State at 13.3 keV in $^{73}$Ge.*

1971Ra22 W.L.Randolph, Jr., R.R.Borchers, R.Michaelsen et al. - Phys.Rev.Lett. 27, 603 (1971).
*Magnetic Moments of the 7/2$^{-}$ Mirror States in $^{37}$Ar and $^{37}$K.*

1971Re06 R.C.Reno, M.Fishbein, C.Hohenemser - Nucl.Phys. A163, 161 (1971).
*Lifetime and g-Factor of the 74.8 keV State of* $^{100}$*Rh.*

1971Re11 R.Repnow, V.Metag, P.von Brentano - Z.Phys. 243, 418 (1971).
*Fission Isomers in Cm and Bk Isotopes.*

1971Ro08 R.G.H.Robertson, R.G.Summers-Gill - Can.J.Phys. 49, 1186 (1971).
*Low-Lying Levels of* $^{58}$*Co.*

1971Ro16 B.Rosner, J.Felsteiner, H.Lindeman, D.Zellermayer - Nucl.Phys. A172, 643 (1971).
*Lifetime and Decay Modes of the 84.6 keV Isomer of* $^{196}$*Au.*

1971Ru03 P.A.Russo, R.Vandenbosch, M.Mehta, J.R.Tesmer, K.L.Wolf - Phys.Rev. C3, 1595 (1971).
*Spin Isomers of the Shape Isomer* $^{237m}$*Pu.*

1971Ru17 S.L.Ruby, R.G.Clark, L.E.Glendenin - Phys.Lett. 36A, 321 (1971).
*Internal-Conversion Coefficients in the Decay of* $^{83}$*Kr-m.*

1971Si19 H.Singh, V.K.Tikku, B.Sethi, S.K.Mukherjee - Nucl.Phys. A174, 426 (1971).
*Decay of* $^{68}$*mCu(3.75 min) and* $^{68}$*gCu(30 sec).*

1971Sm07 F.Smend, I.Borchert, H.Langhoff - Z.Phys. 248, 326 (1971).
*The Influence of the Chemical Environment on the Decay Constant of* $^{90m}$*Nb.*

1971SuZR A.W.Sunyar, P.Thieberger, W.Gelletly, M.Mariscotti - Bull.Amer.Phys.Soc. 16, No.1, 13, AD5 (1971).
$Ni^{62}(\alpha,np\gamma)Cu^{64}$ *Reaction Studies at E = 40 MeV.*

1971Ta07 K.Takahashi, D.L.Swindle, P.K.Kuroda - Phys.Rev. C4, 517 (1971).
$^{122}$*In Isomers.*

1971Te07 J.K.Temperley, J.A.Morrissey, S.L.Bacharach - Nucl.Phys. A175, 433 (1971).
*Spontaneously Fissioning Isomers in* $^{237}$*Pu.*

1971To01 K.S.Toth, R.L.Hahn - Phys.Rev. C3, 854 (1971).
$\alpha$ *Decay from High-Spin Isomers in* $^{153}$*Ho and* $^{154}$*Ho.*

1971To12 D.F.Torgerson, N.S.Oakey, R.D.Macfarlane - Nucl.Phys. A178, 69 (1971).
*Beta-Delayed* $\alpha$*-Emission from* $^{24}$*Al and* $^{24m}$*Al.*

1971Va22 J.van Klinken, S.Goring, D.Habs, M.von Hartrott et al. - Z.Phys. 246, 369 (1971).
$^{139}$*Sm, a New Nuclide of Relevance for the Systematics of the N = 77 Isomers.*

1971Ve04 K.-P.Vester - Z.Phys. 242, 320 (1971).
*K−Conversion Coefficient of the 53.3 keV Transition and K/L+M Ratio of the 13.34 keV Transition in* $^{73}$*Ge.*

1971Wa05 F.W.Walker, T.A.Devito, F.M.Rourke, H.M.Eiland - J.Inorg.Nucl.Chem. 33, 1208 (1971).
$^{148m}$*Pm Half-Life Determination.*

1971Wa16 T.E.Ward, P.E.Haustein - Phys.Rev.Lett. 27, 685 (1971).
*New Five-Quasiparticle Isomeric State in* $^{177}$*Hf.*

1971Wo09 J.L.Wood, D.S.Brenner - Nucl.Phys. A174, 353 (1971).
*The Decay of* $^{162}$*mHo to Levels in* $^{162}$*Dy.*

1972Ad01 A.Adam, D.Horvath, A.Kiss, E.Mayr - Nucl.Phys. A180, 587 (1972).
*Investigation of Short-Lived Activities in* $^{18}$*F,* $^{22}$*Na,* $^{40}$*K,* $^{85}$*Rb and* $^{92}$*Mo Excited by 14.7 MeV Fast Neutrons.*

1972Ad12 B.L.Ader, N.N.Perrin - Nucl.Phys. A197, 593 (1972).
*Niveaux d'Energie du* $^{177}$*Ta.*

1972Al25 M.Alpsten, G.Astner - Phys.Scr. 5, 41 (1972).
*The Decays of* $^{205}$*Po and* $^{203}$*Po.*

1972Al49 G.Albouy, J.M.Lagrange, M.Pautrat, C.Roulet et al. - Phys.Scr. 6, 219 (1972).
*Low-Lying High Spin States in the Neutron Deficient Lead Isotopes.*

1972Am03 B.Amov, T.Vandlik, J.Vandlik, T.Vylov et al. - Yad.Fiz. 16, 881 (1972); Sov.J.Nucl.Phys. 16, 487 (1973).
*Investigation of the Decay of* $Tl^{194}$*.*

1972AnZZ H.R.Andrews, J.S.Geiger, R.L.Graham, S.H.Sie, D.Ward - Bull.Am.Phys.Soc. 17, No.4, 514, EH11 (1972).
*Measurement of E0 Transition Probabilities for* $0^+_1 \rightarrow$ *Transitions in* $^{94}$*Zr,*$^{96}$*Zr and* $^{100}$*Mo* $0^+$*(g.s.).*

1972As04 G.Astner, I.Bergstrom, J.Blomqvist, B.Fant, K.Wikstrom - Nucl.Phys. A182, 219 (1972).
*Effective Quadrupole Charges in the Lead Region.*

1972Ba21 A.Backlin, G.Hedin, V.Berg, S.G.Malmskog - Nucl.Phys. A181, 76 (1972).
*E2 and M1 Transition Probabilities between Low-Energy States in* $^{189}$*Ir.*

1972Ba51 S.A.Bakiev, K.A.Baskova, S.S.Vasilev, M.A.Mokhsen et al. - Izv.Akad.Nauk SSSR, Ser.Fiz. 36, 115 (1972); Bull.Acad.Sci.USSR, Phys.Ser. 36, 109 (1973).
*Levels of* $^{130}$*Xe Excited in* $\beta$ *Decay of* $^{130}$*I and* $^{130}$**I.*

1972Ba65 C.V.K.Baba, T.Faestermann, D.B.Fossan, D.Proetel - Phys.Rev.Lett. 29, 496 (1972).
*Magnetic Moments of the* $7^-$ *and* $5^-$ $(\pi h_{9/2}, g_{9/2})$ *States in* $^{210}$*Bi.*

1972Be37 I.Berka, C.Rolfs, R.E.Azuma - Can.J.Phys. 50, 1682 (1972).
*Excitation Energy of the Fourth Excited State in* $^{18}$*F.*

1972BeWN R.Beeler, L.Balsenc, J.Laplace - Helv.Phys.Acta 45, 929 (1972).
*Methode de Mesure de la Periode d'Elements Radioactifs et DifferencesConstatees.*

1972Bf03 R.Beeler, L.Balsenc, J.Laplace - J.Radioanal.Chem. 10, 257 (1972).

1973Ke07 A.Kerek, J.Kownacki - Nucl.Phys. A206, 245 (1973).
*The Level Structure of the N = 81 and 82 Nucleides* $^{137,138}$*Ba asInvestigated in* $^{136}$*Xe(α,xn) Reactions.*

1973Ke12 A.Kerek, G.B.Holm, S.Borg, P.Carle - Nucl.Phys. A209, 520 (1973).
*Excited States in the Two-Neutron-Hole Nucleus* $^{130}$*Sn Observed in the 0.53 sec* $\beta^-$ *Decay of* $^{130}$*In.*

1973Ke22 A.Kerek, J.Kownacki - Izv.Akad.Nauk SSSR, Ser.Fiz. 37, 1824 (1973); Bull.Acad.Sci.USSR, Phys.Ser. 37, No.9, 30 (1974).
$^{137}$*Ba and* $^{138}$*Ba Levels Excited in (α,xn) Reactions on* $^{136}$*Xe.*

1973Kh02 T.L.Khoo, J.C.Waddington, M.W.Johns - Can.J.Phys. 51, 153 (1973).
*The 1333.1 and 1559.4 keV Isomers in* $^{176}$*Hf.*

1973Ki09 B.T.Kim, W.K.Chu, J.C.Glass - Phys.Rev. C8, 1920 (1973).
*Spin and Magnetic Moment of the 1.094-MeV State in* $^{168}$*Er.*

1973Kl03 W.Klamra, S.A.Hjorth, J.Boutet, S.Andre, D.Barneoud - Nucl.Phys. A199, 81 (1973).
*High-Spin Rotational States in* $^{157}$*Dy.*

1973Kr10 K.Krien, F.Djadali, R.A.Naumann, H.Hubel, E.H.Spejewski - Phys.Rev. C7, 2484 (1973).
*Spectroscopic Investigation of Even-Even Neutron-Deficient Gd Isotopes.*

1973La20 J.C.F.Lau, J.J.Hogan - Phys.Rev. C8, 715 (1973).
*Investigation of Branching Ratios between Isomeric States of* $^{154}$*Tb.*

1973Le09 J.R.Leigh, K.Nakai, K.H.Maier, F.Puhlhofer et al. - Nucl.Phys. A213, 1 (1973).
*Energy-Level Systematics of Odd-Mass Lanthanum Isotopes; a New Coupling Scheme.*

1973Le18 V.E.Lewis, D.Smith, A.Williams - Metrologia 9, 14 (1973).
*Correlation Counting Applied to the Determination of Absolute Disintegration Rates for Nuclides with Delayed States.*

1973LeZJ J.Legrand, J.P.Brethon, F.Lagoutine - CEA-R-4428 (1973).
*Etude de la Desintegration du* $^{137}$*Cs +* $^{137m}$*Ba.*

1973Li17 T.Lindblad, H.Ryde, P.Kleinheinz - Nucl.Phys. A210, 253 (1973); Priv.Comm. (December 1973).
*Effects of the Coriolis Mixing in* $^{181}$*W.*

1973Ma68 B.Martin, D.Merkert - Czech.J.Phys. 23B, 297 (1973).
*Internal Conversion Investigations for the 127.37 keV E3 Transition in* $^{134}$*Cs.*

1973Mc09 J.McDonald, A.Kerek - Nucl.Phys. A206, 417 (1973).
*Transition Probability Measurements in* $^{132}$*Te; The Systematics of E2 Transition Rates in Neutron-Rich Even Te Nuclei.*

1973Mi08 L.D.Miller, F.J.Schima - Int.J.Appl.Radiat.Isotop. 24, 353 (1973).
*The Half-Life of* $^{129m}$*Xe.*

1973Mo07 D.Molzahn, R.Brandt - Phys.Rev. C7, 2596 (1973).
*Production of 62-sec* $^{222m}$*Ac.*

1973Na35 T.Nagy - Magy.Fiz.Foly. 21, 555 (1973).
*Neutronokkal Letrehozott, Izomer Allapotbol Spontan Hasado Magok Keletkezesere Vezeto Reakciok Vizsgalata.*

1973Ne03 D.Nethaway - Priv.Comm. (February 1973).

1973Ni04 A.Nilsson, M.Grecescu - Nucl.Phys. A212, 448 (1973).
*High-Spin States in the Isotones* $^{89}$*Zr,* $^{91}$*Mo and* $^{93}$*Ru Populated in α-Particle-Induced Reactions.*

1973Or03 C.J.Orth, W.R.Daniels, D.C.Hoffman, F.O.Lawrence - Phys.Rev. C8, 718 (1973).
*Decay of* $^{178}$*Yb and the Isomers of* $^{178}$*Lu.*

1973Pa03 M.Pautrat, G.Albouy, J.C.David, J.M.Lagrange et al. - Nucl.Phys. A201, 449 (1973).
*Etude des Isotopes du Plomb de Masse 200, 198, 196 et 194 Formes par Reaction (Ions Lourds, xn).*

1973Pa04 M.Pautrat, G.Albouy, J.M.Lagrange, C.Roulet et al. - Nucl.Phys. A201, 469 (1973).
*Etats Excites des Isotopes de Plomb de Masse 202, 200 et 198.*

1973Pa08 A.Pakkanen, P.Puumalainen, H.Helppi, T.Komppa - Nucl.Phys. A206, 164 (1973).
*Isomeric Transition in* $^{198}$*Au.*

1973Pa21 A.Pakkanen, D.W.Heikkinen - Phys.Fenn. 8, 345 (1973).
*A 6.1 s Isomer and Rotational Bands in* $^{192}$*Os.*

1973PaZV D.G.Parkinson, I.A.Fraser, J.C.Lisle, J.C.Willmott - Priv.Comm. (May 1973).

1973Pr05 I.L.Preiss, J.J.Labrecque - J.Inorg.Nucl.Chem. 35, 1057 (1973).
*Isomerism in* $^{154}$*Pm.*

1973Pr11 D.Proetel, F.Riess, E.Grosse, R.Ley et al. - Phys.Rev. C7, 2137 (1973).
*Gamma Decay of Excited States in* $^{210}$*Bi and an Interpretation with the Shell Model.*

1973Re03 I.Rezanka, I.M.Ladenbauer-Bellis, T.Tamura, W.B.Jones, F.M.Bernthal - Phys.Rev. C7, 1663 (1973); Priv.Comm. (February 1974).
*Study of* $^{173}$*Hf Levels Populated in the Decay of* $^{173}$*Ta.*

1973RoYQ I.A.Romanii, A.P.Klyucharev, Y.N.Rakivnenko, V.A.Lutsik et al. - Program and Theses, Proc.23rd Ann.Conf.Nucl.Spectrosc.Struct.At.Nuclei, Tbilisi, p.113 (1973); See Also 74Ro05.
*Millisecond Isomers in Iridium Isotopes.*

1973Sa10 Z.P.Sawa - Phys.Scr. 7, 5 (1973).
*On* $1f_{7/2}$ *Nuclei. Gamma Ray Spectroscopic Studies on States in* $^{53}$*Mn And* $^{53}$*Fe Using the* $^{50}$*Cr + α Reaction.*

1973Sa22 G.N.Salaita, P.K.Eapen - J.Inorg.Nucl.Chem. 35, 2139 (1973).
*Isomeric Cross Sections for (n,2n) Reactions in Thallium, Lead, and Bismuth at 14.8 MeV.*

1974Ch44 H.C.Cheung, D.G.Burke, G.Lovhoiden - Can.J.Phys. 52, 2108 (1974).
*Odd Proton States in $^{165}$Tm, $^{167}$Tm, $^{169}$Tm, and $^{171}$Tm.*

1974ClZW C.M.Class - ORO-1316-168, p.D4 (1974).
*Re-analysis of Some Neutron Angular Distribution.*

1974ClZX R.G.Clark, L.E.Glendenin, W.L.Talbert,Jr. - Proc.Symp.Phys.Chem.Fission, 3rd, Rochester, N.Y. (1973), Int.At.En.Agency, Vienna, Vol.2, p.221 (1974).
*Fission Fragment Isomers from Spontaneous Fission of $^{252}$Cf.*

1974Co33 W.J.Courtney, C.I.Delaune, J.W.Nelson, J.D.Fox - Phys.Rev. C10, 1922 (1974).
*Half-Life of the 257 keV $1^+ \rightarrow 4^-$ Transition in $^{90}$Nb.*

1974De09 F.W.N.de Boer, P.F.A.Goudsmit, P.Koldewijn, B.J.Meyer - Nucl.Phys. A225, 317 (1974); Priv.Comm. (March 1974).
*Decay Properties of Three $^{166}$Lu Isomers and the Decay of $^{166}$Hf.*

1974De47 F.W.N.de Boer, P.F.A.Goudsmit, B.J.Meijer, P.Koldewijn et al. - Nucl.Phys. A236, 349 (1974).
*The Decays of 21.55 min $^{162g}$Tm and 24.3 s $^{162}$Tm (A New Isomer).*

1974Di03 M.Diksic, L.Yaffe, D.G.Sarantites - Phys.Rev. C10, 1172 (1974).
*Identification and Decay Characteristics of $^{132m}$I.*

1974Di18 F.Dimmling, N.Brauer, B.Focke, T.Kornrumpf et al. - Z.Phys. 271, 103 (1974).
*Lifetime and Magnetic Moment of the $11/2^-$, 731 keV Level in $^{113}$Sn.*

1974Dr02 J.E.Draper, N.S.P.King, W.G.Wyckoff - Phys.Rev. C9, 948 (1974).
*In-Beam Measurement of E0 Matrix Elements in $^{72}$Se and $^{72}$Ge.*

1974Ek03 C.Ekstrom, W.Hogervorst, S.Ingelman, G.Wannberg - Phys.Scr. 10, 301 (1974).
*Nuclear Spins of $^{165}$Lu, $^{168m}$Lu and $^{172}$Lu.*

1974En09 A.Enulescu, I.Piticu, I.Vaviceta - Izv.Akad.Nauk SSSR, Ser.Fiz. 38, 2090 (1974); Bull.Acad.Sci.USSR, Phys.Ser. 38, No.10, 66 (1974).
*K-Forbidden Transitions in $^{169}$Tm.*

1974Er07 H.N.Erten, J.Blachot - Radiochim.Acta 21, 1 (1974).
*Decay of 5.7-Minutes and 40-Minutes $^{130}$Sb Isomers to Levels in $^{130}$Te.*

1974Fi18 A.Filevich, M.Lagos - Nucl.Instrum.Methods 122, 605 (1974).
*Two Simple Circuits for in-Beam μs and ms Time Measurements.*

1974FiZI P.Fintz, A.Gallmann, G.Guillaume, K.W.Jones et al. - Bull.Int.Conf.Gamma-Ray Transition Probabilities, Delhi, Abstr.2C3, p.7 (1974).
*Delayed γ-Rays in the Reaction $^{40}$Ca + $^{16}$O.*

1974Fo01 C.Foin, S.Andre, S.A.Hjorth - Nucl.Phys. A219, 347 (1974).
*Intrinsic and Rotational Excitations in $^{175}$Lu.*

1974Fo06 M.M.Fowler, G.W.Goth, C.-C.Lin, A.C.Wahl - J.Inorg.Nucl.Chem. 36, 1191 (1974); A.C.Wahl, Priv.Comm. (January 1974).
*Half-Lives of Tin and Antimony Fission Products with A = 128-133.*

1974Fo23 B.Fogelberg, T.Nagarajan, B.Grapengiesser - Nucl.Phys. A230, 214 (1974).
*Levels and Transition Probabilities in $^{124}$In as Observed in the Decay of $^{124}$Cd.*

1974FoZY J.Fontanilla, A.L.Prindle, J.H.Landrum, R.A.Meyer - Bull.Amer.Phys.Soc. 19, No.4, 501, EH11 (1974).
*Half-Lives and β-Decay of the 133 and 135 Isomers of Xe.*

1974Gr29 B.Grapengiesser, E.Lund, G.Rudstam - J.Inorg.Nucl.Chem. 36, 2409 (1974).
*Survey of Short-Lived Fission Products Obtained Using the Isotope-Separator-on-Line Facility at Studsvik.*

1974GrZN J.W.Gruter, G.Sadler, K.Sistemich, W.Borgs, H.Lawin - Proc.Int.Conf.Nucl.Structure and Spectrosc., Amsterdam, H.E.Blok, A.E.L.Dieperink, Eds., Scholar′s Press, Amsterdam, Vol.1, p.89 (1974).
*Nuclear Spectroscopy Investigations on Light Fission Fragments with JOSEF.*

1974Ha04 J.H.Hamilton, A.V.Ramayya, W.T.Pinkston, R.M.Ronningen et al. - Phys.Rev.Lett. 32, 239 (1974).
*Evidence for Coexistence of Spherical and Deformed Shapes in $^{72}$Se.*

1974Ha10 J.H.Hamilton, E.H.Spejewski, R.L.Mlekodaj, W.-D.Schmidt-Ott et al. - Izv.Akad.Nauk SSSR, Ser.Fiz. 38, 2036 (1974); Bull.Acad.Sci.USSR, Phys.Ser. 38, No.10, 22 (1974).
*Initial UNISOR Research: New Isotopes $^{186}$Tl, $^{188}$Tl, $^{116}$I; Decays of $^{189,190}$Tl, $^{117}$Xe, and $^{117}$I; and Off-Line Atomic and Nuclear Studies.*

1974Ha39 H.H.Hansen, E.de Roost, W.van der Eijk, R.Vaninbroukx - Z.Phys. 269, 155 (1974).
*The Decay of $^{115m}$In.*

1974Ha41 U.Hagemann, H.F.Brinckmann, W.D.Fromm, C.Heiser, H.Rotter - Nucl.Phys. A228, 112 (1974).
*Observation of Isomeric States and Band Structures in $^{107}$Cd.*

1974Ha48 O.Hausser, D.J.Donahue, R.L.Hershberger, R.Lutter et al. - Phys.Lett. 52B, 329 (1974).
*Particle-Vibration Coupling in $^{107}$Cd.*

1974Ha50 A.Hashizume, H.Kumagai, Y.Tendow, T.Katou - Nucl.Instrum.Methods 119, 209 (1974).
*A Mechanical Beam Chopper System for the Measurement of Half-Lives inthe Millisecond Region.*

1974HeZE R.H.Heffner - Thesis, Univ.Washington (1973); Diss.Abstr.Int. B35, 435 (1974).
*Experimental Study of the Deformation of the Fission Isomer in $^{236}$U.*

1974Ho17 W.Hogervorst, C.Ekstrom, S.Ingelman, G.Wannberg - Phys.Scr. 9, 317 (1974).
*Nuclear Spins of Some Neutron-Deficient Tin Isotopes.*

1974Ho27 P.Hornshoj, P.G.Hansen, B.Jonson - Nucl.Phys. A230, 380 (1974).

1974Va08 J.van Klinken, L.M.Taff - Phys.Rev. C9, 2252 (1974).
*Isomeric State in* $^{74}$*Ga.*

1974Va19 J.Vandlik, N.G.Zaitseva, Z.Mate, I.Mahunka et al. - Izv.Akad.Nauk SSSR, Ser.Fiz. 38, 695 (1974); Bull.Acad.Sci.USSR, Phys.Ser. 38, No.4, 26 (1974).
*Investigation of the Decay* $^{191}Tl \rightarrow {}^{191}Hg \rightarrow {}^{191}Au$.

1974Vi05 A.Visvanathan, E.G.Funk, J.W.Mihelich - Z.Phys. 271, 339 (1974).
*Measurements of Some K-Internal Conversion Coefficients Near Threshold.*

1974ViZS C.L.M.Vieu - CSNSM-T-74-01 (1974).
*Studies of 193 and 195 Gold Isotopes.*

1974Vo03 I.Votsilka, B.Kratsik, Y.Liptak, A.F.Novgorodov, M.Toshev - Izv.Akad.Nauk SSSR, Ser.Fiz. 38, 57 (1974); Bull.Acad.Sci.USSR, Phys.Ser. 38, No.1, 49 (1974).
*Decay of* $^{87}$*Nb. The New Isotope* $^{86}$*Nb.*

1974Wa14 T.E.Ward, P.E.Haustein, J.B.Cumming, Y.Y.Chu - Phys.Rev. C10, 1983 (1974).
*Decay of the* $K^{\pi} = 8^-$ *Isomer in* $^{182}$*Hf.*

1974Wi06 G.Winter, W.Andrejtscheff, L.Funke, P.Manfrass, H.Sodan - Nucl.Phys. A223, 320 (1974).
*The Rotational Bands* $1/2^+[411]$ *and* $1/2^-[541]$ *in* $^{175}$*Lu.*

1974Wi14 D.H.Wilkinson, D.E.Alburger - Phys.Rev. C10, 1993 (1974).
*Half-Lives of* $^{6}$*He,* $^{19}$*Ne, and* $^{42}$*Sc-m.*

1974WoZW K.L.Wolf, J.W.Meadows - Bull.Am.Phys.Soc. 19, No.4, 595, KH1 (1974).
*Measurements on the Fissioning Isomer* $^{238m}$*U with the (n,n′) and (d,pn) Reactions.*

1974Ya02 S.W.Yates, J.C.Cunnane, P.J.Daly, R.Thompson, R.K.Sheline - Nucl.Phys. A222, 276 (1974).
*Levels of* $^{190}$*Os Populated in the Decays of 3.3 h* $^{190m}$*Re and 12 d* $^{190}$*Ir and in the* $^{189}$*Os(d,p)*$^{190}$*Os Reaction.*

1974Ya07 H.Yamamoto, K.Kawade, H.Fukaya, T.Katoh - J.Phys.Soc.Jap. 37, 10 (1974).
*Decay of* $^{154}$*Pm to Levels of* $^{154}$*Sm.*

1974Ya09 Y.Yamazaki, O.Hashimoto, H.Ikezoe, S.Nagamiya et al. - Phys.Rev.Lett. 33, 1614 (1974).
*Blocking Effect of the M1 Core Polarization Studied from the g FactorOf the* $8^+$ *State in* $^{214}$*Ra.*

1975Ad09 I.Adam, K.Y.Gromov, M.Honusek, T.A.Islamov et al. - Nucl.Phys. A254, 63 (1975).
*Levels in* $^{163}$*Tm Excited by the Decay of 11.4 min* $^{163}$*Yb.*

1975Ah05 I.Ahmad, F.T.Porter, M.S.Freedman, R.K.Sjoblom et al. - Phys.Rev. C12, 541 (1975).
*Levels in* $^{237}$*Pu Populated by* $^{237}$*Am (Electron Capture) and* $^{241}$*Cm(*$\alpha$*) Decays.*

1975Al13 A.A.Aleksandrov, G.-Y.Baier, T.Vylov, E.P.Grigorev et al. - Izv.Akad.Nauk SSSR, Ser.Fiz. 39, 458 (1975); Bull.Acad.Sci.USSR, Phys.Ser. 39, No.3, 1 (1975).
*The Excited States of* $^{158}$*Dy.*

1975Al21 V.L.Alexeev, B.A.Emelianov, L.P.Kabina, D.M.Kaminker et al. - Nucl.Phys. A248, 249 (1975).
*Low-Lying* $^{134}$*Cs Excited States from the (n,*$\gamma$*) Reaction.*

1975An08 S.Andre, J.Boutet, J.Rivier, J.Treherne et al. - Nucl.Phys. A243, 229 (1975).
*Excited States of the Transitional* $^{189}$*Ir and* $^{187}$*Ir Nuclei Populated by The (*$\alpha$*,2n) Reaction.*

1975Az01 G.Azuelos, J.E.Kitching - Phys.Rev. C12, 563 (1975).
*Half-Lives of Some* $T = 1/2$ *Mirror Decays.*

1975Ba36 G.Bailleul, J.P.Bocquet, H.Schrader, R.Stippler et al. - Z.Phys. A273, 283 (1975).
*Gamma-Ray Spectroscopy of Short-Lived Fission Products Using the Recoil Separator 'Lohengrin'.*

1975Ba60 H.Bartsch, K.Huber, U.Kneissl, H.Krieger - Nucl.Phys. A252, 1 (1975).
*The Decay of the* $11/2^-$ *Isomeric State in* $^{103}$*Ru.*

1975Be17 V.Berg, R.Foucher, A.Hoglund - Nucl.Phys. A244, 462 (1975).
*Quasi-Prolate and Quasi-Oblate Bands in Soft* $^{189}$*Au.*

1975Be32 H.Bertschat, H.Kluge, U.Leithauser, E.Recknagel, B.Spellmeyer - Nucl.Phys. A249, 93 (1975).
*Low-Lying Levels and the Magnetic Moment of the* $3^+$ *213.5 keV Statein* $^{72}$*As.*

1975Be39 I.Bergstrom, C.J.Herrlander, T.Lindblad, V.Rahkonen et al. - Z.Phys. A273, 291 (1975).
*Evidence for Two-Particle Core Polarization in the Isomeric* $29/2^+$ *and* $21/2^-$ *States of the* $\pi(h^2_{9/2})_8+\ i_{13/2}\nu(j^{-2})_0+$ *and* $\pi(h^3_{9/2})\nu(j^{-2})_0+$ *Configurations in* $^{209}$*At.*

1975Bi01 P.G.Bizzeti, P.R.Maurenzig, G.Poggi, G.Lo Bianco - Lett.Nuovo Cim. 12, 53 (1975).
*The* $6^+$ *Isomer State in* $^{46}$*Ca.*

1975BlZY D.Bloch - Thesis, Fredrich-Alexander Univ., Erlangen-Nurnberg (1975).
*G-Factor-Messungen an Isomeren Zustanden in 126I und 130I, Sowie Vorschlage fur Niveauschemas Dieser Kerne im Niederenergiebereich.*

1975Bo30 E.Bozek, C.Gehringer, C.Jaeger, J.C.Merdinger - Nucl.Phys. A250, 257 (1975).
*Lifetime of the Second Excited State of* $^{43}$*K.*

1975Br01 B.A.Brown, P.M.S.Lesser, D.B.Fossan - Phys.Rev.Lett. 34, 161 (1975).
*High-Spin Isomers in* $^{91}$*Nb and* $^{91}$*Zr via* $^{6}$*Li-Induced Reactions.*

1975Br16 A.R.Brosi, B.H.Ketelle - Nucl.Phys. A245, 243 (1975).
$^{136}$*Nd Decay and the* $^{136}$*Pr Ground State Spin.*

1975BrYM W.Bruchle - Thesis, Johannes Gutenberg-Universitat, Mainz (1975).

*States in N = 80 $^{140}$Nd Populated in the Decay of $^{140}$Pm.*

1975Kh04 T.L.Khoo, F.M.Bernthal, R.A.Warner, G.F.Bertsch, G.Hamilton - Phys.Rev.Lett. 35, 1256 (1975).
*Residual Interactions in Four-Quasiparticle $K^{\pi}$ = $14^{-}$ Isomer in $^{176}$Hf.*

1975Ki06 H.Kirschner, A.Ok - Z.Phys. A272, 287 (1975).
*K-Konversionskoeffizient nahe der K-Bindungsenergie, gemessen am $^{174m}$Lu.*

1975Kl01 P.Kleinheinz, M.R.Maier, R.M.Diamond, F.S.Stephens, R.K.Sheline - Phys.Lett. 53B, 442 (1975).
*Revised Single-Particle Energies in N = 83 Nuclei.*

1975Kl04 W.Klamra, J.Rekstad - Nucl.Phys. A243, 395 (1975).
*Level Structure of $^{103}$Ru Studied by the $^{100}$Mo($\alpha$,n)$^{103}$Ru Reaction.*

1975Kl11 G.Klein, N.Kaffrell, N.Trautmann, G.Hermann - Inorg.Nucl.Chem.Lett. 11, 511 (1975).
*Identification of $^{96}$Y in Fission by a Rapid Chemical Separation Procedure.*

1975Kr08 J.V.Kratz, H.Franz, N.Kaffrell, G.Herrmann - Nucl.Phys. A250, 13 (1975).
*Gamma-Ray Emission from $^{80-86}$As Isotopes.*

1975Ku17 W.Kutschera, B.A.Brown, H.Ikezoe, G.D.Sprouse et al. - Phys.Rev. C12, 813 (1975).
*Lifetimes of the $6_1^+$ States in $^{42}$Ti and $^{46}$Ca and E2 Effective Charges in (1f$_{7/2}$)±2 Nuclei.*

1975Li16 R.M.Lieder, H.Beuscher, W.F.Davidson, A.Neskakis, C.Mayer-Boricke - Nucl.Phys. A248, 317 (1975).
*Excitation of High-Spin States in $^{190,191,192,193,194}$Hg Through ($\alpha$,xn) Reactions.*

1975Ma02 E.Matthias, E.Recknagel, O.Hashimoto, S.Nagamiya et al. - Nucl.Phys. A237, 182 (1975).
*The g-Factor Measurement for the $8^+$ Isomeric State of $^{86}$Sr Produced in Kr Gas.*

1975McZO A.B.McDonald, J.R.Beene, O.Hausser, T.K.Alexander, E.D.Earle - AECL-5121, p.14 (1975).
*Measurements of g-Factors of Isomeric Levels in $^{211}$At.*

1975Me22 M.Meyer, R.Beraud, J.Daniere, R.Rougny et al. - Phys.Rev. C12, 1858 (1975).
*High-Spin States in $^{109}$Cd.*

1975Mi05 J.A.Mirza, K.E.G.Lobner, D.Breitig, H.A.Baader et al. - Z.Phys. A272, 175 (1975).
*The Nuclear Structure of $^{198}$Au from the Reaction $^{197}$Au(n,$\gamma$)$^{198}$Au.*

1975Mo03 J.A.Morman, W.C.Schick,Jr., W.L.Talbert,Jr. - Phys.Rev. C11, 913 (1975).
*Lifetimes of Levels in $^{136}$Xe, $^{140}$Cs, and $^{141}$Cs Populated in the Decays Of Mass-Separated $^{136}$I, $^{140}$Xe, and $^{141}$Xe.*

1975No11 E.Nolte, P.Vogt - Z.Phys. A275, 33 (1975).
*Rotational Bands in $^{77}$Kr.*

1975Pa01 L.Palffy, J.P.Deutsch, L.Grenacs, J.Lehmann, M.Steels - Phys.Rev.Lett. 34, 212 (1975).
*Comparison of the $^{16}$N-m($0^-$) $\beta$-Decay Rate with the Inverse Muon-Capture Rate: The Induced Pseudoscalar Form Factor in Nuclei.*

1975Pl04 D.Plostinaru, E.A.Ivanov, G.Pascovici, A.Iordaschescu - Phys.Lett. 57B, 235 (1975).
*Spin-Parity, Lifetime and Magnetic Moment of a Short-Lived Isomeric State in $^{118}$Sb.*

1975Pl05 W.Plukebaum - Z.Phys. A273, 393 (1975).
*Die Zerfalle des Rh$^{97g}$ und des Rh$^{97m}$.*

1975Pr05 H.S.Pruys, E.A.Hermes, H.R.Von Gunten - J.Inorg.Nucl.Chem. 37, 1587 (1975).
*The Decay of $^{152m1}$Eu and $^{152m2}$Eu and Reaction Cross Sections of$^{153}$Eu for 14 MeV Neutrons.*

1975Ra40 A.V.Ramayya, R.M.Ronningen, J.H.Hamilton, W.T.Pinkston et al. - Phys.Rev. C 12, 1360 (1975).
*Mean life and collective effects of the 937 keV, $0^+$ state in $^{72}$Se: Evidence for nuclear coexistence.*

1975RaYX D.Rabenstein - Proc.Int.Symp.Neutron Capture Gamma-Ray Spectrosc. and Related Topics, 2nd, Petten, p.584 (1974).
*Spins and Lifetimes of New Low-Lying States in $^{204}$Tl.*

1975Re05 I.Rezanka, I.M.Ladenbauer-Bellis, J.O.Rasmussen, W.Ribbe, E.der Mateosian - Phys.Rev. C11, 1767 (1975).
*High-Spin Rotational States in $^{169}$Hf from the $^{159}$Tb($^{14}$N,4n$\gamma$) Reaction and Decay of $^{169}$Ta.*

1975Re06 E.Recknagel - Phys.Scr. 11, 208 (1975).
*Solid State Information from Magnetic Hyperfine Interaction Studied in-Beam.*

1975Ri06 J.Rivier, R.Moret - Radiochim.Acta 22, 27 (1975).
*Mise en Evidence de l'Isotope $^{105}$In et Etude de la Desintegration $^{105}$In → $^{105}$Cd.*

1975Ro05 C.Rolfs, W.S.Rodney, S.Durrance, H.Winkler - Nucl.Phys. A240, 221 (1975).
*A New Method for Measurement of Proton Beam Energies.*

1975Ro25 S.Roodbergen, H.Visser, W.Molendijk, H.S.Bedet, H.Verheul - Z.Phys. A275, 45 (1975).
*Transition Probabilities in the Ni-Zn Region.*

1975Ru03 P.A.Russo, J.Pedersen, R.Vandenbosch - Nucl.Phys. A240, 13 (1975).
*Gamma Decay of the $^{238}$U Shape Isomer.*

1975Ru06 H.Rubinsztein, M.Gustafsson - Phys.Lett. 58B, 283 (1975).
*Nuclear Spin Measurements on Neutron-Deficient Isotopes of the Refractory Elements.*

1975Sa15 G.Sadler, T.A.Khan, K.Sistemich, J.W.Gruter et al. - Nucl.Phys. A252, 365 (1975).
*Studies of the $\beta$-Decay of $^{96}$Y and the Level Scheme of $^{96}$Zr.*

1975Sc17 K.H.Schedl, C.Bercks, E.Hummel - Z.Phys. A272, 417 (1975).
*Untersuchungen isomerer Ubergange in $^{140}$Pr und $^{142}$Pr.*

1975Se12 P.Sen, N.Cue, H.Bakhru - Phys.Rev. C12, 716 (1975).
*32.1 ms Isomer in $^{198}$Tl.*

1975So03 D.C.Sousa, L.L.Riedinger, E.G.Funk, J.W.Mihelich - Nucl.Phys. A238, 365 (1975).

*Decay of the Three Isomers of $^{154}$Tb.*
1975Sq01 G.T.A.Squier, W.E.Burcham, J.M.Freeman, R.J.Petty et al. - Nucl.Phys. A242, 62 (1975).
*The ft Value for the Superallowed Fermi Decay $^{38m}K(\beta^+)^{38}$Ar.*
1975St07 H.Strusny, H.Tyrroff, E.Herrmann, G.Musiol et al. - Nucl.Phys. A242, 30 (1975).
*The Decay of $^{159}$Tm (9.0 min) to $^{159}$Er.*
1975Su02 K.Summerer, N.Kaffrell, N.Trautmann - Z.Phys. A273, 77 (1975).
*The Decay of $^{105}$Tc to Levels in $^{105}$Ru.*
1975Ta06 C.W.Tang, A.Pakkanen, Z.C.Mester, C.D.Coryell et al. - Z.Phys. A272, 301 (1975).
*The Decay of 2.4 h $^{117g}$Cd and 3.4 h $^{117m}$Cd.*
1975To08 H.Toki, A.Faessler - Nucl.Phys. A253, 231 (1975).
*Asymmetric Rotor Model for Decoupled Bands in Transitional Odd-Mass Nuclei.*
1975Uy01 J.Uyttenhove, K.Heyde, H.Vincx, M.Waroquier - Nucl.Phys. A241, 135 (1975).
*Possible Evidence for Microsecond Shape Isomerism in Tl Isotopes.*
1975Va02 M.A.Van Driel, H.Klijnman, G.A.P.Engelbertink, H.H.Eggenhuisen, J.A.J.Hermans - Nucl.Phys. A240, 98 (1975).
*Allowed Electron-Capture Branches in the Decay of $^{34m}$Cl.*
1975Va14 J.van Klinken, S.J.Feenstra - Phys.Rev. C12, 2111 (1975).
*Shape Implications of Unhindered $11/2^- \rightarrow 11/2^-$ $\beta$ Decays in the.*
1975Va16 L.Varnell, I.Rezanka - Nucl.Phys. A255, 75 (1975).
*Magnetic Moment of the $8^+$, 13.9 msec Isomer $^{88m}$Y.*
1975Va20 T.B.Vandlik, J.Vandlik, N.G.Zaitseva, Z.Mate et al. - Yad.Fiz. 22, 873 (1975); Sov.J.Nucl.Phys. 22, 455 (1976).
*Decay of the Isotopes $^{192}$Tl and $^{190}$Tl.*
1975Va21 G.V.Valsky, V.L.Varentsov, G.A.Petrov, Y.S.Pleva et al. - Yad.Fiz. 22, 701 (1975); Sov.J.Nucl.Phys. 22, 363 (1976).
*Formation of the Spontaneously Fissile Isomer $^{242m}$Am in Thermal-Neutron Capture.*
1975Wa19 T.E.Ward, Y.Y.Chu - Radiochem.Radioanal.Lett. 22, 1 (1975).
*Refined Value for the Half-Life of $^{177m}$Lu.*
1975Wa24 T.E.Ward, Y.Y.Chu - Phys.Rev. C12, 1632 (1975).
*Decays of 2.45-h $^{178}$Ta-m and 23.8-min $^{178}$Lu-m to the $K^\pi = 8^-$ Levels In $^{178}$Hf.*
1975Ya03 S.W.Yates, I.Ahmad, A.M.Friedman, F.J.Lynch, R.E.Holland - Phys.Rev. C11, 599 (1975).
*Lifetimes of the $1/2^+[631] \rightarrow 5/2^+[622]$ E2 Transitions in Some Actinide Nuclei.*
1975Ya14 S.W.Yates, F.J.Lynch - Phys.Rev. C12, 1080 (1975).
*Isomerism in $^{194}$Au.*
1975Yo01 N.Yoshikawa - Nucl.Phys. A243, 143 (1975).
*In-Beam Study of $^{136}$Ce and $^{138}$Nd Isotones with $N = 78$.*
1976Ah06 H.Ahrens, N.Kaffrell, N.Trautmann, G.Herrmann - Phys.Rev. C14, 211 (1976).
*Decay Properties of Neutron-Rich Niobium Isotopes.*
1976Ao02 T.Aoki - Nucl.Phys. A272, 47 (1976).
*The Gamma Transition from the $55/22^+(402)$ State to the $77/22^+(404)$ State in $^{177}$Ta.*
1976Ba02 C.V.K.Baba, D.B.Fossan, T.Faestermann, F.Feilitzsch et al. - Nucl.Phys. A257, 135 (1976).
*Lifetime and g-Factor Results for the $13/2^-$ 1985 keV Level in $^{91}$Nb And the $15/2^-$ 2288 keV Level in $^{91}$Zr.*
1976Be11 G.Beyer, A.Jasinski, O.Knotek, H.-G.Ortlepp et al. - Nucl.Phys. A260, 269 (1976).
*The Decay Scheme of $^{127}$Ba and Half-Lives of Some Low-Lying ExcitedStates in $^{123,125,127,129}$Cs.*
1976Be12 H.Beuscher, D.R.Zolnowski, D.R.Haenni, T.T.Sugihara - Phys.Rev.Lett. 36, 1128 (1976).
*High-Spin Isomeric States in $^{202,204,206}$Po.*
1976Be44 I.Bergstrom, J.Blomqvist, C.J.Herrlander, C.G.Linden - Z.Phys. A278, 257 (1976).
*Evidence for a 3.6 Minute Isomeric $12^-$ State of the $\pi h^{-1}_{11}/2$ $\nu i^{-1}_{13}/2$ Configuration in $^{206}$Tl and Effective Two-Particle Interactions.*
1976Bh04 P.Bhattacharyya, R.K.Chattopadhyay, B.Sethi, V.K.Tikku, S.K.Mukherjee - Nuovo Cim. 31A, 519 (1976).
*Decay of the 46.6 s $^{75}$Ge-m.*
1976Bi09 C.R.Bingham, L.L.Riedinger, F.E.Turner, B.D.Kern et al. - Phys.Rev. C14, 1586 (1976).
*Decay of Mass-Separated $^{190}$Tl and $^{190}$Hg.*
1976Bi13 M.Bini, A.M.Bizzeti Sona, P.G.Bizzeti, P.Blasi et al. - Nuovo Cim. 35A, 69 (1976).
*Low-Lying Levels of $^{96}$Tc Excited in the $^{93}$Nb($\alpha$,n)$^{96}$Tc Reaction.*
1976Bo19 M.Bormann, H.-K.Feddersen, H.-H.Holscher, W.Scobel, H.Wagner - Z.Phys. A277, 203 (1976).
*(n,2n) Anregungsfunktionen fur $^{54}$Fe, $^{70}$Ge, $^{74}$Se, $^{85}$Rb, $^{86,88}$Sr, $^{89}$Y, $^{92}$Mo, $^{204}$Hg im Neutronenenergiebereich 13-18 MeV.*
1976Br08 B.A.Brown, D.B.Fossan, P.M.S.Lesser, A.R.Poletti - Phys.Rev. C13, 1194 (1976).
*Isomeric $17/2^-$ Level and Parity Mixing in $^{93}$Tc.*
1976Br14 B.A.Brown, P.M.S.Lesser, D.B.Fossan - Phys.Rev. C13, 1900 (1976).
*Properties of High-Spin States in $^{91}$Nb and $^{91}$Zr via $^{6}$Li-Induced Reactions.*
1976Ch08 Y.Y.Chu, E.-M.Franz - Phys.Rev. C13, 2011 (1976); Erratum Phys.Rev. C14, 772 (1976).
*Conversion Electron Studies on the New High-Spin Isomer in $^{154}$Eu.*
1976Ch17 T.U.Chan, M.Agard, J.F.Bruandet, A.Giorni et al. - J.Phys.(Paris),Lett. 37, L-237 (1976).
*Ferromagnetic Effect and Spin Assignment for the 390 keV State in $^{62}$Cu.*
1976Ch30 S.Chojnacki, T.Morek, A.Korman, Z.Haratym, T.Kempisty - Acta Phys.Pol. B7, 823 (1976).

*Level Properties of the Light Even Sb Nuclei $^{112}$,$^{114}$,$^{116}$Sb.*
1976KaYO N.Kaffrell, G.Franz, G.Klein, K.Summerer et al. - CERN-76-13, p.483 (1976).
*Decay Properties of Neutron-Rich Nuclei in the Mass Region A ≈ 100.*
1976Kh03 T.L.Khoo, F.M.Bernthal, R.G.H.Robertson, R.A.Warner - Phys.Rev.Lett. 37, 823 (1976).
*High-Spin Multiquasiparticle Yrast Traps in $^{176}$Hf.*
1976Ko13 A.Korman, D.Chlebowska, T.Kempisty, S.Chojnacki - Acta Phys.Pol. B7, 141 (1976).
*Excited States in $^{199}$Bi, $^{201}$Bi, $^{203}$Bi Isotopes.*
1976La14 N.Lavi, Y.Nir-El - J.Inorg.Nucl.Chem. 38, 2133 (1976).
*Isomer Yield Ratios of $^{132m}$I and $^{134m}$I.*
1976Le03 W.Leitz, H.-E.Mahnke, W.Semmler, R.Sielemann, T.Wichert - Nucl.Phys. A258, 103 (1976).
*The Half-Lives of the First and Second Excited States in $^{66}$Ga.*
1976Li09 C.G.Linden, I.Bergstrom, J.Blomqvist, K.-G.Rensfelt et al. - Z.Phys. A277, 273 (1976).
*High Spin States in $^{205}$Pb and a Precision Test of the Nuclear Shell Model for Three Nucleons.*
1976Lu07 J.Ludziejewski, J.Bialkowski, Z.Haratym, L.-E.De Geer et al. - Phys.Scr. 14, 133 (1976).
*The Life-Time Measurements of Some High-Spin States in the $^{138}$,$^{139}$Ce and $^{141}$,$^{142}$Nd Nuclei.*
1976Ma20 M.A.J.Mariscotti, M.Behar, A.Filevich, G.Garcia Bermudez et al. - Nucl.Phys. A260, 109 (1976).
*Level Structure of $^{72}$As Studied with the (α,xnp) Reaction.*
1976Ma63 G.C.Martin,Jr. - Int.J.Appl.Radiat.Isotop. 27, 613 (1976).
*The Half-Life of $^{119m}$Sn.*
1976MoZC E.Monnand, J.Blachot, F.Schussler, J.P.Bocquet et al. - Proc.Int.Conf.Nuclei Far from Stability, Cargese, Corsica, p.477 (1976); CERN-76-13 (1976).
*Recent Studies of the Odd-Mass Sr,Y, Zr and Nb Fission Products (A = 95, 97, 99).*
1976NeZT V.G.Nedovesov, Y.V.Kholnov, G.E.Shukin - Program and Theses, Proc. 26th Ann.Conf.Nucl.Spectrosc.Struct.At.Nuclei, Baku, p.109 (1976).
*Half-Life of the Isomers of $^{154}$Tb.*
1976Pi03 M.Piiparinen, S.K.Saha, P.J.Daly, C.L.Dors et al. - Phys.Rev. C13, 2208 (1976).
*$\nu i_{13/2}$ and $\nu h_{9/2}$ Isomers in Odd-A Pt Nuclei.*
1976Pi18 R.G.Pillay, S.H.Devare, P.N.Tandon - Pramana 7, 190 (1976).
*The g-Factor of the 659 keV Level in $^{117}$In.*
1976Po11 L.A.Popeko, G.A.Petrov, Y.P.Rudnev, E.F.Kochubey - Yad.Fiz. 24, 1081 (1976); Sov.J.Nucl.Phys. 24, 567 (1976).
*Investigation of Conversion Electrons in Fission of $^{235}$U by Thermal Neutrons.*
1976Ra37 D.G.Raich, H.R.Bowman, R.E.Eppley, J.O.Rasmussen, I.Rezanka - Z.Phys. A279, 301 (1976); Erratum Z.Phys. A282, 124 (1977).
*Gamma and Alpha Decay from the 2.1-msec Isomer $^{213m}$Ra.*
1976ReZH I.H.Redmount, T.L.Khoo, R.A.Warner - Mich.State Univ., Cyclotron Lab., Ann.Rept., p.66 (1976).
*Three-Quasiparticle States in $^{177}$Hf.*
1976Ri09 N.Rizk, J.Boutet - J.Phys.(Paris), Lett. 37, L-197 (1976).
*Isomeric State and Rotational Band in $^{158}$Ho.*
1976RoYE L.G.Robinson - Thesis, Indiana Univ. (1976).
*A Study of the Levels in Rhenium-179, Tantalum-173, Lutetium-167, andThulium-163 with the (α,6n) Reaction.*
1976RoZE I.A.Romany, A.P.Kliucharev, V.A.Lutsik, Y.N.Rakivnenko - Proc.Int.Conf.Selected Topics in Nucl.Struct., Dubna, Vol.1, p.63 (1976); JINR-D-9682 (1976).
*Decay Schemes of Isomers $^{192m}$Au and $^{194m}$Au.*
1976Sc30 O.Scheidemann - J.Inorg.Nucl.Chem. 38, 1757 (1976).
*The Decay of $^{119}$Cd and $^{119}$In Isomers.*
1976SeZN R.Sellam - Thesis, Univ.Grenoble (1976).
*Etude des Etats Isomeriques de μsec des Produits de Fission de l'$^{235}$U apres Separation en Masse par 'Lohengrin'.*
1976Sh05 T.Shibata, Y.Nagai, M.Fujiwara, H.Ejiri et al. - Nucl.Phys. A257, 303 (1976).
*Low-Lying Levels in $^{145}$Pm and the Effective M2 Transition Moment.*
1976Sj01 T.P.Sjoreen, G.Schatz, S.K.Bhattacherjee, B.A.Brown et al. - Phys.Rev. C14, 1023 (1976).
*High-Spin States and Neutron-Hole Influence in $^{209}$At.*
1976Sl01 G.Sletten, V.Metag, E.Liukkonen - Phys.Lett. 60B, 153 (1976).
*Picosecond Fission Isomers in Even-Even Cm Isotopes.*
1976Sm01 H.A.Smith, Jr., M.E.Bunker, J.W.Starner et al. - Phys.Rev. C13, 387 (1976).
*States in $^{126}$Sb Populated in the β Decay of $10^5$-yr $^{126}$Sn.*
1976SpZQ G.D.Sprouse, O.Hausser, H.R.Andrews, J.R.Beene - AECL-5614, p.17 (1976).
*Quadrupole Moments in Cd Isotopes.*
1976St01 A.M.Stefanini, P.J.Daly, P.Kleinheinz, M.R.Maier, R.Wagner - Nucl.Phys. A258, 34 (1976).
*Isomers in the N = 83 Nucleus $^{149}$Dy.*
1976St08 A.M.Stefanini, P.Kleinheinz, M.R.Maier - Phys.Lett. 62B, 405 (1976).
*A 500 ms Three-Particle Isomer in $^{149}$Dy.*
1976SvZY L.-G.Svensson, A.Backlin, H.Solhed, J.Lindskog - UUIP-873 (1976).
*Transition Rates between the Positive Parity States in $^{101}$Tc.*

1980Ko16 M.Kortelahti, M.Piiparinen, A.Pakkanen, T.Komppa et al. - Nucl.Phys. A342, 421 (1980).
*In-Beam Study of $^{145}$Pm and the Cluster-Vibration Model for Odd Pm Nuclei.*

1980La03 A.N.Larsen, D.Mouchel, H.H.Hansen - Z.Phys. A294, 191 (1980).
*Half-Life of the 23.87 keV Level in $^{119}$Sn.*

1980Le05 V.N.Lebedev, A.I.Levon, O.F.Nemets, G.V.Onishchenko, O.V.Sevastyuk - Izv.Akad.Nauk SSSR, Ser.Fiz. 44, 202 (1980); Bull.Acad.Sci.USSR, Phys.Ser. 44, No.1, 171 (1980).
*Lifetimes and g-Factors of Isomeric States in $^{105}$Ag, $^{111}$In, $^{115}$Sb, $^{117}$Sb.*

1980LeZN K.LeNestour, R.G.Summers-Gill, A.Islam, A.G.Lee - McMaster University, Ann.Rept., 1979-1980, p.56 (1980).
*Conversion Electron Study of the 230 $\mu$s Isomer in $^{146}$Eu.*

1980Li18 C.F.Liang, P.Paris, A.Peghaire, H.Szichman - Z.Phys. A297, 303 (1980).
*New Results in the Decay of $^{150}$Ho and $^{152}$Tm.*

1980Ma16 G.Marguier, A.Charvet, J.Genevey, C.Richard-Serre et al. - Nucl.Phys. A342, 301 (1980).
*Evidence for $^{116}$Xe Quasi-Rotational Bands Populated in the Decays of 0.7 s and 4.1 s $^{116}$Cs States.*

1980Me11 J.C.Merdinger, F.A.Beck, E.Bozek, T.Byrski et al. - Nucl.Phys. A346, 281 (1980).
*Magnetic Moments of the $I(\pi) = 10^+$ Isomers in Ce and Nd N = 80 Isotones.*

1980Mo27 T.Morek, H.Beuscher, B.Bochev, D.R.Haenni et al. - Z.Phys. A298, 267 (1980).
*Isomeric States in $^{134}$Ba.*

1980No06 E.Nolte, G.Korschinek, U.Heim - Z.Phys. A298, 191 (1980).
*Investigation of Very Neutron Deficient Ru and Rh Nuclei with the Help of ($^{40}$Ca; yp + $z\alpha\gamma$) Reactions.*

1980Oh01 S.Ohya, N.Mutsuro, Z.Matumoto, T.Tamura - Nucl.Phys. A334, 382 (1980).
*Lifetimes of the Low-Lying $7/2^-$ States in $^{113,115}$Cd.*

1980Ox01 K.Oxorn, B.Singh, S.K.Mark - Z.Phys. A294, 389 (1980).
*The Decay of 26-s $^{94}$Rh and 71-s $^{94}$Rh Isomers.*

1980Pi03 M.Piiparinen, M.Kortelahti, A.Pakkanen, T.Komppa, R.Komu - Nucl.Phys. A342, 53 (1980).
*Medium-Spin States in the N = 80 Nuclei $^{139}$Pr and $^{141}$Pm.*

1980Po07 A.R.Poletti, G.D.Dracoulis, C.Fahlander - Phys.Rev.Lett. 45, 1475 (1980).
*Low-Lying Yrast States in $^{210}$Rn and $^{211}$Rn and the Competition between Neutron-Hole and Proton Excitations.*

1980Pr02 H.Prade, L.Kaubler, U.Hagemann, H.U.Jager et al. - Nucl.Phys. A333, 33 (1980).
*Structure of High-Spin States in $^{143}$Pm.*

1980RaZM V.Rahkonen - Thesis, Univ. of Jyvaskyla (1980).
*High-Spin Yrast States in the $^{206}$Po, $^{208}$At and $^{210}$At Nuclei.*

1980RoZT I.A.Romany, V.A.Lutsik, Yu.N.Rakivnenko, E.A.Skakun et al. - Program and Theses, Proc.30th Ann.Conf.Nucl.Spectrosc.At.Nuclei, Leningrad, p.141 (1980).
*The Decay Scheme of the $^{192}$Au Isomer.*

1980Sc09 U.J.Schrewe, P.Tidemand-Petersson, G.M.Gowdy, R.Kirchner et al. - Phys.Lett. 91B, 46 (1980).
*Alpha Decay of Neutron-Deficient Isotopes with $78 \leq Z \leq 83$ Including the New Isotopes $^{183,184}$Pb and $^{188}$Bi.*

1980Sc13 F.Schussler, J.A.Pinston, E.Monnand, A.Moussa et al. - Nucl.Phys. A339, 415 (1980).
*Discovery of a Very Low-Lying $0^+$ State in $^{98}$Sr and Shape Coexistence Implication in $^{98}$Sr.*

1980Sj01 T.P.Sjoreen, U.Garg, D.B.Fossan - Phys.Rev. C21, 1838 (1980).
*Spectroscopy of $^{213}$At, $^{212}$Po, and $^{210}$Pb Following $^{208}$Pb + $^{7}$Li.*

1980SuZY R.G.Summers-Gill, A.Islam, A.G.Lee - McMaster University, Ann.Rept., 1979-1980, p.65 (1980).
*The Isomeric State in $^{136}$La.*

1980Te01 B.O.Ten Brink, P.van Nes, C.Hoetmer, H.Verheul - Nucl.Phys. A338, 24 (1980).
*On the Structure of $^{71}$As and $^{73}$As.*

1980Va13 A.Van Poelgeest, J.Bron, W.H.A.Hesselink, K.Allaart et al. - Nucl.Phys. A346, 70 (1980).
*Neutron Excitations in Even Mass Sn Nuclei.*

1980Wa23 P.M.Walker, D.Ward, O.Hausser, H.R.Andrews, T.Faestermann - Nucl.Phys. A349, 1 (1980).
*g-Factors for $^{172,173,174,178}$Hf K-Isomers.*

1980Wi11 J.Wilson, S.R.Faber, P.J.Daly, I.Ahmad et al. - Z.Phys. A296, 185 (1980).
*The $\pi h^3_{11/2}$ Spectrum in the Three Valence Proton Nucleus $^{149}_{67}$Ho$_{82}$.*

1980Wi13 H.S.Wilson, R.W.Kavanagh, F.M.Mann - Phys.Rev. C22, 1696 (1980).
*Gamow-Teller Transitions in Some Intermediate-Mass Nuclei.*

1980Wi20 I.N.Wischnewski, H.V.Klapdor, P.Herges, H.Fromm, W.A.Zheldonozhski - Z.Phys. A298, 21 (1980).
*Investigation of the $\beta^+$-Decay of $^{105,106,108}$In and of $^{100,102}$Ag.*

1980Zo02 D.R.Zolnowski, M.B.Hughes, J.Hunt, T.T.Sugihara - Phys.Rev. C21, 2556 (1980).
*Systematics of Quasirotational States in N = 88 Nuclei.*

1981Ad02 I.Adam, M.Honusek, Z.Hons, V.V.Kuznetsov et al. - Nucl.Phys. A356, 129 (1981).
*Levels in $^{161}$Tm Excited in the Decay of 4.2 Min $^{161}$Yb.*

1981An11 O.Ando, K.Miyano - Int.J.Appl.Radiat.Isotop. 32, 381 (1981).
*Half-Life Determination with High Counting Rate and Half-Lives of $^{202m}$Pb and $^{201}$Pb.*

1981An17 W.Andrejtscheff, L.Zamick, N.A.Lebedev, K.M.Muminov et al. - Nucl.Phys. A368, 45 (1981).
*The Behavior of l-Forbidden M1 Transitions near Z = 50,N = 82.*

1981Av04 V.D.Avchukhov, K.A.Baskova, A.B.Vovk, V.O.Sokolov et al. - Izv.Akad.Nauk SSSR, Ser.Fiz. 45, 2124 (1981).
*Study of the Decay of Millisecond Isomers $^{183m}$Re,$^{180m}$W.*

1981Ba40 G.K.Bavaria, J.E.Crawford, S.Calamawy, J.E.Kitching - Z.Phys. A302, 329 (1981).
*The Decay of $^{78g}$Rb and $^{78m}$Rb to Levels in $^{78}$Kr.*

1981Bo29 H.Bohn, E.Endres, T.Faestermann, P.Kienle - Z.Phys. A302, 51 (1981).
*Spectroscopy of Excited States in $^{212}$Po,$^{210}$Pb,and $^{213}$At Empolying $^{18}$O Induced Few-Nucleon Transfer Reactions.*

1981By04 A.A.Bykov, V.D.Vitman, Yu.V.Naumov, S.Yu.Orlov, V.K.Tarasov - Izv.Akad.Nauk SSSR, Ser.Fiz. 45, 874 (1981).
*β-Strength Function of Neutron Deficient Lu Isotopes (A = 172-162).*

1981Da06 E.Dafni, H.E.Mahnke, J.W.Noe, M.H.Rafailovich, G.D.Sprouse - Phys.Rev. C23, 1612 (1981).
*E2 Moments of the $^{43}$Sc(19/2$^-$) and the $^{43}$Ti(19/2$^-$) States and the $^{40}$Ca Core Deformation.*

1981De38 J.Deslauriers, S.C.Gujrathi, S.K.Mark - Z.Phys. A303, 151 (1981).
*Structure of Even-Even $^{138}$Nd from the Decay of $^{138}$Pm.*

1981Dr07 S.Drissi, S.Andre, J.Genevey, V.Barci et al. - Z.Phys. A302, 361 (1981).
*Identification and Structure of Yrast Levels in the N = 89 Odd-Odd $^{158}$Tm Nucleus.*

1981En02 H.A.Enge, M.K.Salomaa, A.Sperduto, W.Schier, H.E.Wegner - Phys.Rev. C24, 298 (1981).
*Half-Lives of $^{199}$Rn and $^{199m}$Rn.*

1981En05 G.Engler, E.Neeman - Nucl.Phys. A367, 29 (1981).
*Delayed Neutron Emission Probabilities and Half-Lives of Rb,Sr,Y,In,Cs,Ba and La Precursors with A = 93-98,A = 127-131 and A = 142-148.*

1981Fa01 B.Fant, T.Lonnroth, V.Rahkonen - Nucl.Phys. A355, 171 (1981).
*In-Beam Study of $^{211}$Po Excited States.*

1981Fi02 C.A.Fields, F.W.N.De Boer, J.J.Kraushaar, R.A.Ristinen et al. - Nucl.Phys. A363, 311 (1981).
*High-Spin States in $^{90}$Nb.*

1981Fo02 B.Fogelberg, K.Heyde, J.Sau - Nucl.Phys. A352, 157 (1981).
*Energy Levels and Transition Probabilities in $^{130}$Sn.*

1981Fr07 E-M.Franz, S.Katcoff, P.P.Parekh, L.K.Peker - Phys.Rev. C23, 2234 (1981).
*Identification and Decay of New High-Spin Isomer: 5.5-min $^{242}$Np.*

1981FrZY L.H.Fry - Thesis, Florida State Univ. (1981); Diss.Abst.Int. 42B, 264 (1981).
*A Study of the Systematics of Low-Lying Levels in Nuclei in the Mass Region A = 80 with a NIM Controlled Rabbit System.*

1981Ga11 G.Garcia Bermudez, A.Filevich, A.J.Kreiner, M.A.J.Mariscotti et al. - Phys.Rev. C23, 2024 (1981).
*High Spin States in $^{74}$Br.*

1981Go17 B.I.Gorbachev, A.V.Kuznichenko, V.N.Lebedev, A.I.Levon et al. - Izv.Akad.Nauk SSSR, Ser.Fiz. 45, 2116 (1981).
*Nuclear g-Factors for the 6$^+$ State in $^{112}$Sn and 11/2$^-$ State in $^{113}$Sn.*

1981Gu04 W.Gunther, K.Huber, U.Kneissl, H.Krieger et al. - Nucl.Phys. A359, 397 (1981).
*Yield Ratio for the Two $^{241}$Pu Fission Isomers in the $^{242}$Pu(γ,n) Reaction.*

1981Ha17 B.Haas, D.Ward, H.R.Andrews, O.Hausser et al. - Nucl.Phys. A362, 254 (1981).
*Yrast Isomers and very High Spin States in $^{148,149,151,152}$Dy and $^{147}$Gd.*

1981Ha25 D.R.Haenni, H.Beuscher, B.Bochev, T.Kutsarova et al. - Nucl.Phys. A365, 229 (1981).
*A New 5.0 ns Isomer in $^{144}$Eu.*

1981He04 H.Helppi, J.Hattula, A.Luukko, M.Jaaskelainen, F.Donau - Nucl.Phys. A357, 333 (1981).
*In-Beam Study of $^{127}$,$^{129}$Xe and Collective Description of the Level Structures in Odd-A Xe Nuclei.*

1981He07 H.Helppi, S.K.Saha, P.J.Daly, S.R.Faber et al. - Phys.Rev. C23, 1446 (1981).
*High-Spin Level Spectra of the Nuclei $^{195}$Pb,$^{197}$Pb,$^{199}$Pb,and $^{201}$Pb.*

1981Ho05 D.Horn, G.R.Young, C.J.Lister, C.Baktash - Phys.Rev. C23, 1047 (1981); Erratum Phys.Rev. C23, 2798 (1981).
*High Spin Single Particle States in $^{152}$,$^{153}$Er.*

1981Ho10 S.Hofmann, G.Munzenberg, F.Hessberger, W.Reisdorf et al. - Z.Phys. A299, 281 (1981).
*New Neutron Deficient Isotopes in the Range of Elements Tm to Pt.*

1981Ho24 P.Hoff, B.Fogelberg - Nucl.Phys. A368, 210 (1981).
*Properties of Strongly Neutron-Rich Isotopes of Germanium and Arsenic.*

1981Ho29 D.Horn, C.Baktash, C.J.Lister - Phys.Rev. C24, 2136 (1981).
*Spectroscopy of $^{204}$,$^{206}$,$^{208}$Rn and the Systematic Behavior ofZ = 86 Isotopes.*

1981Hs02 H.-C.Hseuh, E.-M.Franz, P.E.Haustein, S.Katcoff, L.K.Peker - Phys.Rev. C23, 1217 (1981).
*Decay of 14.1-h $^{240}$U and 7.22-min $^{240m}$Np.*

1981Io05 M.Ionescu-Bujor, A.Iordachescu, E.A.Ivanov, D.Plostinaru, Chen Yilung - Rev.Roum.Phys. 26, 239 (1981).
*Half-Life and Magnetic Moment of a 7/2$^+$ Level in $^{123}$Te.*

1981Iw04 T.Iwashita - J.Phys.Soc.Jpn. 50, 3191 (1981).
*Gamma-Gamma Directional Correlation in the Decay of $^{168}$Tm.*

1981Ka10 L.Kaubler, H.Prade, F.Donau, U.Hagemann et al. - Z.Phys. A299, 251 (1981).
*Deformation Dependence of Magnetic Moments in the Odd Transitional Nuclei $^{117-125}$Te.*

1981Ko03 J.Konijn, F.W.N.De Boer, A.Van Poelgeest, W.H.A.Hesselink et al. - Nucl.Phys. A352, 191 (1981).
*The Level Structue of $^{156}$Gd Studied by Means of the (α,2nγ) Reaction.*

1981Ko16 M.Kortelahti, M.Piiparinen, A.Pakkanen, T.Komppa, R.Komu - Phys.Scr. 24, 10 (1981).
*Medium-Spin States in the N = 82 Nuclei $^{141}$Pr and $^{143}$Pm.*

*γ-Spectroscopy of* $^{215}_{89}Ac^{126}$ *and Proton Pairing in N = 126 Isotones.*

1983De34 D.J.Decman, J.A.Becker, J.B.Carlson, R.G.Lanier et al. - Phys.Rev. C28, 1060 (1983).
*Electromagnetic Properties of Isomers in* $^{210}Pb$.

1983Dm04 S.V.Dmitriev, G.A.Otroshchenko, S.M.Solovyev - Yad.Fiz. 38, 1394 (1983).
*Yield of Fissionable Isomers from Reactions* $^{234}U(n,n')$, $^{236}U(n,n')$, *and* $^{238}U(n,n')$.

1983Dr05 G.D.Dracoulis, C.Fahlander, A.P.Byrne - Nucl.Phys. A401, 490 (1983).
*Intrinsic States, High-Spin Rotational Bands and Rotation Alignment in* $^{177}Os$ *and* $^{179}Os$.

1983Dr14 J.Drexler, R.Heil, K.Huber, U.Kneissl et al. - Nucl.Phys. A411, 17 (1983).
*The Decay of Uranium Shape Isomers Investigated by Photonuclear Reactions.*

1983Dy02 K.Dybdal, T.Chapuran, D.B.Fossan, W.F.Piel et al. - Phys.Rev. C28, 1171 (1983).
*High-Spin States in* $^{201,203}At$ *and the Systematic Behavior of Z = 85 Isotopes.*

1983Fu21 M.Fujioka, S.Shibuya, N.Kawamura, A.Matsumoto et al. - Hyperfine Interactions 15/16, 1017 (1983).
*Nuclear Level-Crossing Resonance of the 398-keV* $(9/2)^+$ *State of* $^{69}Ge$ *in a Zn Single Crystal.*

1983Ga18 C.A.Gagliardi, G.T.Garvey, J.R.Wrobel, S.J.Freedman - Phys.Rev. C28, 2423 (1983).
$^{16}N(0^-)$ *Beta Decay and Weak Axial Vector Meson Exchange Currents.*

1983Go02 B.I.Gorbachev, A.V.Kuznichenko, V.N.Lebedev, A.I.Levon et al. - Yad.Fiz. 37, 257 (1983).
*Magnetic Moments of* $h(-11/2)$ *States of Even and Odd Isotopes of Te and Xe.*

1983Gr01 H.Grawe, H.Haas - Phys.Lett. 120B, 63 (1983).
*The Lifetime and g-Factor of the* $8^+$ *State in* $^{96}Pd$.

1983Gr33 H.Grawe, H.H.Bertschat, H.Haas - Hyperfine Interactions 15/16, 65 (1983).
*g-Factors of High-Spin States in* $^{93}Ru$ *and* $^{95}Rh$.

1983Gu05 M.Guttormsen, A.V.Grumbkow, Y.K.Agarwal, K.P.Blume et al. - Nucl.Phys. A398, 119 (1983).
*The* $(\nu i^2_{13/2})(8^+,10^+,12^+)$ *Triplets in* $^{190-196}Hg$.

1983Ha06 E.Hagberg, J.C.Hardy, H.Schmeing, E.T.H.Clifford, V.T.Koslowsky - Nucl.Phys. A395, 152 (1983).
*The Decays of the* $T(z) = (-3/2)$ *β-Delayed Proton Precursors* $^{83}Zr$, $^{87}Mo$ *and* $^{91m}Ru$.

1983Ha15 K.Hardt, Y.K.Agarwal, C.Gunther, M.Guttormsen et al. - Z.Phys. A312, 251 (1983).
*Observation of the Weakly Deformed* $\nu i^2_{13/2}$ *Band in* $^{188}Hg$.

1983He08 R.G.Helmer, C.W.Reich - Phys.Rev. C27, 2248 (1983).
*Levels in* $^{204}_{84}Po$ *from the Decay of* $^{204}_{85}At$.

1983Hi08 R.Hingmann, H.-G.Clerc, C.-C.Sahm, D.Vermeulen et al. - Nucl.Phys. A404, 51 (1983).
*Evidence for an Isomeric State in* $^{216}Th$.

1983Ko05 J.Kownacki, J.D.Garrett, J.J.Gaardhoje, G.B.Hagemann et al. - Nucl.Phys. A394, 269 (1983).
*High-Spin Properties of* $^{163}Yb$ *Band-Crossings and Signature-Splitting in Yb Nuclei.*

1983Ko07 P.Komninos, E.Nolte, P.Blasi - Z.Phys. A310, 137 (1983).
*The 2.6 μs* $(21/2)^+$ *Isomer in* $^{93}Ru$.

1983Ko22 V.T.Koslowsky, E.Hagberg, J.C.Hardy, R.E.Azuma et al. - Nucl.Phys. A405, 29 (1983).
*The Half-Lives of* $^{26m}Al$, $^{34}Cl$, $^{38m}K$*: Precision measurements with isotope-separated samples.*

1983Ko42 T.Komppa, R.Komu, M.Kortelahti, J.Muhonen et al. - Z.Phys. A314, 33 (1983).
*Medium-Spin Levels in the N = 83 Nucleus* $^{147}Gd$.

1983Kr11 K.-L.Kratz, H.Ohm, A.Schroder, H.Gabelmann et al. - Z.Phys. A312, 43 (1983).
*The Beta-Decay of* $^{95}Rb$ *and* $^{97}Rb$.

1983Lo01 T.Lonnroth, D.Horn, C.Baktash, C.J.Lister, G.R.Young - Phys.Rev. C27, 180 (1983).
*High-Spin States in* $^{214}Rn$, $^{216}Ra$ *and a Study of Even-Even N = 128 Systematics.*

1983Ma08 H.-E.Mahnke, W.Semmler, H.Grawe, H.Haas, R.Sielemann - Phys.Lett. 122B, 27 (1983).
*Quadrupole Moments of High-Spin Isomers in Astatine Isotopes.*

1983Mi20 T.Minamisono, K.Takeyama, T.Ishigai, H.Takeshima et al. - Phys.Lett. 130B, 1 (1983).
*Beta Decay Rate of the* $^{16}N(0^-) \to {}^{16}O(0^+)$ *Transition for the Study of Meson-Exchange Effects and g(p)/g(A).*

1983Mu15 M.J.Murphy, C.N.Davids - Phys.Rev. C28, 1069 (1983).
*Level Structure of* $^{67}Ge$ *and Its Implications for the General Structure of Nuclei in the 1f-2p Shell.*

1983OxZZ K.W.Oxorn - Diss.Abst.Int. 44, 1501 (1983).
*Structure of Unstable Nuclei in the* $g_{9/2}$.

1983Po14 Yu.S.Popov, P.A.Privalova, G.A.Timofeev, V.B.Mishenev et al. - Radiokhimiya 25, 482 (1983); Sov.Radiochemistry 25, 458 (1984).
*Identification of* $^{246}Pu$, $^{247}Pu$, $^{246m}Am$, *and* $^{247}Am$ *and Determination of Their Half-Lives.*

1983Ra10 S.Rastikerdar, C.Garrett, G.S.Foote, W.Gelletly - J.Phys.(London) G9, 555 (1983).
*The Decay of* $^{100}Ag$.

1983Ra19 D.C.Radford, M.S.Rosenthal, P.D.Parker, J.A.Cizewski et al. - Phys.Lett. 126B, 24 (1983).
*Very High Spin States and Isomerism in* $^{153}Ho$.

1983Ry06 T.B.Ryves, Ma Hongchang, S.Judge, P.Kolkowski - J.Phys.(London) G9, 1549 (1983).
*Indium Cross Sections between 14 and 18 MeV.*

1983Sc24 J.R.H.Schneider, S.Hofmann, F.P.Hessberger, G.Munzenberg et al. - Z.Phys. A312, 21 (1983).
*Alpha Decay of New Neutron Deficient Gold,Mercury, and Thallium Isotopes.*

*High-Spin γ-Ray Spectroscopy in Z = 83 Isotopes:* $^{199,201}$*Bi.*

1985Po06 A.R.Poletti, G.D.Dracoulis, A.P.Byrne, A.E.Stuchbery et al. - Phys.Lett. 154B, 263 (1985).
*Magnetic Moments,E3 Transitions and the Structure of High-Spin Core Excited States in* $^{211}$*Rn.*

1985Po08 A.R.Poletti, G.D.Dracoulis, C.Fahlander, A.P.Byrne - Nucl.Phys. A440, 118 (1985).
*The Yrast Spectroscopy of* $^{209}$*Rn.*

1985Po13 A.R.Poletti, G.D.Dracoulis, S.J.Poletti, A.P.Byrne et al. - Nucl.Phys. A442, 153 (1985).
*Short-Lived Isomers in* $^{211}$*Rn and* $^{210}$*Rn.*

1985Ra18 V.Rahkonen, B.Fant, C.J.Herrlander, K.Honkanen et al. - Nucl.Phys. A441, 11 (1985).
*An Investigation of the Yrast States in* $^{205,207}$*Po.*

1985Ra21 V.Rahkonen, T.Lonnroth - Z.Phys. A322, 333 (1985).
*Low- and Medium-Spin in the N = 124 Isotones* $^{208}$*Po and* $^{209}$*At.*

1985Ro07 U.Rosengard, I.Bergstrom, J.Blomqvist, P.Carle et al. - Phys.Scr. 31, 122 (1985).
*The g-Factors of the* $(25/2)^+$ *Level in* $^{207}$*Po and* $11^-$ *Level in*$^{208}$*Po.*

1985Ru05 E.Runte, K.-L.Gippert, W.-D.Schmidt-Ott, P.Tidemand-Petersson et al. - Nucl.Phys. A441, 237 (1985).
*Decay Studies of Neutron-Rich Isotopes of Manganese, Iron, Cobalt, Nickel, Copper and Zinc.*

1985St02 R.E.Stone, C.R.Bingham, L.L.Riedinger, R.W.Lide et al. - Phys.Rev. C31, 582 (1985).
*Decay of Mass-Separated* $^{199m}$*Po and* $^{199g}$*Po.*

1985St16 C.Stenzel, H.Grawe, H.Haas, H.-E.Mahnke, K.H.Maier - Z.Phys. A322, 83 (1985).
*E2 Moments and g-Factors in Neutron Deficient Lead Isotopes.*

1985Ti02 P.Tidemand-Petersson, R.Kirchner, O.Klepper, E.Roeckl et al. - Nucl.Phys. A437, 342 (1985).
*β-Delayed Particle Emission from Neutron-Deficient Te,I,Xe,Cs and BaIsotopes.*

1985Ur01 W.Urban, T.Morek, Ch.Droste, B.Kotlinski et al. - Z.Phys. A320, 327 (1985).
*In-Beam Study of* $^{127}$*Xe.*

1985We05 T.Weckstrom, B.Fant, T.Lonnroth, V.Rahkonen et al. - Z.Phys. A321, 231 (1985).
*In-Beam Study of High-Spin States in* $^{199,200,201}$*Po and Systematical Features of Z = 84 Polonium Isotopes.*

1985YaZV H.Yamamoto, K.Kawade, H.Miyade, M.Yoshida et al. - Inst.Nucl.Study, Univ.Tokyo, Ann.Rept., 1984, p.43 (1985).
*The Production of Short-Lived Nuclei by the Fast Neutron Irradiation.*

1986Al11 J.Alstad, T.Bjornstad, G.Skarnemark - Nucl.Instrum.Methods Phys.Res. A244, 490 (1986).
*TDD − A Method for On-Line Determination of Short Half-Lives.*

1986Be07 V.Berg, J.Oms, K.Fransson, Z.Hu, and the ISOCELE Collaboration - Nucl.Phys. A453, 93 (1986).
*Half-Life and Electromagentic Deexcitation Rate Measurements for* $(1/2)^+$ *and* $(3/2)^+$ *States in Odd-Proton Nuclei around Z = 82.*

1986Br28 R.A.Braga, P.B.Semmes, W.R.Western, R.W.Fink - Nucl.Phys. A459, 359 (1986).
*Decay of Mass-Separated* $^{201m}$*Po and* $^{201g}$*Po.*

1986By01 A.P.Byrne, G.D.Dracoulis, C.Fahlander, H.Hubel et al. - Nucl.Phys. A448, 137 (1986).
*Spectroscopy of High-Spin States in* $^{211,212,213}$*Fr.*

1986Ch01 T.Chapuran, K.Dybdal, D.B.Fossan, T.Lonnroth et al. - Phys.Rev. C33, 130 (1986).
*High-Spin γ-Ray Spectroscopy in* $^{197}$*Bi.*

1986De35 J.Deslauriers, S.Gujrathi, S.K.Mark, S.P.Sud - Z.Phys. A325, 421 (1986).
*Decay Properties of* $^{139m}$*Sm and* $^{139}$*Eu.*

1986Dr06 S.Drissi, A.Bruder, J.-Cl.Dousse, V.Ionescu et al. - Nucl.Phys. A451, 313 (1986).
*Systematic Study of the Yrast Band in the Doubly-Odd* $^{158,160,162,164,166}$*Tm Nuclei.*

1986Du04 R.Duffait, A.Charvet, K.Deneffe, R.Beraud et al. - Nucl.Phys. A454, 143 (1986).
*High-Spin States in Odd-Odd* $^{100,102,104}$*Rh.*

1986Ek01 B.Ekstrom, B.Fogelberg, P.Hoff, E.Lund, A.Sangariyavanish - Phys.Scr. 34, 614 (1986).
*Decay Properties of* $^{75-80}$*Zn and Q(β)-Values of Neutron-RichZn and Ga Isotopes.*

1986Fa04 B.Fant, T.Weckstrom, V.Rahkonen, C.J.Herrlander, A.Kallberg - Nucl.Phys. A453, 77 (1986).
*High-Spin and Isomeric States in* $^{203}$*Po.*

1986Go10 H.Gokturk, B.Ekstrom, E.Lund, B.Fogelberg - Z.Phys. A324, 117 (1986).
*Measurements of Absolute γ-Ray Intensities in the Decays of very Neutron Rich Isotopes of Cd and In.*

1986Ho22 S.Hofmann, I.Zychor, F.P.Hessberger, G.Munzenberg - Z.Phys. A325, 37 (1986).
*Identification of a 3.2 μs Isomer in* $^{76}$*Rb.*

1986Ja13 H.C.Jain, P.Carle, A.Kallberg, L.O.Norlin et al. - Nucl.Phys. A458, 225 (1986).
*High-Spin Isomers in* $^{202}$*Pb.*

1986Ka20 K.Kawade, H.Yamamoto, M.Yoshida, T.Ishii et al. - J.Phys.Soc.Jpn. 55, 1102 (1986).
*Half-Lives of Levels in* $^{93}$*Sr and* $^{95}$*Sr.*

1986Ka43 N.Kaffrell, P.Hill, J.Rogowski, H.Tetzlaff et al. - Nucl.Phys. A460, 437 (1986).
*Levels in* $^{107}$*Rh.*

1986Ko01 T.Kohno, M.Adachi, S.Fukuda, M.Taya et al. - Phys.Rev. C33, 392 (1986).
*Level Structure and Electromagnetic Properties in* $^{212}$*Ra.*

1986Lh01 G.Lhersonneau, D.Weiler, P.Kohl, H.Ohm et al. - Z.Phys. A323, 59 (1986).
*A New High-Spin Isomer in* $^{97}_{39}Y_{58}$.

*New Shape Isomer in the Self-Conjugate Nucleus $^{72}$Kr.*

2003Cu03 D.M.Cullen, L.K.Pattison, J.F.Smith, A.M.Fletcher et al. - Nucl.Phys. A728, 287 (2003).
*High-spin states, lifetime measurements and isomers in $^{181}$Os.*

2003Fo03 B.Fornal, R.Broda, K.H.Maier, P.J.Daly et al. - Phys.Rev. C 67, 034318 (2003).
*$\gamma$-ray spectroscopy of proton neutron-hole nucleus $^{208}$Bi from deepinelastic heavy ion reactions.*

2003Ge04 J.Genevey, J.A.Pinston, H.R.Faust, R.Orlandi et al. - Phys.Rev. C 67, 054312 (2003).
*High-spin microsecond isomers in $^{129}$In and $^{129}$Sb.*

2003Gi05 T.N.Ginter, K.E.Gregorich, W.Loveland, D.M.Lee et al. - Phys.Rev. C 67, 064609 (2003); Erratum Phys.Rev. C 68, 029901 (2003).
*Confirmation of production of element 110 by the $^{208}$Pb($^{64}$Ni,n) reaction.*

2003Gi10 T.N.Ginter, J.C.Batchelder, C.R.Bingham, C.J.Gross et al. - Phys.Rev. C 68, 034330 (2003).
*Neutron single-particle states populated via proton emission from $^{146}$Tm and $^{150}$Lu.*

2003Gl05 K.A.Gladnishki, Zs.Podolyak, J.Gerl, M.Hellstrom et al. - Acta Phys.Pol. B34, 2395 (2003).
*Isomer spectroscopy in the neutron-deficient lead region following projectile fragmentation.*

2003HeZT M.Hellstrom, M.N.Mineva, A.Blazhev, H.J.Boardman et al. - GSI 2003-1, p.5 (2003).
*Probing neutron-rich In and Cd nuclei with isomer spectroscopy.*

2003KaZP V.G.Kalinnikov, V.P.Volnykh, V.I.Stegailov, D.V.Filosofov et al. - Program and Thesis, Proc.Conf.on Nucl.Physics (53rd Ann.Conf.Nucl.Spectrosc.At.Nuclei), Moscow, p.95 (2003).
*Investigation of Isomers Decay of $^{156}$Ho.*

2003Ke04 H.Kettunen, T.Enqvist, M.Leino, K.Eskola et al. - Eur.Phys.J. A 16, 457 (2003).
*Investigations into the alpha-decay of $^{195}$At.*

2003Ke08 H.Kettunen, T.Enqvist, T.Grahn, P.T.Greenlees et al. - Eur.Phys.J. A 17, 537 (2003).
*Alpha-decay studies of the new isotopes $^{191}$At and $^{193}$At.*

2003Ku26 J.Kurpeta, A.Plochocki, A.N.Andreyev, J.Aysto et al. - Eur.Phys.J. A 18, 31 (2003).
*Isomeric and ground-state decay of $^{215}$Bi.*

2003LaZZ G.J.Lane, A.P.Byrne, G.D.Dracoulis - Proc.Frontiers of Nuclear Structure, Berkeley, California, P.Fallon and R.Clark, Eds., p.386 (2003); AIP Conf.Proc. 656 (2003).
*High-Spin Isomers, Residual Interactions And Octupole Correlations InThe N=128 Isotones: $^{211}$Bi, $^{212}$Po and $^{213}$At.*

2003Lh01 G.Lhersonneau, Y.Wang, R.Capote, J.Suhonen et al. - Phys.Rev. C 67, 024303 (2003).
*Decay of $^{114}$Rh to $^{114}$Pd.*

2003Li30 X.-F.Li, Y.-J.Ma, Y.-Z.Liu, J.-B.Lu et al. - Eur.Phys.J. A 17, 523 (2003).
*Configuration assignments of yrare high-spin structures in $^{126}$Cs.*

2003Li42 Yu.A.Litvinov, F.Attallah, K.Beckert, F.Bosch et al. - Phys.Lett. B 573, 80 (2003).
*Observation of a dramatic hindrance of the nuclear decay of isomeric states for fully ionized atoms.*

2003Me20 A.Melerangi, D.Appelbe, R.D.Page, H.J.Boardman et al. - Phys.Rev. C 68, 041301 (2003).
*Shape isomerism and spectroscopy of $^{177}$Hg.*

2003Mo36 C.-B.Moon, T.Komatsubara, T.Shizuma, Y.Sasaki et al. - Nucl.Phys. A728, 350 (2003).
*Collective bands in odd-odd $^{118}$I.*

2003MuZV G.Mukherjee, T.L.Khoo, D.Seweryniak, I.Ahmad et al. - ANL-03/23 (Physics Division Ann.Rept., 2002), p.51 (2003).
*First Identification of a $\mu$s Isomer in N = 127 $^{217}$Th.*

2003Po21 Yu.S.Popov, L.V.Zakharova, V.N.Kupriyanov, O.I.Andreev et al. - Radiochemistry 45, 537 (2003); Radiokhimiya 45, 481 (2003).
*Half-Lives and Intensities of X-ray and $\gamma$-Ray Emission of $^{113}$Sn, $^{117m}$Sn, and $^{119m}$Sn.*

2003Sa02 M.Sawicka, J.M.Daugas, H.Grawe, S.Cwiok et al. - Eur.Phys.J. A 16, 51 (2003).
*Isomeric decay of $^{67}$Fe – Evidence for deformation.*

2003Sa40 M.Sawicka, R.Grzywacz, I.Matea, H.Grawe et al. - Phys.Rev. C 68, 044304 (2003).
*Low energy levels in $^{72}$Ni.*

2003Sh13 T.Shizuma, Y.Toh, M.Oshima, M.Sugawara et al. - Eur.Phys.J. A 17, 159 (2003).
*Inelastic excitation of $^{187}$Re.*

2003So02 O.Sorlin, C.Donzaud, F.Nowacki, J.C.Angelique et al. - Eur.Phys.J. A 16, 55 (2003).
*New region of deformation in the neutron-rich $^{60}_{24}Cr_{36}$ and $^{62}_{24}Cr_{38}$.*

2003So21 O.Sorlin, C.Donzaud, F.Azaiez, C.Bourgeois et al. - Nucl.Phys. A719, 193c (2003).
*Beta decay studies of neutron-rich $_{21}Sc - _{27}Co$ nuclei at GANIL.*

2003StZX M.A.Stanoiu - Thesis, Universite de Caen/Basse-Normandie (2003).
*Spectroscopie $\gamma$ en ligne de noyaux legers riches en neutrons produits par fragmentation de faisceau radioactif et Mesures de temps de vie des niveaux excites dans des noyaux proches de $^{68}$Ni.*

2003Tu05 A.Turler, Ch.E.Dullmann, H.W.Gaggeler, U.W.Kirbach et al. - Eur.Phys.J. A 17, 505 (2003).
*On the decay properties of $^{269}$Hs and indications for the new nuclide $^{270}$Hs.*

2003Ur01 W.Urban, J.A.Pinston, T.Rzaca-Urban, A.Zlomaniec et al. - Eur.Phys.J. A 16, 11 (2003).
*First observation of the $\nu$9/2[404] orbital in the A ~ 100 mass region.*

2003Va16 K.Van de Vel, A.N.Andreyev, D.Ackermann, H.J.Boardman et al. - Phys.Rev. C 68, 054311 (2003).
*Fine structure in a $\alpha$ decay of $^{188,192}$Po.*

2003Wa13 Y.Wang, S.Rinta-Antila, P.Dendooven, J.Huikari et al. - Phys.Rev. C 67, 064303 (2003).

*Weakly deformed oblate structures in* $^{198}_{76}Os_{122}$.

2009Po04 M.-G.Porquet, A.Astier, Ts.Venkova, A.Prevost et al. - Eur.Phys.J. A 39, 295 (2009).
*High-spin excitations of* $^{81,82,83,85}Se$*; Competing single-particle and collective structures around* $N = 50$.

2009Po10 M.-G.Porquet, A.Astier, Ts.Venkova, I.Deloncle et al. - Eur.Phys.J. A 40, 131 (2009).
*High-spin excitations of the odd-odd* $^{88}_{37}Rb_{51}$ *and* $^{86}_{35}Br_{51}$ *nuclei:* $\pi p_{3/2}\nu d_{5/2}$ *and* $\pi f_{5/2}\nu d_{5/2}$ *residual interactions.*

2009Qi04 J.Qian, A.Heinz, T.L.Khoo, R.V.F.Janssens et al. - Phys.Rev. C 79, 064319 (2009).
*Spectroscopy of* $^{257}Rf$.

2009Rz01 T.Rzaca-Urban, K.Sieja, W.Urban, F.Nowacki et al. - Phys.Rev. C 79, 024319 (2009).
$(h_{11/2}, g_{7/2})^-_9$ *neutron excitation in* $^{92,94,96}Sr$.

2009Rz02 T.Rzaca-Urban, J.Genevey, T.Materna, W.Urban et al. - Phys.Rev. C 80, 064317 (2009).
*Near-yrast structure of* $^{142}Cs$ *and* $^{144}Cs$.

2009Sa09 J.Sauvage, J.Genevey, B.Roussiere, S.Franchoo et al. - Eur.Phys.J. A 39, 33 (2009).
*Nuclear structure of* $^{189}Tl$ *states studied via* $\beta^+$ */EC decay andlaser spectroscopy of* $^{189m+g}Pb$.

2009Sh17 T.Shizuma, T.Ishii, H.Makii, T.Hayakawa, M.Matsuda - Eur.Phys.J. A 39, 263 (2009).
*A 3-quasiparticle isomer in neutron-rich* $^{183}Ta$.

2009Si21 G.S.Simpson, W.Urban, J.Genevey, R.Orlandi et al. - Phys.Rev. C 80, 024304 (2009).
*Two-quasiparticle isomers and bands of* $^{154,156}Nd$ *and* $^{156,158,160}Sm$.

2009St07 I.Stefanescu, D.Pauwels, N.Bree, T.E.Cocolios et al. - Phys.Rev. C 79, 044325 (2009).
*Evidence for a* $\beta$*-decaying 1/2*$^-$ *isomer in* $^{71}Ni$.

2009St28 I.Stefanescu, W.B.Walters, P.F.Mantica, B.A.Brown et al. - Eur.Phys.J. A 42, 407 (2009).
*Spectroscopy of exotic* $^{121,123,125}Ag$ *produced in fragmentation reactions.*

2009Ur01 W.Urban, A.Zlomaniec, G.S.Simpson, H.Faust et al. - Phys.Rev. C 79, 037304 (2009).
*New information on medium-spin structure of* $^{133}Sb$.

2009Ur02 W.Urban, K.Sieja, G.S.Simpson, H.Faust et al. - Phys.Rev. C 79, 044304 (2009).
*New isomers and medium-spin structure of the* $^{95}Y$ *nucleus.*

2009Ur04 W.Urban, J.A.Pinston, G.S.Simpson, A.G.Smith et al. - Phys.Rev. C 80, 037301 (2009).
*The 11/2*$^-$*[505] neutron extruder orbital in* $^{159}Sm$.

2009Wa02 H.Watanabe, G.J.Lane, G.D.Dracoulis, T.Kibedi et al. - Phys.Rev. C 79, 024306 (2009).
*Decay properties of high-spin isomers and other structures in* $^{121}Sb$ *and* $^{123}Sb$.

2009Wa06 P.M.Walker, R.J.Wood, G.D.Dracoulis, T.Kibedi et al. - Phys.Rev. C 79, 044321 (2009).
*Fast decay of a three-quasiparticle isomer in* $^{171}Tm$.

2009Wa11 H.Watanabe, G.J.Lane, G.D.Dracoulis, A.P.Byrne et al. - Phys.Rev. C 79, 064311 (2009).
*Multi-quasiparticle isomers involving proton-particle and neutron-hole configurations in* $^{131}I$ *and* $^{133}I$.

2009Wa24 H.Watanabe, G.J.Lane, G.D.Dracoulis, A.P.Byrne et al. - Eur.Phys.J. A 42, 163 (2009).
*Identification of* $J^\pi = 19/2^+$ *and* $23/2^+$ *isomeric states in* $^{127}Sb$.

2009Zh11 Y.H.Zhang, M.Hasegawa, W.T.Guo, M.L.Liu et al. - Phys.Rev. C 79, 044316 (2009).
*High-spin level structure in* $^{94,95}Mo$.

2010An01 A.N.Andreyev, S.Antalic, D.Ackermann, T.E.Cocolios et al. - J.Phys.(London) G37, 035102 (2010).
*The new isotope* $^{179}Pb$ *and* $\alpha$*-decay properties of* $^{179}Tl^m$.

2010An08 S.Antalic, F.P.Hessberger, S.Hofmann, D.Ackermann et al. - Eur.Phys.J. A 43, 35 (2010).
*Studies of neutron-deficient mendelevium isotopes at SHIP.*

2010As03 A.Astier, P.Petkov, M.-G.Porquet, D.S.Delion, P.Schuck - Eur.Phys.J. A 46, 165 (2010).
*Coexistence of "a +* $^{208}Pb$*" cluster structures and single-particle excitations in* $^{212}$*84Po128.*

2010At03 L.Atanasova, D.L.Balabanski, S.K.Chamoli, M.Hass et al. - Europhys.Lett. 91, 42001 (2010).
*g-factor measurements at RISING: The cases of* $^{127}Sn$ *and* $^{128}Sn$.

2010Ba31 K.Baczynska, J.Billowes, P.Campbell, F.C.Charlwood et al. - J.Phys.(London) G37, 105103 (2010).
*Nuclear spin determination of* $^{100m}Y$ *by collinear laser spectroscopy of optically pumped ions.*

2010Ba43 G.C.Ball, G.Boisvert, P.Bricault, R.Churchman et al. - Phys.Rev. C 82, 045501 (2010).
*Precise half-life measurement of the superallowed* $\beta^+$ *emitter* $^{38}K^m$.

2010Ba48 M.Balodis, I.Tomandl, V.Bondarenko, L.Simonova et al. - Nucl.Phys. A847, 121 (2010).
*Low-lying levels of* $^{188}Re$ *nucleus from* $\gamma\gamma$*-coincidence measurements.*

2010Be16 J.S.Berryman, R.M.Clark, K.E.Gregorich, J.M.Allmond et al. - Phys.Rev. C 81, 064325 (2010); Publishers Note Phys.Rev. C 82, 029906(2010).
*Electromagnetic decays of excited states in* $^{261}Sg$ *(Z=106) and* $^{257}Rf$ *(Z=104).*

2010BeZZ B.R.Beck, J.A.Becker, P.Beiersdorfer, G.V.Brown et al. - Proc.of the 12th Intern.Conf.on Nuclear Reaction and Mechanisms, Varenna, Italy, June 15-19, 2009, F.Cerutti, A.Ferrari Eds.,Vol.1,p.255 (2010).
*Improved Value for the Energy Splitting of the Ground-State Doublet in the Nucleus* $^{229}Th$.

2010Bl13 A.Blazhev, N.Braun, H.Grawe, P.Boutachkov et al. - J.Phys.:Conf.Ser. 205, 012035 (2010).
*High-energy excited states in* $^{98}Cd$.

2010Br15 A.M.Bruce, S.Lalkovski, A.M.D.Bacelar, M.Gorska et al. - Phys.Rev. C 82, 044312 (2010).
*Shape coexistence and isomeric states in neutron-rich* $^{112}Tc$ *and* $^{113}Tc$.

2010Ch16 B.Cheal, E.Mane, J.Billowes, M.L.Bissell et al. - Phys.Rev.Lett. 104, 252502 (2010).

*High-spin structures of $^{124-131}$Te: Competition of proton- and neutron-pair breakings.*

2014As02 A.Astier, T.Konstantinopoulos, M.-G.Porquet, M.Houry et al. - Phys.Rev. C 89, 034310 (2014).
*High-spin structures of $^{138}_{57}La_{81}$.*

2014Au03 K.Auranen, J.Uusitalo, S.Juutinen, U.Jakobsson et al. - Phys.Rev. C 90, 024310 (2014).
*Experimental study of 1/2$^+$ isomers in $^{199,201}$At.*

2014Ba18 J.C.Batchelder, N.T.Brewer, C.J.Gross, R.Grzywacz et al. - Phys.Rev. C 89, 054321 (2014).
*Structure of low-lying states in $^{124,126}$Cd populated by $\beta$ decay of $^{124,126}$Ag.*

2014Bo26 Ch.Bohm, Ch.Borgmann, G.Audi, D.Beck et al. - Phys.Rev. C 90, 044307 (2014).
*Evolution of nuclear ground-state properties of neutron-deficient isotopes around Z=82 from precision mass measurements.*

2014Bu06 I.Budincevic, J.Billowes, M.L.Bissell, T.E.Cocolios et al. - Phys.Rev. C 90, 014317 (2014).
*Laser spectroscopy of francium isotopes at the borders of the region of reflection asymmetry.*

2014Ca03 R.J.Carroll, R.D.Page, D.T.Joss, J.Uusitalo et al. - Phys.Rev.Lett. 112, 092501 (2014).
*Blurring the Boundaries: Decays of Multiparticle Isomers at the Proton Drip Line.*

2014Ch22 Y.Y.Cheng, S.Q.Zhang, X.Q.Li, H.Hua et al. - Phys.Rev. C 89, 054309 (2014).
*High-spin spectroscopy of $^{144}$Tb: Systematic investigation of dipole bands in N=79 isotones.*

2014Ci03 N.Cieplicka, B.Fornal, K.H.Maier, B.Szpak et al. - Acta Phys.Pol. B45, 205 (2014).
*Angular Distributions of $\gamma$-Rays from $^{210}$Bi Produced in $^{208}$Pb+$^{208}$Pb Deep-inelastic Reactions.*

2014De19 M.Devlin, R.O.Nelson, N.Fotiades, J.M.O′Donnell - Nucl.Data Sheets 120, 48 (2014).
*New Millisecond Isomer Lifetime Measurements at LANSCE.*

2014Di03 M.S.Dias, F.Brancaccio, F.Toledo, M.F.Koskinas - Appl.Radiat.Isot. 87, 126 (2014).
*Disintegration rate, gamma-ray emission probabilities and metastable half-life measurements of $^{67}$Ga.*

2014Di08 J.Diriken, N.Patronis, A.N.Andreyev, S.Antalic et al. - Phys.Lett. B 736, 533 (2014).
*Study of the deformation-driving $\nu d_{5/2}$ orbital in $^{67}_{28}Ni_{39}$ using one-neutron transfer reactions.*

2014Dr02 M.C.Drummond, D.O′Donnell, R.D.Page, D.T.Joss et al. - Phys.Rev. C 89, 064309 (2014).
*$\alpha$ decay of the $\pi h_{11/2}$ isomer in $^{164}$Ir.*

2014DrZZ G.D.Dracoulis, G.J.Lane, A.P.Byrne, H.Watanabe et al. - Int.Nuclear Physics Conf. 2013,(IUPAP),Firenze,Italy, June 2-7,2013, S.Lunardi, P.G.Bizzeti, W.S.Kabana, C.Bucci,et al.Eds.p.02033 (2014); EPJ web of Conf.v.66, (2014).
*Deep inelastic reactions and isomers in neutron-rich nuclei across the perimeter of the A = 180-190 deformed region.*

2014Fe03 L.Ferreux, M.-C.Lepy, M.-M.Be, H.Isnard, V.Lourenco - Appl.Radiat.Isot. 87, 101 (2014).
*Photon emission intensities in the decay of $^{108m}$Ag and $^{110m}$Ag.*

2014Ga14 B.S.Gao, X.-H.Zhou, Y.-D.Fang, Y.-H.Zhang et al. - Chin.Phys.C 38, 064001 (2014).
*Identification of a 9/2$^-$[505] isomer in the neutron-rich $^{193}$Os nucleus.*

2014Go20 A.Gottardo, J.J.Valiente-Dobon, G.Benzoni, S.Lunardi et al. - Phys.Rev. C 90, 034317 (2014).
*Isomeric decay spectroscopy of the $^{217}$Bi isotope.*

2014Ha38 H.Hayashi, M.Shibata, M.Asai, A.Osa et al. - Nucl.Instrum.Methods Phys.Res. A747, 41 (2014).
*Performance of a total absorption clover detector for $Q_\beta$ measurements of neutron-rich nuclei far from the $\beta$-stability line.*

2014Hu09 S.-P.Hu, H.-L.Ma, X.-P.Cao, X.-G.Wu et al. - Phys.Lett. B 732, 59 (2014).
*Multiple triaxial bands and abnormal signature inversion in $^{74}_{33}$As.*

2014Io01 M.Ionescu-Bujor, A.Iordachescu, N.Marginean, R.Lica et al. - Phys.Rev. C 90, 014323 (2014), Erratum Phys.Rev. C 90, 029903 (2014).
*Structure of $^{130}$La at low and medium spins.*

2014Is01 M.Ismail, I.A.M.Abdul-Magead - Nucl.Phys. A922, 168 (2014).
*Examples of the failure of proximity approach when the nuclear surface is irregular or has concave regions.*

2014Is04 L.W.Iskra, R.Broda, R.V.F.Janssens, J.Wrzesinski et al. - Phys.Rev. C 89, 044324 (2014).
*Higher-seniority excitations in even neutron-rich Sn isotopes.*

2014Ka23 Z.Kalaninova, S.Antalic, A.N.Andreyev, F.P.Hessberger et al. - Phys.Rev. C 89, 054312 (2014).
*Decay of $^{201-203}$Ra and $^{200-202}$Fr.*

2014Ku20 S.Kumar, V.Singh, K.Singh, S.Sihotra et al. - J.Phys.(London) G41, 105110 (2014).
*Band structures in $^{99}$Rh.*

2014Ku28 J.Kurpeta, W.Urban, A.Plochocki, T.Rzaca-Urban et al. - Phys.Rev. C 90, 064315 (2014).
*Neutron configurations in $^{113}$Pd.*

2014La15 S.Lalkovski, D.Ivanova, E.A.Stefanova, A.Korichi et al. - Phys.Rev. C 89, 064312 (2014).
*Coexisting structures in $^{105}$Ru.*

2014Le20 A.Lennarz, A.Grossheim, K.G.Leach, M.Alanssari et al. - Phys.Rev.Lett. 113, 082502 (2014).
*In-Trap Spectroscopy of Charge-Bred Radioactive Ions.*

2014Li17 R.M.Lindstrom - J.Radioanal.Nucl.Chem. 300, 605 (2014).
*The half-life of $^{111m}$Pd.*

2014Li21 G.S.Li, M.L.Liu, X.H.Zhou, Y.H.Zhang et al. - Phys.Rev. C 89, 054303 (2014).
*In-beam $\gamma$ spectroscopy of the even-even nucleus $^{190}$Pt.*

2014Li31 L.Liu, S.Y.Wang, Z.Q.Chen, C.Liu et al. - Phys.Rev. C 90, 014313 (2014).
*Single-particle structures, high-spin isomers, and a strongly coupledband in odd-odd $^{120}$Sb.*

2014Li32 R.Lica, N.Marginean, D.G.Ghita, H.Mach et al. - Phys.Rev. C 90, 014320 (2014).
*Low-lying isomeric states in $^{80}$Ga from the $\beta^-$ decay of $^{80}$Zn.*

*Impact of Coriolis mixing on a two-quasi-neutron isomer in $^{164}Gd_{100}$ and other N=100 isotones.*

2018Gr01 E.Grodner, J.Srebrny, Ch.Droste, L.Prochniak et al. - Phys.Rev.Lett. 120, 022502 (2018).
*First Measurement of the g Factor in the Chiral Band: The Case of theCs128 Isomeric State.*

2018Ha19 D.J.Hartley, F.G.Kondev, R.Orford, J.A.Clark et al. - Phys.Rev.Lett. 120, 182502 (2018).
*Masses and $\beta$-Decay Spectroscopy of Neutron-Rich Odd-Odd $^{160,162}$EuNuclei: Evidence for a Subshell Gap with Large Deformation at N=98.*

2018Hi07 Y.Hirayama, Y.X.Watanabe, M.Mukai, M.Ahmed et al. - Phys.Rev. C 98, 014321 (2018).
*$\beta$- and $\gamma$-decay spectroscopy of $^{197,198}$Os.*

2018Hu13 T.H.Huang, W.Q.Zhang, M.D.Sun, Z.Liu et al. - Phys.Rev. C 98, 044302 (2018).
*Identification of the new isotope $^{224}$Np.*

2018Ka47 L.Kaya, A.Vogt, P.Reiter, C.Muller-Gatermann et al. - Phys.Rev. C 98, 054312 (2018).
*Millisecond 23/2$^+$ isomers in the N=79 isotones $^{133}$Xe and $^{135}$Ba.*

2018Ku13 J.Kurpeta, A.Plochocki, W.Urban, T.Eronen et al. - Phys.Rev. C 98, 024318 (2018).
*Excited levels in the multishaped $^{117}$Pd nucleus studied via $\beta$ decay of $^{117}$Rh.*

2018La03 N.Lalovic, for the AGATA Collaboration - J.Phys.(London) G45, 035105 (2018).
*Study of isomeric states in $^{198,200,202,206}$Pb and $^{206}$Hg populated in fragmentation reactions.*

2018Le10 M.C.Lewis, E.Parr, R.D.Page, C.McPeake et al. - Phys.Rev. C 98, 024302 (2018).
*Decay of a 19$^-$ isomeric state in $^{156}$Lu.*

2018Mo16 B.Moon, C.-B.Moon, G.D.Dracoulis, R.A.Bark et al. - Phys.Lett. B 782, 602 (2018).
*Occurrence of a chiral-like pair band and a six-nucleon noncollectiveoblate isomer in $^{120}$I.*

2018Mu19 M.Mukai, Y.Hirayama, Y.X.Watanabe, P.Schury et al. - Nucl.Instrum.Methods Phys.Res. A884, 1 (2018).
*High-efficiency and low-background multi-segmented proportional gas counter for $\beta$-decay spectroscopy.*

2018Na20 B.S.Nara Singh, D.M.Cullen, M.J.Taylor, P.C.Srivastava et al. - Phys.Rev. C 98, 024319 (2018).
*Level structure above the 17$^+$ isomeric state in $^{152}_{69}Tm_{83}$.*

2018Pa04 T.Palazzo, G.J.Lane, A.E.Stuchbery, A.J.Mitchell et al. - Phys.Rev. C 97, 014323 (2018).
*High-spin spectroscopy and shell-model interpretation of the $N < 126$ radium isotopes $^{212}$Ra and $^{213}$Ra.*

2018Pa20 J.Park, R.Krucken, D.Lubos, R.Gernhauser et al. - Phys.Rev. C 97, 051301R (2018).
*$\beta$ decays of the heaviest N=Z−1 nuclei and proton instability of $^{97}$In.*

2018Po07 F.Ponce, E.Swanberg, J.Burke, R.Henderson, S.Friedrich - Phys.Rev. C 97, 054310 (2018).
*Accurate measurement of the first excited nuclear state in $^{235}$U.*

2018Po10 S.Pomme, J.Paepen, R.Van Ammel - Appl.Radiat.Isot. 140, 171 (2018).
*Linearity check of an ionisation chamber through $^{99m}$Tc half-life measurements.*

2018Re07 K.Rezynkina, A.Lopez-Martens, K.Hauschild, I.Deloncle et al. - Phys.Rev. C 97, 054332 (2018).
*Influence of octupole vibration on the low-lying structure of $^{251}$Fm and other heavy $N = 151$ isotones.*

2018Rz01 T.Rzaca-Urban, W.Urban, M.Czerwinski, J.Wisniewski et al. - Phys.Rev. C 98, 064315 (2018).
*Low-spin excitations in $^{97}$Zr.*

2018Sa45 A.Saamark-Roth, L.G.Sarmiento, D.Rudolph, J.Ljungberg et al. - Phys.Rev. C 98, 044307 (2018).
*Low-lying states in $^{219}$Ra and $^{215}$Rn: Sampling microsecond $\alpha$-decaying nuclei.*

2018Sh09 H.A.Shugart, E.Browne, E.B.Norman - Appl.Radiat.Isot. 136, 101 (2018).
*Half-lives of $^{101}Rh^{g}$ and $^{108}Ag^{m}$.*

2018Sh23 Y.Shigekawa, Y.Kasamatsu, Y.Yasuda, M.Kaneko et al. - Phys.Rev. C 98, 014306 (2018).
*Variation of half-life and internal-conversion electron energy spectrum between $^{235m}$U oxide and fluoride.*

2018Si28 K.Siegl, K.Kolos, N.D.Scielzo, A.Aprahamian et al. - Phys.Rev. C 98, 054307 (2018).
*B-decay half-lives of $^{134,134m}$Sb and their isomeric yield ratio produced by the spontaneous fission of $^{252}$Cf.*

2018St18 M.Stryjczyk, Y.Tsunoda, I.G.Darby, H.De Witte et al. - Phys.Rev. C 98, 064326 (2018).
*$\beta^-$decay study of the $^{66}Mn - ^{66}Fe - ^{66}Co - ^{66}Ni$ decay chain.*

2018Ya01 H.B.Yang, L.Ma, Z.Y.Zhang, C.L.Yang et al. - Phys.Lett. B 777, 212 (2018).
*Alpha decay properties of the semi-magic nucleus $^{219}$Np.*

2018Zh29 Y.H.Zhang, P.Zhang, X.H.Zhou, M.Wang et al. - Phys.Rev. C 98, 014319 (2018).
*Isochronous mass measurements of $T_z = -1$ fp-shell nuclei from projectile fragmentation of $^{58}$Ni.*

2019Bi04 S.Biswas, A.Lemasson, M.Rejmund, A.Navin et al. - Phys.Rev. C 99, 064302 (2019).
*Effects of one valence proton on seniority and angular momentum of neutrons in neutron-rich $^{122-131}_{51}$Sb isotopes.*

2019Ch24 Z.Q.Chen, Z.H.Li, H.Hua, H.Watanabe et al. - Phys.Rev.Lett. 122, 212502 (2019).
*Proton Shell Evolution below $^{132}$Sn : First Measurement of Low-Lying $\beta$-Emitting Isomers in $^{123,125}$Ag.*

2019Da02 P.J.Davies, J.Park, H.Grawe, R.Wadsworth et al. - Phys.Rev. C 99, 021302 (2019).
*Toward the limit of nuclear binding on the N=Z line: Spectroscopy of $^{96}$Cd.*

2019De24 H.D.Dearmon, K.S.Krane - Eur.Phys.J. A 55, 135 (2019).
*Neutron capture cross sections of $^{74,76,78,80,82}$Se.*

2019Do02 A.C.Dombos, A.Spyrou, F.Naqvi, S.J.Quinn et al. - Phys.Rev. C 99, 015802 (2019).
*$\beta$-decay half-lives of neutron-rich nuclides in the A=100−−110 mass region.*

2019Du12 R.Dunlop, C.E.Svensson, C.Andreoiu, G.C.Ball et al. - Phys.Rev. C 99, 045805 (2019).
*$\beta$ decay and $\beta$-delayed neutron decay of the N=82 nucleus $^{131}_{49}In_{82}$.*

2022Zh45 M.M.Zhang, Y.L.Tian, Y.S.Wang, Z.Y.Zhang et al. - Phys.Rev. C 106, 024305 (2022).
*Fine structure in the $\alpha$ decay of the $8^{+}$ isomer in $^{216,218}U$.*

2022Zh46 W.Q.Zhang, A.N.Andreyev, Z.Liu, D.Seweryniak et al. - Phys.Rev. C 106, 024317 (2022).
*Identification of excited states in $^{188}Bi$ and $^{188}Po$.*